%% file: new.tex
\documentclass[11pt,a4paper,twosided]{book}
\usepackage[italian]{babel}
\usepackage[dvips]{graphics}
\usepackage{amsmath,amsbsy,epsfig,eucal}
\usepackage{amssymb,wasysym}
\usepackage{feynmf}
\usepackage{latexsym}
\usepackage{XThesis}
\usepackage{pennames}
\usepackage{graphicx,graphics}

\begin{document}
\selectlanguage{italian}
\pretolerance=10000 %impedisce la divisione in sillabe T++
\input{front}
\pagenumbering{Roman}\setcounter{page}{1} \tableofcontents \afterpreface
%\clearpage
\pagebreak
%\pagenumbering{Roman}
\newpage
\thispagestyle{empty} \vspace*{5cm}
\newpage
\pagenumbering{arabic}\setcounter{page}{1}
\addcontentsline{toc}{chapter}{Introduzione}
\input{intro}
\input{command}
\input{cap1/cap1}     %cap1
\input{cap2/cap2}     %cap2
\input{cap3/cap3}     %cap3
\input{cap4/cap4}     %cap4
\input{cap5/cap5}     %cap5
\input{cap6/cap6}     %cap6
\input{concl}
\appendix
\input{appendici/glauber.tex}
%\appendix
\input{appendici/V0cinematic.tex}
%\vspace*{10cm}
\newpage
\setcounter{page}{258}
\bibliographystyle{plain}
\input{biblio} 
%\newpage
%\addcontentsline{toc}{chapter}{Indice delle figure}
%\listoffigures 
%\newpage
%\addcontentsline{toc}{chapter}{Indice delle tabelle}
%\listoftables 
\newpage
\addcontentsline{toc}{chapter}{Ringraziamenti}
\input{aknow}
\end{document}

%% file: front.tex
\begin{titlepage}

\pagestyle{empty}

%\renewcommand{\printlandscape}{\special{landscape}}
%\printlandscape
%\sliderotation{right}

\def\bw{1}
\ifnum \bw=0
        \def\Red{\color[named]{Black}}
        \def\Cyan{\color[named]{Black}}
        \def\Blue{\color[named]{Black}}
        \def\Green{\color[named]{Black}}
        \def\Magenta{\color[named]{Black}}
        \def\UMagenta{\underline}
        \def\infn{infn-bw.ps}
\else
        \def\Red{\color[named]{Red}}
        \def\Cyan{\color[named]{Cyan}}
        \def\Blue{\color[named]{Blue}}
        \def\Green{\color[named]{Green}}
        \def\Magenta{\color[named]{Magenta}}
        \def\UMagenta{\color[named]{Magenta}}
        \def\infn{infn.ps}
\fi
%\slidesmag{2}
%\slideframe{none}
%\renewcommand{\footnoterule}{\vspace*{2ex}}
%\renewcommand{\refname}{\begin{center}
%                        \noindent \Large \rm \Red References
%                        \vspace*{2ex}
%                        \end{center}}
%\addtolength{\voffset}{-5cm}
%\addtolength{\topmargin}{-6cm}
%\addtolength{\textheight}{3cm}
%\begin{document}
%\centerslidesfalse
%\begin{slide*}
%\vspace{-20cm}
\begin{center}
%{\huge UNIVERSIT\`A DEGLI STUDI DI BARI}
\huge\sc{UNIVERSIT\`{A} DEGLI STUDI DI BARI} \\
\end{center} 
%\vspace{0.7cm}
\begin{center}
{\Large DOTTORATO DI RICERCA IN FISICA}
\end{center}
\vspace{0.1cm}
\begin{center}
% \includegraphics[scale=1]{LogoUni.ps}
% \hspace{0.8cm}
 \includegraphics[scale=1]{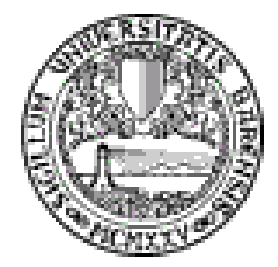}
\end{center}
%\vspace{0.5cm}
\vskip 0.2  truecm
\rule{15cm}{0.3mm}
\begin{center}
%{\Large Tesi di Dottorato}
{\LARGE\sc{Tesi di Dottorato}} \\
\end{center}
%\vspace{3.5cm}
\vspace{1cm}
\begin{center}
 {\huge{\bf Studio della produzione di }} \\ \vspace{0.5cm} 
 {\huge{\bf particelle strane e della dinamica di }}\\ \vspace{0.5cm}
 {\huge{\bf espansione in collisioni nucleari }}\\ \vspace{0.5cm}
 {\huge{\bf ultra-relativistiche all'SPS}}
\end{center}
\vspace{2.0cm}
{\Large Relatore:} \\
{\Large Chia.mo Prof. Bruno Ghidini} \quad\quad\quad\quad\quad 
\quad\quad\quad\quad\quad\quad \quad\quad\quad\quad{\Large Dottorando:} \\
\vspace{1.0cm} 
%\hspace{9.0cm} {\Large Dottorando:} \\
{\hspace{8.0cm} \quad\quad\quad}{\Large Giuseppe Eugenio Bruno} \\
%$ \begin{array}{lr}
% {\rm Relatore:}    \\
% {\rm Chia.mo\: Prof.\: Bruno\:\: Ghidini}  \\
%   &     \hspace{4cm}  {\rm  Dottorando}: \\
%   &     \hspace{4cm}  {\rm Giuseppe\ Eugenio\:\: Bruno}        \\
%\end{array} $
\vspace{1.0cm}
\rule{15cm}{0.3mm}
\begin{center}
\vspace{-0.5cm}
{\Large Novembre 2002}
\end{center}
%\end{slide*}
%\end{document}
\end{titlepage}

%% file: intro.tex
\chapter*{Introduzione}
\indent
\quad  La fisica degli ioni pesanti ultra-relativistici cerca di riprodurre 
in laboratorio le condizioni in cui si pensa che si trovasse l'Universo 
pochi istanti dopo il {\em Big Bang} primordiale. 
\newline
Si ritiene infatti che, prima dell'adronizzazione, l'Universo si trovasse 
in una fase in cui quark e gluoni liberi costituissero un {\em plasma} 
denso e caldo. 
Questa \`e la fase che si vuole ricreare oggi in laboratorio,
facendo collidere nuclei pesanti accelerati ad energie
ultra-relativistiche.
In tale fase, tra l'altro, l'asimmetria tra materia ed anti-materia, cos\`i evidente 
nell'Universo attuale,  era invece una  piccola anomalia. 

L'estremo interesse suscitato da  questa nuova linea di 
ricerca, attivatasi a partire dalla met\`a degli 
anni '80 con la progettazione e realizzazione dei primi 
esperimenti di interazioni tra ioni di massa ed energia 
intermedia all'acceleratore AGS  
({\em Alternate Gradient Synchroton}) di Brookhaven (USA), 
ha dato luogo ad una prima generazione di esperimenti all'SPS 
({\em Super Proton Synchroton}) del CERN con ioni 
relativamente leggeri ($^{16}{\rm O}$, $^{32}{\rm S}$).  
Visti i risultati incoraggianti 
di questi, 
%della prima serie di esperimenti  
%%se pur 
%di tipo esplorativo, 
una seconda generazione, 
condotta con ioni pi\`u pesanti, ha studiato in dettaglio i segnali 
caratteristici proposti come evidenze sperimentali 
dell'avvenuta transizione di fase nello stato di plasma di quark e di 
gluoni (QGP).  

La determinazione sperimentale di queste evidenze \`e una sfida difficile, 
in quanto la sopravvivenza del plasma sarebbe comunque limitata ai primi 
istanti successivi alla collisione ($ \approx 10^{-22}$\ sec).  
La sua rivelazione dipende quindi da misure indirette, o attraverso le 
particelle prodotte nello stato finale o cercando segnali 
caratteristici, capaci di sopravvivere alla transizione stessa 
ed alla successiva evoluzione del sistema collidente.  
\newline
Nonostante queste difficolt\`a, le evidenze sperimentali sin qui 
raccolte dagli esperimenti di seconda generazione del CERN,   
considerate nella loro collegialit\`a, non trovano ragionevoli 
descrizioni in termini di uno scenario adronico (in assenza di QGP).  
Esse, peraltro, hanno ottenuto larga approvazione nella comunit\`a scientifica  
al punto che nel febbraio del 2000, in una conferenza stampa, 
il laboratorio del CERN ha espresso la sua formale convinzione 
che nelle collisioni Pb-Pb pi\`u centrali
sia stato prodotto un nuovo stato della materia, avente  
molte delle caratteristiche attese per il QGP.  

Se dunque l'evidenza del QGP sembra ormai appurata, a molte altre 
domande si deve ancora dar risposta. Eccone alcune:
\begin{itemize}
\item[]
Quali sono le condizioni limite, %della materia, 
in termini di volume e densit\`a di energia iniziale, per far avvenire 
la transizione di fase ? 
\item[]
Che tipo di transizione di fase si ha ? Di quale ordine ? 
\item[]
Che caratteristiche fisiche ha il plasma ?  
\item[]
Si ha, ed a quale livello, il ripristino della simmetria chirale ?
%(e quello tra la materia ed antimateria) ?  
\end{itemize}
Ad alcune di queste domande potranno forse dare una risposta definitiva 
solo i nuovi esperimenti al collisionatore di ioni pesanti RHIC a 
Brookhaven, in funzione dall'anno 2000, e l'esperimento ALICE 
che studier\`a, dal 2007, le collisioni tra ioni di piombo accelerati 
ad un'energia ancora pi\`u elevata ($2.76 + 2.76$\ TeV per nucleone) 
dal {\em Large Hadron Collider} (LHC) del CERN.  

In questa linea di ricerca si inserisce l'attivit\`a descritta in 
questa tesi di dottorato, che affronta alcune tematiche specifiche 
dell'esperimento CERN NA57.   
Questo studia le collisioni su bersaglio fisso di nuclei di piombo 
(e, per confronto, di protoni) accelerati dall'SPS del CERN fino 
a 160 GeV/$c$\ per nucleone e misura, come segnale di QGP, 
la produzione di stranezza. 
Come conseguenza della transizione di fase \`e infatti atteso un 
incremento della produzione di particelle strane, ed in particolare 
di barioni ed anti-barioni multi-strani,  
rispetto alle normali interazioni adroniche.  
\newline
Per individuare le condizioni limite oltre le quali pu\`o 
avvenire la transizione di fase, l'esperimento \`e pensato  
per studiare le collisioni Pb-Pb  
\begin{enumerate}
\item
su un'ampio intervallo  di centralit\`a, 
a partire dalle collisioni pi\`u centrali sino 
a quelle periferiche, per il $60\%$ della sezione 
d'urto anelastica.  
In tal modo \`e possibile esplorare i limiti della transizione 
di fase in termini del volume iniziale coinvolto nella collisione. 
\item
in funzione dell'energia della collisione. L'esperimento 
ha quindi raccolto dati a 160 ed a 40 GeV/$c$\ per nucleone.  
\end{enumerate}
Come interazioni di riferimento, in cui non si attende la 
transizione di fase, si considerano:
\begin{itemize}
\item
a 160 GeV/$c$, le collisioni p-Be e p-Pb raccolte dall'esperimento 
precursore WA97. 
\item
a 40 GeV/$c$, le collisioni p-Be raccolte apposta da NA57.
\end{itemize}

La tesi \`e strutturata nel modo di seguito descritto.    
\newline
Nel primo capitolo si presenta un'introduzione alla fisica 
delle collisioni tra ioni pesanti ultra-relativistici 
ed
%e si presenter\`a  %un'ampia 
una panoramica sullo stato della ricerca di questo campo della fisica.  
\newline
Nel secondo capitolo viene descritto l'esperimento NA57, 
soffermandosi in maniera particolare sull'apparato   
sperimentale utilizzato.   
\newline
Segue, nel terzo capitolo, la descrizione delle procedure sviluppate 
per identificare le particelle strane studiate, \PKzS, \PgL, \PagL, \PgXm, 
\PagXp, \PgOm\ e \PagOp, 
isolandole dalla moltitudine 
di particelle prodotte nelle collisioni tra ioni di piombo (alcune migliaia).  
\newline
Nel quarto capitolo si discute l'analisi dei segnali delle particelle 
strane selezionate come spiegato nel capitolo precedente. 
Si presenta prima uno studio volto a determinare con precisione  
la contaminazione del fondo residuo sotto i segnali di \PKzS, \PgL\ e \PagL;  
vengono quindi affrontati i problemi relativi all'analisi dei segnali  
ottenuti, quali le correzioni per l'accettanza geometrica e per l'efficienza 
di ricostruzione, la determinazione  della finestra di accettanza fiduciale 
delle diverse particelle, nonch\'e la procedura di estrapolazione dei risultati 
ottenuti. In questo capitolo viene anche illustrato il metodo utilizzato 
per la determinazione della centralit\`a della collisione.  
\newline
Nel quinto capitolo si discutono i risultati ottenuti sulla produzione  
di particelle strane nelle collisioni nucleari all'energia dell'SPS,  
in particolare le distribuzioni di massa trasversa,  
i rapporti di produzione tra le varie specie ed i tassi assoluti di produzione   
in funzione della centralit\`a e dell'energia della collisione. 
%Si confronteranno i risultati ottenuti con quelli dell'esperimento WA97 
\newline
Il sesto ed ultimo capitolo \`e dedicato allo studio della interferometria  
HBT tra  particelle di carica negativa. L'analisi \`e  
condotta in funzione della centralit\`a, a partire dai dati raccolti 
dall'esperimento WA97 in collisioni Pb-Pb a 160 GeV/$c$ per nucleone.
Da questo studio si ricava una descrizione della geometria della 
regione di collisione e della dinamica di espansione del sistema 
Pb-Pb a seguito della collisione.  

%% file: command.tex
\newcommand{\lum}[1]{$10^{#1}$ cm$^{-2}$s$^{-1}$ }
\newcommand{\fr}{\mbox{\ $\longrightarrow$\ }}
\newcommand{\pt}{$p_t$ }
\newcommand{\et}{$\eta$ }
\newcommand{\ets}{$\eta$}
\newcommand{\ph}{$\varphi$ }
\newcommand{\phs}{$\varphi$}
\newcommand{\ptc}{$p_t^{cut}$ }
\newcommand{\ptcv}[1]{$p_t^{cut}=#1$ GeV/$c$}
\newcommand{\hz}{Hz/cm$^2$}
\newcommand{\aeta}{$\lvert\eta\rvert$}
\newcommand{\aetal}[1]{$\lvert\eta\rvert <#1$}
\newcommand{\aetag}[1]{$\lvert\eta\rvert <#1$}
\newcommand{\etae}[1]{$\eta =#1$}
\newcommand{\aetae}[1]{$\lvert\eta\rvert =#1$}
\newcommand{\aetai}[2]{$#1<\lvert\eta\rvert <#2$}
\newcommand{\abs}[1]{\mbox{$\lvert#1\rvert$}}
\newcommand{\gv}{GeV/$c$}
\newcommand{\ptg}[1]{$p_t > #1$ \gv }
\newcommand{\ptl}[1]{$p_t < #1$ \gv }
\newcommand{\pte}[1]{$p_t = #1$ \gv }
\newcommand{\bun}{bunch crossing }
\newcommand{\buns}{bunch crossings }
\newcommand{\tr}{trigger }                  
\newcommand{\gvm}{GeV/$c^2$}    
\newcommand{\mut}{$\mu_{tag}$ }

%% file: cap1/cap1.tex
\chapter{Il plasma di quark e di gluoni nelle collisioni nucleari 
         ultra-relativistiche} 
\section{Introduzione}
Tradizionalmente la {\em fisica delle alte energie} viene 
identificata con la {\em fisica delle particelle elementari}:  
l'approccio generalmente seguito \`e quello di descrivere le 
interazioni tra gli oggetti elementari, derivandone le caratteristiche  
da principi primi (le teorie di gauge). Dal punto di vista sperimentale, 
\`e conveniente considerare sistemi in interazione i pi\`u semplici possibile, 
preferibilmente collisioni tra leptoni (elettroni, neutrini) 
e/o, al  pi\`u, adroni (pioni, protoni, etc.).   
\newline  
Nel dominio di pi\`u bassa energia, la fisica nucleare indaga  sistemi 
molto pi\`u estesi (i nuclei, appunto), che sono dotati evidentemente
di caratteristiche ``globali'', 
ed usa pertanto un approccio necessariamente diverso, 
spesso di tipo fenomenologico e statistico-termodinamico.  
\newline
L'evoluzione della fisica delle particelle si accompagna tradizionalmente 
alla possibilit\`a di accelerare le particelle ad energie sempre pi\`u 
elevate, permettendo di raggiungere di volta in volta le soglie per 
la creazione di nuove particelle e per l'innesco di nuovi fenomeni.  
\newline
A partire dalla seconda met\`a degli anni ottanta, con la possibilit\`a 
di accelerare nuclei alle energie ultra-relativistiche (i.e. $v\simeq c$) 
tipiche della moderna fisica delle particelle elementari,  si \`e 
aperto un nuovo campo di indagine dello studio delle propriet\`a della 
materia nucleare, con particolare riferimento alle interazioni forti.    
Esso \`e a cavallo tra la fisica delle particelle e la fisica nucleare  
e si avvale quindi di metodi e concetti propri delle due discipline, 
unificando in un certo senso l'approccio microscopico della  
fisica delle particelle e quello macroscopico della fisica nucleare.  
Si \`e quindi sviluppata una vera e propria ``termodinamica di QCD'', 
dove stati complessi di molte particelle interagenti fortemente 
vengono descritti in termini di variabili quali la {\em temperatura}, 
la {\em densit\`a}, l'{\em{entropia}}, il {\em potenziale chimico} 
e cos\`i via. 
\newline
Perch\'e una simile descrizione sia consistente \`e necessario  che il 
sistema formatosi in seguito alla collisione tra i nuclei sia esteso, 
cio\`e di dimensioni molto maggiori della portata dell'interazione 
forte ($ \gg 1$ fm), riesca a raggiungere l'equilibrio termico, 
cio\`e la sua vita media ecceda i tempi tipici di rilassamento 
($ \gg 1$ fm/$c$), e sia costituito da un numero molto elevato 
di particelle.  
\newline
La normale materia nucleare \`e composta da nucleoni entro cui sono 
confinati i quark ed i gluoni. 
Nel seguito del capitolo si esporr\`a come la termodinamica di QCD ed 
opportuni modelli fenomenologici portino  
a prevedere una transizione di fase della materia nucleare verso uno 
stato di plasma di quark e gluoni (QGP) in corrispondenza di una 
temperatura critica $T_c = 150 \div 200$\ MeV o di una densit\`a barionica 
di $0.5 \div 1.5 \, {\rm nucleoni/fm^3} $. Saranno quindi descritte 
le propriet\`a del QGP e le modalit\`a con cui si cerca di far avvenire la 
transizione di fase agli acceleratori di particelle di altissima energia.  
Seguir\`a una rassegna delle diverse segnature sperimentali attese a seguito 
della transizone di fase  e saranno esposti alcuni dei risultati pi\`u 
recenti e promettenti prodotti dagli esperimenti coinvolti in questa 
attivit\`a di ricerca.  
Si dar\`a particolare rilievo al segnale associato alla produzione di 
particelle strane ed alla correlazione di intensit\`a, argomenti su cui 
verte questo lavoro di tesi.  
\section{Le particelle elementari e le interazioni fondamentali}
Si ritiene oggigiorno che tutta la materia di cui \`e composto l'universo 
sia costituita da pochi costituenti elementari.  
Questi costituenti elementari sono noti come particelle elementari e 
possono  classificarsi in due categorie: i fermioni ed i bosoni.  
\newline
La differenza tra le due famiglie risiede nello spin,   
cio\`e il momento angolare intrinseco  
di una particella, che \`e pari ad un multiplo semi-intero 
di $\hbar$\ per i fermioni, ad un multiplo intero di $\hbar$\ per i bosoni.  
\newline
Le particelle elementari sono poi caratterizzate dal mondo in cui interagiscono 
tra loro. Quattro diverse interazioni fondamentali sono conosciute: % sino ad ora: 
l'elettromagnetica, la debole, la forte e la gravitazionale: 
le prime due sono riconducibili ad un'origine comune (``unificazione 
elettro-debole''). 
Vi sono forti presupposti per ritenere che le interazioni fondamentali 
possano essere correttamente descritte in termini di teorie di campo associate a 
particolari simmetrie di gauge locali~\cite{r1_2}.  
Insieme ad  altri aspetti, tali teorie godono della propriet\`a fondamentale 
di essere teorie rinormalizzabili: in virt\`u di tale propriet\`a, le predizioni 
di queste teorie restano finite ad ogni ordine del calcolo perturbativo, 
pur introducendo solo un numero {\em finito} di parametri, 
quali ad esempio le masse delle particelle e le costanti di accoppiamento.   
Nel contesto delle teorie di campo di gauge quantistiche, i bosoni intervengono 
in quanto associati alle trasformazioni di simmetria che lasciano invariante 
la lagrangiana della teoria. 
\newline
La teoria stessa viene costruita per descrivere una particolare interazione 
tra i fermioni. Per questo motivo, i bosoni sono spesso indicati come i 
``mediatori'' dell'interazione tra i fermioni.  
%(o ``bosoni vettori'' dell'interazione).  
Le interazioni fondamentali conosciute ed i bosoni ad esse associati sono  
elencati nella tabella 1.1 
  \begin{table}[h]
    \label{tab11}
    \begin{center}
      \begin{tabular}{|c|c|c|} \hline
          {\bf Interazione} & {\bf Bosoni vettori} & {\bf Gruppo di gauge}  \\ 
          \hline Elettromagnetica e Debole & $\gamma$, $Z^0$, $W^+$, $W^-$ & $SU(2)_L\times U(1)_Y$  \\
           \hline Forte & 8 diversi gluoni & $SU(3)$ \\
           \hline Gravitazionale & gravitone (?) & ? \\
           \hline
      \end{tabular}
    \end{center}
   \caption{Le interazioni fondamentali note e la loro descrizione in termini 
            di teorie di campo di gauge quantistiche.}
\end{table} 
\newline
Non tutti i fermioni possono interagire mediante ciascuna delle quattro interazioni 
fondamentali. Vi sono infatti due famiglie di fermioni: i quark ed i leptoni.  
I primi interagiscono per mezzo di tutte le forze conosciute, mentre i secondi  
non risentono della %sola 
forza forte.   
Sia i quark che i leptoni si presentano in tre generazioni di coppie, come 
indicato nella tabella 1.2.   
 \begin{table}[h]
    \label{tab12}
    \begin{center}
      \begin{tabular}{|l|c|c|c|} \hline
         \hspace{2cm} {\bf Generazione} & I & II & III \\ 
          {\bf Fermioni} &  &  &  \\
          \hline   Quark  & u \ d & c \ s & t \ b \\
          \hline Leptoni   & e \ $\nu_e$ & $\mu$ \ $\nu_{\mu}$ & $\tau$ \ $\nu_{\tau}$ \\
          \hline
      \end{tabular}
    \end{center}
   \caption{I fermioni elementari suddivisi per generazione.}
\end{table}
\newline
Ad ogni fermione conosciuto, quark o leptone, corrisponde poi un anti-fermione 
caratterizzato da numeri quantici opposti e stessa massa.  
\subsection{L'interazione elettro-debole} 
La forza elettromagnetica viene descritta dalla 
{\em ElettroDinamica Quantistica} (QED) che assume il gruppo abeliano $U(1)$\ 
come gruppo di invarianza di gauge.  
Maggiori difficolt\`a si incontrano nel descrivere in modo completo la sola 
forza debole. Negli anni sessanta Glashow, Weinberg and Salam capirono che 
la forza debole pu\`o trovare una descrizione completa solo all'interno di una 
teoria pi\`u generale, che racchiuda anche l'interazione elettromagnetica. 
La forza elettromagnetica e la forza debole sono state ``unificate'' nell'ambito 
di un'unica teoria di gauge, la teoria elettrodebole, che le descrive entrambe. 
Tale teoria assume come gruppo di simmetria il gruppo di gauge 
$SU(2)_L\times U(1)_Y$~\cite{r1_1}.  Nella teoria, le particelle sono 
inizialmente prive di massa in quanto si assume una perfetta simmetria di gauge. 
I fermioni ed i bosoni acquisiscono le loro masse attraverso il {\em meccanismo 
di Higgs} di rottura spontanea della simmetria di gauge locale~\cite{r1_2}.  
Nella sua formulazione pi\`u semplice ({\em modello minimale}), il meccanismo 
prevede un solo bosone di Goldstone, 
noto in tal caso come particella di Higgs, 
senza spin e dotato di massa. Le uniche particelle della teoria che restano 
prive di massa sono  i fotoni ed i neutrini. La teoria pu\`o essere  
estesa in modo tale che anche i neutrini abbiano una massa non nulla 
(modelli non minimali), come sembrano suggerire i pi\`u recenti 
esperimenti che studiano queste particelle~\cite{NOW}.  
\newline
Col meccanismo di Higgs, si conserva la  
propriet\`a di rinormalizzabilit\`a della teoria, in quanto la rottura della
simmetria \`e di tipo {\em spontaneo}, nel senso che riguarda solo lo stato 
di minima energia e non la lagrangiana della teoria. Considerando tutti i 
possibili stati di minima energia in cui l'universo si pu\`o disporre,  
si riguadagna la completa simmetria di gauge.  

\subsection{L'interazione forte} 
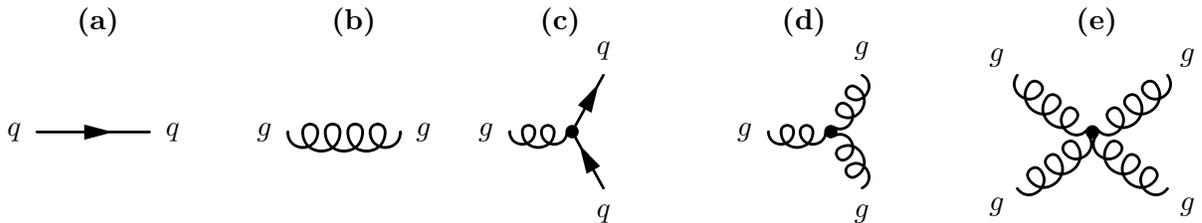
\begin{figure}[b]
%\begin{center}
% \includegraphics[scale=0.30]{cap1/IntQCD.eps} 
% \caption{Schema simbolico dei diversi termini della lagrangiana di QCD.
%  {\bf a)} [{\bf b)}]Termine quadratico nel campo dei quark [gluoni], che descrive la loro 
%                     propagazione. 
%  {\bf c)} Termine, quadratico nel campo dei quark e lineare nel campo gluonico, che 
%           descrive i vertici di interazione dei quark.  
%  {\bf d)} [{\bf e)}] Termine cubico [quartico]  nel campo gluonico, che descrive i 
%  		    vertici di interazione a tre [quattro] gluoni. } 
% \label{IntQCD}
% \end{center}
%\end{figure}
%\begin{figure}[t]
\centering
\hspace{-0.6cm}
\begin{tabular}{ccccc}
{\bf (a)} \quad\quad\quad & {\bf (b)}\quad & 
{\bf (c)} \quad\quad\quad & {\bf (d)} \quad\quad\quad\quad &{\bf (e)} \\ \\
\hspace{-0.5cm}
\setlength{\unitlength}{0.3mm}
\begin{fmffile}{sample11}
\begin{fmfgraph*}(50,50)
\fmfleft{i1} \fmfright{o1} \fmflabel{$q$}{i1}
\fmflabel{$q$}{o1}
\fmf{fermion}{i1,o1}
\end{fmfgraph*}
\end{fmffile} \quad\quad\quad 
&
\setlength{\unitlength}{0.3mm}
\begin{fmffile}{sample21}
\begin{fmfgraph*}(50,50)
\fmfleft{i1} \fmfright{o1} \fmflabel{$g$}{i1}
\fmflabel{$g$}{o1}
\fmf{gluon}{i1,o1}
\end{fmfgraph*}
\end{fmffile} \quad\quad
&
\setlength{\unitlength}{0.5mm}
\begin{fmffile}{sample31}
\begin{fmfgraph*}(50,30)
\fmfleft{i1} \fmfbottom{o1} \fmftop{o2} \fmflabel{$g$}{i1}
\fmflabel{$q$}{o1} \fmflabel{$q$}{o2}
\fmf{gluon}{i1,v1}
\fmf{fermion}{o1,v1}
\fmf{fermion}{v1,o2}
\fmfdot{v1}
\end{fmfgraph*}
\end{fmffile}
&
\setlength{\unitlength}{0.5mm}
\begin{fmffile}{sample41}
\begin{fmfgraph*}(50,30)
\fmfleft{i1} \fmfbottom{o1} \fmftop{o2} \fmflabel{$g$}{i1}
\fmflabel{$g$}{o1} \fmflabel{$g$}{o2}
\fmf{gluon}{i1,v1}
\fmf{gluon}{o1,v1}
\fmf{gluon}{v1,o2}
\fmfdot{v1}
\end{fmfgraph*}
\end{fmffile}
&
\hspace{-0.5cm}
\setlength{\unitlength}{0.5mm}
\begin{fmffile}{sample51}
\begin{fmfgraph*}(50,30)
\fmfleft{i1,i2} \fmfright{o1,o2} \fmflabel{$g$}{i1}
\fmflabel{$g$}{i2} \fmflabel{$g$}{o1}
\fmflabel{$g$}{o2}
\fmf{gluon}{i1,v1,o1}
\fmfdot{v1}
\fmf{gluon}{i2,v1,o2}
\end{fmfgraph*}
\end{fmffile}
\end{tabular}
 \caption{Schema simbolico dei diversi termini della lagrangiana di QCD.
  {\bf a)} [{\bf b)}]Termine quadratico nel campo dei quark [gluoni], che descrive la loro
                     propagazione.
  {\bf c)} Termine, quadratico nel campo dei quark e lineare nel campo gluonico, che
           descrive i vertici di interazione dei quark.
  {\bf d)} [{\bf e)}] Termine cubico [quartico]  nel campo gluonico, che descrive i
                   vertici di interazione a tre [quattro] gluoni. }
 \label{IntQCD}
\end{figure}

La teoria che descrive le interazioni forti prende il nome di 
{\em CromoDinamica Quantistica} (QCD). Cos\`i come nella  
QED la {\em carica elettrica} \`e l'osservabile fisica che si conserva  
in virt\`u della invarianza della lagrangiana di QED per trasformazioni 
del gruppo $U(1)$, cos\`i nella QCD vi \`e una {\em carica di colore} associata 
alla invarianza della lagrangiana di QCD per trasformazioni del gruppo $SU(3)$.  
Pi\`u precisamente, il numero di cariche conservate \`e pari all'ordine del 
gruppo di simmetria ed il numero di bosoni di gauge \`e pari al numero di 
generatori del gruppo. 
\newline
Al gruppo di ordine uno $U(1)$, generato da un unico operatore (di rotazione), 
corrisponde dunque un'unica carica conservata (la carica elettrica) ed un unico 
bosone di gauge (il fotone).   
\newline
Il gruppo $SU(3)$ \`e di ordine tre e dispone  di otto generatori: 
vi sono pertanto tre diversi tipi  
di carica di colore (la carica rossa, la carica blu e la carica verde),   
ed otto bosoni di gauge (i gluoni), mediatori dell'interazione 
di colore tra i quark.   
\newline
La caratteristica peculiare del gruppo di simmetria $SU(3)$\   
associato all'interazione forte \`e quella di essere un gruppo     
non-abeliano. Di conseguenza nella lagrangiana di QCD \`e presente un 
termine di auto-interazione tra i campi gluonici: in altri termini, i gluoni, 
a differenza del fotone che non \`e carico elettricamente, possiedono una 
carica di colore, per cui possono interagire tra loro, oltre che con i quarks.  
La lagrangiana finale di QCD, per uno solo dei sei quark, assume infatti la seguente  
espressione: 
\begin{equation}
 \CMcal{L}=\bar{q}(i\gamma^{\mu}\partial_{\mu} - m)q - 
           g(\bar{q}\gamma^{\mu} T_{a} q )G^{a}_{\mu} -
	   \frac{1}{4}G^a_{\mu\nu}G_a{\mu\nu}
 \label{Lagrangiana}
\end{equation}
dove $q=(q_R,q_B,q_V)$\ \`e il campo dei quark nei tre possibili colori, 
$G_{\mu}^a$\ \`e il campo gluonico ($a=1,2,...,8$), $G_{\mu\nu}^a$\ una 
combinazione di campi gluonici data da 
$G_{\mu\nu}^a=\partial_{\mu}G^a_{\nu} - \partial_{\nu}G^a_{\mu} -
               g f_{abc}G^b_{\mu}G^c_{\nu} $, $g$\ \`e la costante  
di accoppiamneto forte, $T_a$\ sono matrici hermitiane 3$\times$3  
ed $f$\ sono le costanti  
di struttura del gruppo (reali e completamente anti-simmetriche) tali 
che $[T_a,T_b]=if_{abc}T_c$.  
Quanto detto diventa pi\`u evidente se si riscrive la lagrangiana dell'eq.~\ref{Lagrangiana}  
nella forma simbolica  pi\`u semplice 
\newline
\begin{center}
$\CMcal{L}=$``$\bar{q}q$'' + ``$G^2$'' + $g$ ``$\bar{q}qG$'' + $g$ ``$G^3$'' + $g^2$``$G^4$'' 
\label{SimpleLagr}
\end{center}
in cui compaiono solo i campi dei quark e dei gluoni.  
Ciascuno dei cinque termini della espressione %~\ref{SimpleLagr} 
precedente  
pu\`o essere rappresentato da un diagramma di Feynman elementare, 
come schematizzato nella fig.~\ref{IntQCD}.  
%\newline
I primi tre termini di questa espressione descrivono rispettivamente  
la propagazione libera dei quark e dei gluoni e l'interazione quark-gluone; 
essi hanno un corrispettivo in QED con il fotone al posto del gluone.  
I due termini rimanenti comportano la presenza di vertici a tre od a 
quattro gluoni nella QCD e riflettono 
il fatto che i gluoni stessi sono portatori di carica di colore. Essi non hanno un 
corrispettivo in QED e provengono dalla natura non abeliana del gruppo di gauge. \`E importante  
ricordare che l'invarianza di gauge determina in modo univoco la struttura di questi  
termini di auto-accoppiamento per i gluoni e che vi \`e un'unica costante di 
accoppiamento $g$.  

L'insieme delle teorie cui qui si \`e rapidamente accennato
\`e conosciuto come il {\em Modello Standard} delle interazioni 
elettro-deboli e forti.  
Le osservazioni sperimentali hanno sinora confermato le predizioni del Modello 
Standard ad un livello di precisione dell'ordine dello 0.1\%~\cite{r1_7}.  
La teoria si basa su 
%diciannove 
diciotto  
parametri liberi e prevede l'esistenza di una 
particella, il bosone di Higgs, ancora da verificare.
\subsection{Le costanti di accoppiamento in QED ed in QCD}
\begin{figure}[ht]
%\begin{figure}[p]
\begin{center}
 \includegraphics[scale=0.60]{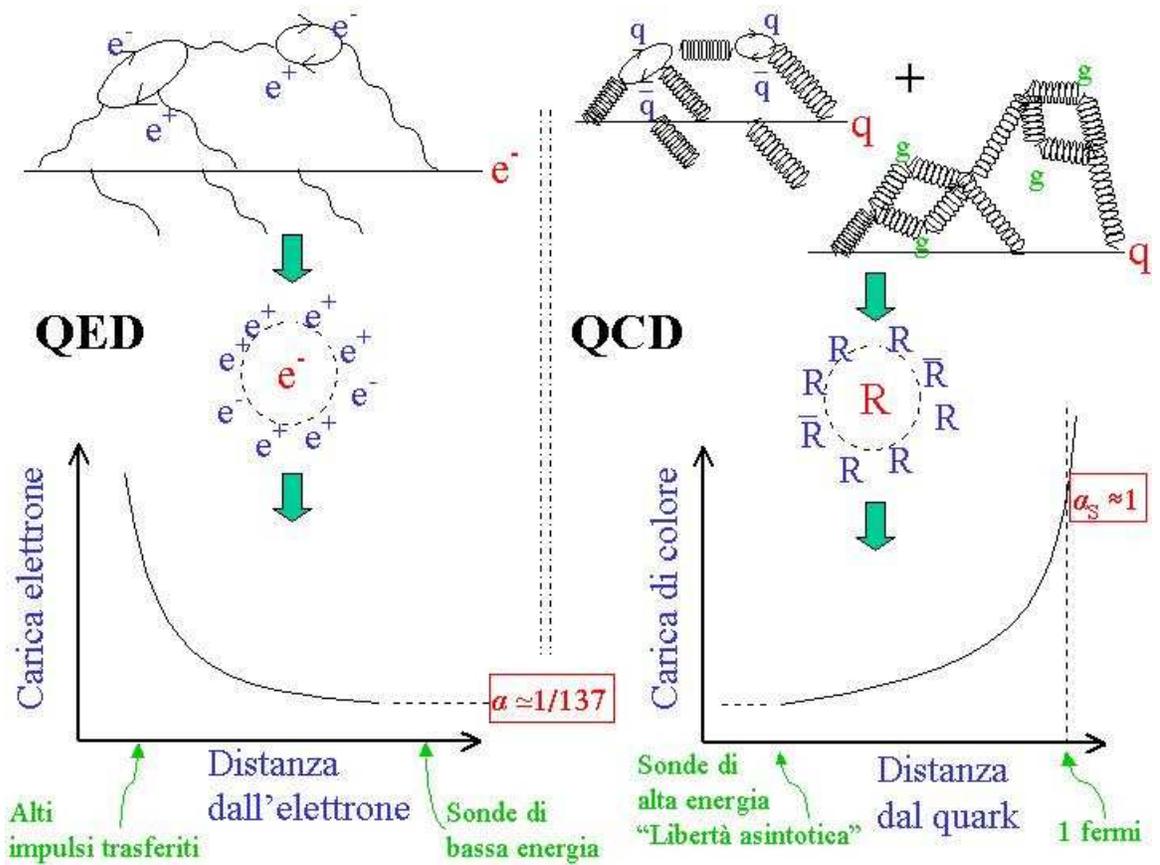}
 \caption{Effetti dello schermaggio della carica elettrica (sinistra) e 
          della carica di colore (a destra) nel vuoto della 
	  teoria di campo quantistica.} 
  \label{Coupling}
\end{center}
\end{figure}

\begin{figure}[h]
\begin{center}
 \includegraphics[scale=0.30]{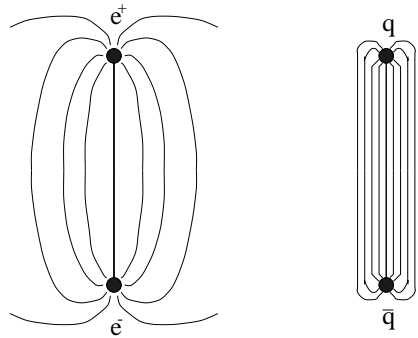}
 \caption{Linee di campo elettrico per il sistema \Pep\ \Pem\  e linee 
           di campo di colore per la coppia $q\bar{q}$.}  
 \label{Linee}
\end{center}
\end{figure}

\begin{figure}[h]
\begin{center}
\includegraphics[scale=0.40]{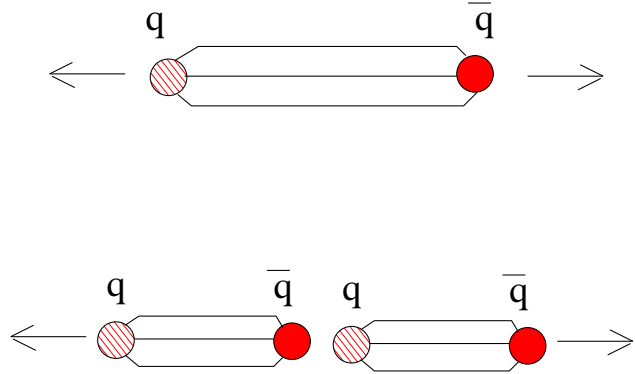}
 \caption{Formazione di una nuova coppia $q \bar{q}$\ per separazione 
           di un quark da un anti-quark.}
\label{fragm}
\end{center}
\end{figure}

La possibilit\`a dell'accoppiamento tra gluoni si manifesta in un diverso 
andamento della costante di accoppiamento effettiva di QCD rispetto a 
quello caratteristico della QED, come mostrato in fig.~\ref{Coupling}.  
Una carica elettrica posta nel vuoto di QED emette e riassorbe continuamente 
fotoni virtuali, i quali possono produrre temporaneamente coppie \Pep\Pem che 
hanno l'effetto di schermare la carica originariamente presente.     
Infatti, delle coppie \Pep\Pem che ``circondano'' l'elettrone libero, i 
positroni gli sono, in media, pi\`u vicini; l'elettrone \`e quindi circondato 
da una nube di carica elettrica polarizzata in modo tale che le 
cariche positive siano preferenzialmente pi\`u vicine; la 
carica negativa di un elettrone \`e dunque schermata dalla polarizzazione 
del vuoto di QED.   
L'effetto di schermaggio, simile al fenomeno di 
polarizzazione di un  dielettrico, si manifesta in una decrescita della 
costante di accoppiamento di QED al crescere della distanza  
dalla carica originaria o, equivalentemente, al decrescere del momento trasferito 
nelle interazioni.  
Considerando i primi due ordini dello sviluppo perturbativo, la dipendenza della 
costante di accoppiamento di QED dall'impulso trasferito $q$\  
(pi\`u precisamente da $Q^2=-q^2$) \`e fornita dalla 
seguente espressione, valida ad elevati impulsi traferiti: 
\begin{equation}
 \alpha(Q^2)=
   \frac{\alpha(\mu^2)}
  {1-\frac{\alpha(\mu^2)}{3\pi}\ln(\frac{Q^2}{\mu^2})}
\label{Alpha}
\end{equation}
dove $\alpha(\mu^2) $\ \`e il valore della costante di accoppiamento al 
valore di riferimento $Q^2=\mu^2$. 
\newline
In maniera analoga, un quark posto nel vuoto di QCD emette e riassorbe 
continuamente gluoni i quali,  per\`o, possono produrre temporaneamente  
anche coppie $gg$, oltre che coppie $q\bar{q}$, in virt\`u 
della loro auto-interazione. 
Poich\'e la probabilit\`a di occorrenza dei vertici gluonici \`e maggiore 
di quella tra quark, il contributo dominante delle coppie $gg$\ produrr\`a, 
nello spazio circostante, una distribuzione della carica di colore originariamente 
posseduta dal quark, secondo un effetto di anti-schermaggio.  
La costante di accoppiamento di QCD assume infatti la seguente espressione: 
\begin{equation}
 \alpha_s(Q^2)=
   \frac{12\pi}
    {(33-2N_f)\ln(\frac{Q^2}{\Lambda^2})}
 \label{AlphaS}
\end{equation}
dove $N_f$\ indica il numero di tipi di quark ({\em ``flavour''})  e 
$\Lambda$\ \`e il parametro di scala della QCD, esprimibile come in QED in 
termini del valore di riferimento di $\alpha_s$\ ad un dato valore del 
momento trasferito, 
$\Lambda^2=\mu^2\exp[\frac{-12\pi}{(33-2N_f)\alpha_s(\mu^2)}]$. 
\newline
La costante di accoppiamento $\alpha_s$\ decresce al crescere di $Q^2$,
purch\'e il numero di quark resti inferiore a  sedici.  
Per valori di $Q^2$\ molto pi\`u grandi di $\Lambda^2$,  $\alpha_s$\
\`e dunque piccola e la descrizione perturbativa in 
termini di quark e gluoni in debole interazione ha senso.   
Pertanto, per interazioni a ``breve distanza'' risulta $\alpha_s \ll 1$ 
e si parla in tal caso di ``libert\`a asintotica'' dei quark.   
\newline 
L'evidenza sperimentale di tale propriet\`a \`e stata ottenuta attraverso 
esperimenti di urti altamente anelastici leptone-nucleone, nei quali la 
misura dell'impulso del leptone prima e dopo la collisione permette di 
sondare le distribuzioni di impulso dei quark e dei gluoni all'interno 
dei nucleoni. Si \`e cos\`i verificato che per grandi impulsi trasferiti,   
i quark all'interno degli adroni possono essere considerati pressocch\'e   
liberi.  
\newline
Per piccoli impulsi trasferiti (per valori di $Q^2$\ dell'ordine 
di $\Lambda^2$), od equivalentemente per distanze di interazione 
superiori ad un fermi, la costante di accoppiamneto diventa 
grande e lo sviluppo perturbativo, basato sull'espansione in serie 
di potenze della costante di accoppiamento, non \`e pi\`u applicabile. 
L'interazione forte tra due cariche di ugual colore diventa dunque tanto 
pi\`u intensa quanto pi\`u grande \`e la loro separazione. 
A causa dell'interazione dei gluoni tra loro, le linee di forza associate 
al campo di colore della coppia $q\bar{q}$ sono concentrate entro un tubo 
di flusso, come mostrato nella fig.~\ref{Linee}. Al contrario, per un dipolo 
elettrico \Pep\ \Pem\, nulla vieta alle linee di forza di occupare tutto 
lo spazio circostante, in quanto i fotoni, non iteragendo tra loro, non sono  
in grado di contenere le linee di campo entro una regione confinata 
dello spazio. 
\newline
Se si assegna al tubo delle linee di colore una densit\`a di energia lineare
costante, l'energia potenziale tra un quark ed un anti-quark aumenta linearmente
con la distanza,  $V(r)\sim \lambda r $, per $r \apprge$ 1 fm. 
Volendo separare una coppia  
$q\bar{q}$, si pu\`o pensare di allontanare il quark dall'anti-quark; in 
tale processo le linee di colore si allungano e l'energia potenziale aumenta 
sino a quando non diventa pi\`u conveniente, da un punto di vista energetico, 
la formazione di una nuova coppia $q\bar{q}$\ piuttosto che un ulteriore 
allontanamento.  La coppia $q\bar{q}$\ prodotta funge da nuovi terminali, 
pozzo e sorgente, delle linee di forza, che cos\`i si spezzano in due nuovi 
pi\`u piccoli tubi, come mostrato in fig.~\ref{fragm}. 
\newline
Questo comportamento a piccoli momenti trasferiti \`e indicato col termine 
di ``schiavit\`u infrarossa'' e si osserva, da un punto di vista 
sperimentale, ad esempio nel processo \Pep \Pem $\longrightarrow q \bar{q}$, 
in cui il quark e l'antiquark non si manifastano mai singolarmente, 
ma adronizzano danno luogo ad una tipica segnatura sperimentale 
caratterizzata da due getti di  adroni formanti tra loro,
nel sistema centro di massa della collisione, un angolo piatto. 
Da un punto di vista sperimentale i quark sono pertanto 
{\em confinati} entro gli adroni, che presentano una carica di colore  
totale nulla, non manifestandosi singolarmente.  
\newline
Considerando il comportamento sia a breve che a lunga distanza,  
il potenziale effettivo del sistema $q\bar{q}$\ si pu\`o esprimere come 
$V_{QCD}(r)=-\frac{4}{3}\frac{\alpha_s}{r} + \lambda r $\ dove il primo 
termine domina per piccole distanze ($r \ll 1$ fm, {\em libert\`a asintotica}) 
ed il secondo a grandi distanze ($r \apprge 1$\ fm, {\em schiavit\`u infrarossa}). 
Il potenziale di QED per il positronio (coppia \Pep\ \Pem) dispone invece  
del solo primo termine, $V_{em}=-\frac{\alpha}{r}$.  
\section{La transizione di fase di QCD} 
Nel paragrafo precedente si \`e descritto come, da un punto di vista 
sperimentale, i quark ed i gluoni risultino confinati entro gli adroni,  
per la natura non abeliana dell'interazione forte.  
In questa sezione si vedr\`a come tanto la termodinamica  
statistica di QCD quanto i modelli fenomenologici prevedano la possibilit\`a 
per sistemi estesi di materia nucleare di una transizione verso una fase  
deconfinata, il plasma di quark e gluoni.  
\subsection{Il modello fenomenologico dell' M.I.T.} 
I modelli fenomenologici forniscono a volte buone indicazioni sul comportamento 
di sistemi di molte particelle  
e possono fornire delle stime sulle variabili termodinamiche. 
Essi si basano su alcune assunzioni fisiche semplificatrici ed hanno il vantaggio 
di evitare laboriosi calcoli numerici, che costituiscono spesso il vero limite 
della QCD su reticolo ({\em cfr.} paragrafo 1.3.4). Si considera, in questo 
paragrafo, uno dei primi modelli proposti per descrivere sistemi estesi di 
materia nucleare, noto come ``modello a {\em bag}'' e sviluppato 
dall'M.I.T. negli anni settanta~\cite{BagModel}, che 
fornisce una equazione di stato approssimata del QGP.   
\newline
Nel modello si assume che i quark abbiano massa nulla all'interno 
degli adroni, o meglio delle {\em bag} (``sacchetti''), mentre assumono  
massa infinita al di fuori di esse. La semplice ipotesi fenomenologica che 
giustifica il confinamento dei quark negli adroni prevede che il moto termico 
dei quark venga bilanciato da una pressione di {\em bag} $B$, diretta 
verso l'interno. 
%Tale pressione \`e conseguenza della maggior densit\`a di energia del vuoto 
%all'interno del {\em bag} rispetto all'esterno. 
Nel modello, considerando $N$\ quark di spin 1/2, senza massa e non interagenti,  
contenuti entro una cavit\`a sferica di raggio $R$, si calcola la seguente 
espressione per l'energia del sistema: 
\begin{equation}
E=\frac{2.04 N}{R} + \frac{4\pi}{3} R^3 B. 
\nonumber
\end{equation} 
Il valore di equlibrio del raggio $R$\ si ricava imponenendo 
${\rm d}E / {\rm d}R = 0$, in corrispondenza del quale \`e possibile 
stimare la pressione $B$:
\begin{equation}
B^{1/4}=\left(\frac{2.04 N}{4\pi}\right)^{1/4} \frac{1}{R} . 
\nonumber
\end{equation}
Assumendo come raggio di confinamento $R=0.8$\ fm per un sistema di 3 quark 
all'interno di un barione, si ottiene la costante di pressione di {\em bag} 
$B^{1/4}=205$\ MeV.  La spettroscopia adronica fornisce 
per $B^{1/4}$\ valori compresi tra 145 
%${\rm MeV/fm^3}$~\cite{Hax80} e 235 ${\rm MeV/fm^3}$~\cite{Has81}. 
${\rm MeV}$~\cite{Hax80} e 235 ${\rm MeV}$~\cite{Has81}.  
\newline
Poich\'e la carica di colore totale all'interno di una {\em bag} 
deve risultare nulla, in virt\`u della legge di Gauss per il flusso, e poich\'e 
vi sono tre possibili cariche di colore, il modello prevede che tra le 
{\em bag} adroniche permesse vi siano gli stati $qqq$\ (barioni) e 
$q\bar{q}$ (mesoni), neutri dal punto di vista della carica di colore.   
\newline
All'interno del modello \`e possibile prevedere anche l'esistenza di uno stato 
in equilibrio in cui, in una {\em bag}, un numero elevato di quark e gluoni siano 
liberi e deconfinati ($\rightarrow$\ plasma).  
La densit\`a di energia e la pressione assumono  
la seguente espressione in termini della temperatura e del potenziale 
chimico\footnote{Il potenziale chimico barionico $\mu_B$
(o del quark/antiquark di dato flavour $f$,  $\mu_{f}$) \`e pari
all'energia minima necessaria per aggiungere un barione
(un quark/antiquark) al sistema.   
%per $T=0$. 
Gli anti-quark (anti-barioni) hanno
potenziale chimico opposto a quello dei quark (barioni), ed il valore del
potenziale chimico \`e differente al variare del flavour del quark considerato.}:  
\begin{subequations}
\begin{equation}
\epsilon=\epsilon_{g}(T,\mu) + \epsilon_{q}(T,\mu) + \epsilon_{\bar{q}}(T,\mu) + B
\label{Energy}
\end{equation}
\begin{equation}
P=P_{g}(T,\mu) + P_{q}(T,\mu) + P_{\bar{q}}(T,\mu) - B 
\label{pression}
\end{equation}
\end{subequations}
dove $g$, $q$ e $\bar{q}$\ indicano, rispettivamente, i contributi dovuti al moto 
dei gluoni, dei quark e degli anti-quark e $B$\ \`e la presione di {\em bag}.  
Il segno negativo per $B$\ nell'eq.~\ref{pression} riflette l'instabilit\`a del vuoto 
entro la {\em bag}, che collassa se non \`e sostenuta dalla pressione dei costituenti 
al suo interno.  Un limite inferiore per 
% la stabilit\`a 
l'esistenza 
dello stato deconfinato 
\`e quindi ottenibile ponendo $P=0$\ nell'eq.~\ref{pression}. I valori critici di 
temperatura e potenziale chimico in corrispondenza dei quali ha luogo la transizione 
di fase si possono quindi ottenere a partire dall'equazione: 
\begin{equation}
P_{g}(T,\mu) + P_{q}(T,\mu) + P_{\bar{q}}(T,\mu) = B
\label{Stability}
\end{equation}
Nell'ipotesi che le interazioni tra i costituenti del plasma siano trascurabili, e che 
le masse siano nulle, le densit\`a di energie per le diverse componenti valgono:  
\begin{subequations}
\begin{equation}
\epsilon_{g} = n_g \int \frac{{\rm d}^3 p}{(2 \pi)^3} \cdot p 
          \left(\frac{1}{e^{p/T}-1} \right)=  n_g \frac{\pi^2}{30}T^4
\end{equation}
\begin{equation}
\begin{split}
\epsilon_{q} + \epsilon_{\bar{q}} = & n_q \sum_f\int 
        \frac{{\rm d}^3 p}{(2 \pi)^3} \cdot p
        \left(\frac{1}{e^{(p-\mu_f)/T}+1} + \frac{1}{e^{(p+\mu_f)/T}+1} \right) = \\
	  & n_q \sum_f\left( \frac{7\pi^2}{120}T^4 + \frac{1}{4}\mu_f^2T^2 + 
	\frac{1}{8\pi^2}\mu_f^4 \right)
\end{split}
\end{equation}
\end{subequations}
dove $n_g=8 \times 2$\ ed $n_q = 3\times2$\ sono il numero di gradi di libert\`a 
dei gluoni e dei quark, la cui massa possa esser trascurata rispetto all'energia 
termica del sistema (8 gluoni ciascuno con 2 stati di elicit\`a, i 3 stati di colore 
con 2 stati di spin per ogni quark ed antiquark di {\em flavour} $f$). 
L'equazione di stato del sistema (eq.~\ref{Energy}) diventa dunque: 
\begin{equation}
\epsilon=\frac{8}{15} \pi^2 T^4 + 3 \sum_f\left( \frac{7}{60}\pi^2 T^4 + 
        \frac{1}{2}\mu_f^2T^2 + \frac{1}{4\pi^2}\mu_f^4 \right) + B
\end{equation}
e, poich\'e per particelle di massa nulla $\epsilon=3 P $\ \`e anch'esso 
nullo, la relazione di stabilit\`a~\ref{Stability}, nel limite inferiore 
di pressione nulla, si riscrive come:
\begin{equation}
\frac{8}{45} \pi^2 T^4 + \sum_f\left( \frac{7}{60}\pi^2 T^4 +
        \frac{1}{2}\mu_f^2T^2 + \frac{1}{4\pi^2}\mu_f^4 \right) = B.
\label{Stability2}
\end{equation}
Si ottiene dunque una relazione tra $\mu$\ e $T$\ che fornisce tutti i possibili 
valori critici per la transizione di fase dallo stato confinato a quello 
deconfinato. 
Per valori di $(\mu,T)$\ superiori a quelli critici  $(\mu_c,T_c)$\ i quark ed 
i gluoni possono esistere nella fase di plasma,
%il plasma di quark e gluoni \`e stabile, 
mentre per valori inferiori 
%il sistema \`e instabile 
ci\`o non \`e possibile 
e la materia nucleare si trova confinata entro {\em bag} contenenti 
coppie $q\bar{q}$ (mesoni) o tripletti $qqq$ (barioni). 
%\newline
%\`E interessante osservare come sia possibile oltrepassare, nel piano $(\mu,T)$, 
%la curva critica data dall'eq.~\ref{Stability2} o riscaldando il sistema ($T$\ 
%elevato) o comprimendolo ($\mu$ elevato). Nel primo caso, la pressione dei 
%costituenti del plasma, necessaria per bilanciare la pressione del vuoto $B$, \`e 
%fornita dal moto di agitazione termica, nel secondo proviene dalla degenerazione 
%delle diverse componenti fermioniche del gas.   
%
\subsection{Il diagramma di fase della materia nucleare}  
\begin{figure}[hbt]
\begin{center}
{\bf a)} \hspace{8.0cm} {\bf b)} \\
\hspace{-1.0cm}
\includegraphics[scale=0.37]{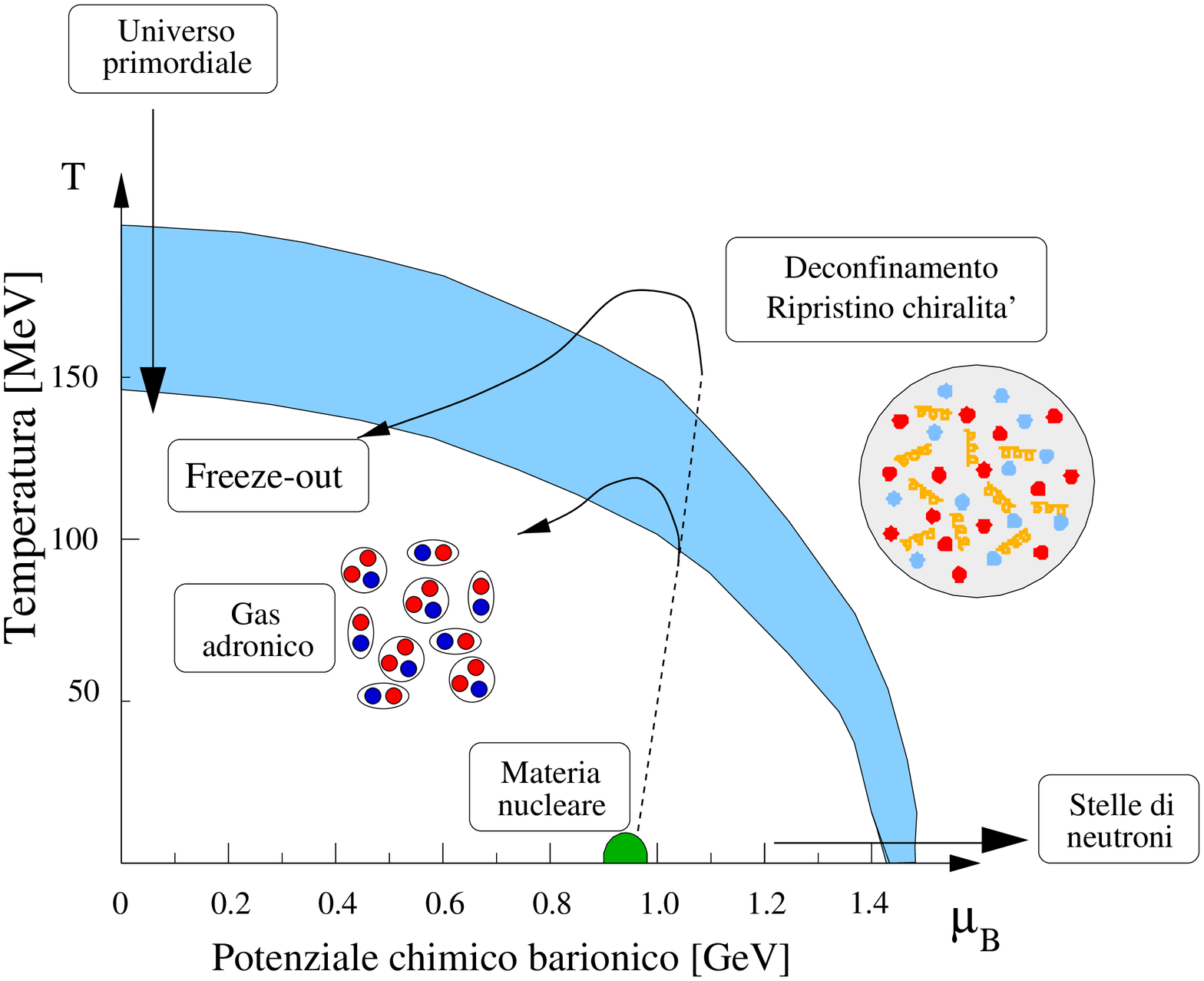}    
\hspace{0.3cm}
\includegraphics[scale=0.28]{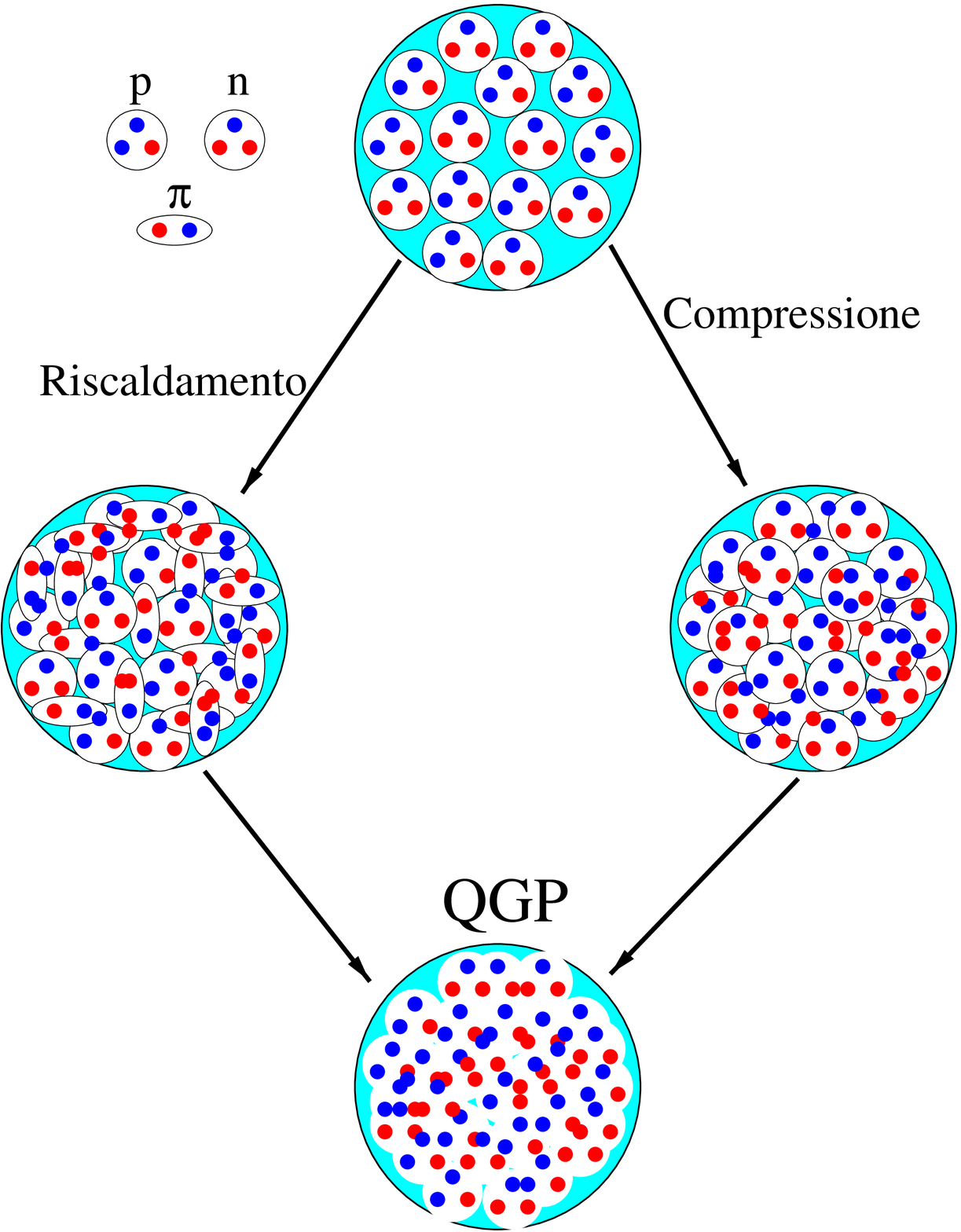}
\caption{ {\bf a)} Diagramma di fase della materia nucleare. Le  tre
       frecce indicano l'evoluzione seguita dall'universo dopo il 
       {\em Big Bang}, dai nuclei pesanti che collidono agli acceleratori, 
       da una {\em supernova} che esplode diventando una stella  di neutroni.  
       {\bf b)} Condizioni favorevoli per la formazione del QGP.  
       Con il {\em riscaldamento}, il contenuto barionico non aumenta, ma 
       la densit\`a di quark \`e incrementata dalla produzione di mesoni. 
       Con la {\em compressione}, aumenta la densit\`a barionica e, di 
       conseguenza, il potenziale chimico barionico.}  
\label{RiscaldaComprimi}
\end{center}
\end{figure}

In fig.~\ref{RiscaldaComprimi}.a \`e mostrato schematicamente il diagramma di 
fase della materia nucleare nel piano $(\mu_B,T)$, dove $T$\ \`e la temperatura 
del sistema e $\mu_B$\ il potenziale chimico barionico\footnote{Vedi nota 1.}, 
parametro che regola la densit\`a barionica del sistema.  
A bassi valori di temperatura e di potenziale chimico barionico la materia 
si trova nello stato confinato (stato adronico), le elevate 
temperature e/o gli elevati potenziali chimici barionici 
corrispondono invece alla fase deconfinata del plasma.  
In particolare, a basse temperature e per $\mu_B \simeq 1$\ si trova la materia 
nucleare stabile. Aumentando l'energia del sistema, per {\em compressione} o 
per {\em riscaldamento}, si raggiunge una fase in cui i nucleoni interagiscono 
anelasticamente. In questa fase, chiamata {\em gas adronico}, si trovano   
all'equilibrio termico anche specie adroniche non originariamente presenti 
nel nucleo, ad esempio pioni o stati eccitati del nucleone, quali la $\Delta$. 
Riscaldando o comprimendo ulteriormente il sistema, la densit\`a di partoni 
aumenta fino al punto che il concetto di appartenenza di un quark ad un adrone 
perde senso (fig.~\ref{RiscaldaComprimi}.b). In queste condizioni la 
materia nucleare non \`e pi\`u confinata e si ha il passaggio alla fase di 
plasma.   
\newline
\`E utile sottolineare come sia possibile oltrepassare, nel piano $(\mu,T)$,
la curva critica data dall'eq.~\ref{Stability2} o {\em riscaldando} il sistema ($T$\
elevato) o {\em comprimendolo} ($\mu$ elevato):  
%come mostrato in  fig.~\ref{RiscaldaComprimi}.b. 
nel primo caso, la pressione dei  
costituenti del plasma, necessaria per bilanciare la pressione del vuoto $B$, \`e
fornita dal moto di agitazione termica, nel secondo proviene dalla degenerazione
delle diverse componenti fermioniche del gas.  
%\`E interessante notare come la transizione dalla fase confinata a quella 
%deconfinata possa avvenire sia per {\em riscaldamento} che per 
%{\em compressione} della materia nucleare (fig.~\ref{RiscaldaComprimi}.b).  
\newline
Sono stati individuati almeno due sistemi in cui vi sono ottimi presupposti 
per ritenere che la materia nucleare 
si sia trovata o si trovi nella fase deconfinata di QGP.  
%abbia effettuato la transizione di fase dalla stato confinato 
%(fase adronica) a quello deconfinato (fase di QGP), o viceversa. 
Il primo sistema \`e costituito dall'intero 
universo nei suoi primi istanti di vita, dopo circa $10^{-5} sec$\ dal 
{\em Big Bang}; in tal caso si pensa che la materia nucleare si 
trovasse in uno stato di QGP caratterizzato da una densit\`a 
barionica pressocch\'e nulla e da 
una elevatissima temperatura; la transizione verso la fase confinata 
\`e quindi avvenuta per raffreddamento a seguito dell'espansione 
dell'universo. 
Lo studio di transizioni con caratteristiche simili, cio\`e prodotte 
per intenso riscaldamento della materia nucleare a bassa densit\`a 
barionica, assume estrema importanza in vari aspetti della Cosmologia, 
in particolare per quanto riguarda la nucleosintesi, la materia oscura 
e la struttura a larga scala dell'universo~\cite{Sch86}. 
\newline
Il secondo sistema, in tal caso di rilevanza astrofisica, \`e costituito 
dalle stelle di neutroni, formatesi in seguito all'esplosione di 
una {\em supernova}. La transizione evolve in tal caso nell'altra 
direzione, dallo stato adronico verso quello di QGP, a bassa temperatura,  
e come conseguenza della compressione cui \`e sottoposta la regione  
pi\`u interna di una stella di neutroni dopo il suo collasso 
gravitazionale. L'interno delle stelle di neutroni potrebbe dunque essere   
costituito da un plasma freddo di quark e gluoni ad elevata densit\`a 
barionica (e quindi con prevalenza di quark $u$\ e $d$, preesistenti dalla fase 
adronica). Surge ulteriore interesse in tale sistema, in quanto la dinamica 
delle esplosioni di supernova, ed in particolare le instabilit\`a di tipo 
idrodinamico, dipendono dalla compressibilit\`a, e quindi dall'equazione 
di stato della materia nucleare~\cite{Gle91}. 
\newline
Le evoluzioni delle due transizioni, qui sommariamente descritte,  sono 
rappresentate in fig.~\ref{RiscaldaComprimi}.a da due frecce il cui 
verso indica l'evoluzione temporale del sistema, durante la transizione. 
\newline
Nella fig.~\ref{RiscaldaComprimi}.a \`e anche schematizzata la 
transizione di fase cui si spera di sottoporre la materia costituente 
i nuclei ultra-relativistici che collidono agli acceleratori 
di particelle. In tal caso si ha inizialmente materia nucleare che, 
in seguito all'energia rilasciata nell'urto, si ritrova nello stato di 
QGP per quindi descrivere la transizione nuovamente verso la fase adronica, 
dove si trovava prima dell'urto. La transizione dal QGP verso il gas 
adronico \`e conseguenza in tal caso sia del raffreddamento che della 
decompressione (con diminuzione della densit\`a barionica, cui 
corrisponde una diminuzione del potenziale chimico barionico).  
Come si esporr\`a nei prossimi paragrafi, le transizioni attesa 
alle energie dei collisionatori RHIC ed LHC hanno caratteristiche che 
pi\`u si avvicinano a quelle della transizione dell'universo primordiale, 
essendo caratterizzate da elevati valori di temperatura e basse 
densit\`a barioniche. Il tipo di transizione che si ritiene di  
raggiungere agli acceleratori SPS ed AGS, \`e invece pi\`u prossimo alla 
transizione delle stelle di neutroni, con densit\`a barionica pi\`u 
elevata e temperatura inferiore alle precedenti.  
\newline
La densit\`a barionica della materia nucleare \`e legata al valore del 
potenziale chimico $\mu$\ dalla relazione:
\begin{equation}
\rho_B = \frac{1}{3}(N_q-N_{\bar{q}}) = \frac{n_q}{3}\sum_{f} \int 
          \frac{{\rm d}^3 p}{(2 \pi)^3} \cdot 
	  \left(\frac{1}{e^{(p-\mu_f)/T}+1} - \frac{1}{e^{(p+\mu_f)/T}+1} \right)
\label{Bariondensity}
\end{equation}
dove $N_q$\ ed $N_{\bar{q}}$\ sono, rispettivamente, le densit\`a di quark ed 
anti-quark del sistema, $n_q=3 \cdot 2$\ \`e pari al numero di degenerazione 
per la carica di colore e per lo spin e la sommatoria sui {\em flavour} pu\`o 
essere estesa al solo numero di quark la cui massa sia trascurabile rispetto
allo stato termico del sistema (tipicamente $u$\ e $d$;  eventualmente anche $s$). 
Calcolando gli integrali, l'eq.~\ref{Bariondensity} diventa:  
\begin{equation}
\rho_B = \sum_{f} \left(\frac{\mu_f}{3}T^2 + \frac{\mu_f^3}{3\pi^2} \right)
\label{Bariondensity2}
\end{equation}
A partire dalle equazioni~\ref{Stability2} e~\ref{Bariondensity2} \`e possibile 
stimare in termini di densit\`a barionica e temperatura i valori critici 
per la transizione di fase. 
Nel caso di formazione di QGP per sola compressione, cio\`e per $T=0$, il valore 
critico di densit\`a barionica risulta pari a $0.5 \div 1.5 \, {\rm nucleoni/fm^3}$, 
cio\`e da 5 a 10 volte superiore alla normale densit\`a barionica 
della materia nucleare ($\rho_B = 0.14 \,{\rm nucleoni/fm^3}$). 

\subsection{Collisioni nucleari ultra-relativistiche}
\subsubsection{Caratteristiche generali}
\begin{figure}[htb]
\begin{center}
\includegraphics[scale=0.40]{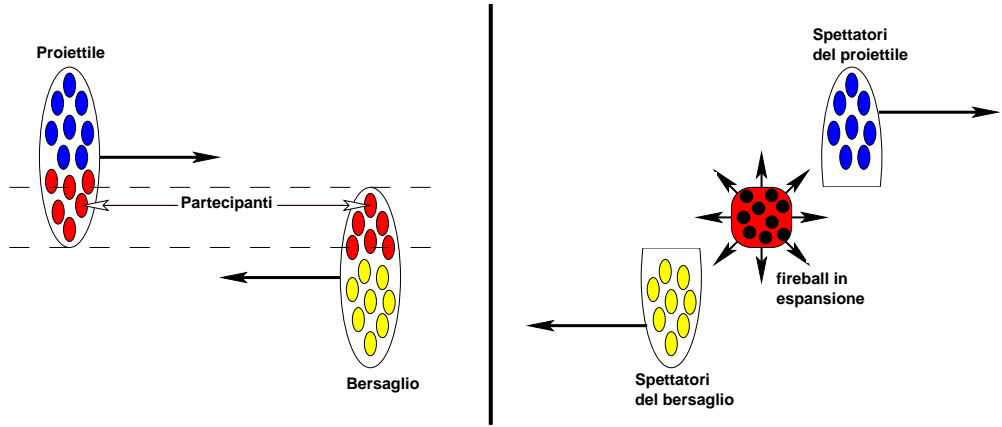}
\caption{Schema di una collisione tra due nuclei nel sistema 
         di riferimento del centro di massa. A sinistra \`e mostrata 
	 la situazione prima della collisione; dopo la collisione 
	 (a destra) i nucleoni spettatori del proiettile e del bersaglio
	 continuano il loro moto sostanzialmente imperturbati, mentre 
	 i partecipanti formano un sistema fortemente eccitato chiamato 
	 {\em fireball}.}
 \label{Collisione}
 \end{center}
\end{figure}

Ad energie ultra-relativistiche, le lunghezze d'onda di {\em De~Broglie}
del nucleo e dei nucleoni sono molto piccole rispetto alle tipiche  
dimensioni nucleari; quindi gli ioni pesanti, ed i nucleoni che li 
costituiscono, si comportano, con ottima approssimazione, come particelle 
classiche.  
Inoltre, poich\'e il raggio d'azione dell'interazione tra nucleoni 
($1 \div 2 $\ fm) 
\`e pi\`u piccolo rispetto alla dimensione dei nuclei, semplici  
modelli geometrici permettono di fornire una buona descrizione nel 
calcolo  di grandezze legate alla centralit\`a di una collisione. 
\newline
Le collisioni tra i nuclei possono quindi essere caratterizzate 
dal parametro di impatto ${\bf b}$, che \`e definito come il vettore 
congiungente i centri dei due nuclei incidenti nel piano trasverso 
all'asse di collisione. Poich\'e normalmente il bersaglio ed il proiettile 
non sono n\`e deformati n\`e polarizzati, la centralit\`a della collisione 
\`e determinata dal modulo del parametro di impatto, $b=|{\bf b}|$. 
Le collisioni pi\`u centrali hanno dunque $b\simeq0$\ e quelle pi\`u 
periferiche $b\simeq R_{proiet} + R_{bers}$.  
Come mostrato in fig.~\ref{Collisione}, solo i nucleoni che si trovano nella 
regione di sovrapposizione geometrica dei nuclei possono interagire nella 
collisione; essi vengono pertanto chiamati nucleoni {\em partecipanti} alla 
collisione.  
Gli altri nucleoni continuano invece lungo la loro traiettoria senza subire 
interazioni e vengono quindi indicati come {\em spettatori} della collisione. 
\newline
Nelle collisioni centrali, in cui il numero di {\em spettatori} viene minimizzato, 
risulta maggiore il trasferimento dell'originale energia longitudinale 
in energia di eccitazione della materia nucleare interagente,  e quindi pi\`u 
favorevoli le condizioni per la transizione di fase nel QGP.  
\newline
Come si esporr\`a nei prossimi paragrafi, la selezione degli urti centrali avviene 
misurando osservabili fisiche correlate (o anti-correlate) al numero di  
{\em partecipanti} ed alla densit\`a di energia raggiunta nella regione centrale, 
quali la molteplicit\`a di particelle secondarie prodotte e l'energia emessa 
in direzione trasversa  (o l'energia residua misurata in avanti lungo la linea 
di fascio). 
\newline
Per parametrizzare il moto delle particelle nella direzione longitudinale conviene 
introdurre la variabile adimensionale {\em rapidit\`a}, definita a partire dall'energia 
$E$\ e dall'impulso longitudinale $p_L$ di una particella: 
\begin{equation}
y=\frac{1}{2} \ln \frac{E+p_L}{E-p_L}. 
\label{Rapidity}
\end{equation}
Il vantaggio di tale variabile cinematica risiede nella sua propriet\`a di 
trasformarsi in maniera additiva per {\em ``boost''} di Lorentz longitudinali: 
\begin{equation}
y=y' + y_{\Sigma'} , 
\nonumber
\end{equation}
dove $y'$\ \`e la rapidit\`a della particella nel sistema $\Sigma'$\ che  
si muove con rapidit\`a $y_{\Sigma'}$\ lungo l'asse longitudinale del sistema $\Sigma$.
Gli {\em intervalli} di rapidit\`a sono quindi invarianti per tali trasformazioni 
del sistema di riferimento, e le distribuzioni di rapidit\`a assumono la stessa 
forma sia nel sistema laboratorio che in quello del centro di massa (CMS), venenedo 
solo traslate della quantit\`a 
\begin{equation}
y_{CM}=\frac{1}{2} \ln \frac{1+\beta_{CM}}{1-\beta_{CM}} \simeq \ln 2\gamma_{CM}, 
\nonumber
\end{equation}
dove $\beta_{CM}$\ e $\gamma_{CM}$\ sono, rispettivamente, la velocit\`a ed il 
fattore di Lorentz del sistema centro di massa rispetto al sistema laboratorio. 
Nel limite ultra-relativistico ($E \gg m$), la rapidit\`a pu\`o essere approssimata 
dalla pseudo-rapidit\`a 
\begin{equation}
\eta=\frac{1}{2} \ln \frac{p+p_L}{p-p_L}=-\ln[\tan(\theta/2)], 
\label{PseudoRapidity}
\end{equation}
dove $\theta$\ \`e l'angolo tra l'impulso della particella ${\bf p}$\ e 
l'asse del fascio, 
espressione usata quando non \`e possibile identificare la particella.  
\newline
Nel CMS, la regione centrale di interazione corrisponde a valori di 
rapidit\`a intorno allo zero, poich\'e i {\em partecipanti} hanno 
perso l'impulso longitudinale originario, mentre i frammenti provenienti 
dal bersaglio e dal proiettile assumono valori opposti di rapidit\`a, in 
corrispondenza dei valori di rapidit\`a del bersaglio e del proiettile prima 
dell'urto. 
Nel sistema laboratorio, la regione centrale corrisponde a valori centrati attorno 
a $y_{CM}$. 
\newline
%I valori di rapidit\`a propri della zona di frammentazione del bersaglio e del 
%proiettile possono essere ricavati a partire dalle relazioni: 
I valori di rapidit\`a di una particella, ed in particolare quelli 
della zona di frammentazione del bersaglio o del proiettile,  
possono essere ricavati a partire dalle relazioni:   
\begin{align}
 &  \sinh y = \frac{p_L}{m_T}  \nonumber \\
 &  \cosh y = \frac{E}{m_T},   \nonumber
\end{align}
dove $m_T=\sqrt{p_T^2 + m^2} $\ \`e la cosiddetta massa trasversa delle particelle.  
\subsubsection{Modello di Glauber} 
Il modello di Glauber delle collisioni nucleari~\cite{Glauber}\cite{Wong} 
\cite{Carrer18}\ si basa sull'ipotesi che i nucleoni di ciascun nucleo 
possano subire pi\`u di una collisione anelastica con i nucleoni 
dell'altro nucleo. Il modello \`e basato sul concetto di cammino libero 
medio e sull'assunzione che i nucleoni interagiscano con la sezione d'urto 
elementare nucleone--nucleone. Nell'urto tra i nuclei, in seguito alle collisioni 
anelastiche, i nucleoni vengono eccitati; la frammentazione e la conseguente 
produzione di 
adroni avvengono, nel sistema di riferimento del nucleone, dopo un tempo 
$  \tau_{prod} $, detto tempo di formazione. 
Le stime teoriche di $ \tau_{prod} $  forniscono un valore compreso tra 
$0.4$\ e  $1.2\,{\rm fm/c}$\ \cite{Wong19}. 
Entro $  \tau_{prod} $, il nucleone eccitato ({\em wounded nucleon}), 
nell'attraversare il nucleo, 
pu\`o subire altre collisioni anelastiche con gli altri nucleoni. 
Nel modello di Glauber si assume che il nucleone eccitato prosegua lungo 
la traiettoria iniziale ed interagisca con la stessa sezione d'urto del 
nucleone incidente. Tale assunzione \`e giustificata dal fatto che la 
sezione d'urto anelastica 
%non diffrattiva 
nucleone--nucleone, 
$ \sigma_{in} $, dipende debolmente dall'energia, per energie nel  
centro di massa $ \sqrt{s} > 3\,{\rm GeV} $, e vale circa 30 mb. Inoltre, 
essendo interessati a processi di produzione di particelle, si considera 
la sezione d'urto anelastica $ \sigma_{in} $\ pari alla sola sezione 
d'urto non diffrattiva. \\
In appendice A si mostra come il modello permetta di calcolare il numero medio 
di collisioni binarie tra i nucleoni, $<N_{coll}>$, ed il numero medio di 
nucleoni partecipanti all'interazione, $<N_{part}>$, in funzione del 
parametro d'impatto, per una data 
distribuzione di densit\`a $ \rho $\ e sezione d'urto anelastica.\\
Come si vedr\`a nei prossimi paragrafi, diversi esperimenti hanno 
mostrato che le osservabili che permettono di determinare la centralit\`a 
di una collisione, quali la molteplicit\`a di particelle cariche 
prodotte o l'energia trasversa della collisione~\footnote{Anche 
l'energia depositata in avanti ad angolo zero entro un calorimetro permette 
di determinare la centralit\`a delle collisioni, ma questa osservabile 
\`e anticorrelata con le precedenti: le collisioni pi\`u centrali corrispondono 
in tal caso ad un piccolo rilascio di energia in avanti, quelle periferiche 
ad elevati valori di tale variabile.},  
%risultino proporzionali 
%(od anti-proporzionali, nel caso dell'energia depositata in avanti) 
%al numero di nucleoni partecipanti piuttosto che al numero di 
%collisioni binarie.  
dipendono dal numero di nucleoni partecipanti secondo una legge di potenza,  
del tipo $<N_{part}>^\alpha$, compatibile o comunque 
molto prossima ad una legge lineare ($\alpha \simeq 1$),  piuttosto che 
dal numero di collisioni binarie, nel 
qual caso la dipendenza  $<N_{coll}>^{\alpha '}$\ si discosta maggiormente 
dalla linearit\`a (i.e. $|\alpha' - 1| > |\alpha - 1| $). 
\newline
Mentre la dipendenza lineare dal numero di collisioni binarie \`e attesa 
in uno scenario di sovrapposizione di collisioni nucleone-nucleone, con 
pur possibili deviazioni dovute agli effetti dello stato iniziale, quella 
dai nucleoni partecipanti \`e pi\`u facilmente associabile ad un sistema 
con intensa diffusione nello stato finale, in cui le particelle entranti 
perdono il ricordo della loro provenienza  e ciascun partecipante 
contribuisce alla produzione delle particelle con circa lo stesso rilascio 
di energia.    
\subsection{Risultati della QCD su reticolo}
\begin{figure}[hbt]
\begin{center}
\includegraphics[scale=0.60]{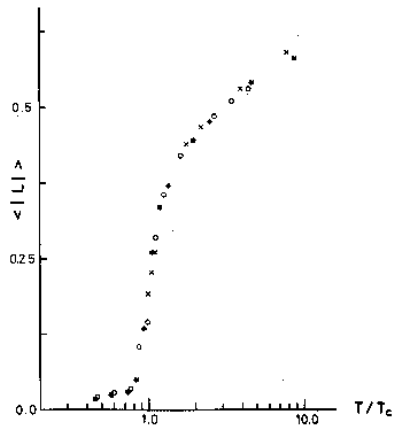}
\caption{Andamento del parametro d'ordine $<L>$\ in funzione del rapporto 
         $T/T_{C}$~\cite{Sat84}.}
\label{trovami}
\end{center}
\end{figure}

\begin{figure}[p]
\begin{center}
\includegraphics[scale=0.40]{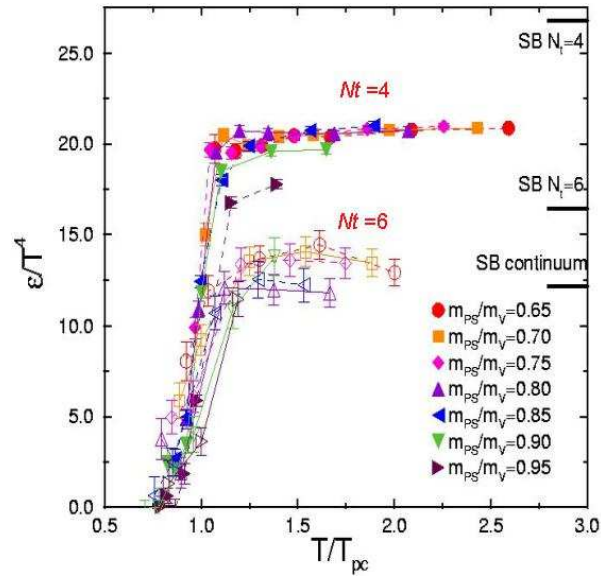}
 \caption{Risultati di un calcolo di QCD su reticolo per la densit\`a di energia 
          in funzione della temperatura del sistema, per diverse 
	  dimensioni del reticolo e per diversi valori delle masse dei 
	  quark~\cite{Condensate}. 
	  I segmenti di linea a tratto pieno rappresentano il limite 
	  statistico del gas pefetto.}
\label{Transit}
\end{center}
\end{figure}

\begin{figure}[p]
\begin{center}
{\bf a)} \hspace{8.0cm} {\bf b)} \\
\hspace{-1.2cm}
\includegraphics[scale=0.60]{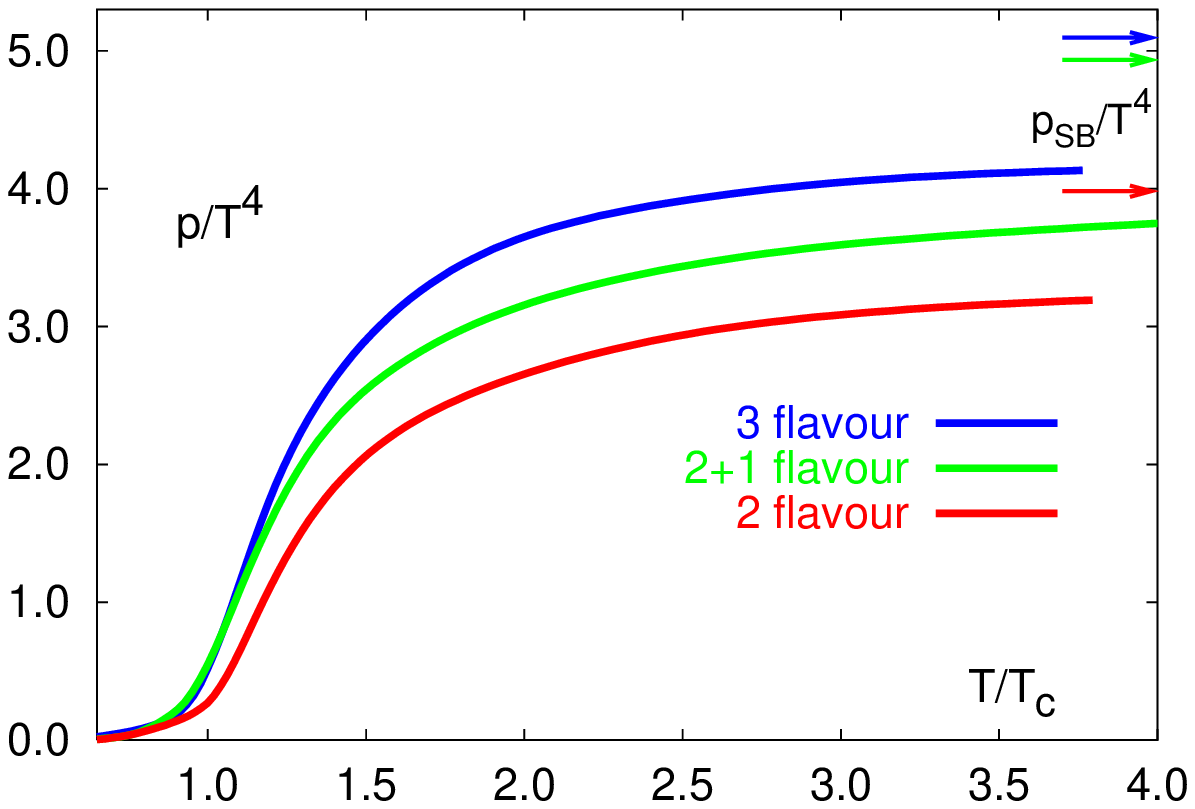}
\includegraphics[scale=0.60]{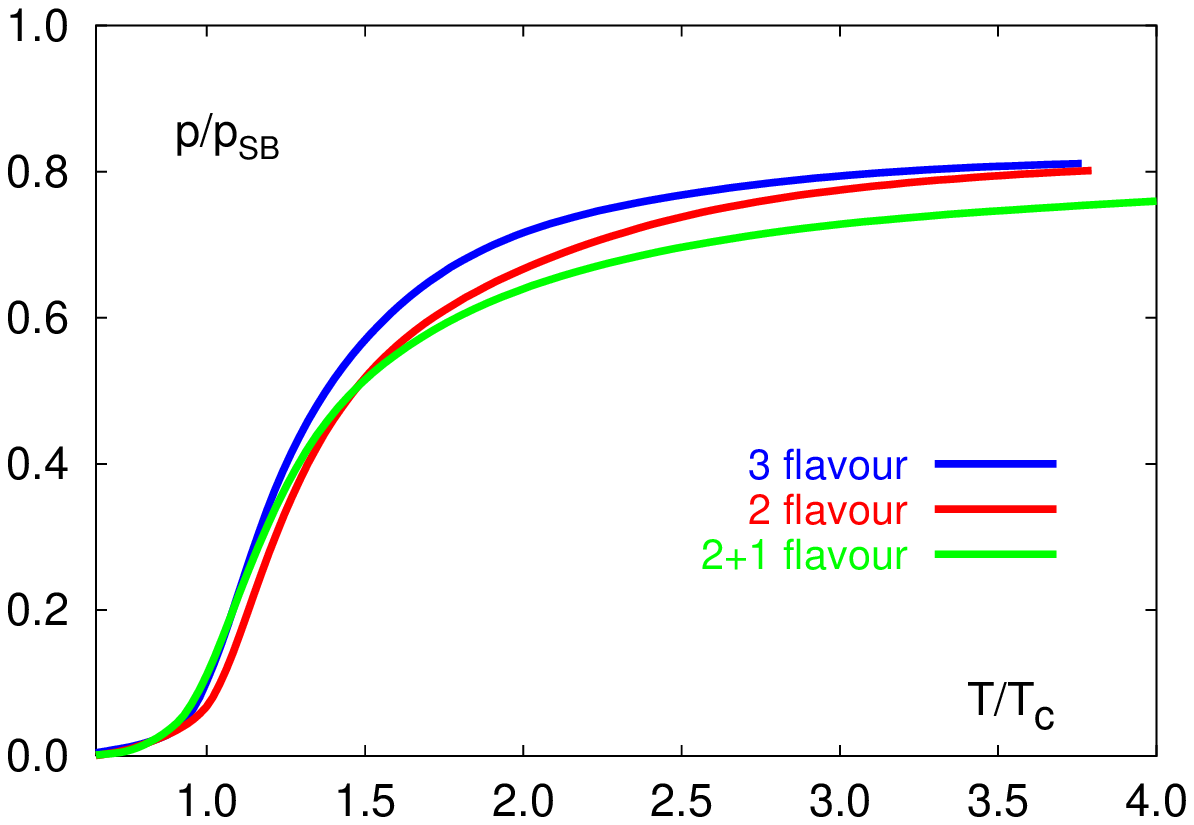}
 \caption{{\bf a)} Risultati di un calcolo di QCD su reticolo per il parametro 
          d'ordine $P/T^4$ per diverse ipotesi sul numero di {\em flavour}.
	  {\bf b)} Dipendenza dalla temperatura del rapporto tra i valori di  pressione 
	  calcolati col reticolo e la pressione del gas ideale di quark e gluoni.}
\label{Lattice}
\end{center}
\end{figure}

Per la materia ordinaria, il fenomeno della transizione di fase \`e 
frequente ed \`e studiato mediante un'approccio statistico-termodinamico. 
Nel caso specifico della materia nucleare, l'approccio statistico deve 
essere formulato all'interno della teoria che descrive l'interazione tra 
i suoi costituenti, vale a dire la QCD.   
Il regime perturbativo di libert\`a asintotica pu\`o determinare le prime 
fasi dell'evoluzione della collisione e fissare le condizioni iniziali 
per la formazione del plasma. 
Tuttavia i quark ed i gluoni interagiscono 
su grandi distanze durante la fase di equilibrazione, di espansione e di 
adronizzazione. Per queste fasi, un trattamento in termini di QCD perturbativa 
\`e inadeguato.  
\newline 
Una trattazione non perturbativa della QCD statistica \`e possibile mediante 
la formulazione di una teoria di gauge su un reticolo rappresentante le 
coordinate spazio-temporali~\cite{Wil74}. La discretizzazione dello 
spazio-tempo permette di eliminare le divergenze che nascono nella QCD in 
seguito all'integrazione dei diagrammi di Feynman per elevati impulsi 
({\em divergenze ultraviolette}). In regime perturbativo tali divergenze sono 
eliminate nella procedura di rinormalizzazione, mentre nella QCD su reticolo 
il limite inferiore della distanza tra i punti del reticolo (il passo) limita 
superiormente la variabile impulso, rendendo finiti gli integrali. Inoltre 
la formulazione su reticolo consente di scrivere la funzione di partizione del 
sistema termodinamico come un integrale di percorso\footnote{ 
In questa espressione $t=i\tau$\ \`e un tempo immaginario e viene utilizzata la 
notazione $\int[dx] \equiv \int\prod_{i=1}^{n_t}dx_i$.}   
\begin{equation}
Z=\sum_{x_a}<x_a|e^{-H/kT}|x_a> = \int [dx]e^{-S(x)}, 
\label{partition}
\end{equation}
dove $S(x)=\int^{\tau_b}_{\tau_a}{L(x,t){\rm d}\tau}$\ 
%($t=i\tau$\ \`e un tempo immaginario)
\`e l'azione associata ad un certo percorso ed $e^{S(x)}$\ la 
sua ampiezza di probabilit\`a. In tal modo \`e  
possibile l'utilizzo di metodi Monte Carlo nel calcolo 
delle variabili termodinamiche del sistema, come la 
densit\`a di energia o la pressione, 
ricavabili a partire da $Z$:
\begin{subequations}
 \begin{equation}
  \epsilon=\frac{T^2}{V}\left(\frac{\partial \ln Z}{\partial T}\right)_V
  \label{Energia}
 \end{equation}
 \begin{equation}
  P=T\left(\frac{\partial \ln Z}{\partial V}\right)_T
  \label{pressione}
  \end{equation} 
\end{subequations}
Per un gas ideale relativistico di quark di massa nulla e di gluoni, servendosi  
delle statistiche di Fermi-Dirac e di Bose-Einstein, si ottiene la seguente 
espressione per la pressione:
\begin{equation}
P=\left[ n_g+\frac{7}{8}(n_q+n_{\bar{q}})\right]  \frac{\pi^2}{90}T^4 ,
\label{IdealGas}
\end{equation}
dove $n_g$, $n_q$\ ed $n_{\bar{q}}$\ rappresentano il grado di degenerazione 
dei gluoni, dei quark e degli anti-quark, rispettivamente. 
Per gli otto gluoni, vi sono due possibili stati di elicit\`a, pertanto 
$n_g=N_g \cdot N_{pol}=$\ $8\times2$. Per i quark bisogna considerare il 
numero dei possibili stati di colore $N_C$, di spin $N_S$\ ed  
i diversi {\em flavour} $N_f$. Per questi ultimi ha senso considerare i soli 
quark $u$, $d$\ ed $s$, che hanno masse dello stesso ordine od inferiori 
alla temperatura del sistema ({\em cfr.} prossimo paragrafo); pertanto 
risulta $n_q=n_{\bar{q}} = N_C \cdot N_S \cdot N_f = 3 \times 2 \times 3 = 18 $.
\newline
Si ricorda inoltre che per un gas ideale la pressione e la densit\`a di 
energia sono legate dalla relazione $P=\frac{\epsilon}{3}$.  
\newline
Le transizioni di fase vengono generalmente esaminate studiando alcuni 
``parametri d'ordine'', vale a dire variabili termodinamiche che sono 
nulle in una fase del sistema e diverse da zero nell'altra. Nello studio 
della transizione di fase di QCD, \`e conveniente usare il valore di 
aspettazione dell'operatore di linea di Wilson $<L>$~\cite{Sat84}. Esso \`e 
legato all'energia libera $F$, indotta dalla presenza di un quark statico 
posto entro un bagno termico di gluoni a temperatura $T$ ($<L>\sim e^{-F/T}$). 
Nella materia adronica confinata, la carica di colore non viene schermata, per 
cui la sua energia libera \`e infinita ed $<L>=0$. Nella fase deconfinata, invece, 
il quark \`e schermato dal mezzo gluonico, la sua energia libera risulta finita ed 
$L$\ diventa diverso da zero. 
In fig.~\ref{trovami} \`e mostrato il risultato di un calcolo di QCD per l'andamento 
del parametro d'ordine $<L>$\ in funzione della temperatura. 
A basse temperature, il valore nullo del parametro indica che i quark si trovano 
nella usuale fase confinata. In corrispondenza di un valore critico di 
temperatura $T_c$, pari a circa 150 MeV, il sistema si porta in una fase 
deconfinata.   
\newline
Altre variabili termodinamiche utili per studiare la transizione di fase sono la 
la densit\`a di energia e la pressione. In fig.~\ref{Transit} 
(fig.~\ref{Lattice}.a) si vede una 
chiara transizione di fase nel parametro d'ordine $\epsilon/T^4$\ ($P/T^4$) alla 
temperatura critica $T_C$. La densit\`a di energia (la pressione) cresce 
improvvisamente dal piccolo valore proprio della fase adronica ad un valore 
vicino a quello previsto per un gas ideale di quark e gluoni. 
Questo riflette l'aumento dei gradi di libert\`a tra la fase adronica  
(per esempio, per un gas di pioni: $n_\pi=3$) ed il QGP 
(per esempio, per un plasma di gluoni e quark $u$\ e $d$: $n_{QGP} =36$). 
N\`e la densit\`a di energia n\`e la pressione raggiungono il limite termodinamico, 
come si envice in dettaglio dalla fig.~\ref{Lattice}.b. 
 
\begin{figure}[htb]
\begin{center}
 {\bf a)} \hspace{8.0cm} {\bf b)} \\
 \includegraphics[scale=0.40]{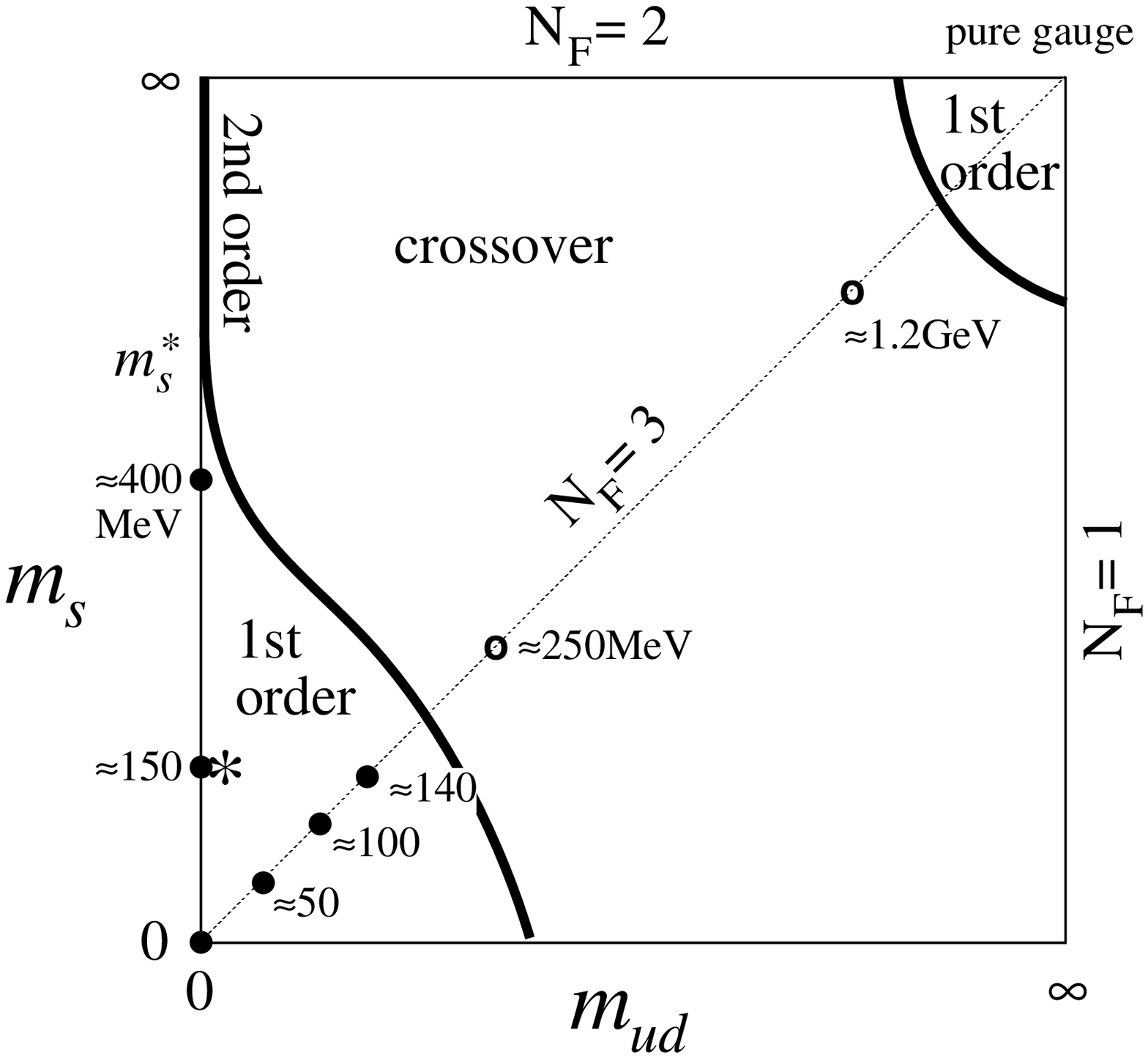}
 \hspace{0.9cm}
 \includegraphics[scale=0.47]{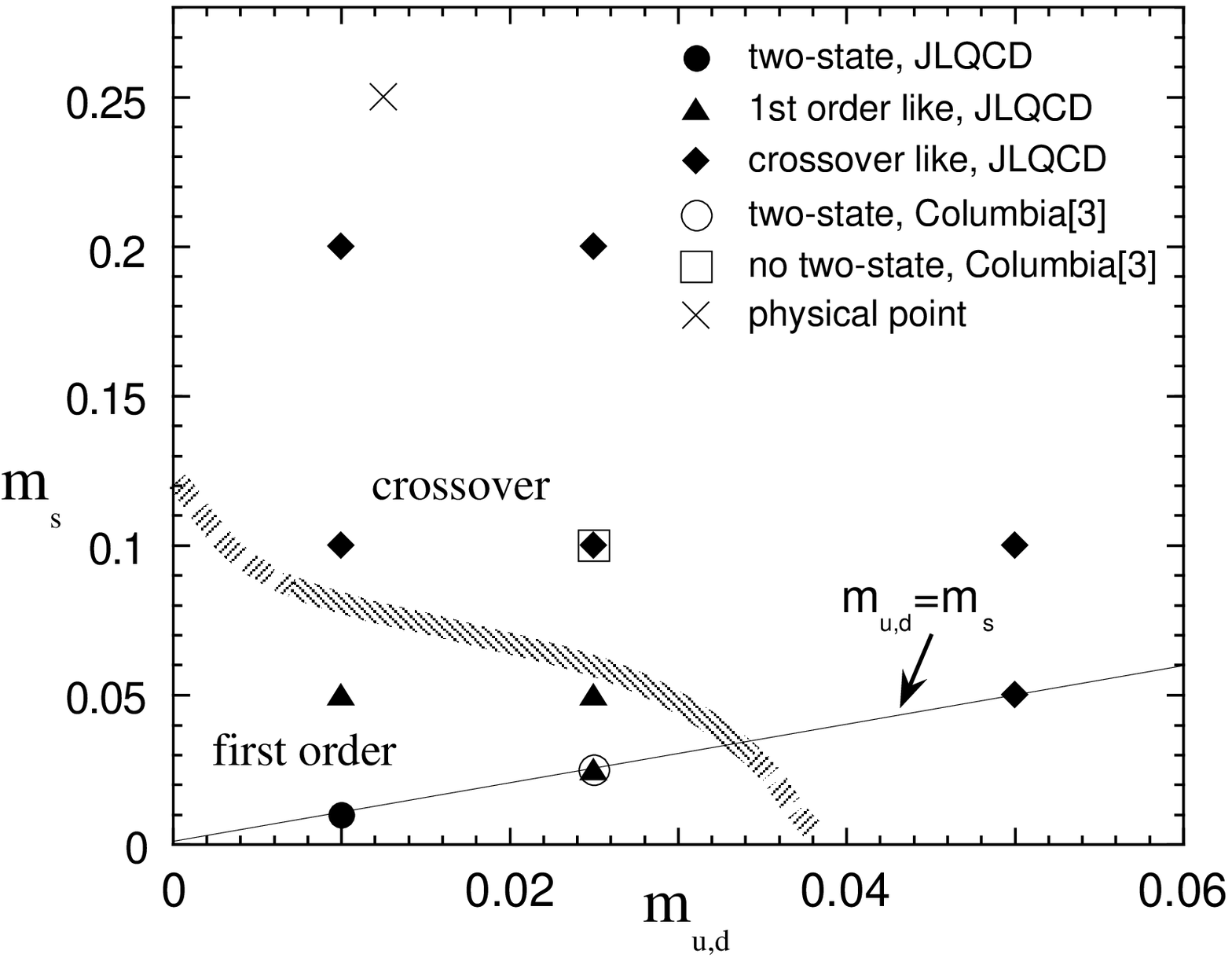}
 \caption{Le regioni delle transizioni di fase nel piano $(m_{u,d},m_s)$.
          Si distinguono una regione detta di {\em pura gauge} in cui le masse dei
         quark sono infinite e la transizione risulta del primo ordine,
         una regione in cui le masse sono nulle ({\em limite chirale}) e
         la transizione \`e ancora del primo ordine, ed una regione intermedia
         ({\em crossover}) in cui la transizione \`e continua. La linea
         corrispondente alla transizione del secondo ordine si discosta
         dall'asse verticale al di sotto di un certo valore di $m_s$.
         {\bf a)} Risultati di Iwasaki et al.~\cite{OrderTransit}; il valore
         fisico della massa dei quark viene determinato dai valori dei rapporti
         delle masse $m_{\phi}/m_{\rho}$\ e $m_{\pi}/m_{\rho}$ e corrisponde
         ad una transizione del primo ordine.
         {\bf b)} Risultati del gruppo JLQCD~\cite{OrderTransit2} ottenuti
         su un reticolo di $8^3 \times 4$ nodi. Il punto fisico della transizione
         viene calcolato assumendo una temperatura critica di 150 MeV, e cade
         nella regione in cui la transizione \`e continua ({\em ``crossover''}).}
\label{PhaseTransit}
\end{center}
\end{figure}

 L'ordine della transizione di fase dipende, cos\`i come il valore
 della temperatura critica, dal numero di {\em flavour} e dalle masse
 dei quark introdotti nelle simulazioni.
 La transizione risulta del primo ordine
 nell'ipotesi di masse dei quark nulle (limite chirale), od infinite
 (limite di gauge pura). Per valori intermedi delle masse dei quark, le
 simulazioni %di QCD su reticolo ({\em cfr.} prossimo paragrafo)
 suggeriscono transizioni di fase continue. Come mostrato in
 fig.~\ref{PhaseTransit}, i risultati della QCD su reticolo a temperatura non
 nulla sono ancora ambigui e necessitano di conferme ottenibili utilizzando
 reticoli di maggiore dimensione.
\subsection{Il ripristino della simmetria chirale nel QGP} 
\begin{figure}[h]
\begin{center}
\includegraphics[scale=0.40]{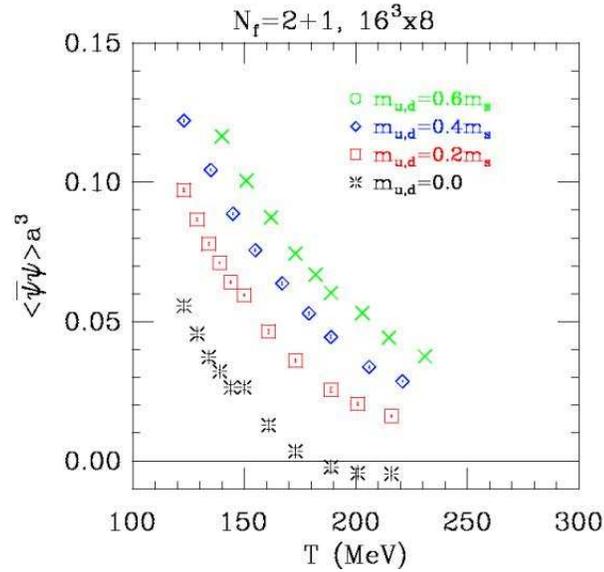}
 \caption{Andamento del parametro d'ordine $<q\bar{q}>$\ in funzione della 
     temperatura per un reticolo di dimensione $16^3\times3$ e per diversi 
     rapporti tra le masse dei quark $u$,$d$ ed $s$~\cite{Condensate}.}
\label{Condensate}
\end{center}
\end{figure}

Si ritiene che la transizione nello stato di QGP venga accompagnata 
da un parziale ripristino della simmetria chirale~\cite{Chirale}. 
La simmetria chirale \`e una simmetria esatta della lagrangiana di QCD, 
nel limite in cui le masse dei quark siano nulle; la presenza dei 
termini di massa, che valgono $m_u \simeq m_d \simeq$ 7 MeV, 
$m_s \simeq$ 156 MeV~\cite{QuarkMass}, 
nella lagrangiana (eq.~\ref{Lagrangiana}) rompe tale simmetria. 
Queste masse non devono essere confuse con le {\em masse costituenti} 
introdotte nei modelli a quark fenomenologici, che hanno valori 
dell'ordine di $m_u \approx m_d \approx $ 330 MeV, $m_s \approx $ 550 MeV, 
e rappresentano piuttosto l'energia dei quark confinati entro gli 
adroni~\cite{QuarkCost}.
La simmetria chirale pu\`o per\`o essere considerata una simmetria approssimata 
per la QCD in quanto le masse dei quark sono piccole se confrontate con la 
scala di energia tipica delle interazioni adroniche. 
\newline
Le grandi differenze tra le masse nello spettro degli adroni, come ad 
esempio tra i mesoni \Pgr e \Pai, non sono per\`o spiegate dalla leggera 
asimmetria tra le masse dei quark~\cite{Chirale}. Le differenze sono invece 
giustificate col meccanismo di rottura spontanea della simmetria chirale:  
la lagrangiana del sistema \`e (approssimativamente) simmetrica, mentre 
lo stato fondamentale (il vuoto) non gode di tale simmetria.  
Questo vuol dire che il valore di aspettazione del condensato di quark 
$<q\bar{q}>$\ nello stato fondamentale \`e diverso da zero.  
\newline
La rottura spontanea della simmetria prevede l'esistenza di mesoni    
pseudo-scalari di massa nulla: i bosoni di Goldstoone. Essi sono 
identificati coi pioni, che hanno infatti una massa di molto inferiore a  
quella di ogni altra particella dello spettro adronico. 
La loro massa non \`e  esattamente nulla in quanto la simmetria 
chirale \`e solo una simmetria approssimata della QCD.
\newline
\`E tuttavia previsto che, all'aumentare della temperatura o della densit\`a, 
lo stato di vuoto divenga anch'esso simmetrico: si parla in tal 
caso di {\em ``ripristino (parziale) della simmetria chirale''}~\cite{Chirale}.  
In tali condizioni il valore di aspettazione del condensato $<q\bar{q}>$\ 
dovrebbe annullarsi. Calcoli su reticolo, quali quello mostrato in 
fig.~\ref{Condensate}, indicano che il ripristino 
della simmetria chirale ($<q\bar{q}=0>$) e la transizione di deconfinamento 
avvengono approssimativamente alla stessa temperatura, sebbene non vi sia 
alcuna ragione fondamentale che giustifichi la loro coincidenza.    
\newline
Questa transizione riveste una notevole importanza nel meccanismo di produzione
dei quark strani. In un sistema deconfinato e quindi simmetrico per chiralit\`a,
la soglia per la produzione di una coppia $s\bar{s}$\ passa dal tipico valore
che assume in interazioni adroniche ($\approx$ 700 MeV) al doppio della massa
del quark strano, $2 \, m_s \, \simeq $ 300 MeV.
\section{I segnali sperimentali di QGP}
In questa sezione verranno illustrati alcuni tra i segnali sperimentali  
e tra i risultati pi\`u recenti e significativi proposti come evidenza del 
raggiungimento del QGP nelle collisioni nucleari ultra-relativistiche.  
\subsection{Esperimenti sugli ioni pesanti ultrarelativistici}
\begin{figure}[ht]
\begin{center}
\includegraphics[scale=0.38,angle=-90]{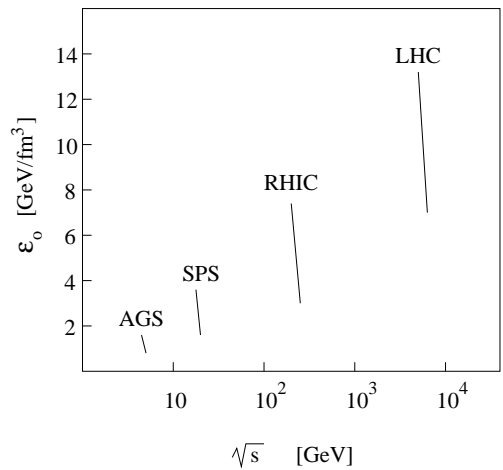}
 \caption{Densit\`a di energia $\epsilon_0$\ raggiunta nelle collisioni 
         per i diversi acceleratori di ioni pesanti.}
\label{Facility}
 \end{center}
\end{figure}

A partire dal 1986, con la disponibilit\`a dei primi fasci di ioni 
$^{16}{\rm O}$ e $^{32}{\rm S}$\ accelerati al sincrotone SPS del CERN e di 
ioni $^{28}{\rm Si}$\ al sincrotone a gradiente alternato (AGS) di Brookhaven, 
si \`e sviluppata la prima attivit\`a sperimentale riguardante la fisica 
degli ioni pesanti relativistici, a bersaglio fisso.  
I primi esperimenti eseguiti, di carattere esplorativo, hanno mostrato 
che vi \`e la possibilit\`a di portare la materia nucleare ad alte densit\`a 
di energia e temperatura.  Ad esempio, nelle collisioni prodotte dal fascio 
di ioni $^{32}{\rm S}$, sono state raggiunte temperature di $\sim 200 $\ MeV 
e densit\`a di energia di $\sim 2 \, {\rm GeV/fm^3}$, molto vicini ai valori 
critici necessari per la formazione del QGP. 
I dati sperimentali non hanno tuttavia consentito una discriminazione tra uno 
scenario di QGP ed uno di normale gas adronico. 
\newline
Questi risultati preliminari hanno suscitato grande interesse nella comunit\`a 
scientifica, ed hanno indotto una seconda generazione di esperimenti, condotta 
con ioni di $^{197}{\rm Au}$\  all'AGS a partire dal 1993 e di $^{208}{\rm Pb}$\ 
all'SPS a partire dal 1994. L'obiettivo di questa seconda fase \`e 
principalmente quello di esplorare le osservabili del QGP individuate durante 
la prima fase, ma in sistemi interagenti pi\`u estesi, dove le condizioni per 
la formazione del plasma sono pi\`u faveroveli. Tra gli esperimenti di seconda 
generazione effettuati al CERN, NA49 \`e stato progettato per consentire uno 
studio generale e simultaneo della dinamica della collisione, disponendo  
di una grande accettanza ed essendo sensibile ad un gran numero di osservabili fisiche. 
Gli altri esperimenti sono invece finalizzati a rivelare, in maniera specifica e con 
maggior precisione, particolari segnali della transizione di fase.  
Il programma sperimentale dell'SPS \`e tuttora in fase di 
svolgimento e 
%svolgimento: molti degli esperimenti dell'Area Nord (NA) del CERN sono in funzione;  
%nuovi esperimenti, quali ad esempio NA57 ed NA60, sono stati proposti e realizzati 
%in un secondo momento, o per finalizzare studi precedenti o per realizzarne di nuovi;  
nuove richieste di prese dati sono state avanzate al Comitato dell'SPS (SPSC) 
per l'anno 2003.  
\newline
A partire dall'anno 2000 il nuovo complesso di accelerazione del RHIC
({\em Relativistic Heavy Ion Collider}), costituito
da un collisionatore di ioni pesanti, \`e entrato in pieno regime di
funzionamento a Brookhaven. Il grande vantaggio dei collisionatori risiede
nella disponibilit\`a di una maggiore energia nel sistema centro di massa:
nelle collisoni Au-Au studiate al RHIC l'energia nel centro di massa
per nucleone ($\sqrt{s_{NN}}$) raggiunge infatti valori di 200 GeV,
oltre dieci volte superiori a quelli tipici dell'SPS.  
Un nuovo collisionatore, il {\em Large Hadron Collider} (LHC),
\`e in fase di realizzazione al CERN; 
\begin{table}[hb]
\label{tab13}
\begin{center}
\begin{tabular}{||c||c|c|c|c||} \hline \hline
     & Ioni & $p_{fascio}/A$ (GeV/$c$)& $\sqrt{s_{NN}}$ (GeV) & Collaborazioni\\\hline\hline
 AGS &Si& 14.6       & 5.4         & E802(E859), E810, E814, E858 \\
 AGS &Au& 10.7       & 4.7         & E866, E877, E878\\
 SPS &O, S& 200      & 19.4        & NA34,35,36,38; WA80,85,94 \\
 SPS &Pb&160, 80, 40 & 17.3, 12.3, 8.8 & NA44,45,49,50,52,57,60; WA97,98\\
 RHIC&Au& 65, 100    & 130, 200    & STAR, PHENIX, BRAHMS, PHOBOS \\ \hline
 LHC &Pb& 2500       & 5500        & ALICE   \\ \hline \hline
\end{tabular}
\end{center}
\caption{Complessi di accelerazione di fasci di ioni relativistici e sigle
     delle principali Collaborazioni coinvolte. LHC
     ({\em Large Hadron Collider}) entrer\`a in funzione come
     collisionatore di ioni di piombo nel 2007.}
\end{table}
esso raggiunger\`a un'energia di circa 5.5 TeV per nucleone nel sistema centro di massa, 
valore cui corrisponde un'energia totale nel centro di massa pari a 
circa 2300 TeV per le collisioni Pb-Pb.  
In tabella 1.3 sono riassunti i dati sul tipo di ioni utilizzati, le energie raggiunte dai 
diversi acceleratori e le sigle dei principali esperimenti coinvolti.  
\newline
In fig.~\ref{Facility} sono mostrate le densit\`a di energia raggiunte nella 
collisione per i diversi acceleratori di ioni pesanti, presenti e futuri. 
Se l'SPS del CERN ha ormai raggiunto, e probabilmente superato,   
il limite di energia critica per la 
transizione di fase, all'energia del RHIC questo limite verr\`a 
abbondantemente oltrepassato, mentre ad LHC esso  
sar\`a ormai molto lontano e pertanto risulta pi\`u difficile 
fare previsioni sul comportamento delle osservabili sin qui studiate 
ed \`e forse lecito aspettarsi anche fenomeni di nuova fisica. 
\subsection{Le osservabili globali}
\subsubsection{Distribuzioni di rapidit\`a}
\begin{figure}[ht]
\begin{center}
\includegraphics[scale=0.58]{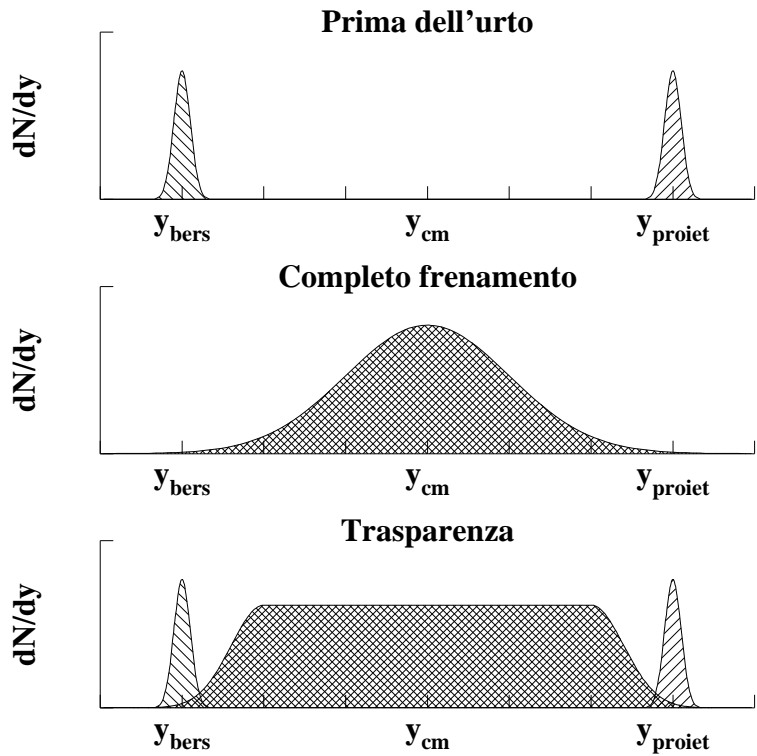}
 \caption{Diversi regimi di frenamento nucleare. Distribuzioni di 
          rapidit\`a prima della collisione (in alto) e dopo la 
	  collisione, nell'ipotesi di frenamento totale (in mezzo) 
	  e di completa trasparenza (in basso).}
 \label{frenamento}
 \end{center}
 \end{figure}

La misura delle distribuzioni di rapidit\`a delle particelle 
nello stato finale apporta importanti informazioni sulla dinamica 
della collisione. 
\newline
Prima della collisione tutta l'energia \`e trasportata dai barioni 
presenti nei nuclei bersaglio e proiettile; dopo la 
collisione il numero barionico netto si distribuisce nello spazio 
delle fasi in un modo che riflette la perdita di energia dei barioni 
iniziali durante la collisione.  
\newline
Passando da energie relativistiche 
%($\sqrt{s_{NN}}=0.1\div1  $\ GeV) 
ad energie ultra-relativistiche 
%($\sqrt{s_{NN}} \apprge 10 $\ GeV) 
cambia la fenomenologia della collisione tra i nucleoni del proiettile e 
del bersaglio. Estremizzando, \`e possibile definire  
due regimi limite per quel che riguarda il processo di decelerazione dei nuclei 
incidenti: il regime di {\em frenamento totale} e quello di 
{\em trasparenza nucleare}.  
Nel regime di frenamento totale i nucleoni incidenti si fermano nel 
sistema di riferimento del centro di massa e formano una zona a 
ricco contenuto barionico nella regione di rapidit\`a centrale. Ad 
energie pi\`u elevate si giunge al regime di trasparenza nucleare, 
in cui i contenuti barionici iniziali del proiettile e del bersaglio 
continuano il loro moto senza una sostanziale decelerazione. Anche 
in questo caso, nella regione di rapidit\`a centrale, viene rilasciata 
una notevole quantit\`a di energia, ma il contenuto barionico \`e 
pressocch\'e nullo. In fig.~\ref{frenamento} sono rappresentate 
schematicamente le due situazioni. 
\newline
Trascurando il ruolo degli spettatori e considerando solo 
collisioni centrali, \`e possibile individuare due distinti 
modelli per i due regimi estremi illustrati sopra: uno, 
sviluppato da Fermi e Landau valido per il caso di frenamento 
totale~\cite{Fer50}\cite{Lan53}, l'altro valido per energie 
superiori (caso di trasparenza totale) dovuto a Bj{\o}rken 
e McLerran~\cite{Bjo83}\cite{Mcl82}.  

Quello di Landau e Fermi \`e un modello idrodinamico in cui 
le condizioni iniziali sono fissate dalla contrazione di Lorentz 
dei nuclei incidenti. Si suppone che tutta l'energia $E_{cm}$\
%cinetica 
nel centro di massa  
%$E_{cm}$\  
dei due nuclei, che qui saranno 
supposti identici per semplicit\`a, sia dissipata nell'urto, e  
che quindi questi si fermino nel centro di massa. Dopo l'urto 
la materia nucleare si espande idrodinamicamente; la produzione di 
particelle \`e massima nei primi istanti dopo l'urto, quando \`e 
massima la densit\`a di energia. I due nuclei, contratti 
relativisticamente nella direzione longitudinale, presentano un 
volume $V=V_0/\gamma$, dove $V_0$ \`e il volume dei nuclei a riposo e 
$\gamma=(1-\beta^2)^{-1/2}$\ \`e il loro fattore di Lorentz nel 
sistema del centro di massa. Un'energia pari a due volte $E_{cm}$\ 
viene perci\`o distribuita entro un volume $V$; la densit\`a di 
energia vale, quindi: 
\begin{equation}
\epsilon=\frac{2E_{cm}}{V_0/\gamma}=\frac{2E_{cm}^2}{\frac{4}{3}\pi R^3 M}
\label{density}
\end{equation}
dove $M$\ ed $R$\ sono la massa ed il raggio dei nuclei.
Come discusso nel paragrafo 1.3 per quanto riguarda le collisioni all'AGS 
e SPS, la transizone di fase verso lo stato di QGP pu\`o avvenire in 
queste condizioni principalmente per compressione. 
\newline
Per energie cinetiche nel centro di massa minori di 1 GeV per nucleone,  
corrispondenti ad urti non relativistici, 
si applicano invece i modelli a {\em fireball} che 
suppongono la simmetria sferica del sistema formatosi in seguito alla 
collisione. Anche in questo caso il sistema \`e inizialmente a riposo 
nel centro di massa ed \`e l'elevata pressione che lo porta ad  
espandersi, con conseguente  raffreddamento. Nel caso del modello di Landau, 
il gradiente di pressione \`e molto pi\`u elevato nella direzione 
longitudinale che in quella trasversale, quindi l'espansione \`e 
pi\`u pronunciata nella direzione lungo l'asse del fascio nei 
primi istanti dopo la collisione.  

\begin{figure}[ht]
\begin{center}
\includegraphics[scale=0.58]{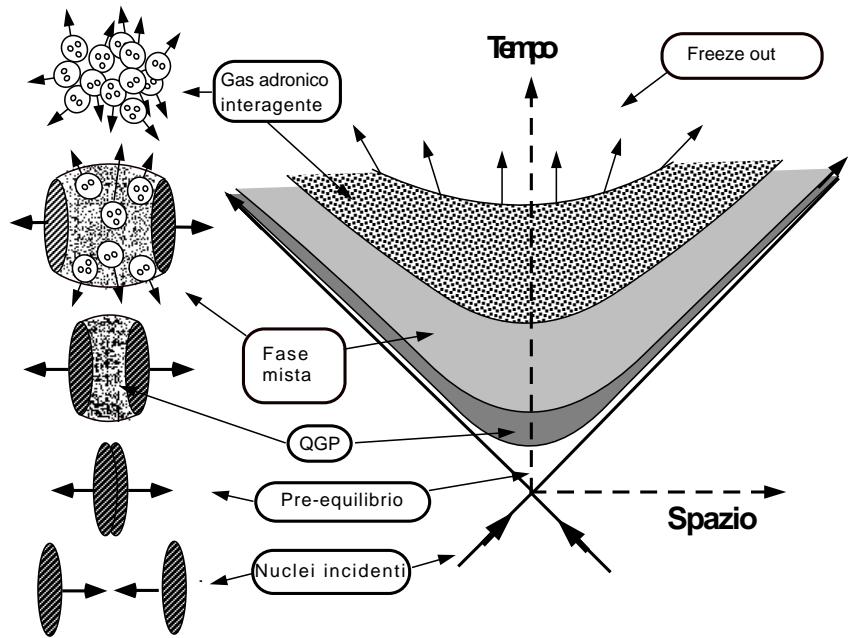}
\caption{Evoluzione spazio-temporale di una collisione centrale 
         nucleo-nucleo ad energie ultra-relativistiche. L'asse 
	 spaziale corrisponde alla direzione del fascio incidente. 
	 Entrambe le coordinate degli assi cartesiani sono valutate 
	 nel sitema centro di massa dei due nuclei.}
\label{spacetime}
\end{center}
\end{figure}

Il modello di Bj{\o}rken si applica invece ad alte energie nel centro 
di massa ($\sqrt{s_{NN}} > 100 $\ GeV) --- raggiunte dall'acceleratore RHIC 
ed, in futuro, da LHC --- dove si prevede che i nuclei, 
fortemente contratti relativisticamente, diventino quasi trasparenti l'uno 
all'altro. La transizione di fase nel QGP \`e conseguenza in tal caso 
del riscaldamento della materia nucleare, piuttosto che della 
compressione, e le condizioni del QGP sono 
simili a quelle proprie dell'universo dopo il {\em Big Bang}.  
I quark di valenza dei nucleoni incidenti conservano 
una frazione della loro energia; nel passaggio dei due nuclei uno  
attraverso l'altro, l'interazione dei nucleoni lascia nella regione di 
rapidit\`a centrale una densit\`a di energia cui corrisponde una carica 
barionica pressocch\'e nulla. 
\newline
L'evoluzione spazio-temporale di una collisione centrale 
\`e schematizzata in fig.~\ref{spacetime}.  
Il sistema di riferimento \`e quello del centro di massa dei nuclei 
incidenti, e la coordinata spaziale \`e stata scelta lungo la linea 
di fascio. Le iperboli rappresentano le linee di tempo proprio costante 
($\tau=\sqrt{t^2-x^2}={\rm cost}$).  
Per $(x,t)=(0,0)$\ i nuclei proiettile e bersaglio collidono e viene creata 
una zona ad alta densit\`a di energia. \`E questo il punto in cui pu\`o 
eventualmente formarsi lo stato di QGP, dopo una fase di pre-equilibrio 
che termina dopo un tempo proprio $\tau_0$. L'espansione provoca il 
raffreddamento del sistema che, di conseguenza, ritorna ad una fase adronica. 
Quando la distanza media tra gli adroni risulta maggiore della portata 
delle interazioni forti, gli adroni sciamano liberamente senza pi\`u 
interagire ({\em ``freeze-out''}). Poich\'e la densit\`a di energia decresce 
rapidamente in seguito all'espansione idrodinamica del plasma, riveste  
particolare interesse il suo valore prima dell'espansione, per $\tau=\tau_0$. 
\newline
Per poter stimare la densit\`a di energia $\epsilon_0$\ prima dell'evoluzione 
idrodinamica occorre determinare l'energia depositata nella reazione ed il 
volume della regione interessata. Poich\'e l'energia depositata nella collisione 
si manifesta infine in produzione di adroni, \`e possibile stimare l'energia 
iniziale ed il rispettivo volume ricostruendo all'indietro  
le traiettorie delle particelle al punto da cui sono originate nello 
spazio-tempo. 
Poich\'e le particelle emergono dal punto di interazione a 
$(x,t)=(0,0)$, il volume che occupano, e quindi la densit\`a di energia, 
dipende dal tempo. Quella che si vuole \`e la densit\`a di energia a rapidit\`a 
centrale ($x=0$) ed al tempo (proprio) $\tau_0$. 
\newline
La velocit\`a longitudinale delle particelle vale $v_x=\frac{p_x}{E}=\tanh y$; una 
particella emessa nell'origine $(0,0)$\ avr\`a coordinate $(x,t)$\ che possono 
esprimersi in funzione della rapidit\`a e del tempo proprio:
\begin{subequations}
 \begin{equation}
  x=\tau \sinh y
 \label{ww1}
 \end{equation} 
 \begin{equation}
  t=\tau \cosh y .
 \label{ww2}
 \end{equation}
 \label{ww}
\end{subequations}     
Per mezzo delle~\ref{ww}, la distribuzione di rapidit\`a pu\`o essere 
riscritta come distribuzione spaziale ad un istante fissato. Considerando 
l'elemento di volume centrato ad $x=0$, cio\`e a rapidit\`a centrale 
al tempo proprio $\tau_0$, di dimensione $\mathcal{A}\Delta x$, dove 
$\mathcal{A}=\pi r_0^2 A^{2/3}$\ \`e l'area trasversa del 
nucleo~\footnote{Si \`e utilizzata la parametrizzazione 
$r_0 A^{1/3}$\ per il raggio nucleare, dove $A$\ \`e il numero di massa 
del nucleo.}, la densit\`a di particelle in questo volume vale: 
\begin{equation}
   \frac{\Delta N}{\mathcal{A} \Delta x}=\frac{1}{\mathcal{A}} 
   \frac{{\rm d}N}{{\rm d}y} \frac{1}{\tau_0 \cosh y} \, .
\label{Donato}
\end{equation}
A partire dall'eq.~\ref{Donato} si calcola la densit\`a di energia a $x=0$\ ed 
al tempo proprio $\tau_0$, mediata sull'area trasversa $\mathcal{A}$:
\begin{equation}
\epsilon_0 = <m_T> \cosh y \frac{\Delta N}{\mathcal{A} \Delta x} = 
     \frac{<m_T>}{\tau_0 \mathcal{A} }\frac{{\rm d}N}{{\rm d}y}. 
\label{EnergDens}
\end{equation} 
Nell'eq.~\ref{EnergDens}, $<m_T> \cosh y $\ rappresenta l'energia media 
di una particella, ed il tempo proprio $\tau_0$\ \`e stimato da Bj{\o}rken 
essere dell'ordine di 1 fm/$c$.  
\newline
All'SPS, trovandosi in un regime intermedio tra quello di Bj{\o}rken e
quello di Landau, la densit\`a di energia calcolata nel modello di
Bj{\o}rken (eq.~\ref{EnergDens}) rappresenta un limite inferiore della
densit\`a di energia effettivamente raggiunta nelle collisioni Pb-Pb.
Ponendo nell'eq.~\ref{EnergDens} ${\rm d}N/{\rm d}y \simeq 500$\  
a rapidit\`a centrale~\cite{NetBaryon49}, $<m_T> \simeq 480$ MeV~\footnote{
Valore ricavato utilizzando le distribuzioni di massa trasversa misurate
dall'esperimento WA97~\cite{mt_WA97}}, un raggio nucleare
$r_0 A^{1/3}=6.62$\ fm~\cite{Carr26} ed un tempo
proprio $\tau_0\simeq1$ fm/$c$, si ottiene un valore della densit\`a
di energia $\epsilon_0 \simeq 1.7 \, {\rm GeV/fm^3}$. Questo valore, pur
essendo un limite inferiore, \`e tuttavia gi\`a compatibile con la
densit\`a di energia a cui ci si aspetta la transizione di fase.
%
%%%%%%%%%
 
\begin{figure}[h]
\begin{center}
\includegraphics[scale=0.90]{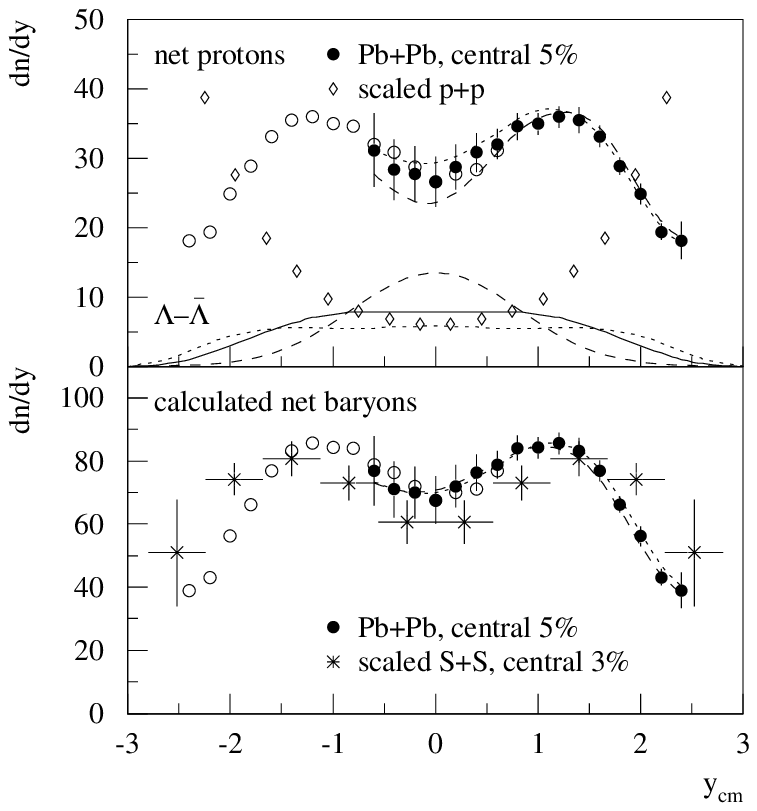}
\includegraphics[scale=0.35]{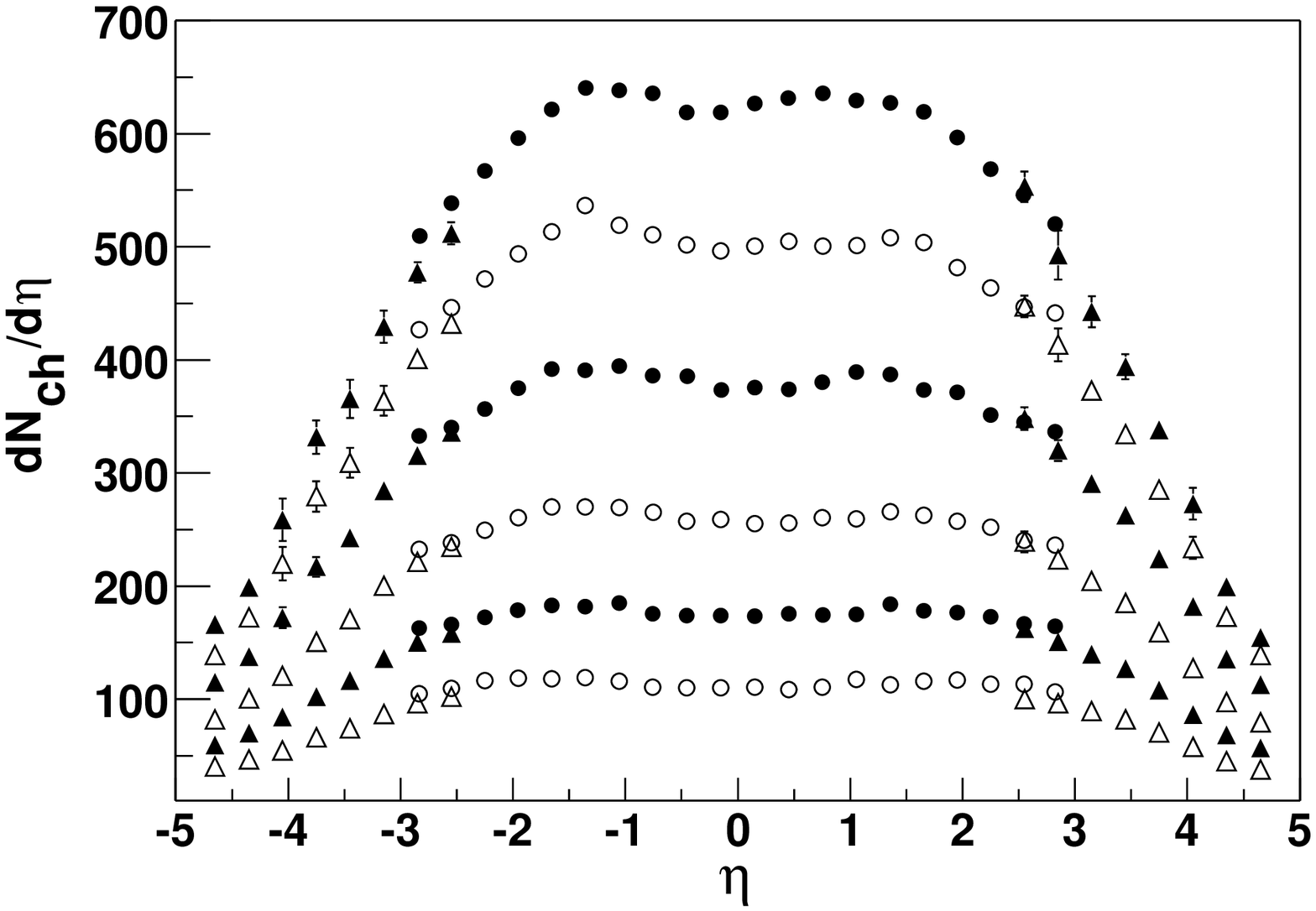}
 \caption{{\em A sinistra:} Risultati dell'esperimento NA49~\cite{NetBaryon49}.  
 In alto: distribuzione di rapidit\`a di $p-\bar{p}$\ in collisioni 
 centrali Pb-Pb a 160 $A$ GeV/$c$; sono anche mostrate la distribuzioni 
 $\Lambda-\bar{\Lambda}$ nelle stesse collisioni (le tre linee in basso, 
  per diversi metodi di correzione) e $p-\bar{p}$\ in collisioni p-p. 
  In basso: distribuzioni di rapidit\`a per i barioni 
  netti ($B-\bar{B}$) in collisioni Pb-Pb e S-S~\cite{NetBaryon49}.
  \newline {\em A destra:} Distribuzione di pseudo-rapidit\`a delle 
        particelle cariche misurata da BRAHMS~\cite{BRHAMS} in collisioni 
	Au-Au a $\sqrt{s_{NN}}$ al RHIC, per diversi intervalli di 
	centralit\`a.}
 \label{NetBaryon}
\end{center}
\end{figure}

Nella parte destra della 
fig.~\ref{NetBaryon} sono mostrate le distribuzioni di rapidit\`a dei 
{\em protoni netti} (cio\`e la differenza tra il numero di protoni e quello 
di anti-protoni) misurate dalla collaborazione NA49~\cite{NetBaryon49} 
in collisioni centrali Pb-Pb e, per confronto in collisioni S-S. Si considera 
la differenza $p-\bar{p}$\ per eliminare gli effetti dovuti alla produzione 
di coppie barione--anti-barione.  Nella stessa fig.~\ref{NetBaryon}, \`e 
mostrata anche la distribuzioni di {\em barioni netti} 
(cio\`e barioni $-$\  anti-barioni),  ricavata dalle distribuzioni di 
{\em protoni netti}, servendosi di modelli che 
calcolano il contributo dei diversi stati di isospin e modelli per il 
calcolo delle distribuzioni di barioni strani. Si distinguono ancora 
i due picchi relativi alla frammentazione del bersaglio ed a quella del 
proiettile; \`e quindi chiaro come il regime dell'SPS sia intermedio tra 
quello di completa trasparenza e quello di totale frenamento, sebbene 
p\`u prossimo al primo. La distribuzione 
di rapidit\`a delle \PgL-\PagL\ presenta invece un unico picco centrale 
molto largo, suggerendo che nel meccanismo di produzione di queste particelle 
si \`e perso ormai il ``ricordo'' del moto iniziale del proiettile 
(e del bersaglio, nel sistema CM). Le distribuzioni di rapidit\`a 
(o di pseudo-rapidit\`a) misurate al RHIC, quale quella misurata da 
BRHAMS~\cite{BRHAMS} e mostrata nella parte destra della fig.~\ref{NetBaryon}, 
sono invece 
%ancora pi\`u vicine a 
quelle attese nel caso di regimi di totale trasparenza.  
\subsubsection{Distribuzioni di energia trasversa}    
\begin{figure}[p]
\begin{center}
{\bf a)} \hspace{8.0cm} {\bf b)} \\   
\includegraphics[scale=0.36]{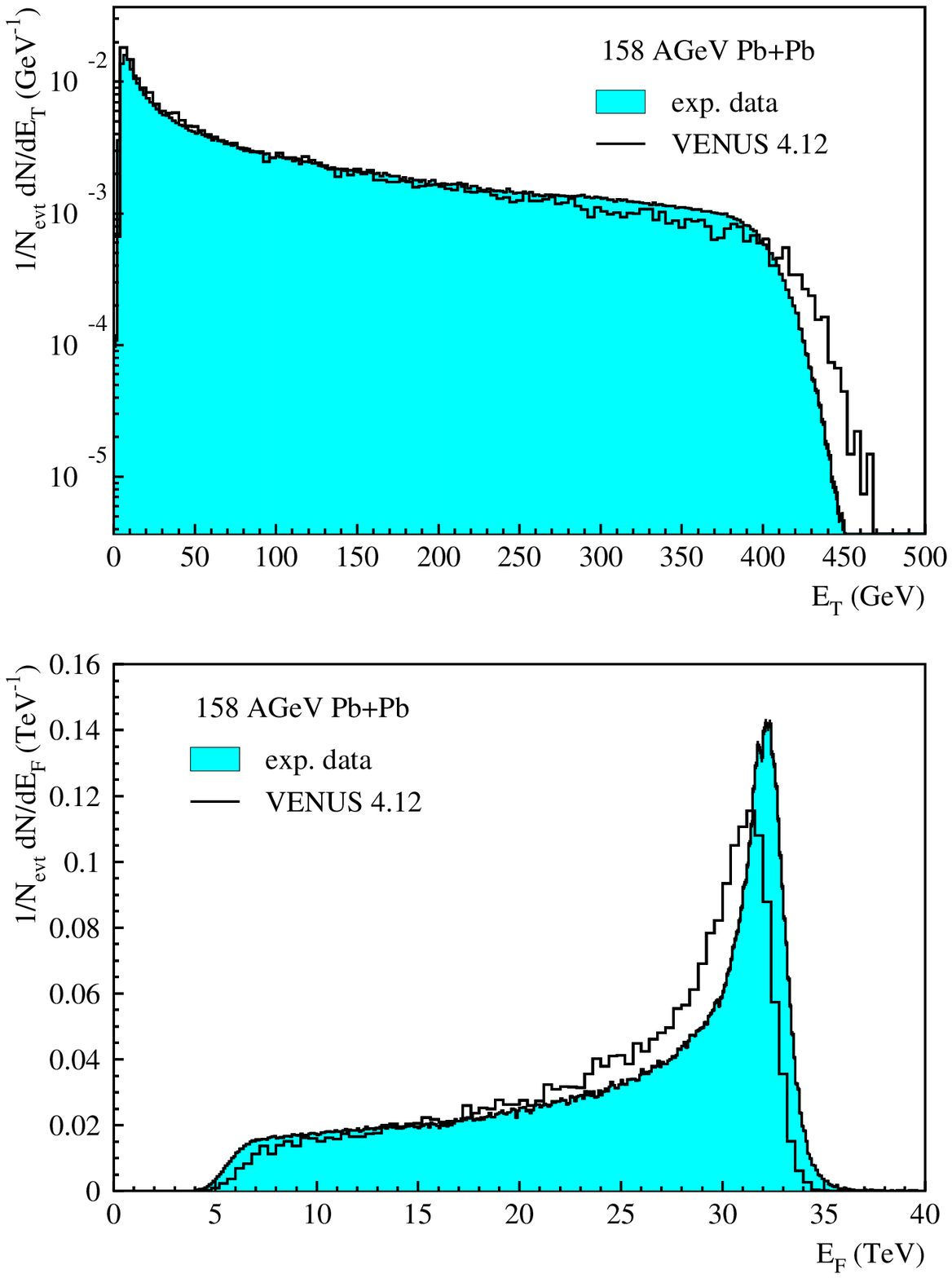}
\includegraphics[scale=0.40]{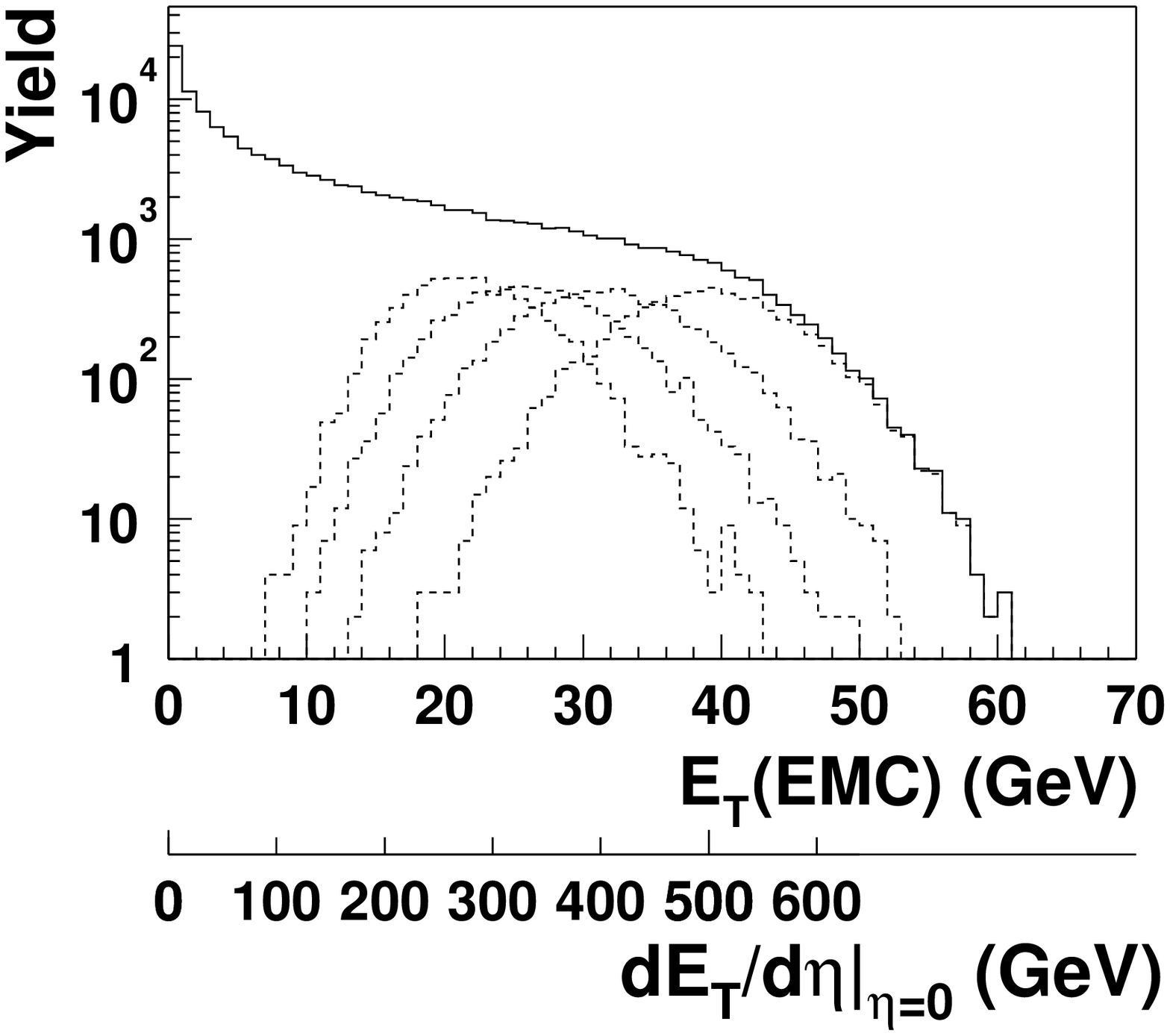}\\
\vspace{0.6cm}
{\bf c)} \\
\vspace{0.5cm}
\includegraphics[scale=0.45]{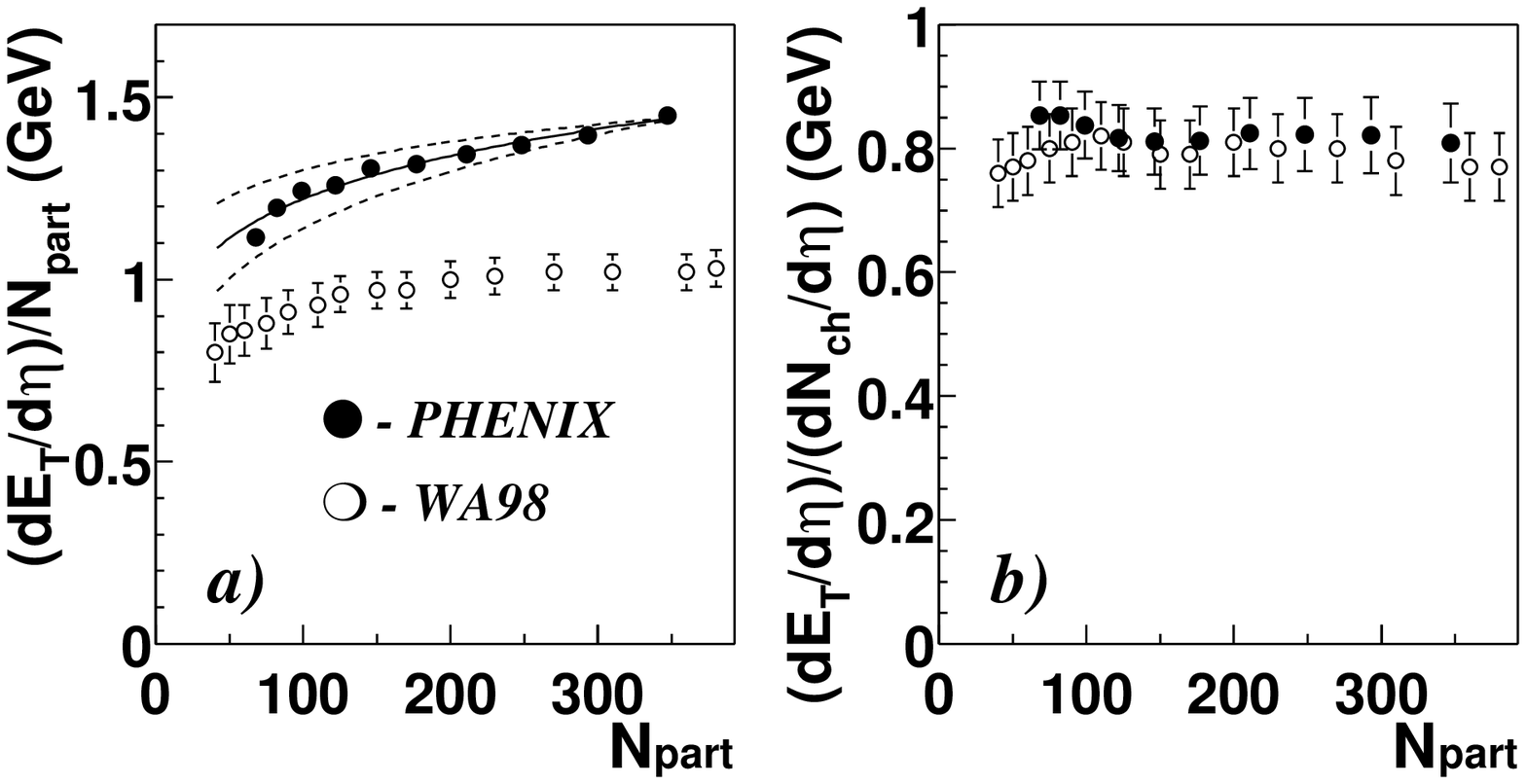}
 \caption{Distribuzione di energia trasversa misurata da  
 WA98 (grafico superiore in {\bf a}, Pb-Pb a 
 $\sqrt{s_{NN}}=17.2$ GeV)~\cite{WA98PRC65}, confrontata con le predizioni 
 del generatore VENUS 4.12~\cite{VENUS}  
 e da PHENIX ({\bf b}, Au-Au a $\sqrt{s_{NN}}=130$ GeV)~\cite{EtPHENIX}, 
 in cui si distinguono i contributi alla 
 distribuzione di {\em ``minimum bias''} (linea intera) delle quattro classi 
 di maggior centralit\`a (linee trattegiate). Nel grafico inferiore 
 dell'inserto {\bf a}) \`e mostrata la distribuzione di energia rilasciata 
 in avanti misurata con uno ZDC ({\em calorimetro ad angolo zero}) da WA98. 
 \newline
 {\bf c}) {\em A sinistra}: densit\`a di energia trasversa per partecipante 
 ($ \left. {\rm d}E_T/{\rm d}\eta \right|_{\eta=0} / N_{part}$) in funzione 
 del numero di partecipanti misurata da PHENIX  
 e da WA98. La linea a tratto continuo 
 rappresenta il miglior fit eseguito sui dati di PHENIX 
 con la funzione $N_{part}^\alpha$\ e fornisce 
 $\alpha= 1.13 \pm 0.05$, le linee trattegiate mostrano di 
 quanto si discosta la curva di potenza se calcolata a $\pm 1 \sigma$.
 {\em A destra}: $ {\rm d}E_T/{\rm d}\eta \left|_{\eta=0} / 
        {\rm d}N_{ch}/{\rm d}\eta \right|_{\eta=0}$\ versus $N_{part}$. }
 \label{EtFig}
\end{center}
\end{figure}

L'energia trasversa \`e definita come 
\begin{equation}
E_T=\sum_{i} E_i \sin \theta_i
\label{Et}
\end{equation}
dove la somma si estende su tutte le particelle prodotte in una collisione,  
$ E_{i}$\ e $\theta_i $\ sono, rispettivamente, l'energia e l'angolo rispetto alla linea 
di fascio della direzione di volo (all'emissione) della particella  $i$-esima. 
\newline  
Operativamente $E_i$\ \`e definita come l'energia depositata in un calorimetro 
che contenga completamente lo sciame in esso generato dalla particella $i$-esima. 
%$E_i$\ \`e quindi l'energia totale per i mesoni ed \`e invece l'energia totale 
%(pi\`u) meno la massa a riposo per gli (anti)barioni.
\newline
Le distribuzioni di energia trasversa, quali ad esempio quelle misurate da 
WA98~\cite{WA98PRC65} all'SPS nelle interazioni Pb-Pb a $\sqrt{s_{NN}}=17.2$ GeV 
(160 $A$\ GeV/$c$\ nel laboratorio) o da PHENIX~\cite{EtPHENIX} al RHIC nelle 
collisioni Au-Au a $\sqrt{s_{NN}}=130$ GeV , mostrate in fig.~\ref{EtFig}, 
risultano interpretabili~\cite{EtPHENIX} in termini del modello 
geometrico della collisione di Glauber~\cite{Carrer18}, a meno dell'introduzione 
di un esponente $\alpha$\ nella dipendenza dal numero di nucleoni partecipanti 
($N_{part}^\alpha$).  WA98 trova infatti una dipendenza del tipo 
$N_{part}^{1.08\pm0.06}$, ancora compatibile con una legge lineare, mentre a pi\`u 
alte energie PHENIX misura $E_T \propto N_{part}^{1.13 \pm 0.05}$, che si 
discosta maggiormente dalla linearit\`a, come mostrato in fig.~\ref{EtFig}.  
\newline
Le collisioni periferiche popolano la regione a bassi valori
di $E_T$\ delle distribuzioni in fig.~\ref{EtFig}.a,b;  
il {\em plateau} nella parte centrale 
della distribuzione corrisponde ad un incremento pi\`u uniforme del numero  
di nucleoni partecipanti; 
infine, dal ginocchio in avanti, emerge il contributo degli eventi pi\`u centrali. 
Questa semplice interpretazione \`e comune anche ad altre variabili globali quali 
la molteplicit\`a di particelle cariche e l'energia residua misurata in avanti 
lungo la linea di fascio. Essa consente di selezionare gli eventi sulla base della 
centralit\`a della collisione.  
L'estensione dello spettro di energia trasversa cresce all'aumentare delle dimensioni 
del sistema interagente, confermando  cos\`i che le condizioni pi\`u idonee per la 
formazione del plasma si hanno per sistemi di collisione pi\`u complessi.   
\newline
Il valore di energia trasversa corrispondente al ginocchio delle distribuzioni pu\`o 
essere utilizzato per stimare la densit\`a di energia raggiunta nelle collisioni. 
Ad esempio nel modello di Bj{\o}rken, riscrivendo l'eq.~\ref{EnergDens} come 
\begin{equation}
\epsilon_0 = \frac{1}{\tau_0 \pi R^2} \left. \frac{{\rm d}E_T}{{\rm d}y} \right|_{max}
\label{EnergDens2}
\end{equation}
dato che la regione centrale, caratterizzata da valori di rapidit\`a nel centro di massa 
intorno a zero, corrisponde ai valori massimi di energia trasversa, si pu\`o 
avere un'ulteriore stima della densit\`a di energia al tempo $\tau_0$. 
Per collisioni Pb-Pb a 160 $A$\ GeV/$c$ si ottiene il valore 
$\epsilon_0=2.8 \, {\rm GeV/fm^3}$, che adesso \`e tuttavia mediato 
sull'intero volume del sistema interagente, mentre nelle regioni pi\`u 
interne della sorgente di particelle pu\`o raggiungere il valore 
$\epsilon_0=3.5 \, {\rm GeV/fm^3}$~\cite{Bla96}, compatibile con la densit\`a 
di energia critica predetto dai calcoli di QCD su reticolo.  
\newline
I dati a pi\`u bassa energia dell'AGS sono perfettamente 
compatibili con una dipendenza di semplice proporzionalit\`a dal 
numero di nucleoni partecipanti~\cite{Abbot}.
\subsubsection{Energia depositata in avanti}
Informazioni sulla centralit\`a della collisione si possono anche 
ottenere misurando il numero di nucleoni spettatori del proiettile. 
Nel caso di reazioni tra nuclei dello stesso numero di massa, ci\`o 
fornisce probabilmente la misura pi\`u diretta della geometria della 
collisione. Allo scopo viene spesso utilizzato un calorimetro posto 
lungo la linea di fascio ({\em Zero Degree Calorimeter}), di dimensioni 
tali da coprire la regione di pseudo-rapidit\`a $\eta \approx \eta_{fascio}$\ 
in cui si manifestano i nucleoni del proiettile che non hanno interagito. 
Gli urti centrali corrispondono dunque agli eventi in cui una piccola frazione 
dell'energia del fascio viene depositata nel calorimetro. L'energia depositata 
in avanti \`e quindi anticorrelata con le osservabili molteplicit\`a  
ed energia trasversa, usate anch'esse per determinare la centralit\`a 
della collisione, come si evince in fig.~\ref{EtFig}.a dove sono 
mostrate le distribuzioni di energia trasversa (in alto) e di energia in avanti 
(in basso) misurate da WA98~\cite{WA98PRC65}.  
\subsubsection{Molteplicit\`a delle particelle cariche}
\begin{figure}[ht]
\begin{center}
{\bf a)} \hspace{8.0cm} {\bf b)} \\
\includegraphics[scale=0.50]{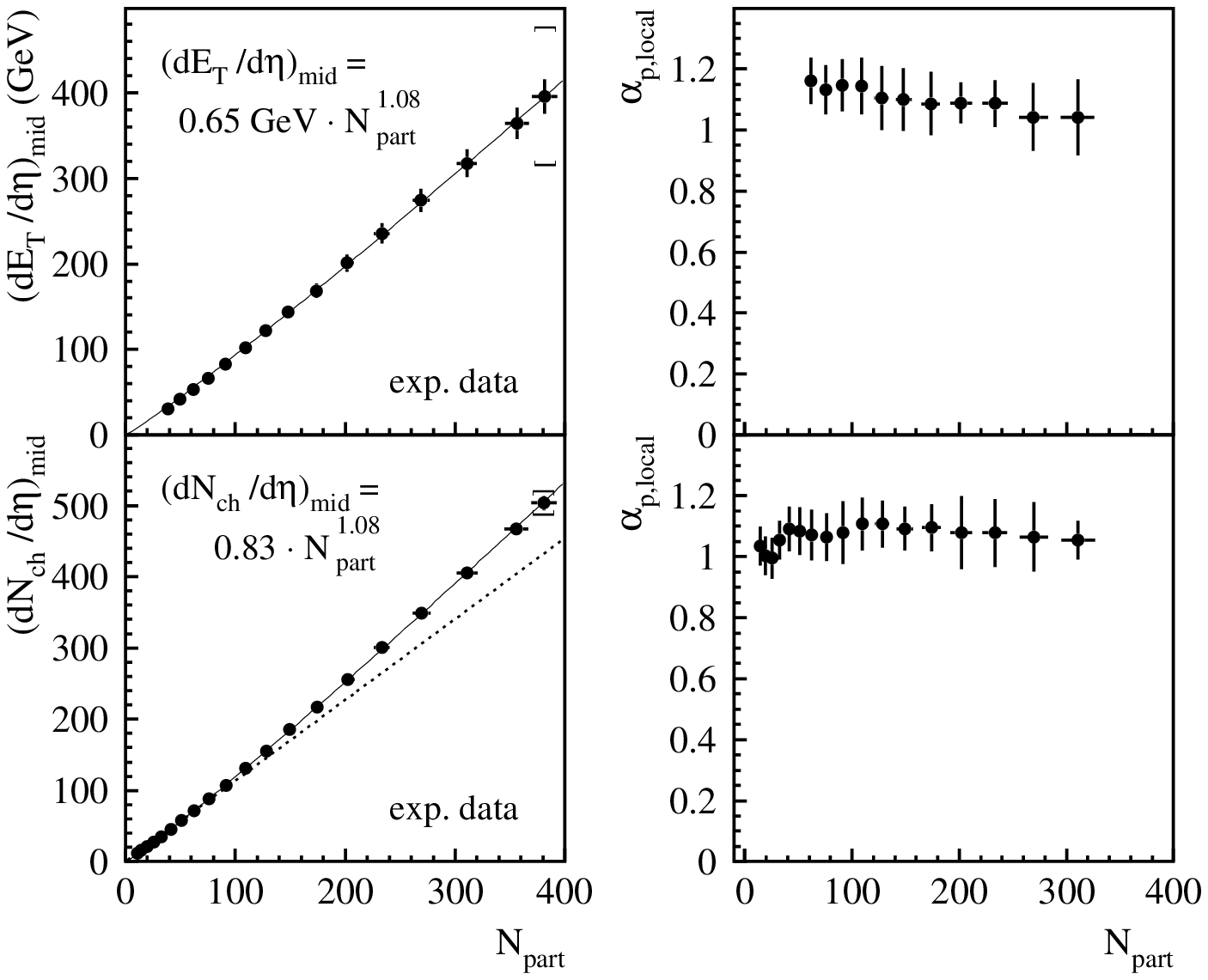}
\includegraphics[scale=0.38]{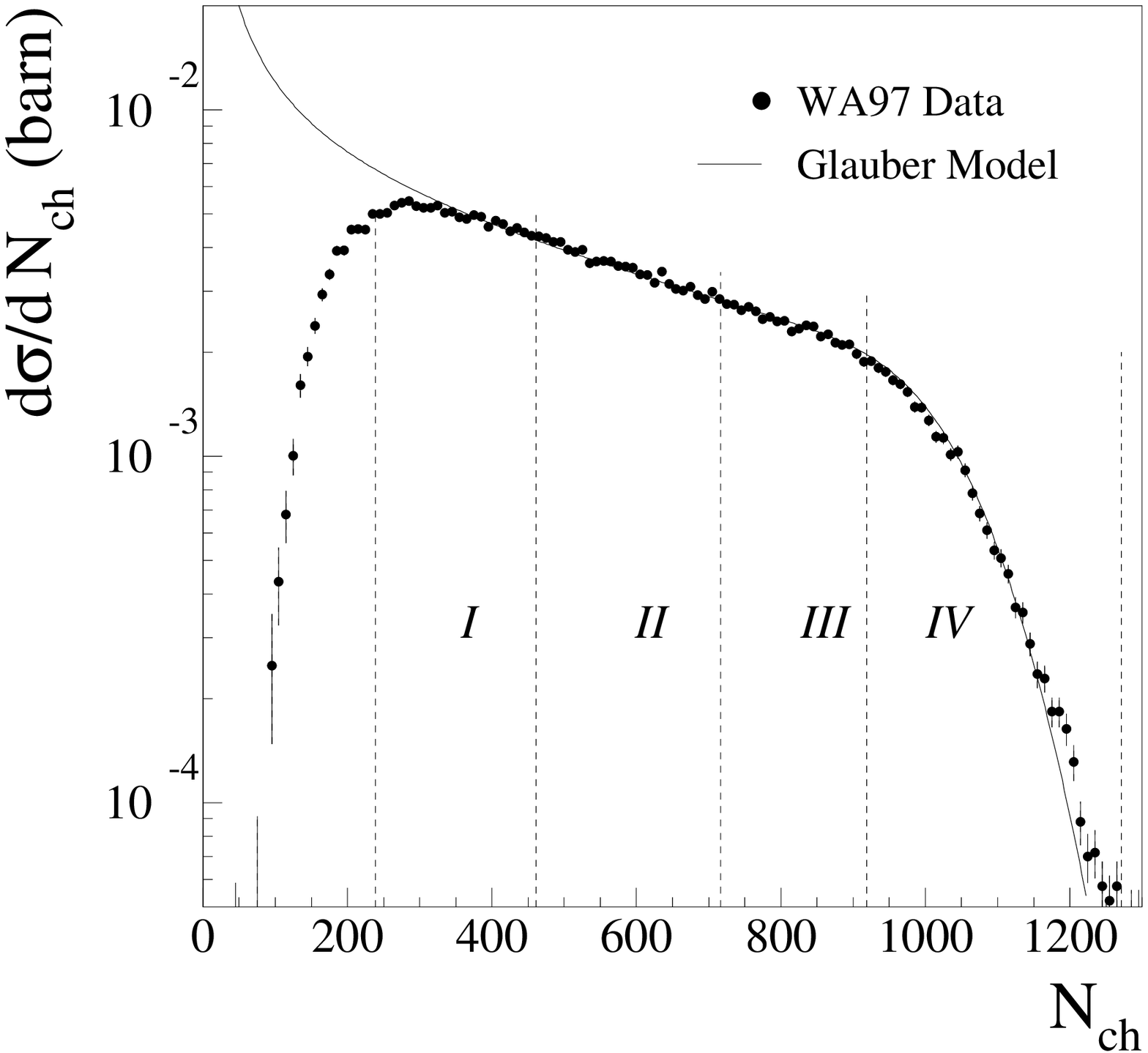}
 \caption{{\bf a)} Negli inserti di sinistra sono mostrate le dipendenze 
  dell'energia trasversa (in alto) e della molteplicit\`a di particelle cariche 
  (in basso) a valori centrali di pseudo-rapidit\`a  
  dal numero di nucleoni partecipanti misurate da WA98~\cite{WA98PRC65}; 
  la linea trattegiata rappresenta 
  l'estrapolazione lineare condotta a partire dalle collisioni pi\`u 
  periferiche ($N_{part} \approx 50$). Per meglio evidenziare il lieve 
  discostarsi dalla linearit\`a, negli inserti di sinistra sono mostrati i valori 
  locali dell'esponente $\alpha$\ in funzione di $N_{part}$, ottenuti 
  eseguendo il fit con i soli cinque punti sperimentali di $E_T$\ ed 
  $N_{ch}$\ pi\`u vicini.
  \newline
   {\bf b)} Fit della distribuzione di molteplicit\`a con il 
   modello dei nucleoni partecipanti eseguito da WA97~\cite{WA97Centr}. 
   Per piccoli valori di $N_{ch}$, la distribuzione sperimentale \`e 
   soppressa a causa del trigger di centralit\`a, che rigetta le interazioni 
   periferiche.}  
  \label{WA97Multiplicity}
 \end{center}
\end{figure}

In fig.~\ref{WA97Multiplicity}.a sono mostrate le distribuzioni di energia 
trasversa (in alto), di cui si \`e gi\`a discusso nella sezione precedente, 
e di molteplicit\`a di particelle cariche (in basso) misurate da 
WA98~\cite{WA98PRC65}. Anche quest'ultima permette di risalire al numero di 
nucleoni partecipanti alla collisione, a partire dalla relazione 
$N_{ch}\propto N_{part}^{1.07\pm0.04}$\ ricavata dal miglior fit con 
una funzione potenza ai dati sperimentali. L'esperimento WA98 ha dunque 
misurato una dipendenza di $E_T$\ ed $N_{ch}$\ da $N_{part}$\ 
leggermente pi\`u rapida della legge lineare, attesa nel modello 
na\"{i}ve dei nucleoni participanti. Le prime misure di $E_T$\ eseguite 
da PHENIX all RHIC, come si \`e visto, si discostano maggiormente dalla 
dipendenza lineare. 
\newline
La distribuzione di particelle cariche misurata da WA97~\cite{WA97Centr}, 
mostrata in fig.~\ref{WA97Multiplicity}.b, \`e invece ancora ben descritta 
in termini di nucleoni partecipanti col modello di Glauber. 
Rilasciando l'ipotesi di linearit\`a col numero di nucleoni partecipanti, 
il fit con la legge di potenza fornisce il valore $\alpha=1.05\pm0.05$, 
compatibile sia con l'ipotesi di proporzionalit\`a ad $N_{part}$\ che con  
i risultati di WA98.   
\subsubsection{Distribuzioni di massa trasversa}
\begin{figure}[hbt]
\begin{center}
\includegraphics[scale=0.38]{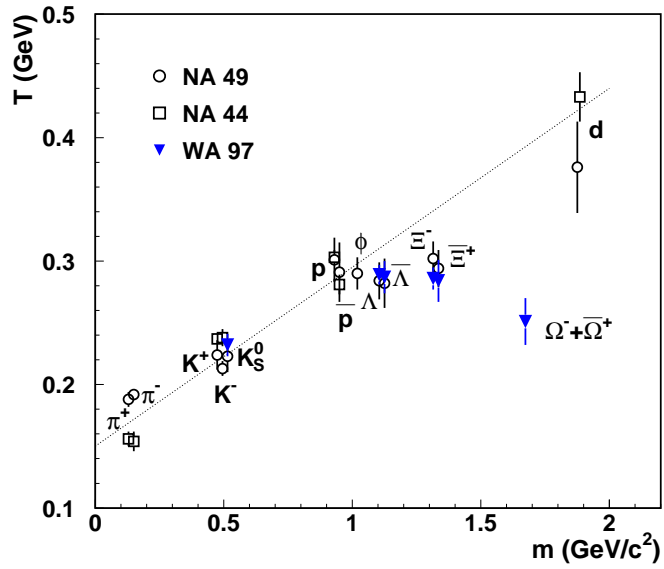}
\caption{Dipendenza dell'inverso della pendenza degli spettri di
  massa trasversa dalla massa della particella nelle collisioni Pb-Pb
   all'SPS. La linea rappresenta l'approssimazione fornita da calcoli
   col modello RQMD per particelle non strane~\cite{RQMD}.}
\label{InverseSlopes}
\end{center}
\end{figure}

Lo studio delle distribuzioni di massa trasversa delle particelle --- 
quantit\`a definita dalla relazione $m_T=\sqrt{p_T^2+m_0^2} $, dove $m_0$\ \`e 
la massa a riposo e $p_T$\ il suo impulso trasverso --- 
prodotte nelle collisioni tra nuclei pesanti 
fornisce indicazioni circa il raggiungimento dell'equilibrio termico 
locale nella sorgente, e sulla dinamica di evoluzione del sistema  
interagente. 
\newline
Indicando con $f_i(x,p)$\ la distribuzione nello spazio delle fasi 
per particelle della specie $i$, dove $x$\ sono le coordinate dello 
spazio-tempo e $p$\ \`e il quadri-vettore energia-impulso, si pu\`o 
mostrare~\cite{Cooper-Frye} che in regime idrodinamico, all'equilibrio 
locale, risulta: 
\begin{equation}
f_i(x,p)= \frac{1}{ \exp \left\{ \left[ p^\mu u_\mu - \sum_k c^{(k)} 
 \mu_i^{(k)}(x)  \right] / T(x) \right\} + \Theta_i }
\end{equation}
dove $T(x)$\ \`e la temperatura locale, $\mu_i^{(k)}(x)$\ e $c^{(k)}$\  
sono, rispettivamente, il potenziale chimico ed il relativo numero 
quantico conservato (di solito stranezza e numero barionico), 
$\Theta_i$\ = +1 per i bosoni e -1 per i fermioni, $u_\mu$\ \`e il 
campo di velocit\`a. 
Nota $f(x,p)$, l'equazione di Cooper-Frye~\cite{Cooper-Frye} 
regola la distribuzione invariante degli impulsi delle particelle 
prodotte al {\em ``freeze-out''}: 
\begin{equation}
E\frac{{\rm d}^3N}{{\rm d}p ^3}=\frac{{\rm d}^2N}{2\pi m_T{\rm d}y{\rm d}m_T}=
       \frac{g}{(2\pi\hbar)^3}\int_{\Sigma_f} f(p,x) p^{\mu}
       {\rm d}^3\Sigma_{\mu}(x)
\label{Cooper-Frye}
\end{equation}
dove  
${\rm d}^3\Sigma_{\mu}(x)$\ \`e l'elemento infinitesimo 
dell'iper-superficie  $\Sigma_f(x)$\ di {\em ``freeze-out''}.  
Due ipotesi fondamentali sono state poste nel ricavare 
l'eq.~\ref{Cooper-Frye}:  l'equilibrio locale per il sistema in regime 
idrodinamico e l'esistenza di una superficie di {\em ``freeze-out''}. 
\newline
L'equilibrio locale ed il regime idrodinamico sussistono fintanto 
che l'intervallo di tempo medio tra due collisioni risulti molto 
pi\`u piccolo dei tempi caratteristici dell'evoluzione macroscopica 
del sistema. Il {\em ``freeze-out''} coincide, in questa descrizione idrodinamica, 
con la rottura dell'equilibrio ed il successivo disaccoppiamento delle 
particelle. Due meccanismi possono portare al {\em ``freeze-out''}: 
l'espansione pu\`o diventare cos\`i rapida che il sistema esce 
dall'equilibrio, oppure il libero cammino medio delle particelle diventa 
maggiore delle dimensioni geometriche del sistema. La combinazione di 
questi due meccanismi fissa la scala dei tempi e definisce 
un'iper-superficie di {\em ``freeze-out''} attraverso la relazione 
\begin{equation}
\tau^{(i)}_{scatt} \lesssim \min \left( 
\tau_{exp},\tau^{(i)}_{escape}\right), 
\label{validity}
\end{equation}
in cui $\tau^{(i)}_{scatt}= \lambda_i/<v_i>$\ \`e il reciproco della frequenza 
di interazione per particelle della specie $i$, con cammino libero medio 
$\lambda_i$\ e velocit\`a media $<v_i>$.  La scala dei tempi che determina 
l'espansione ($\tau_{exp}$) \`e fissata dal profilo della velocit\`a di 
espansione idrodinamica $u(x)$, mentre la scala dei tempi di fuga dalla 
fireball ($\tau^{(i)}_{escape}$) \`e determinata dalle dimensioni 
geometriche della fireball ($R$) e dalla velocit\`a media $<v_i>$: 
\begin{subequations}
\begin{equation}
\tau_{exp}=\frac{1}{\partial_\mu u^\mu(x)}
\nonumber
\end{equation}
\begin{equation}
\tau^{(i)}_{escape}=\frac{R}{<v_i>}
\nonumber
\end{equation}
\end{subequations}
Quando i criteri definiti dall'eq.~\ref{validity} non sono pi\`u 
soddisfatti, termina l'evoluzione idrodinamica e le particelle cessano 
di interagire.
\newline
In~\cite{Cooper-Frye,28} vengono discusse diverse soluzioni relative 
a differenti propriet\`a di simmetria dell'iper-superficie di 
{\em ``freeze-out''} $\Sigma_f$. Qui si prenderanno in considerazione due ipotesi 
di {\em ``freeze-out''}: il caso di una fireball stazionaria in cui non c'\`e moto 
collettivo di espansione, ed il caso di una superficie di {\em ``freeze-out''} a 
simmetria cilindrica in cui il sistema sia soggetto ad un'espansione 
collettiva  longitudinale e trasversale.
\begin{itemize}
\item {\bf Fireball stazionaria} \\
Nel caso limite di una fireball stazionaria in equlibrio, non c'\`e 
correlazione tra la variabile spaziale $x$\ e l'impulso $p$\ nelle 
distribuzioni di spazio delle fasi, che quindi si possono fattorizzare.  
In approssimazione di Boltzmann, si ottiene infatti:
\begin{equation}
f(p,x)=\lambda e^{-(p \cdot u)/T}
\end{equation}
dove $\lambda$\ e la quadri-velocit\`a $u$\ assumono espressione: 
\begin{subequations}
\begin{equation}
\lambda = e^{(\sum_K c^{(k)} \mu^{(k)})/T},
\end{equation}
\begin{equation}
u = (\cosh y_{CM},0,0,\sinh y_{CM}), 
\end{equation}
\end{subequations}
in cui $y_{CM}$\ \`e la rapidit\`a del centro di massa e $\lambda$\ \`e la 
fugacit\`a~\footnote{La fugacit\`a $\lambda$\ di una data specie \`e definita 
a partire dal potenziale chimico e dalla temperatura, 
secondo la relazione: $\lambda = \exp[\mu/kT]$.} 
della specie in questione.
\newline
Lo spettro determinato dall'eq.~\ref{Cooper-Frye} ha quindi la forma: 
\begin{equation}
E\frac{{\rm d}^3N}{{\rm d}p ^3}=
       \frac{g\lambda V}{(2\pi)^3} m_T \cosh(y-y_{CM}) 
       \exp \left[-\frac{m_T \cosh(y-y_{CM}) }{T}  \right], 
\label{Stationary}
\end{equation}
dove $V$\ \`e il volume che si ottiene dall'integrazione 
$\int_{\Sigma_f}p^{\mu} {\rm d}^3\Sigma_{\mu} \longrightarrow EV$. 
Dall'eq.~\ref{Stationary} si ricava che, per dati raccolti in una 
finestra di rapidit\`a ristretta attorno a $y_{CM}$\ 
($\Delta y \lesssim 0.5$), la distribuzione di massa trasversa 
dovrebbe essere interpolata con una funzione della 
forma~\cite{Cooper-Frye}: 
\begin{equation}
\frac{{\rm d}^2N}{{\rm d}y{\rm d}m_T} \propto m_T e^{-m_T/T_{eff}}
\end{equation} 
dove la temperatura efficace $T_{eff}$\ \`e legata alla temperatura 
vera da: 
\begin{equation}
T_{eff}=\frac{T}{\cosh(y-y_{CM})}
\end{equation}
Se invece lo spettro di massa trasversa viene integrato su una grande 
finestra di rapidit\`a attorno a $y_{CM}$, \`e opportuno considerare 
l'integrale nella variabile $y$\ dell'eq.~\ref{Stationary}:
\begin{equation}
\frac{{\rm d}N}{{\rm d}m_T} \propto \sqrt{m_T} e^{-m_T/T} \quad\quad\quad
 {\rm per \; m_T \gg T.}
\end{equation}
\item {\bf Espansione collettiva trasversa e longitudinale}\\
Il parametro $T_{app}$\ determinato dal {\em fit} delle distribuzioni di 
massa trasversa non pu\`o in realt\`a essere interpretato come 
temperatura della sorgente termica a causa dell'espansione cui \`e 
sottoposto il sistema. Infatti, uno stato di equilibrio termico implica 
la presenza di frequenti interazioni tra le particelle del sistema. In 
prossimit\`a della superficie, queste interazioni generano un gradiente 
di pressione che d\`a origine ad un flusso trasverso collettivo. 
Inoltre, a causa del parziale frenamento, una frazione dell'impulso 
longitudinale dei nucleoni incidenti non viene dissipata, ed \`e dunque 
legittimo assumere una notevole espansione longitudinale della sorgente, 
la cui intensit\`a pu\`o essere stimata dallo studio delle distribuzioni 
di rapidit\`a (tipicamente la velocit\`a dell'espansione risulta pari 
a $\beta_L = 0.95$). 
\newline
Per collisioni centrali, o quando si analizzano dati anche per collisioni 
periferiche ma mediate su tutte le possibili orientazioni del parametro 
di impatto ${\bf b}$, si pu\`o assumere una simmetria cilindrica rispetto 
alla linea di fascio. 
\newline
L'espansione longitudinale viene generalmente descritta assumendo un 
profilo di Bj{\o}rken~\cite{Bjo83}, vale a dire una rapidit\`a del 
flusso longitudinale indipendente dal tempo proprio (cio\`e un profilo di 
velocit\`a invariante per {\em ``boost''} di Lorentz longitudinali), 
mentre per l'espansione trasversa si pu\`o assumere una rapidit\`a 
che aumenti linearmente con la distanza dal centro della fireball 
({\em cfr. paragrafo 6.2.7}).  
\newline
Le conclusioni che \`e possibile trarre svolgendo calcoli con questo 
tipo di modelli, possono essere cos\`i riassunte:  
\begin{itemize}
\item[-] La velocit\`a di espansione longitudinale determina la distribuzione 
      di rapidit\`a, mostrando una dipendenza molto debole dalla temperatura 
      $T$\ e dal moto di espansione trasversa;
\item[-] Il flusso trasverso ha l'effetto di innalzare la temperatura 
         misurata rispetto a quella presente al {\em ``freeze-out''}, secondo 
	 un effetto di ``spostamento Doppler verso il blu''. Di fatto 
	 il parametro $T_{app}$\ misura sia il contributo dovuto al moto 
	 termico (quindi la temperatura di {\em ``freeze-out''}) che quello 
	 dovuto al moto collettivo (quindi l'energia cinetica del moto 
	 trasverso) all'energia delle particelle rilevate.
\end{itemize}
Per alti valori della massa trasversa
\begin{equation}
T_{app}\equiv\left[\lim_{m_{T}\rightarrow \infty} \frac{{\rm d}}{{\rm d}m_T}
  (\ln\frac{{\rm dN}}{{\rm d}m^2_T}) \right]^{-1} \approx 
   T \sqrt{\frac{1+<\beta_\perp>}{1-<\beta_\perp>}}
\label{Tapp1}
\end{equation}
dove $<\beta_\perp>$\ \`e la velocit\`a media del flusso trasverso. 
\newline
Nel limite non relativistico invece, cio\`e per $p_T \ll m_0$\ 
($m_0$\ \`e la massa a riposo della particella), la temperatura 
apparente $T'_{app}$\ vale~\cite{Carrer29}: 
\begin{equation}
T'_{app}\equiv\left[\lim_{m_{T}\rightarrow m_0} \frac{{\rm d}}{{\rm d}m_T}
  (\ln\frac{{\rm dN}}{{\rm d}m^2_T}) \right]^{-1} = 
     T + m_0 <\beta^2_\perp>
 \label{Tapp2}
\end{equation}
\end{itemize}
La misura del singolo spettro di massa trasversa, da cui \`e possibile 
estrarre $T_{app}$,  non permette quindi di determinare in modo univoco 
sia la temperatura di {\em ``freeze-out''} $T$\ sia la velocit\`a $<\beta_\perp>$. 
Come si esporr\`a in questo lavoro, \`e possibile estrarre informazioni 
indipendenti sulle variabili $T$\ e $<\beta_\perp>$\ dall'analisi 
%degli spettri di doppie particelle (analisi HBT),  
delle correlazioni fra particelle identiche (analisi HBT), 
in modo tale da poter  
determinare in modo univoco i valori delle due variabili. Un altro metodo 
di analisi, che verr\`a anch'esso sviluppato, \`e quello di studiare gli spettri 
di massa trasversa di {\em diverse} particelle, in modo da poter nuovamente 
separare il contributo termico da quello del flusso trasverso.
\newline
Un'evidenza sperimentale dell'esistenza dell'espansione collettiva trasversale
\`e fornita dal sistematico aumento del parametro $T$\ con la massa della
particella, cio\`e con la sua energia cinetica trasversa, come mostrato
in fig.~\ref{InverseSlopes}.
Nell'espansione idrodinamica, infatti, tutte le particelle si muovono
con la stessa velocit\`a trasversa e, classicamente,  
l'energia cinetica
(collettiva) delle particelle dipende dalla massa delle particelle: pertanto
le particelle di massa pi\`u elevata trasportano maggiore energia,
in accordo con l'eq.~\ref{Tapp2}.  
\subsection{L'interferometria HBT}
Lo studio della correlazione tra particelle identiche, nota come tecnica 
di Hanbury-Brown e Twiss (HBT), riveste un ruolo sempre pi\`u importante 
nella fisica degli ioni pesanti ultra-relativistici. Questa tecnica, 
inizialmente sviluppata negli anni sessanta 
in campo astrofisico per la misura dei raggi 
stellari~\cite{Hanbury}, venne riscoperta in maniera indipendente nella 
fisica delle particelle pochi anni dopo~\cite{Goldhaber}. 
\newline
Sfruttando la propriet\`a di simmetria (anti-simmetria) di cui gode la
funzione d'onda descrivente due bosoni (fermioni) identici, 
questa tecnica permette di ricavare informazioni sulle dimensioni 
geometriche, anche di pochi fermi, 
della sorgente da cui le particelle sono state emesse.  
Nel caso di sorgenti stastiche 
(il caso astrofisico della misura dei raggi stellari), 
\`e possibile ricavare direttamente le dimensioni della sorgente; 
nel caso in cui l'emissione \`e accompagnata da una dinamica di 
espansione della sorgente (il caso delle collisioni tra nuclei pesanti 
ultra-relativistici), \`e anche possibile ricavare informazioni su tale 
dinamica, servendosi di opportuni modelli.
\newline
Il sesto capitolo di questo lavoro sar\`a dedicato completamente  
all'interferometria HBT: si rimanda dunque a quel capitolo 
per i dettagli teorici e sperimentali alla base del metodo.
\newline
Una comprensione dettagliata della geometria spazio-temporale della 
sorgente negli istanti del {\em ``freeze-out''} e dell'evoluzione 
dinamica della zona di reazione pu\`o essere fornita dalla 
interferometria HBT. Queste informazioni forniscono un importante punto 
di partenza sperimentale per un'estrapolazione dinamica, all'indietro, 
verso gli stati iniziali densi e caldi della collisione.  
\newline
Si vedr\`a ad esempio come sia possibile ottenere un'informazione  
indipendente sulla coppia di variabili $T$\ e $<\beta_\perp>$\ che, 
combinata con quella proveniente dagli spettri di massa trasversa 
di singola particella, permette di separare le due quantit\`a.  
\subsection{La soppressione nella produzione degli stati di  charmonio}  
\begin{figure}[h]
\begin{center}
\includegraphics[scale=0.50]{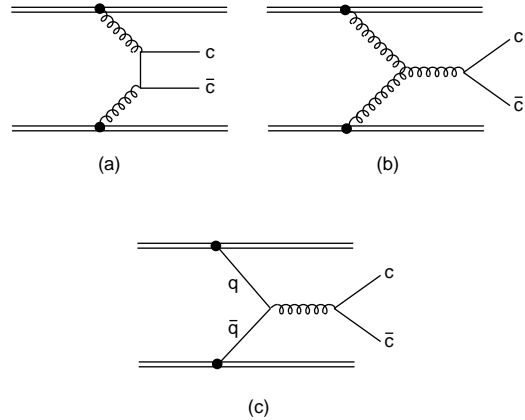}
\caption{Diagrammi di ordine inferiore per la produzione di $c\bar{c}$ 
         in collisioni adroniche, per fusione di 
	 due gluoni ({\bf a}, {\bf b}) o per annichilazione di un quark ed 
	 un anti-quark ({\bf c}).}
\label{Jproduction}
\end{center}
\end{figure}

\begin{figure}[h]
\begin{center}
\includegraphics[scale=0.27,angle=-90]{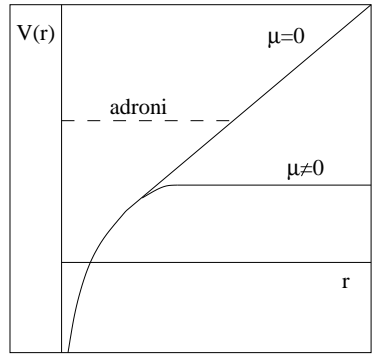}
\caption{Andamento del potenziale effettivo di QCD nel vuoto ($\mu=0$) e 
         in un mezzo ``colorato'' ($\mu \ne 0$).}
\label{Screening}
\end{center}
\end{figure}

\begin{figure}[h]
\begin{center}
{\bf a)} \hspace{8.0cm} {\bf b)} \\
\hspace{-1.2cm}
\includegraphics[scale=0.38]{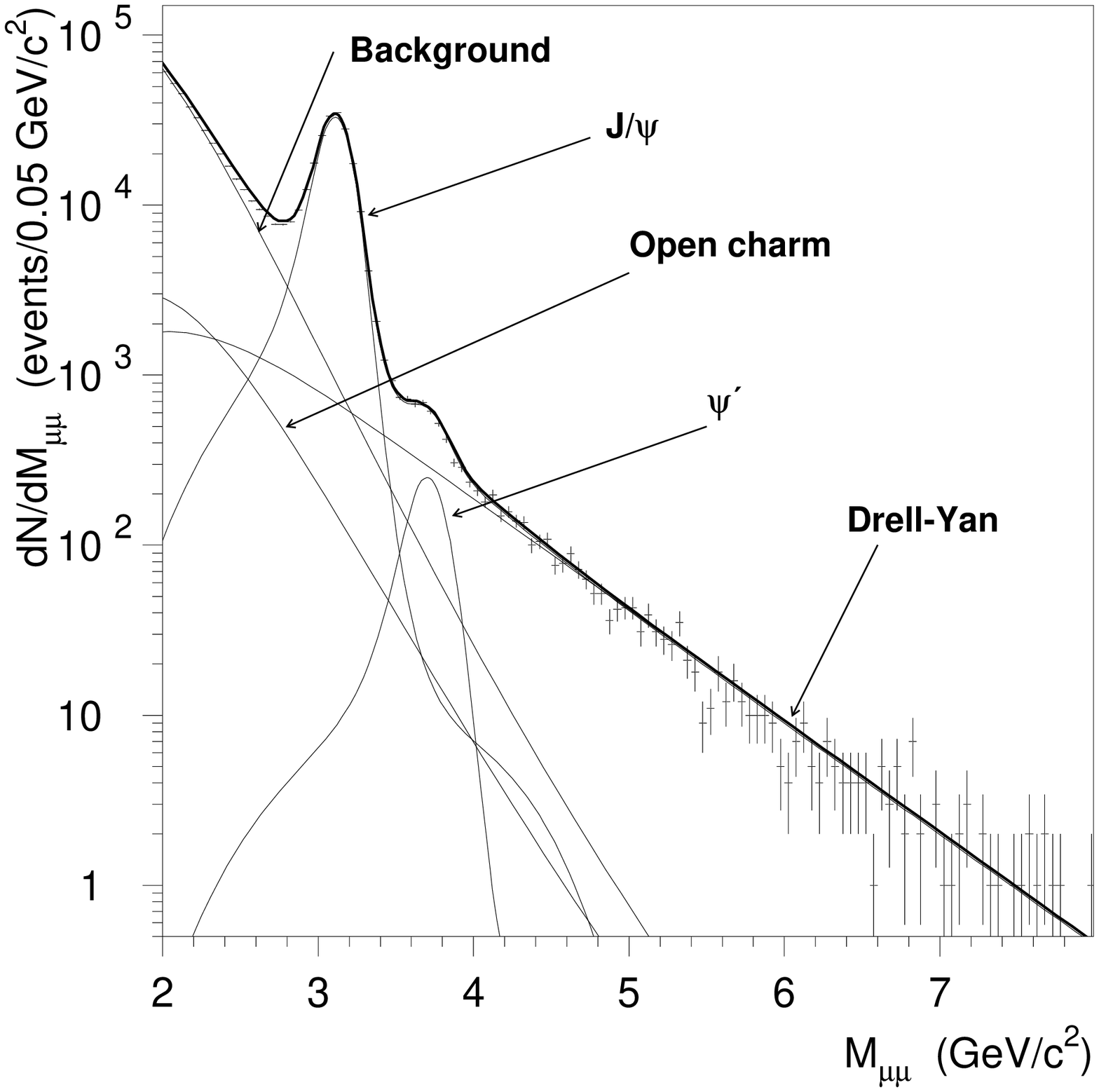}
\includegraphics[scale=0.66]{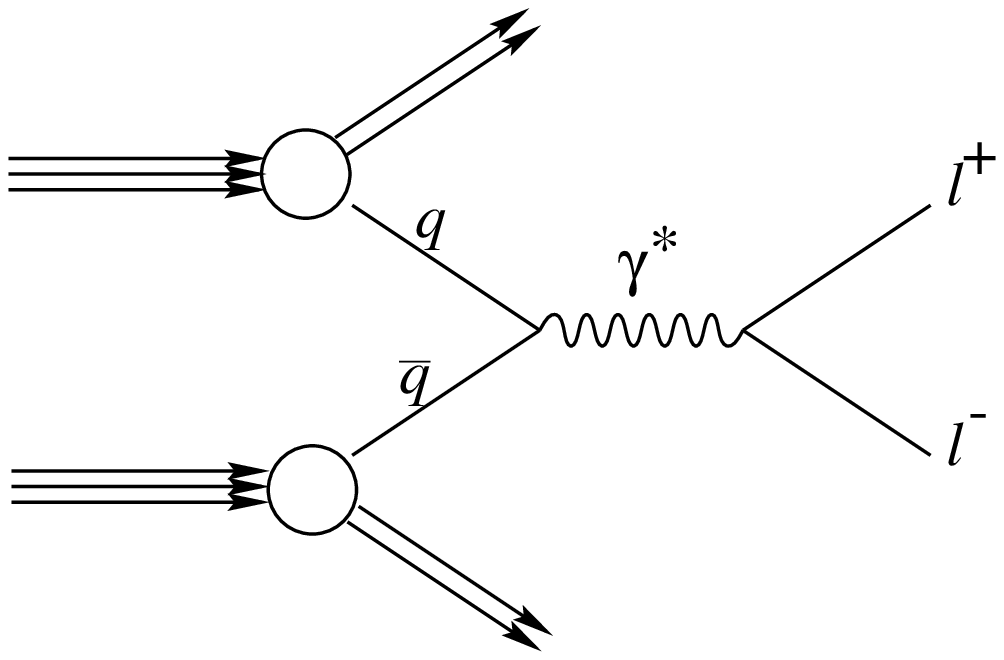}
 \caption{ {\bf a)} Spettro di massa invariante di coppie \Pgmp \Pgmm~\cite{JPSI1}.  
           {\bf b)} Il processo di Drell-Yan di produzione di coppie leptoniche.} 
\label{JPSI}
\end{center}
\end{figure}

\begin{figure}[h]
\begin{center}
\includegraphics[scale=0.37]{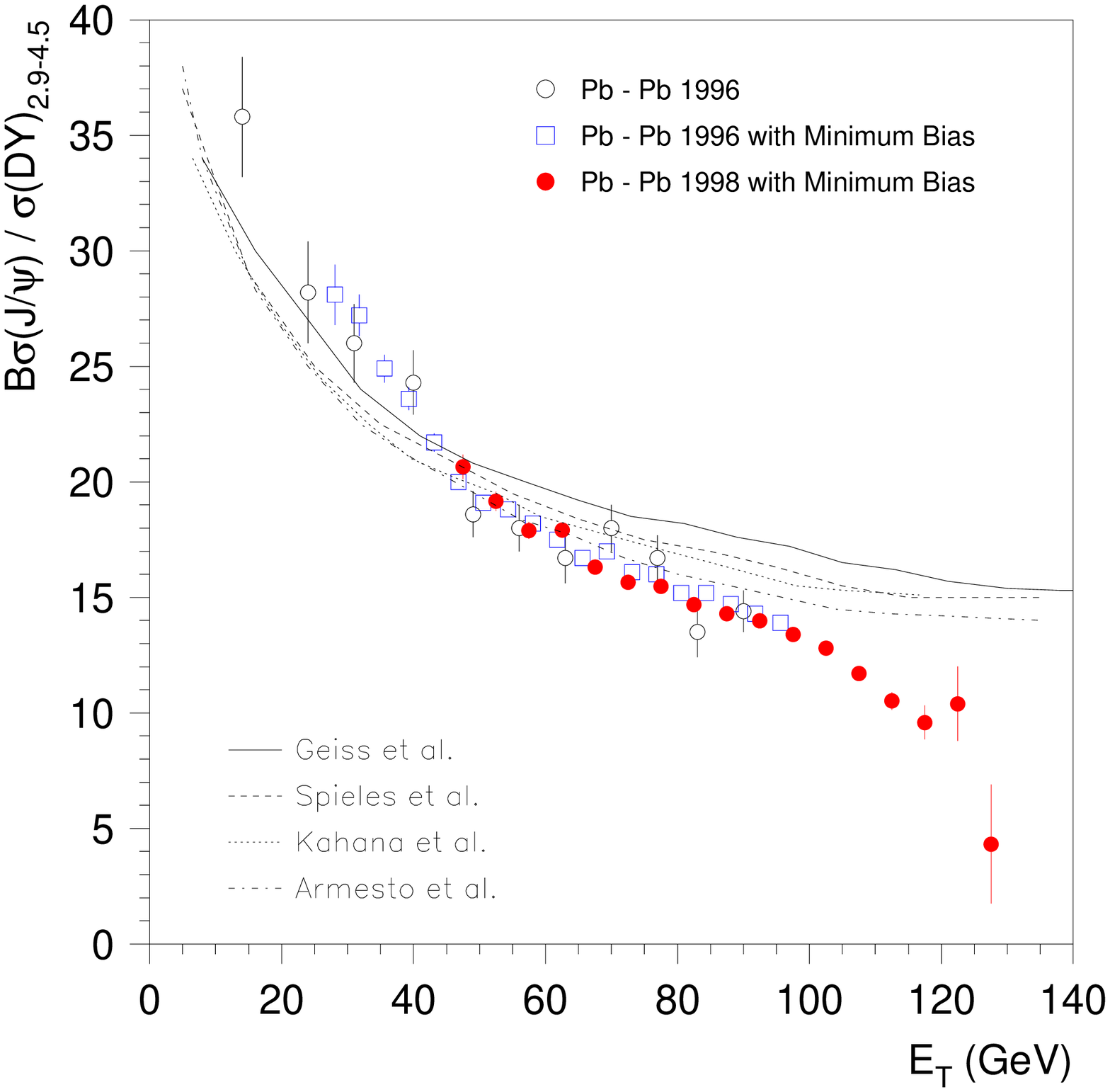}
\includegraphics[scale=0.37]{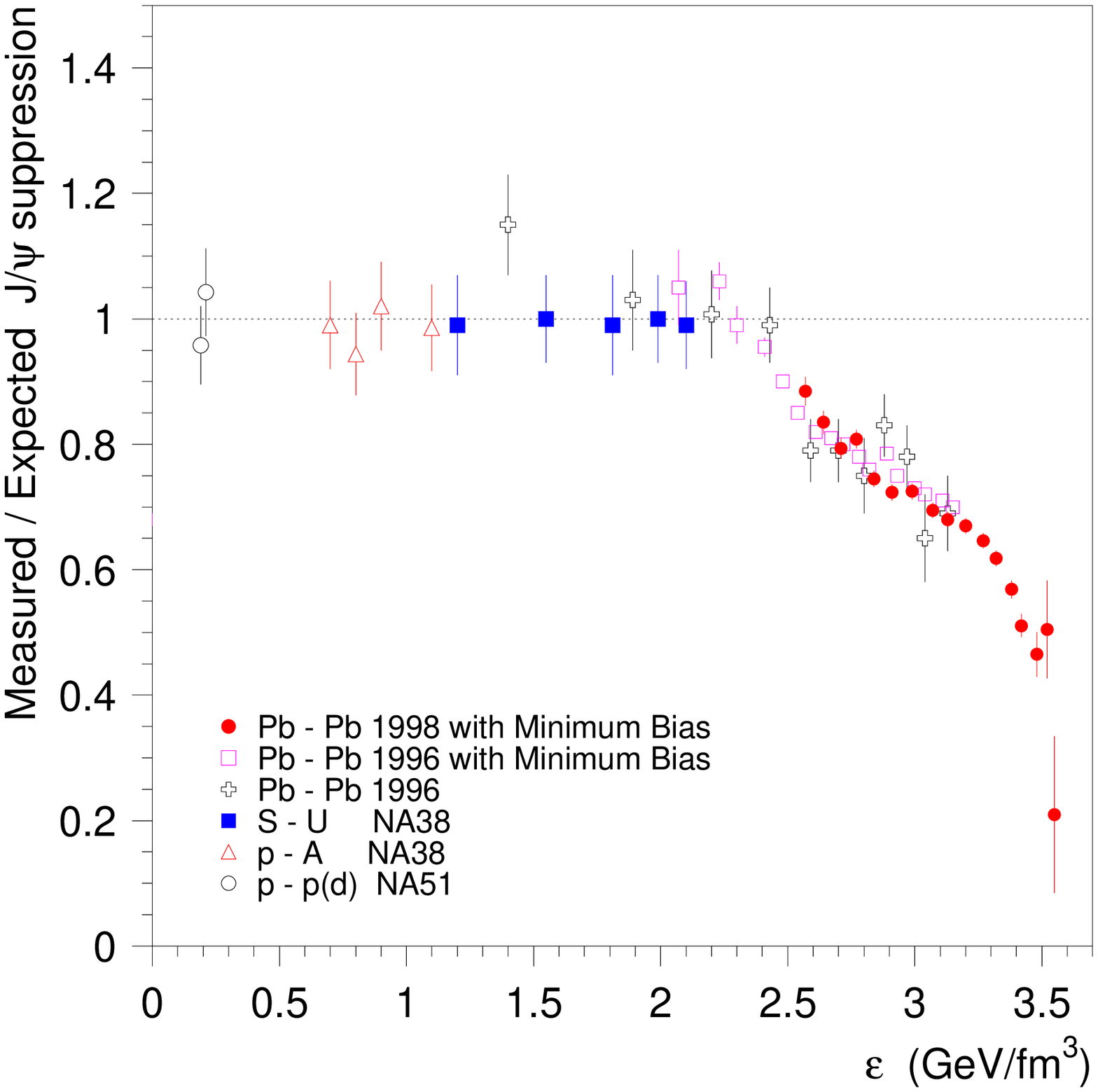}
 \caption{{\bf a)} Rapporto $\sigma_{J/\psi}/\sigma_{DY}$\ in funzione dell'energia 
                 trasversa, confrontato con calcoli convenzionali, in assenza 
		 di QGP.  
         {\bf b)} {\em Yield} misurato della $J/\psi$\  normalizzato allo {\em ``yield''}  
	 atteso nell'ipotesi di soppressione per normale assorbimento del mezzo nucleare, 
	  in funzione della densit\`a di energia raggiunta nelle collisioni, per 
	  diversi sistemi~\cite{JPSI2}.}  
 \label{Evidence}
 \end{center}
\end{figure}

Nel 1986, T.~Matsui e H.~Satz predissero che la produzione della particella 
$J/\psi$\ dovesse essere soppressa all'interno del QGP~\cite{Satz}.  
Il charmonio \`e uno stato legato $c\bar{c}$ che pu\`o esser prodotto 
nelle collisioni nucleari in seguito a processi di fusione del tipo 
$gg \longrightarrow c\bar{c}$\ ed annichilazione  
$q\bar{q} \longrightarrow c\bar{c}$.  
Poich\'e la massa del quark $c$\ \`e elevata, il processo di formazione 
di questi stati si pu\`o descrivere con la QCD perturbativa, come mostrato 
in fig.~\ref{Jproduction}.   
Lo spettro del charmonio pu\`o essere calcolato con buona precisione~\cite{Schr} 
per mezzo dell'equazione di Schr\"{o}dinger, dal momento che il quark 
$c$\ non \`e relativistico per via della sua elevata massa:
\begin{equation}
\left[2m_c+\frac{1}{m_c}\nabla^2 + V(r) \right]\Psi_{n,l} = M_{n,l}\Psi_{n,l}
\label{Schr}
\end{equation}
dove il potenziale $V(r)=-\frac{\alpha}{r} + \lambda r $ contiene 
il termine di interazione a lunga distanza ($\lambda r $) responsabile 
del confinamento ed un termine di interazione a breve distanza di tipo 
coulombiano. Per diversi valori del numero quantico principale $n$ ed 
orbitale $l$, le masse $M_{n,l}$ e le funzioni d'onda $\Psi_{n,l}$\ 
dei diversi stati di charmonio $J/\psi$, $\chi_c$, $\psi'$, ... sono 
calcolabili in termini delle costanti $m_c$, $\lambda$ ed $\alpha$.  
\newline
Nell'ipotesi di formazione del QGP, l'alta densit\`a di cariche di colore 
produce uno schermaggio della forza forte, in analogia  
%all'analogo 
al 
fenomeno di schermaggio di Debye della forza elettrica  
in un mezzo denso. Il potenziale elettrico per due cariche $e_0$ poste 
nel vuoto a distanza r vale infatti $V_0(r)=\frac{e_0^2}{r}$. In un ambiente 
denso di altre cariche elettriche il potenziale viene ``schermato'' e diventa 
$V(r)=\frac{e_0^2}{r} \exp(-\mu r)$, dove $r_D=\mu^{-1}$ \`e il 
{\em raggio di schermaggio di Debye} caratteristico del mezzo; esso diminuisce 
all'aumentare della densit\`a del mezzo  carico.  Se un sistema legato, 
caratterizzato da un raggio medio $r_B$, quale ad esempio
l'atomo di idrogeno, viene posto in un mezzo con densit\`a di carica cos\`i 
elevata da risultare $r_D \ll r_B$, la forza effettiva di legame tra le due 
particelle del sistema diviene cos\`i a corto raggio che il 
sistema si scioglie. Questo implica altres\`i che un mezzo dielettrico isolante 
possa diventare elettricamente conduttore se portato a densit\`a 
sufficientemente elevate: l'effetto dello schermaggio \`e quello di rompere 
i legami negli atomi costituenti il mezzo, ed il mezzo si porta in un nuovo 
stato di {\em plasma} in cui le cariche elettriche sono praticamente libere.  
\newline
Analogamente, nel caso di due cariche di colore immerse in un mezzo ad elevata 
densit\`a di carica di colore, quale il QGP,  il termine di confinamento 
del potenziale --- nel vuoto pari a $V(r)=\lambda r$ --- 
diventa\footnote{La forma della funzione che descrive lo schermaggio 
dipende dal tipo di funzione del potenziale non schermato~\cite{Dixit}; questo 
\`e il motivo del diverso tipo di schermaggio per il potenziale elettrico e 
quello di colore.} $V(r) = \lambda r \left[ \frac{1-\exp(-\mu r)}{\mu r} \right]$. 
Anche qui la massa di schermaggio $\mu$\ \`e pari all'inverso del raggio di 
schermaggio per le cariche di colore.  
L'espressione completa del potenziale effettivo del charmonio assume dunque la 
forma: 
\begin{equation}
V(r)=\frac{\lambda}{\mu} \left[ 1-\exp(-\mu r) \right]-\frac{\alpha}{r}\exp(-\mu r).
\label{VScreened}
\end{equation}
L'andamento  del potenziale effettivo nel vuoto ed in presenza di
schermaggio della carica di colore \`e mostrato in fig.~\ref{Screening}.  
Il termine esponenziale elimina la dipendenza lineare dalla distanza, 
responsabile del confinamento. 
\newline
Il valore del raggio di Debye pu\`o essere calcolato~\cite{DebyePQCD} con 
la QCD perturbativa al primo ordine: 
\begin{equation}
r_D=\frac{1}{\mu}=\frac{1}{\sqrt{(\frac{N_C}{3}+\frac{N_f}{6})g^2}T}; 
\label{Debye}
\end{equation}
il raggio di Debye \`e dunque funzione della temperatura e della costante  
di accoppiamento $g$.  
Un altro approccio per determinare la dipendenza di $\mu$ da T si serve di calcoli 
di QCD su reticolo a temperatura non nulla. Assumendo $T_c=170$ MeV e la relazione approssimata 
$\mu\simeq 3.3 T$~\cite{65}, si ottengono i parametri della dissociazione del charmonio,  
riassunti nella  tabella~1.4.  
\begin{table}[t]
 \label{tab14}
 \begin{center}
  \begin{tabular}{||c||c|c|c||} \hline
     {\bf Stato}            & $J/\psi$(1S)& $\psi'$(2S) & $\chi_c$(1P) \\ \hline \hline
     {\bf $\mu_d$ [GeV]}    & 0.68       & 0.35        & 0.35       \\ \hline
     {\bf $T_d/T_C$    }    & 1.2        & 1           & 1          \\ \hline
     {\bf $\Delta E$ [GeV]} & 0.64       & 0.06        & 0.24       \\ \hline 
\end{tabular}
\end{center}
\caption{Dissociazione del charmonio per schermaggio della carica di colore.} 
\end{table}
Per il mesone $J/\psi$\ --- caratterizzato dalla pi\`u alta energia di legame 
 (nel vuoto) $\Delta E_{J/\psi} = 2 M_D - M_{J/\psi} \simeq 0.64$ GeV, 
 pari alla differenza tra la soglia di produzione di coppie di mesoni 
 con numero di charm non nullo  ({\em ``open charm''}) e la sua massa ---  
si ottiene ad esempio, un raggio di Debye $r_D \simeq 0.29$ fm, e la 
dissociazione pu\`o avvenire ad una temperatura $T_d=1.2\, T_C$, di poco superiore 
alla temperatura critica.  Gli stati eccitati del sistema $c\bar{c}$, 
quali $\psi'$\ e $\chi_c$, possono  
%essere dissociati pi\`u facilmente 
formarsi pi\`u difficilmente   
e dovrebbero essere soppressi non appena la temperatura raggiunge il valore critico. 
\newline
La produzione dei mesoni vettori $J/\psi$, $\psi'$ \`e stata studiata  
attraverso il loro decadimento in coppie \Pgmp \Pgmm\ dagli 
esperimenti NA38 ed NA50 in vari sistemi di collisione. In fig.~\ref{JPSI}.a 
\`e mostrato lo spettro invariante delle coppie muoniche rivelate in 
interazioni Pb-Pb. Esso \`e interpretato come sovrapposizione dei contributi 
relativi al decadimento della $J/\psi$\ e della $\psi'$\ ai processi 
di Drell-Yan (fig.~\ref{JPSI}.b) ed al fondo costituito da coppie muoniche 
provenienti da decadimenti simultanei di pioni e kaoni. \`E quindi possibile 
misurare separatamente la sezione d'urto per i processi di Drell-Yan e per la 
produzione di $J/\psi$\ e $\psi'$ nel canale di decadimento osservato, indicate 
rispettivamente con $\sigma(DY)$, 
$B_{\Pgmp\Pgmm}\sigma(J/\psi)$ e $B_{\Pgmp\Pgmm}\sigma(\psi')$, dove $B_{\Pgmp\Pgmm}$\ \`e la 
frazione di decadimenti nel canale $\Pgmp\Pgmm$. Sebbene lo  
studio della produzione della  $J/\psi$\ sia il pi\`u agevole da un punto di vista sperimentale 
(maggior sezione d'urto di produzione), esso \`e complicato dal 
fatto che gli stati di charmonio di pi\`u alta energia 
(quali la  $\psi'$\ e la $\chi_c$) decadono frequentemente in stati di 
pi\`u bassa energia: 
di tutte le $J/\psi$ ricostruite, circa il 62\%  sono prodotte direttamente, 
mentre $\approx$ 30\% (8\%) proviene dal decadimento del mesone 
$\chi_c$\ ($\psi'$).   
\newline
La sezione d'urto $\sigma(DY)$\ risulta proporzionale al numero di nucleoni
partecipanti alla collisione, poich\'e i processi di Drell-Yan provengono dalle singole
collisioni primarie tra i nucleoni, per cui viene utilizzata come elemento di
normalizzazione nel valutare la produzione di $J/\psi$ in diversi sistemi di collisione.
\newline
La produzione della $J/\psi$ presenta una naturale decrescita in funzione delle dimensioni 
del sistema interagente dovuto all'assorbimento all'interno della materia nucleare. Esso 
\`e interpretabile in termini di assorbimento dello stato pre-risonante $c\bar{c}$, non 
imputabile allo stato di plasma, quanto piuttosto a processi anelastici del tipo 
$\pi \, + \, J/\psi \, \longrightarrow \, D \, \overline{D} \, X$\ innescati da adroni 
di sufficiente energia in presenza del gas adronico~\cite{NormalAbsor}.  
\newline
I pi\`u recenti risultati di NA50~\cite{JPSI2}\cite{JPSI3}, riportati in fig.~\ref{Evidence},  
suggeriscono che vi sia un addizionale meccanismo di soppressione della $J/\psi$\ 
nelle collisioni Pb-Pb, non presente nelle interazioni S-U ({\em soppressione anomala}). 
Inoltre indicano che vi sia un effetto di soglia nella 
{\em soppressione anomala} della $J/\psi$\ che si verifica per le collisioni Pb-Pb pi\`u centrali, 
corrispondenti ad energie trasverse ($E_T$) pi\`u elevate. 
Infine, l'andamento della soppressione sembra suggerire, andando dalle collisioni periferiche 
verso quelle centrali, una  decrescita con una doppia soglia: la prima, meno intensa, potrebbe 
corrispondere alla pi\`u facile dissociazione della $\chi_c$\ nel 
QGP~\footnote{Si ricorda che circa il 30\% delle $J/\psi$\ osservate proviene dal 
decadimento radioattivo della $\chi_c$.}, 
la seconda, pi\`u evidente,  alla dissociazione diretta della $J/\psi$. 
\newline
Questi risultati hanno innescato un'intensa attivit\`a teorica che al momento pu\`o essere 
cos\`i sintetizzata:  precedenti modelli di assorbimento nucleare non sono in grado di predire 
la soppressione anomala, per cui alcuni autori~\cite{Pro1}\cite{Pro2} hanno attribuito il fenomeno 
alla formazione del QGP; altri ~\cite{Capella}~\cite{Gavin} hanno sviluppato modelli pi\`u 
conservativi che riescono a riprodurre la sopressione anomala in collisioni Pb-Pb per mezzo 
delle interazioni con gli adroni, mostrando tuttavia un disaccordo con i precedenti risultati 
relativi a nuclei pi\`u leggeri. 

\subsection{I segnali elettromagnetici}  
I segnali elettromagnetici sono per molti aspetti ideali per la rivelazione del 
QGP in quanto permettono di sondare  i primi e pi\`u caldi stadi dell'evoluzione del 
sistema interagente, non essendo affetti dalle successive interazioni forti. 
La loro intensit\`a \`e tuttavia molto bassa e devono essere distinti dall'abbondante 
fondo proveniente dai processi di decadimento elettromagnetico degli adroni.  
\subsubsection{I fotoni}  
\begin{figure}[htb]
\begin{center}
{\bf a)} \hspace{8.0cm} {\bf b)} \\
\hspace{-1.2cm}
\includegraphics[scale=0.35]{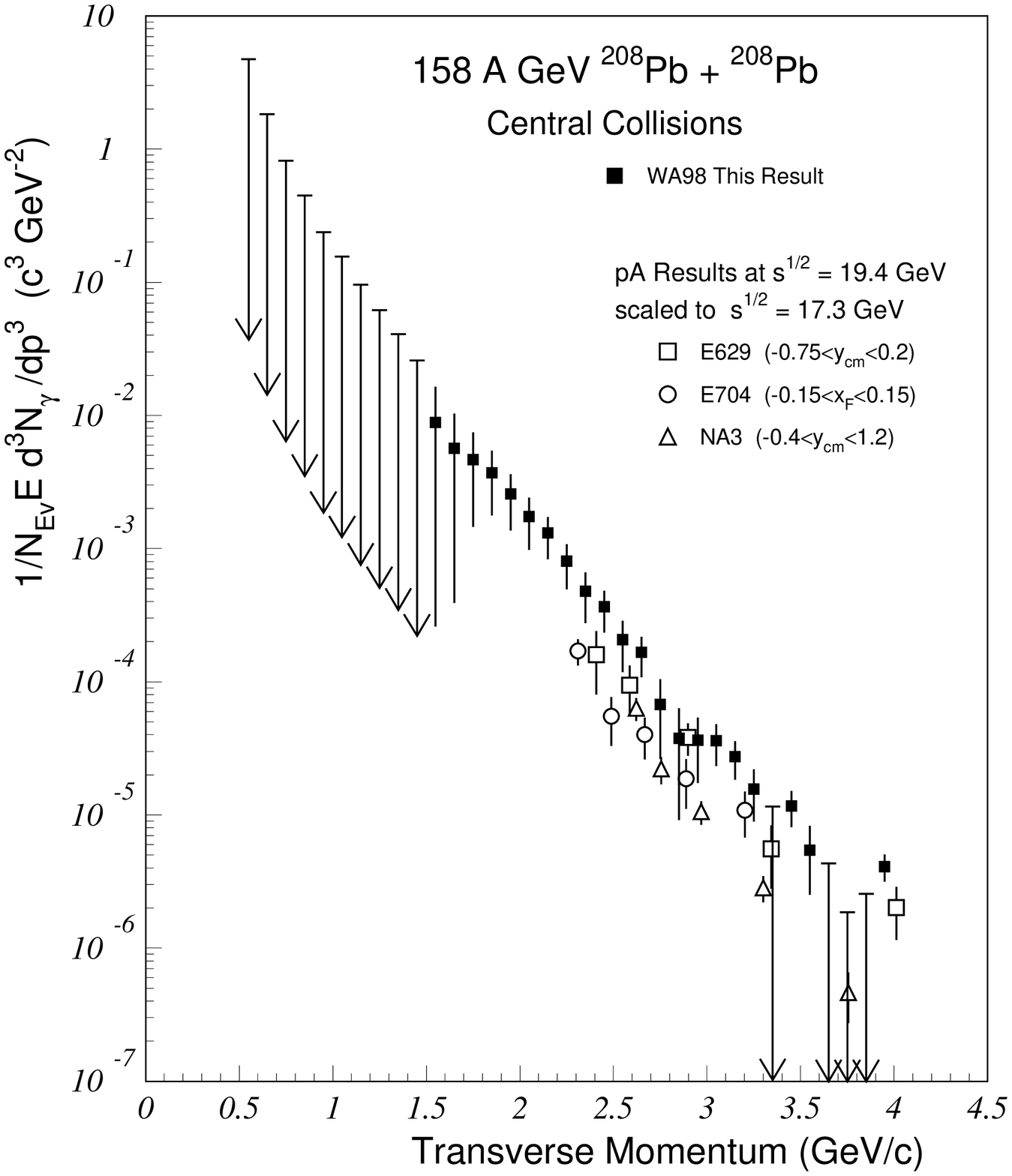}
\includegraphics[scale=0.36]{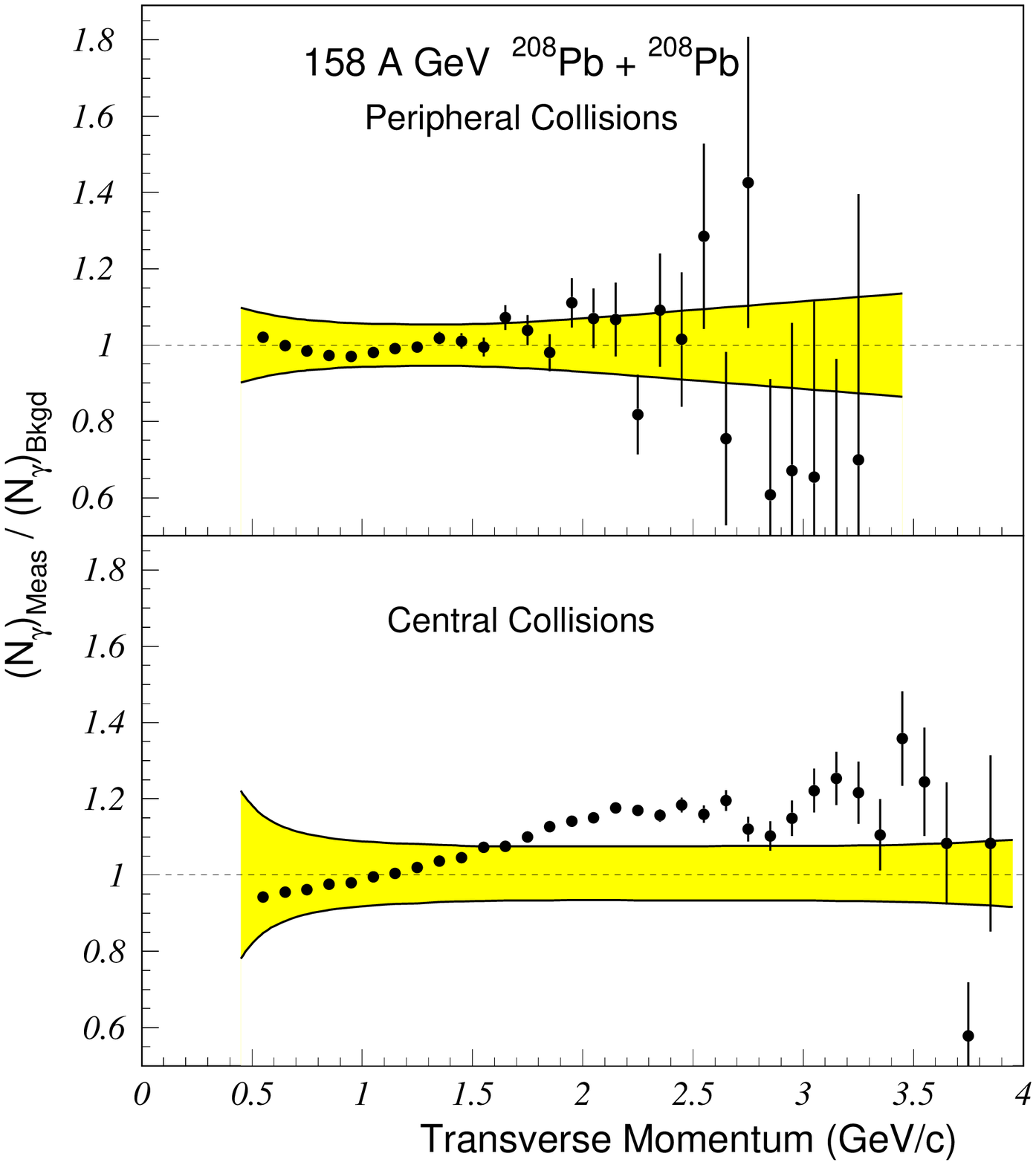} 
 \caption{{\bf a)} Spettro di impulso trasverso dei fotoni diretti misurato dalla 
          collaborazione WA98 in collisioni centrali Pb-Pb a 
	  160 $A$\ GeV/$c$. 
	  {\bf b)} Rapporto  tra il numero di fotoni misurati ed il fondo atteso per 
	           collisioni Pb-Pb periferiche (in alto) e centrali (in basso). 
		   Le bande rappresentano l'errore sistematico.~\cite{WA98} }
 \label{Photons}
 \end{center}
 \end{figure}

L'emissione di fotoni diretti dallo stato di QGP in equilibrio termico avviene
principalmente attraverso il processo d'urto Compton 
$gq \longrightarrow \gamma q$. Questo segnale deve competere con i fotoni provenienti 
dai decadimenti adronici, quali $\pi^0 \longrightarrow \gamma \gamma $ e 
$\eta^0 \longrightarrow \gamma \gamma $, e da quelli formati in seguito a reazioni 
del tipo $\pi \rho \longrightarrow \gamma \rho$\ all'interno del gas adronico 
termalizzato~\cite{Kap93}.   
\newline
Partendo da precedenti stime del tasso di emissione dei 
fotoni~\cite{PhotEm1,PhotEm2,PhotEm3,PhotEm4}, alcuni autori~\cite{Kaputsa}  
confrontarono nei dettagli l'emissivit\`a di fotoni nei due scenari di QGP e di 
gas adronico.  Queste prime stime mostravano come i tassi di emissione termica 
per il gas adronico e per il QGP fossero molto simili e dipendessero solo dalla 
temperatura. Essi concludevano quindi che i fotoni diretti fossero s\`i una buona 
sonda della temperatura del sistema, ma non permettessero di distinguere tra i 
due scenari.  
\newline
Pi\`u recentemente si \`e mostrato~\cite{Aurenche} che il tasso di produzione 
({\em ``yield''}) dei fotoni nel QGP, calcolato utilizzando diagrammi 
sino a due {\em loop}, \`e  
notevolmente pi\`u elevata delle stime precedenti calcolate all'ordine inferiore. In 
particolare, un processo di annichilazione $q\bar{q}$\ con diffusione \`e risultato 
dominante nella {\em ``rate''} di emissione dei fotoni 
%dalla materia a quark 
dal plasma 
per elevate energie dei fotoni. Utilizzando questi nuovi risultati \`e 
possibile ricalcolare la produzione di fotoni nelle collisioni tra ioni pesanti 
e mostrare che, per una temperatura iniziale sufficientemente elevata, 
lo {\em ``yield''} dei fotoni dal QGP sia significativamente superiore a quello 
dalla materia adronica~\cite{Srivastava}.  
\newline
I precedenti risultati sperimentali delle collaborazioni NA34~\cite{Ake90},
WA80~\cite{Alb91} e WA98~\cite{Agg96}, che hanno misurato lo spettro inclusivo 
di fotoni in collisioni centrali S-AU e, pi\`u recentemente Pb-Pb, non rivelavano 
alcun eccesso di fotoni indicante la formazione di QGP.  
\newline
I pi\`u recenti dati di WA98 sembrano invece suggerire un eccesso   
nella produzione di fotoni per le solo collisioni Pb-Pb pi\`u centrali~\cite{WA98}.  
In fig.~\ref{Photons}.a \`e mostrato lo spettro di impulso trasverso 
dei fotoni misurato nelle collisioni centrali Pb-Pb a 160 $A$\ GeV/$c$; nella 
fig.~\ref{Photons}.b si osserva il rapporto tra il numero di fotoni misurati ed il 
fondo di fotoni atteso in funzione dell'impulso trasverso per 
collisioni periferiche (in alto) e per le collisioni centrali (in basso).  Le prime 
(collisioni periferiche) sono compatibili con un rapporto pari ad uno;  
per le seconde (collisioni centrali) il rapporto suggerisce un eccesso di fotoni 
diretti per valori elevati di $p_T$.  
\newline
Poich\'e la produzione di fotoni aumenta fortemente con la temperatura del plasma, 
ad energie pi\`u elevate di quelle dell'SPS  la pi\`u alta temperatura iniziale 
del sistema interagente dovrebbe rendere tale segnale  pi\`u significativo.  
I primi risultati di RHIC a riguardo, presentati dell'esperimento PHENIX alla 
conferenza di Quark Matter 2002~\cite{PHENIXQM02}, non sono ancora conclusivi per via 
degli elevati errori sistematici. 
%\begin{figure}[htb]
%\begin{center}
%\includegraphics[scale=0.40]{cap1/Phenix_photons.eps}
%\caption{Ciccio }
%\label{PHENIXPhotons}
%\end{center}
%\end{figure}
\subsubsection{Le coppie leptoniche}
\begin{figure}[p]
\begin{center}
{\bf a)} \hspace{7.0cm} {\bf b)} \\
%\vspace{-1.2cm}
\includegraphics[scale=0.35]{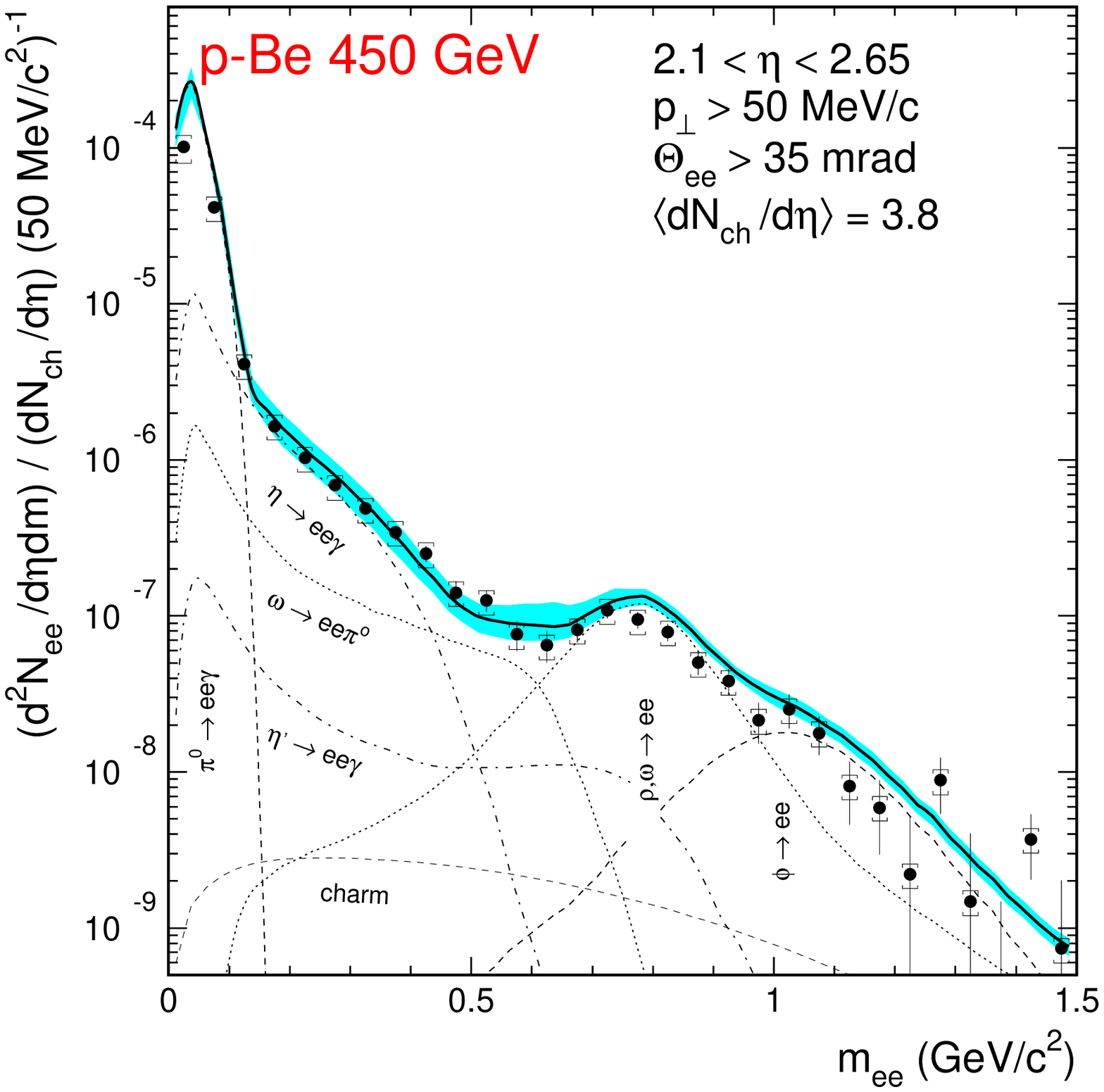}
\includegraphics[scale=0.35]{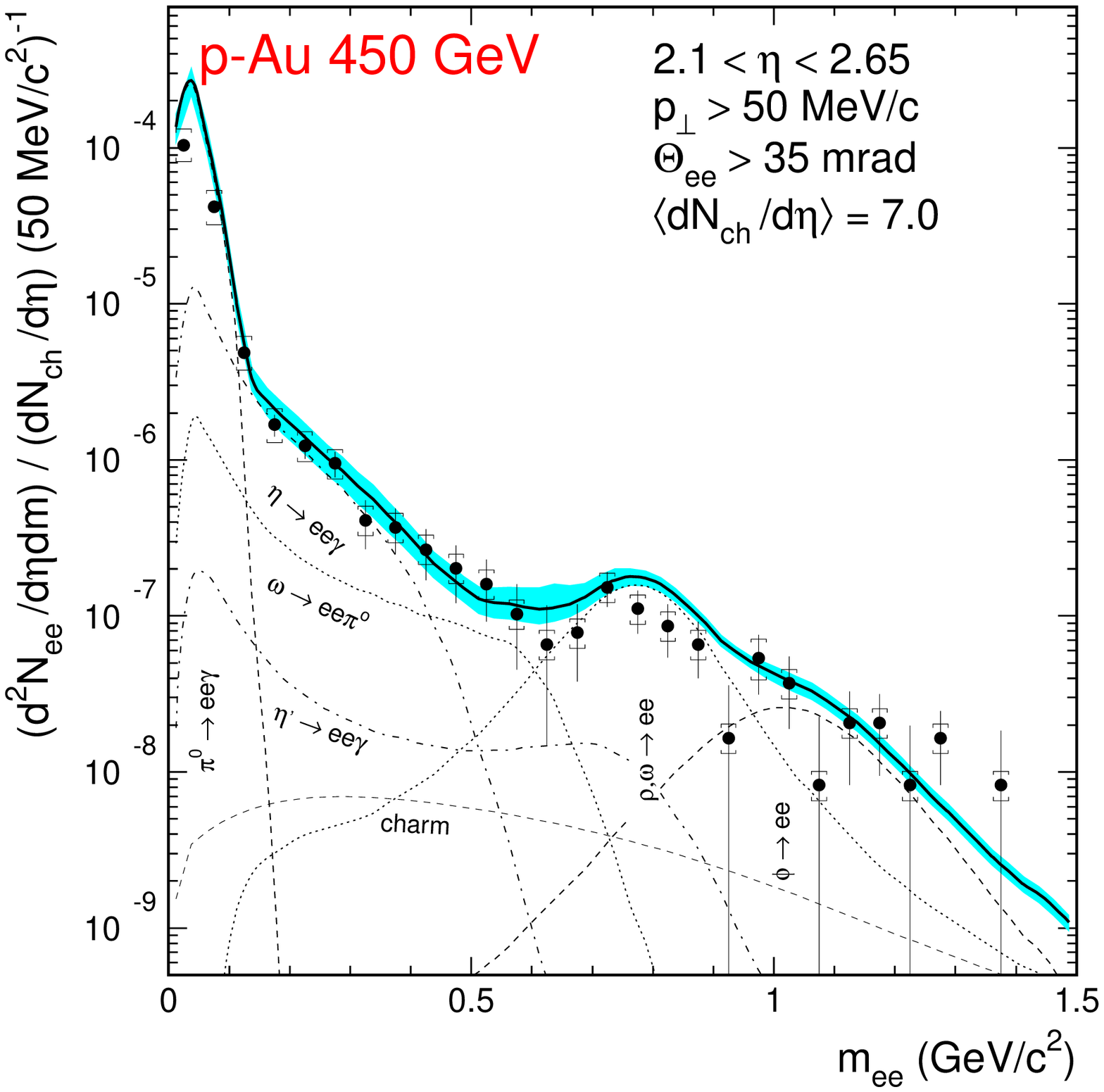} \\
\vspace{0.4cm}
{\bf c)} \hspace{4.6cm} {\bf d)} \hspace{4.6cm} {\bf e)} \\
\vspace{-0.2cm}
\hspace{-0.8cm}
\includegraphics[scale=0.27]{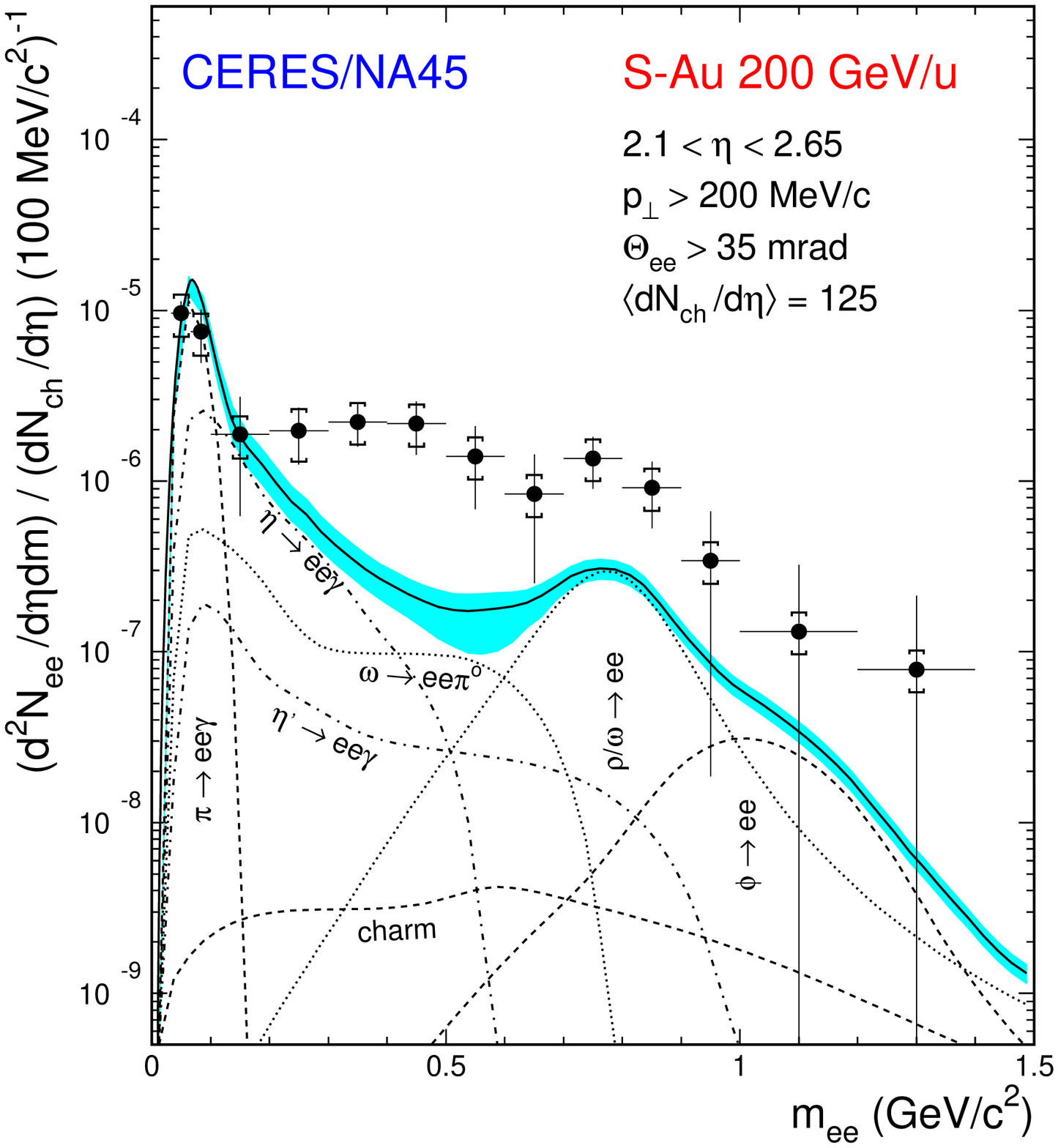}
\includegraphics[scale=0.27]{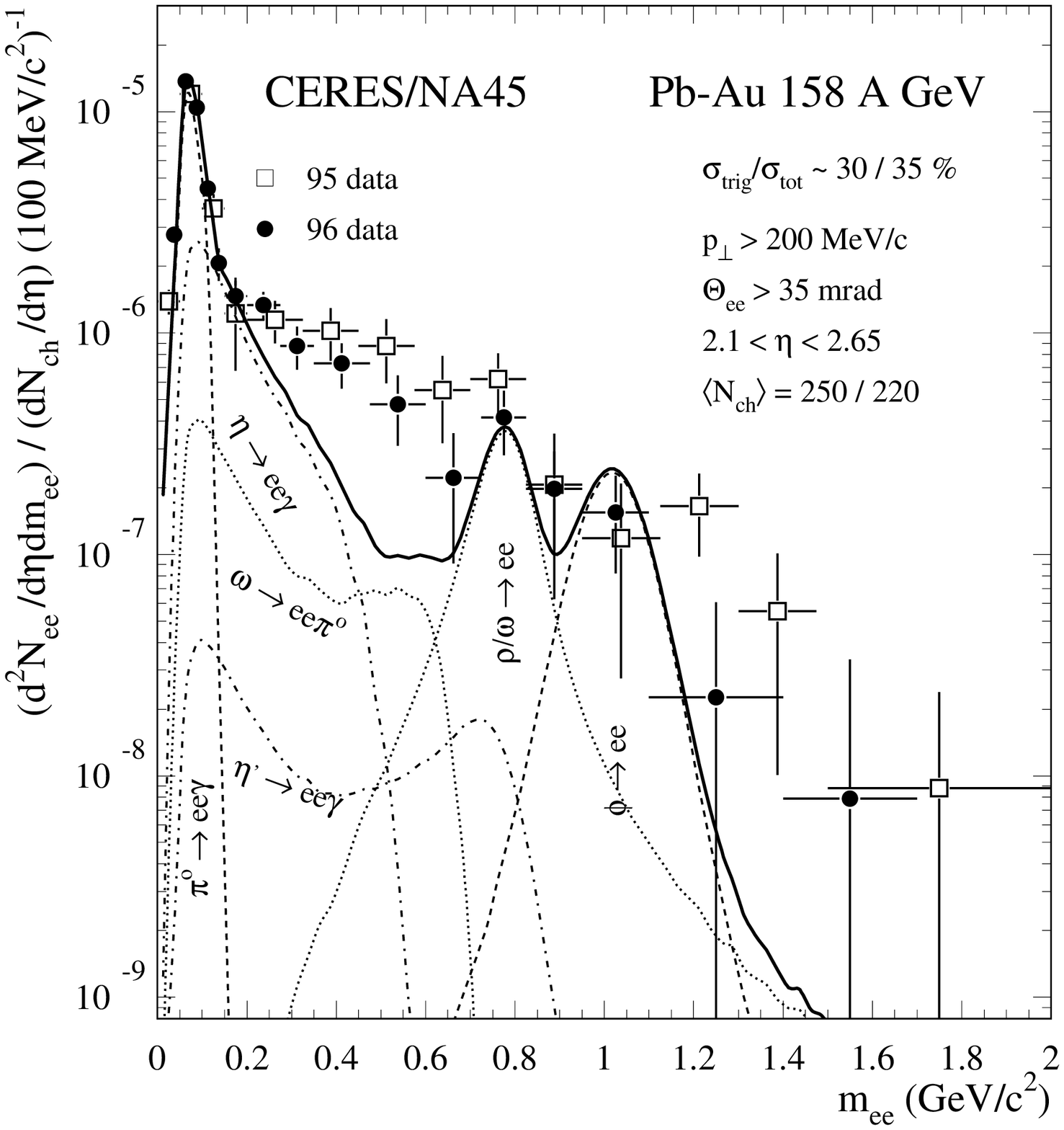}
\includegraphics[scale=0.29]{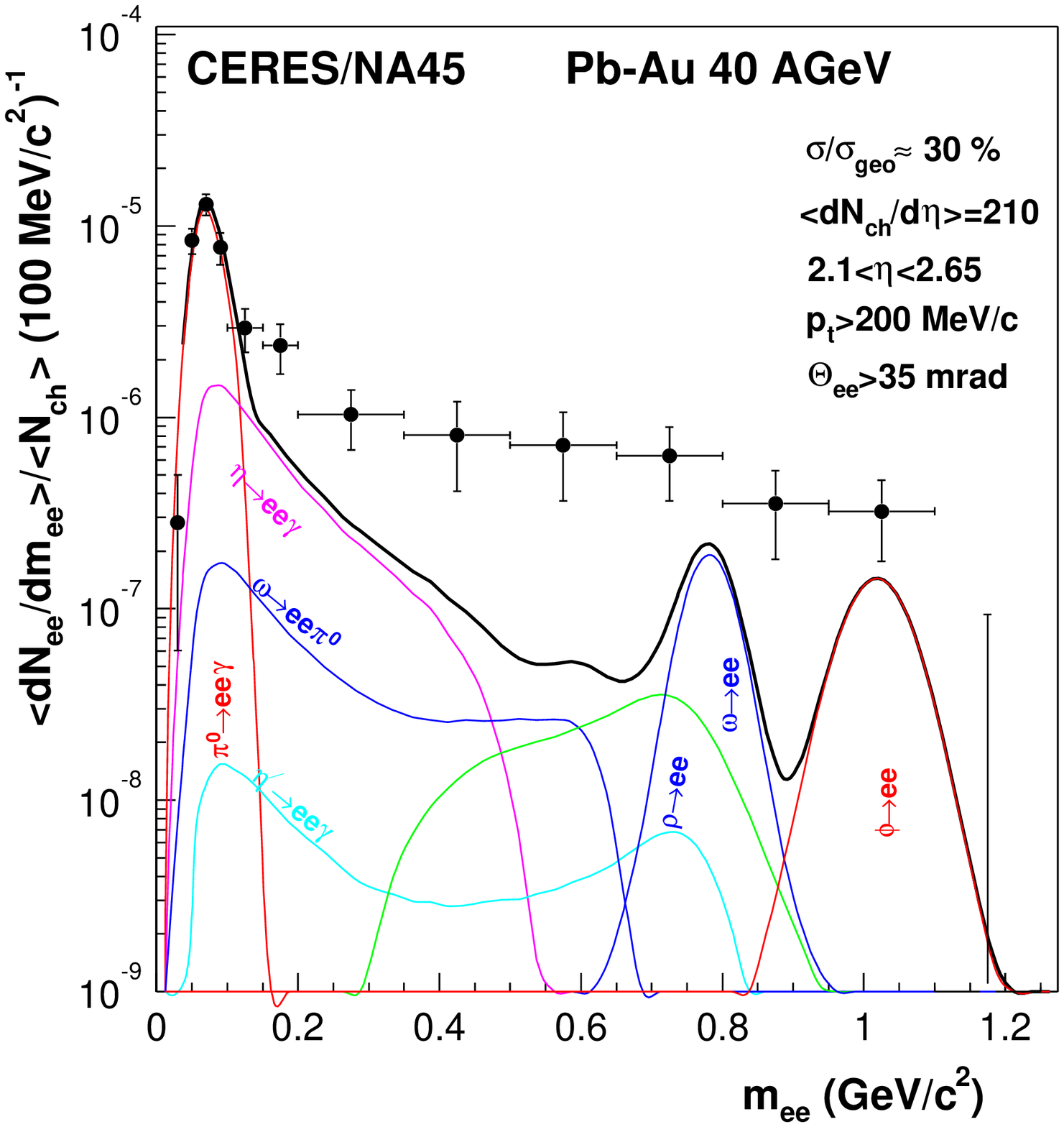} \\
\caption{ Spettri di massa invariante \Pep\Pem misurati
   dall'esperimento NA45/CERES nell'intervallo di
   pseudo-rapidit\`a $2.1 < \eta < 2.65 $. 
   {\bf a)} Collisioni p-Be  a 450 GeV/$c$~\cite{CERESpA}; 
   {\bf b)} collisioni p-Au  a 450 GeV/$c$~\cite{CERESpA}; 
   {\bf c)} collisioni S-Au  a 200 $A$ GeV/$c$~\cite{CERESSAu}; 
   {\bf d)} collisioni Pb-Au a 160 $A$ GeV/$c$~\cite{CERES160}; 
   {\bf e)} collisioni Pb-Au a  40 $A$ GeV/$c$~\cite{CERES40}.}
 \label{Dileptons}
 \end{center}
\end{figure}

La produzione di dileptoni per annichilazione di coppie $q\bar{q}$, 
dovrebbe costituire, al pari della radiazione termica, un segnale non 
influenzato dalla successiva adronizzazione e recante informazioni  
circa le condizioni del sistema interagente nell'istante della sua   
formazione~\cite{Ruu91}.  
Rispetto allo studio dei fotoni, si ha il vantaggio di poter calcolare 
una massa invariante, il che aiuta a distinguere tra diversi processi di 
produzione, ed in particolare a controllare i mesoni vettori tramite il 
loro decadimento diretto. 
\newline
Per estrarre il debole segnale delle prime fasi della collisione, bisogna  
conoscere dettagliatamente i modi di decadimento elettromagnetico degli 
adroni dopo il {\em ``freeze-out''}.  
Anche in tal caso il fondo \`e infatti considerevole: per masse invarianti 
$M_{l^+l^-} < 1.5 $ GeV il contributo dominate del fondo viene dai 
prodotti di decadimento degli adroni, mentre per 
$M_{l^+l^-} > 5 \div 10 $ GeV  diventa dominante il fondo dovuto 
ai processi di Drell-Yan ($q\bar{q} \longrightarrow l^+l^- $,  
fig.~\ref{JPSI}.b)  
a seguito delle violente interazioni tra nucleoni, che avvengono nei primi 
istanti della collisone tra i nuclei, o alla produzione di mesoni 
vettori pesanti che decadono in coppie leptone-antileptone.  
\newline
In fig.~\ref{Dileptons} sono mostrati gli spettri di massa invariante 
\Pep\Pem , misurati nell'intervallo di pseudo-rapidit\`a 
$2.1 < \eta < 2.65 $ dall'esperimento NA45/CERES in interazioni 
p-Be ({\bf a}) e p-Au ({\bf b}) a $450$\ GeV/$c$, 
S-Au ({\bf c}) a 200 $A$\ GeV/$c$, 
Pb-Au a 160 ({\bf d}) e 40  ({\bf e}) 
$A$\ GeV/$c$ \cite{CERESpA}\cite{CERESSAu}\cite{CERES160}\cite{CERES40}.  
Si pu\`o notare come i decadimenti a due ed a tre corpi dei mesoni 
\Pgpz, \Pgh, \Pghpr, \Pgr, \Pgo\ e \Pgf, noti come 
``sorgenti standard'', riescano a riprodurre lo spettro misurato in 
collisioni p-Be e p-Au.  
Nelle caso di collisioni S-Au e Pb-Au, invece, il contributo delle sorgenti 
standard, calcolato assumendo che l'abbondanza relativa di particelle 
nello stato finale sia indipendente dal processo di collisione, 
sottostima vistosamente lo spettro osservato 
nella regione di massa invariante $0.25 < m_{e^+e^-} < 0.7 $ GeV/$c^2$.  
\newline 
L'eccesso osservato per $m_{\Pep\Pem}>0.2$ GeV/$c^2$ nelle interazioni 
nucleo-nucleo pu\`o essere cos\'i quantificato: 
\begin{itemize}
\item[]S-Au  a 200 GeV/Au: 
  $\quad\quad \frac{\rm dati}{\rm sorgenti \; standard}=
        5.0 \pm 0.7 (stat.) \pm 2.0 (sist.) $ 
\item[]Pb-Au a 160 $A$ GeV/$c$:  
  $\quad\quad \frac{\rm dati}{\rm sorgenti \; standard}=
  2.9 \pm 0.3 (stat.) \pm 0.6 (sist.) $
\item[]Pb-Au a  40 $A$ GeV/$c$:  
  $\quad\quad \frac{\rm dati}{\rm sorgenti \; standard}=
  5.1 \pm 1.3 (stat.) \pm 1.0 (sist.) $
\end{itemize}
Lo studio in funzione della centralit\`a~\cite{CERES160}\cite{CERES40} 
suggerisce un incremento dell'eccesso di coppie prodotte 
nell'intervallo $m_{\pi^0} < m_{\Pep\Pem} < m_{\rho,\omega}$\ pi\`u 
rapido di una legge lineare, attesa qualora  
il decadimento degli adroni fosse l'unica sorgente di coppie \Pep\Pem.  
Inoltre l'incremento \`e decisamente pi\`u marcato per coppie di  
piccolo impulso trasverso ($p_T<500$ MeV) che di grande impulso 
trasverso ($p_T>500$ MeV)~\cite{CERES40}.   
\newline
Numerosissimi lavori teorici finalizzati ad una possibile 
interpretazione dei risultati di questi dati sono stati compiuti 
negli ultimi anni (per una {\em review} recente si guardi 
in~\cite{ReviewDilep}). 
Sembra ormai raggiunto un generale consenso: si osserva 
la radiazione diretta dalla fireball, dominata 
dall'annichilazione \Pgpp\Pgpm $\longrightarrow$\ $\rho$\ 
$\longrightarrow$\ \Pep\Pem; la forma dello spettro di massa 
invariante \Pep\Pem pu\`o essere riprodotta richiedendo un 
cambiamento delle propriet\`a della $\rho$\ all'interno del 
mezzo nucleare.  
Due sono i principali candidati, tra loro in competizione,  che tengono 
conto di ci\`o. Il primo \`e il cosidetto {\em ``scaling di 
Brown-Rho''}~\cite{Brown-Rho}, in cui si riduce la massa della 
$\rho$, come fenomeno precursore del ripristino della simmetria chirale. 
Nel secondo si calcola la densit\`a spettrale della $\rho$\ 
all'interno del mezzo adronico denso, allargando la 
larghezza dello spettro~\cite{ReviewDilep}\cite{DiLepThe2}.  
In entrambi i casi si riesce a riprodurre l'intero spettro 
come fosse dovuto ad annichilazione $q\bar{q}$, rispettivamente,  
con una elevata riduzione della (media temporale della) massa 
oppure con un notevole allargamento dello spettro.  
Una descrizione quantitativa dei dati Pb-Au richiede una 
temperatura di 145 MeV\ a 40 $A$\ GeV/$c$ ed una di 170 MeV 
a 160  $A$\ GeV/$c$~\cite{CERES40}. In ogni caso, i dati escludono 
una produzione della $\rho$\ non modificata, cio\`e secondo la 
funzione di densit\`a spettrale nel vuoto.
\newline
Il mesone $\rho$, ha infatti una vita media $\tau$ = 1.3 fm/$c$\ 
pi\`u corta di quelle delle altre sorgenti standard e piccola in 
confronto al tempo di vita del sistema interagente 
($\approx 10 \div 20$ fm/$c$), per cui le coppie leptoniche provenienti 
dal suo decadimento hanno buona probabilit\`a di essere create 
prima del {\em ``freeze-out''}.
\newline
Per riprodurre i risultati sperimentali i modelli teorici richiedono 
quindi un cambiamento delle propriet\`a della \Pgr\ nel mezzo 
oppure una radiazione termica con temperatura $T>T_C$. 
In ogni caso questi risultati forniscono un'indicazione di un 
cambiamento nelle caratteristiche fisiche della materia nucleare, 
in linea con le previste conseguenze della transizione di fase di QCD. 
\subsection{Fluttuazioni evento per evento}
\begin{figure}[h]
\begin{center}
\includegraphics[scale=0.57]{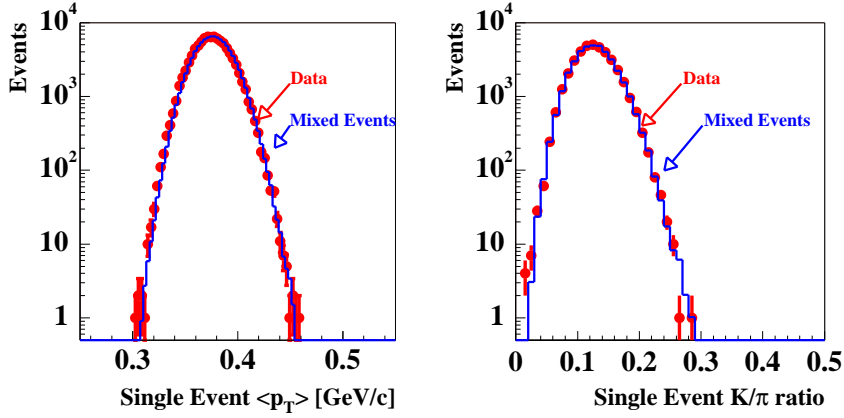}
\caption{ Distribuzioni delle variabili $<p_T>$\ (destra) e 
 $\frac{<K^{\pm}>}{<\pi^{\pm}>}$ (sinistra), calcolate evento per evento, 
 in collisioni centrali Pb-Pb a 160 $A$ GeV/$c$ misurate da NA49. 
 I punti rappresentano i dati degli eventi reali, le linee gli 
 eventi {\em mescolati}.}
\label{Fluct}
\end{center}
\end{figure}

Generalmente tutti i possibili segnali sperimentali vengono analizzati  
mediando su un numero elevato di collisioni, ma \`e tuttavia 
possibile studiare alcune osservabili evento per evento.  
\newline
Infatti, gi\`a nelle collisioni nucleo-nucleo all'energia dell'SPS si 
producono tipicamente $10^3$\ particelle per evento. \`E quindi 
possibile misurare la distribuzione del valor medio --- calcolato 
su base evento per evento --- dell'impulso trasverso delle 
particelle cariche $<p_T>$\ o del rapporto 
$\frac{<K^{\pm}>}{<\pi^{\pm}>}$\ tra il numero di kaoni ed il numero di 
pioni carichi prodotti in un evento.  
\newline
Se lo stato di QGP venisse raggiunto solo in una frazione degli eventi 
raccolti, essi mostrerebbero caratteristiche globali differenti, ed in 
tal caso lo studio di alcune osservabili su base evento per evento 
permetterebbe di isolare l'insieme degli eventi di tale frazione.  
Inoltre, altre segnature sperimentali della fase di QGP potrebbero essere 
correlate con le fluttuazioni evento per evento, e quindi, isolando una certa 
frazione di eventi sulla base di 
una simile analisi, 
%un'analisi evento per evento, 
si renderebbe pi\`u pronunciato il segnale di tali segnature.  
\newline
L'esperimento dell'SPS pi\`u idoneo  a studiare le fluttuazioni 
\`e NA49, 
caratterizzato da una estesa accettanza e dalla possibilit\`a di 
identificazione delle particelle per perdita di energia nella TPC  
e dalla misura del tempo di volo~\cite{NA49pID}. 
\newline
In fig.~\ref{Fluct} sono mostrati i dati sperimentali (punti) e gli 
eventi di fondo calcolati mescolando eventi diversi (linee a tratto pieno) 
per le distribuzioni di $<p_T>$\ (a destra) e del rapporto 
$\frac{<K^{\pm}>}{<\pi^{\pm}>}$\ (a sinistra) in collisioni 
centrali Pb-Pb~\cite{Fluct1}\cite{Fluct2}.
Per il rapporto $\frac{<K^{\pm}>}{<\pi^{\pm}>}$, gli eventi di fondo 
sono calcolati prendendo i kaoni carichi da un evento e confrontando 
con i pioni carichi di un altro evento. Per la distribuzione in 
$<p_T>$\, il fondo viene calcolato formando nuovi eventi {\em mescolati}, 
ottenuti scegliendo a caso le particelle da pi\`u eventi diversi, 
assicurandosi tuttavia di conservare la distribuzione di molteplicit\`a 
degli eventi reali. In entrambi i casi \`e pertanto possibile stimare 
l'entit\`a  di eventuali fluttuazioni non statistiche, definendo 
l'intensit\`a della fluttuazione non statistica: 
\begin{equation}
\sigma_{non \; stat}=\sqrt{\sigma^2_{data}-\sigma^2_{mixed}}.
\label{NSFluctua}
\end{equation}
\newline
La collaborazione NA49 trova che in tutti gli eventi risulta 
$\sigma_{non \; stat}<4.0\%$\ con livello di confidenza del 90\% 
per il rapporto $\frac{<K^{\pm}>}{<\pi^{\pm}>}$~\cite{Fluct1}. 
Questo valore \`e piccolo rispetto al valore di $\sim$0.2 
misurato per il rapporto ed indica che il meccanismo responsabile 
dell'incremento di produzione di stranezza \`e presente in tutte
le collisioni centrali Pb-Pb. 
\newline
In maniera simile, $\sigma_{non \; stat}<1.2\%$ con livello di 
confidenza del 90\% per le fluttuazioni della distribuzione 
di $<p_T>$ in ciascun evento~\cite{Fluct2}. 
\newline
In entrambi i casi non vi \`e evidenza di significative fluttuazioni 
statistiche su base evento per evento.  
\section{La produzione di stranezza come segnale di QGP} 
La produzione di particelle strane rappresenta una vera e propria pietra 
miliare per la scoperta sperimentale del QGP ed in particolare 
per lo studio della formazione ed evoluzione della materia deconfinata. 
Questo campo di indagine \`e stato sviluppato intensamente da quando 
venne proposto, pi\`u di vent'anni or sono, sino ad 
oggi~\cite{Raf82-86}\cite{Raf86}\cite{Raf92}\cite{Raf96}\cite{Raf2000}. 
Dal punto di vista teorico, l'interesse suscitato dal segnale della produzione 
di particelle strane proviene dalle semplici ed attendibili differenze tra 
uno scenario di produzione di stranezza entro la materia confinata 
(gas adronico) ed uno entro il QGP: 
\begin{itemize}
\item Nella fase di QGP, la densit\`a di particelle \`e sufficientemente alta 
      e la soglia per la produzione di quark ed anti-quark di {\em flavour} strano 
      sufficientemente bassa affinch\'e si riesca a produrre una elevata 
      popolazione di quark (ed anti-quark) strani entro la durata di vita del 
      plasma~\cite{Raf82-86}\cite{Raf86}\cite{Raf96}\cite{Bilic}; 
      nella fase di gas adronico, invece, una popolazione cos\`i 
      abbondante non \`e raggiungibile~\cite{Raf86}\cite{Koch}, a meno di non 
      stravolgere i parametri dei modelli al punto tale che molti altri risultati 
      sperimentali non vengano pi\`u riprodotti.  
\item \`E possibile raggiunge, od addirittura superare, la popolazione 
      di equilibrio chimico nello spazio delle fasi degli adroni solamente 
      quando la fase di QGP, caratterizzata da un alto contenuto 
      entropico, 
%      si disintegra 
      condensa   
      rapidamente ed in modo esplosivo 
      in adroni~\cite{Raf86}\cite{Raf96}\cite{Letessier}.
\end{itemize}
Come si discuter\`a nei prossimi paragrafi, vi 
sono diverse ed importanti caratteristiche riguardanti la stranezza,  
che sono attese nell'ipotesi di deconfinamento; tre in particolare possono 
essere considerate come i pilastri alla base dell'idea di considerare 
l'osservabile ``stranezza'' come segnale del QGP: 
\begin{enumerate}
\item La simmetria tra materia ed anti-materia, cio\`e tra i barioni ed 
      i rispettivi anti-barioni strani emessi direttamente, negli spettri 
      di massa trasversa e nella fugacit\`a~\footnote{Vedi nota 7.} 
      del quark strano;  
\item L'incremento nel tasso di produzione (lo {\em ``yield''}, cio\`e il numero 
      di particelle prodotte per collisione) di barioni 
      ed anti-barioni strani che aumenta con il contenuto di stranezza dei quark; 
\item l'incremento del rapporto tra il  tasso di produzione  di un dato 
     iperone (od anti-iperone) ed il numero di nucleoni partecipanti,  
     nelle collisioni nucleo-nucleo rispetto a quelle protone-nucleo, 
     usate come riferimento.  
\end{enumerate}
Ciascuna di queste tre predizioni \`e stata  confermata dagli esperimenti 
dell'SPS che  studiano collisioni indotte da nuclei 
di piombo all'energia di 160 GeV per nucleone, ed alcune anche dai precedenti 
esperimenti col fascio di zolfo a 200 GeV per nucleone. 
Pi\`u in particolare: 
\begin{enumerate}
\item La collaborazione WA97 ha studiato in dettaglio gli spettri di massa
      trasversa dei barioni ed anti-barioni strani, che presentano una  
      notevolissima simmetria, alquanto inusuale, tra il settore barionico 
      e quello anti-barionico~\cite{mt_WA97}.  
\item Un'analisi dettagliata dei risultati Pb-Pb condotta dalla collaborazione 
      WA97 ha mostrato, confrontando i risultati p-Be e p-Pb con Pb-Pb, un notevole 
      incremento dello {\em ``yield''} dei barioni ed anti-barioni strani che 
      aumenta col contenuto di 
      stranezza~\cite{WA97Nucl}\cite{WA97Eur}\cite{WA97PhysLett}\cite{WA97Lietava}.  
      I risultati della collaborazione NA49~\cite{NA49PLB444} sono consistenti con 
      quelli di WA97. 
      Anche la collaborazione WA85
      misura un incremento dei barioni ed anti-barioni strani che  
      aumenta col numero di stranezza, nelle reazioni indotte da S~\cite{WA85PLB447}.
\item[3.a] L'incremento nella produzione di stranezza, per valori centrali di 
	   rapidit\`a, \`e stato osservato in collisioni indotte da S 
	   dagli esperimenti NA35~\cite{NA35}, 
	   WA85 e WA94~\cite{WA85-WA94}, ed NA44~\cite{NA44}. 
	   Nelle collisioni Pb-Pb, l'incremento \`e stato misurato 
	   dagli esperimenti WA97~\cite{WA97Nucl}\cite{WA97Eur}\cite{WA97PLB449}, 
	   NA49~\cite{NA49SiKler} ed NA44~\cite{NA44PLB471}. In particolare, come si 
	   discuter\`a nei prossimi paragrafi, i risultati 
	   di WA97 suggeriscono che, entro l'intervallo di centralit\`a 
	   corrispondente a circa il 40\% delle collisioni Pb-Pb pi\`u centrali, 
	   l'incremento ha ormai raggiunto un valore di 
	   saturazione~\cite{WA97PLB449}. I dati dell'esperimento NA52 per i kaoni 
	   sembrano indicare ulteriormente che la saturazione venga raggiunta molto rapidamente, 
	   non appena la centralit\`a della collisione, e quindi la dimensione 
	   della materia nucleare interagente, 
	   oltrepassi un limite corrispondente ad un numero barionico 
	   $B=40 \div 50$\ per la materia partecipante alla collisione~\cite{NA52}. 
\item[3.b] Un incremento globale nella produzione di particelle strane 
	   \`e stato osservato in reazioni indotte da S e da Pb dagli 
	   esperimenti NA35~\cite{NA35} ed NA49~\cite{NA49SiKler}.
\end{enumerate}
\subsection{Produzione di stranezza in fase adronica} 
Nello scenario adronico, in assenza cio\`e di QGP, la produzione di particelle 
strane avviene per mezzo di interazioni secondarie nella fase di gas adronico 
successiva alla collisione tra i nuclei, ed \`e in genere sfavorita rispetto 
alla produzione di adroni contenenti solo i quark leggeri $u$\ e $d$. 
Dal punto di vista teorico, ci\`o \`e riconducibile alla rottura della simmetria 
chirale ed al conseguente innalzamento della massa del quark $s$\ nello stato 
confinato.
\newline
Le tipiche reazioni in cui si generano iperoni sono quelle {\em dirette}
o di {\em produzione associata}, 
%ad esempio per $Y$ = $\Lambda$\ o $\Sigma$:  
%\begin{equation}
%\pi\pi \rightarrow K \bar{K}; \quad \quad \pi N  \rightarrow  K Y, \quad 
%NN \rightarrow  N K Y
%\nonumber
%\end{equation}
%I processi diretti non giocano praticamente alcun ruolo poich\'e hanno una 
%soglia molto elevata rispetto alla temperatura del sistema.  
Nell'interazione primaria tra i nucleoni vengono infatti prodotti principalmente 
pioni; questi possono interagire con i nucleoni o con altri adroni prodotti 
nell'interazione dando luogo alla {\em produzione associata}. 
Nella seguente tabella sono riportate le reazioni che dominano la produzione 
di particelle strane in funzione del contenuto di stranezza dell'adrone 
prodotto ($Y$ = $\Lambda$\ o $\Sigma$):  
\begin{align}
& |s| = 1 :  \quad \quad \pi N \rightarrow K Y \quad \pi \bar{N} 
	\rightarrow \bar{K} \bar{Y}  \quad  \pi \pi \rightarrow K \bar{K} \nonumber \\
& \quad \quad \quad \quad \; \,
 	\quad N N \rightarrow N K Y \quad N N \rightarrow N N Y \bar{Y} \nonumber \\
& |s| = 2 :  \quad \quad \pi Y \rightarrow \Xi  K \quad 
	\pi \bar{Y} \rightarrow \bar{K} \bar{\Xi} 
	\quad N N \rightarrow N N \Xi \bar{\Xi} 
	\quad \pi \pi \rightarrow  \Xi \bar{\Xi} \\
& |s| = 3 :  \quad \quad \pi  \Xi \rightarrow \Omega  K \quad 
	\pi \bar{\Xi} \rightarrow \bar{K} \bar{\Omega}   
	\quad  N N \rightarrow N N \Omega \bar{\Omega} 
	\quad \pi \pi \rightarrow  \Omega \bar{\Omega}
\nonumber
\end{align}
Tutte queste reazioni sono caratterizzate da sezioni d'urto piuttosto piccole, 
dell'ordine dei 100 $\mu$b~\cite{Raf86}, e da soglie molto elevate rispetto 
alla temperatura del sistema. 
\newline
Ad esempio, per produrre una $\Lambda$:
\begin{equation}
p p  \rightarrow p + \Lambda + K^+
\nonumber
\end{equation}
\`e richiesta un'energia di soglia nel centro di massa di 671 MeV, e
per produrre un $\bar{\Lambda}$ ($p p  \rightarrow p p \Lambda \bar{\Lambda}$)
un'energia di 2231 MeV. La produzione diretta di un'antiomega
($\bar{\Omega}[\bar{s}\bar{s}\bar{s}]$) pu\`o avvenire mediante la reazione
$\pi\pi \rightarrow \Omega \bar{\Omega}$\ con un'energia di soglia di 3.3 GeV.
\newline
La produzione di barioni ed anti-barioni  
con pi\`u di un'unit\`a di stranezza \`e particolarmente sfavorita: essa avviene 
infatti attraverso reazioni dirette caratterizzate da soglie particolarmente elevate 
oppure deve procedere attraverso una catena di reazioni nelle quali la stranezza 
dell'adrone aumenta progressivamente, il che richiede comunque tempi molto pi\`u 
lunghi rispetto alla vita della {\em fireball}.  
\newline
Una volta creata, la stranezza viene ridistribuita tra le varie specie adroniche   
tramite le seguenti reazioni:
\begin{align}
&\bar{K} N \rightarrow Y \pi \quad\quad K \bar{N} \rightarrow \bar{Y}\pi 
   \nonumber \label{StrangeDistrib}\\ 
&\bar{K} Y \rightarrow \Xi \pi \quad\quad K \bar{Y} \rightarrow \bar{\Xi}\pi 
    	     \\
&\bar{K} \Xi \rightarrow \Omega \pi \quad\quad K \bar{\Xi} \rightarrow \bar{\Omega}\pi
   \nonumber 
\end{align} 
Tutte queste reazioni sono esotermiche e le relative sezioni d'urto sono circa un 
ordine di grandezza maggiori della sezione d'urto di produzione di stranezza. 
Tuttavia, come evidenziato dalle eq.~\ref{StrangeDistrib}, la produzione di anti-barioni strani 
\`e penalizzata poich\'e avviene attraverso interazioni con anti-nucleoni, la cui 
presenza \`e particolarmente scarsa nel gas adronico.  
Al crescere della densit\`a di particelle strane cominciano a diventare 
importanti le reazioni di annichilazione, di cui si pu\`o anche tener conto per 
calcolare con maggiore precisione l'evoluzione temporale delle abbondanze 
delle diverse specie:
\begin{align} 
& N \bar{N} \rightarrow  \sim 5\pi             \nonumber \\
& Y \bar{N} \rightarrow \bar{K} + \sim 4\pi \quad\quad  
  \bar{Y} N \rightarrow K + \sim 4\pi          \nonumber \\ 
& \Xi  \bar{N} \rightarrow 2\bar{K} + \sim 3\pi \quad\quad  
   \bar{\Xi} N \rightarrow 2K + \sim 3\pi      \nonumber \\ 
& \Omega  \bar{N} \rightarrow 3\bar{K}+ \sim 2\pi \quad\quad 
  \bar{\Omega} N \rightarrow 3K + \sim 2\pi \nonumber .
\end{align}

\begin{figure}[hb]
\begin{center}
\includegraphics[scale=0.60]{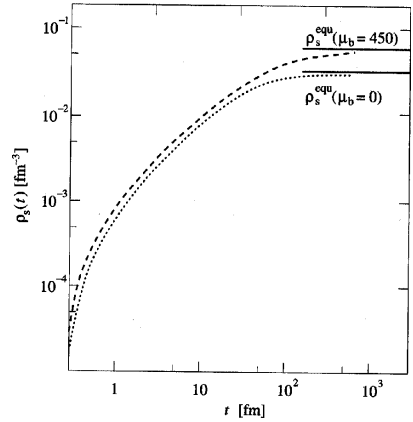}
  \caption{Produzione di stranezza in funzione del tempo in un gas adronico 
  a $T=160$\ MeV. Sono mostrati i risultati per due valori del potenziale chimico 
  barionico ($\mu_b=$\ 0 e 450 MeV)~\cite{Raf86}.}  
 \label{DaAggiungere}
\end{center}
\end{figure}

D'altra parte \`e stato dimostrato che, in un gas adronico ad elevata 
densit\`a, qualora si riesca a raggiungere l'equilibrio chimico, la produzione di 
stranezza pu\`o anche diventare confrontabile con quella relativa al 
plasma~\cite{Cle91}. Diventa quindi cruciale valutare il tempo in cui l'equilibrio 
chimico di stranezza pu\`o essere raggiunto nelle interazioni adroniche. 
In fig.~\ref{DaAggiungere} \`e mostrato l'andamento della produzione di stranezza in 
funzione del tempo per un gas adronico avente temperatura $T = 160$\ MeV e potenziale 
barionico nullo ($\mu_B =0$, linea punteggiata) o pari a 450 MeV (linea trattegiata).  
\newline
La saturazione nella produzione di stranezza viene raggiunta, nello scenario di 
gas adronico, dopo tempi dell'ordine dei 100 fm/$c$, molto pi\`u lunghi rispetto 
al tempo di vita del sistema interagente ($\sim$ 10 fm/$c$).
Non vi \`e dunque il tempo necessario per raggiungere la popolazione di 
particelle strane attesa all'equilibrio chimico. 
\subsection{Produzione di stranezza nel QGP} 
\begin{figure}[htb]
\begin{center}
\includegraphics[scale=0.55]{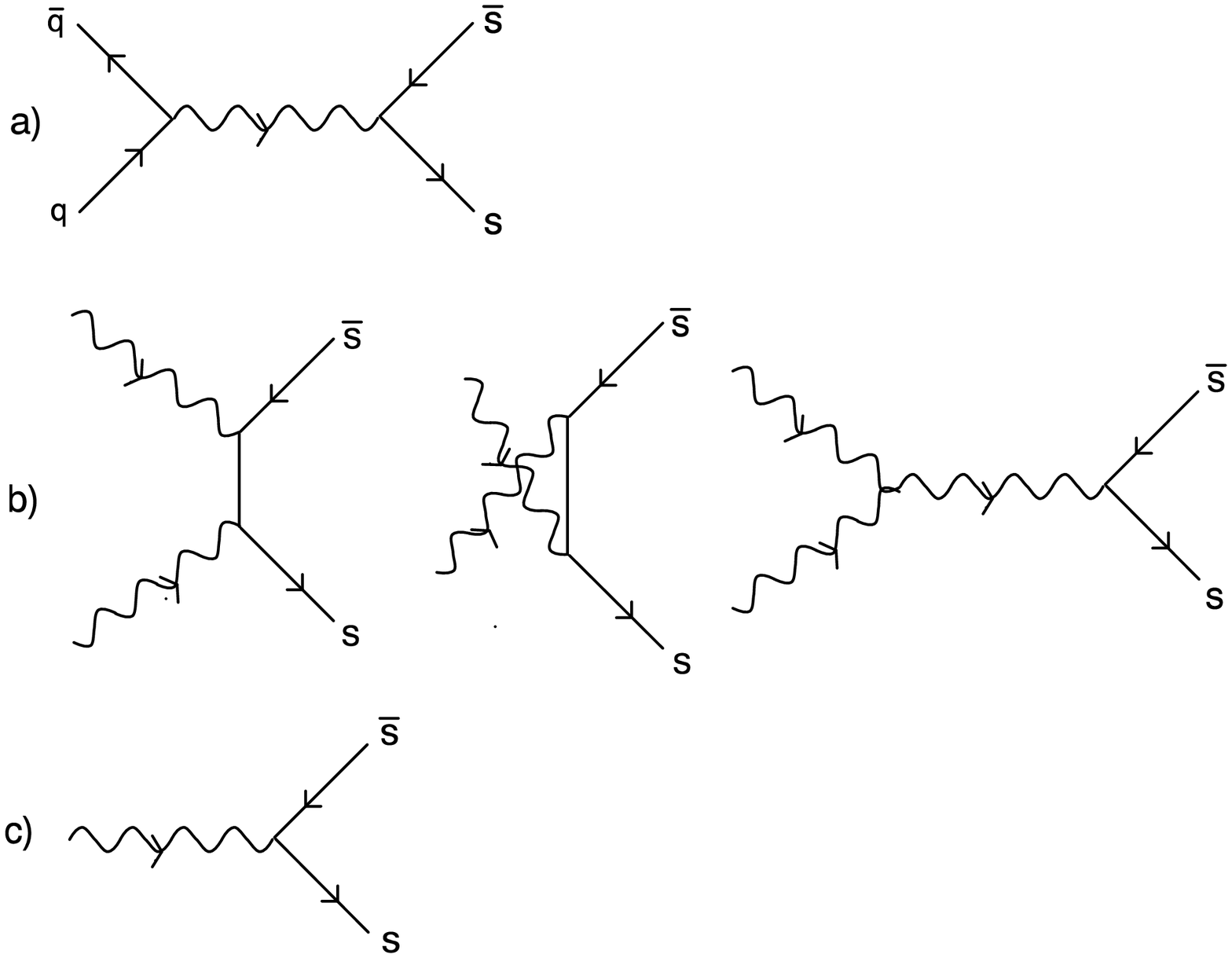}
  \caption{Diagrammi di QCD all'ordine pi\`u basso per i processi 
           $q\bar{q} \longrightarrow s\bar{s}$\ {\bf (a)},   
	   $gg \longrightarrow s\bar{s}$\ {\bf (b)} e 
	   $g^* \longrightarrow s\bar{s}$.}  
 \label{DiagrStran}
\end{center}
\end{figure}

\begin{figure}[htb]
\begin{center}
\includegraphics[scale=0.75]{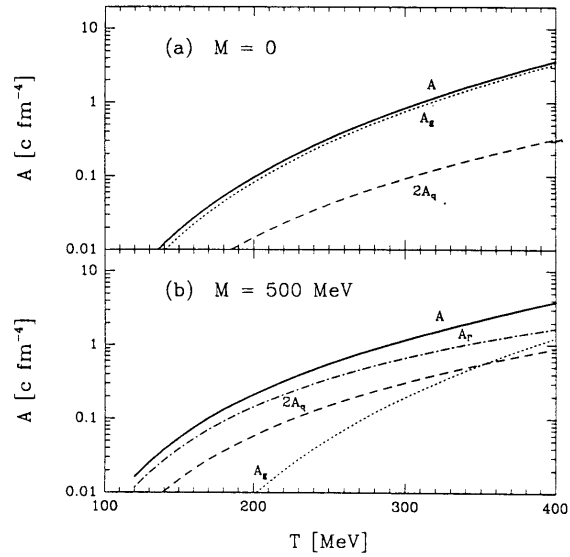}  
  \caption{Frequenza di produzione ($A={\rm d}^4N/{\rm d}x^4$\ parziale e totale 
  dei quark strani, per 
  i processi $q\bar{q} \longrightarrow s\bar{s}$ (linea trattegiata, $A_q$), 
  $gg \longrightarrow s\bar{s}$ (linea a puntini, $A_g$) e 
  $g^* \longrightarrow  s\bar{s}$ (linea tratto-punto, $A_r$), 
  in funzione della temperatura. $M$ indica il valore usato per 
  la massa termica dei gluoni.}  
\label{Heinz2}
\end{center}
\end{figure}

All'interno del QGP la produzione di quark strani \`e limitata  
unicamente dalla conservazione della stranezza; la frequenza di 
produzione e la scala dei tempi per il raggiungimento dell'equilibrio chimico 
possono essere calcolate per mezzo della teoria di campo all'equilibrio termico oppure, 
se la temperatura \`e sufficientemente elevata, anche con la QCD perturbativa. 
Si accenner\`a brevemente a questo secondo tipo di approccio, meno  
consolidato rispetto al primo, volendo soffermarsi sui diversi processi elementari di 
produzione.  
\newline
Da un punto di vista microscopico, la produzione di stranezza avviene per mezzo dei  
processi elementari riportati in fig.~\ref{DiagrStran}: annichilazione   
quark anti-quark ({\bf a}) e fusione gluonica ({\bf b}). 
Inoltre vi \`e il contributo dovuto al decadimento degli stati massivi di eccitazione 
collettiva, i  {\em ``plasmoni''}, del tipo $g^* \rightarrow s\bar{s}$\ 
(fig~\ref{DiagrStran}.c). 
A temperatura finita, infatti, tutte le teorie di gauge con particelle a massa nulla
presentano complicazioni dovute alle singolarit\`a infrarosse; per superare
queste difficolt\`a si pu\`o adoperare una tecnica per ricalcolare le somme dei
contributi a piccoli impulsi per temperature e densit\`a non nulle~\cite{Pisarski};
l'idea alla base di questo approccio consiste nell'assumere un termine per il
propagatore delle particelle leggere, con impulso $\lesssim gT$ ($g \ll 1$), che
contenga correzioni dovute alla auto-interazione con il mezzo. In tal modo,
all'ordine pi\`u basso dello sviluppo perturbativo in $g$, le particelle di piccolo
impulso, in un ambiente a temperatura finita, acquistano una massa termica ($m_{ter}$) 
pari a 
\begin{align}
& m_g=(N_c + N_f/2) g^2 T^2/9 \\
& m_q=m_q^0 + g^2 T^2/6 ,
\end{align}
rispettivamente, per i gluoni e per i quark.  
Questi ``modi'' massivi per i quark ed i gluoni sono l'equivalente per la QCD 
a temperatura finita dei modi plasmonici nel gas di elettroni o nel plasma 
elettromagnetico. All'ordine perturbativo successivo, la massa di questi ``modi'' 
acquisisce  anche una larghezza termica ($\Delta m_{ter}$), cui corrisponde  
una vita media finita. Il processo in fig.~\ref{DiagrStran}.c,  cinematicamente vietato 
per un propagatore gluonico senza auto-interazione col mezzo ($m_g=0$), \`e ora 
permesso in virt\`u della massa finita e della larghezza termica del gluone.
\newline  
Con questa nuova tecnica \`e stato possibile ricalcolare la frequenza di produzione 
dei quark strani entro il QGP~\cite{Bilic}\cite{Altherr93}, con maggior precisione 
rispetto ai primi calcoli inizialmene eseguiti~\cite{Raf82-86}\cite{Raf86}  
per i processi $q\bar{q} \longrightarrow s\bar{s}$\ e  $gg \longrightarrow s\bar{s}$, 
ma soprattutto \`e possibile considerare il nuovo processo 
$g^* \longrightarrow s\bar{s}$. A ragion del vero, il decadimento di un 
gluone termico in una coppia di quark strani era gi\`a stata 
considerata in~\cite{Biro}, l\`i  
dove, per\`o, la motivazione di introdurre una massa finita per il gluone proveniva 
da quanto suggerito dai risultati della QCD su reticolo~\cite{Ukawa}. 
\newline  
I risultati di un calcolo svolto con questa tecnica,  
assumendo per i quark le loro masse costituenti e per i gluoni una massa  
costante ed indipendente dalla temperatura, sono mostrati in fig.~\ref{Heinz2}. 
Il grafico nel riquadro superiore \`e il risultato dei primi
calcoli~\cite{Raf82-86}\cite{Raf86} svolti con $m_g=0$\ (propagatore
gluonico ``nudo''), quello nel riquadro inferiore assume $m_g=500$\ MeV.  
Nel primo caso, il processo di fusione gluonica (fig.~\ref{DiagrStran}.b) \`e 
la sorgente dominante per la produzione di quark strani, superando 
di un ordine di grandezza la produzione per annichilazione di quark leggeri, 
a causa del maggior numero di gradi di libert\`a a disposizione dei gluoni.  
Nel secondo caso, una massa non nulla per i gluoni provoca una diminuzione della 
frequenza delle fusioni tra gluoni, in quanto si riduce notevolmente il fattore 
statistico di Bose per i gluoni entranti nel diagramma di fig.~\ref{DiagrStran}.b, 
ed un aumento della frequenza di produzione per annichilazione $q\bar{q}$\ dei 
quark leggeri, in quanto il gluone intermedio \`e ora solo leggermente 
{\em ``off mass shell''} (fuori dalla massa fisica), in virt\`u della sua 
massa termica. Il processo pi\`u importante diventa, per valori elevati di $m_g$\ 
quali quelli qui considerati, il decadimento del gluone termico. Tuttavia, sebbene 
con l'introduzione della massa termica dei gluoni vi siano cambiamenti anche 
sostanziali nelle frequenze parziali dei diversi processi, la frequenza totale  
$A=R_{tot}=R_{q\bar{q}\rightarrow s\bar{s}}+R_{gg \rightarrow s\bar{s}} 
  +R_{g^* \rightarrow s\bar{s}}$ rimane quasi invariata~\cite{Biro}.  

\begin{figure}[htb]
\begin{center}
\includegraphics[scale=0.95]{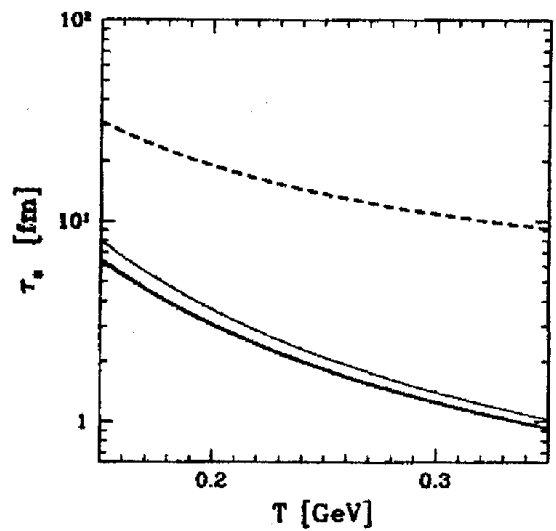}
% \caption{Costante di tempo di rilassamento chimico per la stranezza 
%           nel QGP in funzione della temperatura, per 
%	   $\alpha_S(M_Z)=0.118$,  $m_s(M_Z)=90$\ MeV e $ $. 
%	   Il contributo dei quark }
\caption{Costante di tempo di rilassamento chimico per la stranezza
         nel QGP in funzione della temperatura. Il contributo dei quark 
	 (linea tratteggiata), dei gluoni (linea sottile) e totale 
	 (linea spessa) sono calcolati per $m_s=160$\ MeV ed 
	 $\alpha_s=0,6$~\cite{Raf96}.}
\label{DaAgg}
\end{center}
\end{figure}

Si pu\`o dimostrare~\cite{Raf96} che la densit\`a di quark strani nel plasma 
($\rho_s$) raggiunge l'equilibrio in maniera asintotica, secondo la legge: 
\begin{equation}
\rho_s(t)=\rho_s(\infty)\tanh(t/2\tau_s) \sim \rho_{s}(\infty)(1-2e^{-t/\tau_s}) 
    \quad {\rm per \, t \gg \tau_s } 
\label{StrangeTime}
\end{equation}
dove $\rho_s(\infty)$\ indica la densit\`a di quark strani all'equilibrio e $\tau_s$, 
detta costante di tempo di rilassamento chimico, fornisce una stima della scala di 
tempo necessaria per la saturazione della produzione di stranezza. 
In fig.~\ref{DaAgg} \`e mostrato l'andamento di tale costante in funzione della   
temperatura del QGP. 
Il contributo dei processi gluonici (linea sottile), fusione 
od annichilazione, risulta dominante  rispetto a quello proveniente 
dall'annichilazione dei quark (linea trattegiata) e la somma dei contributi 
(linea spessa) indica che l'equilibrio chimico nella produzione di quark strani 
\`e raggiunto entro $\sim 3$\ fm/$c$\ se la temperatura del plasma si mantiene 
al di sopra dei 200 MeV. Il calcolo della costante di tempo di rilassamento dipende 
dai valori usati per $m_s$\ ed $\alpha_s$\  
ma, anche nei casi pi\`u sfavorevoli, risulta confrontabile col tempo di vita del QGP. 
\newline
Il parziale ripristino della simmetria chirale contribuisce a favorire la produzione 
di stranezza nel QGP, riducendo la soglia di produzione di coppie $s\bar{s}$. Inoltre, 
la produzione dei quark $u$\ e $d$\ risulta sfavorita in regimi di alta densit\`a 
barionica, quale quello creato in seguito alle collisioni all'SPS. Infatti, la 
grande abbondanza di quark $u$\ e $d$\ inizialmente pre-esistenti nel plasma si 
traduce in un alto valore del potenziale bariochimico $\mu_B$ del plasma, ed in 
una riduzione di un fattore $e^{-\frac{\mu_B}{3}/T}$\ dell'ulteriore produzione 
di quark leggeri ({\em ``Pauli blocking''}).   
Questi meccanismi fanno s\`i che nel plasma all'equilibrio chimico, 
l'abbondanza di quark $\bar{u}$, $\bar{d}$, e $\bar{s}$\ sia simile e, di 
conseguenza, sia fortemente favorita la produzione di antibarioni strani e 
multi-strani rispetto a quanto avviene nelle normali interazioni adroniche.
\subsection{Risultati sperimentali di WA97}
\begin{figure}[htb]
\begin{center}
\includegraphics[scale=0.28]{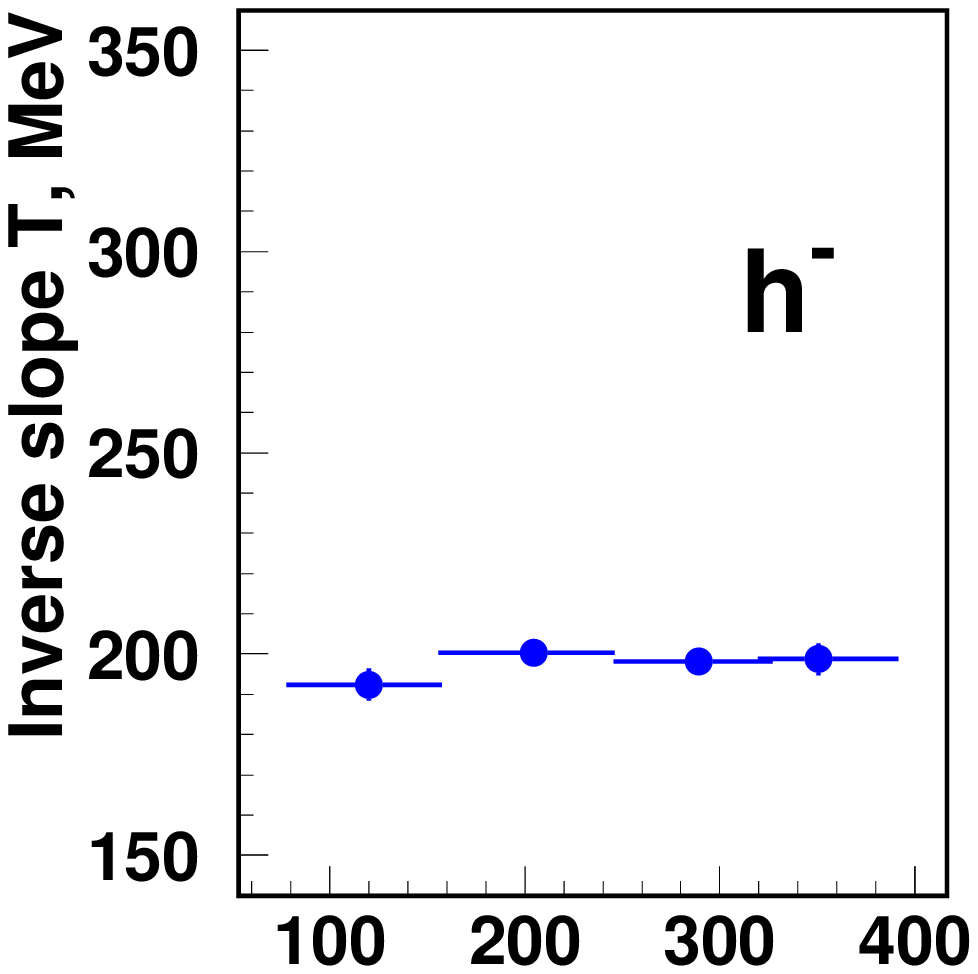}
\includegraphics[scale=0.28]{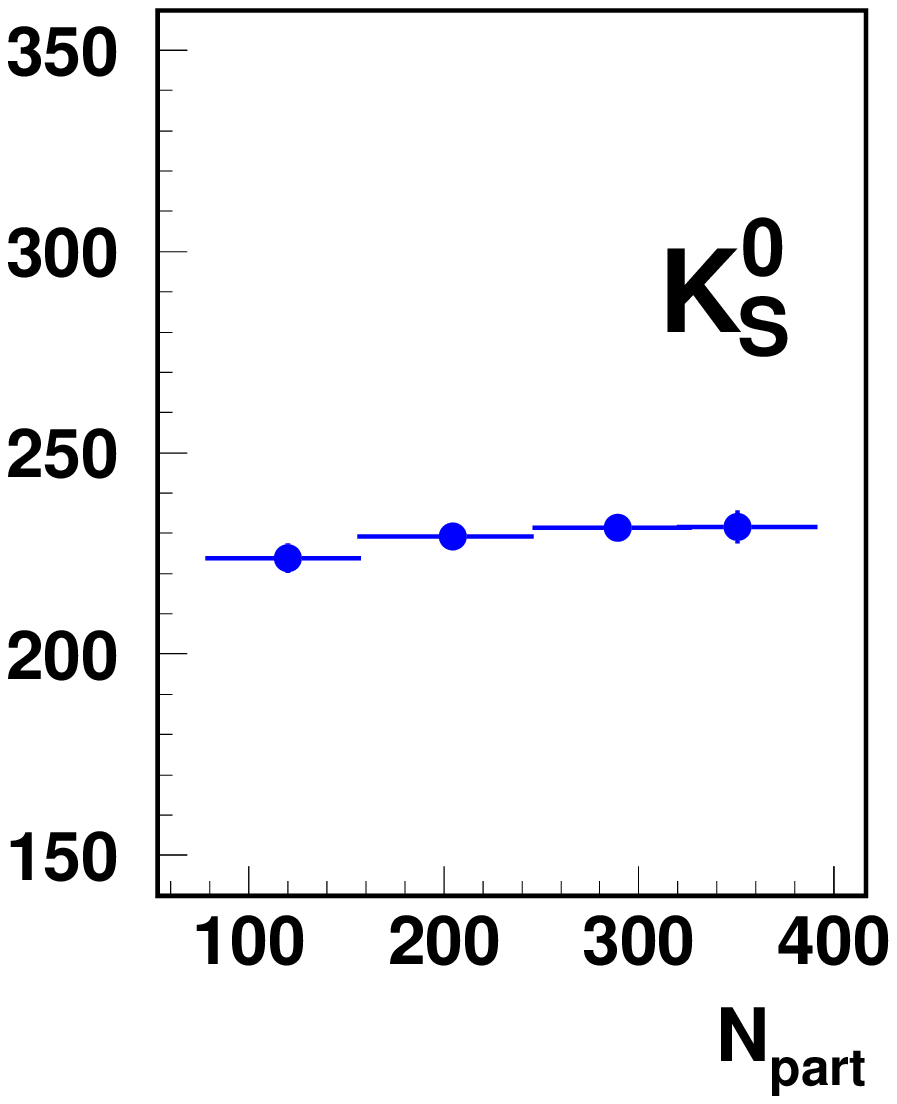}
\includegraphics[scale=0.28]{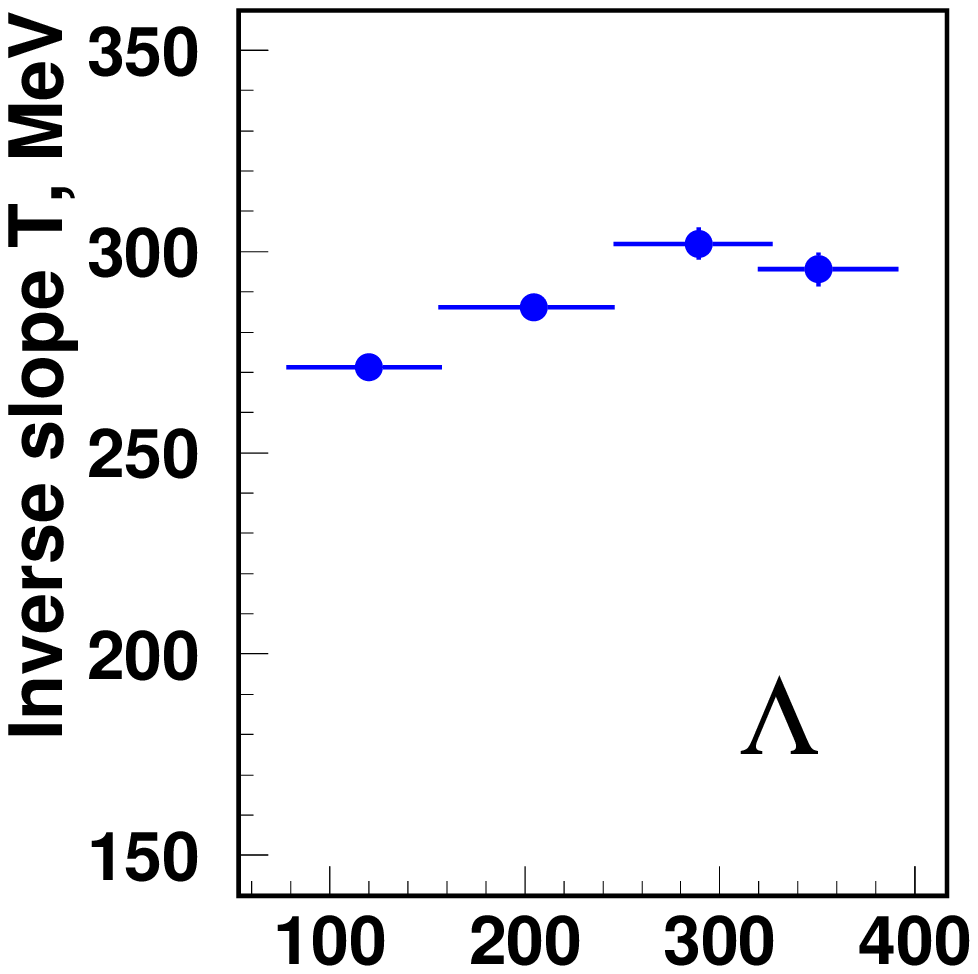}
\includegraphics[scale=0.28]{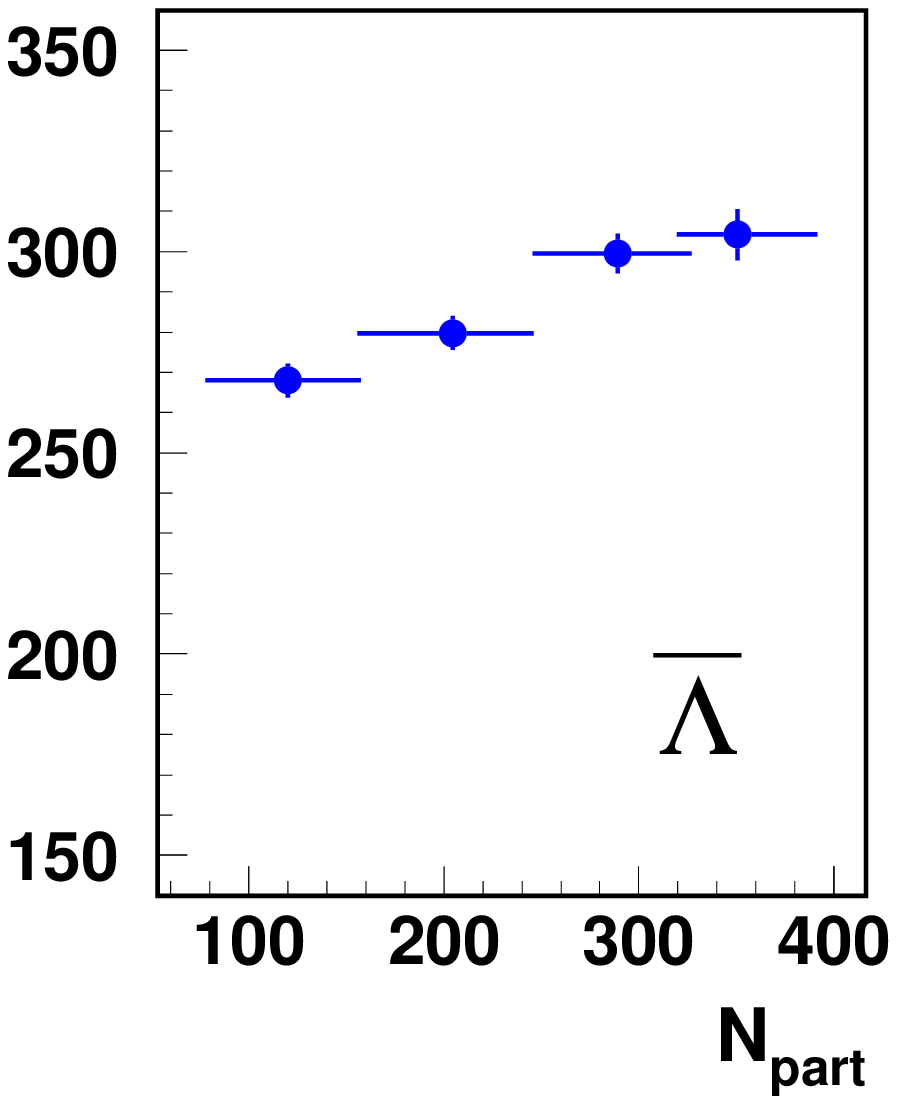}\\
\includegraphics[scale=0.28]{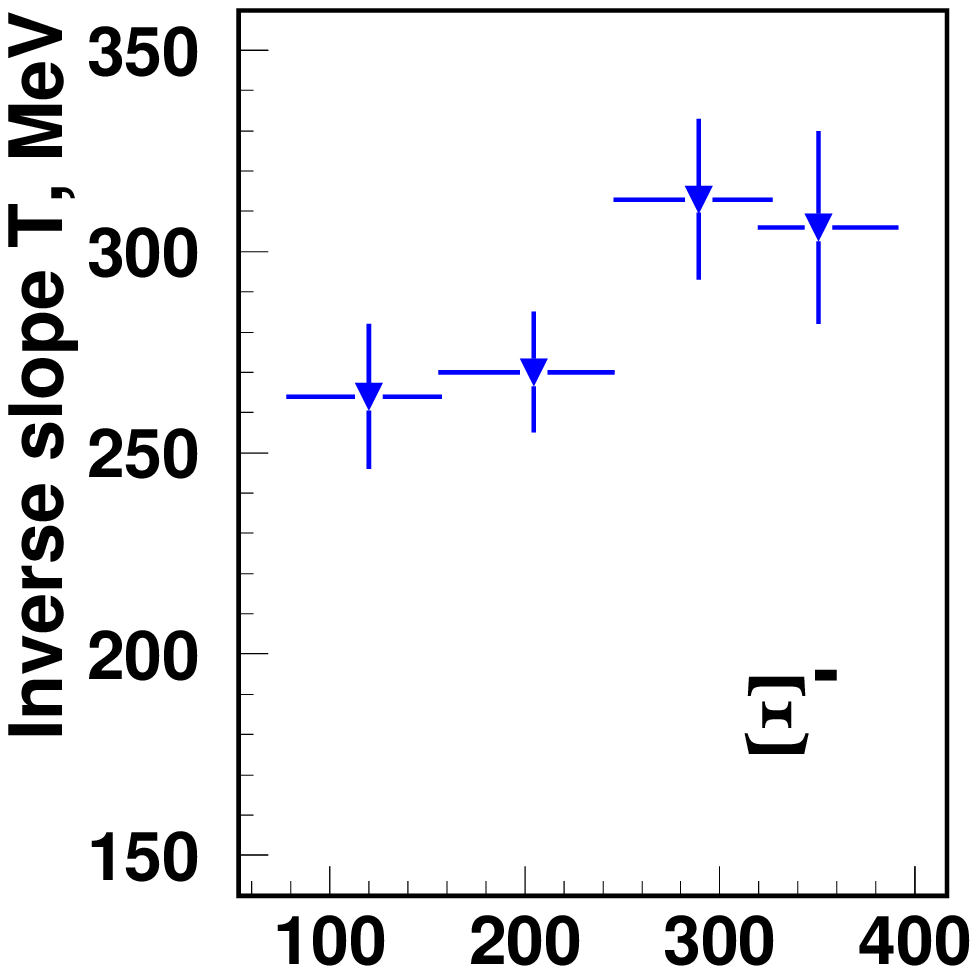}
\includegraphics[scale=0.28]{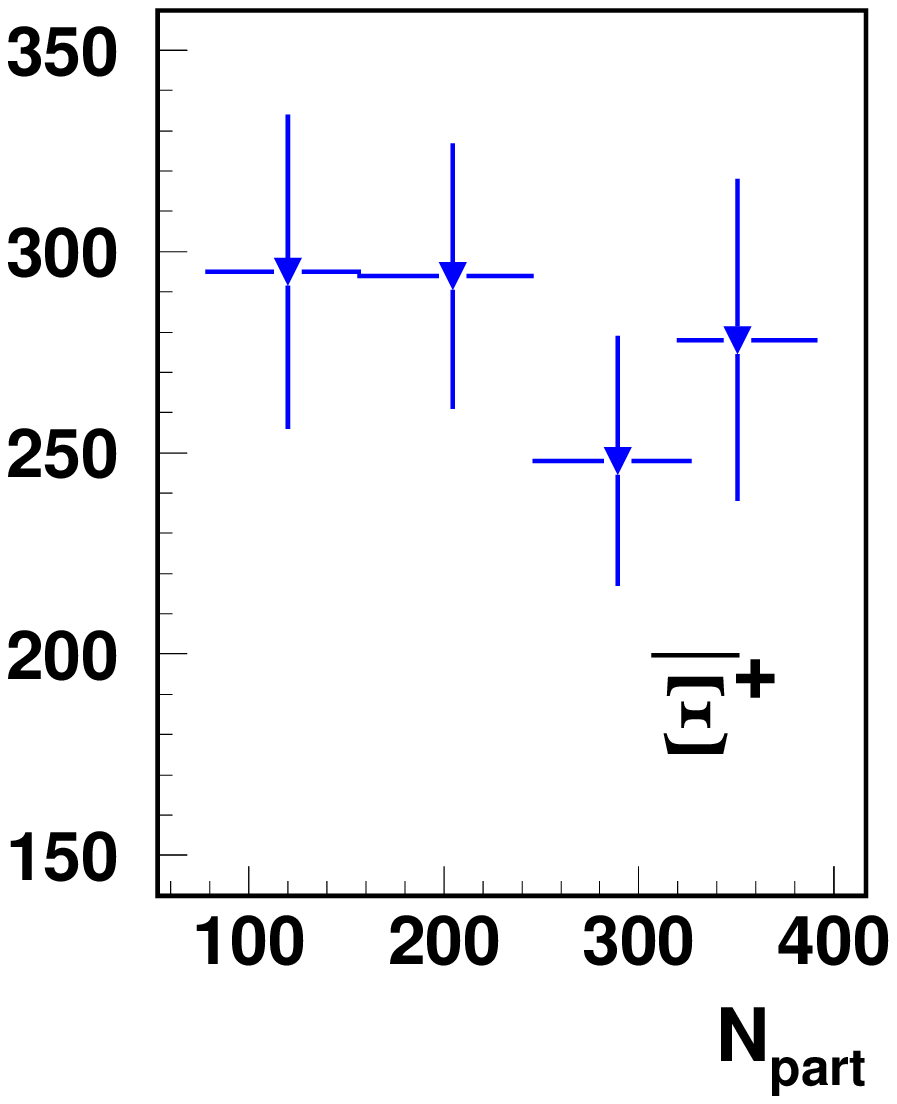}
\includegraphics[scale=0.28]{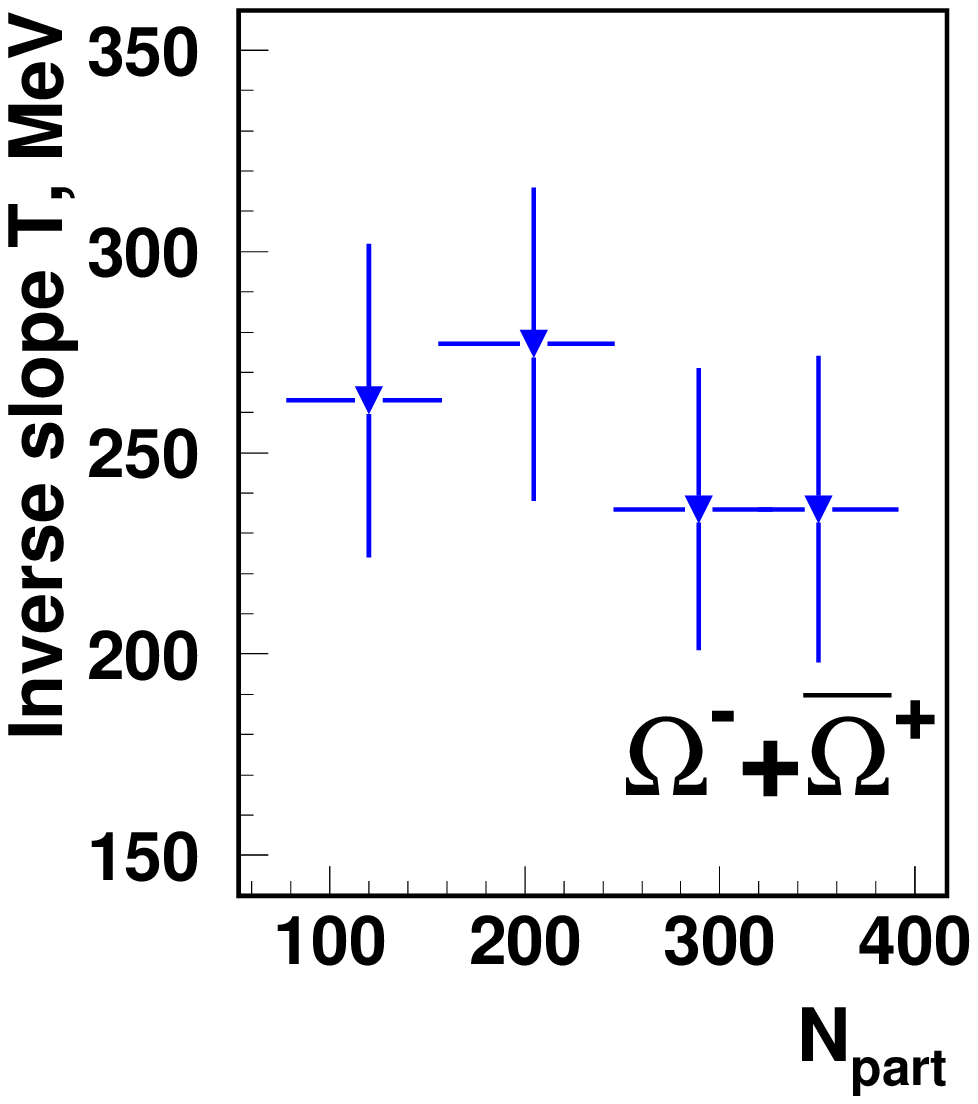}
\caption{Parametro $T_{app}$\ degli spettri di massa trasversa in funzione dalla 
 centralit\`a della collisione, come misurato da WA97~\cite{mt_WA97}.}
\label{WA97mt_fig}
\end{center}
\end{figure}

L'esperimento che sino ad ora ha studiato in maggior dettaglio la produzione 
di particelle strane nelle collisioni tra ioni ultra-relativistici   
\`e WA97~\cite{WA97}. Esso ha studiato la produzione di iperoni $\Lambda$, 
$\Xi^-$, $\Omega^-$ ed i rispettivi anti-iperoni, di $K_S^0$\ e di particelle 
cariche negativamente ($h^-$) a rapidit\`a centrale ed impulso trasverso 
medio-alto ($p_T \sim 0.5 \div 3 $ GeV/c) in collisioni Pb-Pb a 158 Gev/$c$\ 
per nucleone ed  in collisioni p-Be e p-Pb alla stessa energia. 
In questo paragrafo si esporanno brevemente i principali risultati della 
collaborazione WA97. 
\newline
La centralit\`a delle collisioni Pb-Pb viene determinata~\cite{WA97Centr}  
misurando la molteplicit\`a di particelle cariche prodotte nell'interazione 
ed \`e  parametrizzata in termini di nucleoni partecipanti alle collisioni, 
valutati con il modello di Glauber. In termini di centralit\`a, sono state definite 
quattro classi, indicate con i numeri romani da I a IV, cui corrispondono 
i numeri medi ($N_W$) e le $FWHM$\ della distribuzione di nucleoni partecipanti    
riportate in tabella 1.5. L'insieme delle quattro classi corrisponde alle 
interazioni pi\`u centrali, per circa il 40\% della sezione d'urto d'interazione  
anelastica.  
\begin{table}[h]
 \label{tab15}
  \begin{center}
  \begin{tabular}{ccccccc} \hline
    & p-Be & p-Pb  & Classe I & Classe II & Classe III & Classe IV   \\
\hline $<N_W>$ & $2.5$ & $4.75$ & 
                     $120.1^{+5.7}_{-6.1}$ & $204.6^{+4.1}_{-4.4}$ & 
                     $289.0^{+2.5}_{-2.9}$ & $350.6^{+0.9}_{-1.1}$\\
       $FWHM$  & & & $79.7$ & $90.3$ & $81.3$ & $72.1$ \\ \hline
\end{tabular}
\end{center}
\caption{Valor medio e $FWHM$\ delle distribuzioni di nucleoni 
	 partecipanti  per le collisioni p-Be, p-Pb e Pb-Pb nelle 
	 quattro classi di centralit\`a di WA97.}
\end{table}
\newline
Dalla misura 
sperimentale della distribuzione degli impulsi delle particelle di una data 
specie $ \frac{{\rm d}^2N(m_T,y) }{{\rm d}m_T {\rm d}y}$,  si determina 
il numero di particelle prodotte per evento ({\em ``yield''}) nell'unit\`a 
di rapidit\`a intorno al valore centrale ed estrapolato all'intero spettro 
di impulso trasverso, 
secondo l'espressione: 
\begin{equation}
 Yield = \int_m^{\infty} {\rm d}m_T \int_{y_{CM}-0.5}^{y_{CM}+0.5}
       {\rm d}y \, \frac{{\rm d}^2N(m_T,y) }{{\rm d}m_T {\rm d}y}
\label{Yield}
\end{equation}
Poich\'e si considera un piccolo intervallo di rapidit\`a (un'unit\`a) 
attorno al valore di rapidit\`a centrale, \`e ragionevole assumere una 
parametrizzazione per la distribuzione sperimentale 
$\frac{{\rm d}^2N(m_T,y) }{{\rm d}m_T {\rm d}y} $\ 
--- che viene misurata generalmente in una finestra di accettanza pi\`u 
piccola rispetto a quella in cui si valuta lo {\em ``yield''} --- 
indipendente da $y$\ ({\em cfr.} paragrafo 1.4.2),  per cui:
\begin{equation} 
\frac{{\rm d}^2N(m_T,y) }{{\rm d}m_T {\rm d}y} = 
    {\rm cost} \, m_T \exp(-\frac{m_T}{T_{app}})
\nonumber
\end{equation}
I valori del parametro $T_{app}$, pari all'inverso della pendenza degli spettri 
di massa trasversa in coordinate logaritmiche, misurati da WA97 sono 
riportati in tabella 1.6~\cite{mt_WA97}:  
\begin{table}[h]
 \label{tab16}
 \begin{center}
\begin{tabular}{|c|c|} \hline
{Particella} & Pendenza Inversa $T_{app}$ (MeV)   \\
\hline   $h^-$                 & $197 \pm 2$  \\
\hline   $K^0_S$               & $230 \pm 2$  \\
\hline   $\Lambda$             & $289 \pm 3$  \\
\hline   $\bar{\Lambda}$       & $287 \pm 4$  \\
\hline   $\Xi^-$               & $286 \pm 9$  \\
\hline   $\bar{\Xi}^+$         & $284 \pm 17$ \\
\hline   $\Omega^- + \Omega^+$ & $251 \pm 19$ \\ \hline 
\end{tabular}
\end{center}
\caption{Inverso della pendenza degli spettri di massa trasversa per le
         particelle misurate da WA97~\cite{mt_WA97}.}
\end{table}
L'andamento del parametro $T_{app}$\ in funzione della centralit\`a della 
collisione \`e mostrato in fig.~\ref{WA97mt_fig}.
Si osserva dalla tabella~1.6 e dalla fig.~\ref{WA97mt_fig} che, entro gli errori 
sperimentali, le particelle e le rispettive antiparticelle strane 
(i.e. le $\Lambda$\ e le $\Xi$) presentano 
stessi valori per il parametro $T_{app}$. L'esigua statistica a disposizione per la 
particella $\Omega$, che dispone di tre unit\`a di stranezza ed \`e quindi molto 
pi\`u rara, non ha permesso di separare le $\Omega^-$\ dalla rispettiva 
antiparticella $\bar{\Omega}^+$. 
Una conferma di un simile risultato di simmetria tra il settore
barionico e quello anti-barionico, anche per la $\Omega$,  
sarebbe dunque estremamente interessante. 

\begin{figure}[htb]
\begin{center}
\includegraphics[scale=0.45]{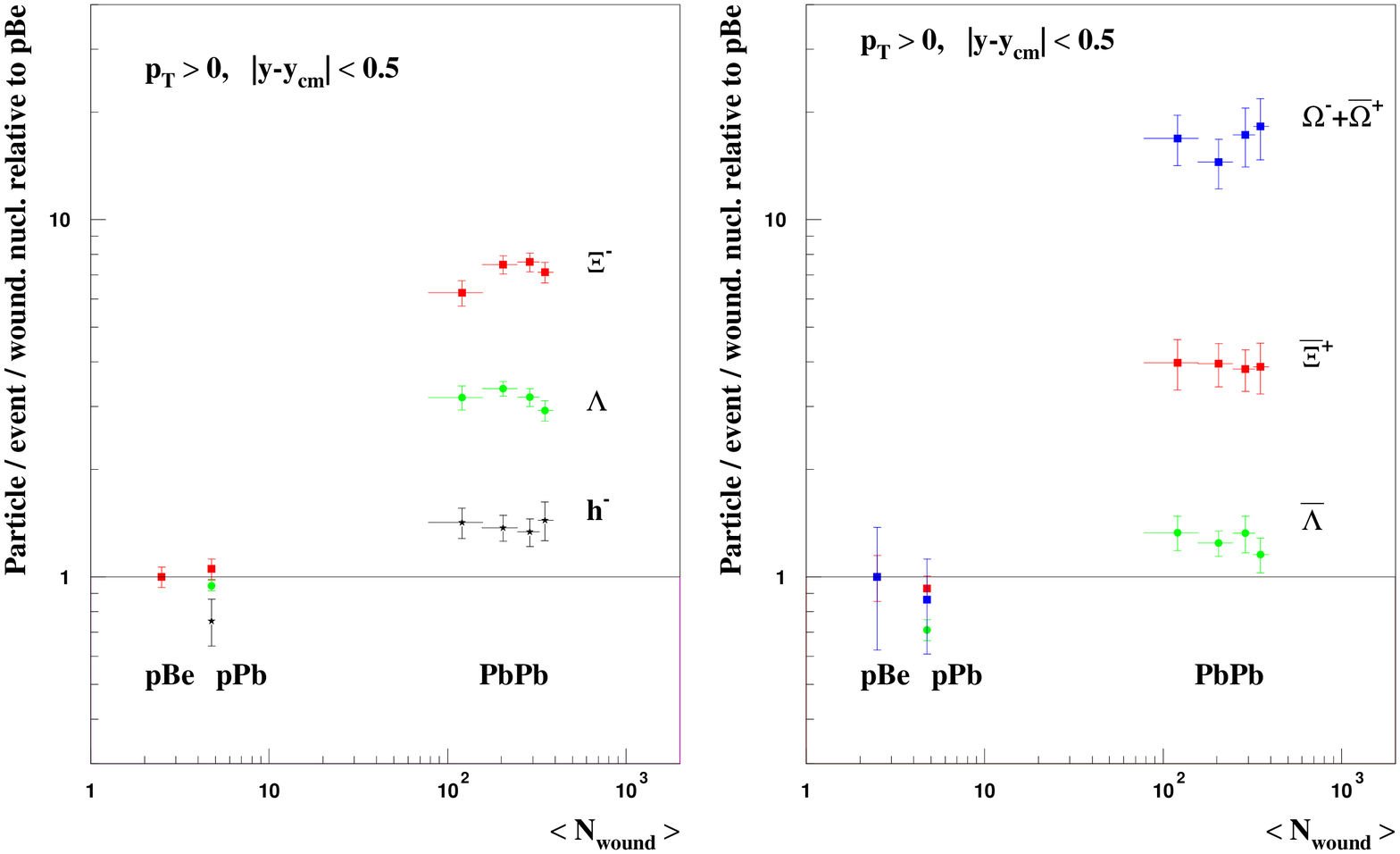}
\caption{Incremento nella produzione di stranezza misurato da WA97
        (si veda il testo per i dettagli).}
\label{EnWA97}
\end{center}
\end{figure}

Una volta noto lo {\em ``yield''}, l'incremento nella produzione di una 
data particella tra le interazioni Pb-Pb e quelle di riferimento (p-Be) 
viene definito come: 
\begin{equation}
E = \left( \frac{<Yield>}{<N_{part}>}\right)_{{\rm Pb-Pb}} / 
    \left( \frac{<Yield>}{<N_{part}>}\right)_{{\rm p-Be}} 
\label{Enhanc}
\end{equation}
dove $<Yield>$\ \`e lo {\em ``yield''} di produzione nell'interazione o 
classe di centralit\`a cui corrisponde un numero medio di nucleoni 
partecipanti pari a $<N_{part}>$.  
In fig.~\ref{EnWA97} \`e mostrata la dipendenza dell'incremento $E$\  
dal numero di nucleoni partecipanti alla collisione~\cite{WA97PLB449}. Le 
particelle sono state divise tra quelle che non contengono alcun quark 
di valenza in comune con quelli originariamente presenti nel sistema 
interagente ($\bar{\Lambda}[\bar{u}\bar{d}\bar{s}]$, 
$\bar{\Xi}^+[\bar{d}\bar{s}\bar{s}]$, $\Omega^-$[sss]\ e 
$\bar{\Omega}^+[\bar{s}\bar{s}\bar{s}]$), e quelle che ne contenengono 
($h^-[\sim d\bar{u}] $, $\Lambda[uds]$\ e $\Xi^-[dss]$)~\footnote{Tra le particelle 
di carica negativa $h^-$, costituiti prevalentemente da $\pi^-[d\bar{u}]$, 
vi sono per\`o anche 
$K^-[\bar{u}s]$\ e $\bar{p}[\bar{u}\bar{u}\bar{d}]$\ che non contengono alcun 
quark.}. 
La linea orizzontale corrisponde al caso di uno {\em ``yield''} che aumenti 
linearmente col numero di nucleoni partecipanti, andamento atteso 
nell'ipotesi che la collisione Pb-Pb possa interpretarsi semplicemente 
come una sovrapposizione di collisioni elementari $N$-$N$. In quest'ultima 
ipotesi, infatti, non dovrebbe osservarsi alcun incremento tra le collisioni  
p-Be, che meglio approssimano le collisioni $N$-$N$, o p-Pb, cui corrisponde  
un numero di nucleoni partecipanti di poco superiore a p-Be, e la collisione 
Pb-Pb. Il discostarsi dell'incremento dall'unit\`a, od equivalentemente 
la dipendenza dello {\em ``yield''} da $<N_{part}>$\ pi\`u rapida di una legge 
lineare, suggerisce che differenti processi di produzione delle particelle 
si instaurano nelle collisioni Pb-Pb.
\newline
Si osserva che l'incremento aumenta con il contenuto di stranezza dell'adrone, 
come atteso nello scenario di QGP:  
\begin{align}
& E(\Omega^- + \bar{\Omega}^+) > E(\bar{\Xi}^+) > E(\bar{\Lambda}) \nonumber \\
& E(\Xi) > E (\Lambda)  > E(h^-)  \, .\nonumber
\end{align}
L'incremento pi\`u elevato osservato, in corrispondenza dell'iperone $\Omega$, 
\`e pari ad un fattore circa quindici rispetto alle interazioni p-Be, cio\`e 
allo scenario di normali interazioni adroniche. 
Questi risultati costituiscono una delle evidenze sperimentali 
pi\`u significative del raggiungimento dello stato di QGP al SPS. 
\newline
I dati in fig.~\ref{EnWA97} suggeriscono inoltre che, nelle collisioni Pb-Pb, entro 
l'intervallo di centralit\`a coperto da WA97, l'incremento della produzione 
di stranezza abbia ormai raggiunto un valore di saturazione. Se questo incremento 
\`e imputabile al raggiungimento della fase deconfinata, ci\`o implica che  
in tutte le classi di centralit\`a definite in WA97 pu\`o aver luogo 
la transizione. 
\newline
\`E dunque di notevole interesse poter innanzitutto confermare con un nuovo 
esperimento i risultati di WA97 qui brevemente esposti,  
ed ancor di pi\`u poter esplorare la regione 
di fig.~\ref{EnWA97} compresa tra $N_{part}\approx 5$ (corrispondente alle 
interazioni p-Pb) ed $N_{part}\simeq 120$ (corrispondente alla I classe di WA97, 
cio\`e quella delle  collisioni pi\`u  periferiche). Infine, se la 
materia nucleare effettua 
davvero la transizione nel QGP, si pu\`o pensare di determinare i limiti per cui 
questa pu\`o avvenire anche diminuendo l'energia inizialmente a disposizione per il 
sistema; ci\`o vuol dire utilizzare fasci di ioni meno energetici rispetto 
a quello di Pb a 160 $A$ GeV/$c$\ usato da WA97, oppure sistemi pi\`u leggeri.  
\newline
%Agendo in queste direzioni, qualora i nuovi risultati confermassero 
%l'interpretazione qui esposta dei dati di WA97, si potrebbero anche determinare, 
%in termini di centralit\`a e di energia dei nuclei collidenti, 
%le condizioni richieste per la transizione nel QGP.
Questo \`e il programma sperimentale di NA57, come sar\`a descritto nei prossimi 
capitoli.  
\section{Conclusioni}
In questo capitolo si sono dunque descritte le maggiori 
evidenze sperimentali di QGP all'SPS,  
dovute agli esperimenti che hanno studiato (e studiano tuttora) 
le collisioni tra nuclei pesanti ultra-relativistici.  
\newline
%Considerate 
Considerando  
{\em globalmente} tutte queste evidenze,  
%e con maggior enfasi  
e con maggior rilevanza quelle sulla soppressione della produzione 
degli stati di charmonio ({\em paragrafo} 1.4.4)
e sull'incremento della produzione di stranezza ({\em paragrafo} 1.5.3), diventa 
un'impresa ardua, se non impossibile,   
giustificarle simultaneamente in uno 
scenario adronico senza la transizione nella fase di QGP. 
\newline
Nel febbraio del 2000 il laboratorio del CERN ha infatti formalmente 
annunciato~\cite{PressRelease} 
di ritenere conclusive  le evidenze sperimentali sin qui raccolte dai 
%sette  
suoi 
esperimenti coinvolti in questo campo: 
%della fisica: 
%sulle collisioni tra ioni relativistiche: 
\newline
``{\em ... A common assessment of the collected data leads us to conclude
that we now have compelling evidence that a new state of matter has indeed
been created, at energy densities which had never been reached over
appreciable  volumes in laboratory experiments before and which exceed
by more than a factor 20 that of normal nuclear matter. The new state of
matter found in heavy ion collisions at the SPS features many of the
charateristics of the theoretically predicted quark-gluon 
plasma...''} (dal discorso del prof. L.~Maiani, direttore generale del CERN).
\newline
Restano tuttavia tantissimi interrogativi cui urge fornire risposta.   
Il primo e pi\`u ovvio \`e il seguente:  
a partire da quali condizioni avviene la transizione di fase ? 
%Tutti i segnali sperimentali proposti sono davvero interpretabili in 
%termini di QGP? 

%% file: cap2/cap2.tex
\chapter{L'esperimento NA57 al CERN }

\section{Introduzione}
L'esperimento NA57 si propone di studiare la produzione di particelle  
strane e multi-strane nelle collisioni tra nuclei di piombo ultra-relativistici  
ed in collisioni di riferimento tra protone e berillio~\cite{NA57p}.  
\newline
Nell'ambito del programma di studio delle interazioni tra nuclei pesanti 
di altissima energia portato avanti dal CERN, l'esperimento NA57 rappresenta 
la naturale evoluzione dei precedenti esperimenti WA85~\cite{WA85}, 
WA94~\cite{WA94} e WA97~\cite{WA97}. Questi hanno studiato, nell'ordine, le 
interazioni zolfo-tungsteno (protone-tungsteno) a 200 $A$ GeV/$c$, 
zolfo-zolfo (protone-zolfo) a 200 $A$ GeV/$c$ e 
piombo-piombo (protone-piombo e protone-berillio) a 160 $A$ GeV/$c$ 
nello spettrometro OMEGA~\cite{OMEGA} del CERN. 
I primi due esperimenti, WA85 e WA94, utilizzavano un sistema di camere 
proporzionali multifili per la ricostruzione dei decadimenti di particelle 
strane. In WA97 si \`e usato un telescopio di rivelatori a microstrip di 
silicio in cui 
sono stati introdotti i primi rivelatori a pixel di silicio, 
gli {\em Omega2} ({\em cfr}. paragrafo 2.4.3); questi sono stati poi 
ripresi e sviluppati ulteriormente ({\em Omega3}) in NA57, sostituendo 
completamente le microstrip.  
\newline
Come esposto nel capitolo primo, WA97 ha studiato la produzione di particelle 
strane in funzione della centralit\`a della collisione Pb-Pb confrontando con 
la produzione nelle interazioni di riferimento p-Be e p-Pb. 
Il principale risultato di questo esperimento consiste nell'aver osservato che 
la produzione di particelle strane non scala linearmente col numero di nucleoni che 
prendono parte all'interazione. Al contrario,  
il numero di particelle strane di una data specie prodotte per evento 
(il ``tasso'', o {\em yield} in inglese), diviso il numero di 
nucleoni interagenti, \`e incrementato nelle interazioni Pb-Pb rispetto a quelle 
p-Be e p-Pb.  
Tale incremento presenta una gerarchia che dipende dal numero di stranezza dell'iperone: 
l'incremento aumenta con il contenuto di stranezza della particella. 
Un simile comportamento \`e atteso nell'ipotesi di formazione del QGP 
nelle collisioni Pb-Pb e 
di normali interazioni adroniche nelle collisioni di riferimento. 
\newline
Il secondo pi\`u importante risultato di WA97 \`e quello che l'incremento  
di produzione di stranezza sembra aver ormai raggiunto un valore di saturazione 
nell'intervallo di centralit\`a coperto dall'esperimento. 
Ci\`o dovrebbe dunque suggerire che, nell'ipotesi che l'incremento osservato sia 
conseguenza della transizione di fase nello stato di QGP, tale transizione avvenga  
in tutte  le collisioni Pb-Pb osservate da WA97, indipendentemente dalla loro 
centralit\`a. 
\newline 
L'esperimento NA57 si prefigge di estendere e completare le misure di WA97, in particolare 
si vuole:  
\begin{itemize}
\item[-]
 investigare la dipendenza dall'energia del fascio incidente dell'incremento nella produzione 
 di particelle strane e multi-strane; 
\item[-]
 misurare il tasso di produzione di particelle strane e multi-strane su di un intervallo di 
 centralit\`a pi\`u esteso (verso la regione periferica) rispetto a quello coperto da WA97.  
\end{itemize}
Per rispondere al primo obiettivo, si sono raccolti dati utilizzando il fascio di 
ioni di piombo a due diverse energie: 160 $A$\ GeV/$c$ e 40 $A$\ GeV/$c$.  
Come interazione di riferimento all'energia pi\`u bassa (40 GeV/$c$ per nucleone), 
si sono considerate le collisioni p-Be a 40 GeV/$c$. 
Non si prevede di raccogliere nuovi dati sulle interazioni di 
riferimento protone-nucleo all'energia di 160 GeV/$c$, 
poich\'e si possono utilizzare i dati gi\`a raccolti da WA97. 
\newline
Per raggiungere il secondo obiettivo, ci si \`e  impegnati per 
ridurre al minimo tutte le interazioni del fascio al di fuori del bersaglio e per 
migliorare le  caratteristiche del fascio stesso, quali la stabilit\`a e la 
focalizzazione. 

\section{Il fascio}
\begin{figure}[hbt]
\begin{center}
 \includegraphics[scale=0.40]{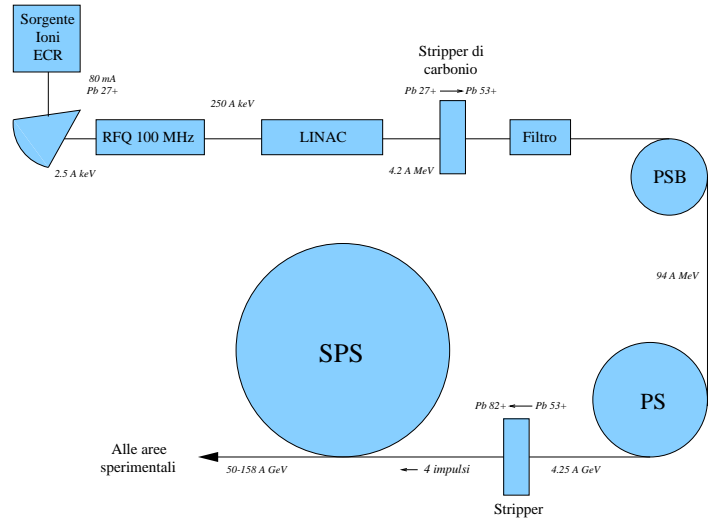}
 \caption{Schema degli stadi di accelerazione del fascio di piombo.}
 \label{SPS_1}
 \end{center}
\end{figure}

\begin{figure}[b]
\begin{center}
 \includegraphics[scale=0.35]{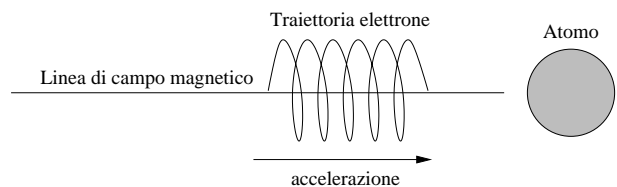} 
  \caption{Principio di funzionamento dell'ECR.}
  \label{ECR}
  \end{center}
\end{figure}

Dalla sua costruzione nel 1976 il {\em Super Proton Synchrotron} 
(SPS) del CERN ha subito varie modifiche prima di essere in
grado di accelerare ioni pesanti~\cite{SPS}.
\newline
L'esperimento NA57 utilizza la linea di fascio H4 estratta nell'area Nord  
dell'SPS. In fig.~\ref{SPS_1} sono mostrati i vari stadi di accelerazione  
degli ioni di piombo.  
Il primo stadio della produzione del fascio di ioni piombo prevede 
una sorgente a risonanza elettronica di ciclotrone (ECR). Nell'ECR gli atomi 
di piombo vengono sublimati da un campione di 
%$^{208}{\rm Pb}$; 
piombo;   
essi sono quindi 
ionizzati a seguito di urti con degli 
elettroni liberi, che provengono da molecole di ossigeno, 
e che vengono accelerati lungo un campo magnetico applicato alla sorgente  
(fig.~\ref{ECR}).
Dall'ECR gli ioni vengono immessi nella {\em linea di trasporto di bassa energia} 
e quindi nello spettrometro ad alta risoluzione  che seleziona gli ioni 
${\rm Pb}^{27+}$. Gli ioni passano quindi attraverso il campo di un 
quadrupolo a radiofrequenza (RFQ) da 100 MHz dove vengono focalizzati, 
suddivisi in pacchetti ed accelerati da 2.5 $A$ keV/$c$ a 250 $A$ keV/$c$. 
A questo punto la {\em linea di trasporto di energia intermedia} guida il fascio 
fino ad un acceleratore lineare (LINAC). Nel LINAC gli ioni di piombo vengono 
accelerati sino a 4.2 $A$ MeV/$c$; il fascio passa quindi attraverso un foglio 
di carbonio nel quale perde parte dei rimanenti elettroni. Una serie di magneti 
seleziona lo stato di carica ${\rm Pb}^{53+}$. Da questo momento gli ioni 
vengono portati ai successivi stadi di accelerazione tramite la 
{\em linea di trasporto di alta energia}. 
\newline
Il ${\rm Pb}^{53+}$\ viene iniettato senza ulteriore ionizzazione nel 
{\em Proton Synchrotron Booster} (PSB). Esso \`e stato appositamente 
modificato per il fascio di piombo in modo da poter raggiungere un vuoto con 
pressioni sino a soli $10^{-7}$\ Pa. Le radiofrequenze del PSB sono 
impulsate in modo tale da preservare la struttura a pacchetti del fascio ed 
impartiscono agli ioni una accelerazione tale che la loro velocit\`a passi dal 
10\% al 20\% della velocit\`a della luce.  
Il processo di accelerazione viene ripetuto una seconda volta, a seguito della 
quale la velocit\`a degli ioni diventa pari al 42 \% della velocit\`a della luce 
e gli ioni lasciano il PSB con un momento di 94 $A$ MeV/$c$. 
\newline
Lo stadio successivo si serve del {\em Proton Synchrotron (PS)}. Il processo di 
accelerazione nel PS \`e organizzato in super-cicli di durata di 19.2 $s$, che 
corrispondono a quattro cicli del PSB, ciascuno della durata di 1.2 $s$. Il tempo 
rimanente \`e dedicato alla iniezione di elettroni e positroni: il complesso 
dell'SPS \`e infatti in grado di fornire contemporaneamente un fascio di ioni 
pesanti ed uno di elettroni e positroni da iniettare nel {\em Large Electron 
Positron collider} (LEP). Dopo aver raggiunto un'energia di 4.5 $A$ GeV/$c$\ i 
quattro pacchetti di ioni sono consecutivamente iniettati nell'SPS, dopo esser 
passati attraverso un foglio che li priva dei rimanenti elettroni, portandoli 
nello stato di carica ${\rm Pb}^{82+}$. Nell'SPS gli ioni vengono quindi accelerati 
sino ad un momento di 40 o 156 $A$ GeV/$c$\ ed invati verso i vari esperimenti 
situati nella {\em Nord Area} e nella {\em West Area}. 
Il tempo totale di estrazione di un pacchetto \`e di 4.8 $s$\  e due pacchetti sono 
intervallati tra loro di 19.2 $s$, mentre la tipica intensit\`a del fascio di piombo 
nella linea H4  \`e di circa $10^6$\ ioni per estrazione. 

Per lo studio dell'interazione p-Be a 40 $A$ GeV/$c$\ viene utilizzato un fascio 
secondario, costituito cio\`e da protoni non direttamente accelerati 
dall'SPS, in quanto quest'ultimo \`e stato progettato per accelerare 
protoni a pi\`u elevato momento (450 GeV/$c$). 
Il fascio primario di protoni di energia pari a 450 GeV, subito dopo la sua 
estrazione dall'SPS, interagisce con un bersaglio di berillio; le particelle 
secondarie positive aventi momento di 40 GeV/$c$\ ($\pm 0.5 \%$) sono selezionate 
ed introdotte nella linea di fascio H4 per mezzo di collimatori e magneti 
bipolari; esse sono infine focalizzate sul bersaglio per mezzo di magneti 
quadrupolari. L'intensit\`a del fascio primario estratto \`e di circa 
$10^9$\ protoni per ciclo, alla quale corrisponde un'intensit\`a del fascio 
secondario di circa $10^6$\ protoni per ciclo. La durata del ciclo di accelerazione 
per i protoni \`e di 14.4 $s$\ e la durata dell'estrazione \`e di 2 $s$.  
\newline
Il fascio estratto \`e composto prevalentemente da pioni e kaoni, oltre che da  
protoni (circa il 30\%). La contaminazione delle altre particelle deve quindi essere 
eliminata; a tale scopo si  
utilizzano due rivelatori \v{C}erenkov ($C1$\ e $C2$) 
%differenziali a gas, denominati CEDAR, 
a soglia 
posti lungo la linea di fascio H4. 
La regolazione della pressione del gas 
consente di ottimizzare l'efficienza di identificazione dei protoni  
($\simeq 100 \%$), con una reiezione di questi  pari a  
circa il 30\%  delle particelle incidenti~\footnote{La soglia (cio\`e la 
pressione del gas) \`e impostata in modo tale che i soli pioni e kaoni 
rilasciano un segnale, per effetto \v{C}erenkov.}.  
Il segnale proveniente da questi rivelatori \`e  
incluso nella logica di trigger per scartare interazioni non attivate da protoni. 
\section{Il bersaglio}
NA57 \`e un esperimento a bersaglio fisso. Il fascio incidente viene focalizzato  
sul bersaglio, caratterizzato dalla massa atomica e dallo spessore. 

\noindent
%\subsubsection{Il bersaglio col fascio di piombo}
Nel caso del fascio di ioni di piombo, 
per quanto riguarda la scelta del materiale 
si \`e optato per il Pb ($A$ = 207.19), 
in quanto si ottiene un sistema simmetrico nella collisione: in tal modo
si pu\`o estendere l'intervallo di rapidit\`a coperto dall'esperimento
riflettendo la regione di accettanza intorno al valore di rapidit\`a del
centro di massa. Inoltre, 
l'alto numero atomico del piombo comporta il raggiungimento di 
elevate densit\`a di energia e di estesi volumi di interazione.  
Per quanto concerne lo spessore del bersaglio, si deve trovare 
un compromesso tra la necessit\`a di aumentare la frequenza delle interazioni e quella 
di render trascurabili le probabilit\`a di interazioni multiple e di conversioni 
di fotoni.  La probabilit\`a di interazione col bersaglio di una particella incidente 
\`e pari a 
\begin{equation}
P_I(L) = 1 - \exp(-L/\lambda_I) 
\label{eq2_1}
\end{equation}
dove $L$\ \`e lo spessore del bersaglio e $\lambda_I$\ la lunghezza di interazione, pari 
a 3.99 $cm$ per il sistema Pb-Pb. 
La probabilit\`a di conversione dei fotoni in coppie \Pep \Pem si calcola a partire 
dalla seguente formula, analoga alla~\ref{eq2_1},  
\begin{equation}
P_{\gamma con}(L)=1 - \exp(-L/X_0) 
\label{eq2_2}
\end{equation}
in cui $X_0$\ prende il nome di lunghezza di radiazione, ed \`e pari a 0.56 $cm$\ per il 
piombo.
Il bersaglio 
utillizzato coi fasci di Pb sia a 160 che a 40 $A$\ GeV/$c$\ ha uno spessore pari 
a 0.4 $mm$, corrispondente a circa l'1\% di lunghezza di interazione  
ed a 0.071 lunghezze 
di radiazione.  
Alcuni run speciali sono stati eseguiti utilizzando  bersagli pi\`u spessi, 
pari al 2\%, al 4\% ed all'8\% 
di lunghezza di interazione.
%, o senza bersaglio. 
Confrontando i dati con bersaglio di spessore %pari all'8\% di lunghezza di interazione 
maggiore  
con quelli con bersaglio standard si possono avere informazioni sulle doppie interazioni. 
Run speciali eseguiti senza bersaglio ({\em ``empty target''}) sono stati utilizzati 
invece per determinare il contributo delle interazioni (di fondo) che non avvengono 
entro il bersaglio.  
%La misura dei parametri geometrici del bersaglio \`e stata condotto con la 
%seguente procedura:
%\begin{itemize}
%\item
%\item
%\end{itemize}
%\subsubsection{Il bersaglio col fascio di protoni}

\noindent
Il fascio di protoni a 40  GeV/$c$\  \`e stato invece indirizzato su un 
bersaglio di Be, di spessore pari all'8\% della lunghezza d'interazione. 
%Davanti a questo bersaglio, di elevate \`e posto 
\section{L'apparato sperimentale}
\begin{figure}[bt]
  \begin{center}
\includegraphics[scale=0.50]{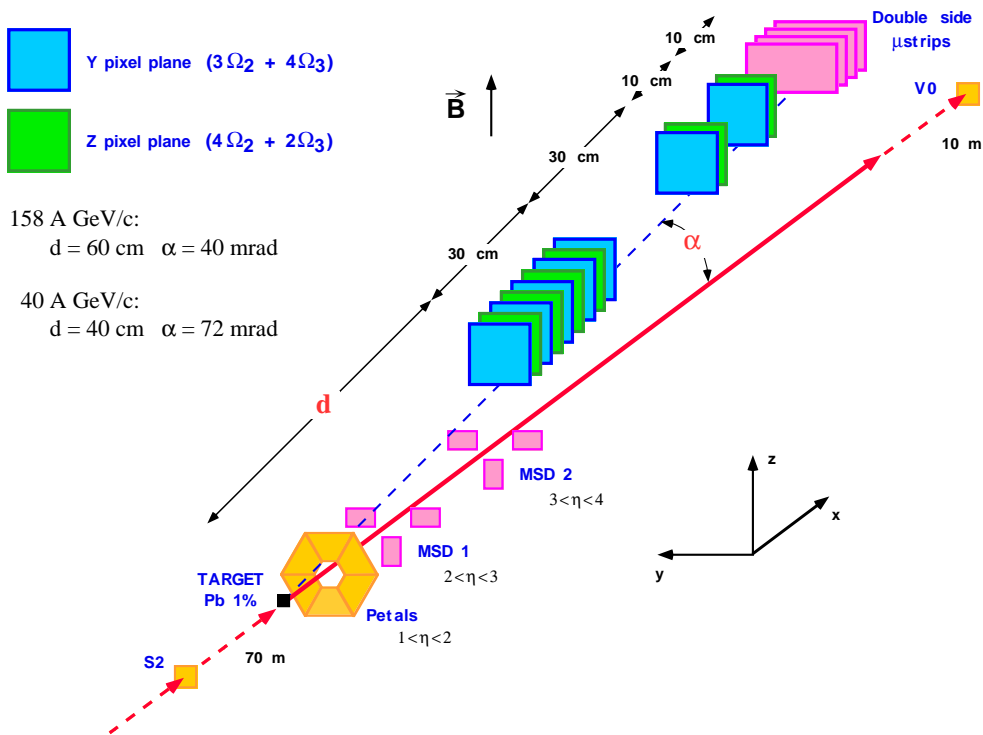}
    \caption{L'insieme dei rivelatori usati nell'esperimento NA57.}
    \label{cap2_1}
  \end{center}
\end{figure}

In fig.~\ref{cap2_1} \`e mostrato uno schema dell'insieme dei rivelatori 
utilizzati nell'esperimento NA57. Esso \`e composto, in sequenza,  
%nella direzione del fascio incidente, 
da una  schiera di sei scintillatori disposti a corona (i ``petali''), da due piani 
di rivelatori di molteplicit\`a di tipo microstrip di silicio 
(le ``MSD''), da un  telescopio realizzato esclusivamente con piani di rivelatori 
a pixel di silicio, e da rivelatori a microstrip in silicio con doppia faccia.  
Tali rilevatori sono stati opportunamente montati su un banco ottico di alluminio 
in modo da poter variare, a seconda dell'energia del fascio, 
{\em (i)} l'angolo di inclinazione $\alpha$\ del telescopio rispetto 
alla linea del fascio, {\em (ii)} la distanza dei rivelatori dal bersaglio.   
Il telescopio \`e posto in una regione di campo magnetico 
di cui \`e nota l'intensit\`a punto per punto (la ``mappa'') con grande accuratezza 
(all'interno del magnete GOLIATH).  
%e disposti , a seconda dell'energia del 
%fascio o del tipo di bersaglio, come si illustrer\`a in dettaglio nei 
%prossimi paragrafi, in una regione di campo magnetico approsimativamente uniforme 
%(magnete GOLIATH).  
\subsection{Il magnete}
\begin{figure}[bt]
  \begin{center}
  \includegraphics[scale=0.60]{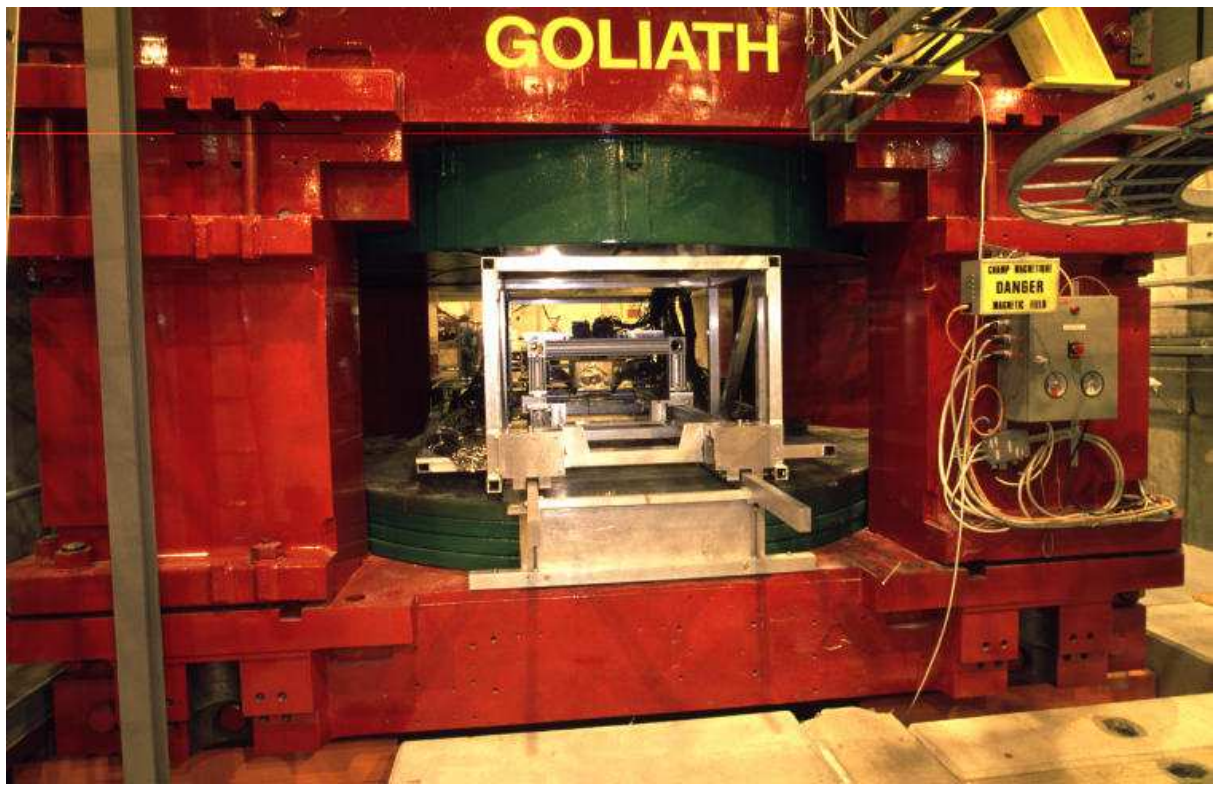}
 \caption{Fotografia del magnete ''GOLIATH''. Al suo interno si 
  riconosce il banco ottico di alluminio.} 
 \label{cap2_2}
 \end{center}
\end{figure}

L'esperimento NA57 utilizza 
%in maniera esclusiva 
il magnete GOLIATH,  una cui
fotografia \`e riportata in fig.~\ref{cap2_2}, 
situato lungo la linea di fascio H4 nella zona PPE134.  
Esso \`e composto in realt\`a da due bobine montate in serie. 
%come schematizzato nella fig.~\ref{}. 
La prima (GOLIATH) \`e alimentata da due convertitori in parallelo, la 
seconda (DAVID) da un convertitore e da un diodo di anti-ritorno.  
Il campo magnetico prodotto, sufficientemente uniforme, \`e diretto 
verticalmente ed  ha un'itensit\`a massima di circa 1.4 Tesla. 
Il sistema di riferimento del laboratorio ha origine nel centro del magnete,
l'asse $x$\ diretto lungo la direzione del fascio, l'asse $z$\ diretto verso
l'alto, perpendicolarmente alle espasioni polari (quindi lungo la direzione del
campo $\vec{B} $\ del magnete), e l'asse $y$\ tale da formare una terna destrorsa. 
Il verso del campo magnetico, determinato da quello della corrente entro le
bobine, viene invertito nel corso delle misure per
eliminare eventuali asimmetrie dovute a piccoli errori
sistematici nella misura 
della mappa di $\vec{B} $, e per verificare l'allineamento dei rivelatori 
del telescopio, posti in posizione simmetrica rispetto al piano x--z.
Il banco ottico sul quale sono disposti i diversi rivelatori, che saranno 
descritti nei prossimi paragrafi, pu\`o scorrere su se stesso, per mezzo di 
apposite rotaie in lega di alluminio, ed esser introdotto completamente  
all'interno del magnete GOLIATH.  
\subsection{I rivelatori di molteplicit\`a}
La misura della molteplicit\`a delle particelle cariche prodotte in un 
evento Pb-Pb consente una determinazione della  
centralit\`a dell'urto, come esposto nel capitolo precedente.  
Nell'esperimento NA57 vengono utilizzati due sistemi di rivelatori di 
centralit\`a per studiare le collisioni Pb-Pb. 
\subsubsection{Gli scintillatori a petali}
Una schiera di sei scintillatori di forma trapezoidale (i ``petali''), 
disposti in modo tale da formare un corona attorno alla linea di fascio, 
\`e posta posteriormente al bersaglio a $10$\ cm da questo. Essa copre la 
regione di pseudorapidit\`a 1 $<$\ $\eta$\ $<$\ 2. Tale regione cinematica non 
si sovrappone all'angolo solido coperto dal telescopio di tracciamento, pertanto 
la diffusione multipla nei petali non deteriora la risoluzione in momento delle 
tracce ricostruite nel telescopio. 
\newline
I ``petali'' vengono utilizzati per selezionare, a livello di trigger, gli eventi 
a pi\`u elevata molteplicit\`a, cui corrispondono gli urti pi\`u 
centrali tra i nuclei, e per rigettare quelli a pi\`u bassa molteplicit\`a, 
conseguenti da urti periferici. Trattandosi di scintillatori, 
la loro risposta \`e infatti estremamente rapida ed il segnale \`e reso disponibile 
per formare la condizione di trigger in $\approx$\ 200 ns.  
I segnali dei petali vengono equalizzati, quindi la somma dei segnali in uscita 
da questi rivelatori \`e 
proporzionale all'energia rilasciata dalle particelle cariche che li hanno 
attraversati; trattandosi in gran parte di particelle al minimo di ionizzazione, 
questo segnale \`e circa proporzionale al numero di particelle.  
\newline
Il limite inferiore di molteplicit\`a  oltre cui non \`e opportuno spingersi dipende 
principalmente dalla percentuale di contaminazione d'interazioni che non 
avvengono nel bersaglio (``empty target''). Queste ultime sono infatti caratterizzate 
da basse molteplicit\`a di particelle cariche, rispetto alle collisioni 
centrali Pb-Pb.  
La soglia al di sopra della quale i singoli scintillatori forniscono un segnale 
viene scelta in corrispondenza delle collisioni Pb-Pb 
pi\`u periferiche che si possono studiare senza essere dominati dal contributo 
d'interazioni da ``empty target''.  
A livello di trigger, il sistema dei petali interviene fornendo esito positivo 
per un evento qualora almeno cinque scintillatori su sei siano sopra soglia.
La percentuale di eventi selezionati dai petali corrisponde a circa il 60\% 
della sezione d'urto d'interazione anelastica nelle interazioni Pb-Pb a 160 
e 40 $A$ GeV/$c$. 
\subsubsection{I rivelatori di molteplicit\`a a micro-strip}
\begin{figure}[h]
\begin{center}
  {\Large \bf{a)}}
  \includegraphics[scale=0.50]{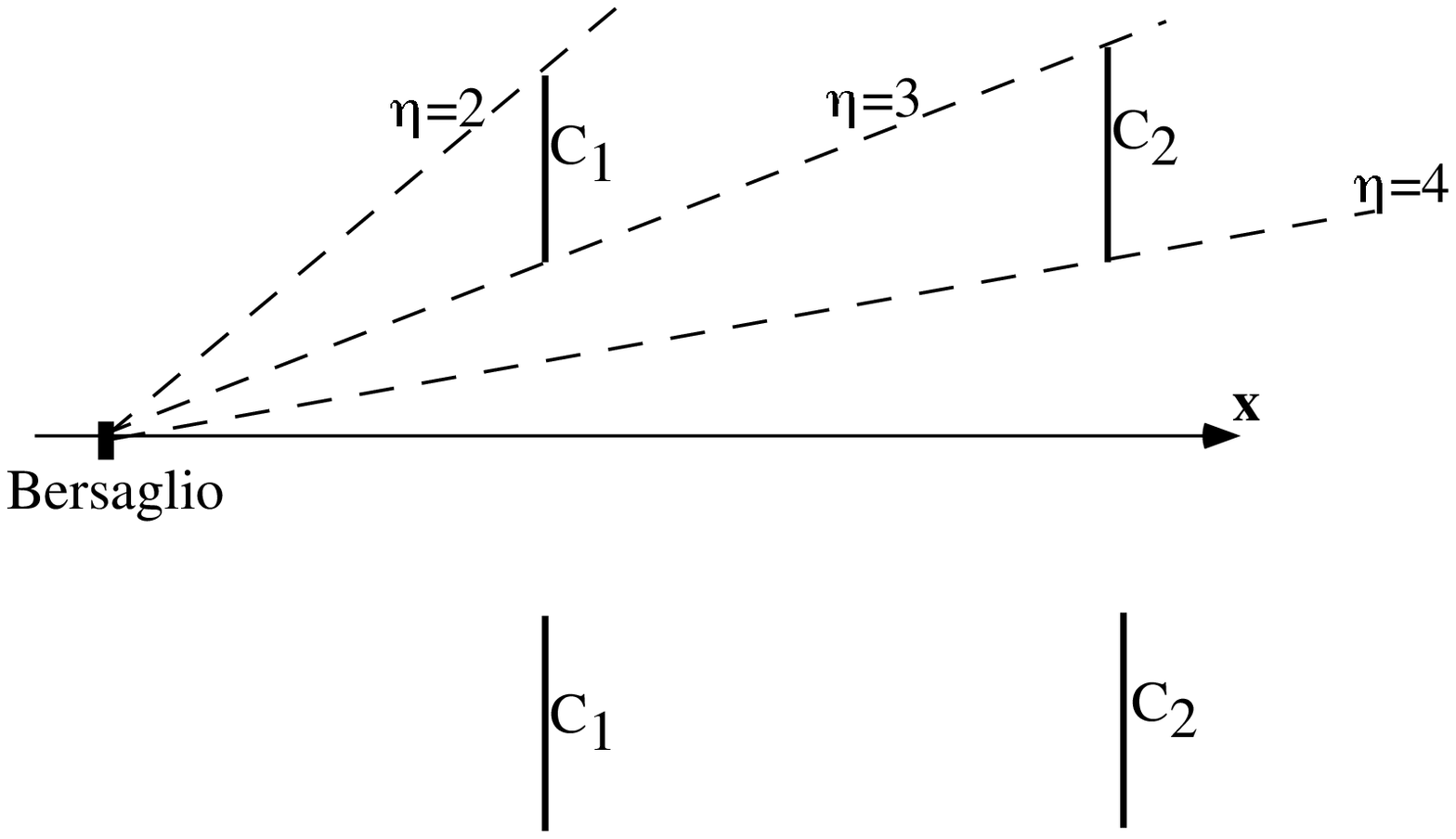} \\
  \vspace{1.6cm}
  {\Large \bf{b)}}
  \includegraphics[scale=0.40]{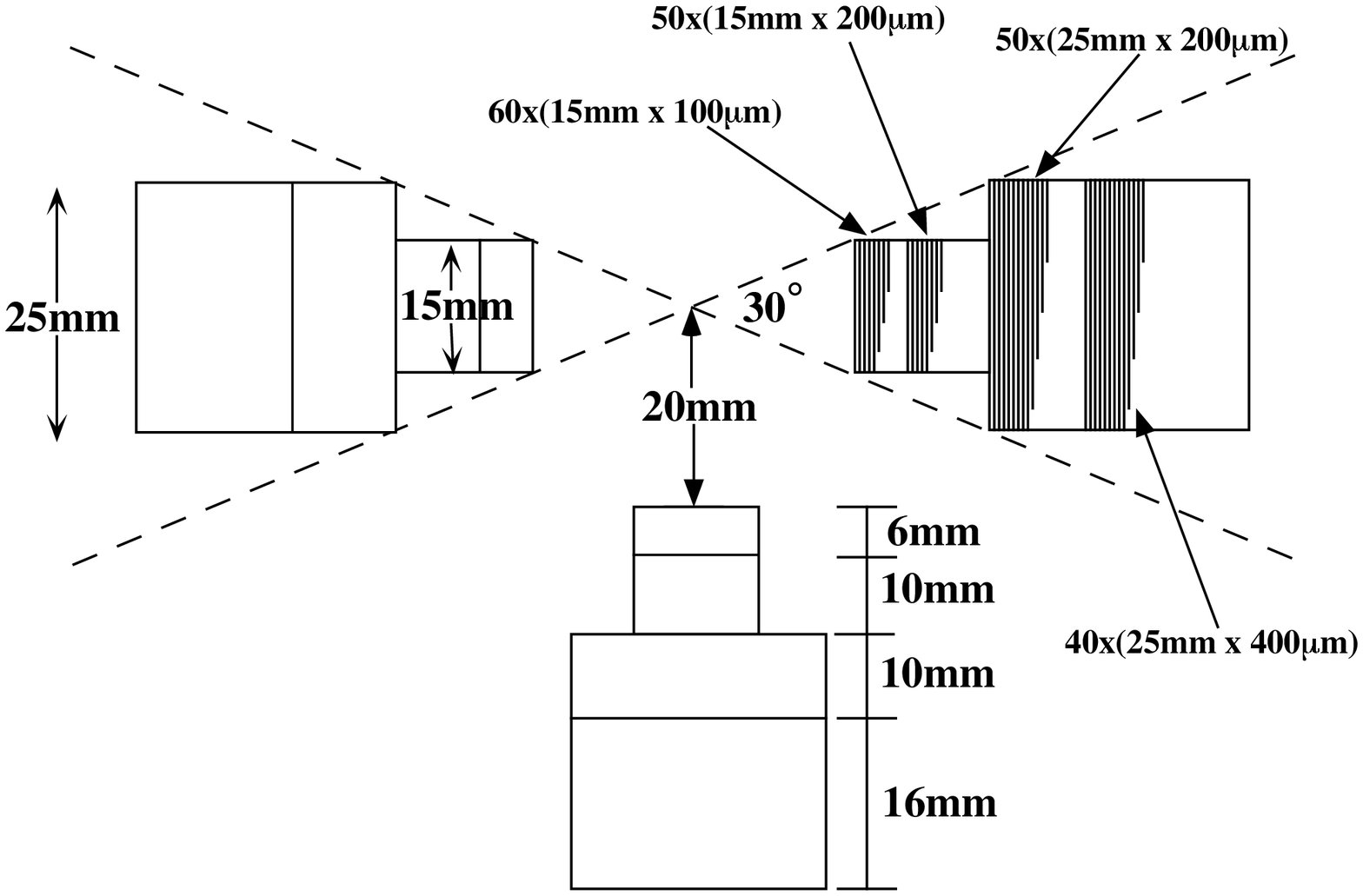}
  \caption{Rivelatori
   di molteplicit\`a. {\bf a)} Vista laterale: le stazioni C1 e C2, tra loro
   identiche, coprono insieme
   l'intervallo di pseudorapidit\`a 2 $<$\ $\eta$\ $<$\ 4.
   {\bf b)} Vista frontale: struttura delle singole stazioni.}
 \label{cap2_3}
 \end{center}
\end{figure}

Nelle interazioni Pb-Pb a 160 $A$\ GeV/$c$\  
la molteplicit\`a delle particelle cariche prodotte viene 
misurata per mezzo di due stazioni  di rivelatori a microstrip di silicio,  
campionando il primo nell'intervallo di pseudorapidit\`a 2 $<$\ $\eta$\ $<$\ 3 
ed il secondo entro 3 $<$\ $\eta$\ $<$\ 4. Le due stazioni coprono quindi regioni 
cinematiche quasi simmetriche rispetto alla rapidit\`a del centro di massa e praticamente 
disgiunte (la sovrapposizione \`e $\lesssim$ 10\%). 
\newline
Le due stazioni sono disposte subito dopo gli scintillatori a petali, lungo 
la linea di fascio;   
ciascuna stazione consta di tre rivelatori a microstrip, montati su un
telaio con tre bracci, vuoto in corrispondenza della zona di passaggio del
fascio, come mostrato nelle figg.~\ref{cap2_1},~\ref{cap2_3}.a. 
Nel piano$(y,z)$\ la copertura azimutale per ciascuna stazione \`e circa 
del 34\%. 
La mancanza di un quarto braccio, che completerebbe la simmetria azimutale, 
\`e dovuta alla necessit\`a di limitare la diffusione multipla delle particelle 
nell'angolo solido sotteso dal telescopio, per non degradare la risoluzione 
delle tracce riscostruite. 
\newline
Come schematizzato nella fig.~\ref{cap2_3}.b, ciascuno dei tre piani \`e realizzato 
con microstrip di differente lunghezza (da 15 a 22 $mm$) e passo 
(da 100 a 400 $\mu m$), per un totale di 200 canali, e disposti in modo tale da 
assicurare un'occupazione uniforme di particelle nell'ipotesi di distribuzione 
costante in pseudo-rapidit\`a. Tale ipotesi \`e infatti applicabile in 
prossimit\`a  della regione centrale di rapidit\`a, come discusso nel primo 
capitolo, ed inoltre la rapidit\`a del centro di massa nel sistema laboratorio 
($y_{cm}\simeq2.9$) 
cade al centro dell'intervallo di rapidit\`a coperto dalle due stazioni. 
L'occupazione di particelle per strip \`e dunque ragionevolmente 
uniforme su ciascun piano di 
rivelatori, risultando inferiore a 0.4, anche per le interazioni  
Pb-Pb a 160 $A$ GeV/$c$\ pi\`u centrali~\cite{WA97p}. 
\newline
Nelle interazioni Pb-Pb a 40 $A$ GeV/$c$\ le due stazioni sono state disposte 
in modo tale da coprire ancora un intervallo pari a quasi due unit\`a di rapidit\`a 
centrato sul valore di rapidit\`a del centro di massa ($y_{cm}\simeq2.2$), sebbene 
vi sia ora una regione di sovrapposizione di cui si tiene conto con una correzione 
Monte Carlo.  
\newline
Questi rivelatori permettono una determinazione della molteplicit\`a di 
particelle cariche pi\`u precisa rispetto a quella fornita dagli scintillatori 
a petali. Essi non possono essere inclusi nel trigger in quanto il tempo di 
lettura \`e dell'ordine dei 500 $\mu sec$, ma vengono usati nell'analisi ``off-line'' 
degli eventi.  
\newline
Ciascuna strip fornisce infatti un segnale analogico proporzionale alla 
energia rilasciata dalla particella. In un evento, si definisce {\em cluster} 
un insieme di  
strip contigue in cui vi \`e stato rilascio di energia al di sopra 
di un valore di soglia. Dal numero di cluster si risale alla molteplicit\`a di 
particelle cariche entro il rivelatore utilizzando un algoritmo che tiene conto 
dell'energia totale depositata nel cluster. L'algoritmo prevede anche una 
correzione per i doppi {\em ``hit''}, che avvengono con maggior frequenza negli eventi 
a pi\`u alta molteplicit\`a. 
\subsection{Il telescopio a piani di pixel}
\begin{figure}[bt]
 \begin{center}
 \includegraphics[scale=0.30,angle=270]{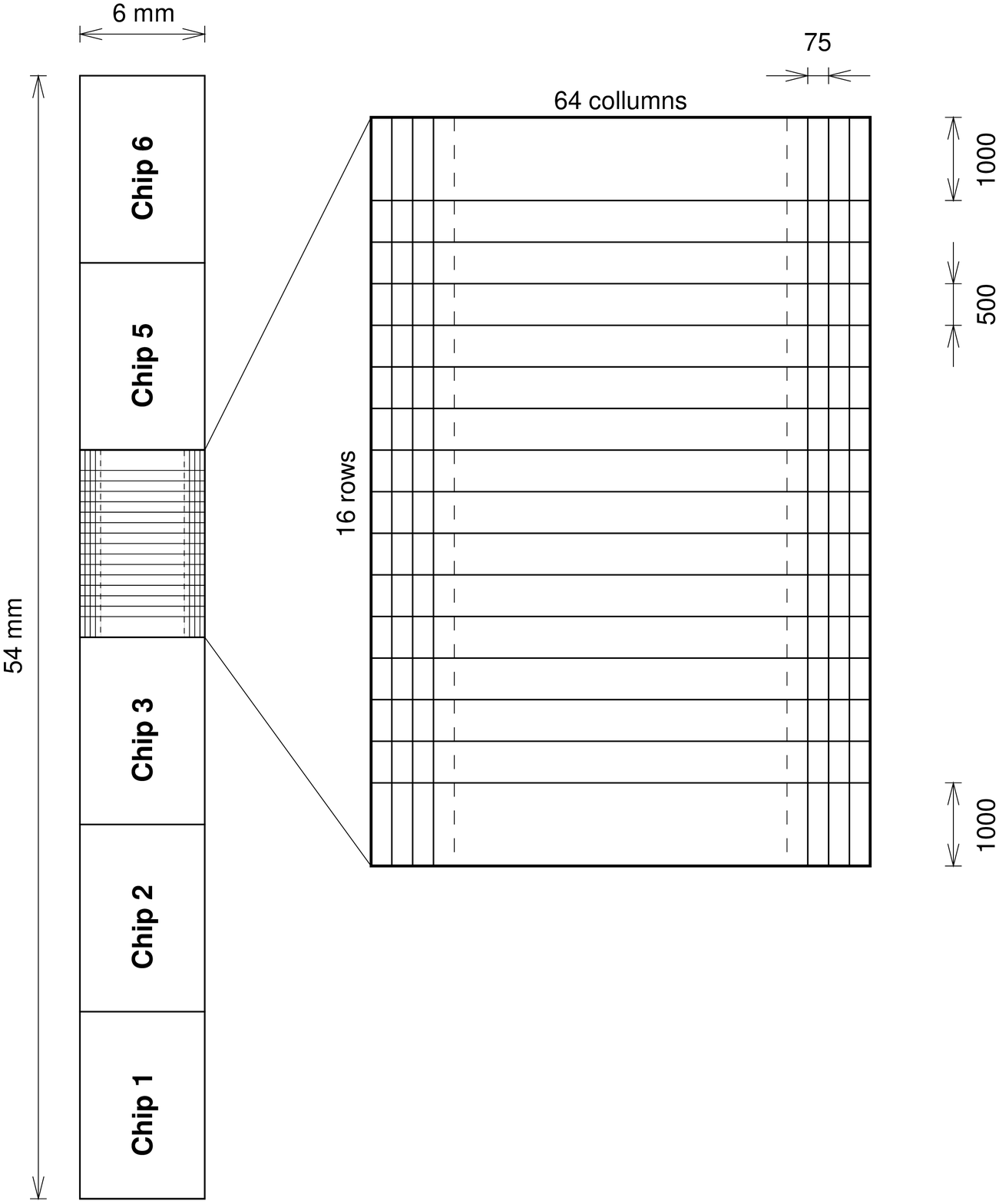}
  \caption{Schema di un {\em ladder} di un {\em Omega2}, costituito da sei 
           {\em chip} affiancati, con ingrandimento del singolo {\em chip}.}  
 \label{Ladder}
 \end{center}
\end{figure}

\begin{figure}[h] 
\begin{center}
 {\Large \bf{a)}}
 \includegraphics[scale=0.35]{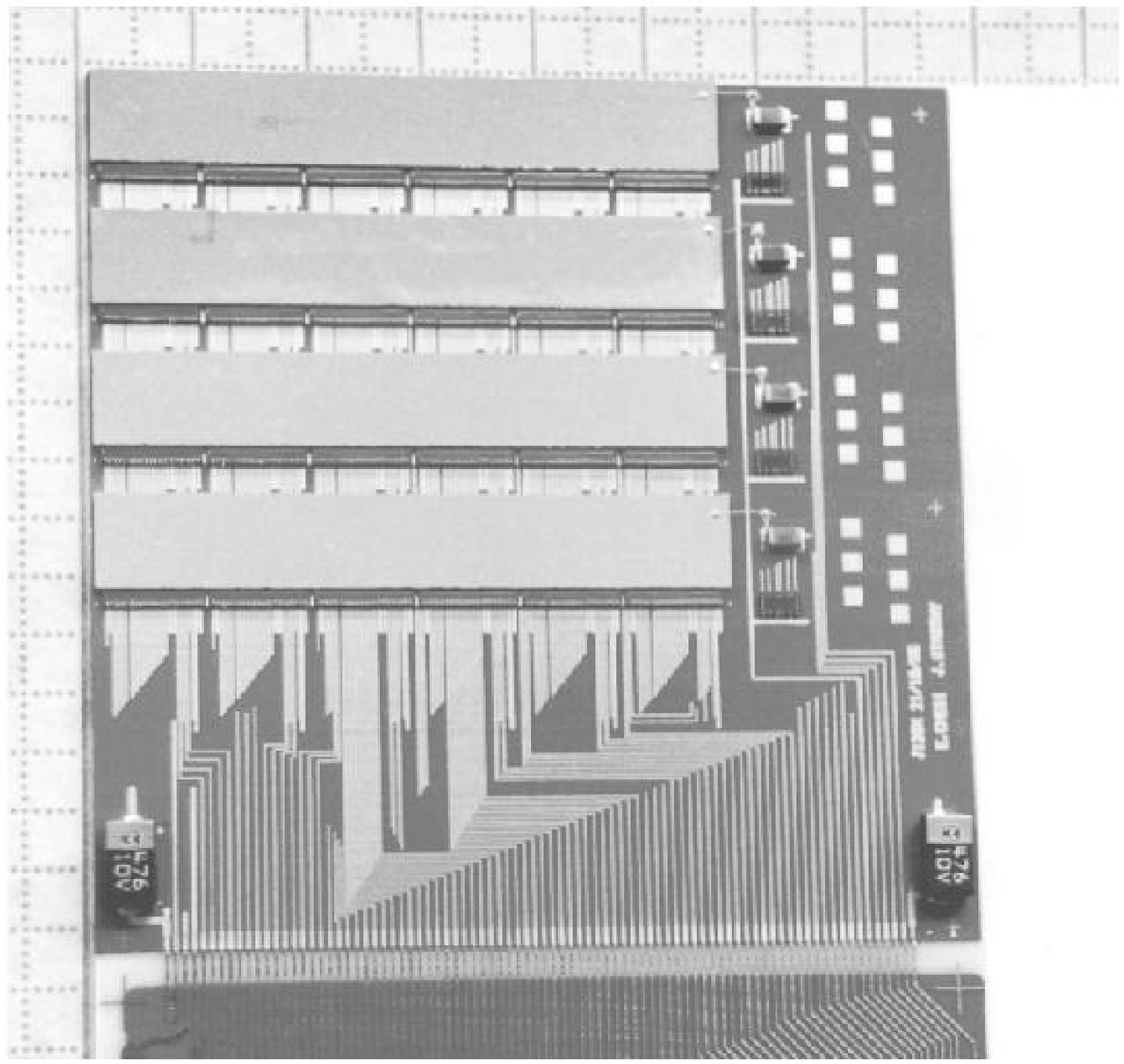} 
 \hspace{0.3cm}
 {\Large \bf{b)}}
 \includegraphics[scale=0.35]{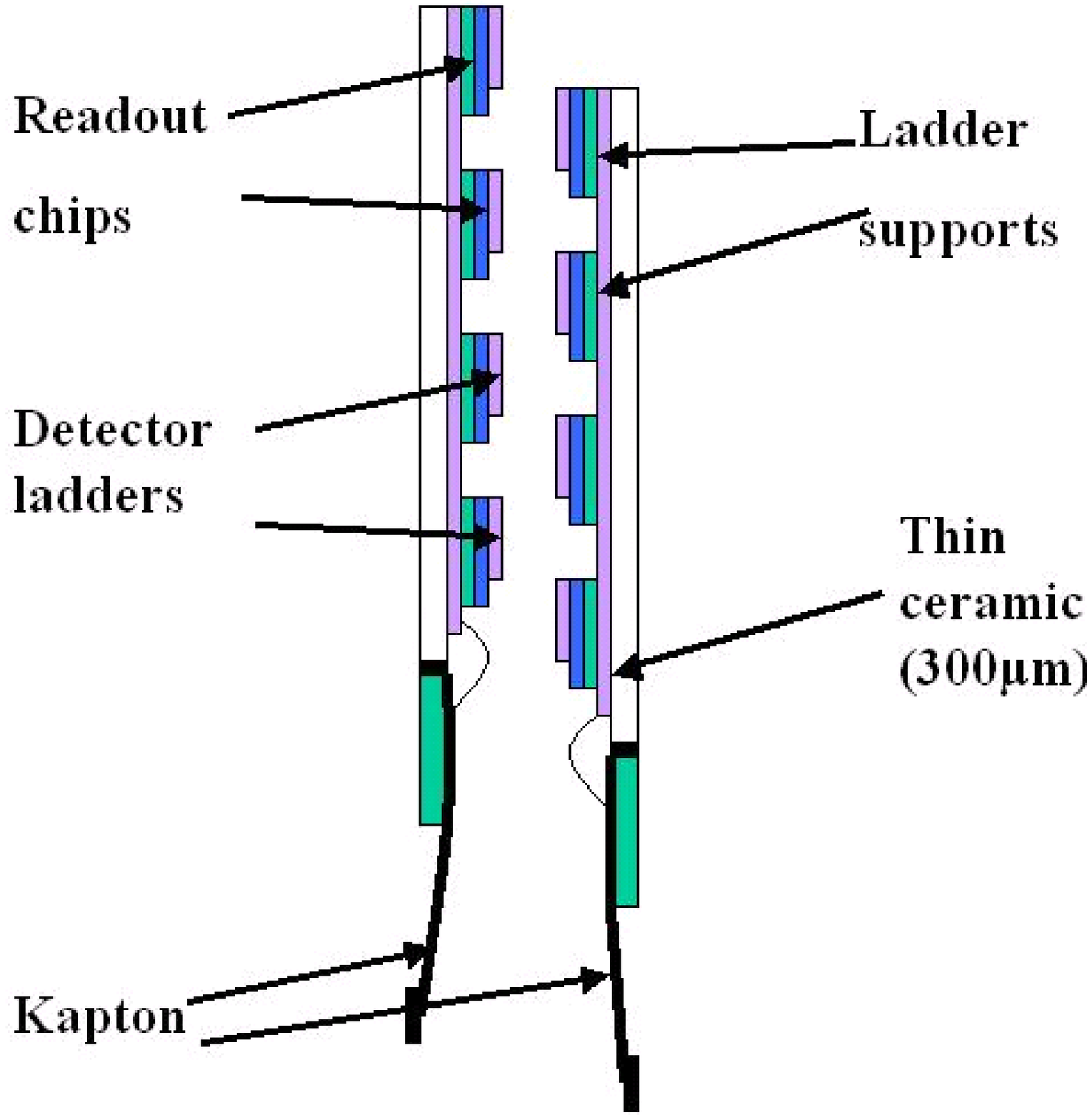}\\
% \vspace{0.6cm}
% {\Large \bf{c)}}
%  \includegraphics[scale=0.45]{cap2/pixlay.eps}
 \caption{Rivelatori a piani di pixel. 
 {\bf a)} Semi-piano {\em (array)} di un {\em Omega3} montato sul supporto ceramico.
 {\bf b)} Schema di un piano completo, formato da due {\em array} affacciati.
% {\bf c)} Fotografia dell'elettronica associata a ciascun pixel.
 }
\label{cap2_4}
\end{center}
\end{figure}

Il cuore dell'apparato \`e un insieme di piani di rivelatori a pixel 
di silicio disposti consecutivamente a formare un {\em ``telescopio''}, nel 
quale avviene la rivelazione e ricostruzione delle tracce cariche.  
\newline
La tecnica alla base dei rivelatori a pixel di silicio ibridi \`e stata 
sperimentata con successo ed in maniera pionieristica dall'esperimento 
WA97~\footnote{Nell'Universit\`a di Bari, oltre all'attuale gruppo della Fisica  
degli Ioni Pesanti, ha dato un importante contributo allo sviluppo dei rivelatori a 
pixel di silicio il gruppo della prof.~ssa~Muciaccia (N.~Armenise, M.G.~Catanesi, 
M.T.~Muciaccia e S.~Simone) ed in particolare il prof.~Simone.}
%il prof.~N.~Armenise, la prof.~ssa~M.T.~Muciaccia, 
%il prof.~S.~Simone e la dott.sa~M.Catanesi.}
in stretta  collaborazione col gruppo RD19 del CERN~\cite{RD19}. 
Il loro sviluppo \`e stato possibile grazie al progresso raggiunto 
nelle densit\`a delle 
diverse componenti nei {\em chip} microelettronici di tipo CMOS e nelle tecniche di 
assemblaggio superficiale a passo fine (assemblaggi di tipo ``flip-chip''). 
\newline
Un rivelatore a pixel ibridi \`e una matrice di diodi in silicio polarizzati 
inversamente, saldati con la tecnica detta del ``bump-bonding'', direttamente sul 
{\em chip} di front-end dell'elettronica.  
Ciascuna singola cella della matrice \`e collegata con una micro-saldatura in  
Sn-Pb ad una cella di pari dimensione del {\em chip} di front-end, che contiene 
gran parte dell'elettronica necessaria per la lettura del rivelatore. 
\newline
Due generazioni di rivelatori a pixel in silicio sono stati sviluppati con 
questa tecnica: gli {\em Omega2}~\cite{Omega2} e gli {\em Omega3}~\cite{Omega3}. 
\newline
Negli {\em Omega2} la dimensione dei pixel \`e pari a 75$\times$500 $\mu m^2$\ e 
ciascun {\em chip} contiene 1024 rivelatori indipendenti (pixel), disposti secondo una 
matrice di 64 righe e 16 colonne.  Un segmento di 
sei {\em chip} a colonne adiacenti costituisce un {\em ladder}, composto dunque 
da 64 righe e 96 colonne di pixel, come mostrato in fig.~\ref{Ladder}.  
\noindent
Sei {\em ladder} vengono quindi incollati per formare una struttura rigida piana 
su un supporto ceramico, spesso 
630 $\mu m$; i {\em ladder} sono spaziati tra loro di alcuni millimetri 
dove trovano spazio le connessioni elettroniche. 
Tale struttura rigida prende il nome di {\em array}.  
\newline
Negli {\em Omega3} 
la dimensione dei pixel \`e 50$\times$500 $\mu m^2$\ 
e ciascun {\em chip} contiene 2032 pixel. In tal caso un {\em chip} \`e costituito 
da una matrice di 127 righe e 16 colonne, un {\em ladder} da  127 righe e 96 colonne 
di pixel, e 4 {\em ladder} paralleli alloggiano sulla 
lastra di ceramica per formare un {\em array} (fig.~\ref{cap2_4}.a).  
\newline
Ciascun {\em piano} del telescopio di NA57 \`e formato da due {\em array}, entrambi  
{\em Omega2} od entrambi {\em Omega3}, montati l'uno di fronte all'altro su uno 
stesso telaio; essi presentano lo strato  
ceramico rivolto verso l'esterno   
e sono allineati in modo tale che le regioni sensibili dell'uno, i ladder, si 
affaccino sulle regione inattive dell'altro, che contengono le linee  
elettroniche (fig.~\ref{cap2_4}.b).  
\newline
Un piano logico di {\em Omega2} ({\em Omega3}) consiste dunque di 
%72576 
73728 (97536) {\em pixel} concentrati su un'area sensibile di 
$5 \times 5\,{\rm cm^2}$.  Il telescopio di NA57 dispone di sette piani 
di  {\em Omega2} e di sei piani di {\em Omega3} per un totale di 
%1.1 milioni  
oltre un milione  
di canali elettronici.
\newline
Si definisce {\em piano z} un piano logico con i ladder disposti in 
posizione orizzontale, in modo che i lati da 50 o 75 ${\rm \mu m}$\ dei pixel siano
paralleli all'asse $z$, e {\em piano y} quello ottenuto ruotando di
$90^{\circ}$\ un piano $z$, attorno alla normale al piano.
Nel telescopio, i piani sono alternativamente di tipo $y$\ e di tipo $z$, per
garantire una migliore risoluzione spaziale nelle due direzioni trasverse, come 
mostrato in fig.~\ref{cap2_1}.  
\newline
Come gi\`a detto, su ciascun chip \`e saldato direttamente un  
microchip di elettronica di lettura frazionato in celle elementari 
(una per pixel). 
%Come mostrato in fig.~\ref{cap2_4}.c, 
Ciascuna comprende un preamplificatore a  
basso rumore, un comparatore, un'unit\`a di ritardo ed una di memoria che permettono 
di registrare i segnali in concomitanza con la selezione operata dalla logica
di trigger. L'informazione fornita da ciascun pixel \`e di tipo binario, nel senso 
che esso segnala se \`e stato attaversato, o meno, da una particella
carica, ma non informa sulla quantit\`a di carica rilasciata.  

\subsection{I rivelatori a microstrip doppia faccia}
\begin{figure}[p] 
\begin{center}
  {\Large \bf{a)}}  \vspace{0.4cm}
  \includegraphics[scale=0.50]{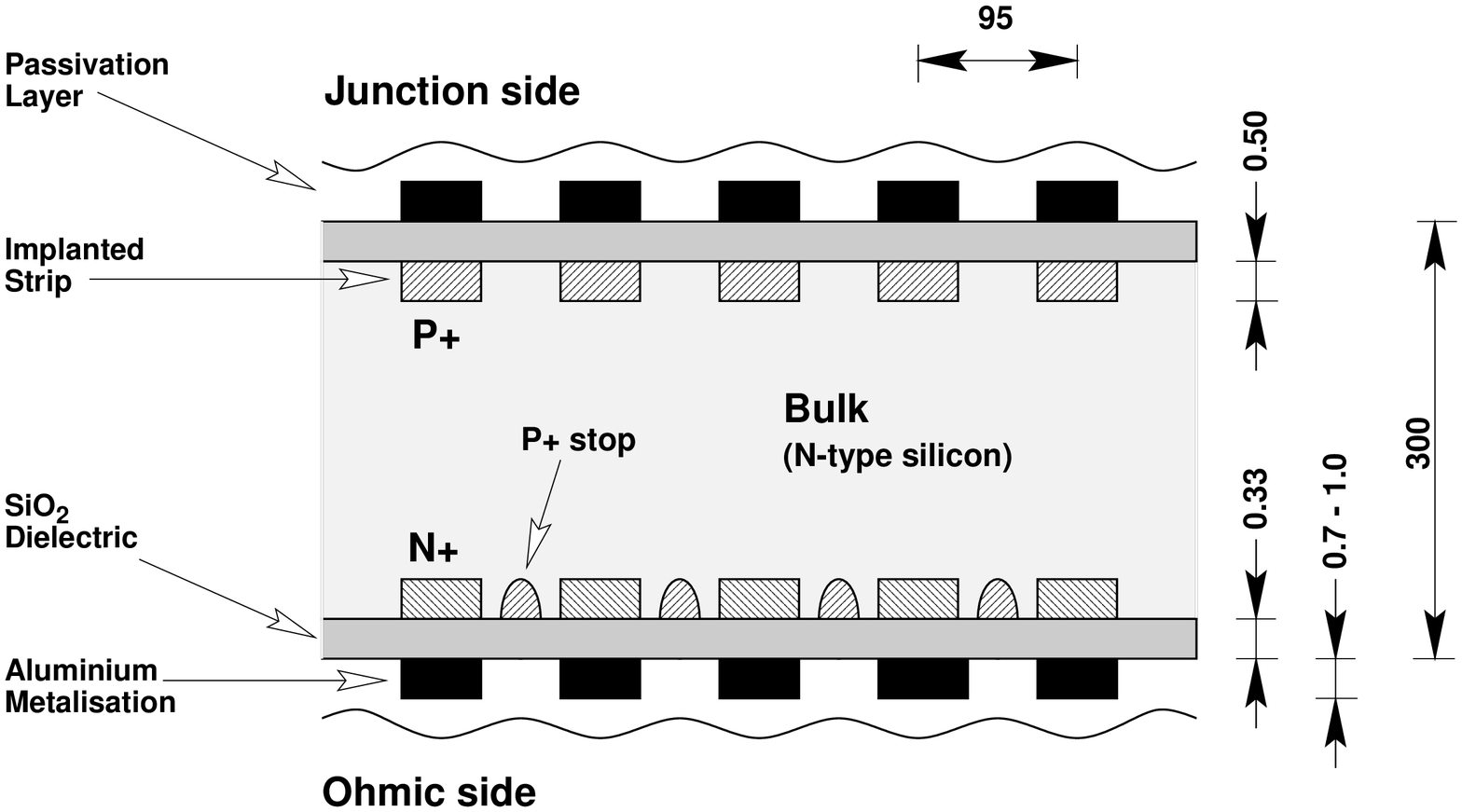} \\
  \vspace{0.9cm}
  {\Large \bf{b)}}  \vspace{0.4cm}
  \includegraphics[scale=0.60]{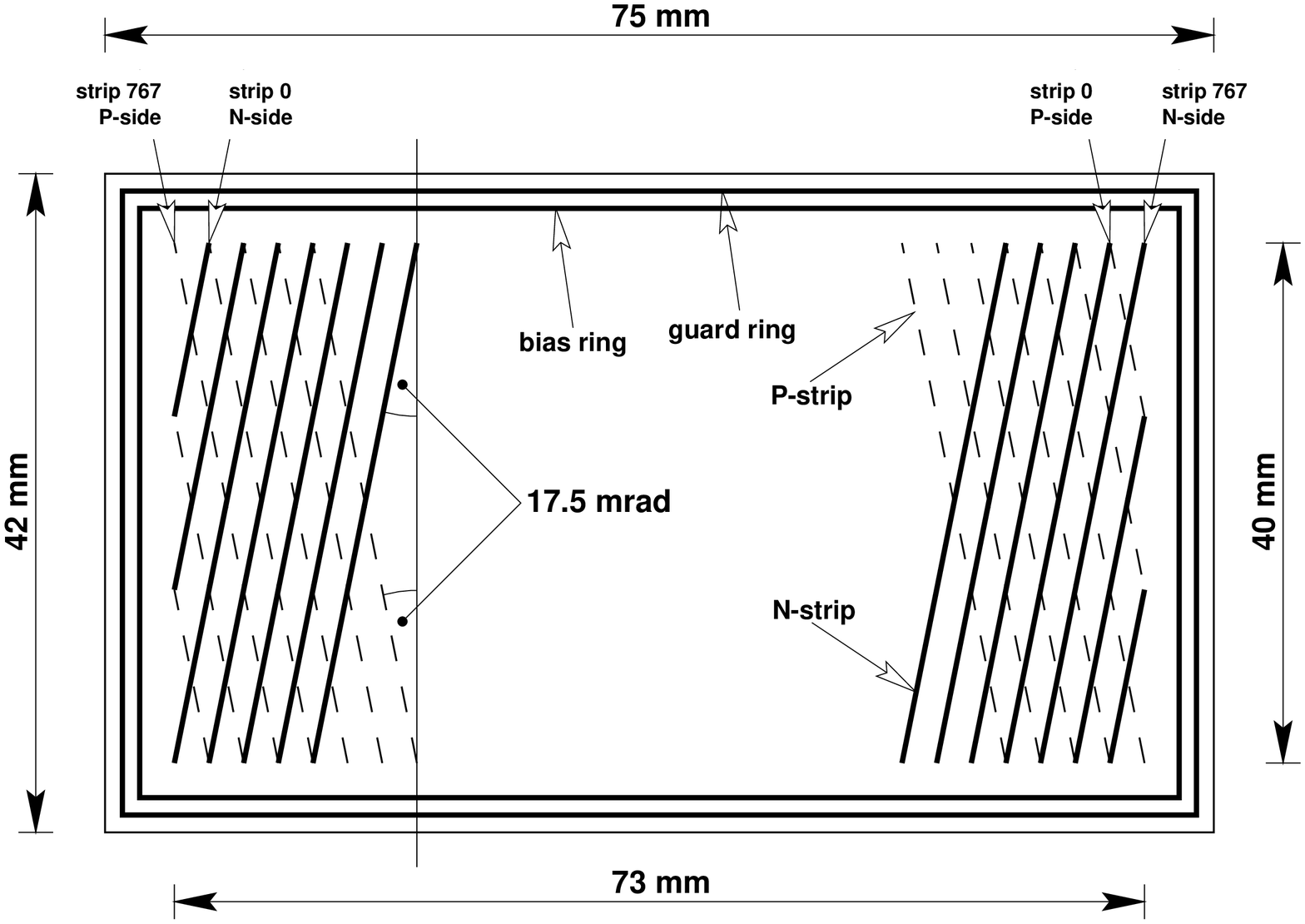}
  \caption{Rivelatori a $\mu$-strip doppia faccia.
  {\bf a)} Sezione di un piano di $\mu$-strip nella direzione del  
           lato pi\`u lungo. Le distanze sono espresse in $\mu m$\ 
	   e la figura non \`e in scala. 
  {\bf b)} Schema di un piano, non in scala, visto dal lato $N$. 
           Le strip di lettura del lato $P$\ sono trattegiate.   }
 \label{Strip1}
 \end{center}
\end{figure}

Quattro moduli di rivelatori a microstrip in silicio doppia faccia  
sono stati utilizzati a partire dalla presa dati del 1998. Una descrizione 
dettagliata di questi rivelatori pu\`o esser trovata in~\cite{DeRijke}.
I moduli sono montati nella parte finale del banco ottico di alluminio,  
dieci centimetri dietro l'ultimo piano di pixel, per 
rendere pi\`u precisa la misura dell'impulso  
delle tracce pi\`u veloci (braccio di leva o ``lever arm'').  
\newline
La parte attiva del rivelatore \`e costituita da uno strato (``wafer'') di 
silicio drogato negativamente. Su entrambe le superfici del wafer sono 
impiantate delle strip in silicio di spessore 0.5 $\mu m$, come 
mostrato in fig.~\ref{Strip1}.a; esse sono drogate 
con ${\rm N}^+$\ sul lato ohmico (lato di tipo N) del wafer e 
con ${\rm P}^+$\ sul lato della giunzione (lato di tipo P). La superficie 
attiva del wafer in silicio \`e quindi ricoperta da uno strato di dielettrico 
${\rm SiO_2}$\ di spessore pari a 0.3 $\mu m$. Al di sopra dello strato di 
isolante, in corrispondenza delle sottostanti strip impiantate nel silicio,  
sono depositate delle strip di alluminio di spessore di 1 $\mu m$. 
\newline
Ciascun piano ha una dimensione di $75 \times 42$\ $mm^2$\, con una regione sensibile 
pari a $73\times 40$\ $mm^2$, come mostrato in fig.~\ref{Strip1}.b. 
Il rivelatore dispone di 768 strip di lettura lunghe 
40 $mm$\ e con passo di 95 $\mu m$, impiantate su entrambe le facce del rivelatore. 
Le strip di lettura sono inclinate rispetto al lato pi\`u corto del rivelatore 
di un angolo pari a $\pm 17.5$\ $mrad$, in modo tale che l'angolo stereoscopico tra 
le strip del lato N e quelle del lato P sia di  35 $mrad$.  
\newline
Combinando i segnali provenieti dalle due facce del rivelatore \`e possibile 
ricostruire la posizione bidimensionale dei punti di impatto delle particelle. 
Con un simile angolo stereoscopico si raggiunge una 
risoluzione di $\approx$\ 27 $\mu m$\ nella 
direzione orizzontale ($y$), in cui curvano le particelle 
per effetto del campo   
magnetico. La risoluzione nella direzione verticale ($z$) \`e invece pari 
a circa 1 $mm$. 
\newpage
\section{Selezione degli eventi ed acquisizione dei dati}
\subsection{Il trigger} 
Le condizioni di trigger sono molto diverse a seconda che si usi il fascio  
di ioni di piombo od il fascio di protoni.  
Nelle interazioni p-Be (caratterizzate da basse molteplicit\`a),  
al fine di aumentare la probabilit\`a che l'evento contenga  un 
decadimento tra quelli oggetto della ricerca, si richiede che almeno due 
tracce potenzialmente ricostruibili attraversino il telescopio.  
Con il fascio di ioni di piombo, invece, si vogliono 
selezionare le collisioni pi\`u centrali, 
senza preoccuparsi dell'occupazione di tracce nel telescopio, 
comunque cospicua. 
\newline
%A partire dalla presa dati del 1998 ({\em cfr.} paragrafo 2.6), il segnale 
%di trigger \`e stato generato servendosi di logica NIM.  
Il segnale di trigger  \`e costruito servendosi di moduli basati sulla logica NIM.  
Parallelamente al trigger NIM, \`e stato sviluppato un secondo sistema di 
trigger di tipo VME~\cite{trigger1,trigger2}, studiato in previsione del 
futuro esperimento ALICE ad LHC.  
Sin dalla prima presa dati del 1998 ({\em cfr. paragrafo 2.6}) 
il trigger VME \`e stato utilizzato per etichettare gli 
eventi raccolti, al fine di studiarne l'efficienza a posteriori. 
A partire dalla presa dati del 2001 il sistema VME \`e stato utilizzato come 
trigger principale per la selezione degli eventi, sostiduendo quello NIM.  
\newline
Vi sono alcuni elementi comuni al trigger per protoni  
ed a quello per gli ioni: i {\em contatori di fascio} ed il {\em tempo morto}. 
Si inizier\`a col descrivere i primi, mentre conviene rimandare la discussione 
del {\em tempo morto}, introducendo prima la logica del trigger per  
il fascio di Pb. Infine si accenner\`a al trigger col fascio di protoni.  
\subsubsection{Contatori di fascio}
\begin{figure}[h]
 \begin{center}
  \includegraphics[scale=0.50]{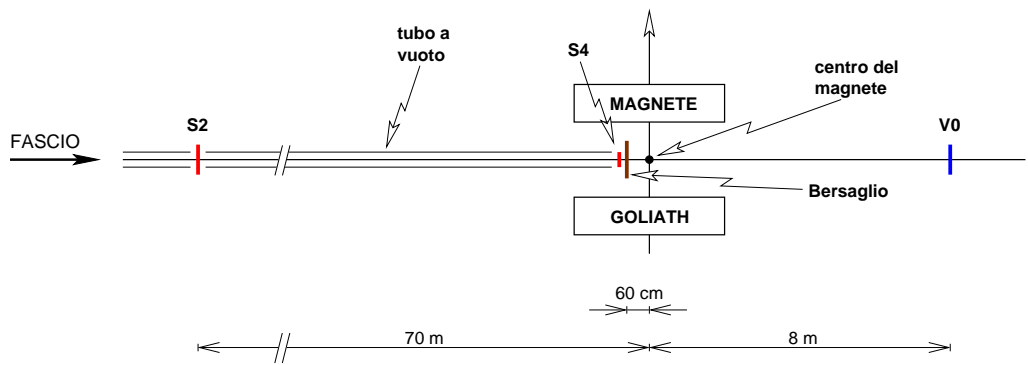}
 \caption{Disposizione degli contatori di fascio ($S2$, $S4$, $V0$) e del bersaglio 
          rispetto al centro del magnete. Le distanze cui sono posizionati i 
	  contatori in questo schema si riferiscono al caso del fascio di ioni di piombo.}
 \label{Contatori}
 \end{center}
\end{figure}

Sia per il fascio di protoni che per  quello di piombo, vengono
adoperati dei rivelatori per definire l'arrivo delle particelle proiettile: 
i contatori di fascio. 
Nella fig.~\ref{Contatori} \`e schematizzata 
la loro disposizione rispetto al centro del 
magnete~\footnote{Le distanze cui sono posti i contatori, in questo schema, si 
riferiscono alla particolare disposizione per il fascio di piombo. Nel caso 
del fascio di protoni, $S2$\ \`e posto $8$\ metri prima del bersaglio.}.   
Il bersaglio \`e disposto a 60 cm dal centro del 
magnete, il contatore $S2$\ circa 70 m prima, il contatore $S4$\ pochi centimetri prima 
del bersaglio. Quest'ultimo, uno scintillatore di dimensioni 8 $\times$ 8 $mm^2$,   
\`e usato solo per aggiustare il posizionamento del fascio di ioni di piombo 
e viene rimosso durante la presa dati con questo tipo di fascio. 
Il tubo a vuoto che conduce il fascio termina proprio davanti ad $S4$; in tal 
modo si riducono al minimo le interazioni in aria del fascio.  
\noindent
Il contatore V0 \`e posizionato infine circa 9 m dietro al centro del magnete.  
Esso pu\`o svolgere una funzione di veto, antiselezionando i proiettili incidenti che 
non hanno avuto interazioni nel bersaglio. 
\newline
Nelle prese dati col fascio di piombo, $S2$\ e $V0$\ sono due rivelatori \v{C}erenkov al 
quarzo, mentre col fascio di protoni si utilizzano due scintillatori.  
\newline 
Gli scintillatori non sono adatti al fascio di piombo in quanto i valori elevati di carica 
elettrica propri di tali ioni inducono fenomeni di saturazione, pregiudicando la 
loro efficienza e linearit\`a di risposta. Per le loro qualit\`a di sensibilit\`a, sono 
invece indicati per risolvere i piccoli segnali prodotti dai protoni. 
\newline
Nei rivelatori \v{C}erenkov al quarzo, la quantit\`a di luce prodotta al passaggio di una 
particella \`e proporzionale al quadrato della sua carica (ed alla sua velocit\`a), 
inoltre offrono appropriati requisiti di resistenza a fasci di considerevole intensit\`a. 
Essi permettono pertanto un'adeguata identificazione degli ioni di piombo, discriminandoli 
dai frammenti di minor carica.  
%\subsubsection{Tempo morto}
\subsubsection{Condizioni di trigger in eventi Pb-Pb}
\begin{figure}[h]
 \begin{center}
 \includegraphics[scale=0.55]{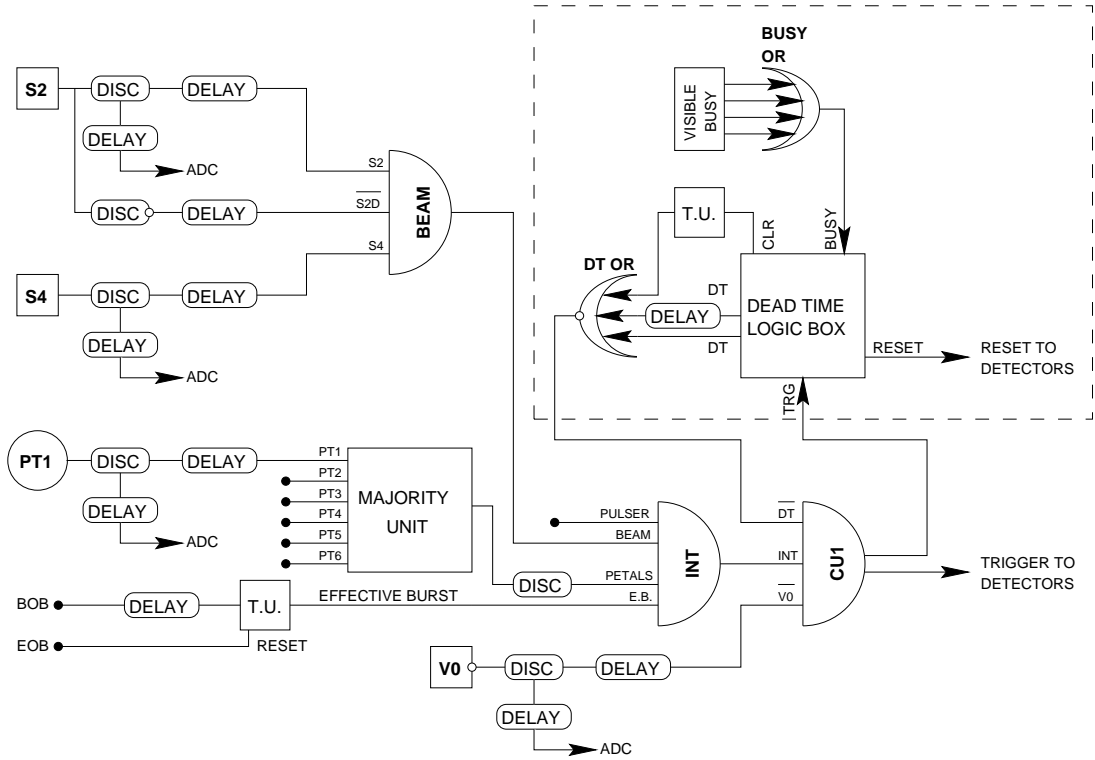}
  \caption{Diagramma del trigger di NA57 nella configurazione per la presa dati 
   Pb-Pb a 160 $A$ GeV/$c$\ del 1998. La zona racchiusa entro il riquadro trattegiato 
   si riferisce alla generazione del segnale di {\em tempo morto}. Il contatore 
   $S4$\ \`e stato utilizzato solo per l'allineamento del fascio. 
   BOB: segnale di inizio {\em burst}; 
   EOB: segnale di fine  {\em burst}. }
 \label{PbTrig}
\end{center}
\end{figure}

Lo schema completo del trigger per il fascio di piombo \`e mostrato in 
fig.~\ref{PbTrig}.  
\noindent
La logica di trigger pu\`o essere classificata in tre livelli:  
\begin{itemize}
\item identificazione del proiettile incidente e reiezione degli 
      eventi concomitanti; 
\item selezione degli eventi in cui vi \`e stata un'interazione 
      anelastica tra il proiettile ed il bersaglio; 
\item selezione degli eventi Pb-Pb sufficientemente centrali. 
\end{itemize}
Il contatore $S2$\ \`e utilizzato per identificare gli ioni di piombo e rigettare 
gli eventi concomitanti~\footnote{Il segnale di ciascun contatore verr\`a indicato con lo
stesso nome dato al rivelatore.}.  
Il singolo ione di piombo \`e  
identificato per mezzo di un discriminatore, che genera il segnale $S2$. 
Nel caso in cui due ioni di piombo attraversino $S2$\ a breve distanza temporale,  
il segnale del contatore $S2$ sar\`a pi\`u ampio. In tal caso anche un secondo 
discriminatore, con una soglia pi\`u elevata rispetto a quella del discriminatore  
che ha generato $S2$, fornisce un segnale, $S2D$. Il primo livello di trigger  
corrisponde dunque alla condizione $ BEAM = S2 \cdot \overline{S2D}$, 
in cui si ha un solo ione di piombo per volta nel contatore $S2$. 
Come detto, il contatore $S4$\ viene utilizzato 
solo per definire il corretto posizionamento del fascio, ed in tali condizioni 
viene posto in ``and'' logico con $BEAM$.  
\newline
Il contatore di veto $V0$\ svolge la seconda funzione nella logica di trigger,  
misurando la carica dei frammenti del nucleo incidente  
(o dell'intero nucelo se questo non ha interagito). 
La soglia di tale rivelatore \`e impostata in modo tale da fornire un segnale 
quando viene attraversato da grossi frammenti nucleari. 
Esso pu\`o quindi antiselezionare gli ioni incidenti che, dopo il bersaglio, si trovano 
ancora lungo la direzione del fascio o quelli che hanno subito sola una 
piccola frammentazione nell'interazione.   
L'assenza del segnale da $V^0$\ ($\overline{V0}$) conferma quindi 
l'interazione del proiettile entro il bersaglio.  
\newline
I rivelatori a petali $PT$  forniscono il segnale 
che seleziona gli eventi pi\`u centrali ({\em cfr}. paragrafo 2.4.2). 
Il segnale di ciascuno dei sei petali viene prima inviato ad un discriminatore, 
essi confluiscono quindi in una unit\`a di ``logica maggioritaria'' 
({\em ``majority unit''}) che d\`a segnale quando almeno cinque dei sei petali 
scintillatori sono sopra soglia ($PT(5/6)$).   
Durante la presa dati del 1998 uno dei sei scintillatori, che non funzionava  
perfettamente, \`e stato escluso dalla logica di trigger. In tali condizioni 
il sistema dei petali forniva la terza condizione di trigger quando tutti i 
cinque rimanenti scintillatori si trovavano sopra soglia, $PT(5/5)$.  
Per consistenza, anche durante le prese dati col fascio di Pb a 
40 $A$\ GeV/$c$ si \`e mantenuta la stessa configurazione nel trigger, 
sebbene ormai tutti i sei scintillatori funzionassero regolarmente.
\newline
Il segnale dei petali ($PT(5/6)$) e quello del contatore $V0$\ ($\overline{V0}$) 
sono in realt\`a complementari. Infatti, uno ione che non abbia interagito nel bersaglio 
o che abbia subito una piccola frammentazione (per cui dunque $\overline{V0}=FALSO$), 
non riesce a produrre una molteplicit\`a di particelle cariche tale da superare la 
condizione richiesta dai petali (cio\`e anche $PT(5/6)=FALSO$).  
Il segnale di $V0$\ \`e stato pertanto escluso dalla logica finale di trigger.   
\newline
La condizione del segnale di $BEAM$\ e dei petali ($PT(5/6)$) seleziona l'interazione 
desiderata  $ INT = BEAM \cdot PT(5/6) $.  

%\subsubsection{Tempo morto}
\noindent
Il segnale di {\bf tempo morto} $DT$ inibisce l'acquisizione di un'ulteriore evento qualora 
si stia  ancora processando l'evento precedente. 
L'intervallo del {\em tempo morto} viene determinato con la stessa procedura negli 
eventi Pb-Pb ed in quelli p-Be. A riguardo, conviene riferirsi alla zona della 
fig.~\ref{PbTrig} racchiusa entro il rettangolo tratteggiato. 
\newline
Quando un evento \`e accettato da tutti i livelli di trigger per essere acquisito, il 
segnale di trigger di livello pi\`u alto $CU1$ 
%({\em cfr.} prossimo paragrafo) 
(che sar\`a di seguito definito)  
viene distribuito a tutti i rivelatori. 
Non appena il segnale $CU1$\ raggiunge un rivelatore, il processo di lettura viene avviato 
ed un segnale di {\em busy} viene generato per indicare che per quel rivelatore 
(o per  una parte del rivelatore, a seconda dei casi) 
\`e in corso la lettura da parte  
dell'elettronica di acquisizione. Tutti i segnali di {\em busy} 
($VISIBLE\_BUSY$\ in fig.~\ref{PbTrig}) sono posti in $OR$ logico dando luogo al 
segnale $BUSY\_OR$, che viene inviato all'ingresso di {\em busy} dell'unit\`a logica 
di tempo morto ($DEAD\_TIME\_LOGIC\_BOX$  in fig.~\ref{PbTrig}). Tale unit\`a genera 
il segnale di {\em tempo morto} $DT$ che agisce come veto per gli eventi successivi.  
\newline
Poich\'e il segnale $BUSY\_OR$\ impiega circa 500 ns per essere formato e giungere 
all'unit\`a logica di tempo morto, il segnale di trigger viene inviato 
direttamente all'ingresso $TRG$\ dell'unit\`a per generare immediatamente il 
segnale $DT$.  
\newline
Il segnale di {\em busy} $BUSY\_OR$ \`e mantenuto sino a quando per tutti i 
rivelatori \`e completato il processo di lettura, terminato il quale rilasciano 
il proprio segnale di {\em busy}. A questo punto un segnale di azzeramento 
($RESET$) viene inviato ai rivelatori, ed il segnale di {\em clear} ($CLR$) \`e 
attivato. 
\newline
La frequenza del trigger dipende principalmente dal tempo morto dei rivelatori. 
I rivelatori pi\`u critici sotto questo punto di vista sono i piani di pixel 
ed i piani di $\mu$-strip per i quali il tempo morto si aggira sul millisecondo 
per evento.   
%
%\newline

\noindent
Riassumendo la condizione finale di trigger per l'interazione Pb-Pb \`e data da 
\begin{equation}
% CU1 = \overline{V0} \cdot  INT \cdot\ \overline{DT}
 CU1 = INT \cdot\ \overline{DT}
 \label{CU1_Pb}
\end{equation}
%dove $INT$\ \`e definito come $ INT = BEAM \cdot PT(5/6) $.
\subsubsection{Condizioni di trigger in eventi p-Be} 
Nel caso delle collisioni p-Be si \`e interessati a raccogliere quegli 
eventi in cui almeno due tracce attraversino la regione del telescopio, 
in modo da essere potenzialmente ricostruibili.   
Si intende infatti studiare la produzione di particelle strane 
che decadono in stati finali contenenti solo 
(almeno due) particelle cariche. 
\newline
Due scintillatori, $SPH1$\ ed $SPH2$, di dimensioni simili a quelle dei 
piani di pixel ($5\times5\, {\rm cm^2}$) sono posti immediatamente 
avanti al primo piano di pixel del telescopio. Entrambi hanno una soglia tale 
da selezionare principalmente il passaggio di due particelle cariche. Posti 
in coincidenza ($SPH1 \cdot SPH2$), si riduce drasticamente la contaminazione 
di segnali di elevata ampiezza ma di singola traccia 
(le code delle distribuzioni di Landau).  
\newline
Dopo l'ultimo piano di pixel, \`e posto un terzo scintillatore $ST2$, di 
maggiori dimensioni, che accerta se almeno una delle due tracce abbia 
attraversato la parte compatta del telescopio, per una buona frazione 
della sua lunghezza.  
\newline 
La condizione che definisce un evento potenzialmente 
interessante, in base al numero di tracce atteso nel telescopio,  
\`e dunque $TEL INT = ( SPH1 \cdot SPH2 \cdot ST2 ) \cdot BEAM $. 
\newline
Il fascio \`e ora definito dalla condizione  
$BEAM=S2\cdot S4$~\footnote{Nel caso del fascio di Pb, \`e 
indispensabile rimuovere $S4$\ dalla linea di fascio  
per evitare interazioni fuori dal bersaglio.}, a cui si deve per\`o aggiungere 
l'antiselezione dei pioni (e dei kaoni) operati dai rivelatori   
\v{C}erenkov: $\overline{C1} \cdot \overline{C2} $.
\newline
La condizione finale di trigger \`e quindi  
$CU1= \overline{C1} \cdot \overline{C2} \cdot TEL INT 
      \cdot \overline{DT}$.  
\subsection{L'acqusizione dei dati}
\begin{figure}[h]
 \begin{center}
 \includegraphics[scale=0.52]{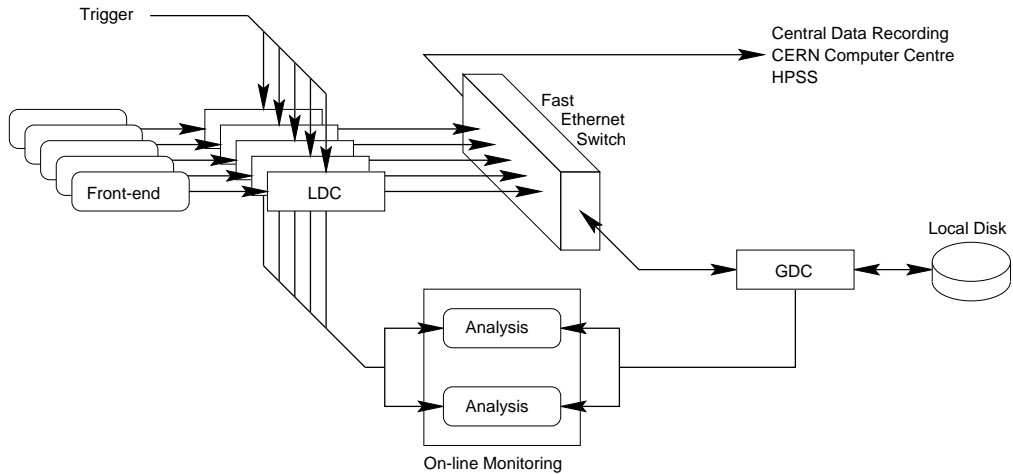}
 \caption{L'architettura del sistema DATE di acqusizione dati nell'esperimento NA57.}
 \label{DATE}
\end{center}
\end{figure}

L'acqusizione finale dei dati raccolti dai diversi rivelatori dell'esperimento  
avviene servendosi del sistema DATE ({\em Data Acqusition and Test Environment}).  
Tale sistema \`e stato sviluppato per l'esperimento ALICE ed \`e in grado di 
trasportare ed immagazzinare un'elevatissima quantit\`a di dati in breve tempo.  
Nelle interazioni Pb-Pb, le pi\`u critiche per quel che riguarda l'aspetto 
dell'acqusizione, si selezionano sino a $3000$\ eventi per pacchetto 
({\em burst}); ciascun evento ha una dimensione di circa 5 kByte. Ne risulta che
la velocit\`a di trasferimento dei dati debba essere di circa 2 MByte al 
secondo~\cite{NA57p1}, ed il volume totale dei dati di qualche TByte. 
\newline
L'architettura del sistema DATE \`e disegnata  in modo schematico nella 
fig.~\ref{DATE}. Si tratta di un sistema software che si serve di sette  
processori VME in parallelo.  
\noindent
I processori VME, chiamati LDC
({\em Local Data Concentrator}), acquisiscono,
indipendentemente gli uni dagli altri, i dati provenienti da differenti
rivelatori, o parte di questi. In particolare  al trigger, alle MSD ed
alle $\mu$-strip sono dedicate tre LDC, una per ciscuno, mentre
ai tredici piani di pixel sono riservati i rimanenti quattro LDC.
\newline
Ciascun LDC esegue la lettura dell'elettronica di front-end e organizza    
i dati in sotto-eventi. La memoria degli LDC \`e capace di contenere i dati 
relativi ad almeno un intero {\em burst}. 
La sincronizzazione dell'acqusizione dei dati \`e operata dal sistema di trigger. 
\newline
Tramite un interruttore di tipo {\em Fast Ethernet} i processori LDC sono connessi 
al GDC ({\em Global Data Concentrator}). Solo al termine di un {\em burst} i dati 
di tutti gli eventi raccolti vengono trasferiti dagli LDC verso il GDC. 
Quest'ultimo raccoglie tutte le informazioni, assemblando i sotto-eventi 
provenienti da uno stesso trigger fisico in un unico evento. 
\newline
Gli eventi cos\`i riassemblati, spesso indicati col termine di {\em raw data} per 
distinguerli da quelli ricostruiti, vengono quindi memorizzati su un disco locale, 
in un unico file. L'insieme degli eventi contenuti all'interno di questo singolo file 
prende il nome di {\em run}. Tipicamente ciascun {\em run} contiene sino a 
2 $\times$ $10^5$\ eventi Pb-Pb, numero imposto  dalla dimensione  tipica di un  
file di {\em raw data}, pari a circa 1 GByte. Dunque, quando sono stati raccolti 
dati per circa 1 GByte di informazioni, il {\em run} viene automaticamente 
interrotto e ne inizia uno nuovo associato ad un nuovo file.  
\newline
Attraverso la stessa porta Ethernet, i file contenenti i {\em raw data} vengono  
trasferiti al {\em centro calcolo} del CERN, dove vengono salvati in modo 
definitivo e permanente su nastro magnetico. Il sistema HPSS 
({\em High Performance Storage System})~\cite{HPSS}, che controlla  
l'accesso ai dischi ed ai nastri magnetici e pu\`o gestire milioni di file di 
grandi dimensioni, \`e stato utilizzato sin dall'inizio;  
a partire dall'anno 2000 ci si \`e serviti del nuovo sistema denominato CASTOR  
({\em {\bf\em C}ERN {\bf\em A}dvanced {\bf\em STOR}age Manager})~\cite{CASTOR}. 
\newline
Durante la raccolta dei dati, col sistema DATE \`e possibile monitorare all'istante  
l'acquisizione dei dati. Diversi programmi di analisi possono essere quindi applicati 
{\em on line} al flusso di dati per controllare la qualit\`a degli eventi raccolti.  

\section{La ricostruzione degli eventi}
Il primo passo dell'analisi degli eventi consiste nella ricostruzione dei punti 
spaziali in cui le particelle hanno attraversato i rivelatori e nella successiva 
correlazione tra questi punti per la ricostruzione delle tracce. Questo compito 
\`e assolto dal programma ORHION 
({\em Omega Reconstruction code for Heavy ION experiments})~\cite{ORHION}, 
sviluppato inizialmente per l'esperimento WA97 nello spettrometro OMEGA. 
I compiti principali di ORHION sono il riconoscimento delle tracce e la 
determinazione dei loro parametri cinematici. 
\subsubsection{Riconoscimento delle tracce}
La ricostruzione delle tracce viene effettuata nella parte compatta 
del telescopio (i primi 30 cm) composta da un numero cospicuo di piani 
di pixel: nella presa dati del 1998  la parte compatta era costituita da 
nove piani, in quella del 1999 da dieci, ed in quella del 2000 da undici.  
\newline
I punti delle tracce sui due piani pi\`u esterni vengono inizialmente 
associati in maniera combinatoria e vengono cos\`i a determinare le prime 
tracce candidate. I punti dei piani pi\`u interni vengono quindi 
utilizzati per confermare o rigettare la scelta della coppia iniziale 
di punti. 
\newline 
Con questa prima ricerca non si sono ancora ricostruite quelle tracce che 
non hanno rilasciato un segnale sui piani esterni, o 
per l'inefficienza dei rivelatori o perch\'e non vi sono passate attraverso.  
Vengono dunque considerate nuove coppie di piani iniziali, p\`u interne, 
uno nella parte iniziale del telescopio e l'altro nella parte posteriore. 
\newline 
La ricerca delle tracce pu\`o inoltre trarre vantaggio dal fatto che la 
maggior parte delle tracce ha origine nel vertice di interazione primaria 
(tracce {\em primarie}). In questo caso i punti dell'ultimo piano possono 
essere associati alla posizione del bersaglio invece che ai punti del 
primo piano. In tal modo, il numero di combinazioni \`e proporzionale 
alla molteplicit\`a dell'ultimo piano, mentre col procedimento precedente 
era proporzionale al prodotto delle molteplicit\`a sul primo e sull'ultimo 
piano. La ricerca delle tracce primarie \`e quindi relativamente rapida. 
\newline
Dopo aver trovato con questo metodo tutte le possibili tracce primarie, 
si ripete l'operazione rilasciando il vincolo del bersaglio e si cercano 
le tracce secondarie, quelle cio\`e che non originano dal vertice di 
interazione.
\newline
Le tracce vengono cercate prima nella proiezione $(x,z)$, dove sono 
rappresentabili con delle linee rette (il campo magnetico \`e orientato 
lungo l'asse $z$), quindi nel piano di curvatura $(x,y)$, dove sono 
approssimate a degli archi di parabola, ed alla fine le proiezioni 
vengono combinate per formare la traccia tridimensionale.  
\newline
Per tutte le tracce ricostruite si definiscono dei parametri di qualit\`a, 
i quali stabiliscono infine se una traccia possa essere accettata o meno.  
Tra questi vi \`e ovviamente il numero di punti associato alle tracce; esso 
sar\`a oggetto di discussione nel prossimo capitolo. 
\subsubsection{Determinazione dei parametri cinematici delle tracce}
Ogni traccia \`e completamente definita da cinque parametri. La scelta 
pi\`u conveniente di questi parametri dipende dalla geometria 
dell'esperimento (ad esempio se si tratta di un esperimento a bersaglio 
fisso o di uno ad un collisionatore) e dalla configurazione del campo 
magnetico. Nel caso di NA57, una scelta conveniente \`e quella di 
considerare l'intersezione della traccia con un piano $\Pi$\ ortogonale 
all'asse $x$\ e definire i seguenti parametri (fig.~\ref{TrkParam}): 
\begin{figure}[h]
 \begin{center}
  \includegraphics[scale=0.52]{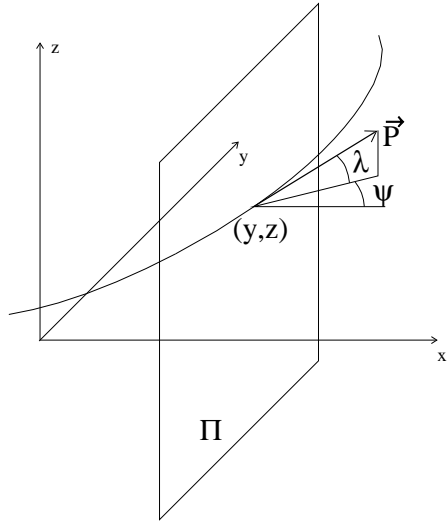}
   \caption{Rappresentazione dei parametri cinematici che definiscono una traccia.}
  \label{TrkParam}
 \end{center}
\end{figure}
\noindent
\begin{itemize}
\item le coordinate $y$\ e $z$\ della traccia all'intersezione col piano $\Pi$; 
\item $1/\rho$, dove $\rho$\ \`e il raggio di curvatura della traccia 
      all'intersezione col piano $\Pi$; 
\item l'angolo $\psi$\ tra la proiezione dell'impulso nel piano di curvatura 
      $(x,y)$\ e la normale al piano $\Pi$\ (asse $x$); 
\item l'angolo di {\em ``dip''} $\lambda$\ tra il vettore impulso ed il 
	piano di curvatura $(x,y)$.
\end{itemize}
Come piano di riferimento $\Pi$\ viene utilizzato un piano la cui ascissa 
corrisponda al primo punto misurato della traccia. 
\newline
La determinazione dei parametri delle tracce viene effettuata da ORHION  
iterativamente: prima vengono utilizzati solo i punti nella parte compatta 
del telescopio, dove le tracce vengono ricostruite, poi vengono cercati 
eventuali ulteriori punti della traccia sui piani del telescopio che 
formano il braccio di leva (gli eventuali piani di pixel rimanenti 
e le $\mu$-strip).  
L'aggiunta di punti appartenenti al braccio di leva permette di ridurre 
l'errore sulla determinazione del momento, in particolare per le tracce 
pi\`u veloci. 
\newline
Operativamente, la procedura di interpolazione viene effettuata coi seguenti 
passi: 
\begin{itemize}
\item nella ricerca delle tracce entro la parte compatta del 
      telescopio, esse vengono interpolate con un modello 
      {\em ``spline''} di quinto grado~\cite{SPLINE} 
      servendosi della mappa misurata del campo magnetico. In questo modo si 
      ottiene la prima stima dei parametri cinematici; 
\item completata la ricerca delle tracce nella parte compatta del
      telescopio, le tracce vengono quindi estrapolate e ne vengono 
      determinate le eventuali intersezioni sui piani del braccio di leva. 
      I punti su questi rivelatori, che si trovino entro determinate tolleranze 
      dalla posizione della traccia estrapolata, vengono associati ad essa; 
\item utilizzando anche questi nuovi punti, vengono quindi ricalcolati i 
      parametri cinematici. 
\end{itemize}
\section{Il campione dei dati}
Un primo run dell'esperimento, non finalizzato alla raccolta dati per analisi 
di fisica,   \`e stato eseguito nel 1997 
per testare l'apparato sperimentale e sviluppare il codice per 
la ricostruzione degli eventi.  
La prima presa dati \`e avvenuta invece nei mesi di Ottobre e Novembre del 1998, 
in cui si sono studiate le interazioni Pb-Pb a 160  $A$ GeV/$c$, raccogliendo un 
totale di 230 milioni di trigger. Nel Luglio del 1999 si sono raccolti i primi 
dati per l'interazione  p-Be di riferimento a 40 GeV/$c$. Nel periodo 
Ottobre-Novembre 1999 si \`e passati quindi ad utilizzare il fascio di ioni 
di piombo ad un energia di 40 GeV per nucleone (Pb-Pb a 40 $A$ GeV/$c$). 
Un secondo campione di dati Pb-Pb a 160  $A$ GeV/$c$ \`e stato quindi 
raccolto nell' autunno dell'anno 2000. Infine, nell'estate del 2001, si \`e 
completata la presa dati per il sistema p-Be a  40  GeV/$c$. 
Tutti gli eventi dei campioni di dati raccolti sono stati ricostruiti, 
ad eccezione di quelli relativi all'anno 2001, tuttora in fase di 
completamento.  
\newline
Nella tabella~2.1 sono riassunte le informazioni sui diversi 
campioni di dati sin qui raccolti dall'esperimento.  
\begin{table}[h]
  \label{tab2_1}
  \begin{center}
  \begin{tabular}{|c|c|c|c|c|} \hline
    {\bf Interazione}                  & 
  \begin{tabular}{c} 
    {\bf Impulso} \\
    {\bf del fascio}
  \end{tabular}                      &  
  \begin{tabular}{c}
    {\bf Dimensione} \\ 
    {\bf del campione} \\
    {\bf (trigger $\times$ $10^6$)}  
  \end{tabular}                      &
  \begin{tabular}{c}
    {\bf Anno di} \\
    {\bf presa dati} 
  \end{tabular}                      &
  \begin{tabular}{c}
    {\bf Ricostruzione } \\
    {\bf degli eventi}
  \end{tabular}                \\
\hline Pb-Pb & 160 $A$ GeV/$c$  & \begin{tabular}{c} 230 \\ 230 \end{tabular}  & 
  \begin{tabular}{c} 1998 \\ 2000 \end{tabular} & 
  \begin{tabular}{c} completata \\ completata \end{tabular}  \\
\hline  Pb-Pb & 40 $A$ GeV/$c$  & 260 & 1999   & completata  \\
\hline  p-Be  & 40  GeV/$c$     & \begin{tabular}{c} 60 \\ 110 \end{tabular}  &
  \begin{tabular}{c} 1999 \\ 2001 \end{tabular} & 
  \begin{tabular}{c} completata \\ in corso \end{tabular}  \\
\hline
\end{tabular}
\end{center}
\caption{Campioni di dati raccolti dall'esperimento NA57.}
\end{table}
\newline
Nell'anno 2002 l'esperimento NA57 non ha raccolto dati, ma tuttavia 
non \`e stato  dismesso. 
Si \`e infatti avanzata richiesta~\cite{NA57Mem} di un'ulteriore  
presa dati per studiare le collisioni di  
un sistema pi\`u leggero del Pb-Pb, quale l'In-In.  
L'SPS dovrebbe infatti fornire un fascio di tali ioni a  partire dal 2003.  
Vi \`e grande interesse nello studio di un simile sistema in  
quanto collisioni centrali tra due nuclei di indio   
corrisponderebbero, in termini di volumi coinvolti nell'interazione  
(o pi\`u precisamente in termini di nucleoni  che prendono parte  
alle collisioni iniziali, {\em cfr.  paragrafo 1.3.3}),  alle collisioni 
pi\`u periferiche che si possono studiare nel sistema Pb-Pb. 
Inoltre, considerando anche le collisioni periferiche 
del sistema In-In, vi sarebbe la possibilit\`a di studiare  
interazioni con un numero ancora minore di partecipanti.  
%rispetto aquanto possibile nel sistema Pb-Pb. 
Infine, a parit\`a di nucleoni partecipanti,  sarebbe molto importante 
poter confermare le misure periferiche del sistema Pb-Pb nelle nuove interazioni,   
con il grande vantaggio che lo studio di collisioni pi\`u centrali 
comporta rispetto a quelle periferiche: un miglior controllo sul trigger 
di centralit\`a, una misura pi\`u precisa della centralit\`a stessa, etc.

%% file: cap3/cap3.tex
\chapter{Ricostruzione e selezione delle particelle strane}

\section{Introduzione}
\begin{figure}[hbt]
\begin{center}
 \includegraphics[scale=0.50]{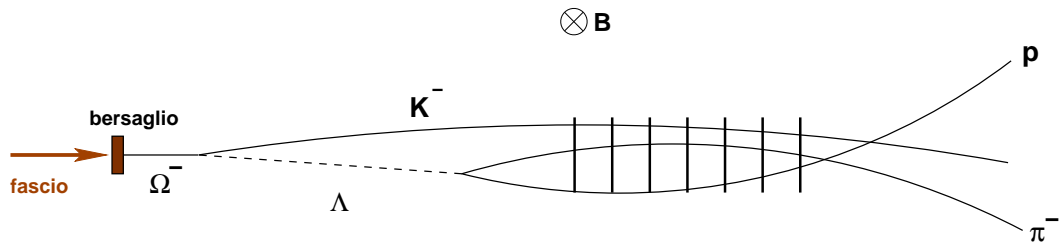}
  \caption{Schema del decadimento a ``cascata'' di una $\Omega^-$\ in particelle 
           cariche  entro un campo magnetico. Il decadimento viene ricostruito 
	   a partire dalla misura delle tracce entro il telescopio, i cui piani 
	   sono schematizzati dai segmenti trasversali.}
\label{OmDecay}
\end{center}
\end{figure}

Nel capitolo precedente si \`e esposta in maniera sintetica 
la logica di funzionamento del programma ORHION per la ricostruzione 
delle tracce di un evento. In questo capitolo si discuter\`a come sia possibile 
ricostruire ed identificare le particelle strane che decadono per interazione 
debole in uno stato finale contenente solo particelle cariche,  
a partire dai parametri cinematici dei prodotti di decadimento. 
\newline
I decadimenti di particelle strane neutre che si possono ricostruire 
con l'apparato sperimentale di NA57, con le relative probabilit\`a di 
decadimento nel canale considerato ({\em Branching Ratio}), sono i seguenti: 
\begin{align}
K_S^0 \longrightarrow  \; & \pi^+ + \pi^- & BR=68.6 \pm 0.3 \% \nonumber \\
\Lambda \longrightarrow \;& p + \pi^-     & BR=63.9 \pm 0.5 \% \nonumber \\
\bar{\Lambda} \longrightarrow \;& \bar{p} + \pi^+  & BR=63.9 \pm 0.5 \% \nonumber
\end{align}
Questi decadimenti carichi formano
delle configurazioni chiamate $V^0$, intendendo con ci\`o una coppia di
tracce di curvatura opposta nel campo magnetico che si diramano da uno
stesso punto di cuspide distinto dal vertice primario di interazione.
\newline
\`E inoltre possibile ricostruire i decadimenti a ``cascata'' delle particelle 
multi-strane, del tipo: 
\begin{align}
\Xi^- \longrightarrow \; & \Lambda + \pi^- \quad BR=99.89 \pm 0.04 \% \nonumber \\
                      & 
		      \begin{picture}(20,20)(0,0) \put(4,2){\vector(1,0){13}} 
		      \put(4,2){\line(0,1){12}}
		      \end{picture} \; p + \pi^- \nonumber \\ \nonumber \\
\Omega^- \longrightarrow \; & \Lambda + K^- \quad BR=67.8 \pm 0.7 \% \nonumber \\
		      &
                      \begin{picture}(20,20)(0,0) \put(4,2){\vector(1,0){13}}
                      \put(4,2){\line(0,1){12}}
                      \end{picture} \; p + \pi^- \nonumber
\end{align}
e quelli corrispondenti alle relative antiparticelle. Come si vedr\`a, 
questi ultimi decadimenti vengono ricostruiti combinando le $V^0$\ riscostruite 
con un'altra traccia carica dell'evento, ed operando opportuni tagli geometrici 
e cinematici.   
\newline
Nel caso di un decadimento a due corpi --- sia essa una $V^0$\ che decade in due 
particelle cariche, od una cascata ($\Xi$\ od $\Omega$) che decade in una $\Lambda$\ 
ed un mesone carico ($\pi$\ o $K$) --- in particelle 
di massa $m_i$\ ed impulso 
$\vec{p}_i$\ ($i=1,2$), dalla misura dell'impulso delle tracce al vertice 
di decadimento \`e possibile risalire alla massa della particella madre:
\begin{equation}
M^2 = m_1^2+m_2^2+2\sqrt{(m_1^2+p_1^2)(m_2^2+p_2^2)} - 2 \vec{p}_1\cdot\vec{p}_2
\label{V0M}
\end{equation}
Lo schema del decadimento di una $\Omega^-$\ \`e riportato in fig.~\ref{OmDecay}.  

\section{Ricostruzione delle $V^0$}
\begin{figure}[hbt]
\begin{center}
\includegraphics[scale=0.60]{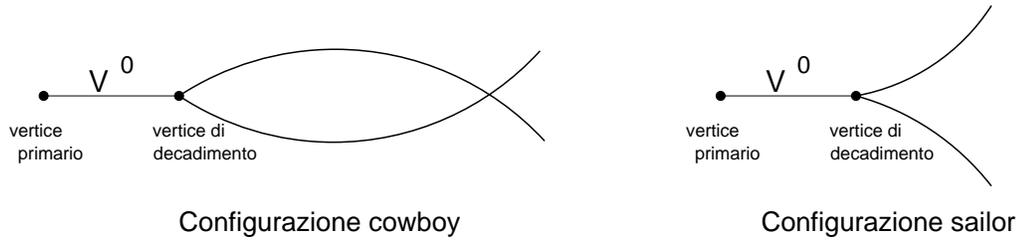}
\caption{Possibili configurazioni di decadimento delle $V^0$.} 
%         {\bf a)} configurazione ``cowboy'', {\bf b)} configurazione ``sailor''.}
\label{cowsai}
\end{center}
\end{figure}

\begin{figure}[htb]
\begin{center}
\includegraphics[scale=0.42]{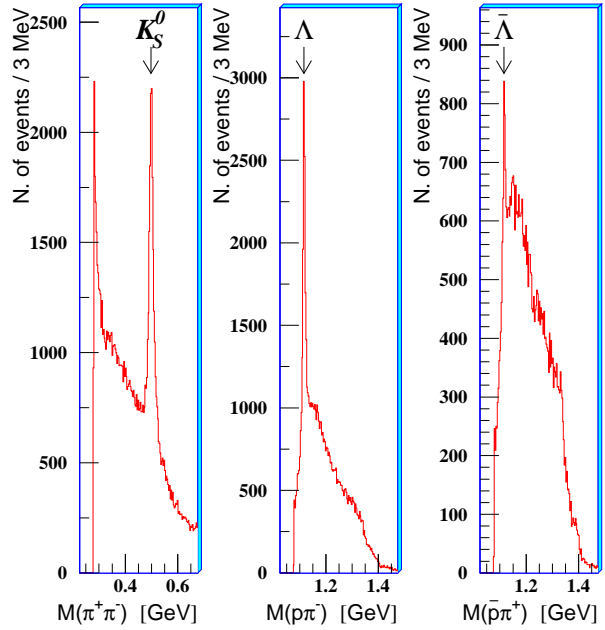}
 \caption{ Spettri di massa invariante $M(\pi^+,\pi^-)$\ (a sinistra), 
 $M(p,\pi^-)$\ (al centro) e $M(\bar{p},\pi^+)$\ per un campione 
 di candidate $V^0$\ ricostruite da STRIPV0.}
\label{V0candidates}
\end{center}
\end{figure}

Nel programma di ricostruzione ORHION vi \`e un sottoprogramma,
STRIPV0, che \`e incaricato di individuare tutte le candidate $V^0$\ contenute in
un evento. STRIPV0 ricostruisce i vertici di decadimento di tipo $V^0$\ come
intersezione di due tracce di carica opposta estrapolate all'indietro. 
Il programma associa in modo combinatorio le tracce di carica opposta e 
calcola la minima distanza tra le due traiettorie; coppie per le quali tale 
distanza \`e superiore ad $ 1\,{\rm mm}$\ vengono subito rigettate. 
Le coordinate del vertice della $V^0$\ sono calcolate come i valori medii delle 
coordinate dei punti di minima distanza tra le tracce abbinate. L'impulso 
della particella $V^0$\ \`e la somma vettoriale degli impulsi delle sue due 
tracce, valutati nei punti di minima distanza.
Come illustrato in fig.~\ref{cowsai} si distinguono due topologie
per una candidata $V^0$\, a seconda della
disposizione della proiezione delle tracce cariche nel piano $xy$:
\begin{itemize}

\item[{\bf a)}]
la topologia di tipo ``cowboy'', in cui le proiezioni delle due tracce 
si intersecano una seconda volta dopo il vertice di decadimento.
\item[{\bf b)}] 
quella di tipo ``sailor'',  
in cui le proiezioni subito divergono senza  pi\`u intersecarsi.
\end{itemize}
La rivelazione dei vertici ``cowboy'', ben pi\`u probabile di quelli
``sailor'' per la piccola sezione trasversa del telescopio, conduce
ad una determinazione dei parametri della $V^0$\  pi\`u precisa.  
Infatti, la determinazione del vertice di decadimento, in 
cui si calcolano i parametri cinematici della $V^0$, \`e molto 
pi\`u precisa per i vertici {\em cowboy}.  
Ad esempio, un contributo all'errore sperimentale nella misura di $M$\  
\`e legato a quello dell'angolo di apertura $\phi$\ della $V^0$\ 
dalla relazione $ dM\,=\,\frac{p_1p_2}{M}\sin(\phi)\,d\phi$, ottenuta
differenziando la eq.~\ref{V0M}, e tale angolo \`e pi\`u critico da
valutare nella topologia ``sailor''.   
%\`E anche pi\`u difficile separare i decadimenti fisici con topologia  
%{\em sailor} dal fondo combinatorio di tracce primarie, rispetto a 
%quelli di tipo ``cowboy''.  
Risulta quindi pi\`u conveniente  
considerare i soli vertici ``cowboy'', che corrispondono al 70\% dei vertici 
ricostruiti. 
In appendice B si mostra come, per la configurazione ``cowboy'', la distanza 
tra il vertice di decadimento ed il secondo punto di intersezione delle 
proiezioni delle due tracce nel piano $xy$\ sia invariante per trasformazioni 
di Lorentz ({\em ``boost''}) ed assuma, all'interno del campo del magnete 
GOLIATH, il massimo valore 
$ L_{MAX} \approx \; 100 $\ cm per i $K_S^0 $\ ed 
$ L_{MAX} \approx \; 50  $\ cm per le $\Lambda$\ ed $\bar{\Lambda} $. Allo 
stesso tempo la massima distanza trasversa tra le due tracce di 
decadimento, valutata nel piano $xy$\ rispetto alla direzione $L$\ della 
$V^0$, risulta piuttosto piccola, dell'ordine del centimetro per $K_S^0 $\ e 
$\Lambda$. La rivelazione di queste particelle \`e pertanto possibile  
all'interno del telescopio dell'esperimento NA57, caratterizzato 
da una sezione trasversa 
%solo {\em apparentemente} 
limitata ($5\times5 \; {\rm cm^2}$) 
ma in realt\`a appropriata per la cinematica di decadimento delle $V^0$, 
e di cui si pu\`o sfruttare tutta la lunghezza per la misura delle traiettorie 
delle particelle di decadimento.  
\newline
Il massimo valore che pu\`o assumere l'impulso trasverso $q_T$\ delle tracce 
di decadimento rispetto alla linea di volo della particella madre \`e fornito 
da: 
\begin{equation}
q_{T}^{MAX}=\frac{1}{2M}\sqrt{M^2-(m_1+m_2)^2}\,
         \sqrt{M^2-(m_1-m_2)^2}, 
\label{qtMax}
\end{equation}
ed \`e pertanto pari a $0.206$\ GeV/$c$\ per i $K^0_S$\ e 
$0.101$\ GeV/$c$\ per le $\Lambda$\ e le $\bar{\Lambda}$. Nel programma 
STRIPV0, al fine di ridurre le combinazioni spurie di tracce che non 
corrispondono al decadimento di una $V^0$, si richiede pertanto la condizione 
$q_T<0.4$\ GeV/$c$.  
\newline
Come ultima condizione, al fine di ridurre drasticamente il numero di candidate 
$V^0$\ ottenute abbinando tracce {\em primarie} prodotte al vertice dell'interazione 
Pb-Pb (o p-Be), si richiede che il vertice di decadimento della candidata $V^0$\ 
(coincidente, come si \`e detto, con il punto medio del segmento di minima distanza 
tra le due tracce cariche) disti pi\`u di $15$\ cm dal bersaglio. Come si vedr\`a nel 
prossimo paragrafo, la scelta finale del volume fiduciale entro il quale una $V^0$\ deve  
decadere per non essere rigettata dai criteri di selezione 
\`e ben pi\`u restrittiva di questa condizione.  
\newline
In fig.~\ref{V0candidates} sono mostrati gli spettri di massa inavariante 
$M(\pi^+,\pi^-)$, $M(p,\pi^-)$\ e $M(\bar{p},\pi^+)$\ corrispondenti ad un 
piccolo campione, pari a circa lo 0.5\% dell'intera statistica dei dati raccolti 
nel 1998, delle candidate $V^0$\ selezionate da STRIPV0. I segnali delle diverse 
specie di particelle sono gi\`a evidenti nei rispettivi spettri, in corrispondenza 
della loro massa nominale, in misura statisticamente significativa rispetto al fondo.  
Il fondo \`e costituito da false $V^0$, in cui si distinguono due  
contributi, entrambi da eliminare;  
il primo \`e dovuto a casuali intersezioni geometriche delle 
tracce presenti nell'evento ($V^0$\ ``geometriche''), 
il secondo, dovuto ad un'errata assegnazione delle masse alle particelle 
del decadimento, si manifesta nello spettro con picchi, detti ``riflessi'', 
in corrispondenza di valori non fisici per 
le masse delle $V^0$\ (fondo cinematico). 
Il lavoro di analisi eseguito sulle candidate $V^0$\ ricostruite da STRIPV0 
ha lo scopo di isolare il segnale fisico di $K_S^0$, $\Lambda$\ ed $\bar{\Lambda}$\ 
dagli eventi di fondo e dai riflessi.  
\section{Selezione delle $V^0$}
Per selezionare ed isolare le particelle $K_S^0$, $\Lambda$\ ed $\bar{\Lambda}$\ 
si possono applicare una serie di criteri di selezione, di natura sia 
geometrica che cinematica, finalizzati a minimizzare la contaminazione 
geometrica delle false $V^0$\ e quella cinematica dei riflessi.  
La determinazione dei criteri di selezione \`e stato oggetto di uno studio 
sistematico, volto a render massimo il rapporto segnale su fondo e tenendo 
anche conto dei risvolti che l'introduzione dei diversi criteri comporta    
nelle successive fasi di analisi, allorch\'e si vogliono misurare quantit\`a 
quali le distribuzioni di impulso trasverso o il tasso di produzione di una 
data specie di particelle. Tali criteri sono dunque stati finalizzati 
differentemente per i $K^0_S$\ e per le $\Lambda$-$\bar{\Lambda}$; per 
queste ultime \`e preferibile adottare gli stessi criteri di selezione, in 
modo tale da controllare la presenza di eventuali errori sistematici 
considerando la completa simmetria dei due decadimenti quando il campo 
magnetico viene invertito.  
\subsection{Selezioni generali}  
Si considerano in questo paragrafo i criteri di selezione comuni alle 
diverse specie, finalizzati alle interazioni Pb-Pb a 160 A GeV/$c$ ed 
idonei ad eliminare la contaminazione geometrica e 
parte di quella cinematica dovuta alla presenza dei $\gamma$\ che convertono in 
coppie \Pep-\Pem. La selezione finale \`e il risultato di un processo 
iterativo che, partendo da tagli approssimati su una o pi\`u variabili, li 
affina via via alla luce dell'effetto combinato su altre variabili. 
\newline
Non verr\`a discusso in dettaglio tutto l'iter delle successive approssimazioni, 
ma  a scopo illustrativo si mostreranno alcune  
distribuzioni ottenute applicando alle candidate $V^0$\ 
tutti i criteri finali di selezione, ad esclusione di quello che si verr\`a 
considerando di volta in volta e di quello sul valore della massa invariante 
della specie considerata.  
\begin{itemize}
\item {\bf Volume fiduciale per il decadimento: $x_{V^0}$} \\
In fig.~\ref{XV0fig} sono mostrate le distribuzioni 
della posizione del vertice 
di decadimento della $V^0$\ lungo l'asse $x$\ del sistema di riferimento 
del magnete per un campione di candidate \PgL+\PagL\ (fig.~\ref{XV0fig}.a) 
ed uno di candidate \PKzS\ (fig.~\ref{XV0fig}.b).   
\begin{figure}[hbt]
\begin{center}
\includegraphics[scale=0.45]{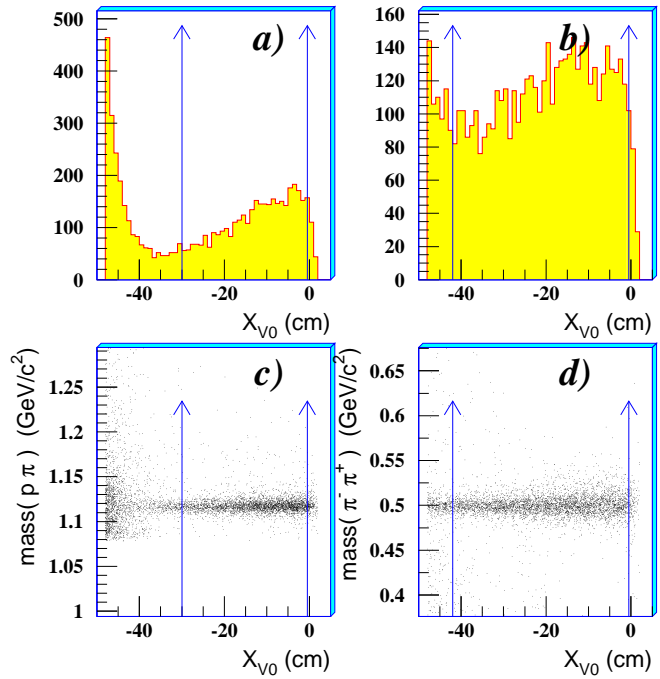}
\caption{{\em In alto}: 
         distribuzioni della posizione $x_{V^0}$\ del vertice di 
	 decadimento delle $V^0$\ per le $\Lambda$\ ed $\bar{\Lambda}$ ({\bf a})  
	 e per i $K^0_S$\ ({\bf b}).  
	 {\em In basso}: 
	 correlazioni della variabile $x_{V^0}$ (in ascissa),  
	 relativa ai grafici soprastanti, con le masse invarianti (in ordinata)  
	 $M(p,\pi)$\ ({\bf c}) ed $M(\pi^+,\pi^-)$\ ({\bf d}).}
\label{XV0fig}
\end{center}
\end{figure}
Questi campioni sono stati ottenuti applicando tutti gli altri criteri 
di selezione, descritti in seguito, per isolare i segnali di $\Lambda$\ e 
\PKzS, tranne quello della posizione del vertice di decadimento, che si vuole 
studiare, e quello sull'intervallo finale di massa invariante ammesso 
per la specie considerata.  
La scelta del taglio \`e stata affinata osservando le correlazioni tra la 
posizione del vertice lungo l'asse $x$\ e la massa invariante $M(p,\pi)$\ od 
$M(\pi^+,\pi^-)$, a seconda che si studino le \PgL+\PagL\ od i \PKzS, 
mostrate rispettivamente nella fig.~\ref{XV0fig}.c e nella fig.~\ref{XV0fig}.d.    
Nel sistema di ascisse in fig.~\ref{XV0fig}, il centro del primo piano 
di pixel nel telescopio \`e posto ad $x\simeq 0.5$ cm, mentre il bersaglio 
si trova ad $x \simeq -60$\ cm. Le frecce indicano il criterio di selezione 
operato sulla variabile $x_{V^0}$: le candidate \PgL\ o \PagL\ sono accettate 
se decadono nell'intervallo $[-30,-0.5]$\ cm, mentre per le candidate \PKzS\ 
l'intervallo fiduciale \`e  $[-42,-0.5]$\ cm. Il limite superiore viene 
imposto per eliminare la contaminazione residua di false $V^0$\ create 
utilizzando tracce prodotte nell'interazione col materiale del primo piano di 
rivelatori, contaminazione peraltro gi\`a ampiamente ridotta dall'azione  
degli altri criteri di selezione. Il valore del limite inferiore viene scelto 
con l'obiettivo di eliminare i vertici di decadimento posti troppo in 
prossimit\`a del bersaglio, l\`i dove diventa elevata la concentrazione di coppie di 
tracce primarie, che solo apparentemente si incrociano al di fuori di esso a causa
degli errori di misura, e non provengono dunque dal decadimento 
di una $V^0$.  
La necessit\`a di richiedere un volume fiduciale pi\`u esteso   
(nella direzione verso il bersaglio) nel caso  
dei \PKzS\ discende in parte dalla loro minor vita media ($c\tau=2.68$\ cm)  
rispetto a quella delle \PgL\ ($c\tau=7.89$\ cm),  
e quindi dallo loro maggior probabilit\`a di decadere pi\`u vicino al bersaglio. 
Inoltre, la forma della distribuzione di $x_{V^0}$\ viene  
modulata dall'accettanza del telescopio, in maniera differente 
per le diverse specie di particelle, e non \`e quindi direttamente 
riconducibile alla distribuzione temporale, di tipo esponenziale, del 
decadimento della ${V^0}$.  
\item {\bf Minima distanza  tra le tracce di decadimento: $close_{V^0}$} \\
In fig.~\ref{CloseV0fig} sono mostrate le distribuzione della minima distanza tra 
le due tracce che formano la $V^0$\ per i campioni di candidate 
\PgL+\PagL\ e di candidate \PKzS. Anche in questo caso i campioni sono stati 
ottenuti applicando tutti gli altri tagli tranne quello sull'intervallo di 
massa invariante ammesso per la specie considerata.  
Nei grafici in basso della stessa 
fig.~\ref{CloseV0fig} sono anche mostrate le correlazione tra tale variabile 
e la massa invariante della specie che si vuole selezionare.  
Il criterio di selezione indicato dalle 
frecce \`e stato scelto in modo da eliminare le $V^0$\ con tracce 
di decadimento aventi minima distanza maggiore di $0.030$\ cm per le \PgL\ ed \PagL\ 
e di $0.035$\ cm per i \PKzS. Ci\`o riduce la contaminazione dovuta a false 
$V^0$\ generate dall'associazione di due tracce non correlate.  
\begin{figure}[htb]
\begin{center}
\includegraphics[scale=0.45]{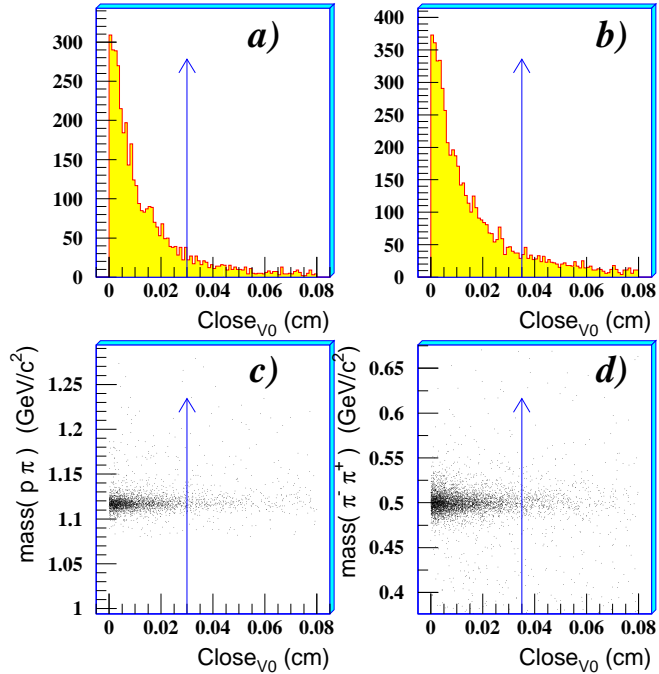}
\caption{{\em In alto}: distribuzioni della minima distanza, $close_{V^0}$, 
         tra le tracce della $V^0$\ per un campione di $\Lambda$\ ed
         $\bar{\Lambda}$ selezionate ({\bf a}), e  per un campione di $K^0_S$\ ({\bf b}).
         {\em In basso}:  
	 correlazioni della variabile $close_{V^0}$\ (in ascissa), relative ai grafici 
	 soprastanti,  
	 con le masse invarianti (in ordinata) $M(\pi^+,\pi^-)$\ ({\bf c}) e  
	 $M(\pi^+,\pi^-)$\  ({\bf d}).} 
\label{CloseV0fig}
\end{center}
\end{figure}
Come gi\`a discusso nel paragrafo precedente, si accettano unicamente i 
decadimenti con topologia di tipo {\em cowboy}. 
%Tecnicamente si definisce il vertice primario 
\item {\bf Parametro d'impatto della $V^0$: ${b_y}_{V^0}$, ${b_z}_{V^0}$} \\
Il parametro d'impatto della $V^0$\ si definisce, nel piano $\Sigma$\ normale 
all'asse del fascio (cio\`e parallelo al piano $yz$\ del riferimento nel 
laboratorio) e contenente il bersaglio, come il vettore congiungente   
il vertice primario dell'interazione con il punto d'intersezione tra la linea 
di volo della $V^0$\ ed il piano $\Sigma$. 
\newline
Nelle interazioni Pb-Pb, caratterizzate da elevate molteplicit\`a di tracce 
cariche entro il telescopio, \`e quasi sempre possibile 
(oltre il $95\%$\ dei casi)   
determinare la posizione del vertice primario dell'interazione Pb-Pb 
con buona precisione (qualche centinaia di micron). 
%Quest'ultimo 
Esso   
(vertice {\em ``evento per evento''})  
viene determinato con la seguente procedura iterativa. 
Si considerano inizialmente tutte le tracce cariche e le loro  
intersezione col piano $\Sigma$;  
%(parallelo al piano $yz$\ del riferimento del laboratorio); entro $\Sigma$, 
il vertice 
viene definito come il ``baricentro'' di queste intersezioni. 
%All'iterazione successiva, 
Vengono quindi eliminate quelle tracce la cui distanza nel piano $\Sigma$\ 
dal vertice precedentemente determinato ecceda un dato limite e si 
determina il ``baricentro'' delle intersezioni rimanenti; e cos\`i via. 
In tal modo si eliminano le tracce non provenienti dal vertice primario, ma generate 
ad esempio dal decadimento di una $V^0$. La procedura termina quando la posizione 
del vertice diventa stabile rispetto ad un'ulteriore iterazione.  
\newline 
Tuttavia, nelle interazioni
pi\`u periferiche, caratterizzate da una minor molteplicit\`a di particelle
cariche, la determinazione del vertice primario in un dato evento
\`e meno efficiente e precisa. Si preferisce pertanto definire differentemente,
uno per ciascun {\em run}~\footnote{Si ricorda, come discusso nel par.~2.5.2, che
un {\em run} \`e un insieme di eventi temporalmente contigui, tipicamente
pari a $2 \, \times \, 10^5$\ interazioni Pb-Pb.},
il vertice primario dell'interazione (vertice {\em ``run per run''})
utilizzato per calcolare i parametri d'impatto delle particelle prodotte.
All'interno di ciascun {\em run} si considerano tutti gli eventi in cui si
riesce a ricostruire il vertice primario {\em evento per evento} 
dell'interazione Pb-Pb.  
Il vertice {\em run per run} \`e quindi definito come il valore medio della 
distribuzione dei vertici {\em evento per evento} all'interno di un dato 
{\em run} e viene usato per tutti gli eventi di quel {\em run}.  
\newline
Il vertice {\em run per run} cos\`i determinato \`e quello che viene adoperato 
in tutte le interazioni studiate da NA57, indipendentemente 
dalla centralit\`a della collisione.
\newline
Considerando le proiezione sull'asse $y$\ e sull'asse $z$\ 
di queste distribuzioni, oltre alle quantit\`a $y_{run}$\ e $z_{run}$\ che 
forniscono le coordinata $y$\ e $z$\ del vertice 
{\em ``run per run''}~\footnote{La coordinata $x$\ del vertice dell'interazione 
viene assunta coincidente con la posizione nominale del bersaglio, il cui 
spessore \`e  molto minore rispetto alla precisione con cui sarebbe possibile 
determinare la $x$\ del vertice a partire da un estrapolazione all'indietro delle
tracce delle particelle prodotte.}, \`e possibile definire la precisione  
sulla determinazione di tali coordinate, $\sigma_{y_{run}}$\ e 
$\sigma_{z_{run}}$, come 
%lo scarto quadratico medio di tali proiezioni.  
la sigma della gaussiana di {\em ``best fit''} a tali proiezioni.  
Con tale definizione, risulta tipicamente $\sigma_{y_{run}} \simeq 350 \mu$m  
e $\sigma_{z_{run}} \simeq 600 \mu$m.  
\newline
La possibilit\`a di definire con buona precisione, su base {\em run per run},  
la posizione $(y_{run},z_{run})$\ del vertice d'interazione \`e conseguenza 
della notevole stabilit\`a spaziale del fascio di Pb e delle 
sua ottima focalizzazione sul bersaglio.  Ci\`o permette di analizzare le 
collisioni pi\`u periferiche con gli stessi strumenti di analisi usati per 
le collisioni pi\`u centrali, senza introdurre la possibilit\`a di errori 
sistematici differenti per le due classi.   
\newline
Nelle figg.~\ref{ImpactV0fig}.a, \ref{ImpactV0fig}.b sono mostrate le 
distribuzioni della componente $y$\ del parametro d'impatto della $V^0$\ per un 
campione di candidate $\Lambda$\ ed \PagL\ ({\bf a}) e per un campione 
di candidate \PKzS\ ({\bf b}), normalizzate alla precisione sulla determinazione 
del vertice {\em run per run} $\sigma_{y_{run}}$. La correlazione di tale variabile 
${b_y}_{V^0} / \sigma_{y_{run}}$\ con  
la massa invariante $M(p,\pi)$\ e  $M(\pi^+,\pi^-)$\ \`e riportata, 
rispettivamente, nella fig.~\ref{ImpactV0fig}.c e nella fig.~\ref{ImpactV0fig}.d.  
Gran parte del fondo geometrico viene eliminato richiedendo che la $V^0$\ 
provenga dal vertice d'interazione, entro qualche unit\`a di 
$\sigma_{y_{run}}$. 
Negli inserti {\em e}), {\em f}), {\em g}) e {\em h}) della fig.~\ref{ImpactV0fig} 
sono mostrate le analoghe distribuzioni per la proiezione $z$\ del 
parametro d'impatto delle $V^0$.  
\begin{figure}[p]
\begin{center}
\includegraphics[scale=0.45]{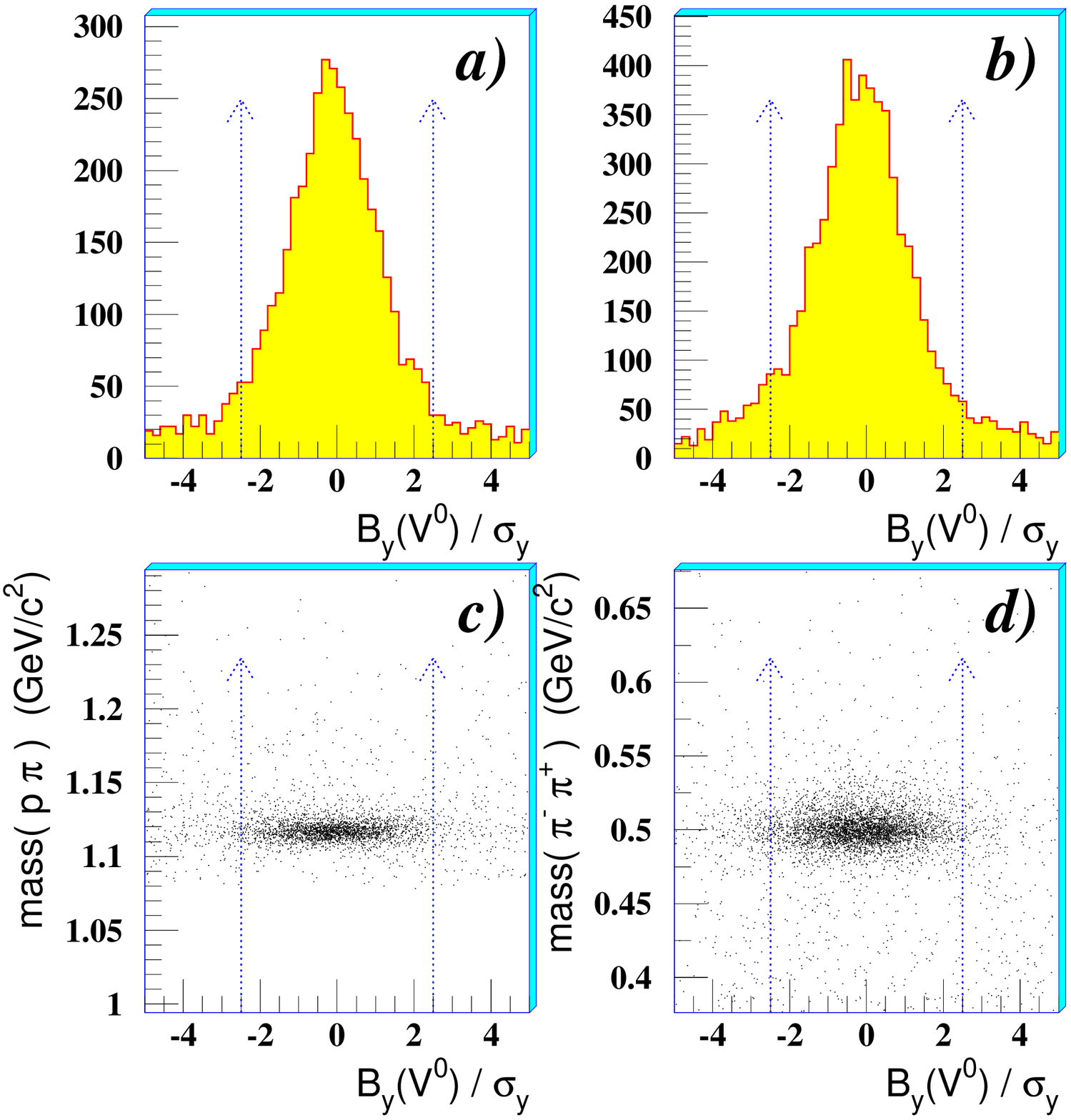}
\includegraphics[scale=0.45]{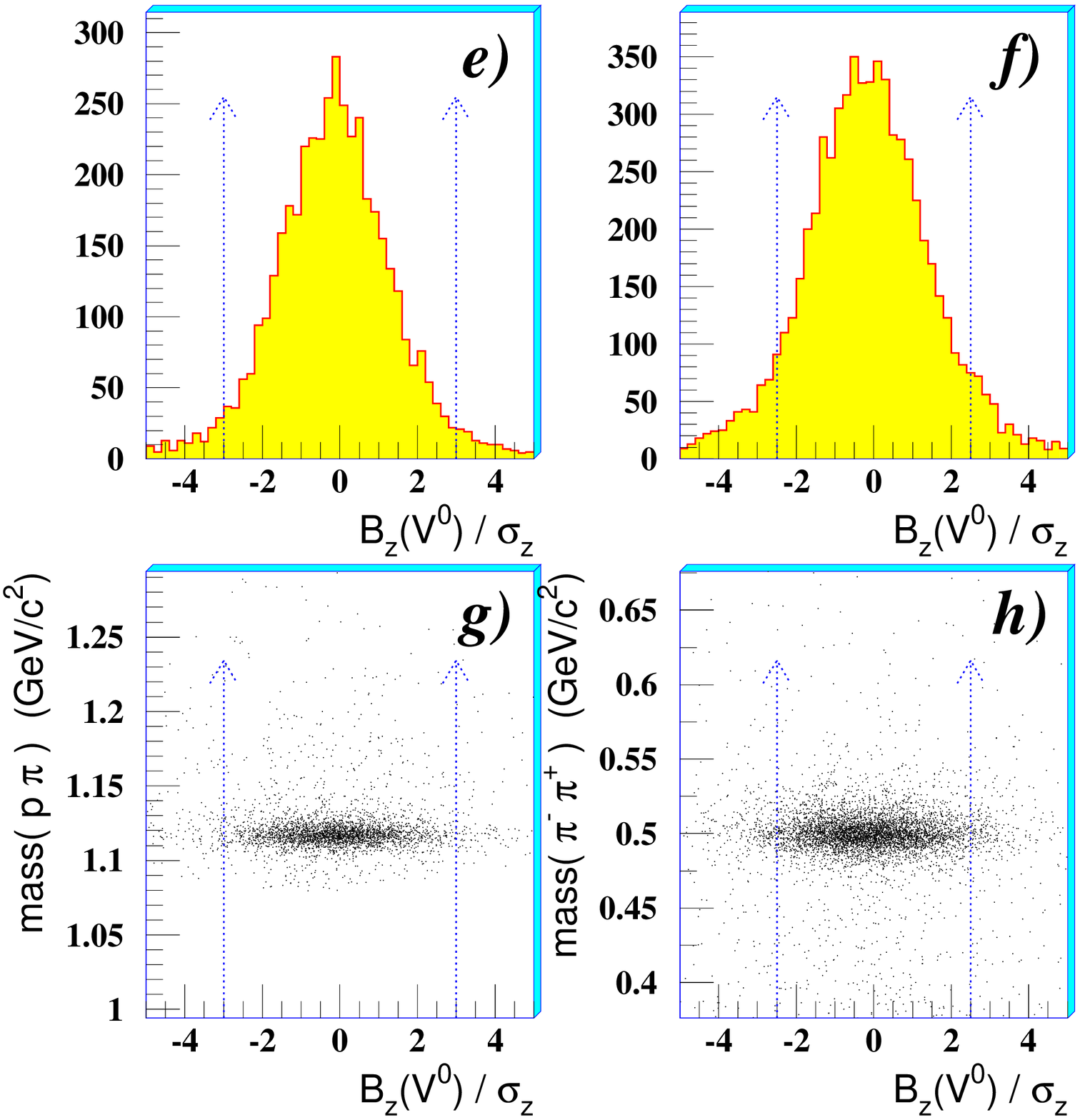}
\caption{{\em Nei quattro grafici in alto}: 
         Distribuzioni della proiezione lungo l'asse $y$\ del parametro
         d'impatto della $V^0$\  per un campione di candidate
         $\Lambda$-$\bar{\Lambda}$ ({\bf a}) e sua 
         correlazione con la massa invariante $M(p,\pi)$\ ({\bf c});    
	 distribuzione della proiezione $y$\ del parametro 
	 d'impatto della $V^0$\ per un campione di candidate \PKzS\ 
	 ({\bf b}) e sua   
	 correlazione con la massa invariante $M(\pi^+,\pi^-)$\  ({\bf d}). 
	 {\em Nei quattro grafici in basso}: 
	 analoghe distribuzioni per la proiezione $z$\ del parametro d'impatto.} 
\label{ImpactV0fig}
\end{center}
\end{figure}
\newline
Dallo studio del rapporto segnale su fondo, risulta che la scelta pi\`u  
conveniente per il criterio di selezione sul parametro d'impatto richiede un 
taglio correlato sulle due proiezioni ${b_z}_{V^0}$\ e ${b_y}_{V^0}$ . 
La correlazione tra le due variabili normalizzate alle rispettive 
$\sigma_{run}$\ \`e mostrata nella fig.~\ref{ImpactV0fig2} per il campione 
di \PgL\ ed \PagL ({\bf a}) e di \PKzS ({\bf b}).  
\begin{figure}[bt]
\begin{center}
\includegraphics[scale=0.45]{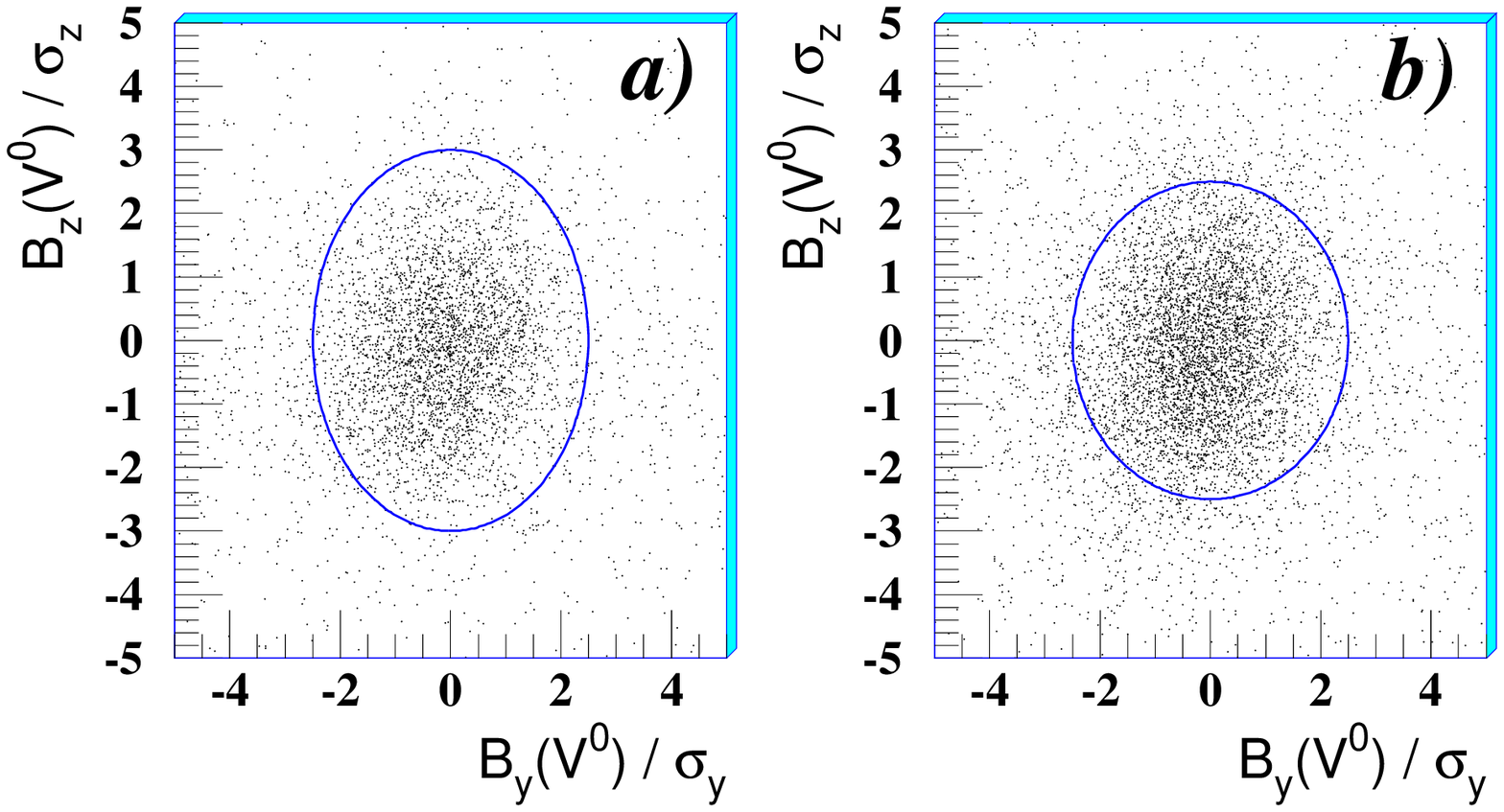}
\caption{Distribuzioni bidimensionali delle proiezioni del parametro 
         d'impatto ${b_z}_{V^0}$, ${b_y}_{V^0}$\ per un campione di 
	 candidate $\Lambda$\ ed 
	 $\bar{\Lambda}$ ({\bf a}), e di \PKzS ({\bf b}).} 
\label{ImpactV0fig2}
\end{center}
\end{figure}
La scelta finale sul criterio di selezione \`e dunque la seguente: 
per \PgL\ ed \PagL\, la $V^0$\ deve possedere un parametro d'impatto 
compreso entro un'ellisse di semiassi pari a $2.5 \sigma_{y_{run}}$\ 
e $3 \sigma_{z_{run}}$; nel caso dei \PKzS\ si \`e definita un'ellisse 
limite con semiassi pari a  $2.5 \sigma_{y_{run}}$\ e  $2.5 \sigma_{z_{run}}$. 
Le ellissi cos\`i definite sono riportate in blu nella fig.~\ref{ImpactV0fig2}.  
Le frecce in blu della fig.~\ref{ImpactV0fig} forniscono invece il limite massimo 
ammesso per una delle due componenti quando l'altra assume valore nullo. 
\newline
Come si discuter\`a nel prossimo capitolo, 
nell'operare questa scelta si \`e tenuto conto del fatto che nella procedura 
di correzione per accettanza ed efficienza \`e necessario introdurre 
la distribuzione dei parametri d'impatto delle $V^0$\ nelle simulazioni; la 
forma di queste distribuzioni non \`e nota esattamente a priori, ed inoltre, 
a causa degli errori di estrapolazione, c'\`e la possibilit\`a che 
vengano introdotte delle distorsioni in queste distribuzioni.  
Una taglio troppo severo, dell'ordine di $1 \div 2 \; \sigma_{run}$, potrebbe 
quindi introdurre degli errori sistematici nella procedura di correzione 
per efficienza ed accettanza. A riguardo di ci\`o, si discuter\`a nel prossimo 
capitolo l'entit\`a degli effetti introdotti dalla scelta qui operata  
per il taglio sul parametro d'impatto, nel caso delle \PgL, \PagL\ e \PKzS.  
\item {\bf Componente $y$\ del parametro d'impatto dei prodotti di 
           decadimento: ${b_y}^{neg},{b_y}^{pos}$} \\
Come si \`e definito il parametro d'impatto della $V^0$, cos\`i si pu\`o considerare il 
prolungamento all'indietro, verso il bersaglio, delle tracce dei prodotti di decadimento di 
una $V^0$. In tal caso si deve %per\`o 
inseguire nel campo magnetico la traiettoria delle  
particelle, sino al piano $\Sigma$\ perpendicolare all'asse $x$\ e contenente 
il bersaglio;  
anche in tal caso conviene riferire il parametro d'impatto al vertice {\em run per run}  
dell'interazione.  
\newline
Prima dell'applicazione degli altri tagli geometrici, una cospicua frazione del fondo delle 
false $V^0$, costituite da tracce provenienti dal vertice primario, presenta valori 
del parametro d'impatto delle tracce di decadimento concentrati attorno allo zero.  
L'applicazione degli altri tagli geometrici rimuove quasi completamente tale frazione;  
tuttavia, nel caso dei \PKzS, \`e possibile rimuovere una piccola contaminazione residua 
di fondo geometrico operando su queste variabili.  
In fig.~\ref{DecayImpacts} sono mostrate le distribuzioni della componente $y$\ del parametro 
d'impatto della particella di carica positiva (fig.~\ref{DecayImpacts}.a) e di quella 
di carica negativa (fig.~\ref{DecayImpacts}.b) provenienti dal decadimento di un campione 
di candidate $V^0$, ottenute applicando tutti i restanti criteri di selezione finalizzati 
alla ricostruzione dei \PKzS.  
\begin{figure}[h]
\begin{center}
\includegraphics[scale=0.45]{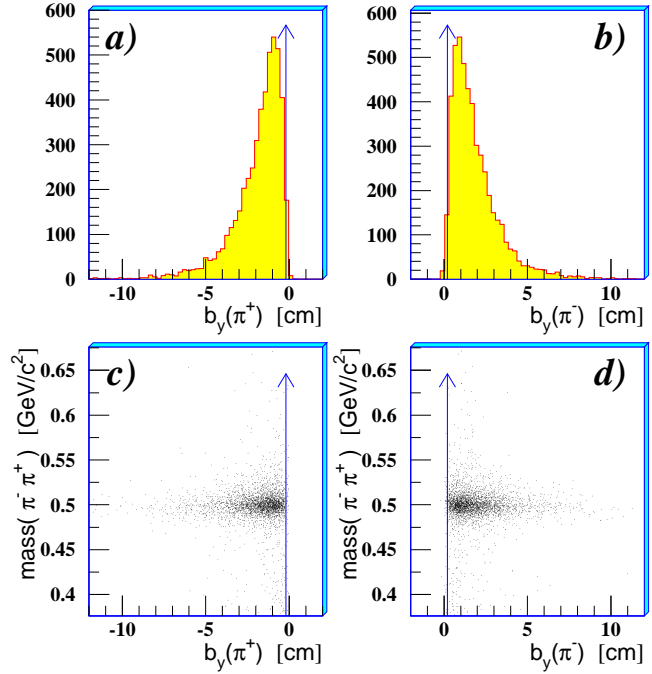}
\caption{{\em In alto}: distribuzioni della componente $y$\ del parametro d'impatto della 
         particella di decadimento di carica positiva ({\bf a}) e di 
	 carica negativa ({\bf b}), per un campione di candidate $V^0$\ selezionate 
	 con i tagli dei \PKzS. 
	 {\em In basso}: correlazioni con la massa invariante $M(\pi^+,\pi^-)$ 
	                (in ordinata).}
\label{DecayImpacts}
\end{center}
\end{figure}
Nelle figg.~\ref{DecayImpacts}.c e \ref{DecayImpacts}.d sono mostrate le 
correlazioni delle due variabili ${b_y}^{neg}$\ e ${b_y}^{pos}$\  
con la massa invariante $M(\pi^+,\pi^-)$. Il taglio impostato su queste variabili,  
indicato dalle frecce nella fig.~\ref{DecayImpacts}, elimina quelle 
candidate $V^0$\  le cui tracce di decadimento presentino un parametro 
d'impatto superiore a $-0.2$\ cm, od 
inferiore a $0.2$\ cm, a seconda che si tratti della particella di carica negativa 
o di quella di carica positiva e tenendo opportunamente conto dell'orientazione, 
rispetto all'asse $z$, del campo magnetico.  
\newline
Nel caso della selezione finalizzata ad isolare i segnali di \PgL\ e \PagL, 
invece, n\`e l'applicazione di un simile taglio, n\`e quello 
sulla componente $z$\ del parametro d'impatto dei prodotti di 
decadimento, risulta di giovamento e quindi non \`e stato applicato.   
\item {\bf Impulso trasverso delle tracce di decadimento: $q_T$} \\
Nella fig.~\ref{qtV0fig} sono riportate le distribuzioni del momento trasverso 
$q_T$, valutato rispetto alla linea di volo della $V^0$, 
per il campione di \PgL\ ed \PagL\ (fig.~\ref{qtV0fig}.a) e di \PKzS\ 
(fig.~\ref{qtV0fig}.b) ottenuti al solito applicando tutti gli altri tagli, ad 
esclusione di quello sulla massa nominale della $V^0$.   
\begin{figure}[h]
\begin{center}
\includegraphics[scale=0.45]{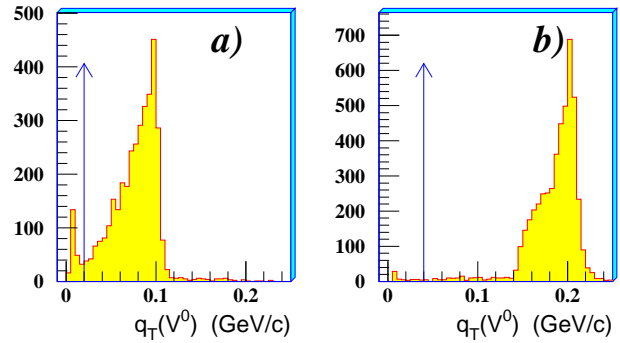}
\caption{Distribuzione del momento trasverso $q_T$\ per un campione di 
candidate $\Lambda$\ ed $\bar{\Lambda}$ (a sinistra), e di \PKzS (a destra).}
\label{qtV0fig}
\end{center}
\end{figure}
I picchi a bassi valori di $q_T$\ sono dovuti prevalentemente alla 
contaminazione di $\gamma$\ che convertono in coppie \Pep-\Pem\ nell'aria 
o nel materiale del primo piano di pixel. Tale contaminazione, di carattere 
cinematico, \`e lievemente pi\`u problematica per le \PgL\ ed \PagL, in quanto 
la distribuzione del momento trasverso $q_T$\ per i decadimenti di queste 
particelle si estende sino a poche decine di MeV/$c$; nel caso dei \PKzS, 
invece, la distribuzione ad essi relativi \`e nettamente separata dal picco, 
peraltro pi\`u ridotto per l'azione degli altri tagli.  
La distribuzione attesa, calcolabile nell'ipotesi di isotropia del 
decadimento nel sistema centro di massa, \`e infatti la seguente:
\begin{equation}
\frac{{\rm d}N}{{\rm d}q_T}=\frac{q_T}{2q_T^{MAX}\sqrt{(q_T^{MAX})^2-(q_T)^2}}. 
\label{qtDistr}
\end{equation}
Essa ha un massimo in corrispondenza di $q_T^{MAX}$\ (picchi principali in 
fig.~\ref{qtV0fig}), ma si riduce a zero per bassi impulsi trasversi, l\`i 
dove \`e stato operato il taglio. Nel caso delle \PgL\ ed \PagL\ si \`e scelto 
dunque il valore di 20 MeV/$c$, mentre per i \PKzS\ si pu\`o assumere un taglio 
pi\`u conservativo a 40 MeV/$c$, senza perdere alcun segnale fisico.  
\item {\bf Angolo interno azimuthale del decadimento: $\phi_{V^0}$} \\
Si consideri il sistema di riferimento $x'y'z'$\ ottenuto ruotando il 
sistema del laboratorio $xyz$\ in modo tale che l'asse $x'$\ sia diretto 
lungo la direzione di volo della $V^0$, l'asse $y'$\ giaccia ancora nel 
piano $xy$\ di curvatura delle tracce cariche,  
e $z'$\ in modo tale da formare una terna destrorsa con $x'$\ ed $y'$.  
\`E possibile definire due angoli interni di decadimento nel sistema 
$x'y'z'$\ appena specificato: 
\begin{enumerate}
\item
l'angolo azimuthale $\phi$ formato  dall'impulso trasverso 
della particella di decadimento di carica 
negativa~\footnote{l'angolo formato dalla particella 
di carica positiva \`e supplementare all'angolo $\phi$ qui definito.} 
e l'asse $y'$\ 
(cio\`e, nel caso della $\Lambda$, l'angolo tra la proiezione dell'impulso 
 $\vec{p}_{\pi^-}$\ del $\pi^-$\ nel piano $y'z'$\ e l'asse $y'$, 
 come schematizzato in fig.~\ref{phiDef_best}); 
\begin{figure}[h]
\begin{center}
\includegraphics[scale=0.30]{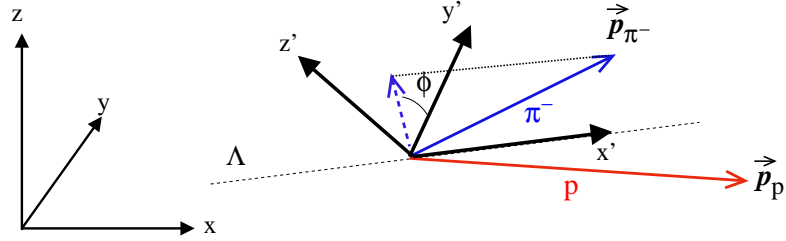}
\caption{Definizione dell'angolo interno $\phi$\ nel caso del decadimento di 
una $\Lambda$.}
\label{phiDef_best}
\end{center}
\end{figure}
\item l'angolo $\theta^*$\ compreso tra l'impulso della particella 
      di carica negativa, valutato nel sistema di riferimento ottenuto 
      applicando a $x'y'z'$ un {\em ``boost''} di Lorentz lungo $x'$\ in modo 
      tale da porsi nel sistema a riposo della $V^0$, 
      e la direzione di volo della $V^0$\ 
      (direzione dell'asse $x'$)~\footnote{In tale sistema, l'angolo $\theta^*$\ 
      ottenuto considerando la particella di carica positiva \`e supplementare a 
      quello relativo alla particella di carica negativa che, convenzionalmente, 
      viene preso come riferimento.}.  
      \`E opportuno tener presente che per porsi nel sistema a riposo della $V^0$, 
      cio\`e per effettuare il {\em ``boost''} di Lorentz, bisogna assegnarle una massa; per 
      una stessa $V^0$\ \`e quindi possibile definire due angoli: 
      $\theta^*_{K^0_s}$ nell'ipotesi che la $V^0$\ sia un $K^0_s$\, oppure 
      $\theta^*_{\Lambda}$ nell'ipotesi che la $V^0$\ sia una $\Lambda$\ od una 
       $\bar{\Lambda}$. L'angolo $\phi$\ \`e invece indipendente dal fattore 
       $\gamma$\ di Lorentz associato al {\em ``boost''} e verr\`a indicato come 
       $\phi_{V^0}$\ per distinguerlo, quando si considereranno i decadimenti 
       delle cascate ($\Xi$\ ed $\Omega$), dagli analoghi angoli interni di 
       decadimento definibili per le cascate.  
\end{enumerate}
In fig.~\ref{PhiV0fig} \`e mostrata la distribuzione dell'angolo $\phi$\ 
%riferito alla particella di decadimento di carica negativa 
per un campione di candidate 
\PgL\ ed \PagL, per le due possibili orientazioni del campo magnetico.  
Poich\'e si accettano solo i decadimenti con configurazione di tipo {\em cowboy}, 
l'angolo $\phi$\ pu\`o assumere i valori compresi nell'intervallo  $[-90^o,90^o]$\ 
quando il campo \`e diretto verso l'alto (parallelo a $z$), 
o negli intervalli esplementari $[-180^o,-90^o]$, $[90^o,180^o]$ 
quando il campo magnetico \`e diretto verso il basso (antiparallelo a $z$). 
\begin{figure}[h]
\begin{center}
\includegraphics[scale=0.45]{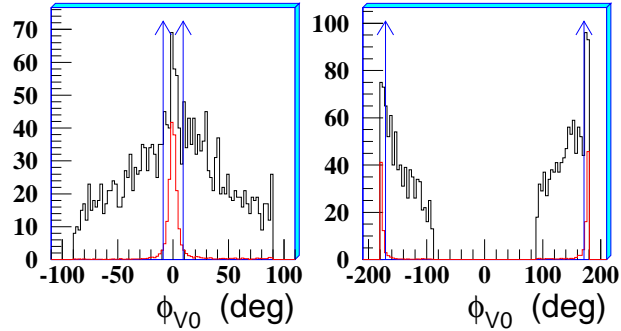}
\caption{Distribuzione dell'angolo interno di decadimento $\phi$\ per un 
         campione di $\Lambda$\ ed $\bar{\Lambda}$\ 
	 ottenuto applicando gli altri criteri di selezione (ad esclusione 
	 di quello sulla massa invariante $M(p.\pi)$). L'istogramma in rosso 
	 rappresenta il fondo geometrico valutato con la tecnica degli 
	 eventi mescolati (si veda il {\em paragrafo 4.2}). 
	 A sinistra \`e mostrata la distribuzione ottenuta   
	 con campo magnetico rivolto verso l'alto, a destra quella con campo 
	 magnetico invertito.}   
\label{PhiV0fig}
\end{center}
\end{figure}
Piccoli angoli $\phi$\ ($\sim 0^o$) per campo magnetico verso l'alto, od angoli 
$\phi$\ prossimi all'angolo piatto ($\sim \pm180^o$) per campo magnetico 
verso il basso, 
corrispondono al caso in cui le due tracce di decadimento abbiano circa    
lo stesso angolo di {\em dip} $\lambda$\ 
%tra il proprio vettore impulso ed il piano di curvatura $(x,y)$ 
(fig.~\ref{TrkParam} nel {\em paragrafo 2.6}),  
angolo $\lambda$\ dunque anche simile a quello  della $V^0$. 
\newline
Il segnale fisico dei veri decadimenti 
delle $V^0$\ ha un massimo (che fortunatamente non \`e eccessivamente 
pronunciato)  in tale regione a causa della sezione trasversa finita  
del telescopio di NA57: per un decadimento di tipo {\em cowboy} 
completamente contenuto nel piano $x'y'$\ ($\phi=0^o\; {\rm o} \; 180^o$), 
infatti, l'azione del campo magnetico  
\`e quella di contenere le tracce rispetto alla loro direzione di emissione 
($y'$\ appunto); mentre per un decadimento nel piano $x'z'$\  
perpendicolare ad $x'y'$\    
(cio\`e per $\phi \sim \pm 90^o$),  
%per campo magnetico verso l'alto o verso il basso), 
l'azione di contenimento delle tracce del campo magnetico 
\`e quasi nulla, e le tracce escono pi\`u facilmente dal 
telescopio\footnote{Si ricorda che il telescopio \`e inclinato di 
soli 40 mrad rispetto alla direzione $x$\ del fascio; pertanto l'angolo 
tra l'asse $z'$\ e l'asse $z$, lungo cui \`e diretto il campo magnetico, 
\`e anch'esso dell'ordine  
di qualche decina di~ mrad.}.  
\newline
Allo stesso tempo anche il fondo geometrico di false $V^0$, ottenute abbinando tracce   
primarie, si concentra in questa regione sia per la maggior accettanza geometrica, come 
nel caso del decadimento delle vere $V^0$, sia in quanto \`e maggiore l'incertezza sulla  
determinazione del punto di minor distanza tra le due tracce  
(punto di apparente intersezione), 
%che viene ora determinato inseguendo le tracce nel piano $x'y'$, 
%in cui le tracce vengono maggiormente curvate dall'effetto 
%del campo magnetico,  
perch\'e esse sono quasi coincidenti nella proiezione $xz$, 
ed \`e dunque pi\`u probabile per esse superare il taglio
sul volume fiduciale di decadimento, a causa di errori di estrapolazione.
\newline
La determinazione del miglior taglio sulla variabile $\phi$\   
per discriminare il segnale fisico dal fondo geometrico non \`e pertanto 
delle pi\`u agevoli.  Un notevole aiuto a riguardo si ottiene considerando la 
distribuzione del solo fondo geometrico, ottenuta sviluppando una tecnica 
di {\em mescolamento degli eventi} che verr\`a discussa in dettaglio nel 
prossimo capitolo. Tale distribuzione \`e riportata in rosso nella 
fig.~\ref{PhiV0fig}. Il criterio di selezione adottato, sia per le \PgL\ e le 
\PagL\ che per i \PKzS, \`e quello di rigettare le candidate $V^0$\ quando le 
particelle di decadimento risultano quasi complanari, entro $\pm 9^o$, 
con il piano $x'y'$, come indicato dalle frecce in blu nella fig.~\ref{PhiV0fig}.  
\end{itemize}
\subsection{Identificazione di $K_S^0$, $\Lambda$\ ed $\bar{\Lambda}$}
\begin{figure}[tb]
\begin{center}
\includegraphics[scale=0.45]{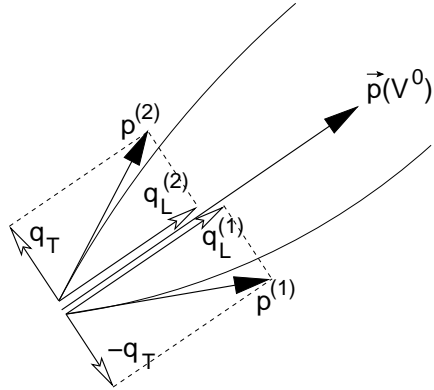}
\caption{Schema del decadimento a due corpi di una $V^0$: 
         si definiscono, rispetto alla linea  di volo della $V^0$, l'impulso 
	 trasverso $q_T$\ delle tracce di decadimento ed i due impulsi longitudinali 
	 $q_L^{(1)}$\ e $q_L^{(2)}$.}
\label{V0Decay2}
\end{center}
\end{figure}

\begin{figure}[htb]
\begin{center}
\includegraphics[scale=0.37]{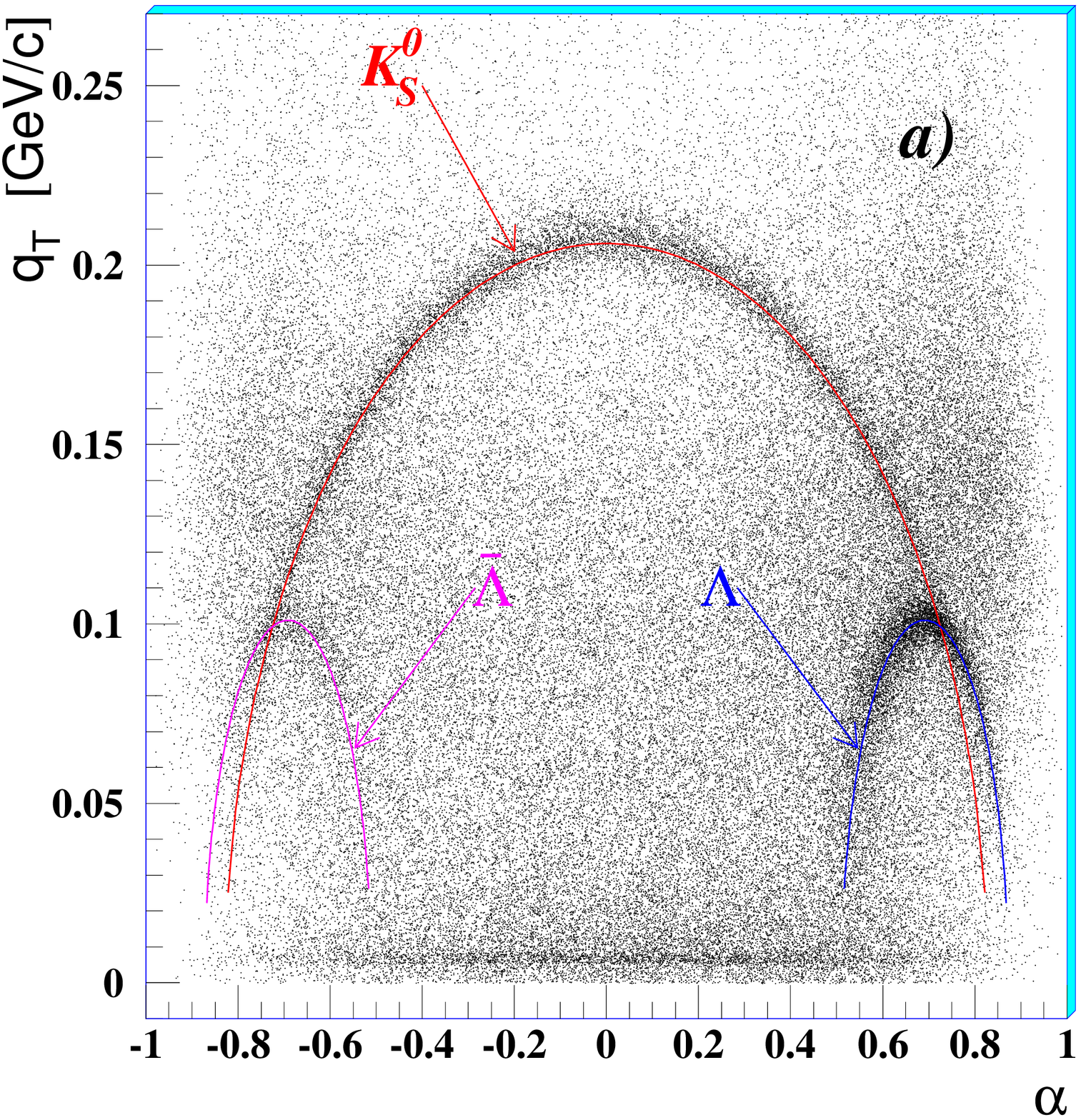}
\includegraphics[scale=0.37]{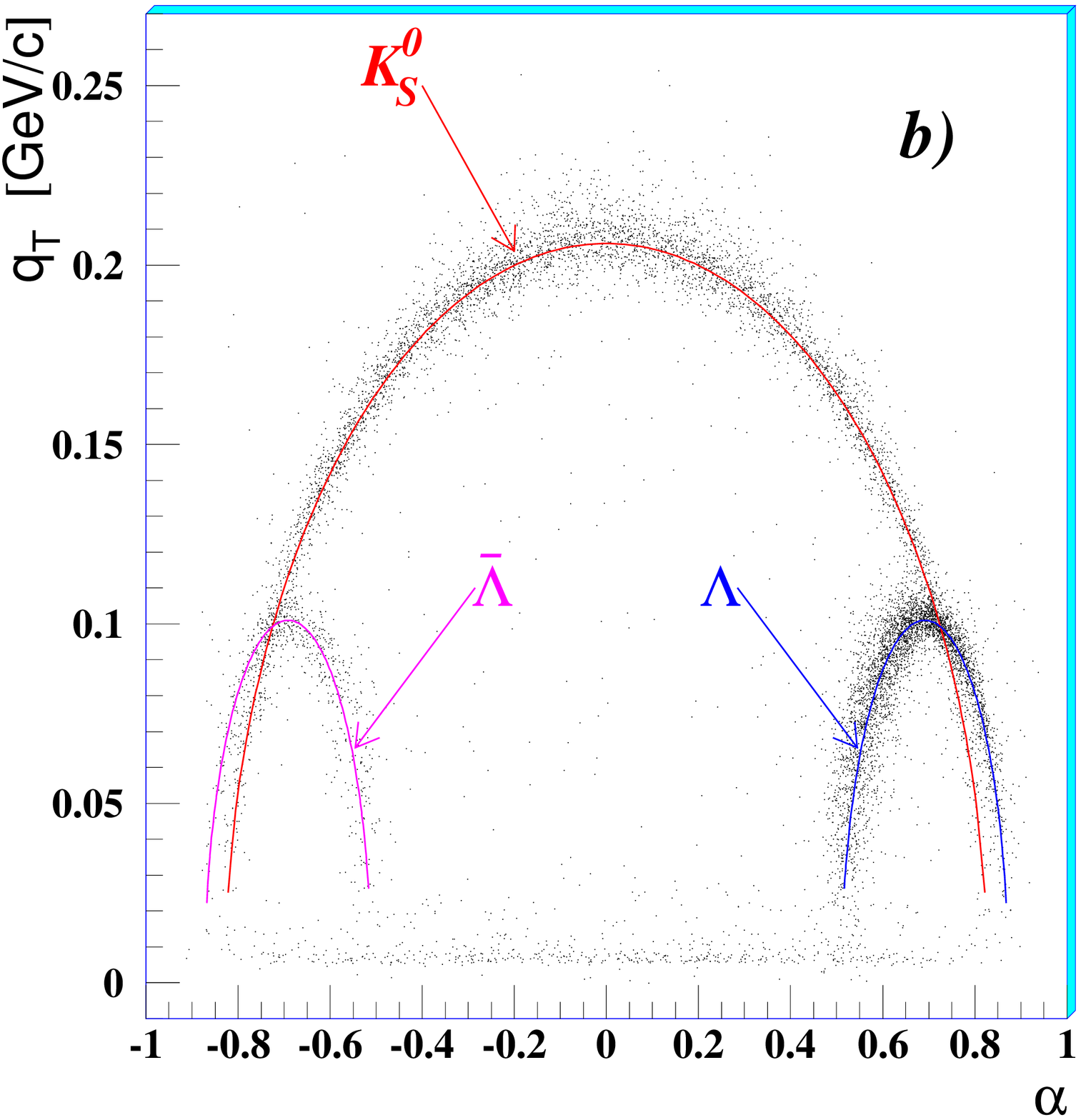}
\caption{Distribuzione dei decadimenti $V^0$\ nel piano $(\alpha,q_T)$\ per 
         un campione di candidate $V^0$\ selezionate da STRIPV0 ({\bf a}) e dopo 
	 l'applicazione dei tagli geometrici ({\bf b}).}
\label{Arment-plot}
\end{center}
\end{figure}

Una volta operati i tagli comuni, la selezione pu\`o essere affinata per 
distinguere i vari tipi di particelle considerando la distribuzione 
dei decadimenti ricostruiti nello spazio delle fasi.   
%\subsubsection{Grafico di Podolanski-Armenteros}
Riferendosi alla fig.~\ref{V0Decay2}, 
le variabili cinematiche pi\`u utili a tal fine sono quelle di 
Armenteros-Podolanski~\cite{Pod54} e precisamente: 
l'impulso trasverso $ q_T $, gi\`a 
definito, e la grandezza $\alpha$,  definita in termini dei 
momenti longitudinali delle particelle di decadimento, rispetto alla 
direzione di volo della $V^0$: 
\begin{equation}
\alpha\,=\, \frac{{q_L}^{(1)}-{q_L}^{(2)}}{{q_L}^{(1)}+{q_L}^{(2)}}
\label{alpha-arment}
\end{equation}
Detto
$<\alpha>=\frac{m_{1}^{2} - m_{2}^{2}}{M_{V^0}^2}$ il suo valore medio,
si pu\`o dimostrare~\cite{Pod54} che, nello spazio delle fasi cos\`i
determinato, i punti corrispondenti ad un dato schema di decadimento
si dispongono lungo la curva di equazione:
\begin{equation}
\left(\frac{q_T}{q_T^{MAX}}\right)^2 +
\left(\frac{\alpha - <\alpha>}{2q_T^{MAX}\sqrt{\frac{1}{M_{V^0}^2}+\frac{1}{P^2}}}
\right)^2=1
\label{Arment-esatta}
\end{equation}
dove $P=q_{L}^{(1)}+q_{L}^{(2)}$\ \`e l'impulso della particella primaria. Per
decadimenti relativistici ($P \gg M_{V^0}$), l'eq.~\ref{Arment-esatta} diventa
l'equazione di un'ellisse nel piano $(\alpha,q_T)$:
\begin{equation}
 \left(\frac{q_T}{q_T^{MAX}}\right)^2 +
 \left(\frac{\alpha - <\alpha>}{\alpha_L}\right)^2=1 \quad \quad
 \text{dove:} \quad \alpha_L=\frac{2q_T^{MAX}}{M_{V^0}}
 \label{Arment-ellis}
\end{equation}
La forma dell'ellisse dipende quindi, per $V^0$\ relativistiche,  
soltanto dalle masse delle particelle coinvolte nel decadimento, 
mentre ciascun punto dell'ellisse corrisponde ad  
un diverso angolo di decadimento nel sistema centro di massa
(dalla relazione $ q_T=q_T^{MAX}\,\sin\theta^*$).
In fig.~\ref{Arment-plot}.a \`e
mostrata la distribuzione, nel piano $(\alpha,q_T)$, di un campione
di candidate $V^0$\ selezionate da STRIPV0 (cio\`e senza le ulteriori selezioni
descritte in questo paragrafo).\\
Si riconoscono agevolmente tre archi di ellisse: i \PKzS, col loro
decadimento simmetrico,
si dispongono lungo l'ellisse centrale con $<\alpha>=0$,
le \PgL\ sono invece concentrate nella regione $ \alpha > 0.45 $,  le \PagL,
meno abbondanti delle \PgL, nella regione $ \alpha < -0.45 $. Si nota anche
l'addensamento a bassi $q_{t}$\ corrispondente alle conversioni dei \Pgg\ in 
coppie \Pep-\Pem, che vengono
eliminati con la condizione sul minimo valore di $q_{t}$\  gi\`a discussa.
\newline
Nella fig.~\ref{Arment-plot}.b \`e mostrata la stessa distribuzione, dopo 
l'applicazione dei tagli geometrici, ottimizzati per la selezione delle 
\PgL\ e \PagL. Per giungere ai segnali fisici finali, bisogna dunque 
operare la selezione delle diverse specie 
operando sui parametri $\alpha$\ e $q_T$.  
%ed applicare il taglio sul minimo valore di impulso trasverso $q_T$.  
\subsubsection{Selezione delle \PgL\ ed \PagL}
\begin{figure}[htb]
\begin{center}
\includegraphics[scale=0.55]{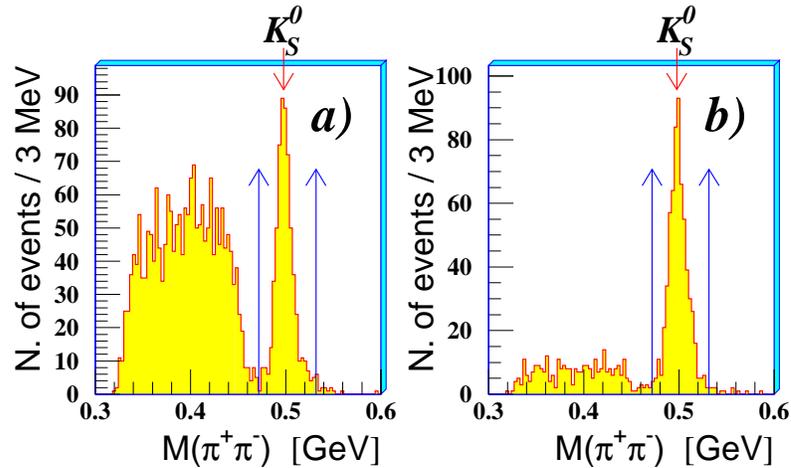}
\caption{Spettri di massa invariante $M(\pi^+,\pi^-)$\ per un campione 
         $V^0$\ cui sono stati applicati i tagli geometrici  
%	 finalizzati alla selezione di \PgL\ e \PagL,  
	 nelle regioni $0.45 < \alpha <0.65 $\ ({\bf a}) e 
	 $-0.65 < \alpha <-0.45 $\ ({\bf b}).}
\label{AntiSelectKaon}
\end{center}
\end{figure}

Si pu\`o restringere la regione cinematica dello spazio delle fasi  
in cui sono concentrate le \PgL\ (\PagL) con la condizione  
$\alpha > 0.45 $\ ($\alpha < -0.45 $). 
Bisogna comunque eliminare la regione di spazio delle fasi in cui vi \`e 
ambiguit\`a tra \PgL\ (\PagL) ed i \PKzS. Rispetto alle variabili $q_T$\ 
ed $\alpha$, tale regione corrisponde all'intersezione delle ellissi 
associate alle \PgL\ (\PagL)  ed ai \PKzS. La miglior condizione per 
eliminare l'ambiguit\`a, massimizzando il rapporto segnale su fondo residuo, 
\`e quella di eliminare le candidate $V^0$\ che abbiano una massa invariante 
$M(\pi^+,\pi^-)$\ vicina alla massa nominale dei \PKzS, entro opportuni 
limiti. Per determinare questi limiti si \`e considerato lo spettro 
di massa invariante $M(\pi^+,\pi^-)$\ nella regione di spazio delle fasi 
$0.45 < |\alpha| <0.65 $, in cui non vi \`e ambiguit\`a con le $\Lambda$\ ed 
\PagL, applicando i rimanenti criteri di selezione ottimizzati per   
le \PgL\ ed \PagL, che qui si vogliono isolare.  
In fig.~\ref{AntiSelectKaon} sono mostrati tali   
spettri relativamente alla regione $0.45 < \alpha <0.65$ 
(fig.~\ref{AntiSelectKaon}.a) ed a quella $-0.65 < \alpha <-0.45$\ 
(fig.~\ref{AntiSelectKaon}.b). Il segnale dei \PKzS, centrato sul 
valore nominale di massa indicato dalle frecce rosse, \`e ora 
ben distinto dal riflesso del segnale delle \PgL\ ed \PagL, distribuito 
su tutto lo spettro, ma concentrato  in questa particolare regione di 
spazio delle fasi ($0.45 < | \alpha |<0.65$) a masse invarianti minori.   
Le frecce blu indicano il criterio   
di selezione adottato per anti-selezionare i \PKzS\ dal campione 
finale di \PgL\ ed \PagL: si richiede che le $V^0$\ abbiano una massa 
invariante $M(\pi^+,\pi^-)$\ superiore di 34 MeV/$c^2$\ 
od inferiore di 26 MeV/$c^2$\ alla massa nominale del \PKzS.  
%Come si vedr\`a considerando il campione finale di \PKzS\ 
%selezionati (e ci\`o \`e vero allo stesso modo per le altre specie di 
%particelle che si considereranno), essi presentano una distribuzione di  
%massa invariante che non corrisponde ad una gaussiana: basti ossservare, 
%per esempio, che gli eventi non hanno tutti la stessa precisione di misura. 
%\`E questo il motivo alla base del taglio sulla massa invariante 
%$M(\pi^+,\pi^-)$\ non simmetrico rispetto al valore nominale di massa, intorno 
%cui, peraltro, \`e centrata la distribuzione sperimentale.   
\subsubsection{Selezione dei \PKzS}  
\begin{figure}[htb]
\begin{center}
\includegraphics[scale=0.55]{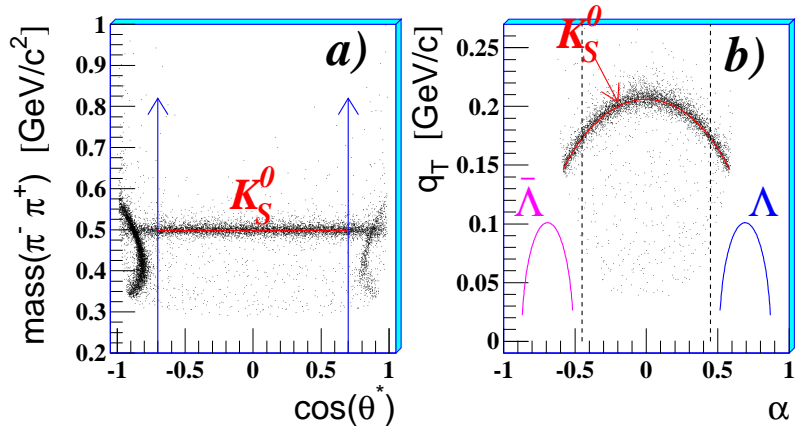}
\caption{{\bf a)} Correlazione tra la massa invariante $M(\pi^+,\pi^-)$\ ed il coseno 
         dell'angolo interno di decadimento $\theta_{\PKzS}^*$\ per un campione 
	 di candidate $V^0$\ cui sono stati applicati i tagli geometrici 
	 finalizzati alla selezione dei \PKzS. 
	 {\bf b)} Effetto del taglio $|\cos \theta_{\PKzS}^*| < 0.7 $\ sul grafico 
	 di Armenteros per lo stesso campione di candidate $V^0$.}  
\label{KaonSelection}
\end{center}
\end{figure}

L'estrazione dei \PKzS\ dal campione di $V^0$\ selezionate coi tagli geometrici 
\`e pi\`u agevole di quella richiesta per le \PgL\ e \PagL, 
in quanto nella gran parte di spazio delle fasi in cui il loro decadimento 
\`e distribuito non vi \`e ambiguit\`a con \PgL\ e \PagL.  
\newline
Un primo criterio, spesso adottato a tal fine, potrebbe essere quello di 
selezionare la regione $ |\alpha| < 0.45$\ in cui non \`e ammesso alcun 
decadimento delle \PgL\ ed \PagL. In tal modo si scarterebbero per\`o 
anche i \PKzS\ concentrati nella regione 
tra $ 0.45 < | \alpha | \apprle 0.65$, 
che sono comunque cinematicamente distinguibili dalle \PgL\ ed \PagL. Per 
recuperare questi ultimi, 
il criterio pi\`u semplice sarebbe quello di operare un taglio 
sulla variabile $q_T$, ad un valore sui 120 MeV/$c$.   
Per\`o questa variabile \`e fortemente correlata con la massa invariante della specie 
considerata (eq.~\ref{V0M}),  ed il taglio su $q_T$\  si tradurebbe quindi in un taglio 
sulla massa invariante $M(\pi^+,\pi^-)$. Un simile criterio di selezione 
presenta dunque il rischio di mascherare la presenza del fondo residuo 
nel segnale dei \PKzS, che quindi non sarebbe agevolmente quantificabile.  
\newline 
Il compromesso tra questi due criteri qui adottato \`e quello di considerare la 
variabile cinematica $\cos \theta_{\PKzS}^*$, dove $\theta_{\PKzS}^*$\ \`e 
l'angolo interno di decadimento formato 
tra l'impulso della particella di decadimento di 
carica negativa e la linea di volo della $V^0$\ 
nel sistema di riferimento in cui il \PKzS\ \`e a riposo, secondo la  
definizione data nel par.3.3.1.  
In fig.~\ref{KaonSelection}.a \`e mostrata la correlazione tra la massa invariante 
$M(\pi^+,\pi^-)$\ e la variabile $\cos \theta_{\PKzS}^*$, dopo aver applicato i 
soli tagli geometrici.  
Rispetto a questa nuova coppia di variabili dello spazio delle fasi, 
le \PgL\ (\PagL), che  si concentrano nella regione 
$\cos \theta_{\PKzS}^* \apprle - 0.75 $ ($\cos \theta_{\PKzS}^* \apprge 0.75 $), 
risultano pi\`u concentrate e quindi pi\`u efficacemente separabili dai 
\PKzS. Il criterio di selezione adottato per la selezione dei 
\PKzS\ \`e quello di accettare le candidate $V^0$\ quando $| \cos \theta_{\PKzS}^*| < 0.7 $\ 
(frecce blu nella fig.~\ref{KaonSelection}.a).  
Questo criterio di selezione permette di accettare anche parte di quei decadimenti di 
\PKzS, che si dispongono nel grafico di Armenteros (fig.~\ref{Arment-plot}) nella parte 
di ellisse corrispondente a $ 0.45 < |\alpha| \apprle 0.65$, e 
che sarebbero invece scartati dalla condizione pi\`u restrittiva 
$|\alpha| < 0.45 $.  
In fig.~\ref{KaonSelection}.b \`e mostrato, nello spazio delle fasi 
$(\alpha,q_T)$, la distribuzione delle $V^0$\ selezionate con questo criterio. 
La  regione in cui \`e cinematicamente permesso  il decadimento delle  \PgL\ 
e \PagL\ \`e ancora completamente rimossa, mentre \`e ora pi\`u esteso l'arco di 
ellisse relativo ai \PKzS\ selezionati, rispetto a quanto possibile col taglio 
$|\alpha|<0.45$\  (linee tratteggiate in fig.~\ref{KaonSelection}.b): il guadagno 
dei \PKzS\ selezionati \`e del 27\%.  

Di seguito sono riassunti i criteri di selezione per le $\Lambda$, 
le $\bar{\Lambda}$\ ed i $K_S^0$\ nella presa dati del 1998 (Pb-Pb a 
160 A GeV/$c$).   
%Quest'ultimi 
Quelli relativi ai \PKzS\   
sono ancora da considerarsi preliminari in quanto si deve   
verificare la loro stabilit\`a rispetto alla procedura di correzione   
per efficienza ed accettanza, procedura  avviata solo per un 
campione molto ridotto di $K_S^0$. 

\begin{center}
{\bf Criteri di selezione per le \PgL\ e \PagL\ in Pb-Pb a 160 A GeV/$c$\ (1998)}
\end{center}
\begin{itemize}
\item[{\bf $\Lambda$.a)}] tutte le tracce del decadimento attraversano la
                      parte compatta del telescopio (primo e nono piano)
\item[{\bf $\Lambda$.b)}] decadimento della $V^0$\ di tipo {\em cowboy}
\item[{\bf $\Lambda$.c)}] -30. cm $<x_{V^0}<$\ -0.5 cm
\item[{\bf $\Lambda$.d)}] $close_{V^0} < $\  0.030 cm
\item[{\bf $\Lambda$.e)}] 0.020 GeV/$c$ $< {q_T}_{V^0} < $\ 0.40 GeV/$c$
\item[{\bf $\Lambda$.f)}] $\left[ \frac{{b_y}_{V^0}}{2.5\cdot\sigma_y} \right]^2 +
                           \left[ \frac{{b_z}_{V^0}}{3\cdot\sigma_z} \right]^2 < 1 $
\item[{\bf $\Lambda$.g)}] $ \alpha  > 0.45$\ ($\Lambda$)
                	  \hspace{0.6cm} {\em OR} \hspace{0.6cm}
		          $\alpha  < -0.45$\ ($\bar{\Lambda}$)
\item[{\bf $\Lambda$.h)}] $ M(\pi^+,\pi^-) - m_{K^0} >$\ 0.034  GeV/$c^2$\
                          \hspace{0.6cm} {\em OR} \hspace{0.6cm}
			  $ m_{K^0} - M(\pi^+,\pi^-) >$\ 0.026  GeV/$c^2$
\item[{\bf $\Lambda$.i)}] $|\phi_{V^0}| > $\ 0.157 rad
\item[{\bf $\Lambda$.j)}] $\left| M(p,\pi) - m_{\Lambda} \right|  < $\
                          0.010 GeV/$c^2$\
\end{itemize}

\begin{center}
{\bf Criteri di selezione preliminari per i $K_s^0$\ in Pb-Pb a 160 A GeV/$c$\ (1998)}
\end{center}
\begin{itemize}
\item[{\bf $K_S^0$.a)}] tutte le tracce del decadimento attraversano la
                        parte compatta del telescopio (primo e nono piano).
\item[{\bf $K_S^0$.b)}] decadimento della $V^0$\ di tipo {\em cowboy}
\item[{\bf $K_S^0$.c)}] -42. cm $<x_{V^0}<$\ -0.5 cm
\item[{\bf $K_S^0$.d)}] $close_{V^0} < $\  0.035 cm
\item[{\bf $K_S^0$.e)}] 0.040 GeV/$c$ $< {q_T}_{V^0} < $\ 0.40 GeV/$c$
\item[{\bf $K_S^0$.f)}] $\left[ \frac{{b_y}_{V^0}}{2.5\cdot\sigma_y} \right]^2 +
                         \left[ \frac{{b_z}_{V^0}}{2.5\cdot\sigma_z} \right]^2<1$
\item[{\bf $K_S^0$.g)}] $ | \cos\theta^*_{K^0_S} | < 0.7 $
\item[{\bf $K_S^0$.h)}] $|\phi_{V^0}| > $\ 0.157 rad
\item[{\bf $K_S^0$.i)}] ${b_y}_{\pi^+}*Segno(B) < -0.2 $
\item[{\bf $K_S^0$.j)}] ${b_y}_{\pi^-}*Segno(B) >  0.2 $
\item[{\bf $K_S^0$.k)}] $ M(\pi^+,\pi^-) - m_{K^0} <$\ 0.0166  GeV/$c^2$\
                          \hspace{0.6cm} {\em AND} \hspace{0.6cm}
                        $ m_{K^0} - M(\pi^+,\pi^-) <$\ 0.0134  GeV/$c^2$
\end{itemize}
\subsection{Spettri di massa finali}
\begin{figure}[htb]
\begin{center}
\includegraphics[scale=0.50]{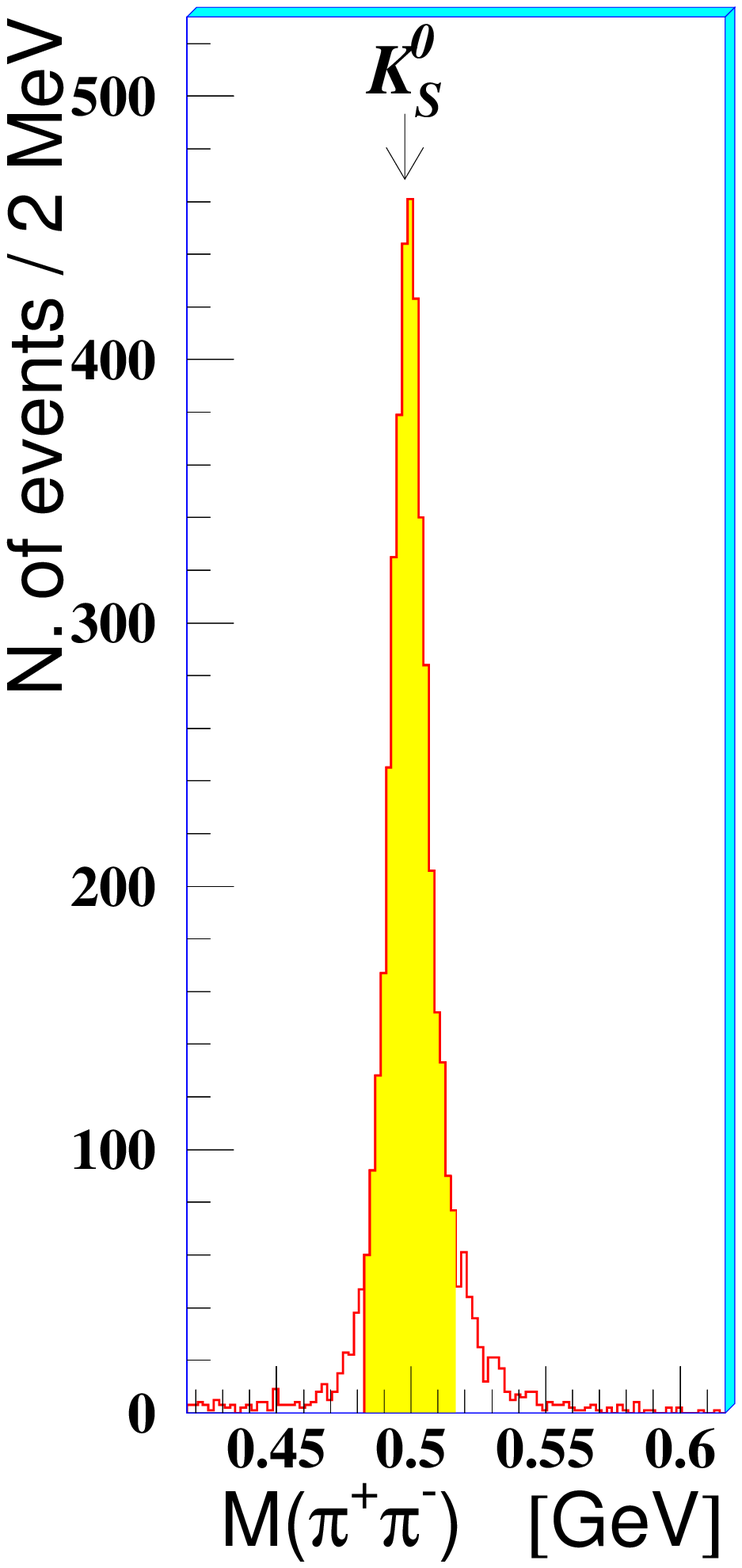}
\hspace{-0.5cm}
\includegraphics[scale=0.50]{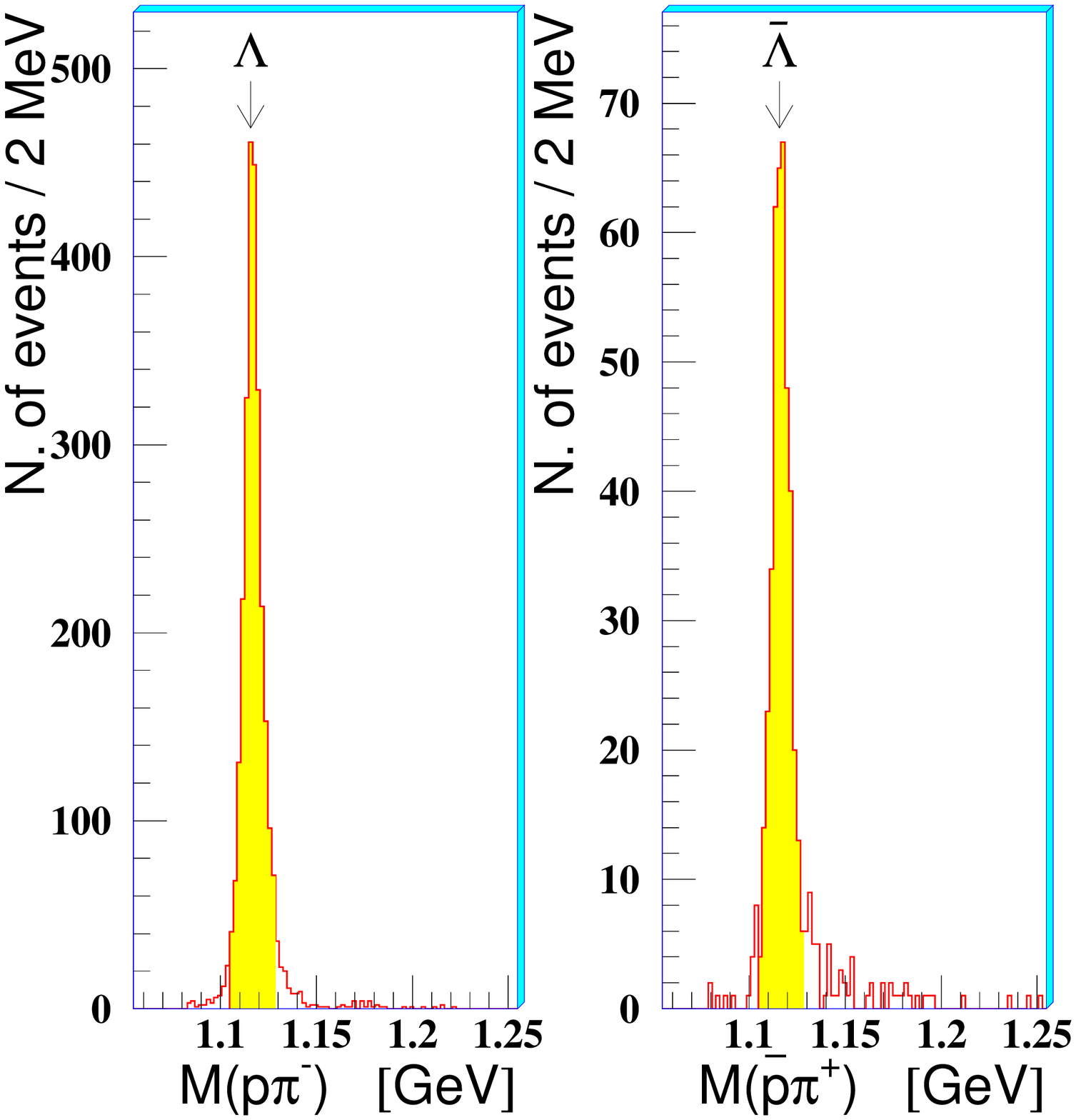}
\caption{Spettri finali di massa invariante $M(\pi^+,\pi^-)$\ (a sinistra),
	 $M(p,\pi^-)$\ (al centro) e $M(\bar{p},\pi^+)$\ (a destra) per un 
	 campione di $V^0$, corrispondente a circa 
	 lo 0.5\% dell'intera statistica raccolta nella presa dati del 1998. 
	 Le frecce indicano il valore nominale di massa delle diverse specie 
	 selezionate.} 
\label{FinalSpectra}
\end{center}
\end{figure}

In fig.~\ref{FinalSpectra} sono mostrati gli spettri di massa invariante per \PKzS, 
\PgL\ e \PagL\ che si ottengono a partire da un campione di candidate $V^0$, 
pari a circa $1/200$\ dell'intera statistica raccolta nella presa dati del 1998, 
applicando i criteri di selezione discussi nel paragrafo precedente per 
le diverse specie di particelle.  
\newline
Per meglio valutare la qualit\`a dei segnali selezionati si \`e eseguito un {\em ``best fit''} 
degli spettri di massa invariante in fig.~\ref{FinalSpectra}, utilizzando una gaussiana 
per descrivere il segnale ed un polinomio di terzo grado per descrivere il fondo. 
%Come  gi\`a osservato, 
Le distribuzioni di massa invariante non corrispondono a semplici gaussiane 
e ci\`o \`e riflesso dai valori del $\chi^2$\ dei {\em ``best fit''}, che non \`e ottimale; 
tuttavia l'approssimazione \`e sufficiente per gli scopi qui prefissi.
I risultati del fit per i parametri della gaussiana (valor medio $\mu$\ e larghezza $\sigma$) 
sono riportati in tab. 3.1 insieme ai valori del $\chi^2$\ e della larghezza a met\`a altezza 
delle distribuzioni ($FWHM$).
\begin{table}[h]
\label{tab3.1}
\begin{center}
\begin{tabular}{|c|c|c|c|c|} \hline
        & $\mu$\ (MeV/$c^2$) & $\sigma$\ (MeV/$c^2$) & $FWHM$\ (MeV/$c^2$) & $\chi^2/ndf$ \\ \hline
   \PKzS    & $499.3 \pm 0.1$ & $7.3 \pm 0.2 $ & $ 15 $ & $2.15 $\\ 
   \PgL     & $1116.1\pm 0.1$ & $4.6 \pm 0.1 $ & $ 8 $ & $3.46 $\\ 
   \PagL    & $1116.7\pm 0.1$ & $4.8 \pm 0.1 $ & $ 9 $ & $1.17 $\\ \hline
\end{tabular}
\end{center}
\caption{Parametri del fit e larghezza del picco a met\`a altezza ($FWHM$) degli 
	 spettri di massa invariante dei \PKzS, \PgL\ e \PagL\ 
	 nell'interazione Pb-Pb a 160 A GeV/$c$; $\mu$\ \`e il valore centrale e 
	 $\sigma$\ la larghezza della gaussiana.}
\end{table}
\newline
I valori medii delle distribuzione sono compatibili entro $1\div2 $\ MeV/$c^2$\ coi 
valori nominali delle rispettive particelle, mentre la larghezza della gaussiana e 
la $FWHM$\ forniscono una stima della risoluzione in massa degli spettri, dell'ordine 
dei $5 \div 10$\ MeV/$c^2$.
\newline
La procedura di fit permette, inoltre, di ottimizzare la scelta degli intervalli di 
massa da considerare per isolare le $V^0$\ che vengono infine selezionate per l'analisi 
fisica che seguir\`a (spettri di massa trasversa, tasso di produzione delle diverse specie, 
etc.). Sono stati cos\`i selezionati un intervallo di ampiezza 
20 MeV/$c^2$\ centrato attorno al valore nominale di massa per le \PgL\ e le \PagL, 
e l'intervallo 
$\left[ m_{\PKzS} - 26\, {\rm MeV}/c^2 \, , \, m_{\PKzS} + 34\, {\rm MeV}/c^2 \right]$\ 
per i \PKzS.  
Gli intervalli cos\`i selezionati sono evidenziati in giallo negli spettri di massa della 
fig.~\ref{FinalSpectra}. 

\section{Ricostruzione e selezione di $\Xi$\ ed $\Omega$}
\begin{figure}[htb]
\begin{center}
\includegraphics[scale=0.42]{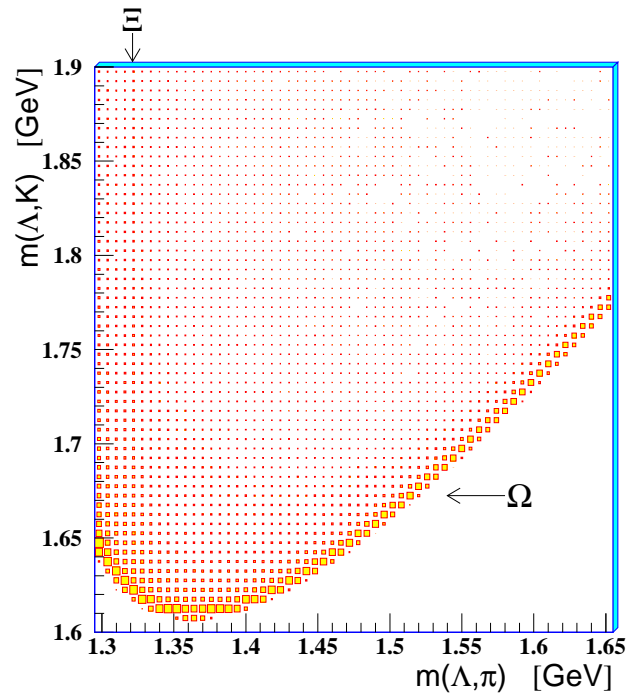}
\caption{Correlazione tra lo spettro di massa invariante 
	$M(\Lambda,K)$\  e quello $M(\Lambda,\pi)$\ per le candidate cascate 
	ricostruite in interazioni Pb-Pb a 160 A GeV/$c$.}
\label{CandScatter}
\end{center}
\end{figure}

La ricostruzione delle ``cascate'', cio\`e dei decadimenti  di 
$\Xi$\ ed $\Omega$\ in una $\Lambda$\ ed un mesone ($\pi$\ o 
$K$, rispettivamente per $\Xi$\ o $\Omega$) con successivo 
decadimento della $\Lambda$\ in $p$\ e $\pi^-$\ 
(od in $\bar{p}$\ e $\pi^+$ nel caso degli anti-iperoni), 
avviene a partire dall'abbinamento di tutte le 
candidate $V^0$\ trovate in un evento con tutte le tracce 
di particelle cariche ricostruite nello stesso evento, ad 
esclusione di quelle relative alla $V^0$\ considerata.  
\newline
Come nel caso delle candidate $V^0$, si considerano inizialmente 
le ``candidate cascate'', ottenute nel modo appena descritto, richiedendo 
tuttavia alcuni tagli preliminari, meno selettivi di quelli infine usati, 
che tuttavia riducono notevolmente il numero di candidate iniziali.  
A tal fine, le $V^0$\ delle ``candidate cascate'' devono  
soddisfare gli stessi criteri preliminari introdotti per definire le 
candidate $V^0$\ prodotte nell'interazione primaria. 
Ulteriormente, sia per quanto riguarda la $V^0$\ che per i parametri della 
cascata stessa, si richiede che: 
\begin{enumerate}
\item
la minima distanza spaziale tra la traccia del mesone carico e la 
traiettoria della $V^0$\ sia minore di $0.1$\ cm; 
\item 
la coordinata $x$\ del vertice di decadimento della cascata, 
valutato come il punto medio del segmento congiungente i punti 
di minima distanza tra la traccia del 
mesone e la traiettoria della $V^0$, sia compresa nell'intervallo 
$[-45,3]$\ cm;
\item
la massa invariante $M(p,\pi)$, calcolata a partire dalle particelle 
componenti la $V^0$, sia 
compresa entro $\pm 25$\ MeV/$c^2$ rispetto al valore nominale della massa 
della $\Lambda$,
\item
la cascata punti verso il vertice primario dell'interazione, in modo 
tale che il suo parametro d'impatto sia compreso entro 
$\pm \,6 \,\sigma_{run}$, sia nella componente $y$\ che in quella $z$, 
dove $\sigma_{run}$\ \`e la precisione sulla determinazione del vertice 
primario {\em run per run}, 
come gi\`a definito discutendo i criteri di selezione per le $V^0$. 
\end{enumerate}
A partire da questi criteri preliminari si ottengono le candidate 
cascate sulle quali bisogna operare dei tagli geometrici e cinematici 
per isolare il segnale fisico dei decadimenti delle $\Xi^-$, delle 
$\Omega^-$\ e delle anti-particelle relative.  
Riferendosi all'intero campione dei dati raccolti nell'anno 2000, 
la correlazione per le candidate cascate 
tra la massa invariante $M(\Lambda,\pi)$\  e quella $M(\Lambda,K)$\  
\`e mostrata in fig.~\ref{CandScatter}. 
\`E gi\`a possibile riconoscere il segnale delle $\Xi$, 
%in corrispondenza del valore nominale $M_{\Xi} = 1.321$\ GeV/$c^2$\  
%nello spettro di massa invariante $M(\Lambda,\pi)$, che si distingue dalle 
in corrispondenza della massa invariante $M(\Lambda,\pi)$\ intorno al  
valore nominale $M_{\Xi} = 1.321$\ GeV/$c^2$, 
che si distingue dalle regioni di accumulazione di punti poste ai bordi 
delle regioni permesse dello spazio delle fasi; invece, il segnale della  
particella $\Omega$, molto pi\`u rara, \`e ancora completamente 
sommerso dal fondo di cascate geometriche e quindi non \`e visibile 
in corrispondenza della massa $M(\Lambda,K)$\ intorno al valore nominale 
$M_{\Omega}=1.672$\ GeV/$c^2$\ nel diagramma bidimensionale.     
\newline
Poich\'e la selezione delle $\Omega^-$\ e delle $\bar{\Omega}^+$\ risulta 
molto pi\`u delicata di quella per le $\Xi^-$\ e le $\bar{\Xi}^+$, a causa 
del minor rapporto iniziale segnale su fondo, si esporranno qui nei dettagli 
i criteri di selezione adottati per le prime, mentre ci si 
limiter\`a ad elencare i valori finali usati per le $\Xi^-$\ e 
le $\bar{\Xi}^+$, soffermandosi invece solo su quei criteri di selezione 
studiati {\em ad hoc} per queste ultime particelle. 
\newline
Poich\'e la disposizione dei piani del telescopio \`e differente per la 
presa dati dell'anno 1998 e per quella del 2000, la selezione delle diverse 
specie di cascate \`e stata affinata separatamente per i due diversi 
campioni di dati. Nell'analizzare i dati dell'anno 2000,  si \`e data 
la precedenza alle $\Omega^-$\ ed alle $\bar{\Omega}^+$, volendo 
ridurre in maniera pi\`u efficace gli errori statistici associati al 
solo campione di dati dell'anno 1998, precedentemente analizzato.  
\newline 
Dal momento che l'analisi dei due campioni presenta, in prima 
approssimazione, stesse problematiche e stessi modi d'approccio 
al problema, ci si riferir\`a, in questa discussione, ad uno solo dei 
due campione di dati disponibili, quello raccolto nell'anno 2000. 
Per una discussione dettagliata sulla scelta finale dei 
valori dei tagli utilizzati per i dati del 1998, che comunque sono qui 
riportati, ci si pu\`o riferire a~\cite{Omega98Bruno} nel caso 
delle $\Omega$, ed a~\cite{XiCarrer} nel caso delle $\Xi$. 
\newline
Si discuteranno dunque i criteri di selezione adottati per isolare il segnale 
delle \PgOm\ ed \PagOp\ dal campione di dati raccolto nell'anno 2000; come 
nel caso della selezione delle $V^0$\ si applicheranno dei criteri puramente 
geometrici, che saranno preceduti nella loro discussione dal simbolo 
$\bullet$, ed altri di tipo cinematico, che agiscono dunque direttamente 
nello spazio delle fasi, e che verranno preceduti dal simbolo $\blacktriangle$. 
Come nella discussione dei criteri di selezione delle $V^0$, si mostreranno 
solo le distribuzioni finali, ottenute applicando i restanti criteri di 
selezione ad eccezione di quello in studio e la selezione finale dell'intervallo 
di massa invariante $M(\Lambda,K)$\ per la definizione del campione di $\Omega$.
\begin{itemize}
\item 
     {\bf Minima distanza  tra le tracce di decadimento della 
          cascata: $close_{cas}$} \\
     {\bf Volume fiduciale per il decadimento della cascata: $x_{cas}$} \\ 
In modo analogo a quanto visto per il decadimento delle $V^0$, 
si considerano i parametri geometrici $close_{cas}$\ --- pari alla minima 
distanza di approccio tra la linea di volo della $V^0$\ e la traiettoria 
del mesone carico, inseguito all'indietro nel campo magnetico --- ed $x_{cas}$\ 
---  pari alla coordinata $x$\ del punto di decadimento della candidata cascata, 
     definito come il punto medio del segmento congiungente le traiettorie della 
     $V^0$\ e del mesone carico quando sono al loro minimo approccio. 
In fig.~\ref{Om8_3} sono mostrate le distribuzioni di queste due variabili e la 
correlazione con la massa invariante $M(\Lambda,K)$, riportata in ordinata.
\begin{figure}[htb]
\begin{center}
\includegraphics[scale=0.42]{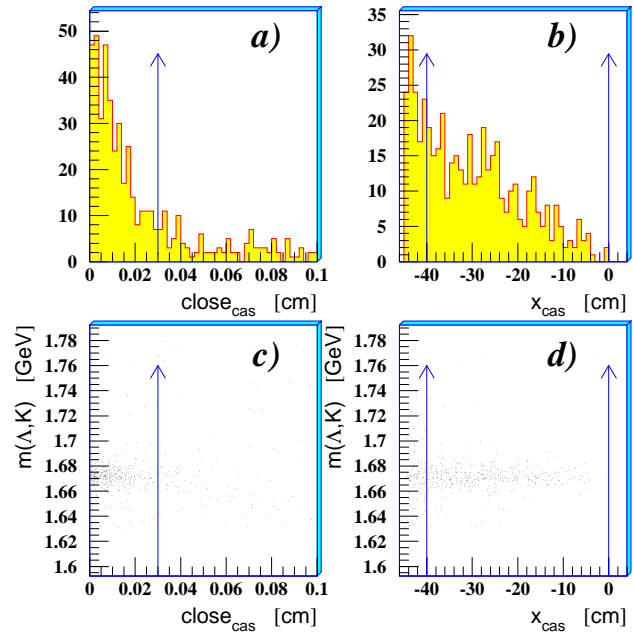}
\caption{{\bf a)} Distribuzione della variabile $close_{cas}$\ per 
	 le candidate cascate, selezionate per isolare il segnale delle 
	 $\Omega$. {\bf b)} Distribuzione della coordinata $x_{cas}$\ del 
	 punto di decadimento delle candidate cascate.
	 {\bf c)} e {\bf d)} Correlazioni con la massa invariante $M(\Lambda,K)$\ 
	 (in ordinata) delle variabili delle distribuzioni soprastanti  
	 (in ascissa).}
\label{Om8_3}
\end{center}
\end{figure}
I criteri di selezione adottati (frecce blu nella fig.~\ref{Om8_3}) escludono 
le candidate cascate che presentino un $close_{cas}>0.030$\ cm ed una $x_{cas}$\ 
esterna all'intervallo $[-40,0]$\ cm. In tal modo si elimina parte di 
quel fondo geometrico costituito da false cascate formate prendendo una o 
pi\`u, delle tre tracce cariche che formano la cascata, proveniente dal 
vertice primario, anzich\`e da un decadimento fisico.
\item
{\bf Minima distanza  tra le tracce di decadimento della \PgL: $close_{V^0}$} \\
{\bf Volume fiduciale per il decadimento della \PgL: $x_{V^0}$} \\
Anche per il decadimento della $V^0$\ associata alla cascata 
\`e possibile agire sulle variabili $close_{V^0}$\ ed $x_{V^0}$, gi\`a 
definite nella discussione sulla selezione delle $V^0$\ primarie, per 
eliminare parte della contaminazione geometrica imputabile ad una $V^0$\ 
geometrica associata ad una traccia carica dell'evento.    
In fig.~\ref{Om8_4} sono mostrate le distribuzioni di queste variabili  
e la loro correlazione con la massa invariante $M(\Lambda,K)$, che viene 
sempre riportata in ordinata.  
\begin{figure}[htb]
\begin{center}
\includegraphics[scale=0.42]{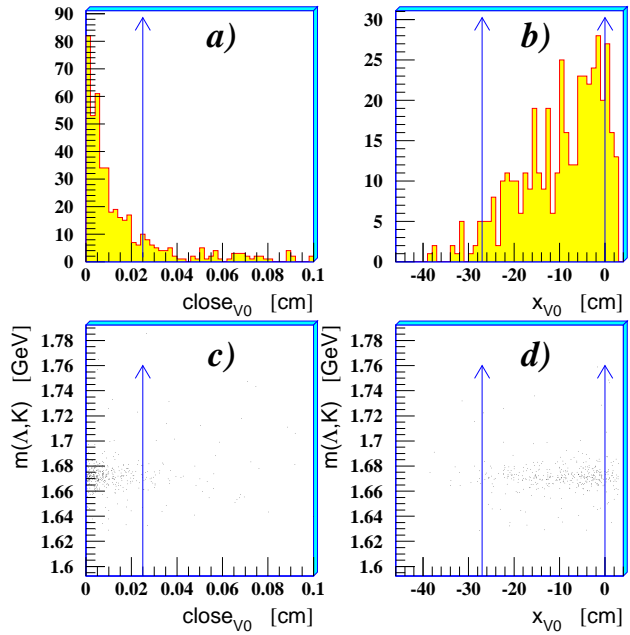}
\caption{{\bf a)} Distribuzione della variabile $close_{V^0}$\ per
         le $V^0$\ delle candidate cascate, selezionate per isolare il segnale 
	 delle $\Omega$. {\bf b)} Distribuzione della coordinata $x_{V^0}$\ del
         punto di decadimento delle $V^0$\ associate alle candidate cascate.
         {\bf c)} e {\bf d)} Correlazioni con la massa invariante $M(\Lambda,K)$\
         (in ordinata) delle variabili delle distribuzioni soprastanti
         (in ascissa).}
\label{Om8_4}
\end{center}
\end{figure}
Il tagli imposti su queste variabili richiedono che il $close_{V^0}$\ sia inferiore 
a $0.025$\ cm, ed il ``volume'' fiduciale per il decadimento della $V^0$\ 
corrisponda all'intervallo $[-27,0]$\ cm.  
\item
{\bf Differenza tra le coordinate $x$ dei due vertici di decadimento: 
$x_{V^0}-x_{cas}$} \\
Nel decadimento a cascata di una $\Xi$\ o di una $\Omega$, il  
decadimento della $\Lambda$\ (cio\`e della $V^0$) deve seguire, pi\`u 
avanti nello spazio, il precedente decadimento della $\Xi$\ o della $\Omega$.
La distribuzione della differenza $x_{V^0}-x_{cas}$ delle candidate cascate 
\`e mostrata in fig.~\ref{Om8_14}, in cui \`e anche mostrata la correlazione 
con la massa invariante $M(\Lambda,K)$.  
\begin{figure}[htb]
\begin{center}
\includegraphics[scale=0.42]{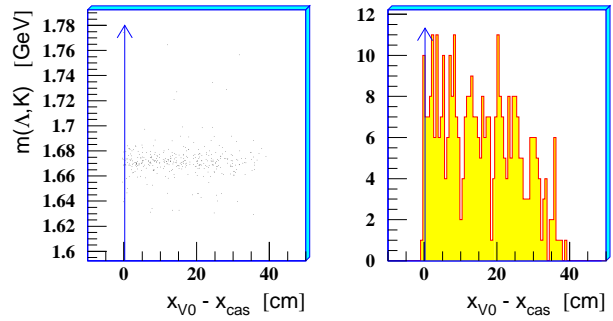}
\caption{Distribuzione della differenza tra le coordinate $x$\ del
         vertice di decadimento della $V^0$\ e quella della cascata.
         La condizione imposta richiede che la $V^0$\ decada almeno $0.5$\ cm
         dopo la cascata.}
\label{Om8_14}
\end{center}
\end{figure}
Il taglio imposto, leggermente pi\`u selettivo della semplice consequenzialit\`a 
dei due decadimenti, richiede che  $x_{V^0}-x_{cas}>0.5$\ cm, per tener conto 
della possibilit\`a di errori di estrapolazione.  
\item
{\bf Parametro d'impatto della cascata: ${b_y}_{cas}$, ${b_z}_{cas}$} \\
Il parametro d'impatto della candidata cascata si definisce in modo analogo 
al parametro d'impatto delle candidate $V^0$. A partire dall'impulso della 
cascata nella posizione di decadimento --- impulso valutato come la somma 
vettoriale  degli impulsi della $V^0$\ e del mesone carico 
nei punti di minima distanza delle loro traiettorie --- 
si estrapola all'indietro, inseguendo nel campo 
magnetico, sino al piano $\Sigma$\ parallelo al piano $yz$\ e contenente il 
bersaglio. Il parametro d'impatto viene valutato in tale piano $\Sigma$, rispetto 
alla posizione {\em run per run} del vertice primario, come nel caso 
dello studio delle $V^0$.  
In fig.~\ref{Om8_5} sono mostrate le distribuzioni delle coordinate $y$\ 
(fig.~\ref{Om8_5}.a) e $z$\ (fig.~\ref{Om8_5}.b) del parametro d'impatto delle 
candidate cascate, normalizzate alla rispettiva $\sigma_{run}$. 
\begin{figure}[p]
\begin{center}
\includegraphics[scale=0.42]{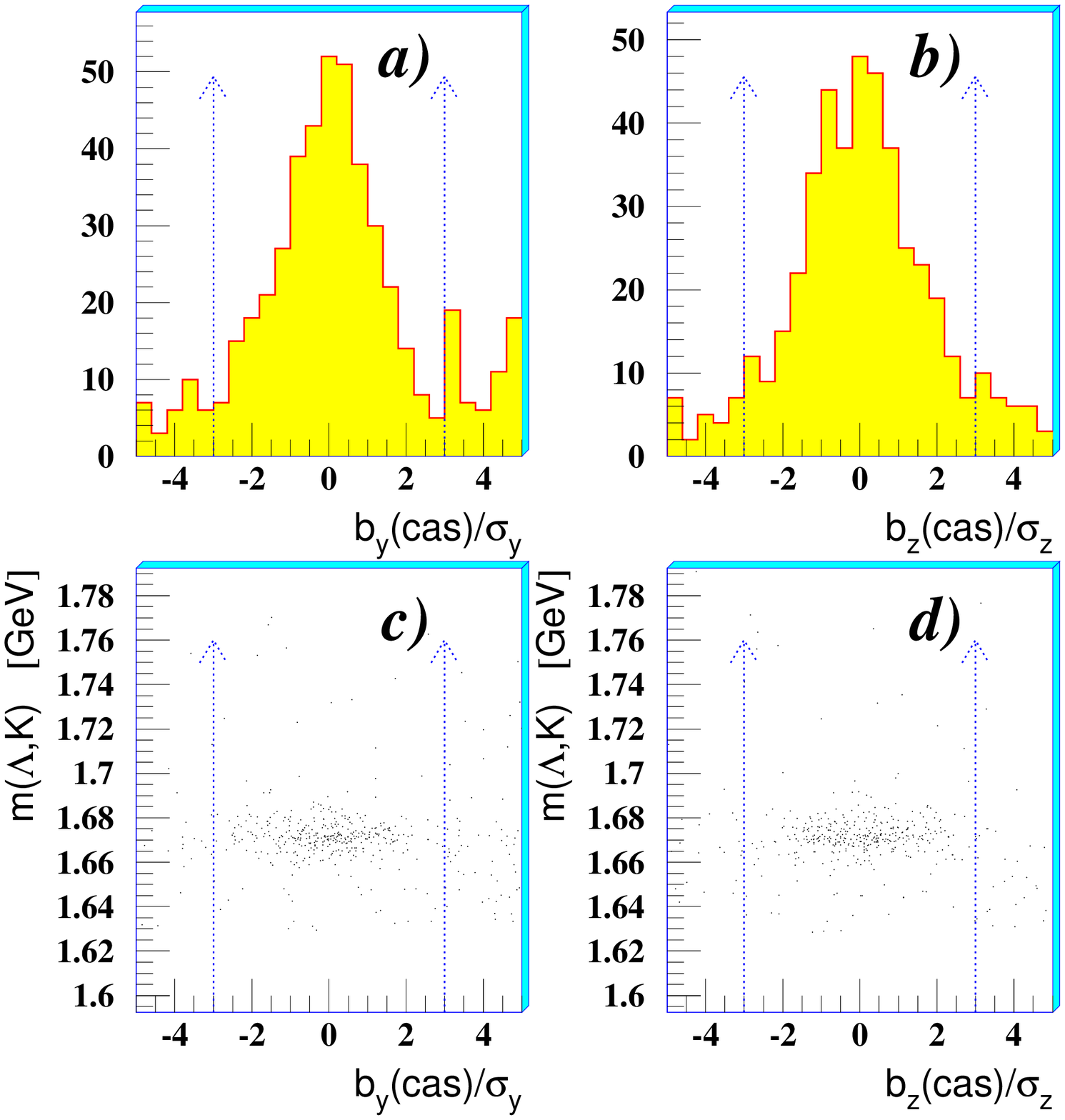} \\
\includegraphics[scale=0.21]{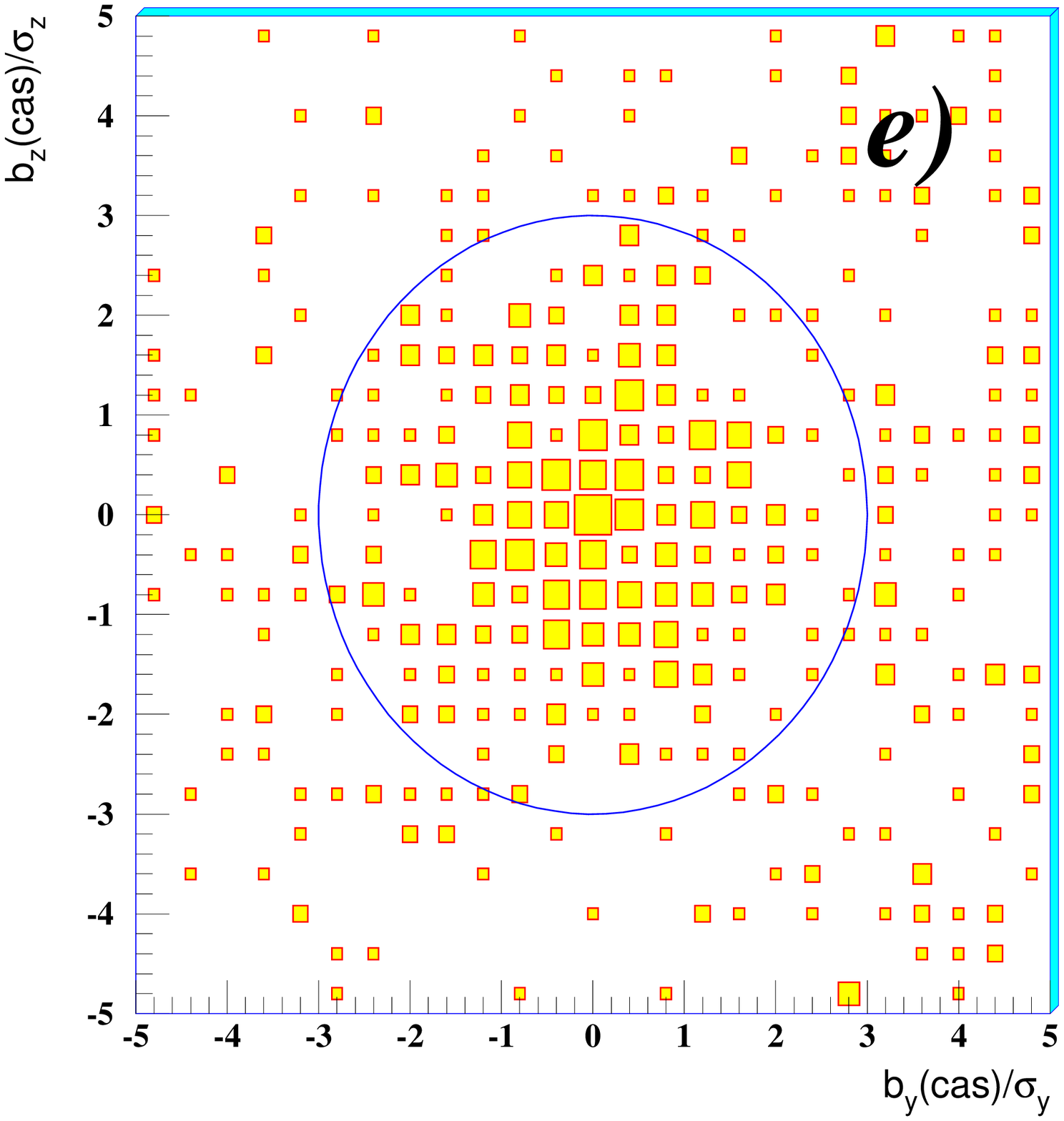}
\caption{Distribuzione dei parametri d'impatto delle candidate 
	 cascate, selezionate per isolare il segnale delle $\Omega$, 
	 normalizzate alla relativa $\sigma_{run}$. In {\bf a)} \`e 
	 mostrata la componente $y$, in {\bf b)} quella $z$. 
	 \newline
	 {\bf c)} e {\bf d)} Correlazione della massa invariante 
	 $M(\Lambda,K)$\ con le due proiezioni del parametro 
	 d'impatto delle cascate. 
	 \newline
	 {\bf e)} Correlazione delle due componenti $z$\ e $y$\ del 
	 	  parametro d'impatto delle cascate, normalizzate 
		  rispettivamente a $\sigma_{z}$\ e $\sigma_{y}$.}
\label{Om8_5}
\end{center}
\end{figure}
Nelle figg.~\ref{Om8_5}.c, \ref{Om8_5}.d sono mostrate le correlazioni 
della massa invariante $M(\Lambda,K)$\ con le due proiezioni. 
Per selezionare le cascate fisiche bisogna richiedere che esse puntino 
al vertice primario dell'interazione; come nel caso delle $V^0$\ primarie, 
si \`e adottato un taglio combinato sulle due proiezioni, richiedendo che 
il parametro d'impatto normalizzato,  
$\vec{b}_{cas}/\sigma_{run}=({b_y}_{cas}/\sigma_{y}, {b_z}_{cas}/\sigma_{z})$, 
sia compreso entro una circonferenza di raggio 3.  Le candidate cascate 
selezionate da questo taglio sono dunque quelle comprese entro la 
circonferenza della fig.~\ref{Om8_5}.e. 
Nell'impostare tale criterio di selezione si \`e tenuto conto delle successive 
fasi di analisi: in particolare nella procedura del calcolo delle correzioni 
per efficienza ed accettanza, la variabile parametro d'impatto  viene 
generata nelle simulazioni e deve essere parametrizzata.  
Valgono dunque le stesse considerazioni espresse discutendo i tagli sul parametro 
d'impatto delle $V^0$\ primarie: anche in tal caso si preferisce adottare 
un taglio non eccessivamente selettivo (a tre sigma, in tal caso) in modo tale che 
eventuali discrepanze tra le distribuzioni reali e quelle delle simulazioni 
influiscano in modo trascurabile sulle correzioni che si vogliono calcolare.  
\item
{\bf Piani del telescopio attraverso cui devono passare le tracce di decadimento} \\
Nella disposizione sperimentale della presa dati del 1998 
({\em cfr. paragrafo 2.4.3}), i piani di pixel del telescopio erano distribuiti 
in una {\em ``parte compatta''}, costituita da nove piani entro $30$\ cm, 
e  nel {\em ``lever arm''} (``braccio di leva''), in cui erano posti  
quattro piani entro dieci centimetri,  a $30$\ cm 
dall'ultimo piano della parte compatta.  Come condizione sulla qualit\`a delle 
tracce, si richiede che tutte le tracce attraversino la parte compatta 
del telescopio. Pi\`u precisamente, la traiettoria di ogni traccia deve 
passare entro il profilo rettangolare, che racchiude la parte sensibile per 
la rivelazione, del primo e dell'ultimo piano di pixel della parte compatta 
del telescopio. 
\newline
Nella disposizione dell'anno 2000, la funzione di {\em ``lever arm''} viene  
invece completamente assolta dai rivelatori a $\mu$strip doppia faccia, 
mentre undici piani di pixel sono distribuiti, con buona uniformit\`a,  
nel telescopio. 
Con tale disposizione, diventa importante studiare in maniera sistematica il 
criterio sulla qualit\`a delle tracce, in termini dei piani entro cui 
le traiettorie delle tracce vanno vincolate.  
In vista di ci\`o, le candidate cascate sono state inizialmente selezionate 
richiedendo come condizione, poco selettiva,  
l'attraversamento del primo e dell'ottavo piano. 
La condizione geometrica che pi\`u si avvicina a quella adoperata nel 1998, 
impone che la traiettoria delle tracce passi per il primo ed il decimo piano 
di pixel, ed \`e stata  adottata come condizione di riferimento.  
Senza dilungarsi nei dettagli di questa analisi, si dir\`a unicamente che, 
tanto per le $\Xi$\ quanto per le $\Omega$, 
la condizione pi\`u vantaggiosa per isolare il segnale delle cascate 
--- condizione determinata rilasciando ad uno ad uno il vincolo 
sull'ultimo piano per le tre tracce costituenti una cascata --- 
richiede che le traiettorie delle tre tracce della cascate attraversino 
tutte il primo ed il nono piano del telescopio.  
In fig.~\ref{Om2} \`e mostrato il segnale delle \PgOm\ e delle \PagOp\ 
selezionate con la condizione di riferimento 
(primo e decimo piano, fig.~\ref{Om2}.a), e quelle 
guadagnate rilasciando il vincolo dal decimo piano al nono piano 
(fig.~\ref{Om2}.b).   
\begin{figure}[htb]
\begin{center}
\includegraphics[scale=0.42]{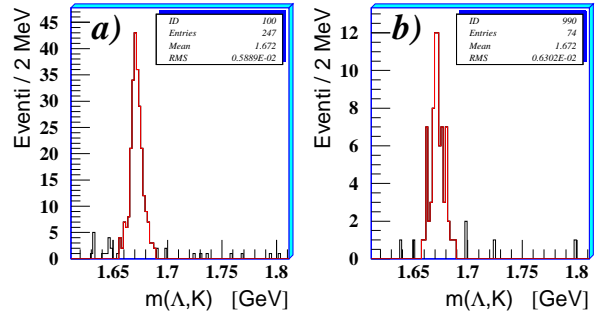}
\caption{{\bf a)} Spettro di massa invariante $M(\Lambda,K)$\ per 
	 le \PgOm\ e le \PagOp\ selezionate richiedendo che tutte le tracce 
	 di decadimento passino attraverso primo e decimo piano del telescopio.
	 {\bf b)} Segnale guadagnato rilasciando la condizione del decimo piano 
	 e portandola al nono.}
\label{Om2}
\end{center}
\end{figure}
Il guadagno sul segnale, calcolato entro l'intervallo di $\pm 17$\ MeV/$c^2$ attorno 
a  $M_{\Omega}=1.672$\ GeV/$c^2$, \`e del 30\%, con un rapporto segnale su fondo 
paragonabile a quello di riferimento. 
Rilasciando ulteriormente la condizione sull'ultimo piano, 
portandolo all'ottavo, si ottiene un ulteriore guadagno dell'8\%, ma con un 
sensibile deterioramento del rapporto segnale su fondo.  
\newline
La situazione \`e molto simile nel caso delle $\Xi$: il guadagno passando 
dal decimo al nono piano \`e del 32\% con stessa qualit\`a del segnale, 
quello passando dal nono all'ottavo del 7\%, ma con un deterioramento.  
\item[$\blacktriangle$]
{\bf Massa invariante $M(p,\pi)$} \\
In fig.~\ref{Om8_6} \`e mostrato lo spettro di massa invariante $M(p,\pi)$\ 
per le $V^0$\ associate al decadimento delle candidate cascate, selezionate  
coi criteri finalizzati ad isolare il segnale delle \PgOm\ ed \PagOp. Nel 
realizzare tale spettro, \`e stato applicato anche il taglio finale sulla massa 
invariante $M(\Lambda,K)$, corrispondente alle 
linee trattegiate nella correlazione tra le masse invarianti $M(\Lambda,K)$\ e 
$M(p,\pi)$\ che \`e altres\`i riportata nella fig.~\ref{Om8_6}. 
Tale spettro, perfettamente centrato sul  
valore nominale $m_{\Lambda}=1.1157$\ MeV/$c^2$, risulta pi\`u stretto di 
quello ottenuto studiando il segnale delle $\Lambda$\ primarie, 
probabilmente in quanto adesso esse decadono mediamente pi\`u in prossimit\`a 
del telescopio e quindi vengono ricostruite con maggiore precisione, 
a riprova del fatto che la precisione sulla misura della 
massa invariante dipende dalla cinematica di decadimento della particella. 
Eseguendo il {\em ``best fit''} con una gaussiana, si ottiene una sigma di soli 
$3.7$\ MeV/$c^2$. Il criterio di selezione sulla massa invariante 
$M(p,\pi)$\ \`e stato scelto in modo da accettare le candidate cascate se 
tale variabile cade nell'intervallo $[m_{\Lambda}-8 \, , \, m_{\Lambda}+8 ]$\ 
MeV/$c^2$, corrispondente dunque a circa due sigma della gaussiana di 
{\em ``best fit''}.  
\begin{figure}[t]
\begin{center}
\includegraphics[scale=0.42]{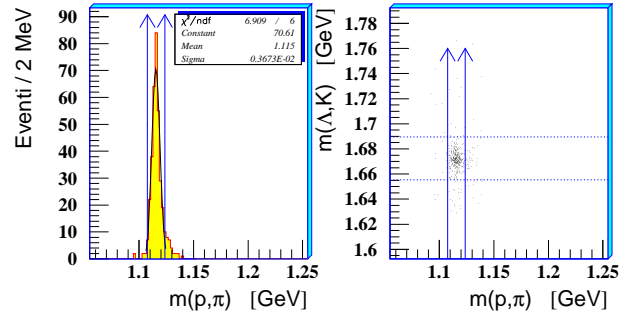}
\caption{{\em A sinistra:} spettro di massa invariante $M(p,\pi)$\
         relativo alle $V^0$\  provenienti dal decadimento delle cascate
         selezionate per isolare il segnale delle \PgOm\ ed \PagOp. Nell'inserto
         \`e riportato il risultato del {\em ``best fit''} con una gaussiana.
         {\em A destra:} correlazione tra le masse invarianti $M(\Lambda,K)$\
         e $M(p,\pi)$.}
\label{Om8_6}
\end{center}
\end{figure}
\item[$\blacktriangle$]
{\bf Angoli interni di decadimento azimuthali: $\phi_{cas}$, $\phi_{V^0}$} \\
Riferendosi alle definizioni gi\`a date discutendo la selezione delle $V^0$\ 
primarie, si considera in fig.~\ref{Om8_8-10} la correlazione tra l'angolo 
interno $\phi_{V^0}$\ nel decadimento della $V^0$\ e quello $\phi_{cas}$\ nel 
decadimento della cascata. L'angolo riportato si riferisce al mesone carico 
nel caso del decadimento della cascata, o alla particella di carica negativa 
nel caso del decadimento della $V^0$. 
Anche nel caso del decadimento della cascata si accettano unicamente le 
topologie di tipo {\em cowboy}~\footnote{Nel decadimento della cascata solo 
la traiettoria del mesone ($\pi$\ o $K$, a seconda che si abbia 
una $\Xi$\ od un $\Omega$) \`e incurvata dall'azione del campo magnetico. La 
definizione di topologia {\em cowboy} o {\em sailor} \`e per\`o identica a quella 
data nel caso della $V^0$.}: pertanto, a seconda dell'orientazione del campo 
magnetico, e nel caso di $\phi_{cas} $\  anche a seconda che si tratti di 
cascata o di anti-cascata,  gli angoli $\phi_{cas}$\ e  $\phi_{V^0}$\ variano 
o entro l'intervallo $[-\pi,+\pi]$\ oppure entro $[-2\pi,-\pi] \cup [\pi,2\pi]$, 
come nel caso delle $V^0$\ primarie.  Nella fig.~\ref{Om8_8-10}, gli angoli 
al di fuori dell'intervallo $[-\pi,+\pi]$\ sono stati ribaltati, sommando o 
sattraendo $2\pi$, riportandoli entro tale intervallo.  
\newline
Il miglior taglio, determinato studiando la variazione del rapporto segnale su 
fondo, \`e quello combinato sulle due variabili. Le candidate cascate vengono 
rigettata se, simultaneamente, risulta: $|\phi_{cas}|<0.15$\ rad e 
$|\phi_{V^0}|<0.15$\ rad. Tale condizione esclude i punti compresi entro il 
quadrato in blu della fig.~\ref{Om8_8-10}.a.  
\begin{figure}[p]
\begin{center}
\includegraphics[scale=0.21]{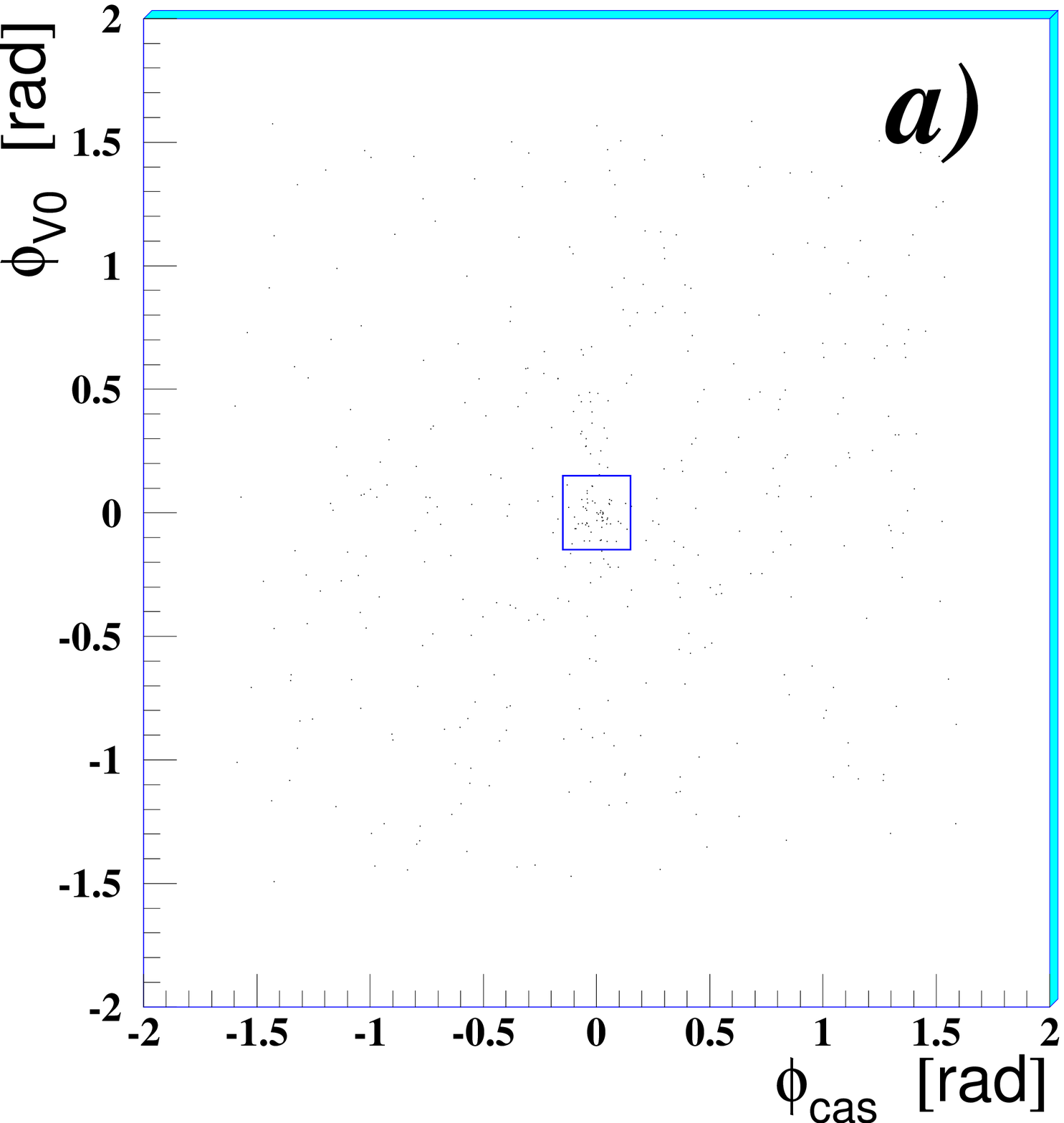} \\
\includegraphics[scale=0.42]{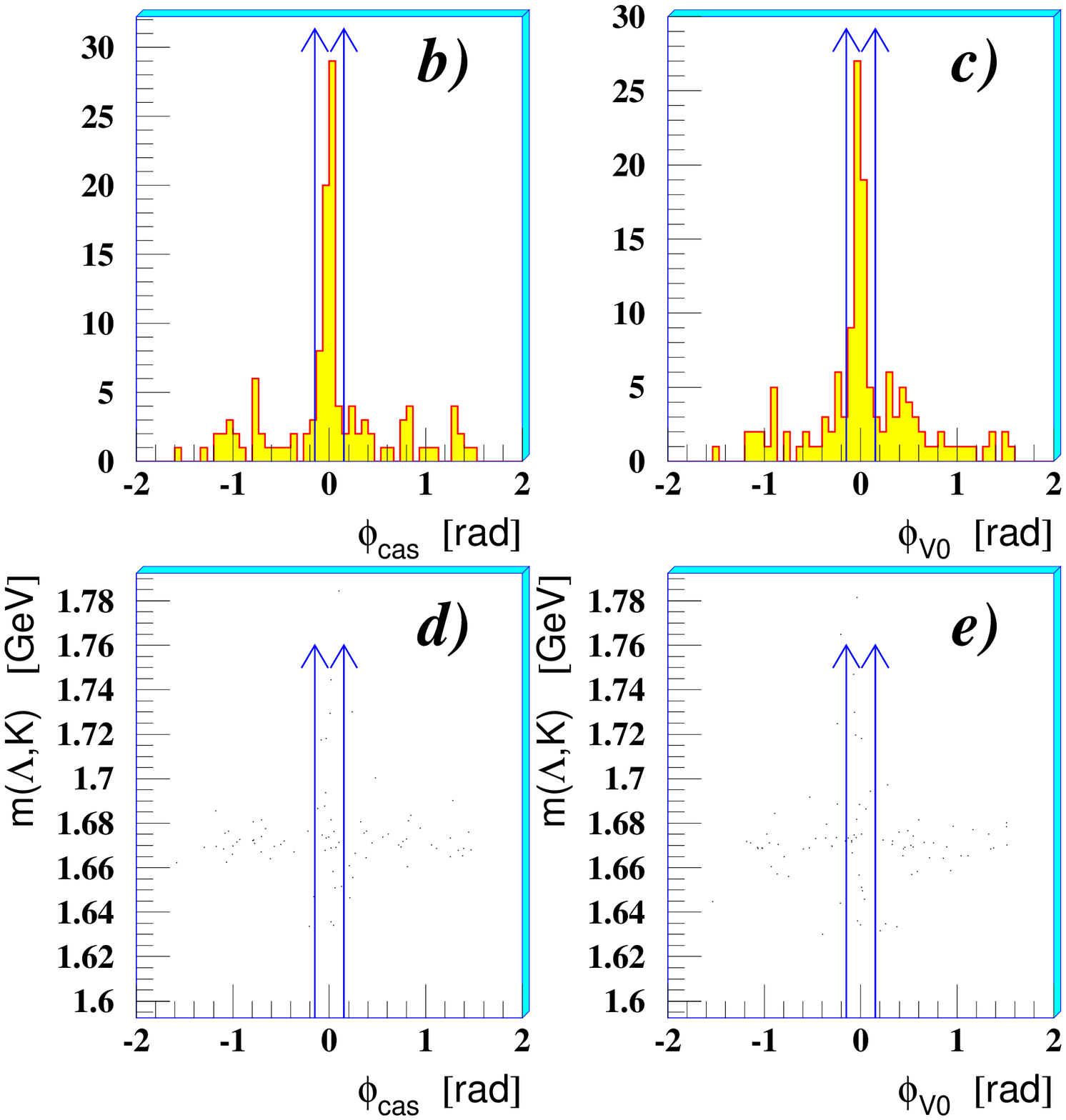}
\caption{{\bf a)} Correlazione tra l'angolo $\phi_{V^0}$\ della $V^0$\ e l'angolo 
	 $\phi_{cas}$\  della cascata (o dell'anti-cascata).
	 {\bf b)} Distribuzione dei valori di $\phi_{cas}$, per le candidate 
	 	  con $|\phi_{V^0}|< 0.30$\ rad.
	 {\bf c)} Distribuzione dei valori di $\phi_{V^0}$, per le candidate
	                   con $|\phi_{cas}|< 0.30$\ rad. 
	 {\bf d)} ({\bf e}) 
	 Correlazione della massa invariante $M(\Lambda,K)$\ con 
	 la variabile $\phi_{cas}$\ ($\phi_{V^0}$) per gli eventi 
	 in {\bf b)}  ({\bf c}).}
\label{Om8_8-10}
\end{center}
\end{figure}
Se si considera la distribuzione della variabile $\phi_{cas}$\  
(fig.~\ref{Om8_8-10}.b) e la sua correlazione con la massa invariante  
$M(\Lambda,\pi)$\ (fig.~\ref{Om8_8-10}.d), per le candidate per le quali 
l'altra variabile $\phi_{V^0}$\ \`e confinata entro una piccola banda  
$[-0.30,+0.30]$\ rad, \`e evidente come  in tal modo si riduca gran parte 
del fondo geometrico, completamente scorrelato dalla variabile 
$M(\Lambda,\pi)$. 
Analoghe conclusioni si traggono considerando 
la distribuzione (fig.~\ref{Om8_8-10}.d) 
e correlazione (fig.~\ref{Om8_8-10}.e) per l'angolo $\phi_{V^0}$, 
limitando l'altra variabile all'intervallo $|\phi_{cas}|<0.30$\ rad. 
La scelta del criterio di selezione combinato \`e dunque pienamente 
giustificata.  
\item[$\blacktriangle$]
{\bf Angolo interno di decadimento longitudinale: ${\theta^*}_{\Omega}$} \\
Si consideri nel sistema di riferimento $x'y'z'$\ --- dove l'asse $x'$\ 
\`e preso parallelo all'impulso della cascata nel punto in cui essa decade, 
l'asse $y'$\ giace nel piano $xy$\ del riferimento del laboratorio, 
l'asse $z'$\ \`e normale agli altri due --- in cui la cascata, 
cui viene assegnata la massa di una $\Omega$, \`e a riposo. Come definito per 
le $V^0$\ primarie, si considera ora l'angolo longitudinale 
${\theta^*}_{\Omega}$\ formato dall'impulso della particella di decadimento 
neutra (la $V^0$) e la direzione $x'$.  
In fig.~\ref{Om8_15} \`e mostrata la distribuzione del 
$\cos({\theta^*}_{\Omega})$\ e la sua correlazione con la massa invariante 
$M(\Lambda,K)$. Nel misurare queste distribuzioni, si \`e rilasciato il 
taglio su ${q_T}_{cas}$\ il quale, altrimenti, altererebbe parte dello 
spazio dello spazio delle fasi del decadimento.   
\begin{figure}[htb]
\begin{center}
\includegraphics[scale=0.42]{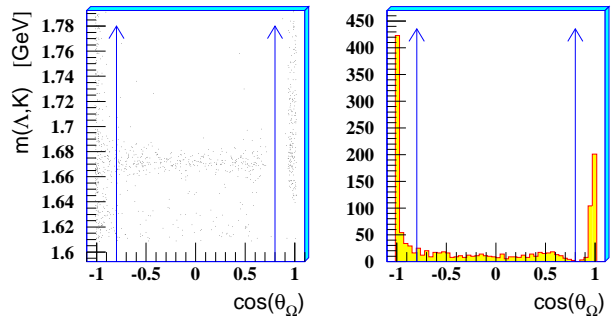}
\caption{{\em A destra:} correlazione tra la massa invariante $M(\Lambda,K)$\ ed il 
         coseno dell'angolo di decadimento ${\theta^*}_{\Omega}$, valutato 
	 nel sistema a riposo della $\Omega$. 
	 Si riconosce l'azione dell'anti-selezione delle $\Xi$, che rimuove 
	 le cascate per le quali $\cos({\theta^*}_{\Omega}) \approx 0.8$.
	 {\em A destra:} distribuzione di $\cos({\theta^*}_{\Omega})$.}
\label{Om8_15}
\end{center}
\end{figure}
Da tali grafici si deduce che la maggior parte del fondo geometrico delle 
cascate \`e concentrata nella regione $\cos({\theta^*}_{\Omega}) \approx 
\pm 1$. Il taglio impostato, indicato dalle frecce blu nella 
fig.~\ref{Om8_15}, elimina le candidate cascate per le quali 
$| \cos({\theta^*}_{\Omega}) | > 0.8$. L'effetto di questo taglio pu\`o essere 
dedotto confrontando il grafico bidimensionale delle masse invarianti 
$M(\Lambda,\pi)$--$M(\Lambda,K)$\ in fig.~\ref{Om8_7}.a, ottenuto 
senza l'applicazione di questo criterio, con quello in fig.~\ref{Om8_7}.c, 
dove il criterio \`e stato applicato.  
\item[$\blacktriangle$]
{\bf Antiselezione delle $\Xi$: $M(\Lambda,\pi)$} \\
In fig.~\ref{Om8_7}.a \`e mostrata la correlazione tra la massa invariante 
$M(\Lambda,K)$\ e quella $M(\Lambda,\pi)$\ per le candidate cascate cui 
sono applicati tutti i criteri di selezione per isolare il segnale delle 
$\Omega$, ad esclusione di quello sulle variabili $\cos({\theta^*}_{\Omega})$, 
${q_T}_{cas}$\ e $M(\Lambda,\pi)$\ che si vuole studiare. 
La distribuzione della massa invariante $M(\Lambda,\pi)$\ \`e mostrata in 
fig.~\ref{Om8_7}.b: il segnale delle $\Xi$\ domina completamente rispetto al 
debole segnale delle $\Omega$, distributo su tutto lo spettro di $M(\Lambda,\pi)$.  
L'effetto dell'applicazione del taglio sulla variabile $\cos({\theta^*}_{\Omega})$, 
gi\`a discussa, \`e visibile nelle fig.~\ref{Om8_7}.c, \ref{Om8_7}.d in cui 
tale taglio \`e adesso applicato: esso rimuove dunque le candidate vicine ai bordi 
dello spazio delle fasi $\left[  M(\Lambda,\pi) \;, \; M(\Lambda,K) \right]$. 
In particolare rimuove gran parte delle $\Xi$, che bisogna comunque anti-selezionare 
per isolare il segnale delle $\Omega$, e rimuove soprattutto l'accumulo 
di fondo geometrico presente, per un dato valore di $M(\Lambda,K)$, in 
corrispondenza del massimo valore di $ M(\Lambda,\pi) $\ ammesso 
cinematicamente.  
\begin{figure}[htb]
\begin{center}
\includegraphics[scale=0.42]{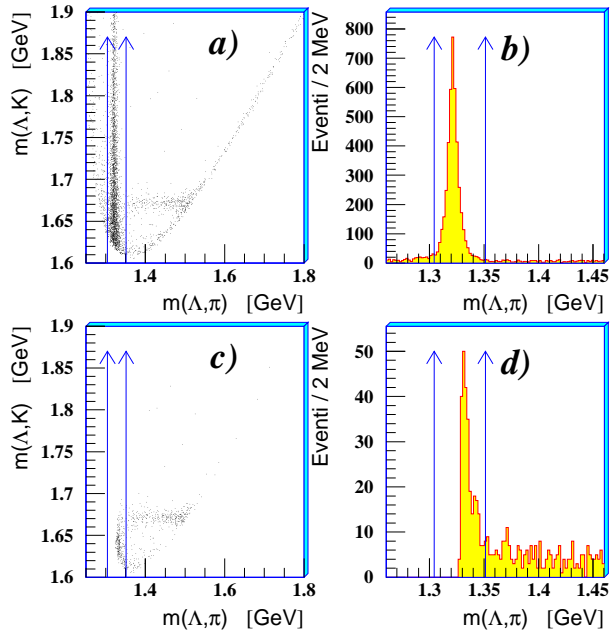}
\caption{Correlazione tra le masse invarianti $M(\Lambda,K)$\ e $M(\Lambda,\pi)$\ 
         ({\bf a}) e distribuzione di $M(\Lambda,\pi)$ ({\bf b}) per un 
	 campione di candidate cascate cui sono applicati i tagli geometrici 
	 per la selezione delle $\Omega$. 
	 In basso ({\bf c} e {\bf d}), si mostra l'effetto che l'applicazione 
	 del criterio di selezione $|\cos({\theta^*}_{\Omega})|<0.8$\ ha 
	 sulle distribuzioni in {\bf a)} ed in {\bf b)}.} 
\label{Om8_7}
\end{center}
\end{figure}
Per anti-selezionare le $\Xi$\ residue, gi\`a ridotte dall'applicazione del taglio 
sulla variabile $\cos({\theta^*}_{\Omega})$, si \`e imposto che la massa invariante 
$M(\Lambda,\pi)$\ cada al di fuori dell'intervallo $[m_{\Xi}-20 \, , \, m_{\Xi} + 30]$\ 
MeV/$c^2$\ ($m_{\Xi}=1321.3$\ MeV/$c^2$\ \`e la massa nominale della $\Xi$). Il limite 
inferiore di quest'ultimo criterio \`e 
%ridondante 
ineffettivo in quanto ridondante con il cretirio di selezione imposto 
su $\cos({\theta^*}_{\Omega})$.  Le frecce blu nella fig.~\ref{Om8_7} sono, al solito, 
poste in corrispondenza dei valori limite ammessi dal criterio di selezione. 
\item[$\blacktriangle$]
{\bf Impulso trasverso minimo delle  tracce di decadimento della cascata: 
	${q_T}_{cas}$} \\
Anche per il decadimento a due corpi di una $\Omega$\ (o di una $\Xi$) \`e possibile 
definire le variabili di Armenteros, $\alpha_{cas}$\ e ${q_T}_{cas}$, come fatto 
nel caso delle $V^0$. Il grafico di Armenteros per le candidate cascate selezionate 
applicando i tagli ottimizzati per le \PgOm\ e \PagOp\ \`e mostrato in fig.~\ref{Om8_17}.a. 
Si riconoscono i due archi di ellisse corrispondenti alle \PgOm\  
(per valori positivi di $\alpha_{cas}$) ed alle \PagOp\ (per valori negativi), mentre 
le \PgXm\ e le \PagXp\ sono gi\`a state anti-selezionate.    
\begin{figure}[htb]
\begin{center}
\includegraphics[scale=0.42]{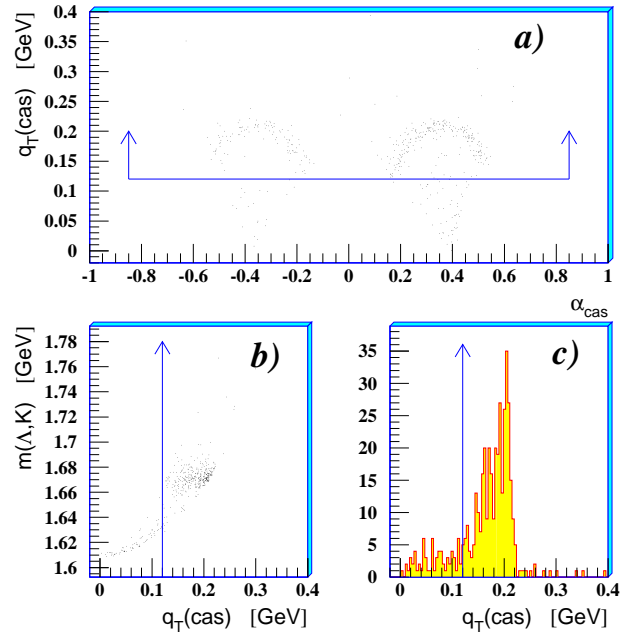}
\caption{{\bf a)} Grafico di Armenteros per il decadiemto delle candidate 
         cascate, selezionate per isolare il segnale delle \PgOm\ 
	 ($\alpha_{cas}\approx 0.15 \div 0.55$) e delle \PagOp\ 
	 ($\alpha_{cas}\approx -0.55 \div -0.15$). 
	 {\bf b)} Correlazione tra la massa invariante $M(\Lambda,K)$\ e 
	 l'impulso trasverso dei prodotti di decadimento della cascata, 
	 rispetto alla sua direzione (${q_T}_{cas}$). 
	 {\bf c)} Distribuzione della variabile ${q_T}_{cas}$.}
\label{Om8_17}
\end{center}
\end{figure}
Gran parte del fondo residuo \`e accumulato in corrispondenza di piccoli valori della 
variabile ${q_T}_{cas}$, la cui correlazione con la massa invariante $M(\Lambda,K)$\ \`e 
mostrata in fig.~\ref{Om8_17}.b, e la cui distribuzione \`e riportata 
in fig.~\ref{Om8_17}.c.  
Si \`e applicato un taglio sulla variabile ${q_T}_{cas}$, accettando le candidate 
cascate per le quali questa variabile risulti superiore a $120$\ MeV/$c$.  
Nel momento in cui si applica il taglio sulla massa invariante $M(\Lambda,K)$, 
che sar\`a di seguito discusso, l'applicazione della condizione 
${q_T}_{cas}>120$\ MeV/$c$\ diventa praticamente ineffettiva, perch\'e 
corrisponde ad eliminare candidate cascate con bassa massa invariante $M(\Lambda,K)$, 
al di sotto del limite inferiore infine utilizzato per isolare il segnale delle 
$\Omega$.
Tale taglio ha dunque ragion d'essere unicamente perch\'e ha permesso di meglio 
definire i limiti degli altri criteri di selezione, rimuovendo gran parte del 
fondo che si trova al di fuori dell'intervallo di massa invariante  $M(\Lambda,K)$, 
il quale, se altrimenti non rimosso, avrebbe artificialmente alterato il rapporto 
segnale su fondo, che \`e il vero parametro di qualit\`a nella definizione dei criteri di 
selezione geometrici.  
\end{itemize}
Di seguito sono riassunti i criteri di selezione per le \PgOm\ e le \PagOp\ del campione 
di dati raccolto nell'anno 2000, appena discussi, 
per quelle del campione di dati del 1998, e per 
le \PgXm\ e \PagXp\ estratte dal campione di dati del 1998.  

\begin{center}
 {\bf Criteri di selezione per le \PgOm\ e \PagOp\ in Pb-Pb a 160 A GeV/$c$\ (2000)}
\end{center}
\begin{itemize}
\item[{\bf $\Omega$.a)}] tutte le tracce del decadimento attraversano il primo,
                          l'ottavo ed il nono piano di pixel.
\item[{\bf $\Omega$.b)}] decadimento della cascata di tipo {\em cowboy}
\item[{\bf $\Omega$.c)}] -40. cm $<x_{cas}<$\ 0. cm
\item[{\bf $\Omega$.d)}] $close_{cas} < $\  0.030 cm  
\item[{\bf $\Omega$.e)}] ${q_T}_{cas} > $\ 0.120 GeV/$c$
\item[{\bf $\Omega$.f)}] $\left[ \frac{{b_y}_{cas}}{3\cdot\sigma_y} \right]^2 + 
                 \left[ \frac{{b_z}_{cas}}{3\cdot\sigma_z} \right]^2 < 1 $
\item[{\bf $\Omega$.g)}] $| \cos\theta^*_{\Omega} | < 0.8 $
\item[{\bf $\Omega$.h)}] $ M(\Lambda,\pi) - m_{\Xi} >$\ 0.030  GeV/$c^2$\ 
		\hspace{0.6cm} {\em OR} \hspace{0.6cm}
                $ m_{\Xi} - M(\Lambda,\pi) >$\ 0.020  GeV/$c^2$ 
\item[{\bf $\Omega$.i)}] decadimento della $V^0$\ di tipo {\em cowboy}
\item[{\bf $\Omega$.j)}] -27. cm $<x_{V^0}<$\ 0. cm
\item[{\bf $\Omega$.k)}] $close_{V^0} < $\  0.025 cm
\item[{\bf $\Omega$.l)}] 0.030 GeV/$c$ $< {q_T}_{V^0} < $\ 0.40 GeV/$c$
\item[{\bf $\Omega$.m)}] $ x_{V^0} - x_{cas} > $\ 0.2 cm
\item[{\bf $\Omega$.n)}] $ \alpha  > 0.45$\ ($\Omega^-$)  
		\hspace{0.6cm} {\em OR} \hspace{0.6cm}
		$\alpha  < -0.45$\ ($\bar{\Omega}^+$)
\item[{\bf $\Omega$.o)}] $|M(p,\pi) - m_{\Lambda}| < $\ 0.008 GeV/$c^2$
\item[{\bf $\Omega$.p)}] $|\phi_{V^0}| > 0.15 $\ rad  
		\hspace{0.6cm} {\em OR} \hspace{0.6cm}
		$|\phi_{cas}| > 0.15 $\ rad
\item[{\bf $\Omega$.q)}] $ |M(\Lambda,K) - m_{\Omega}| <$\ 0.017  GeV/$c^2$
\end{itemize}

\begin{center}
{\bf Criteri di selezione per le \PgOm\ e \PagOp\ in Pb-Pb a 160 A GeV/$c$\ (1998)}
\end{center}
\begin{itemize}
\item[{\bf $\Omega$.a')}] tutte le tracce del decadimento attraversano la
                      parte compatta del telescopio (primo e nono piano).
\item[{\bf $\Omega$.b')}] decadimento della cascata di tipo {\em cowboy}
\item[{\bf $\Omega$.c')}] -40. cm $<x_{cas}<$\ 0. cm 
\item[{\bf $\Omega$.d')}] $close_{cas} < $\  0.038 cm 
\item[{\bf $\Omega$.e')}] ${q_T}_{cas} > $\ 0.120 GeV/$c$
\item[{\bf $\Omega$.f')}] $-2.0<\frac{{b_y}_{cas}*Segno(B)}{\sigma_y}< 2.5 $\
                     ($\Omega^-$)  \hspace{0.4cm} {\em OR} \hspace{0.4cm} 
		     $-2.5<\frac{{b_y}_{cas}*Segno(B)}{\sigma_y}< 
		     2.0 $~($\bar{\Omega}^+$) 
\item[{\bf $\Omega$.g')}] 
              $   \left| {b_z}_{cas}\right|< {2.5 \sigma_z} $
\item[{\bf $\Omega$.h')}] $| \cos\theta^*_{\Omega} | < 0.75 $
\item[{\bf $\Omega$.i')}] $ M(\Lambda,\pi) - m_{\Xi} >$\ 0.030  GeV/$c^2$\ 
		\hspace{0.6cm} {\em OR} \hspace{0.6cm}
                $ m_{\Xi} - M(\Lambda,\pi) >$\ 0.020  GeV/$c^2$ 
\item[{\bf $\Omega$.j')}] decadimento della $V^0$\ di tipo {\em cowboy}
\item[{\bf $\Omega$.k')}] -27. cm $<x_{V^0}<$\ 0. cm
\item[{\bf $\Omega$.l')}] $close_{V^0} < $\  0.025 cm
\item[{\bf $\Omega$.m')}] 0.030 GeV/$c$ $< {q_T}_{V^0} < $\ 0.40 GeV/$c$
\item[{\bf $\Omega$.n')}] $ x_{V^0} - x_{cas} > $\ 0
\item[{\bf $\Omega$.o')}] $ \alpha  > 0.45$\ ($\Omega^-$)  
		\hspace{0.6cm} {\em OR} \hspace{0.6cm}
		$\alpha  < -0.45$\ ($\bar{\Omega}^+$)
\item[{\bf $\Omega$.p')}] $M(p,\pi) - m_{\Lambda} < $\ 0.010 GeV/$c^2$\
                \hspace{0.6cm} {\em AND} \hspace{0.6cm}
		                $m_{\Lambda}- M(p,\pi) < $\ 0.0072 GeV/$c^2$
\item[{\bf $\Omega$.q')}] $|\phi_{V^0}| > 0.15 $\ rad  
		\hspace{0.6cm} {\em OR} \hspace{0.6cm}
		$|\phi_{cas}| > 0.15 $\ rad
\item[{\bf $\Omega$.r')}] $ |M(\Lambda,K) - m_{\Omega}| <$\ 0.017  GeV/$c^2$
\end{itemize}

\begin{center}
 {\bf Criteri di selezione per le \PgXm\ e \PagXp\ in Pb-Pb a 160 A GeV/$c$\ (1998)}
\end{center}
\begin{itemize}
\item[{\bf $\Xi$.a)}] tutte le tracce del decadimento attraversano la
                      parte compatta del telescopio (primo e nono piano).
\item[{\bf $\Xi$.b)}] decadimento della cascata di tipo {\em cowboy}
\item[{\bf $\Xi$.c)}] -40. cm $<x_{cas}<$\ -3. cm
\item[{\bf $\Xi$.d)}] $close_{cas} < $\  0.050 cm  
\item[{\bf $\Xi$.e)}] $\left[ \frac{{b_y}_{cas}}{3\cdot\sigma_y} \right]^2 + 
                 \left[ \frac{{b_z}_{cas}}{3\cdot\sigma_z} \right]^2 < 1 $
\item[{\bf $\Xi$.f)}] $ \cos\theta^*_{\Xi}  > -0.9 $
\item[{\bf $\Xi$.g)}] decadimento della $V^0$\ di tipo {\em cowboy}
\item[{\bf $\Xi$.h)}] -30. cm $<x_{V^0}<$\ 0. cm
\item[{\bf $\Xi$.i)}] $close_{V^0} < $\  0.035 cm
\item[{\bf $\Xi$.j)}] 0.020 GeV/$c$ $< {q_T}_{V^0} < $\ 0.40 GeV/$c$
\item[{\bf $\Xi$.k)}] $ x_{V^0} - x_{cas} > $\ 0
\item[{\bf $\Xi$.l)}] $ \alpha  > 0.45$\ ($\Xi^-$)  
		\hspace{0.6cm} {\em OR} \hspace{0.6cm}
		$\alpha  < -0.45$\ ($\bar{\Xi}^+$)
\item[{\bf $\Xi$.m)}] $M(p,\pi) - m_{\Lambda} < $\ 0.015 GeV/$c^2$\ 
                \hspace{0.6cm} {\em AND} \hspace{0.6cm}
		$m_{\Lambda}- M(p,\pi) < $\ 0.012 GeV/$c^2$
\item[{\bf $\Xi$.n)}] $|\phi_{V^0}| > 0.20 $\ rad  
		\hspace{0.6cm} {\em OR} \hspace{0.6cm}
		$|\phi_{cas}| > 0.20 $\ rad
\item[{\bf $\Xi$.o)}] $({b_y}_{V^0}-{b_y}_{cas})*Segno(B) < -0.200 $\ cm ($\Xi^-$)
                \hspace{0.6cm} {\em OR} \hspace{0.6cm}
		$({b_y}_{V^0}-{b_y}_{cas})*Segno(B) >  0.200 $\ cm ($\bar{\Xi}^+$)
\item[{\bf $\Xi$.p)}] $({b_y}_{t}-{b_y}_{cas})*Segno(B) > 0 $\ ($\Xi^-$)
		\hspace{0.6cm} {\em OR} \hspace{0.6cm}
		$({b_y}_{t}-{b_y}_{cas})*Segno(B) < 0 $\ ($\bar{\Xi}^+$)
\item[{\bf $\Xi$.q)}] $ M(\Lambda,\pi) - m_{\Xi} <$\ 0.020  GeV/$c^2$\
                \hspace{0.6cm} {\em AND} \hspace{0.6cm}
                $ m_{\Xi} - M(\Lambda,\pi) <$\ 0.015  GeV/$c^2$
\end{itemize}
\newpage
\subsection{Spettri di massa invariante di \PgOm, \PagOp, \PgXm\ ed \PagXp }
\begin{figure}[p]
\begin{center}
\includegraphics[scale=0.42]{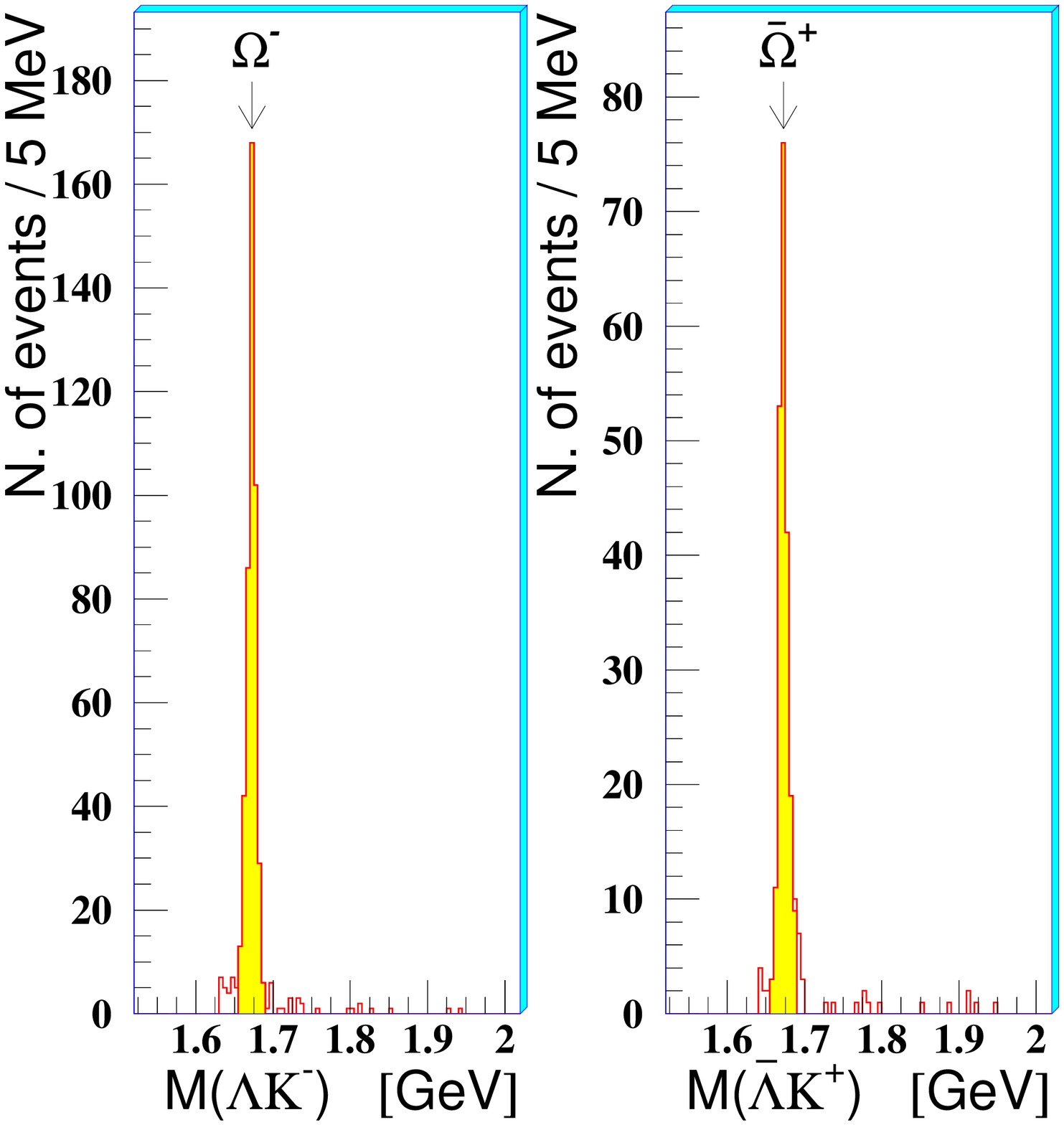}\\
\includegraphics[scale=0.42]{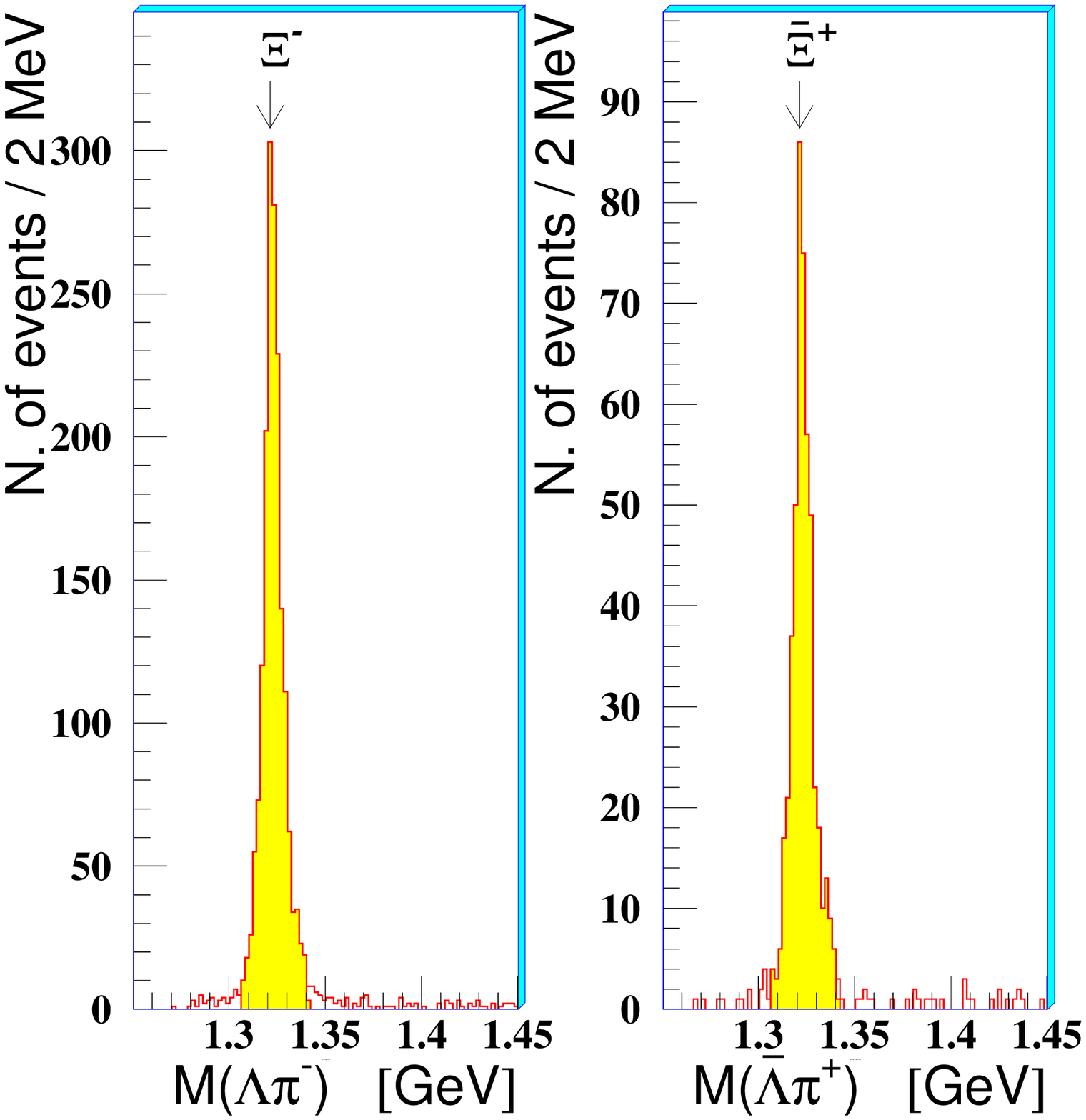}
\caption{Spettri finali di massa invariante delle diverse specie di 
         cascate nelle interazioni Pb-Pb a 160 A Gev/$c$. 
         Per le $\Xi^-$\ e le $\bar{\Xi}^+$\ si \`e considerato il 
         campione di dati raccolto nell'anno 1998, per le $\Omega^-$\ ed  
         $\bar{\Omega}^+$\ entrambi i campioni del 1998 e del 2000. La parte in 
	 giallo degli spettri rappresenta il segnale selezionato per l'analisi 
	 successiva, discussa nei prossimi capitoli.}
\label{CascadeSpectra}
\end{center}
\end{figure}

In fig.~\ref{CascadeSpectra} sono mostrati gli spettri finali di massa invariante 
$M(\Lambda,K^-)$, $M(\bar{\Lambda},K^+)$, $M(\Lambda,\pi^-)$\ e $M(\bar{\Lambda},\pi^+)$\ 
per i campioni di, rispettivamente,  \PgOm, \PagOp, \PgXm\ ed \PagXp\ selezionati con le  
tecniche discusse nel paragrafo precedente. 
Il segnale delle  \PgOm\ ed \PagOp\ \`e stato ottenuto considerando 
tutta la statistica disponibile, unendo quindi i dati raccolti nell'anno 1998 e 
nell'anno 2000; il segnale delle \PgXm\ ed \PagXp, prodotte pi\`u cospicuamente, 
proviene invece dal solo campione del 1998. Sebbene la ricostruzione e selezione delle 
$\Xi$\ dell'anno 2000 sia ormai ultimata, il calcolo delle correzioni per accettanza ed 
efficienza \`e ancora agli inizi e pertanto nel seguito si considerer\`a il solo  
campione di $\Xi$\ del 1998.
\newline
Gli spettri sono centrati sul valore nominale di massa delle rispettive particelle, 
presentano un fondo residuo inferiore al 5\% per tutte le specie, ed hanno una 
larghezza a met\`a altezza di $9 \div 12 $\ MeV/$c^2$.  
In tab. 3.2  sono riassunti i parametri degli spettri delle cascate di 
fig.~\ref{CascadeSpectra}
\begin{table}[h]
\label{tab3.2}
\begin{center}
\begin{tabular}{|c|c|c|c|c|} \hline
    & Particelle & $\mu$\ (MeV/$c^2$) & $\sigma$\ (MeV/$c^2$) & $FWHM$\ (MeV/$c^2$)  \\ \hline
\PgOm  & $432$  & $1672.3$ & $5.7 $ & $ 12 $ \\
\PagOp & $193$  & $1672.7$ & $5.4 $ & $ 12 $  \\
\PgXm  & $1744$ & $1322.7$ & $4.7 $ & $ 9 $  \\
\PagXp & $484$  & $1322.6$ & $4.8 $ & $ 9 $  \\ \hline
\end{tabular}
\end{center}
\caption{Numero di particelle selezionate, valor medio e  
         sigma del {\em ``best fit''} con una gaussiana sul picco, e $FWHM$\ degli 
	 spettri di massa invariante di fig.~\ref{CascadeSpectra}.}
\end{table}

\section{Selezione delle particelle strane in interazioni Pb-Pb a 40 A GeV/$c$}
\begin{figure}[hbt]
\begin{center}
\includegraphics[scale=0.42]{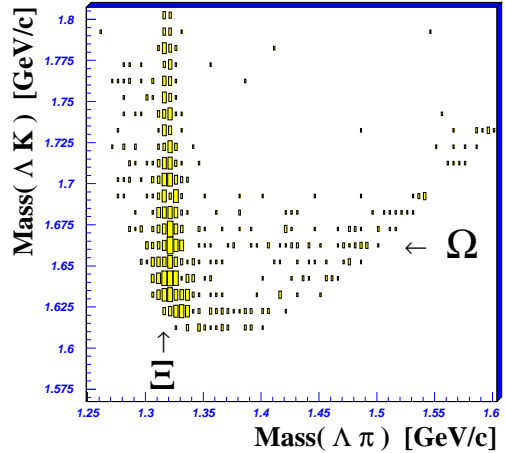}
\caption{Correlazione tra le masse invarianti $M(\Lambda,K)$\ e $M(\Lambda,\pi)$\  
         per le candidate cascate prodotte nell'interazione Pb-Pb a 40 A GeV/$c$,  
	 cui sono stati applicati i criteri di selezione delle $\Xi$.} 
\label{ScatterMass40}
\end{center}
\end{figure}

\begin{figure}[p]
\begin{center}
\includegraphics[scale=0.36]{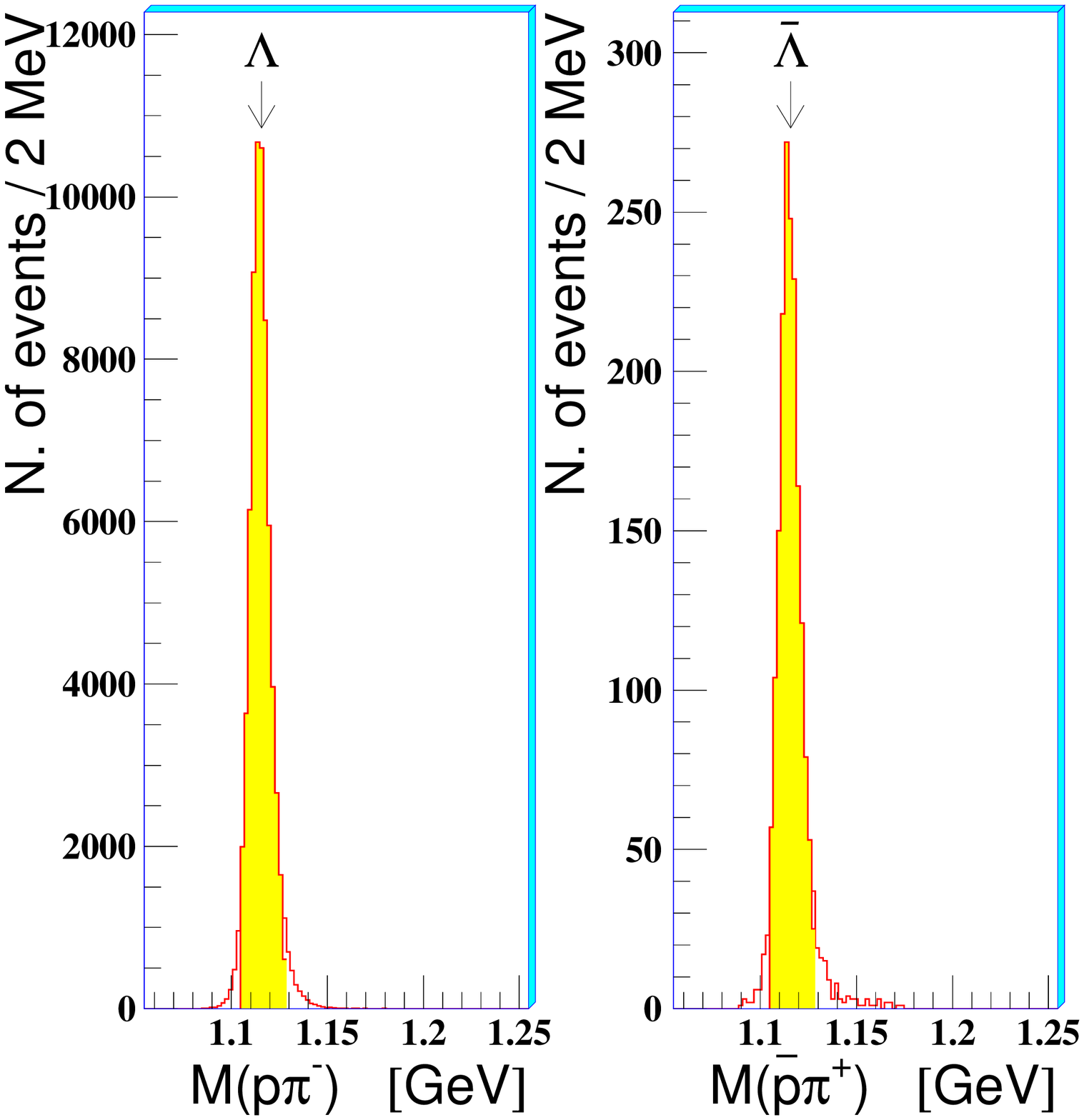}\\
\includegraphics[scale=0.36]{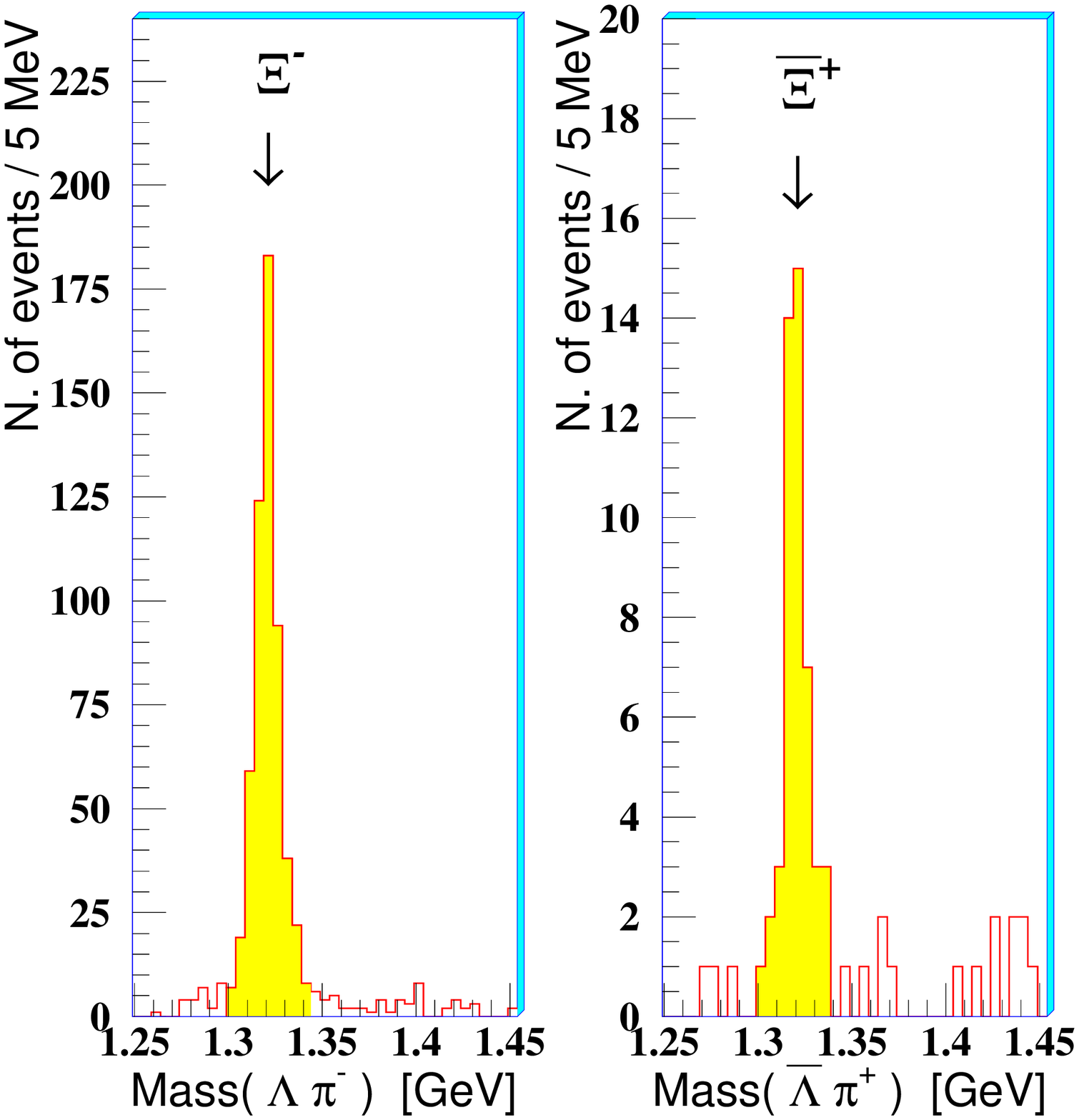}
\includegraphics[scale=0.36]{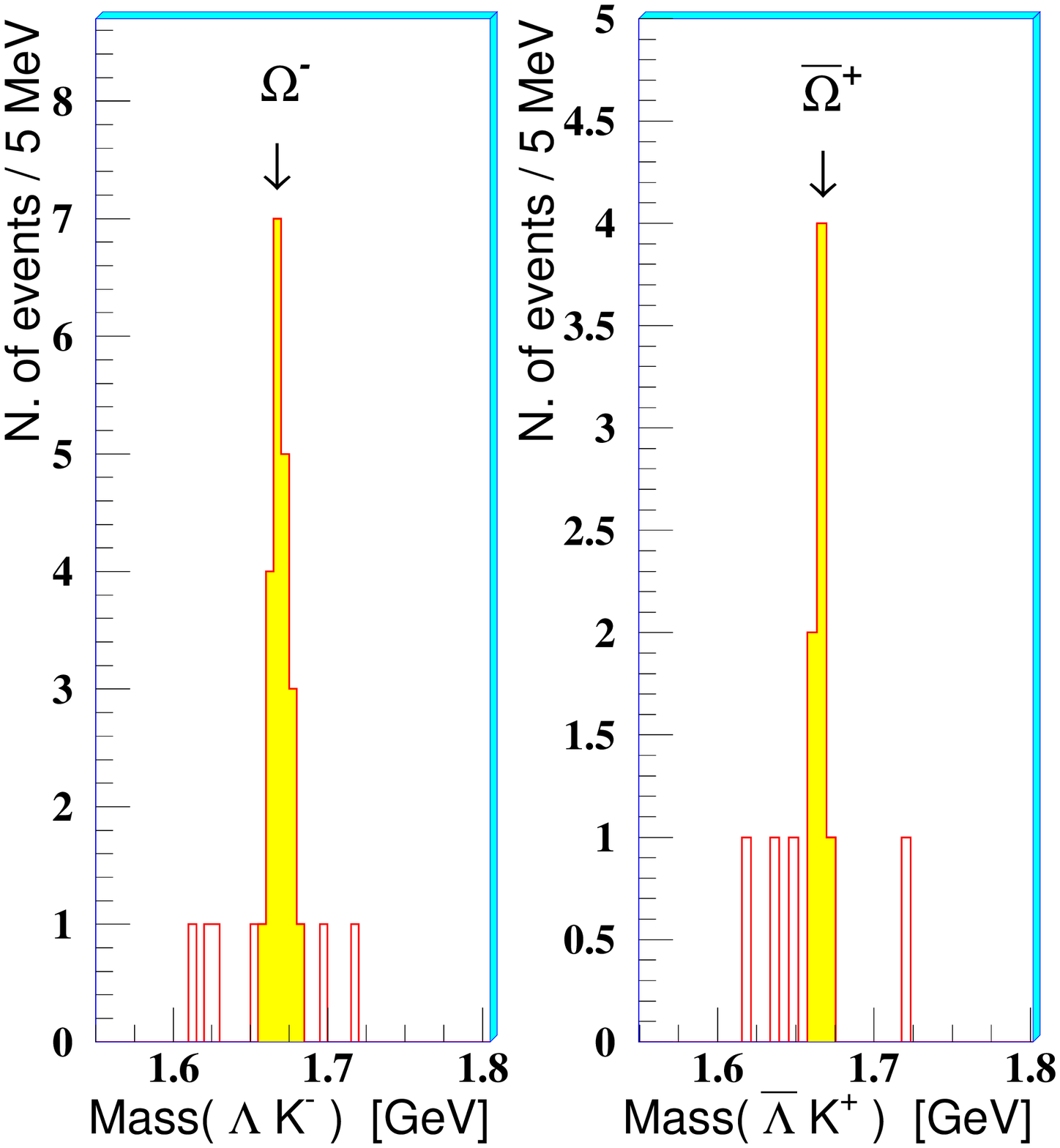}
\caption{Spettri di massa invariante degli iperoni ricostruiti 
	 nell'interazione Pb-Pb a 40 A GeV/$c$. La parte in
	 giallo degli spettri rappresenta il segnale selezionato per l'analisi
         successiva, discussa nei prossimi capitoli.}
\label{InvMass40}
\end{center}
\end{figure}

Non si intende affrontare anche la discussione sulla selezione delle cascate e delle $V^0$\ nelle  
collissioni a 40 A GeV/$c$, ma ci si limiter\`a ad evidenziare le analogie e le differenze 
con le tecniche di selezione utilizzate a pi\`u alta energia e si riporter\`a sullo 
stato dell'analisi.  
\newline
Come esposto nel {\em paragrafo 2.7}, nel corso dell'anno 1999 l'esperimento NA57 ha 
raccolto dati all'energia di 40 GeV/$c^2$\ per nucleone sia nell'interazione Pb-Pb sia  
in quella di riferimento p-Be. La presa dati col fascio di protone, avvenuta in realt\`a solo 
per un breve arco  temporale della presa dati del 1999, 
%che era inizialmente dedicata allo studio  delle sole collisioni Pb-Pb, 
\`e stata completata da una seconda presa dati terminata 
nel dicembre 2001. 
Per le collisioni p-Be dunque, a parte qualche analisi molto preliminare 
condotta sulle $V^0$\ dell'anno 1999, si sta al momento 
ultimando la ricostruzione degli eventi del secondo campione di dati 
(2001)  prima di passare all'analisi completa di queste interazioni. Non saranno quindi 
presentati risultati per le collisioni p-Be a 40 GeV/$c$.  
\newline
L'analisi delle collisioni Pb-Pb a 40 A GeV/$c$\ \`e invece in via di completamento. 
Per quanto riguarda le cascate ($\Xi$\ ed $\Omega$), i primi risultati, gi\`a corretti 
per le perdite dovute all'accettanza ed all'efficienza dei rivelatori e dei programmi 
di ricostruzione, sono stati presentati recentemente~\cite{EliaQM02}. Per le $V^0$, invece, 
l'analisi sin qui condotta dalla Collaborazione ha riguardato la sola selezione  
del segnale delle \PgL\ e \PagL, mentre il calcolo delle correzioni deve essere ancora 
affrontato.  
\newline
Per quanto riguarda la disposizione sperimentale dei rivelatori, le condizioni rispetto 
al caso del fascio di Pb a 160 GeV/$c$\ per nucleone cambiano in quanto si vuole 
mantenere l'accettanza delle particelle strane concentrata intorno a valori di rapidit\`a 
prossimi alla rapidit\`a 
del centro di massa ($y_{cm} \simeq 2.22$ in Pb-Pb a 40 A GeV/$c$) e con impulsi trasversi 
medio-alti non troppo discosti da quelli misurati a 160 A GeV/$c$:  
il primo piano di pixel \`e quindi posto a circa $40$\ cm  
dal bersaglio, che \`e lo stesso utilizzato nelle collisioni a 160 GeV,  
ed il telescopio \`e inclinato di $72$\ mrad rispetto 
alla linea del fascio. Inoltre nel telescopio {\em dieci} piani di pixel sono distribuiti 
entro $30$\  cm  e nel {\em ``lever arm''} operano solo i rivelatori a $\mu$strip 
doppia faccia, con una logica simile a quella adottata poi anche nell'anno 2000.  
\newline
Per quanto riguarda le tecniche di identificazione delle diverse specie, le stesse variabili 
introdotte e gli stessi strumenti di analisi sviluppati a $160$\ GeV sono ancora validi ed 
applicabili per queste collisioni; ovviamente cambiano i valori ottimali dei diversi tagli, 
come del resto si \`e anche trovato analizzando separatamente i campioni di dati a 160 GeV
relativi alla presa dati del 1998 ed a quella del 2000.  
\newline
Sei poi si entra nei dettagli, la prima osservazione \`e  che le particelle prodotte 
sono mediamente pi\`u lente, rispetto al caso dei 160 GeV, e curvano quindi di pi\`u nel 
telescopio, sotto l'azione del campo magnetico; nonostante ci\`o, la risoluzione sperimentale 
sulla misura degli spettri di massa invariante delle diverse particelle strane \`e 
leggermente peggiore, forse in quanto le tracce pi\`u lente risentono maggiormente 
della diffusione multipla lungo la loro traiettoria.  In compenso,   
la molteplicit\`a di particelle cariche \`e pi\`u bassa, e quindi la ricostruzione delle 
particelle strane avviene, entro il telescopio, in un ambiente relativamente meno affollato 
e quindi pi\`u favorevole. 
Infine, la determinazione del vertice primario dell'interazione, calcolato 
ancora su base {\em run per run}, presenta una precisone confrontabile con quella 
trovata a 160 GeV, dato che, in entrambi i casi, un numero molto elevato di tracce 
viene utilizzato per la sua determinazione.  
\newline
Alla fine di questo paragrafo sono elencati i criteri di selezione adottati per isolare 
i segnali delle \PgL, \PgXm, \PgOm\ e delle relative anti-particelle; in fig.~\ref{ScatterMass40} 
\`e mostrata la correlazione tra la massa invariante $M(\Lambda,K)$\ e la massa invariante 
$M(\Lambda,\pi)$\ per le candidate cascate dopo l'applicazione dei criteri di selezione di 
tipo geometrico~\footnote{Si mantengono anche qui le definizioni di {\em tagli geometrici} 
e {\em tagli cinematici}, come introdotto discutendo la selezione delle cascate 
nell'interazione Pb-Pb a 160 A GeV/$c$.}; in fig.~\ref{InvMass40}, infine, sono mostrati 
gli spettri di massa invariante delle diverse specie di particelle strane selezionate 
nelle collisioni Pb-Pb a 40 A GeV/$c$.
Gli spettri sono centrati ancora sui valori nominali delle masse delle diverse 
particelle e presentano una percentuale di fondo residuo che, sebbene debba essere 
ancora stimata con la dovuta precisione, risulta di poco superiore a quella degli 
spettri a 160 GeV, specialmente per le cascate, probabilmente in quanto la 
resa di produzione di queste particelle \`e pi\`u bassa a minor energia.  

\begin{center}
 {\bf Criteri di selezione preliminari per le \PgL\ e \PagL\ in Pb-Pb a 40 A GeV/$c$\ (1999)}
\end{center}
\begin{itemize}
\item[{\bf $\Lambda$.a)}] tutte le tracce del decadimento attraversano il primo
                      ed il nono piano di pixel.
\item[{\bf $\Lambda$.b)}] decadimento della $V^0$\ di tipo {\em cowboy}
\item[{\bf $\Lambda$.c)}] -30. cm $<x_{V^0}<$\ -20. cm
\item[{\bf $\Lambda$.d)}] $close_{V^0} < $\  0.020 cm
\item[{\bf $\Lambda$.e)}] 0.040 GeV/$c$ $< {q_T}_{V^0} < $\ 0.15 GeV/$c$
\item[{\bf $\Lambda$.f)}] $\left[ \frac{{b_y}_{V^0}}{1.5\cdot\sigma_y} \right]^2 +
                           \left[ \frac{{b_z}_{V^0}}{2\cdot\sigma_z} \right]^2 < 1 $
\item[{\bf $\Lambda$.g)}] $ 0.45 <\alpha  < 0.70$\ ($\Lambda$)
                	  \hspace{0.6cm} {\em OR} \hspace{0.6cm}
		          $ -0.70 <\alpha  < -0.45$\ ($\bar{\Lambda}$)
\item[{\bf $\Lambda$.h)}] $ |M(\pi^+,\pi^-) - m_{K^0}| >$\ 0.040  GeV/$c^2$\
\item[{\bf $\Lambda$.i)}] $|\phi_{V^0}| > $\ 0.200 rad
\item[{\bf $\Lambda$.j)}] $\left| M(p,\pi) - m_{\Lambda} \right|  < $\
                          0.012 GeV/$c^2$\
\end{itemize}

\begin{center}
 {\bf Criteri di selezione per le \PgXm\ e \PagXp\ in Pb-Pb a 40 A GeV/$c$\ (1999)}
\end{center}
\begin{itemize}
\item[{\bf $\Xi$.a)}] tutte le tracce del decadimento attraversano il primo
                      ed il nono piano di pixel.
\item[{\bf $\Xi$.b)}] decadimento della cascata di tipo {\em cowboy}
\item[{\bf $\Xi$.c)}] -50. cm $<x_{cas}<$\ -20. cm
\item[{\bf $\Xi$.d)}] $close_{cas} < $\  0.075 cm  
\item[{\bf $\Xi$.e)}] $\left[ \frac{{b_y}_{cas}}{3\cdot\sigma_y} \right]^2 + 
                 \left[ \frac{{b_z}_{cas}}{2\cdot\sigma_z} \right]^2 < 1 $
%\item[{\bf $\Xi$.e)}] $ | \cos\theta^*_{\Xi} | < 0.75 $
\item[{\bf $\Xi$.f)}] decadimento della $V^0$\ di tipo {\em cowboy}
\item[{\bf $\Xi$.g)}] -30. cm $<x_{V^0}<$\ -20. cm
\item[{\bf $\Xi$.h)}] $close_{V^0} < $\  0.045 cm
\item[{\bf $\Xi$.i)}] 0.040 GeV/$c$ $< {q_T}_{V^0} < $\ 0.120 GeV/$c$
\item[{\bf $\Xi$.j)}] $ x_{V^0} - x_{cas} > $\ 1.25 cm
\item[{\bf $\Xi$.k)}] $ \alpha  > 0.45$\ ($\Xi^-$)  
		\hspace{0.6cm} {\em OR} \hspace{0.6cm}
		$\alpha  < -0.45$\ ($\bar{\Xi}^+$)
\item[{\bf $\Xi$.l)}] $| M(p,\pi) - m_{\Lambda}| < $\ 0.016 GeV/$c^2$\ 
\item[{\bf $\Xi$.m)}] ${b_y}_{V^0}*Segno(B) < -0.15 $\ cm ($\Xi^-$)
                \hspace{0.6cm} {\em OR} \hspace{0.6cm}
		${b_y}_{V^0}*Segno(B) >  0.15 $\ cm ($\bar{\Xi}^+$)
\item[{\bf $\Xi$.n)}] ${b_y}_{t}*Segno(B) > 0.25 $\ cm ($\Xi^-$)
		\hspace{0.6cm} {\em OR} \hspace{0.6cm}
		${b_y}_{t}*Segno(B) < -0.25 $\ cm ($\bar{\Xi}^+$)
\item[{\bf $\Xi$.o)}] $ |M(\Lambda,\pi) - m_{\Xi}| <$\ 0.020  GeV/$c^2$\
\end{itemize}

\begin{center}
{\bf Criteri di selezione per le \PgOm\ e \PagOp\ in Pb-Pb a 40 A GeV/$c$\ (1999)}
\end{center}
\begin{itemize}
\item[{\bf $\Omega$.a)}] tutte le tracce del decadimento attraversano il primo
                         ed il nono piano di pixel.
\item[{\bf $\Omega$.b)}] decadimento della cascata di tipo {\em cowboy}
\item[{\bf $\Omega$.c)}] -47.5 cm $<x_{cas}<$\ -20. cm
\item[{\bf $\Omega$.d)}] $close_{cas} < $\  0.075 cm
\item[{\bf $\Omega$.e)}] $\left[ \frac{{b_y}_{cas}}{3\cdot\sigma_y} \right]^2 +
                 \left[ \frac{{b_z}_{cas}}{2\cdot\sigma_z} \right]^2 < 1 $
\item[{\bf $\Omega$.f)}] $ | \cos\theta^*_{\Omega} | < 0.75 $\
\item[{\bf $\Omega$.g)}] $ {p_t}_{cas} > $  1 GeV/$c$
\item[{\bf $\Omega$.h)}] $ M(\Lambda,\pi) - m_{\Xi} >$\ 0.029  GeV/$c^2$\
\item[{\bf $\Omega$.i)}] decadimento della $V^0$\ di tipo {\em cowboy}
\item[{\bf $\Omega$.j)}] -30. cm $<x_{V^0}<$\ -20. cm
\item[{\bf $\Omega$.k)}] $close_{V^0} < $\  0.045 cm
\item[{\bf $\Omega$.l)}] 0.040 GeV/$c$ $< {q_T}_{V^0} < $\ 0.120 GeV/$c$
\item[{\bf $\Omega$.m)}] $ x_{V^0} - x_{cas} > $\ 1.25 cm
\item[{\bf $\Omega$.n)}] $ \alpha  > 0.45$\ ($\Omega^-$)
                \hspace{0.6cm} {\em OR} \hspace{0.6cm}
                $\alpha  < -0.45$\ ($\bar{\Omega}^+$)
\item[{\bf $\Omega$.o)}] $| M(p,\pi) - m_{\Lambda}| < $\ 0.016 GeV/$c^2$\
\item[{\bf $\Omega$.p)}] ${b_y}_{V^0}*Segno(B) < -0.25 $\ cm ($\Omega^-$)
                \hspace{0.6cm} {\em OR} \hspace{0.6cm}
                ${b_y}_{V^0}*Segno(B) >  0.25 $\ cm ($\bar{\Omega}^+$)
\item[{\bf $\Omega$.q)}] ${b_y}_{t}*Segno(B) > 0.25 $\ cm ($\Omega^-$)
                \hspace{0.6cm} {\em OR} \hspace{0.6cm}
                ${b_y}_{t}*Segno(B) < -0.25 $\ cm ($\bar{\Omega}^+$)
\item[{\bf $\Omega$.r)}] $|{b_z}_{V^0} |>$\ 0.1 cm
\item[{\bf $\Omega$.s)}] $ |M(\Lambda,K) - m_{\Omega}| <$\ 0.015  GeV/$c^2$\
\end{itemize}

%% file: cap4/cap4.tex
\chapter{Analisi dei dati} 
\section{Introduzione}
In questo capitolo  si affronter\`a lo studio dei segnali fisici per le 
particelle strane ottenute applicando i criteri di selezione discussi 
nel capitolo precedente. Si discuter\`a inizialmente una tecnica 
sviluppata per misurare con precisione il fondo geometrico residuo sotto 
i segnali delle $V^0$\ (\PgL, \PagL\ e \PKzS), studio motivato principalmente 
dalla necessit\`a di controllare precisamente gli eventuali errori sistematici 
per queste particelle, per le quali gli errori statistici risultano molto piccoli. 
%di quelli per le cascate (\PgXm, \PgOm\ e rispettive particelle).  
\newline
Si discuter\`a la procedura di correzione per tenere conto delle perdite
di particelle strane dovute all'accettanza geometrica dell'apparato
sperimentale, all'efficienza dei rivelatori e dei programmi di ricostruzione,
ed infine ai tagli introdotti nella procedura di identificazione dei
diversi tipi di particelle.
\newline
Verrano 
%riportati il numero di particelle strane identificate analizzando 
%l'intera statistica, al momento disponibile per l'analisi, accumulata 
%dall'esperimento NA57 nelle collisioni Pb-Pb a 160 ed a 40 $A$\ GeV/c, e 
discusse 
le regioni cinematiche scelte per lo studio dei differenti segnali. 
Particolare enfasi si dar\`a alla determinazione della finestra di accettanza 
per le particelle con pi\`u bassa statistica, le $\Omega$, per le quali 
\`e stato compiuto uno studio Monte Carlo dedicato, ed alla stabilit\`a 
delle quantit\`a misurate (tasso di produzione e spettri di massa trasversa) 
per le altre specie.  
\newline
All'interno delle finestre di accettanza determinate per le diverse specie, 
con una procedura di {\em ``best fit''} si misureranno gli spettri di massa 
trasversa ed il tasso di produzione, applicando le correzioni precedentemente 
calcolate. I parametri del {\em ``best fit''}~\footnote{In questo capitolo la 
pendenza inversa degli spettri di massa trasversa, ricavata 
dal {\em ``best fit''}, verr\`a sempre indicata con $T_a$\ 
(``temperatua apparente'').} 
verranno quindi utilizzati per 
estrapolare il tasso di produzione delle diverse particelle 
ad una regione cinematica pi\`u facilmente confrontabile con quella degli altri 
esperimenti. 
\newline
Le possibili sorgenti di errori sistematici intodotte nelle varie fasi di 
analisi del segnale saranno evidenziate di volta in volta e saranno discusse 
le procedure che hanno portato al calcolo dei relativi contributi.  
\section{Studio del fondo geometrico residuo nel segnale delle $V^0$}
\subsection{Introduzione}
Nel paragrafo 3.3 sono state discusse le tecniche di selezione ed identificazione 
delle particelle \PgL, \PagL\ e \PKzS; gli spettri di massa per un campione di 
queste particelle, selezionate con tali tecniche, sono state mostrate nella 
fig.~\ref{FinalSpectra}; infine, nella tabella 3.1 sono state riassunte le 
caratteristiche degli spettri.  Nel determinare i criteri di selezione per 
le tre specie di particelle considerate, si \`e cercato di mantenere il rapporto 
segnale su fondo residuo il pi\`u alto possibile, tenendo tuttavia in 
considerazione gli effetti che i diversi tagli introdotti comportano nelle 
successive fasi di analisi, che verranno descritte in questo capitolo.  
\newline
Il fondo residuo pu\`o essere descritto con una funzione analitica, 
tipicamente un polinomio, ma questa descrizione \`e insoddisfacente poich\'e 
l'adattamento della funzione ai punti sperimentali (il {\em ``best fit''}) 
viene eseguito sulle code delle distribuzioni di massa invariante, 
l\`i dove si giudica che il segnale fisico sia ormai diventato trascurabile. 
Vi \`e dunque un aproccio di tipo soggettivo in questa descrizione: 
sebbene il segnale non richieda alcuna
parametrizzazione\footnote{Volendo quantificare la contaminazione residua col  
rapporto $\frac{B}{T-B}$, il fondo $B$\ viene stimato come l'integrale della 
funzione che lo descrive nell'intervallo considerato, mentre $T$\ \`e 
pari al numero di eventi sperimentali trovati.}, si 
assume che esso vada a zero, oltre certi limiti anch'essi arbitrari, e che quindi 
gli estremi delle code degli spettri siano imputabili al solo fondo. 
Inoltre, il vero limite del metodo consiste nell'assumere che la stessa 
funzione (il polinomio) che ben descrive le code della distribuzione sia anche 
adatta a descrivere la parte di spettro in cui domina il segnale 
(la regione sotto il picco del segnale), che poi \`e la regione in cui 
si ha reale interesse a stimare il fondo stesso. 
\newline
La principale motivazione per avere una stima precisa del rapporto segnale 
su fondo nella regione selezionata degli spettri di massa discende 
dall'intenzione di misurare il tasso assoluto di produzione 
(lo {\em ``yield''}) delle diverse specie nelle collisioni Pb-Pb, 
misura che sar\`a oggetto di discussione di questo capitolo. 
In questo paragrafo si descriver\`a dunque la tecnica di mescolamento degli 
eventi sviluppata per riprodurre il fondo  di $V^0$\ geometriche.   
\newline
Tale studio ha permesso al tempo stesso una migliore definizione dei 
criteri si selezione, come discusso ad esempio a riguardo dell'angolo 
interno di decadimento $\phi_{V0}$\ (vedi fig.~\ref{PhiV0fig}).  
\newline
La conoscenza dettagliata delle diverse distribuzioni per i parameteri 
geometrici e/o cinematici del solo fondo di $V^0$\ geometriche render\`a 
anche possibile determinare con precisione l'entit\`a di eventuali errori 
sistematici nella procedura di correzione per efficienza ed accettanza.
Una discussione di questo metodo di analisi pu\`o esser trovata 
in~\cite{BrunoMoriond}.  
\newline
Nel caso delle cascate, cio\`e delle particelle $\Xi$\ ed $\Omega$, 
lo studio dettagliato del fondo residuo non \`e indispensabile in quanto 
l'errore statistico sulle osservabili 
che si intende misurare (tasso di produzione, pendenza dello spettro di massa 
trasversa, etc.), risulta dominante rispetto a qullo introdotto dalla 
presenza di eventi di fondo.  
\subsection{La tecnica di mescolamento degli eventi}
Come esposto nel capitolo precedente, le candidate 
$V^0$\ vengono selezionate,  durante la fase di ricostruzione degli eventi, 
provando ad abbinare tutte le tracce di carica  negativa con tutte 
quelle di carica positiva ricostruite nell'evento ed applicando dei 
criteri di selezione preliminari molto conservativi. 
Nelle interazioni Pb-Pb, per contenere la quantit\`a di informazioni da 
riversare su dischi e nastri magnetici,  
vengono memorizzati, per le fasi successive di analisi, solamente quegli eventi 
che contengono una o pi\`u candidate $V^0$.  
\newline
L'idea alla base del metodo sviluppato per la descrizione del fondo 
geometrico \`e quella di eliminare ogni segnale fisico proveniente 
dal decadimento a due corpi di una particella (\PgL $\rightarrow$ \Pp \Pgpm, 
\PagL $\rightarrow$ \Pap \Pgpp, \PKzS $\rightarrow$ \Pgpp \Pgpm, conversione 
di \Pgg $\rightarrow$ \Pep \Pem) formando nuove candidate $V^0$, ottenute 
abbinando le tracce di carica negativa di un evento con quelle di 
carica positiva di un'altro evento.  
\newline
Il codice sviluppato legge le informazioni sui parametri cinematici delle tracce 
ricostruite e cerca di formare le candidate $V^0$\ con la stessa procedura 
usata nel programma STRIPV0 per la ricostruzione delle $V^0$\ reali.  
Il programma \`e stato messo alla prova accertandosi che, quando applicato ad eventi 
reali non mescolati, fornisca le stesse candidate $V^0$\ di STRIPV0 e con gli 
stessi parametri cinematici.  
\newline
Per porsi nelle stesse condizioni degli eventi reali, si \`e ripetuta la 
ricostruzione di un campione rappresentativo dell'intera statistica dei dati 
raccolti, ottenuto selezionando gli eventi uno ogni duecento, 
salvando adesso su nastro magnetico {\em tutti} gli eventi ricostruiti, 
anche quelli non contenenti candidate $V^0$.  
\newline
Su questo campione di dati si \`e applicato il programma sviluppato 
per costruire il campione di $V^0$\ geometriche. 
Per mantenere le condizioni degli ``eventi mescolati'' il pi\`u vicino possibile 
a quelle degli eventi reali, due eventi sono stati combinati solo se:
\begin{enumerate}
\item provengono dallo stesso {\em run}
\item differiscono in molteplicit\`a di 
%      particelle cariche per meno di venti tracce 
      {\em ``hit''} per meno di venti {\em ``hit''}
      (molteplicit\`a misurata dalle $MSD$\ non estrapolata, 
      {\em cfr. paragrafo 4.5}).  
\end{enumerate}
Con la prima condizione ci si assicura che i due eventi siano temporalmente vicini, 
in modo tale che le caratteristiche del fascio e dei rivelatori  
siano simili tra loro ed a quelle degli eventi reali; inoltre in tal modo 
si \`e certi di combinare eventi in cui l'orientazione del campo magnetico 
sia la stessa ed a cui si possa associare lo stesso vertice primario 
dell'interazione (il vertice {\em run per run}).  
Con la seconda condizione si vogliono riprodurre quanto pi\`u \`e possibile le 
condizioni cinematiche dell'evento reale.   
Inoltre, per ciascuna coppia 
di eventi $A$,$B$\ si sono abbinate le negative di $A$\ con le positive di $B$, e
le positive di $A$\ con le negative di $B$. 
Per render trascurabile l'errore statistico associato alle distribuzioni 
delle $V^0$\ mescolate, ciascun evento \`e stato combinato 
%circa 17 volte 
parecchie volte con altri eventi (in media circa 17), 
prima di essere abbandonato.  
In tal modo l'incertezza (statistica) 
sulla misura del rapporto segnale su fondo residuo \`e determinata 
principalmente dalla consistenza statistica del campione di particelle reali 
identificate.  
\newline
Nel confrontare le distribuzioni delle $V^0$\ reali e delle ``$V^0$\ mescolate''
si \`e applicato a queste ultime un unico fattore di normalizzazione 
$\mathcal{N}$, pari a:
\begin{equation}
\mathcal{N}= \frac{{\underset{A}{\sum}}\,  N_{A} \cdot P_{A} }
 {{\underset{\substack{A,B \\ A \ne B}}{\sum}}\, 
  (N_{A} \cdot P_{B} + N_{B} \cdot P_{A})}
\label{NormBKG}
\end{equation}
dove $N_{A}$\  ($P_{A}$) \`e il numero di tracce negative (positive) contenute 
nell'evento $A$, e la somma si estende a tutti gli eventi reali nel numeratore ed 
a tutte le combinazioni $A$,$B$ nel denominatore.  
Il fattore di normalizzazione, dunque, tiene unicamente  conto 
del numero di tentativi fatti, negli eventi reali ed in quelli ''mescolati'', 
di formare una $V^0$.  
\subsection{Confronto tra le distribuzioni reali ed il fondo}
\begin{figure}[hbt]
\begin{center}
\includegraphics[scale=0.42]{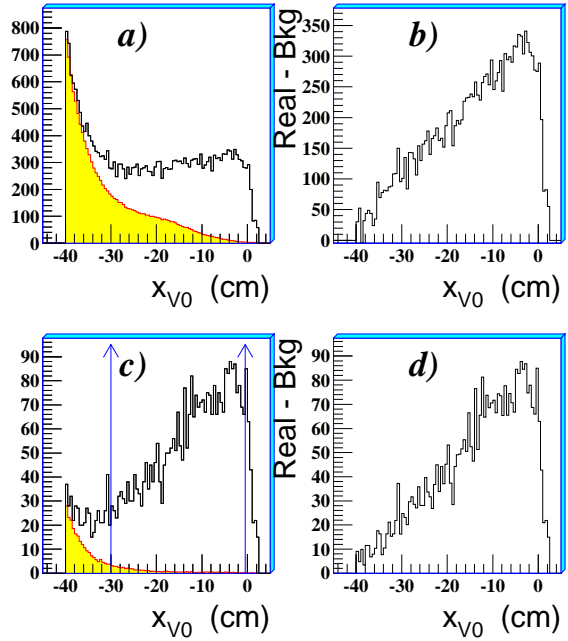}
\caption{{\bf a)} Distribuzione della variabile $x_{V^0}$\ per le candidate 
	 $V^0$\ reali (in chiaro) e per il fondo geometrico (in giallo). 
	 {\bf b)} Distribuzione di $x_{V^0}$\ per il solo segnale fisico, ottenuto 
	 sottraendo dalla distribuzione delle $V^0$\ reali il fondo delle $V^0$\ 
	 geometriche.  
	 {\bf c)} e {\bf d)} Distribuzioni analoghe a quelle soprastanti, 
	 ma ottenute dopo l'applicazione dei criteri di selezione per isolare 
	 il segnale delle \PgL\ e \PagL.}
\label{xvcBKG}
\end{center}
\end{figure}

\begin{figure}[hbt]
\begin{center}
\includegraphics[scale=0.42]{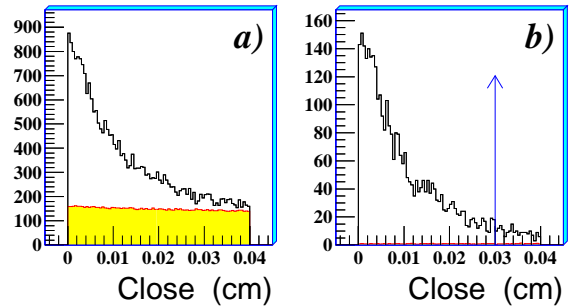}
\caption{Distribuzione della variabile $close_{V^0}$\ per le 
	 candidate $V^0$\ ({\bf a}) e per le  $V^0$\ selezionate 
	 coi tagli delle \PgL\ e \PagL\ ({\bf b}). Le distribuzioni 
	 in chiaro si riferiscono agli eventi reali, quelle in giallo 
	 agli eventi mescolati.}
\label{CloseBKG}
\end{center}
\end{figure}

\begin{figure}[hbt]
\begin{center}
\includegraphics[scale=0.42]{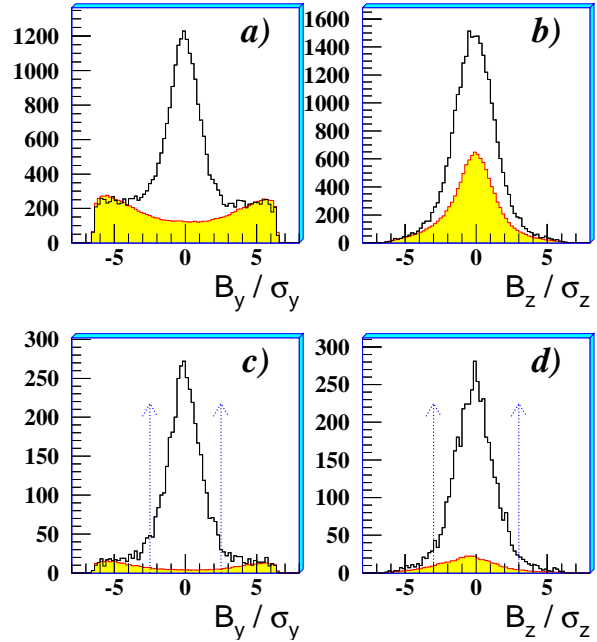}
\caption{Distribuzioni delle proiezione $y$\ e $z$\ del parametro di impatto 
	 delle candidate $V^0$\ ({\bf a} e {\bf b}) e delle $V^0$\ selezionate
	 coi tagli delle \PgL\ e \PagL\ ({\bf c} e {\bf d}). Le distribuzioni
         in chiaro si riferiscono agli eventi reali, quelle in giallo
         agli eventi mescolati.}
\label{V0impactBKG}
\end{center}
\end{figure}

Le distribuzioni delle ``$V^0$\ reali'' e di quelle ``mescolate'' sono state 
confrontate in maniera sistematica sia a livello di candidate --- ottenute 
applicando dei criteri di selezione preliminari poco selettivi, ma 
leggermente pi\`u severi di quelli di STRIPV0 ({\em cfr. paragrafo} 3.2), 
tanto agli eventi reali quanto a quelli mescolati  --- sia dopo l'applicazione  
dei criteri di selezione per isolare i segnali delle diverse specie. 
Un secondo confronto, anch'esso con esito positivo ma di cui si accener\`a 
soltanto, \`e stato eseguito selezionando gli eventi nella regione 
centrale del grafico di Armenteros (fig.~\ref{Arment-plot}) 
in cui nessun segnale fisico \`e presente.  
Questi confronti hanno permesso di accertarsi che il campione di 
``$V^0$\ mescolate'' ben descrive il fondo geometrico degli eventi reali, 
nelle regioni in cui non \`e atteso alcun segnale fisico. 
Inoltre hanno permesso di determinare  in modo ottimale la definizione 
dei criteri di selezione, come gi\`a discusso. 
A conferma di quest'ultimo punto, nei grafici che verranno mostrati  
(solo alcuni di quelli studiati) si riporter\`a l'intervallo fiduciale 
selezionato per la variabile in studio, delimitandolo al solito con delle 
frecce di colore blu.  
\newline
Come primo esempio, in fig.~\ref{xvcBKG} \`e mostrata la distribuzione 
della coordinata $x$\ del vertice di decadimento delle $V^0$\ reali 
(in chiaro) con sovrapposta la distribuzione delle $V^0$\ ``mescolate'' 
(in giallo), prima ({\bf a}) e dopo 
({\bf c}) l'applicazione degli altri criteri di selezione.  
Sottraendo le due distribuzioni, si ottiene la distribuzione del solo segnale 
fisico, mostrato nella fig.~\ref{xvcBKG}.b (candidate) e nella 
fig.~\ref{xvcBKG}.d (particelle selezionate).  
\newline
Come secondo esempio, in fig.~\ref{CloseBKG} si \`e considerata la distribuzione 
di $close_{V^0}$.  L'intensit\`a del fondo geometrico ha una piccola dipendenza 
da tale variabile, mentre il segnale fisico decresce molto rapidamente 
all'aumentare di $close_{V^0}$\ sino a scomparire del tutto.   
\newline
Il fondo nella distribuzione delle variabile $\phi_{V^0}$\ \`e stato 
gi\`a discusso nel {\em paragrafo 3.3}.    
Come ultimo esempio si presentano 
in fig.~\ref{V0impactBKG} le distribuzioni delle proiezioni $y$\ e $z$\ 
del parametro d'impatto delle $V^0$. Il fondo geometrico presenta in tal 
caso una struttura difficilmente prevedibile a priori; queste distribuzioni 
saranno usate pi\`u avanti per stimare l'entit\`a di un errore 
sistematico introdotto dall'uso, nelle simulazioni Monte Carlo, di una 
parametrizzazione di tipo gausssiano per tali distribuzioni. 
Si noti ancora una volta come, sulle code delle distribuzioni, 
vi sia buon accordo tra eventi ``mescolati'' ed eventi ``reali'', sia a 
livello di candidate che per le $V^0$\ selezionate~\footnote{
Si ricordi che la normalizzazione non \`e effettuata sulle singole variabili 
ma \`e ``assoluta'', cio\`e calcolata a priori (eq.~\ref{NormBKG}).}.  
Per completezza  
si osservi che il criterio di selezione su queste due variabili 
\`e di tipo correlato (ellittico, {\em cfr. paragrafo} 3.3) e le frecce, 
apposta tratteggiate, forniscono il limite massimo per una proiezione 
quando l'altra \`e nulla.  
\subsection{Stima del fondo geometrico residuo per \PKzS, \PgL\ e \PagL }
\begin{figure}[p]
\begin{center}
\includegraphics[scale=0.35]{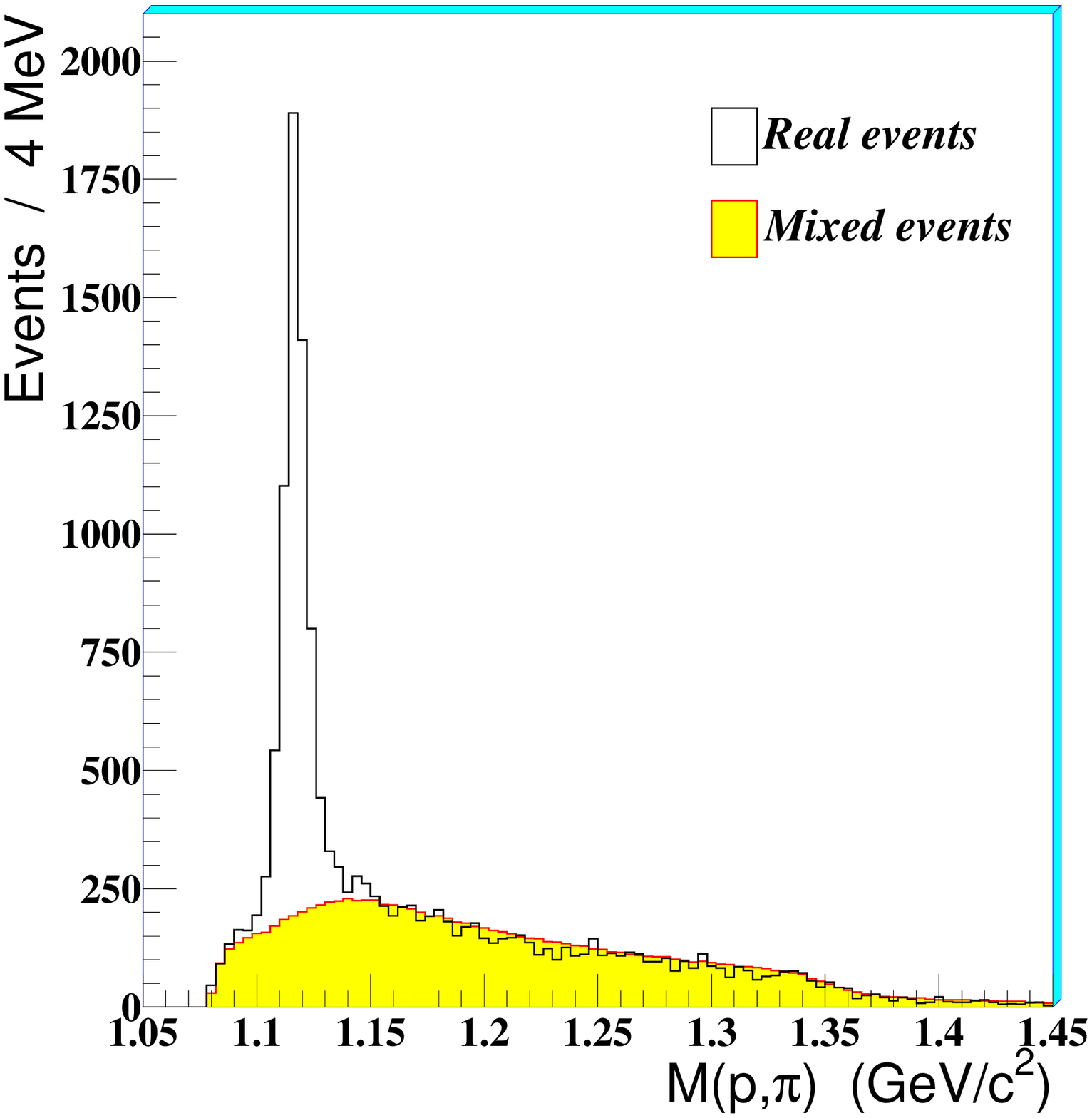}
\includegraphics[scale=0.35]{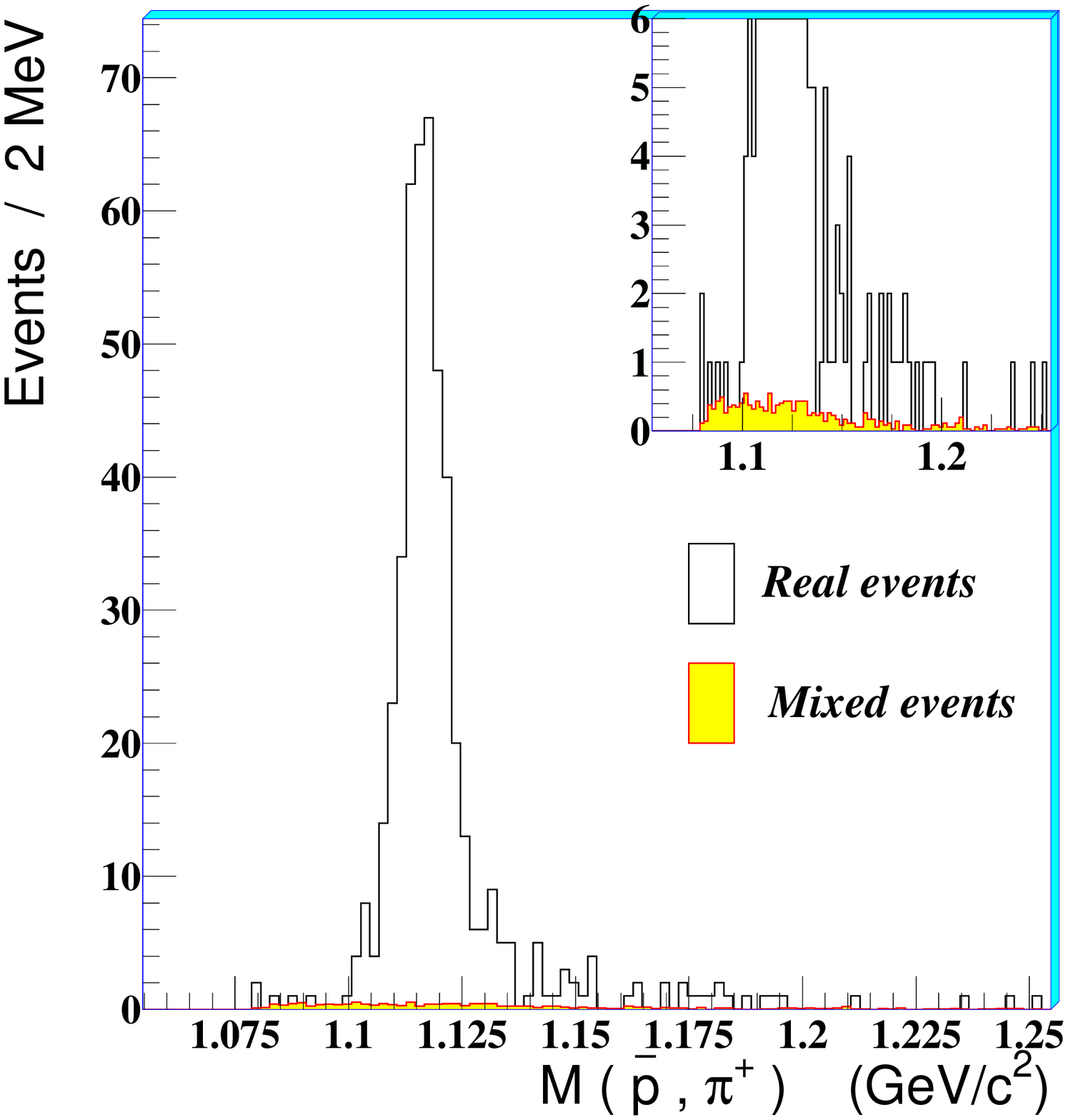}\\
\includegraphics[scale=0.35]{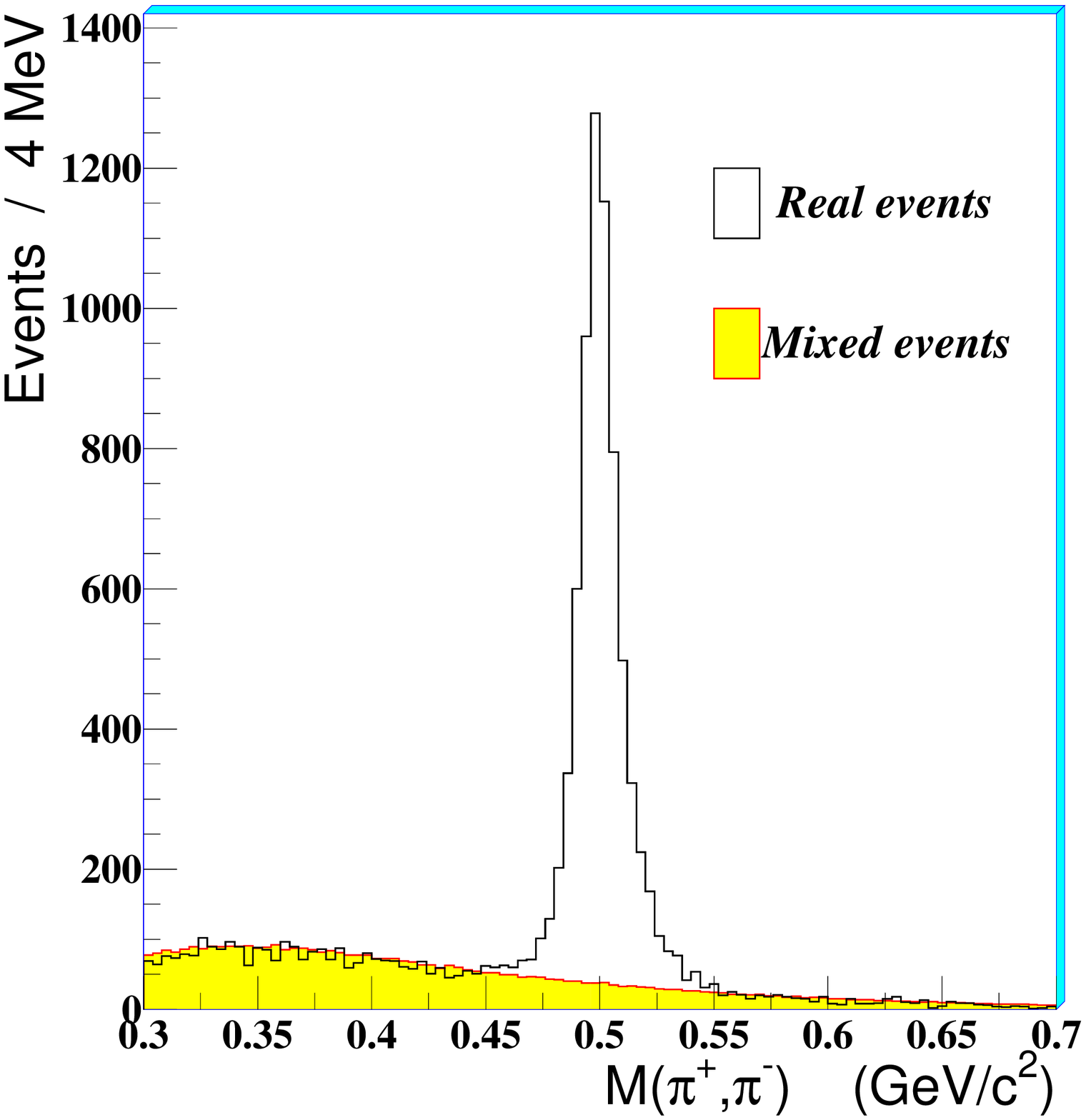}
\includegraphics[scale=0.35]{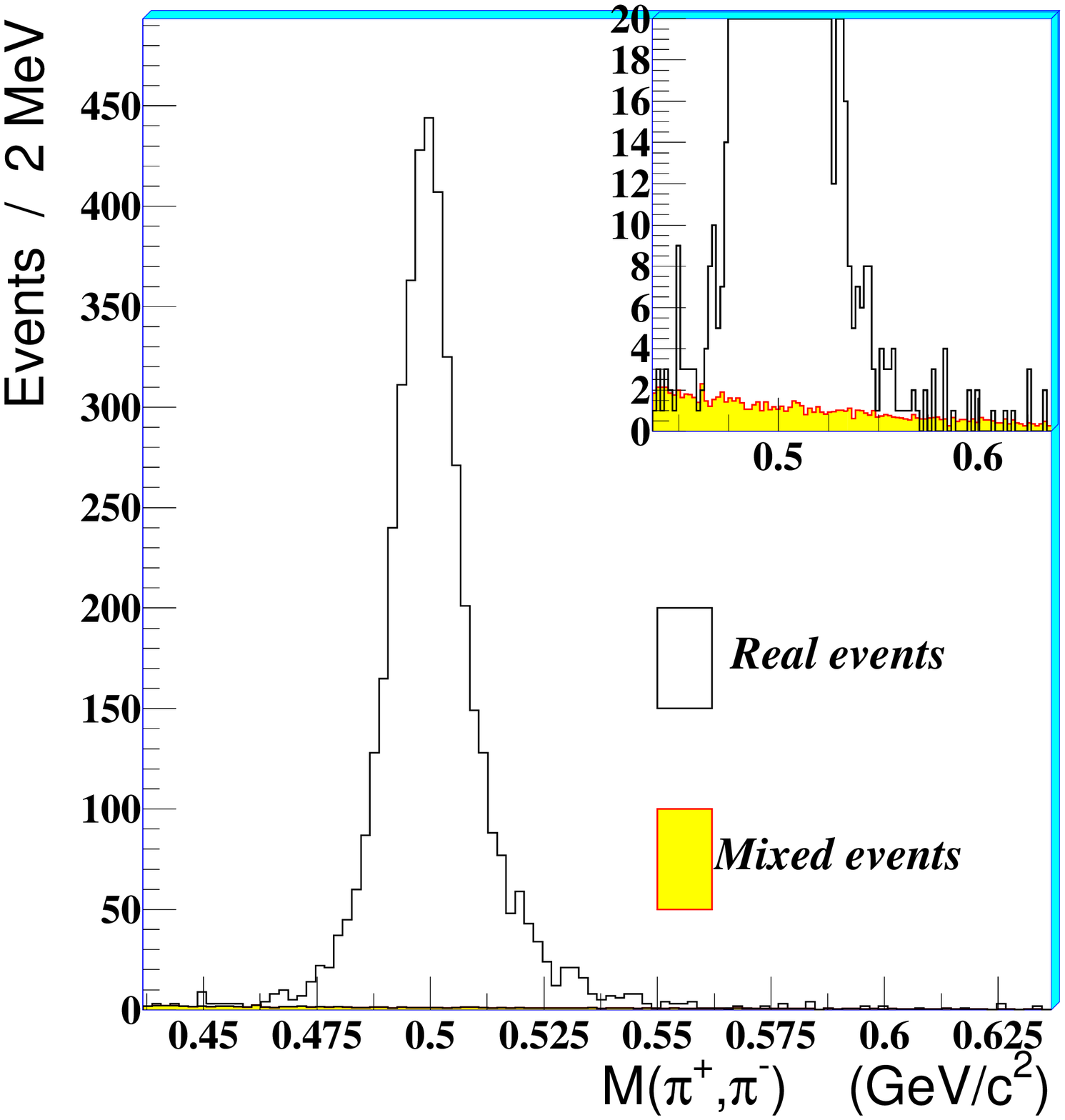}
\caption{\newline
	{\bf In alto:} \newline
	{\em A sinistra:} distribuzione della massa invariante $M(p,\pi)$\ per
        un campione di {\em candidate} $V^0$\ (\PgL+\PagL) 
	di eventi reali (in chiaro) e di eventi mescolati (in giallo).  
        {\em A destra:} spettro di massa invariante $M(\bar{p},\pi^+)$\ per
        un campione di \PagL\ ricostruite in eventi reali (in chiaro) ed 
        in eventi mescolati (in giallo); l'inserto in alto \`e un ingrandimento
        lungo l'asse delle ordinate. \newline
	{\bf In basso:} \newline
	{\em A sinistra:} distribuzione della massa invariante $M(\pi^+,\pi^-)$\
        per un campione di {\em candidate} $V^0$\ di eventi reali (in chiaro) e di
        eventi mescolati (in giallo).
        {\em A destra:} spettro di massa invariante $M(\pi^+,\pi^-)$\ per
        un campione di \PKzS\ ricostruiti in eventi reali (in chiaro) ed
        in eventi mescolati (in giallo); l'inserto in alto \`e un ingrandimento
        lungo l'asse delle ordinate.}
\label{V0MassBKG}
\end{center}
\end{figure}

%In fig.~\ref{LambdaMassBKG}
%\begin{figure}[hbt]
%\begin{center}
%\includegraphics[scale=0.35]{cap4/Mor1a.eps} 
%\includegraphics[scale=0.35]{cap4/Mor1b.eps}
%\caption{{\em A destra:} distribuzione della massa invariante $M(p,\pi)$\ per 
%	 un campione di candidate $V^0$\ di eventi reali (in chiaro) e di  
%	 eventi mescolati (in giallo). 
%	 {\em A sinistra:} spettro di massa invariante $M(\bar{p},\pi^+)$\ per 
%	 un campione di \PagL\ ricostruite in eventi reali (in chiaro) ed 
%	 in eventi mescolati (in giallo); l'inserto in alto \`e un ingrandimento 
%	 lungo l'asse delle ordinate.}
%\label{LambdaMassBKG}
%\end{center}
%\end{figure}
%\begin{figure}[hbt]
%\begin{center}
%\includegraphics[scale=0.35]{cap4/K0a_Mor.eps}
%\includegraphics[scale=0.35]{cap4/K0b_Mor.eps}
%\caption{{\em A destra:} distribuzione della massa invariante $M(\pi^+,\pi^-)$\ 
%	 per un campione di candidate $V^0$\ di eventi reali (in chiaro) e di
%         eventi mescolati (in giallo).
%         {\em A sinistra:} spettro di massa invariante $M(\pi^+,\pi^-)$\ per
%         un campione di \PKzS\ ricostruite in eventi reali (in chiaro) ed
%         in eventi mescolati (in giallo); l'inserto in alto \`e un ingrandimento
%         lungo l'asse delle ordinate.}
%\label{K0MassBKG}
%\end{center}
%\end{figure}
In fig.~\ref{V0MassBKG} sono mostrati gli spettri di massa invariante  
$M(p,\pi)$\ e $M(\pi^+,\pi^-)$\ per il campione di candidate $V^0$\ 
reali (in chiaro) e ``mescolate'' (in giallo). Per entrambi gli spettri 
il fondo geometrico degli eventi reali \`e ben descritto dagli eventi mescolati 
a livello delle candidate.  Nella stessa fig.~\ref{V0MassBKG} sono anche 
mostrate le $V^0$\ selezionate applicando, sia alle particelle reali sia a 
quelle ``mescolate'', i criteri di selezione delle \PagL\ (in alto) e 
dei \PKzS\ (in basso)~\footnote{Un'analoga distribuzione per le \PgL\ pu\`o 
trovarsi in~\cite{BrunoBKG}; qui si \`e preferito soffermarsi sulle \PagL\ per 
le quali si attende una maggior contaminazione di fondo, in quanto il segnale 
delle \PagL\ \`e di un ordine di grandezza inferiore a quello delle \PgL.}.  
Nel caso delle \PagL\ il fondo geometrico residuo \`e concentrato sotto il picco 
del segnale fisico, principalmente nel suo versante sinistro. 
Una simile distribuzione conferma 
la fondatezza dei timori avanzati precedentemente, allorch\'e si considerava 
la difficolt\`a di stimare il fondo utilizzando un polinomio adattato ai 
punti sperimentali sulle code dello spettro del segnale fisico.  Simili 
considerazioni valgono anche nel caso dei \PKzS\ in cui il fondo geometrico 
aumenta monotonicamente al diminuire della massa invariante 
$M(\pi^+,\pi^-)$\ ed al diminuire di ${q_T}_{V^0}$\ 
(e delle \PgL, il cui fondo \`e distribuito in modo molto  
simile a quello delle~\PagL).  
\newline
Dopo l'applicazione dei criteri di selezione, \`e evidente dalla 
fig.~\ref{V0MassBKG} come il fondo geometrico non descriva completamente 
le code, specialmente ad alte masse invarianti, nello spettro delle \PagL. 
Questa discrepanza \`e stata oggetto di un ulteriore studio che ha condotto ad 
attribuire l'eccesso di eventi nella distribuzione reale a due fattori: 
\begin{enumerate}
\item presenza dei \PKzS\ residui non completamente anti-selezionati dal 
      criterio sulla massa invariante $M(\pi^+,\pi^-)$;  
\item $V^0$\ fisiche cui viene attribuita una massa invariante alterata a causa 
	di errori nell'associazione degli {\em ``hit''} alle tracce nel 
	programma ORHION.
\end{enumerate}
Di entrambi i contributi il metodo sviluppato non pu\`o, n\`e vuole (in quanto 
si intende valutare il solo fondo geometrico), tener conto. Uno studio che 
ha messo in risalto come il segnale dei \PKzS\ si riflette nello spettro 
di massa invariante $M(\Pap,\Pgpp)$\ al variare dello spazio delle fasi del 
decadimento (pi\`u precisamente, per  diversi settori dell'ellisse 
dei decadimenti delle \PKzS\ nel grafico di Armenteros) 
conduce a concludere che il residuo dei \PKzS\ si concentra ad elevate masse 
invarianti  $M(\Pap,\Pgpp)$. 
La contaminazione residua dei \PKzS\ all'interno dell'intervallo di massa 
invariante selezionato per l'analisi delle \PagL\ ($\pm 10$\ MeV/$c^2$\ rispetto 
al valore nominale) \`e invece trascurabile ($\lesssim 2\% $).  
\newline
Il problema dell'ambiguit\`a con altre particelle non \`e invece presente nel caso 
dei \PKzS, come discusso nel paragrafo 3.3.2; nel caso delle \PgL, infine, 
il problema \`e marginale in quanto, sebbene il numero assoluto di \PKzS\ 
residui sia confrontabile con quello del caso delle \PagL\ per la simmetria 
del decadimento dei \PKzS,  la contaminazione percentuale \`e un'ordine di 
grandezza inferiore al caso delle \PagL.  
\newline
Per quanto riguarda il secondo contributo, un'indagine condotta allo scopo ha 
accertato che, in una percentuale piccola ma non irrilevante di casi 
($1\div3 \, \%$), il programma ORHION pu\`o distorcere lievemente la traiettoria di 
una delle due tracce della $V^0$\ vincolandola a passare per l'ultimo {\em ``hit''} 
associato all'altra traccia. Il problema si presenta solamente per le tracce 
ricostruite in uno stesso evento e non per gli ``eventi mescolati''. 
Questo aspetto non desta tuttavia ulteriori preoccupazioni in quanto si presenta, 
con la stessa frequenza, anche nelle simulazioni sviluppate per calcolare 
le correzioni per accettanza ed efficienza dei rivelatori  e dei programmi 
di ricostruzione ed identificazione. 
Pertanto, anche le perdite di particelle dovute a questo   
problema vengono tenute nel giusto conto nelle correzioni.  
%\newline
%Nella tabella~\ref{tab4.1} sono stati riassunti i risultati sulla stima del fondo 
%geometrico per le diverse specie di $V^0$\ identificate: 
%il numero di particelle reali usate per la stima, che ne determina la precisione, 
%la percentuale di fondo nell'intero spettro, ed infine quella all'interno degli 
%intervalli di massa selezionati, alla cui misura questa analisi \`e stata 
%finalizzata.  
%
\newline
Nella tabella~\ref{tab4.1} sono stati riassunti i risultati sulla stima del fondo
geometrico per le diverse 
\begin{table}[h]
\begin{center}
\begin{tabular}{|c|c|c|c|} \hline
  & N. di particelle 
  & \begin{tabular}{c} 
      $\frac{B}{T-B}$\ \quad  \% \\ 
      (intero spettro)
    \end{tabular} 
  & \begin{tabular}{c}
      $\frac{B}{T-B}$\   \quad  \% \\
      (intervallo selezionato)
    \end{tabular} \\ \hline
 \PgL  & $2800$  & $1.35$ & ${\bf 0.7} $  \\
 \PagL & $480$   & $3.45$ & ${\bf 2.3} $  \\
 \PKzS & $4650$  & $9.0$  & ${\bf 0.5} $  \\ \hline
\end{tabular}
\end{center}
\caption{Consistenza statistica del campione di $V^0$\ reali identificate 
	 ed usate per la stima del rapporto segnale  su fondo 
	 geometrico;  
	 percentuale di fondo geometrico residuo sull'intero spettro e 
	 negli intervalli di massa invariante selezionati per l'analisi.
\label{tab4.1}}
\end{table}
\newline
%Nella tabella~\ref{tab4.1} sono stati riassunti i risultati sulla stima del fondo
%geometrico per le diverse 
specie di $V^0$\ identificate:
il numero di particelle reali usate per la stima, che ne determina la precisione,
la percentuale di fondo nell'intero spettro, ed infine quella all'interno degli
intervalli di massa selezionati, alla cui misura questa analisi \`e stata
finalizzata.
\section{Correzione per accettanza ed efficienza}
Fra gli obiettivi primari dell'esperimento vi \`e la misura 
%Per poter misurare con precisione la 
della 
sezione d'urto inclusiva (doppiamente) 
differenziale~\footnote{Si ricorda, come discusso nel {\em paragrafo 1.5.3}, 
che la ``massa trasversa'' di una particella di massa $m$\ viene 
definita come $m_T=\sqrt{m^2+p_T^2}$, mentre $y$\ rappresenta la rapidit\`a
della particella.} 
\begin{equation}
\frac{{\rm d}^2N(m_T,y)}{{\rm d}m_T{\rm d}y}
\label{Differential}
\end{equation}
di ciascuna delle diverse specie di particelle osservate e, a partire da 
questa, del loro tasso di produzione integrale 
$Y=\iint_V \frac{{\rm d}^2N(m_T,y)}{{\rm d}m_T{\rm d}y}{\rm d}m_T{\rm d}y$\ 
in un certo volume $V$\ dello spazio delle fasi $(m_T,y)$. 
A questo scopo 
\`e necessario conoscere il modo 
con cui la particolare geometria dei rivelatori limita la loro misura 
(``accettanza geometrica'') e l'efficienza con la quale le particelle sono 
rivelate e ricostruite.
\newline
Come discusso nel secondo capitolo, l'esperimento NA57 dispone di un 
telescopio di altissima risoluzione ma che copre una porzione limitata dello 
spazio delle fasi permesso alle particelle nello stato finale; il calcolo 
dell'accettenza riveste dunque un ruolo cruciale. La precisione richiesta nella 
misura del tasso di produzione di particelle strane impone altres\`i che la loro 
efficienza di rivelazione (cio\`e l'efficienza dei diversi piani di rivelatori 
nel telescopio) e di ricostruzione (cio\`e l'efficienza dei programmi di 
ricostruzione degli eventi e quelli sviluppati per la selezione ed identificazione 
delle diverse particelle) siano note con grande accuratezza.  
\newline
\`E stato dunque sviluppato un metodo di calcolo basato su una simulazione Monte 
Carlo che consente di associare ad ognuna delle particelle identificate una 
correzione che contiene un contributo dovuto all'accettanza geometrica ed uno 
dovuto all'efficienza di rivelazione e ricostruzione. 
La procedura consiste, {\em per ciascuna particella identificata}, nei seguenti  
passi:
\begin{enumerate} 
\item {\bf Generazione di particelle con metodo Monte Carlo e 
	    loro tracciamento nel telescopio}\\
Si generano delle particelle Monte Carlo dello stesso tipo, di impulso trasverso 
$p_T$\ e di rapidit\`a $y$\ pari a quelli della particella reale in questione, cui 
si vuole applicare la correzione. 
Tutti gli altri parametri del decadimento sono invece privi di vincoli,  
ad eccezione dei due seguenti, modificati unicamente per velocizzare il calcolo:  
({\em i}) si considera solo il canale di decadimento  nel quale sono state ricostruite 
le particelle misurate (ad esempio: \PgL $\rightarrow$ \Pp \Pgpm, 
\PgXm$\rightarrow$ \PgL \Pgpm); la percentuale di particelle perse perch\'e, nei 
decadimenti reali, decadute in un altro canale \`e fornita da $1 - BR$, dove $BR$\ 
\`e la frequenza di decadimento ({\em ``Branching Ratio''}) nel canale considerato.
({\em ii}) Le particelle sono generate in modo tale che l'angolo $\phi$\ di azimuth 
che il loro impulso trasverso $\vec{p}_T$\ forma con l'asse $y$, nel piano $yz$ del 
riferimento del laboratorio, sia compreso tra $35^o$\ e $145^o$, in modo da includere  
un'angolo solido ben pi\`u esteso di quello sotteso dal telescopio, ma comunque 
ridotto rispetto ai $4\pi$.  
\newline 
%Le particelle vengono generate in punti dello spazio, 
Il vertice primario d'interazione \`e generato 
entro il piano $\Sigma$\ contente il bersaglio e parallelo al piano $yz$\ del 
riferimento del laboratorio (quindi a coordinata $x$\ fissata), 
%ma distribuiti secondo una 
con una distribuzione 
gaussiana bidimensionale nel piano $\Sigma$\ di valor medio $(y_{run},z_{run})$, 
pari alle coordinate del vertice {\em run per run} dell'evento in cui la particella 
reale \`e stata misurata, e larghezze $({\sigma_y}_{run},{\sigma_z}_{run})$\ pari 
%agli scarti quadratici medii (in $y$\ ed in $z$) della distribuzione dei vertici 
alle larghezze delle gaussiane di {\em ``best fit''} (in $y$\ ed in $z$) alle distribuzioni 
dei vertici
{\em evento per evento} utilizzati per determinare il vertice {\em run per run}.  
Si ricorda che, negli eventi reali, i parametri d'impatto delle diverse particelle,  
su cui si sono operati criteri di selezione, sono stati calcolati riferendosi 
al vertice {\em run per run}. Questo aspetto della simulazione sar\`a oggetto di 
discussione nei prossimi paragrafi.  
%e nel confronto coi risultati dell'esperimento WA97 nel prossimo capitolo.
\newline
Il decadimento delle particelle generate viene quindi simulato rispettando la 
vita media di ciascuna specie ed i prodotti di decadimento sono tracciati all'interno 
dell'apparato sperimentale per mezzo del programma GEANT3~\cite{GEANT3}. Durante 
la generazione vengono registrate tutte le informazioni relative alle particelle 
simulate le cui tracce di decadimento attraversano il telescopio.  
All'interno di GEANT \`e stata inserita la struttura a pixel dei piani del telescopio 
e quella relativa alle $\mu$strip a doppia faccia del {\em ``lever arm''}; 
ai pixel ed alle {\em ``strip''} interessati dal passaggio dei prodotti carichi di decadimento 
sono assegnati degli {\em ``hit''} elettronici, secondo l'efficienza dei rivelatori.  
L'efficienza dei pixel e quella delle 
$\mu$strip viene misurata ed assegnata per ciascun {\em ``chip''} di lettura  
({\em cfr. paragrafo 2.4.3}), e per differenti periodi di funzionamento 
durante la presa dati.  
Questa fase termina quando il numero di particelle  che  
risultano potenzialmente ricostruibili nelle successive fasi di analisi, 
in quanto i loro prodotti di decadimento sono  geometricamente accettati 
dal telescopio, diventa tale da assicurare un numero cospicuo di particelle infine 
ricostruite (qualche centinaio).  
In tal modo l'errore sulla correzione \`e reso indipendente dall'accettanza 
della particella sotto esame. Tipicamente  
si richiede di avere, a seconda della specie in esame, $2500 \div 5000$\ 
particelle potenzialmente ricostruibili sulla base della loro accettanza geometrica.  
\item {\bf Impiantazione degli {\em ``hit''} di GEANT in eventi di fondo}\\
Per riprodurre la situazione reale in cui ogni particella strana \`e stata 
ricostruita --- in presenza cio\`e di altre tracce e del rumore elettronico dei 
rivelatori --- gli {\em ``hit''}  elettronici delle tracce di GEANT vengono 
impiantati entro eventi reali (detti ``eventi di fondo''), uno per per ciscuna 
delle particelle simulate; gli eventi di fondo vengono campionati sull'intera 
statistica accumulata, e sono quindi rappresentativi delle condizioni operative 
durante tutta la presa dati.  
Si costruisce in tal modo, per ciascuna delle particelle generate da GEANT, un 
evento ``ibrido'' costituito dagli {\em ``hit''} delle tracce reali, 
dagli {\em ``hit''} di quelle simulate e dagli {\em ``hit''} dovuti al rumore elettronico.  
Gli eventi di fondo vengono scelti in modo tale che la loro molteplicit\`a 
di {\em ``cluster di hit''}~\footnote{Un cluster \`e un un insieme di pi\`u 
{\em ``hit''} tra loro adiacenti. 
La molteplicit\`a di {\em cluster} \`e generalmente 
correlata alla molteplicit\`a di {\em ``hit''}, ma si \`e preferito utilizzare 
il numero di {\em cluster} come riferimento perch\'e tale variabile si \`e dimostrata 
pi\`u rappresentativa della reale occupazione del telescopio soprattutto 
rispetto ad eventi con rumorosit\`a localizzata solo in alcuni chip.}  
nei piani di pixel del telescopio sia 
prossima, entro certi limiti opportunamente definiti, a quella dell'evento 
contenente la particella da correggere.  Questo criterio 
%\`e verificato in media una volta ogni 100 eventi di fondo ed 
ha lo scopo di ricreare, nell'evento ibrido, 
la stessa situazione fisica presente in quello originale.  
\item {\bf Ricostruzione degli eventi ibridi col programma ORHION} \\
Gli eventi ibridi sono poi processati col programma di ricostruzione ORHION, 
come nel caso degli eventi reali. Le candidate $V^0$\ vengono  al solito 
preventivamente individuate e selezionate in STRIPV0. In tutte le fasi della 
ricostruzione il programma ignora se un dato {\em ``hit''} \`e originariamente 
associato ad una traccia della particella simulata od all'evento di fondo.  
Gli eventi cos\`i processati sono registrati per le succesivi fasi di analisi  
nel formato usuale, conservando tuttavia l'informazione sugli {\em ``hit''}  
associati a ciascuna delle tracce ricostruite.  
\item {\bf Identificazione e ricostruzione delle particelle strane}\\
A questo punto gli eventi ricostruiti da ORHION vengono analizzati 
dal programma sviluppato in fase di analisi per la selezione e  
l'identificazione delle particelle strane, esattamente come nel caso degli 
eventi reali. Si applicano dunque anche agli eventi ibridi gli stessi criteri 
di selezione usati per isolare ed identificare i diversi campioni di particelle.  
Il programma, utilizzato tanto in fase di analisi quanto in fase di 
correzione, \`e diverso a seconda che si tratti di una $V^0$\ (\PgL\ o \PKzS) 
od una cascata ($\Omega$\ o $\Xi$), ma in entrambi i casi 
fornisce in uscita un sotto-insieme degli eventi processati da ORHION, formato 
soltanto da quegli eventi che contengono una $V^0$\ od una cascata, a seconda del 
programma, che abbia superato i tagli dell'analisi.  
\item {\bf Riconoscimento delle particelle generate}\\
\`E necessario infine determinare se e quale, tra le particelle strane ricostruite 
nei precedenti passi, ve ne sia una che corrisponde a quella originariamente generata 
con GEANT. Vi \`e infatti la possibilit\`a che la particella generata non sia stata 
affatto ricostruita (e ci\`o a causa dell'inefficienza stessa, che si intende misurare), 
e quella osservata corrisponda ad un particella presente nell'evento di fondo. 
Per ciascun evento ricostruito --- che pu\`o contenere dunque, al pi\`u, una sola 
particella Monte Carlo, di cui si conoscono tutti i parametri iniziale ed anche  
gli {\em ``hit''} nei rivelatori --- si confrontano gli {\em ``hit''} associati alle tracce di 
ciascuna particella strana che ha superato i criteri di selezione 
(tre tracce nel caso di una cascata, due nel caso di una $V^0$) 
con quelli della particella Monte Carlo originariamente generata. Se le tracce 
sono tra loro compatibili, entro determinate tolleranze che tengono conto degli errori 
di estrapolazione e della possibilit\`a di perdere parte degli {\em ``hit''} originari 
durante la fase di ricostruzione, la particella viene identificata con quella 
originariamente generata.
\end{enumerate}
Vengono cos\`i calcolati, per ogni particella da correggere, il numero di particelle 
Monte Carlo generate ($N_{gen}$), quello delle particelle Monte Carlo accettate 
geometricamente ($N_{acc}$), numero generalmete mantenuto costante per una data 
specie ($2500\div5000$), e quello delle particelle ricostruite ($N_{rec}$) 
ed identificate come corrispondenti alla particella Monte Carlo originaria.  
\newline
Il fattore di correzione per ciascuna particella  (il ``peso'') ---  
che dunque tiene conto, simultaneamente, del contributo delle perdite
per l'accettanza geometrica e di quelle per l'efficienza   
di rivelazione e di ricostruzione --- \`e dato da:
\begin{equation}
w=
  \frac{360^o}{\Delta\phi} \times \frac{N_{gen}}{N_{rec}} \times \frac{1}{BR}
\label{Weight}
\end{equation}
dove $\Delta\phi$\ corrisponde all'intervallo di angolo azimuthale $\phi$\ 
entro il quale le particelle sono state generate nella simulazione (ed al di 
fuori del quale nessuna sarebbe stata comunque ricostruita) ed il  
fattore $\frac{1}{BR}$\ tiene conto che  si \`e imposto, nella simulazione,  
il decadimento delle particelle generate secondo un unico canale, cui 
l'esperimento \`e sensibile, di cui \`e noto il {\em ``branching ratio''} $BR$.  
Poich\'e la simulazione \`e espletata per mezzo di generatori di numeri a caso, 
l'errore sul peso \`e calcolabile a partire dalla 
distribuzione binomiale ed, indicando con $P$\ il rapporto 
$\frac{N_{rec}}{N_{gen}}$, esso \`e fornito 
da~\footnote{L'incertezza sulla probabilit\`a di decadimento nel canale considerato 
($BR$) \`e del tutto trascurabile.}:  
\begin{equation}
\delta w = w \times \frac{\delta P}{P} = \frac{w}{P} 
           \sqrt{\frac{P\cdot(1-P)}{N_{gen}}}
\label{ErrWeight}
\end{equation}
\newline
Nella tabella~4.2 sono riportati i valori medii della correzione  ($<w>$)  
per i segnali studiati nelle interazioni Pb-Pb.  
%i valori medii della correzione globale,
%si sono riportati anche quelli dell'efficienza di ricostruzione
%per i segnali studiati nelle interazioni Pb-Pb.
\begin{table}[h]
\label{tab4.2}
\begin{center}
\begin{tabular}{|l|c|l|c|}  \cline{2-2}\cline{4-4}
 \multicolumn{1}{ c|}{} &
        \multicolumn{1}{|c|}{
        \begin{tabular}{c}
          Pb-Pb \\
          160 A GeV/$c$
         \end{tabular}}  &     & \multicolumn{1}{|c|}{
                               \begin{tabular}{c}
                                Pb-Pb \\
                                40 A GeV/$c$
                                \end{tabular}} \\ \cline{2-2}\cline{4-4}
\multicolumn{1}{ c|}{}   & $<w>$ & & $<w>$  \\  \hline
\PgOm  & $20740\pm1360$ & 
               \multicolumn{1}{ ||c|}{\PgOm+\PagOp}& $63600\pm25100$  \\
\PagOp & $19245\pm1770$  &
               \multicolumn{1}{ ||c|}{ }    &   \\
\PgXm  & $26090\pm719 $ &
               \multicolumn{1}{ ||c|}{\PgXm} & $28530\pm1550$ \\
\PagXp & $25590\pm1365$ & 
               \multicolumn{1}{ ||c|}{\PagXp}& $26100\pm4600$ \\ \cline{3-4}
\PgL   & $1517\pm38$  \\
\PagL  & $1552\pm36$  \\
\PKzS  & $ 986\pm52$  \\ \cline{1-2}
\end{tabular}
\end{center}
\caption{Valori medii delle correzioni calcolate per le particelle strane nelle 
	 collisioni Pb-Pb.}  
\end{table}
\newline
La necessit\`a di generare e processare migliaia di eventi per ogni particella 
rivelata rende il calcolo delle correzioni per accettanza ed efficienza molto 
dispendioso in termini di tempo di esecuzione al computer. Nel caso delle 
$V^0$\ non \`e quindi possibile correggere con questa tecnica  
l'intero campione raccolto, costituito tipicamente da decine o centinaia di 
migliaia di particelle~\footnote{Si pensa allo scopo di sviluppare metodi di correzione 
pi\`u veloci, da applicarsi direttamente alle diverse distribuzioni, e non 
particella per particella, del tipo di quello sperimentato da WA97 in~\cite{mt_WA97}.}.  
Per le cascate, invece, che nelle collisioni Pb-Pb (in cui sono pi\`u 
abbondantemente prodotte) ammontano a poche centinaia ($\Omega$) o migliaia ($\Xi$) di 
esemplari, questa tecnica \`e ottimale ed applicabile all'intero campione.  
\newline
Nella tab. 4.3 sono riassunti i dati sulla consistenza statistica dei diversi 
campioni di particelle strane identificate, e la frazione di queste corrette 
con la tecnica esposta.  
Come discusso nel terzo capitolo, nel calcolare le correzioni per le cascate 
raccolte nella presa dati dell'anno 2000 
si \`e data precedenza alle $\Omega$, volendo diminuire in modo pi\`u significativo 
l'errore statistico delle osservabili legate a questa particella, la pi\`u rara 
tra quelle studiate; pertanto le $\Xi$\ corrette sino ad ora provengono dall'intero 
campione del 1998, mentre un secondo campione delle $\Xi$\ raccolto nell'anno 
2000 raddoppier\`a la statistica sino ad ora analizzata. I dati sulle $V^0$\ a 
160 A GeV/$c$\ si riferiscono invece al solo campione di dati del 1998.  
I dati p-Be a 160 GeV/$c$\ sono invece quelli raccolti dall'esperimento WA97 e 
verranno utilizzati come %interazione di 
riferimento a questa energia nel prossimo capitolo. 
Nell'ultima colonna della tab. 4.3, infine, \`e riportata la stima --- 
fatta a partire dal campione del 1999 ---  
delle particelle attese nelle collisioni p-Be a 40 GeV/$c$, utilizzabili per 
l'analisi.  
\begin{table}[h]
\label{tab4.3}
\begin{center}
\begin{tabular}{|c|lr|lr|lr|lr|} \hline
  &  
	\multicolumn{2}{|c|}{ 
	\begin{tabular}{c}
         Pb-Pb \\ 
	 160 A GeV/$c$  
	\end{tabular}} & \multicolumn{2}{|c|}{
			 \begin{tabular}{c}
			  Pb-Pb \\
			  40 A GeV/$c$ 
			 \end{tabular}} & \multicolumn{2}{|c|}{
			 		\begin{tabular}{c}
			 		   p-Be (WA97)\\
					   160 GeV/$c$ 
				        \end{tabular}} & \multicolumn{2}{|c|}{
							\begin{tabular}{c}
					   		      p-Be \\
							      40 GeV/$c$
						       \end{tabular}} \\ \hline 
       &raccolte& \% pesate&raccolte& \% pesate&raccolte& \% pesate&
        \multicolumn{2}{|c|}{stima attese} \\ \cline{2-9}
%\PgOm  & 432 & 100 & 35  & 100  & 17  & 100 & \multicolumn{2}{|c|}{---}  \\ 
%\PagOp & 193 & 100 &  4  & 100  & 16  & 100 & \multicolumn{2}{|c|}{---} \\
\PgOm  & 432 & 100 & 35  & 100  &\multicolumn{2}{||c||}
            {\PgOm\ + \PagOp}&\multicolumn{2}{|c|}{---} \\ 
\PagOp & 193 & 100 &  4  & 100  & \multicolumn{1}{||c}{33}& \multicolumn{1}{c||}{100} & 
             \multicolumn{2}{|c|}{---}  \\ \cline{6-7}
\PgXm  &$\approx 2\times 2234$ 
             & $\approx$ 50  &   550 & 100 & 397 & 100 & \multicolumn{2}{|c|}{$30$}\\% \div40$}\\
\PagXp &$\approx 2\times 601 $  & 
        $\approx$ 50  &           47 & 100 & 157 & 100 & \multicolumn{2}{|c|}{---} \\
\PgL   & 470000 & 0.5 &78000& --- & 119300 & 4   & \multicolumn{2}{|c|}{ 25000} \\
\PagL  & 67950  &  4  & 2100& --- &  34350 & 15  & \multicolumn{2}{|c|}{ 1500 } \\
\PKzS  & 706000 & 0.08&$\approx 60000$& --- &\multicolumn{2}{|c|}{da analizzare}&
	\multicolumn{2}{|c|}{10000} \\\hline
\end{tabular}
\end{center}
\caption{Statistica delle particelle strane e frazione di particelle corrette   
	 contenute nell'intero campione di dati di NA57 relativi all'interazione 
	 Pb-Pb a 160 e 40 A GeV/$c$, ed in quello   
	 dell'esperimento WA97 relativo all'interazione p-Be che 
	 sar\`a usata come riferimento a 160 GeV/$c$;  stima delle particelle 
	 attese nelle collisioni p-Be a 40 GeV/$c$.}
\end{table}

\subsection{Confronto tra dati e Monte Carlo}
Il calcolo delle correzioni riveste un ruolo di primo piano per come \`e 
concepito l'esperimento NA57, che si prefigge di studiare, tra l'altro, 
i tassi di produzione delle diverse specie di particelle strane. 
\`E pertanto fondamentale accertarsi che la simulazione Monte Carlo ben descriva le 
distribuzioni dei dati reali, ed in modo particolare quelle sulle quali si sono  
applicati i tagli dell'analisi. 
\newline
Si consideri, ad esempio, per una cascata ($\Xi$\ od $\Omega$) 
la distribuzione della variabile $x_{V^0}$, pari alla coordinata $x$\ del vertice di 
decadimento della $\Lambda$\ prodotta nel decadimento della cascata. La forma di questa 
distribuzione \`e il risultato della convoluzione della distribuzione esponenziale 
dei decadimenti della cascata, secondo la propria vita media 
($c\tau=4.91$\ cm per la $\Xi$, $c\tau=2.461$\ cm per la $\Omega$) e   
di quella dei decadimenti della $\Lambda$\ stessa ($c\tau=7.89$\ cm), il tutto 
modulato dall'accettanza ed efficienza dei rivelatori. Un buon accordo tra la 
distribuzione dei dati reali e quella ottenuta al Monte Carlo suggerisce che  
la vita media osservata per le particelle reali \`e effettivamente quella 
nominale {\em introdotta} nella simulazione e che nella simulazione \`e ben 
riprodotta la modulazione dovuta all'accettanza ed all'efficienza dei rivelatori.  
\newline
Per tutte le specie si sono effettuati confronti sistematici tra le distribuzioni 
delle particelle generate al Monte Carlo e quelle dei dati reali.  

\begin{figure}[t]
\begin{center}
\includegraphics[scale=0.190]{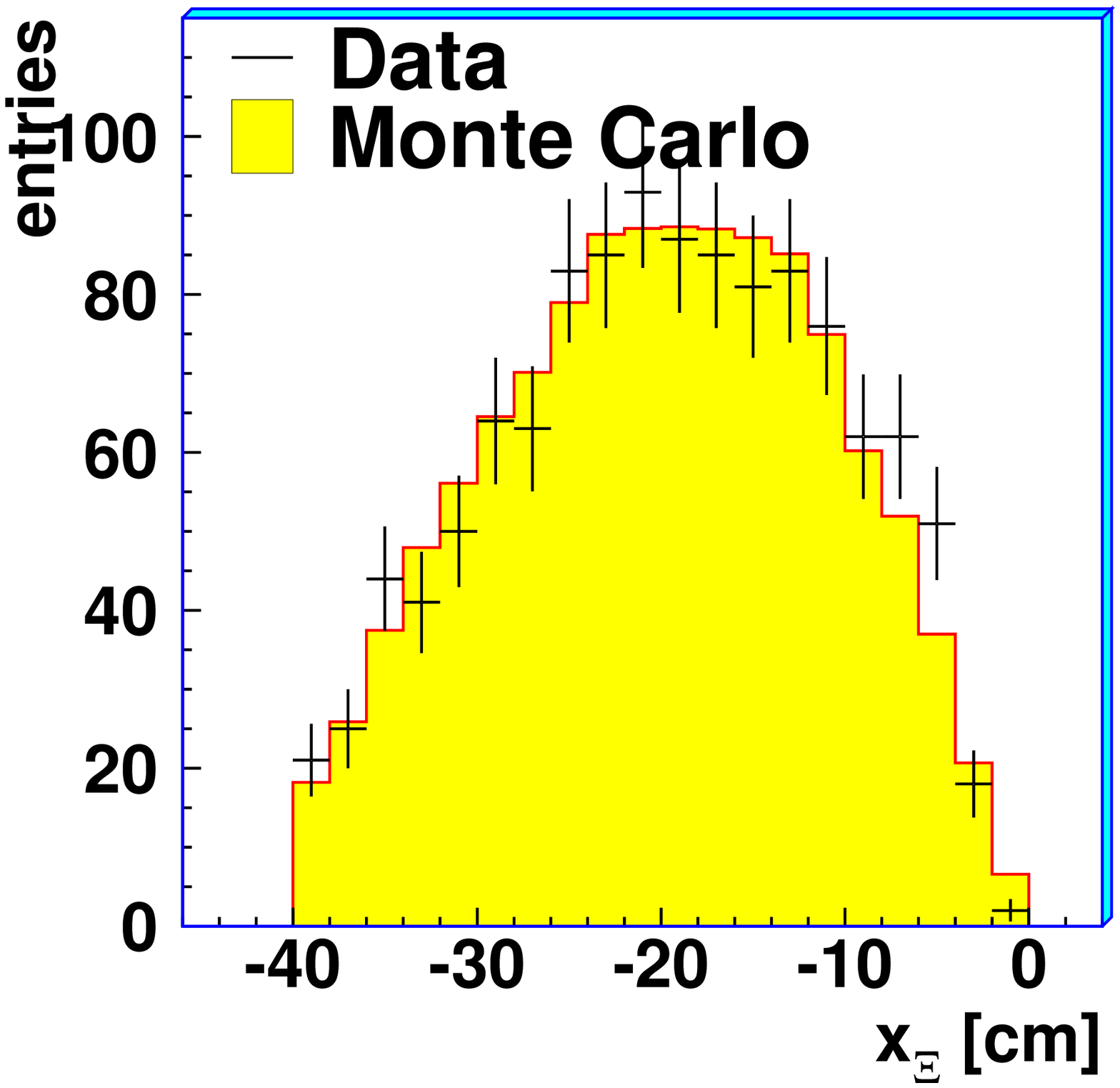}
\includegraphics[scale=0.190]{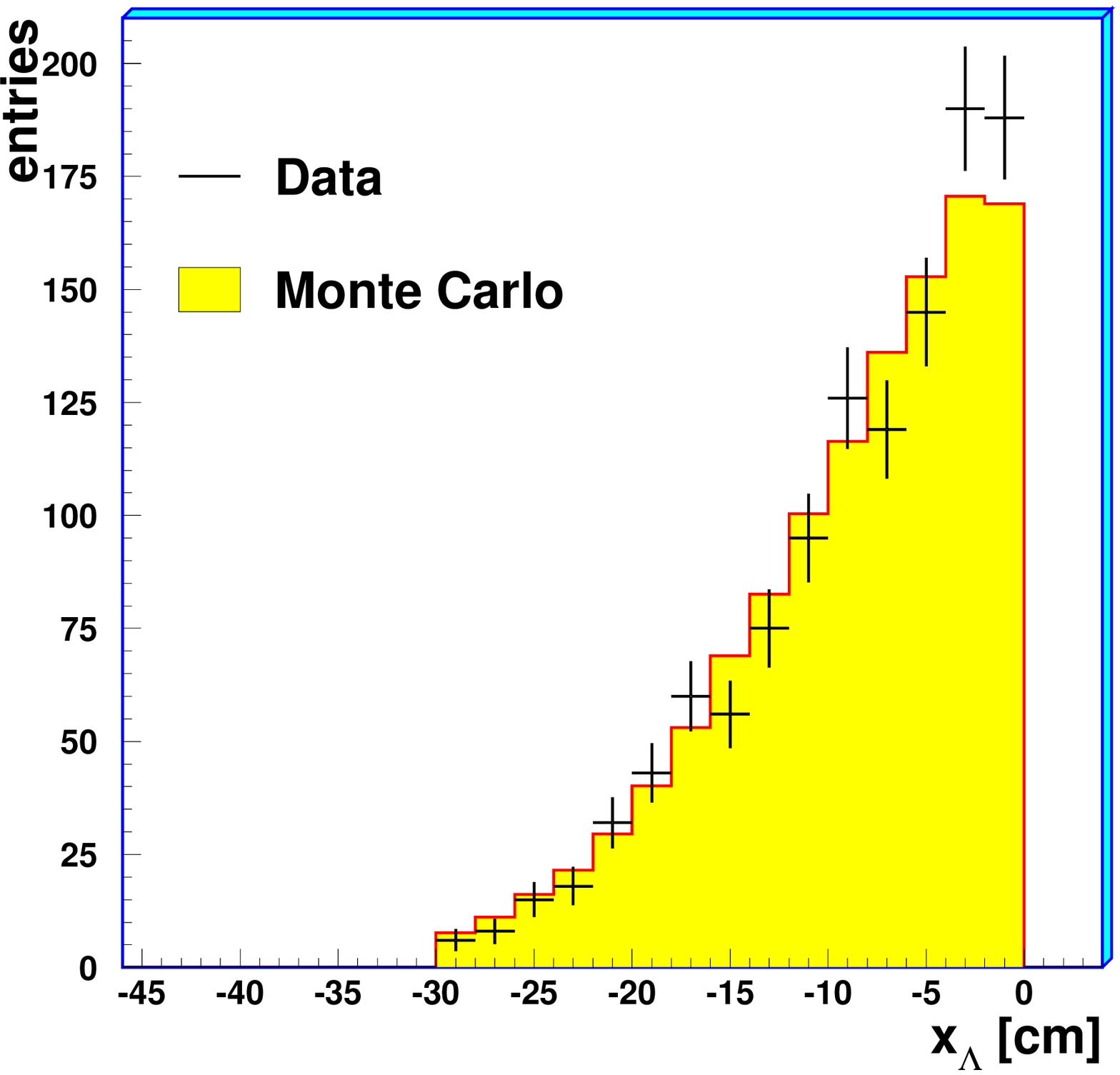}
\includegraphics[scale=0.190]{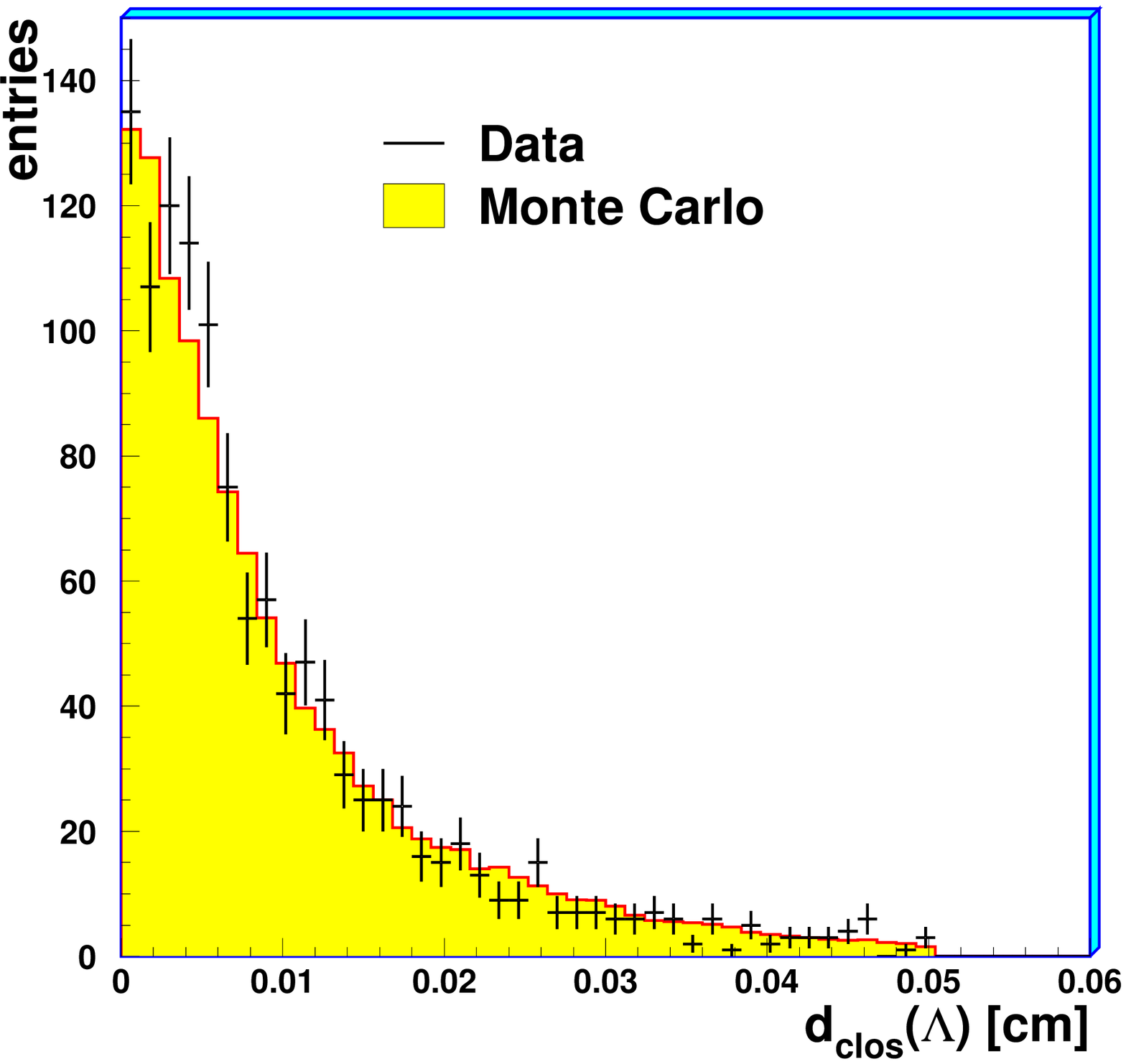} 
\includegraphics[scale=0.190]{cap4/xiMC.dir/xi_byLa.eps} \\
\includegraphics[scale=0.200]{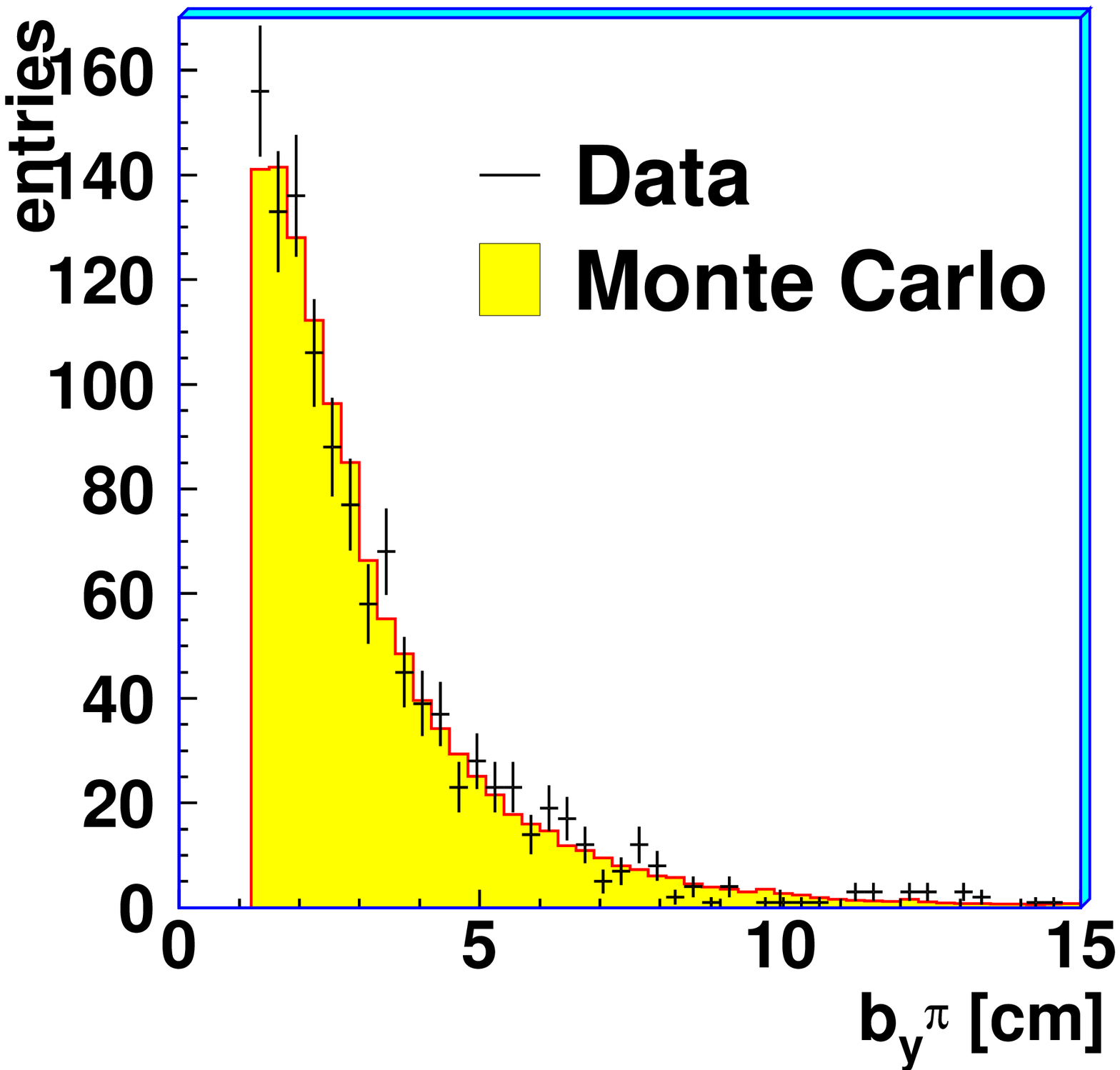} 
\includegraphics[scale=0.195]{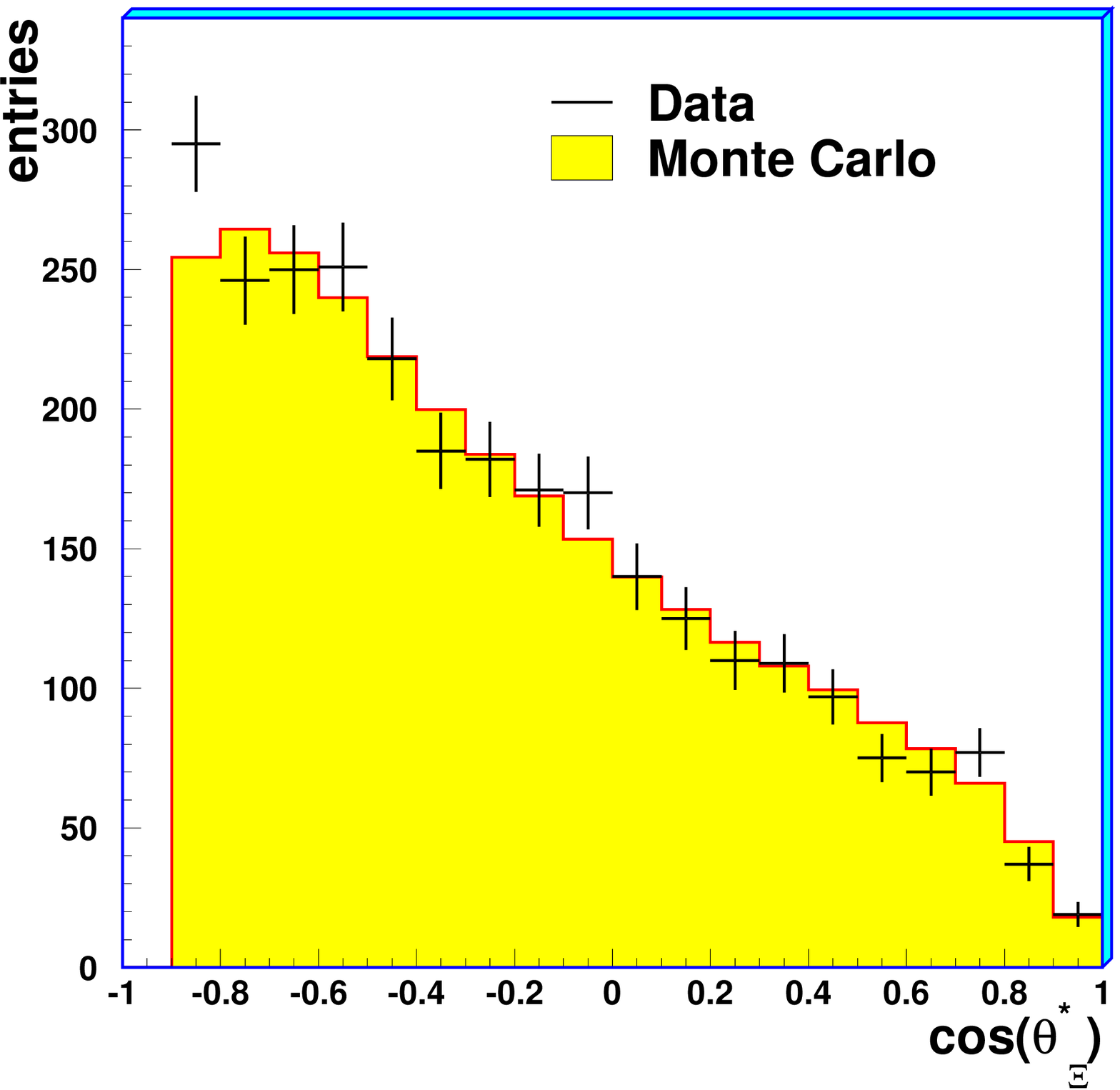}
\includegraphics[scale=0.195]{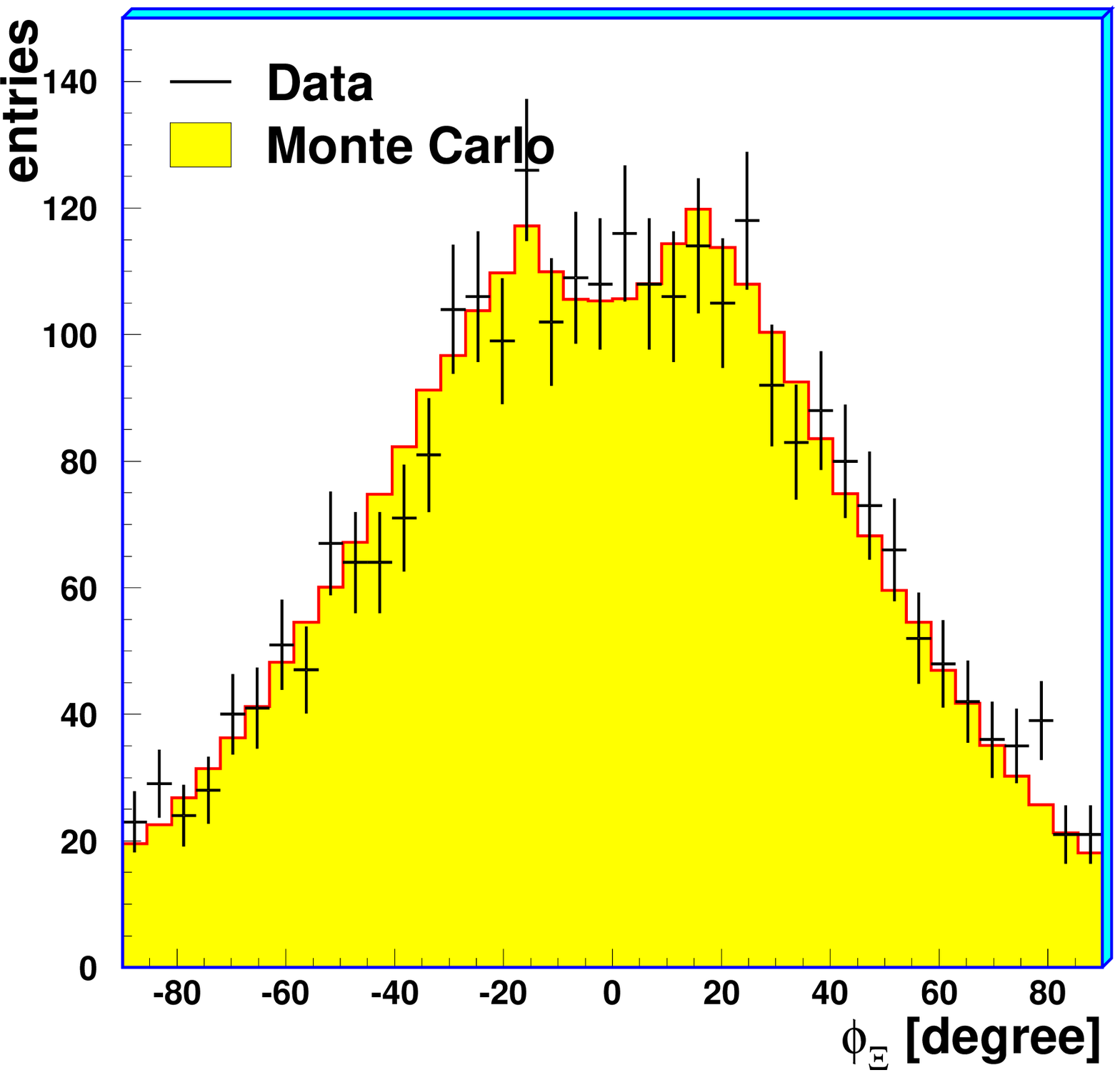}
\caption{Confronto tra varie distribuzioni delle \PgXm\ e \PagXp\ reali
         (punti con errori statistici) %o distribuzioni bidimensionali in chiaro) 
	 e le relative distribuzioni ottenute col Monte Carlo
         (distribuzioni in giallo).}
\label{XiComparison}
\end{center}
\end{figure}

In fig.~\ref{XiComparison}  sono mostrati alcuni dei confronti per le \PgXm\ e le 
\PagXp; da sinistra verso destra, e dall'alto in basso, si osservano le 
distribuzioni per: $x_{cas}$, $x_{V^0}$, $close_{V^0}$, $b_y(\Lambda)$; $b_y(\pi)$, 
$\cos(\theta^*_{\Xi})$, $\phi_{cas}$.
Nei grafici della fig.~\ref{XiComparison}, e cos\`i in tutti quelli relativi 
alle altre particelle considerate, le distribuzioni Monte Carlo, caratterizzate 
da un pi\`u piccollo errore statistico, sono rappresentate per mezzo di istogrammi in 
giallo, mentre per i dati reali si sono adoperati i punti con le barre di errore 
(e le distribuzioni in chiaro per le correlazioni bidimensionali).  

\begin{figure}[p]
\begin{center}
\includegraphics[scale=0.185]{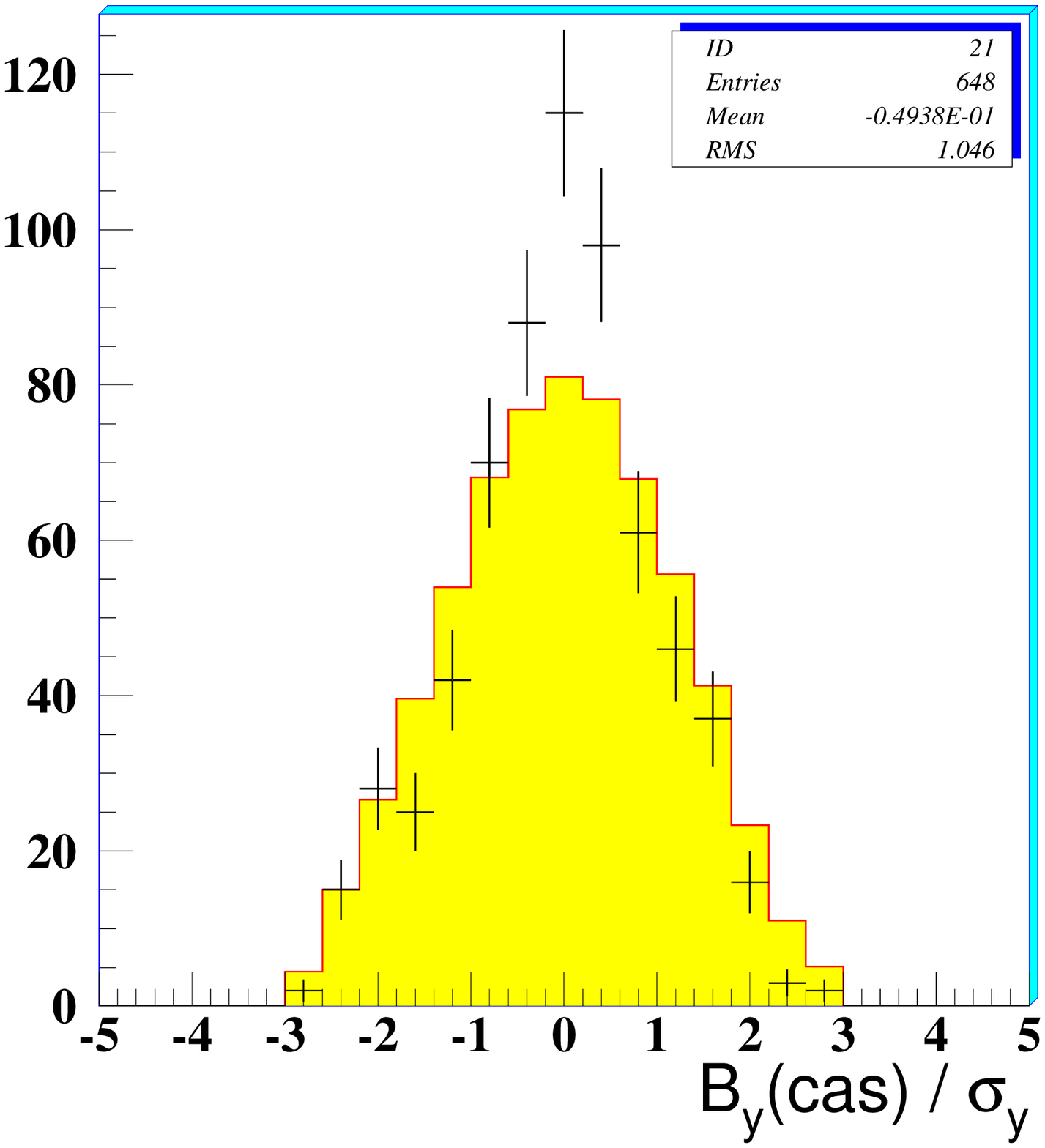}
\includegraphics[scale=0.185]{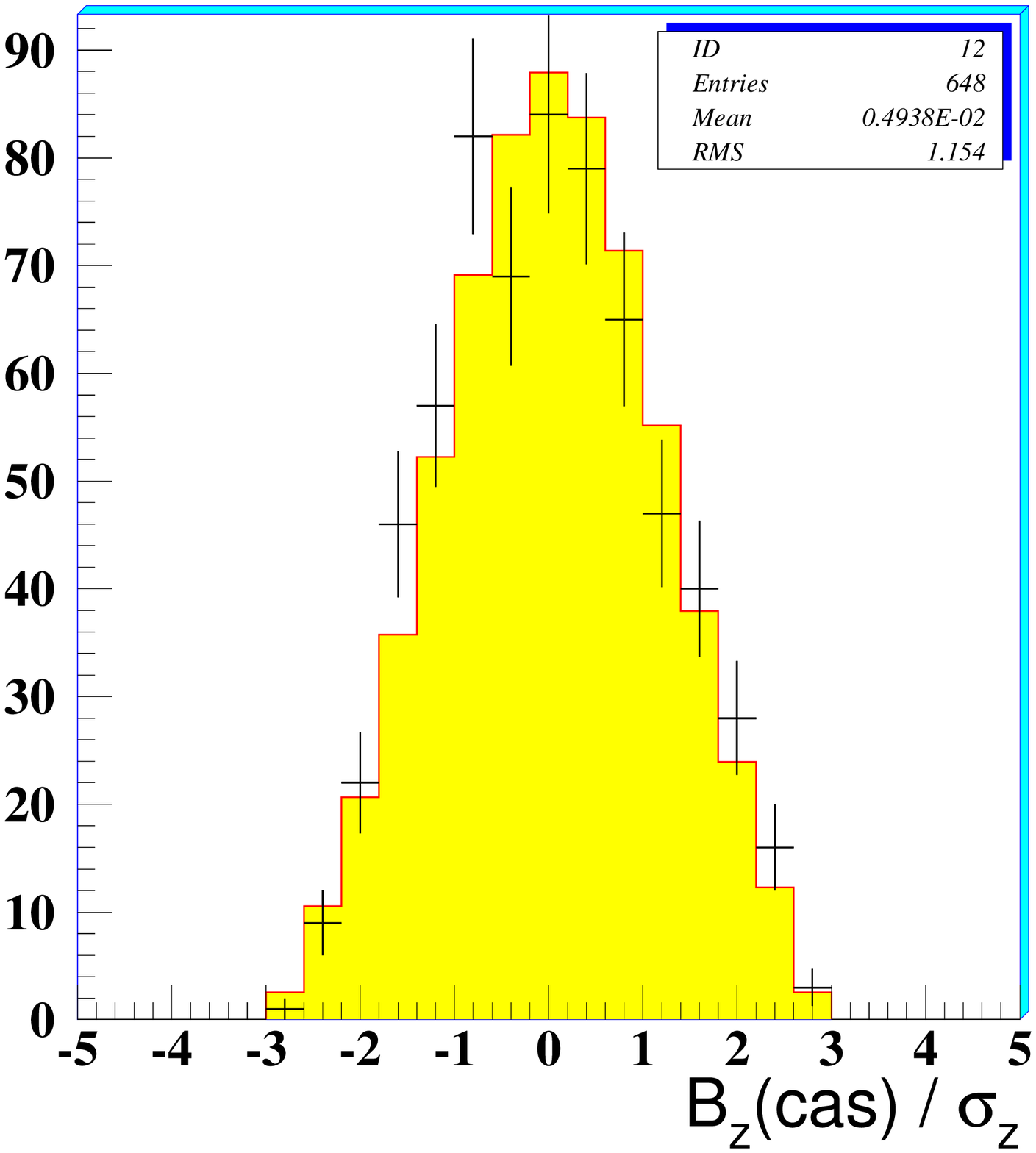}
\includegraphics[scale=0.185]{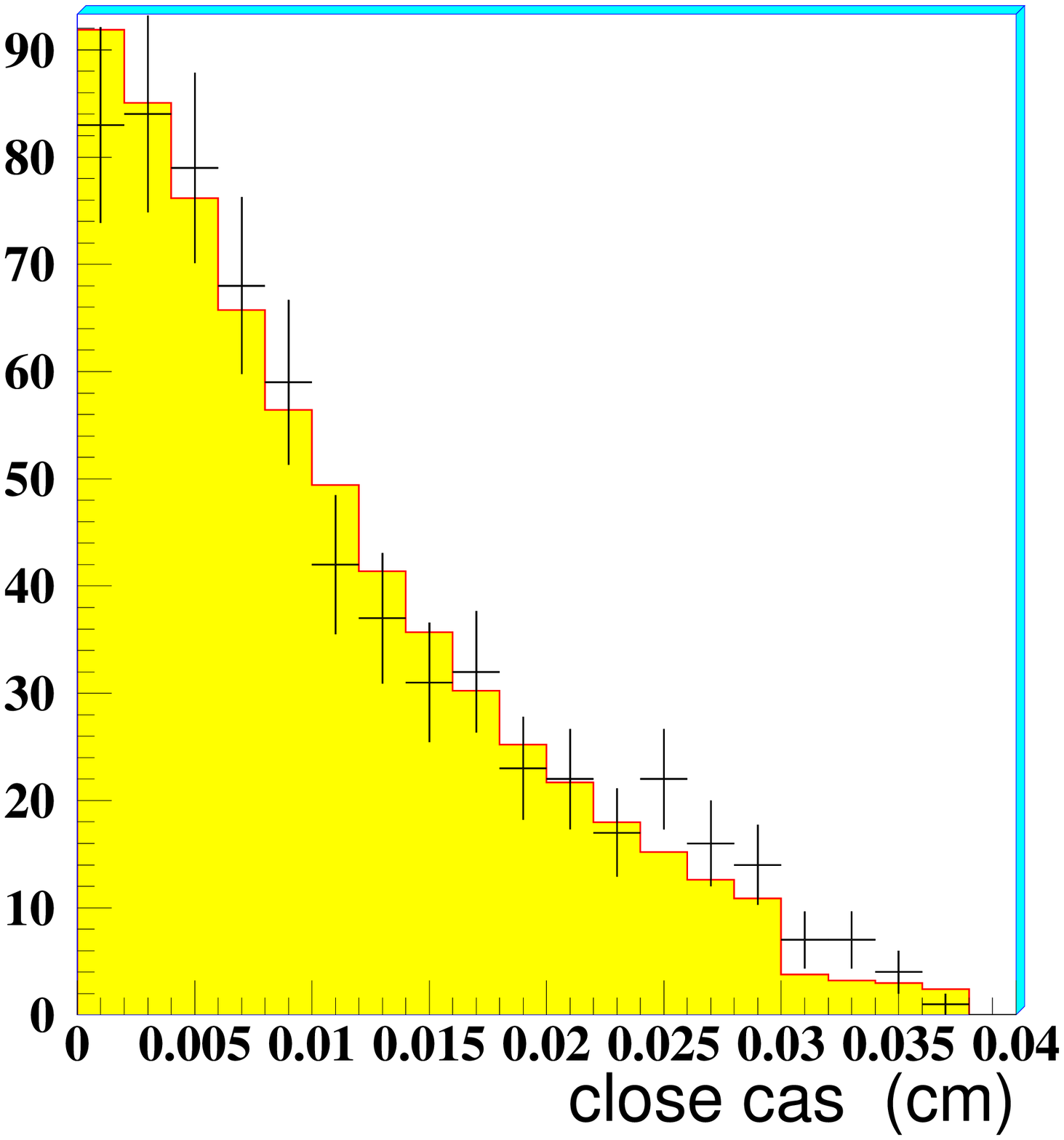}
\includegraphics[scale=0.185]{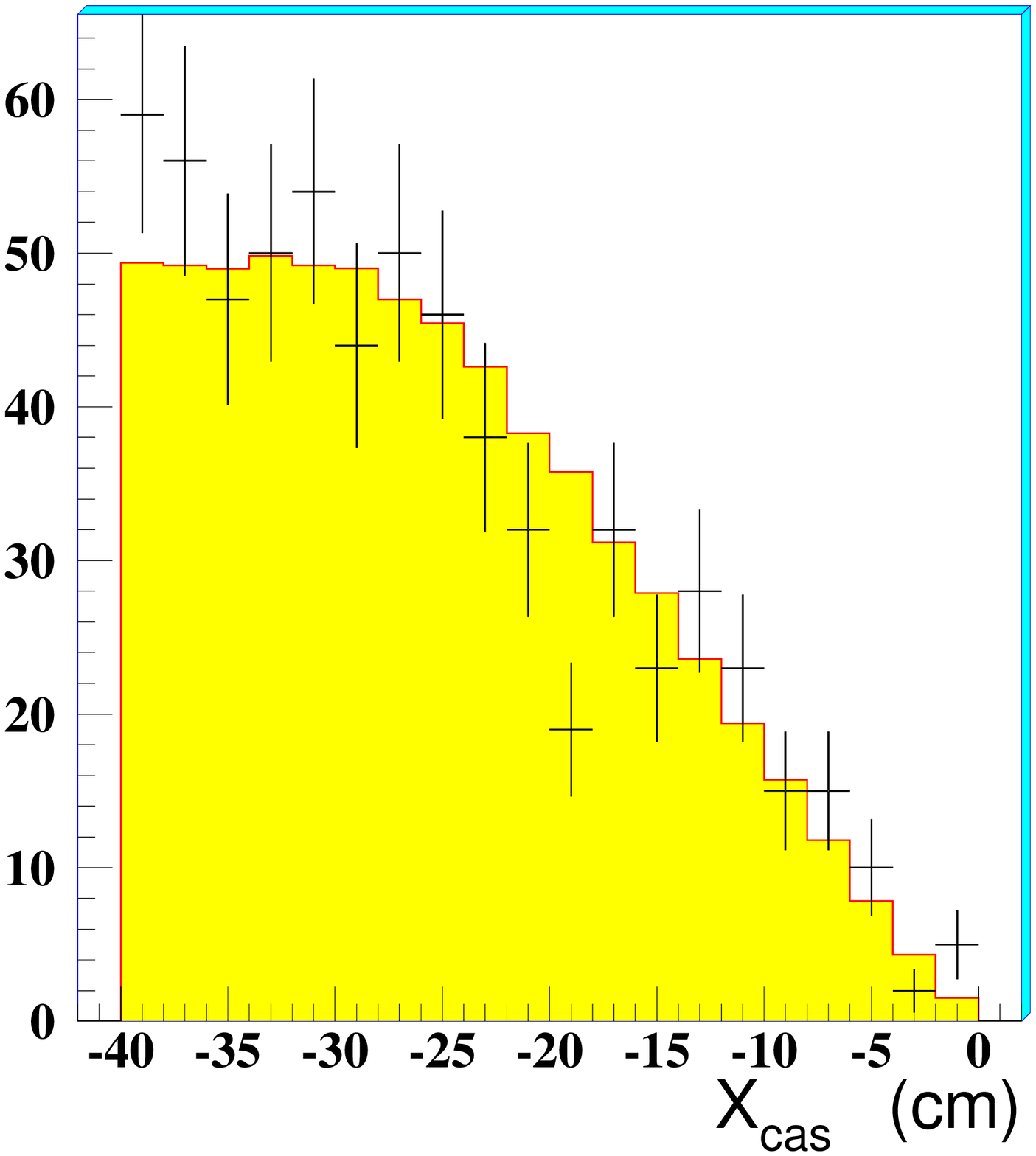} \\
\includegraphics[scale=0.185]{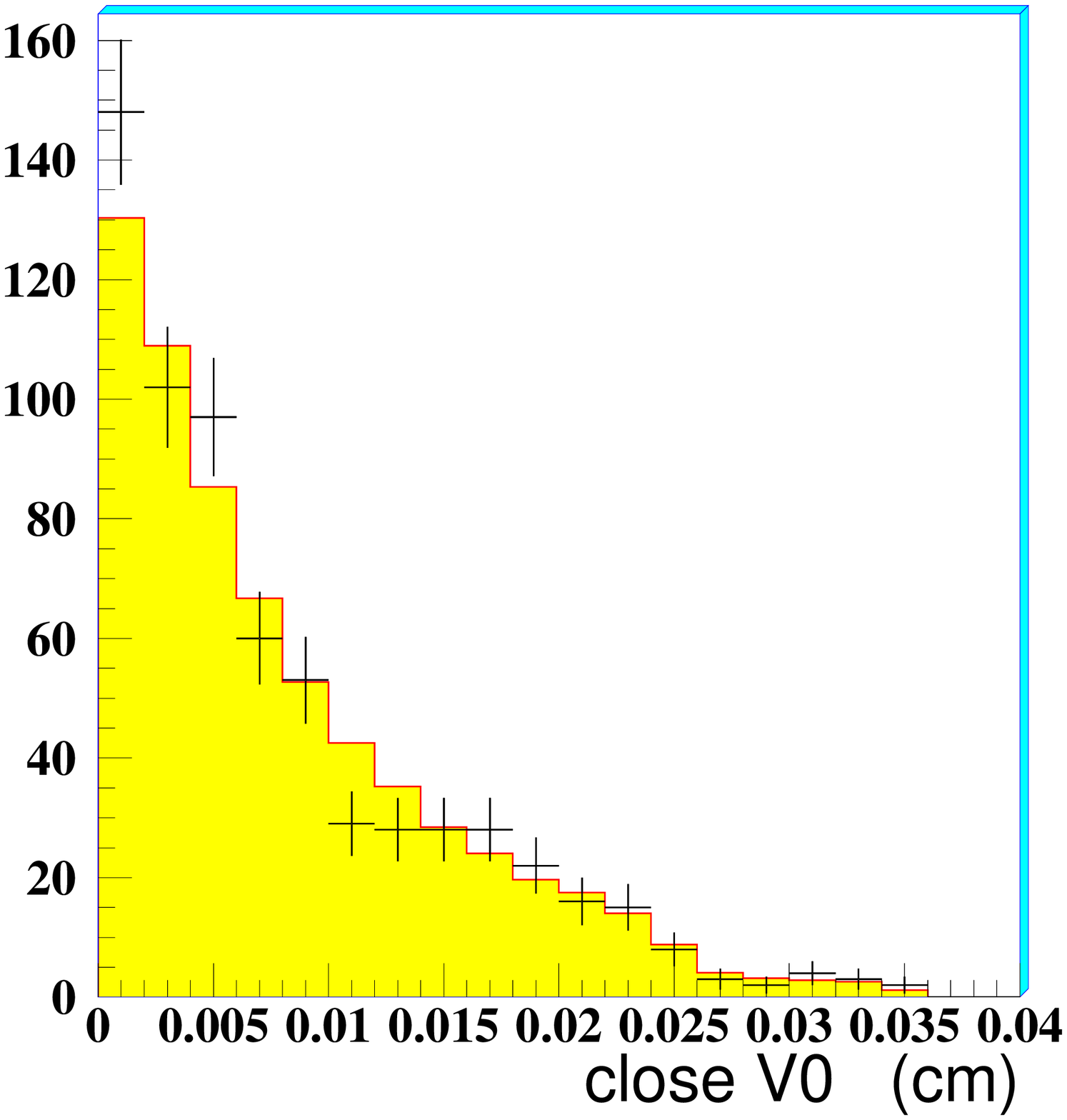}
\includegraphics[scale=0.185]{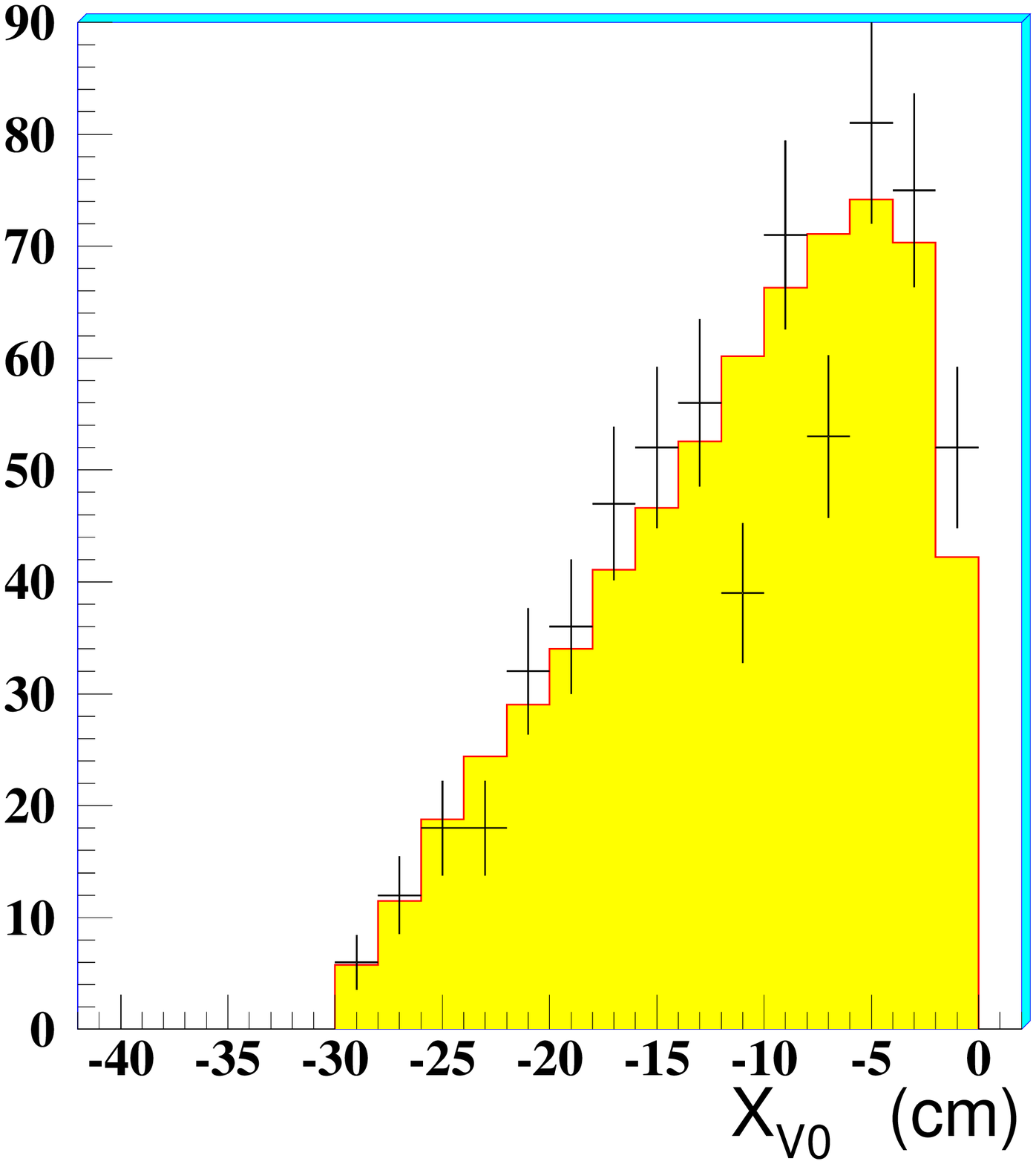}
\includegraphics[scale=0.185]{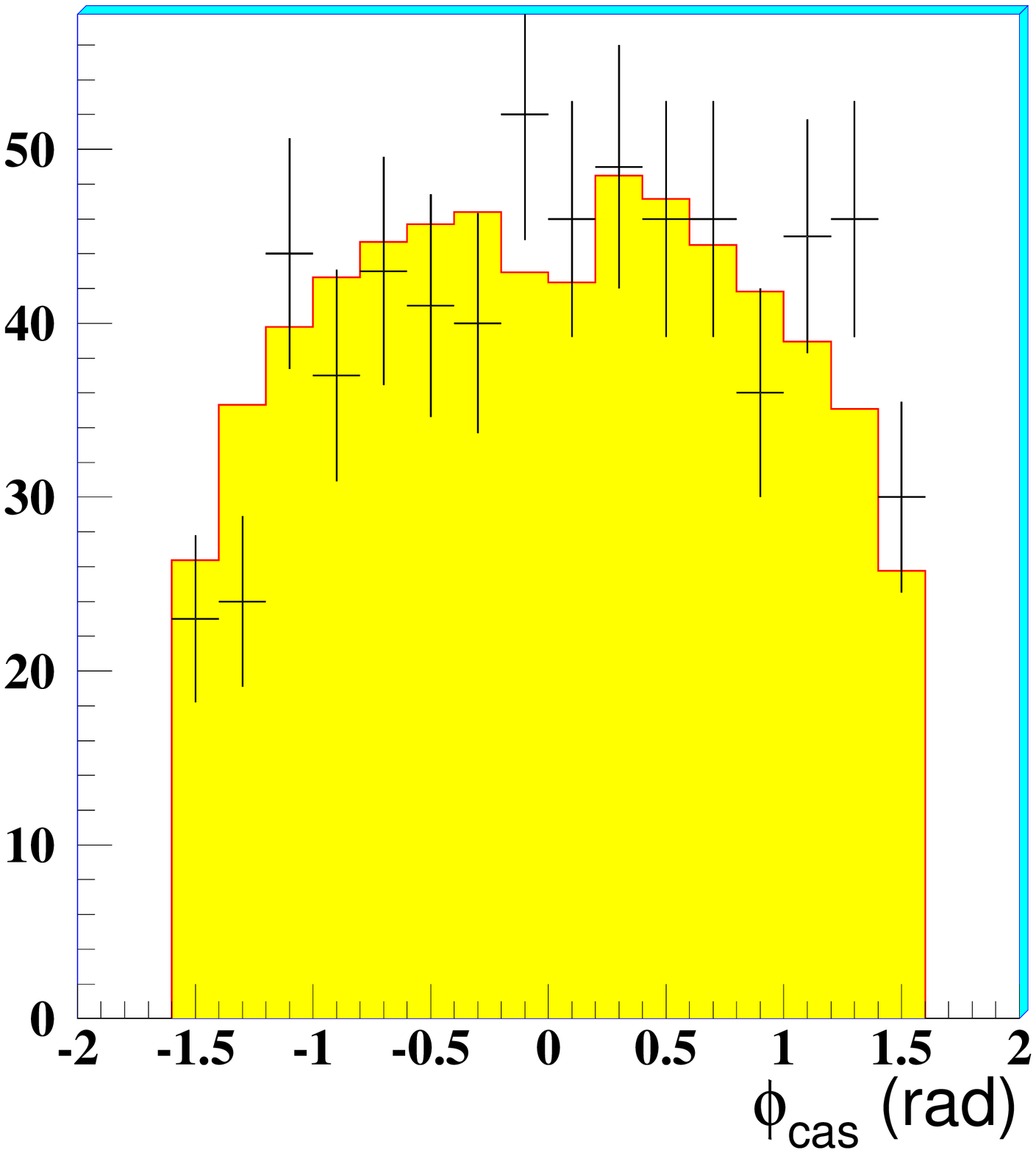}
\includegraphics[scale=0.185]{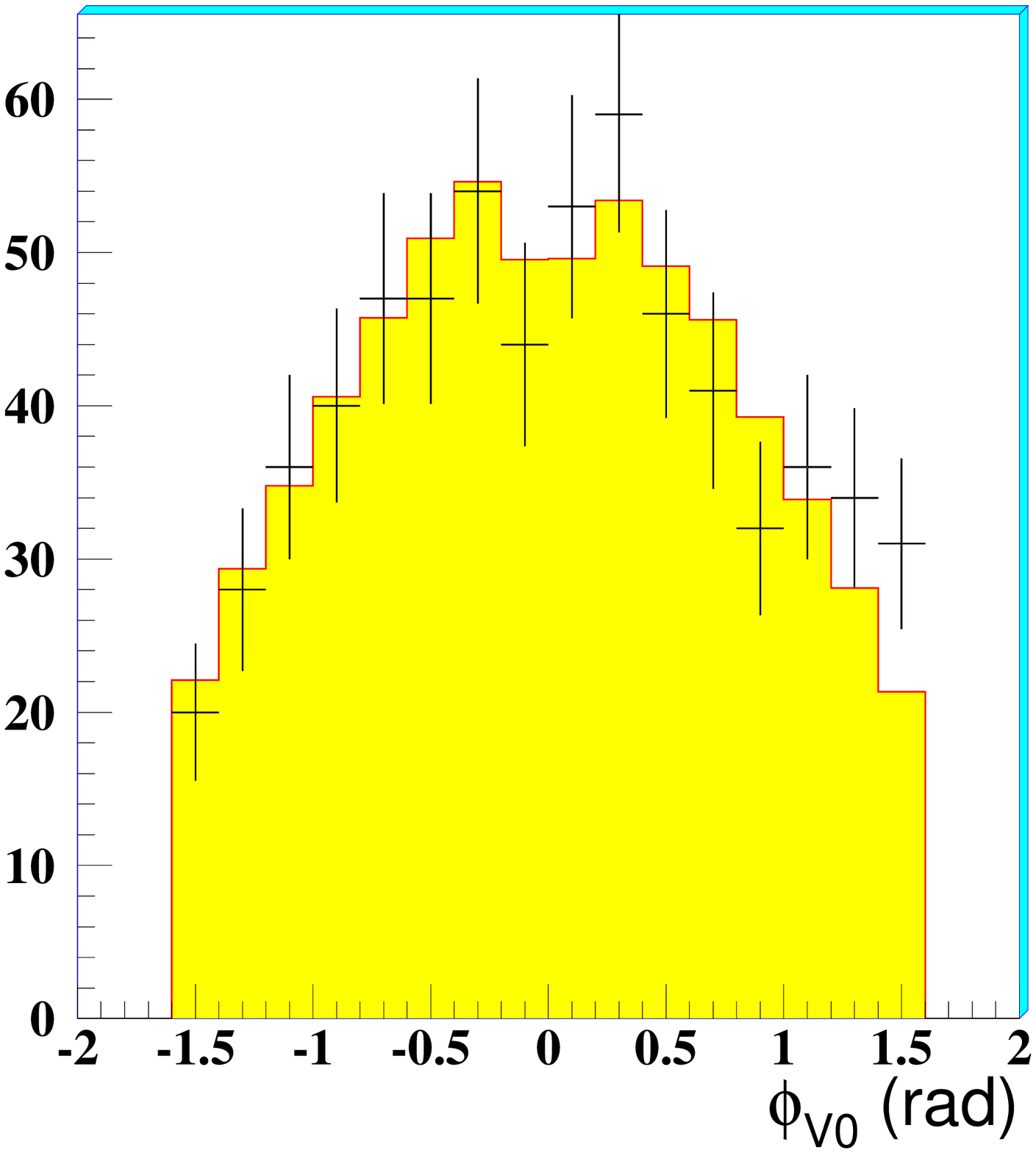}\\
\includegraphics[scale=0.185]{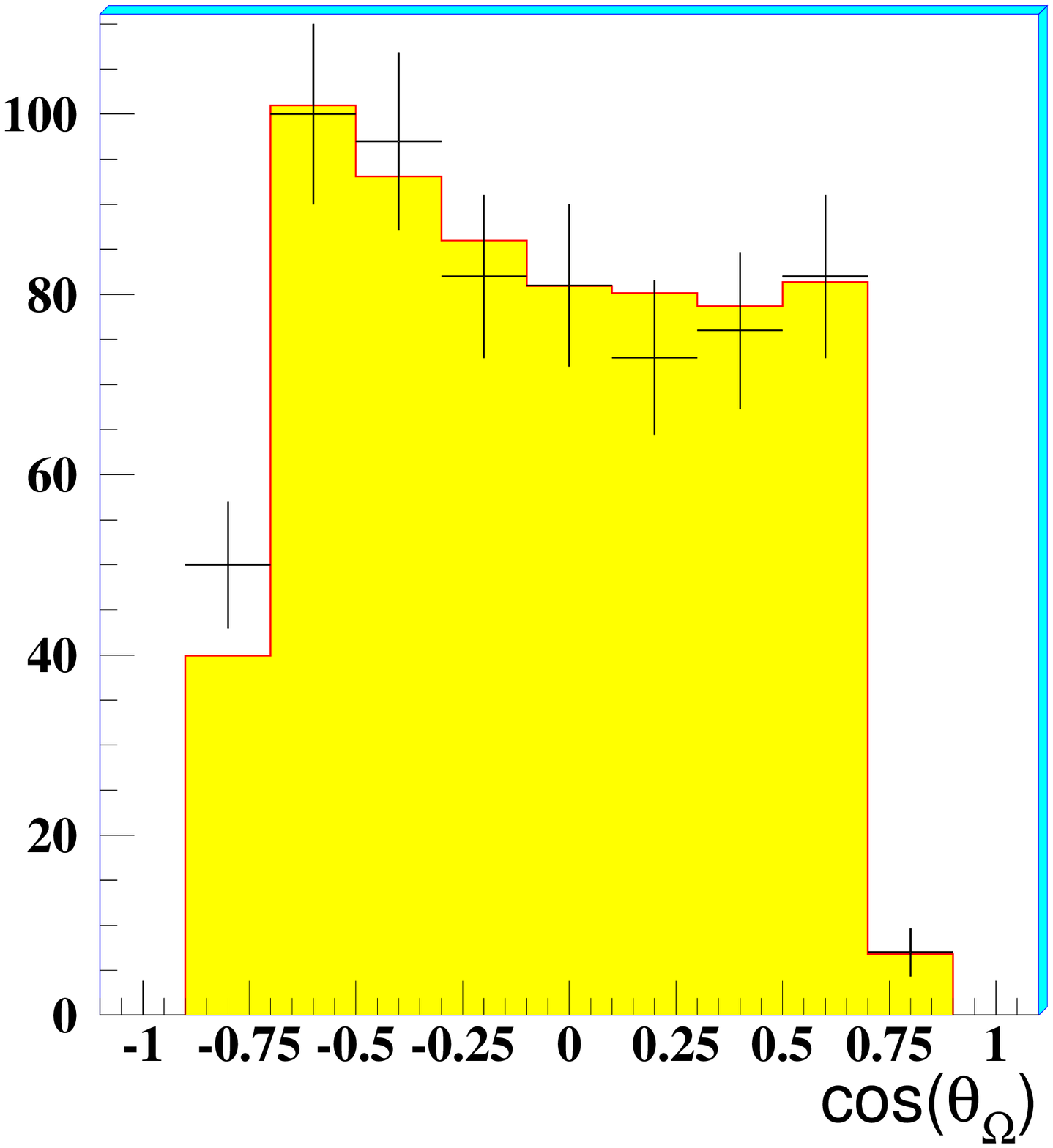}
\includegraphics[scale=0.185]{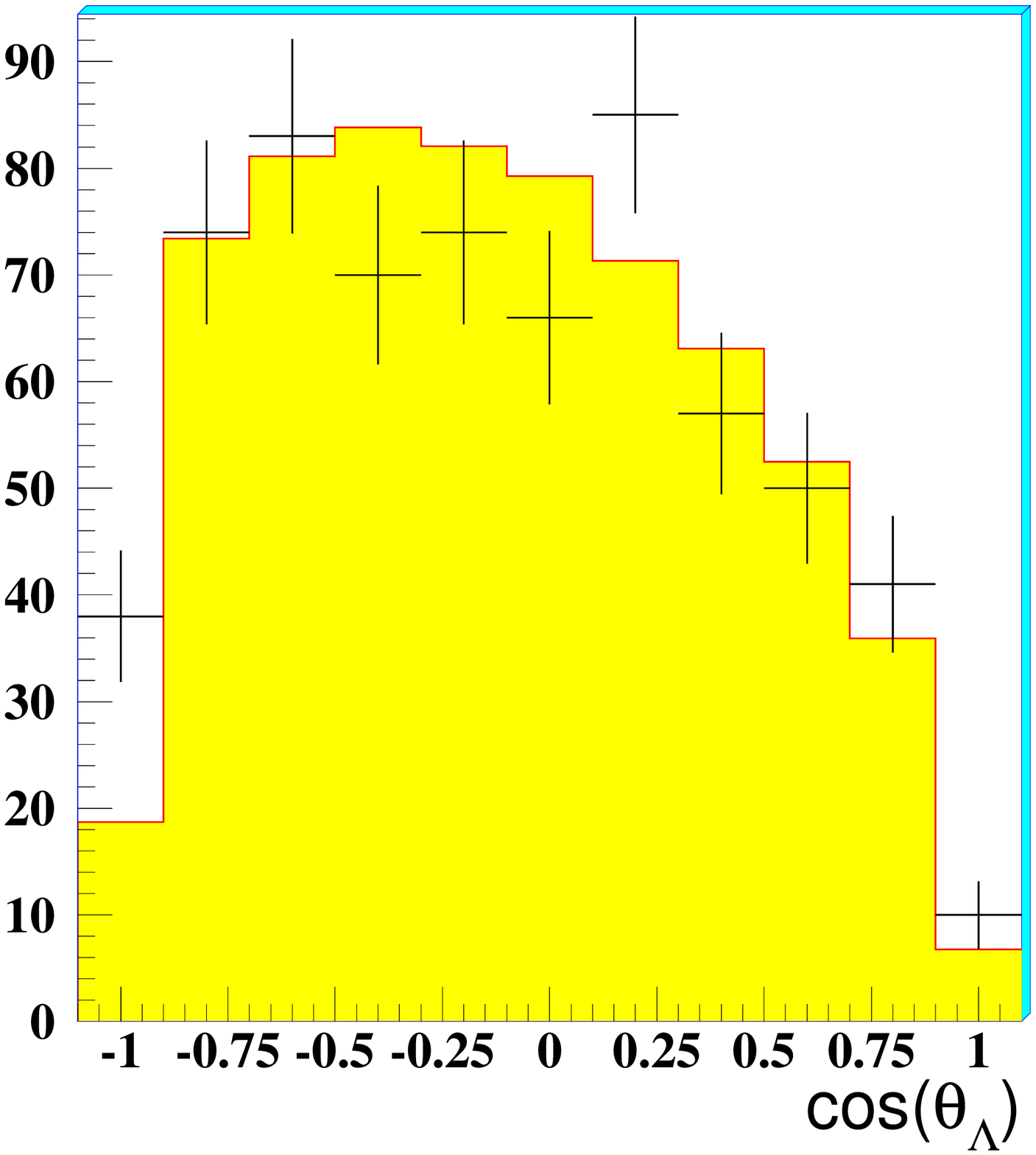}
\includegraphics[scale=0.185]{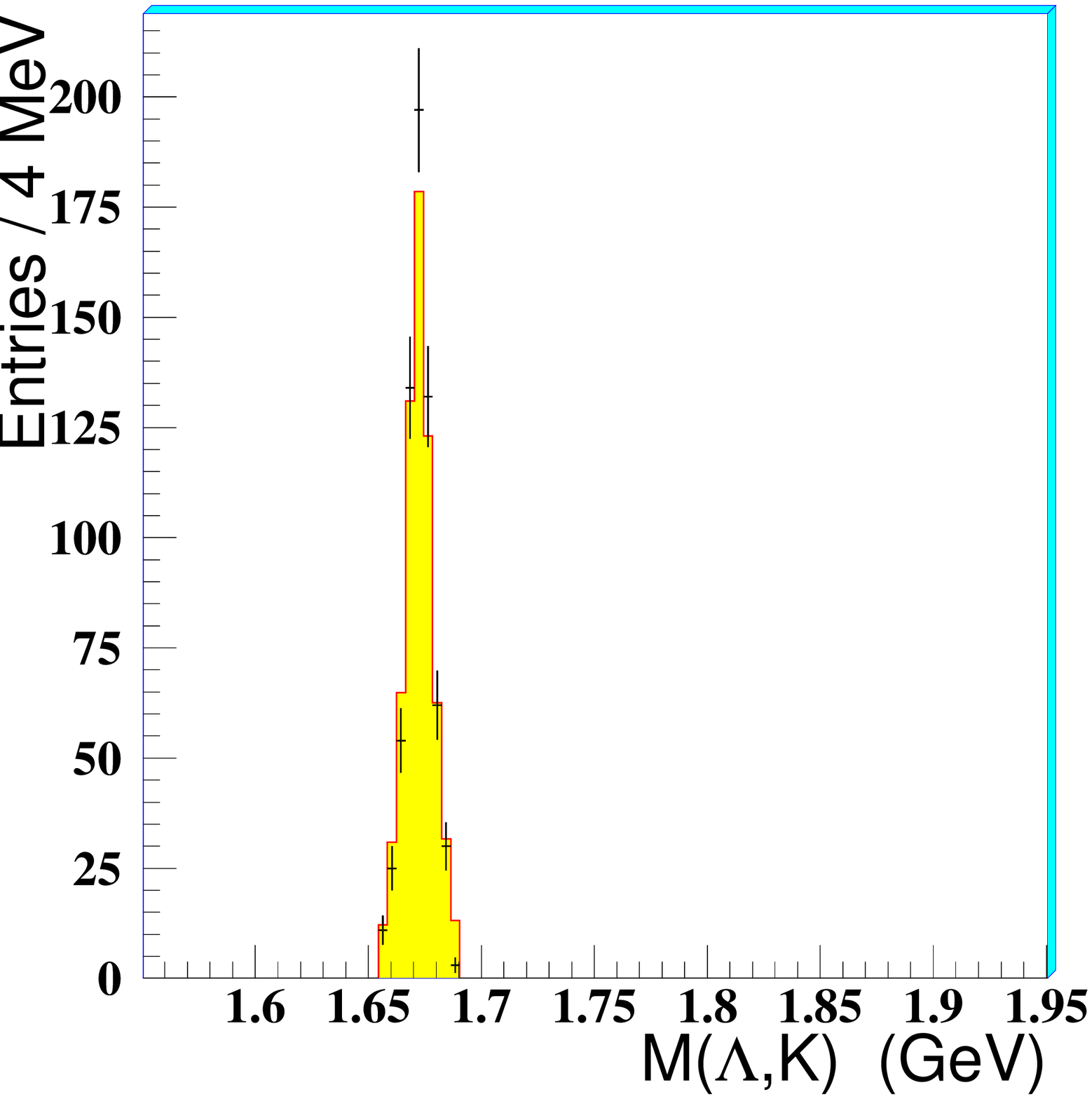}
\includegraphics[scale=0.185]{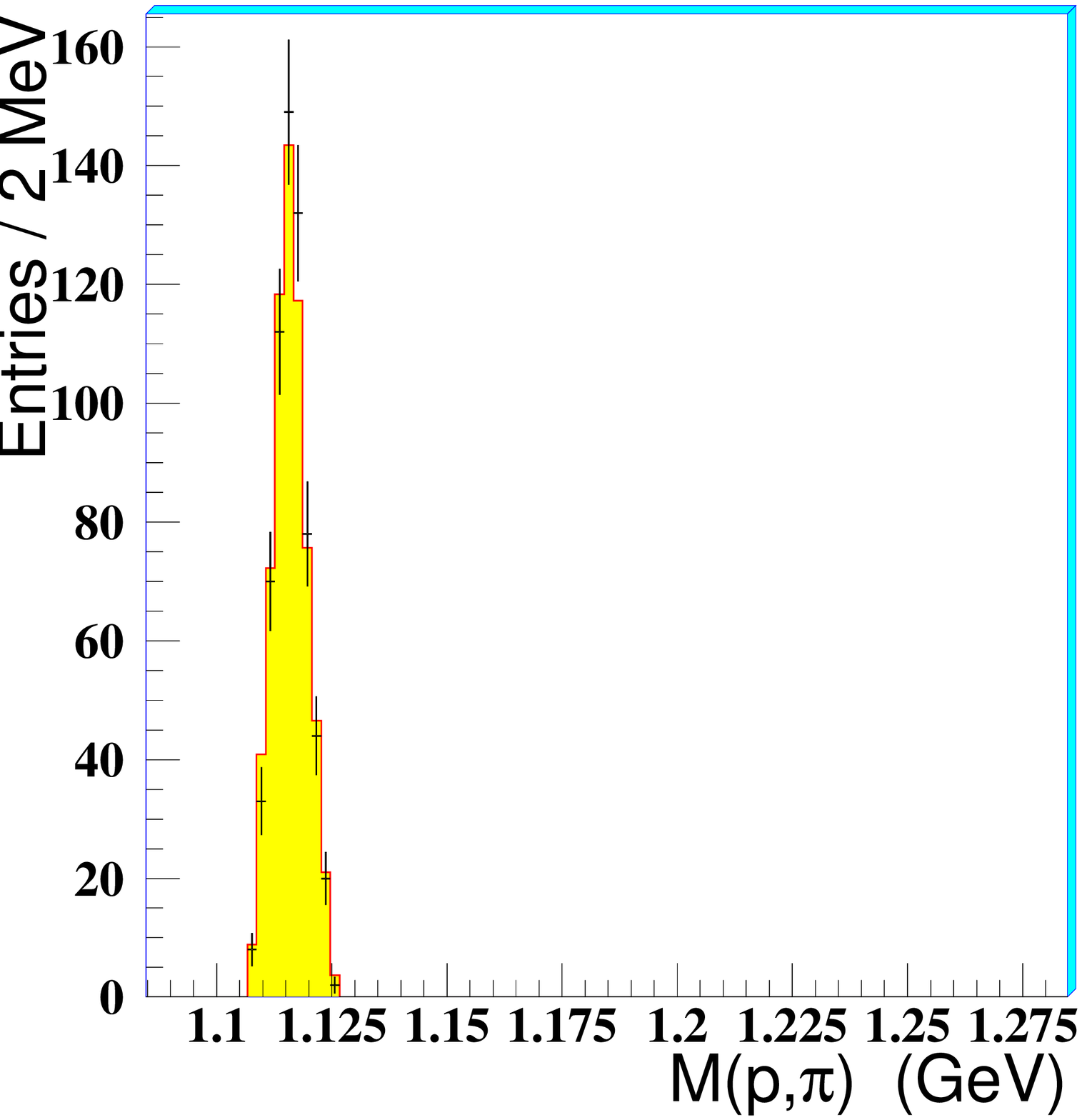} \\
\includegraphics[scale=0.185]{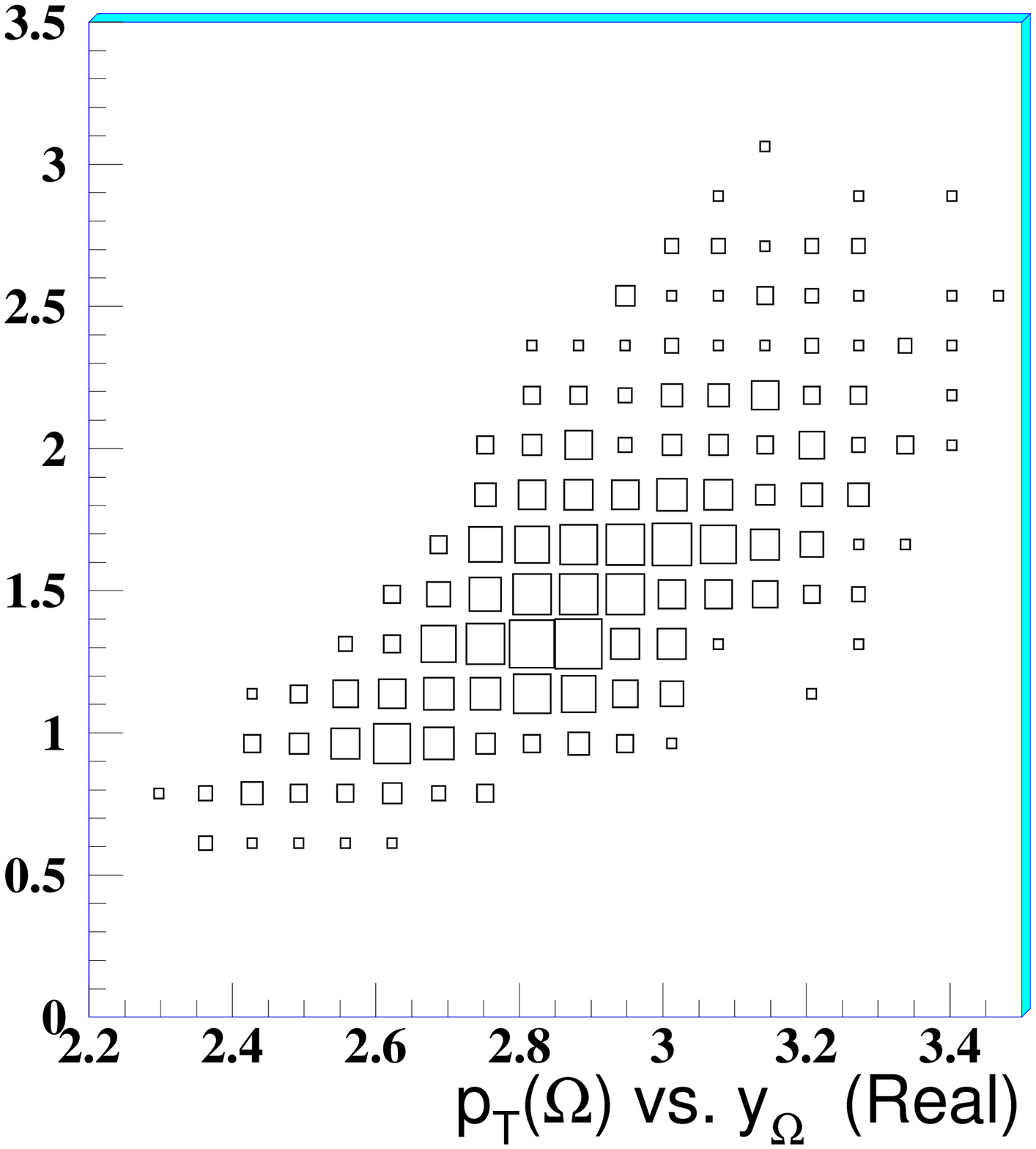}
\includegraphics[scale=0.185]{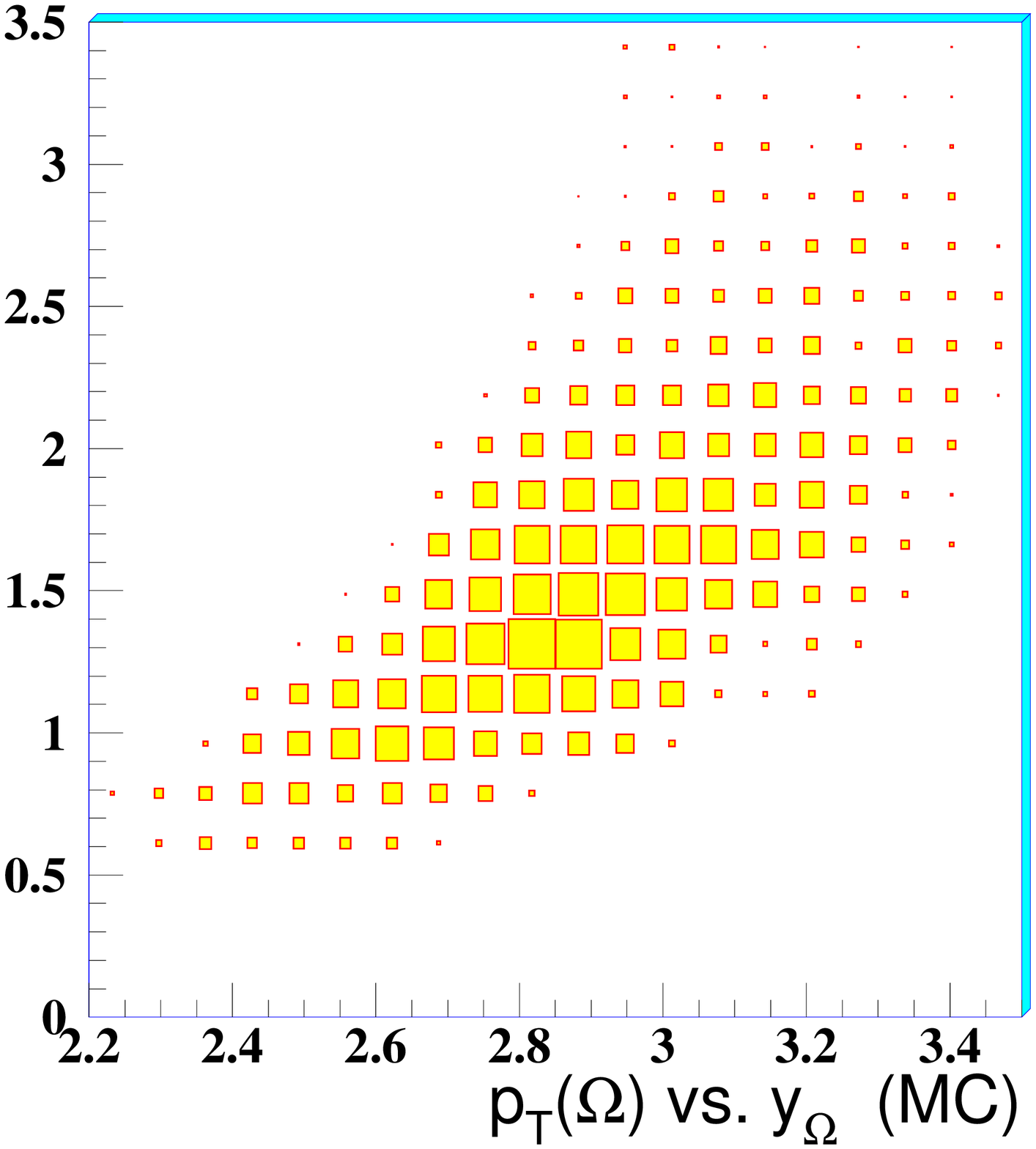}
\includegraphics[scale=0.185]{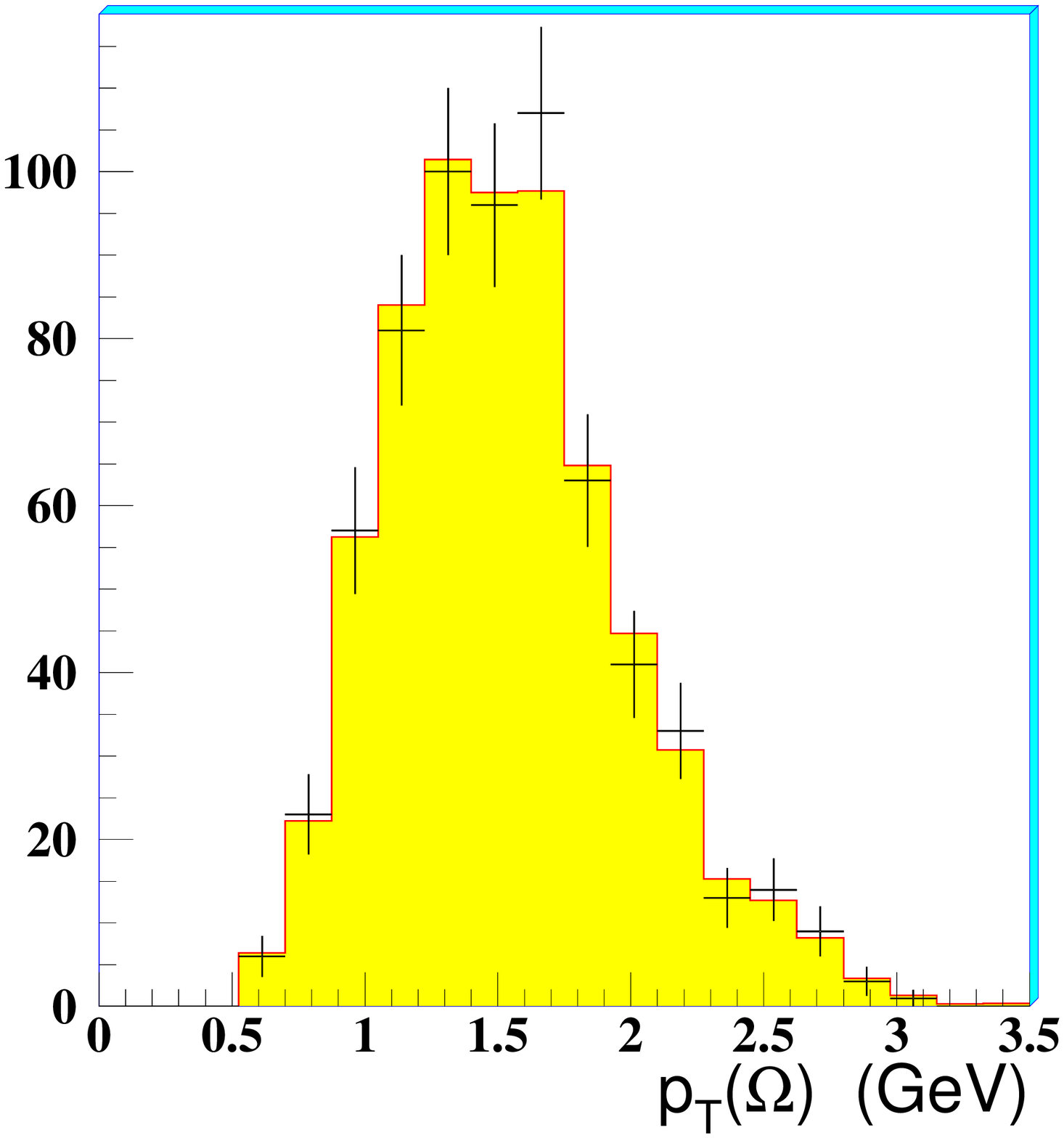}
\includegraphics[scale=0.185]{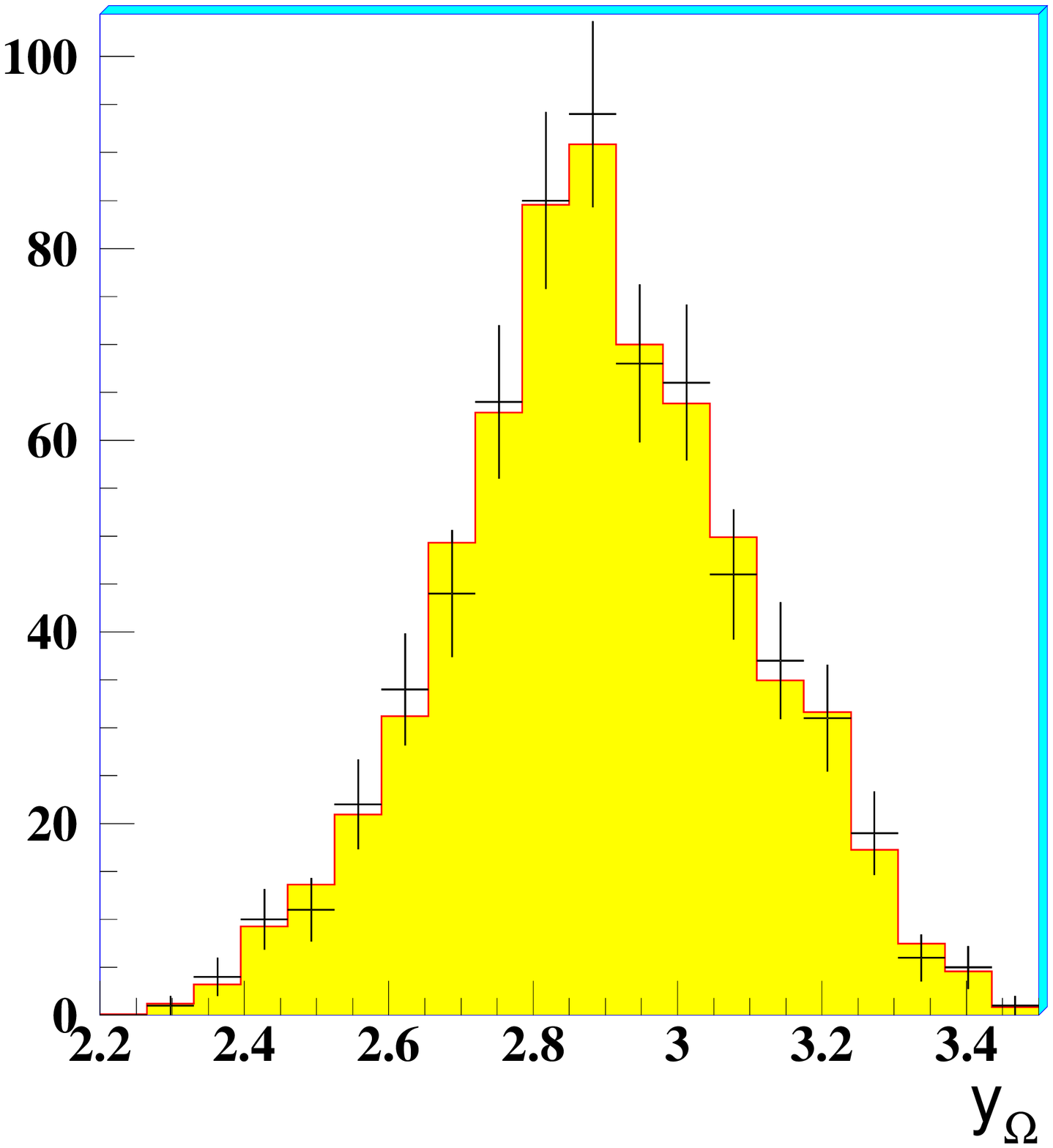}\\
\includegraphics[scale=0.185]{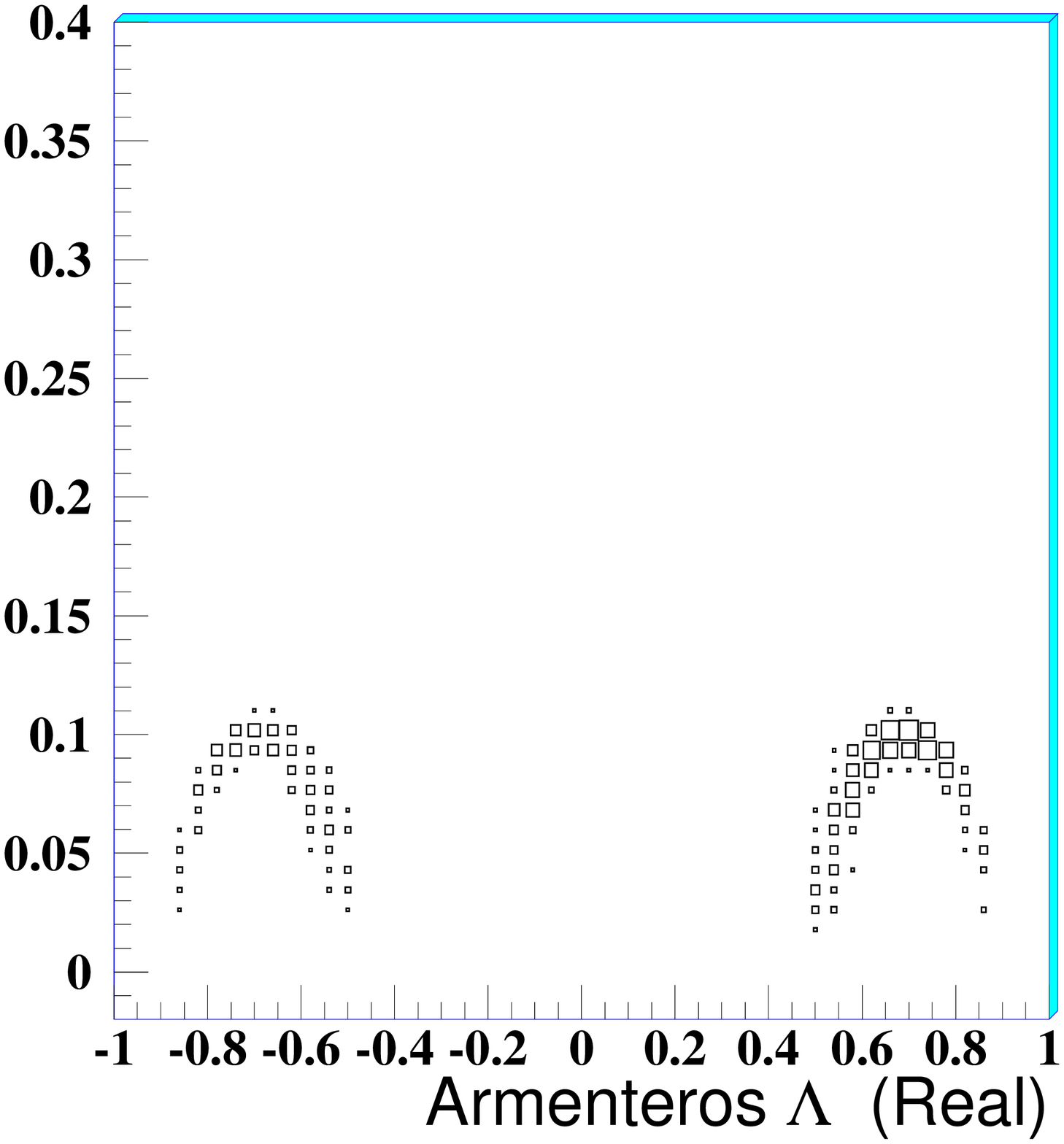}
\includegraphics[scale=0.185]{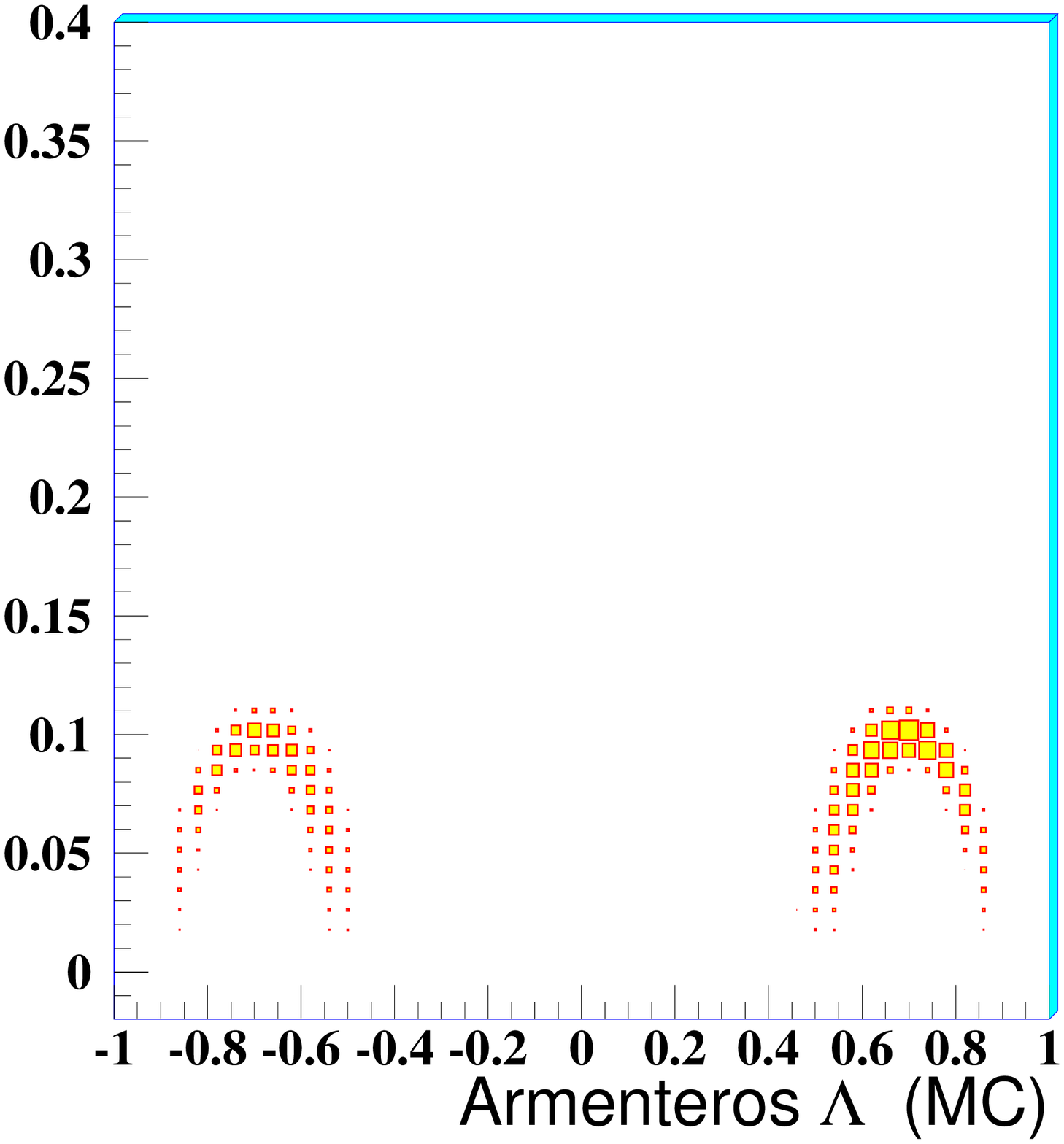}
\includegraphics[scale=0.185]{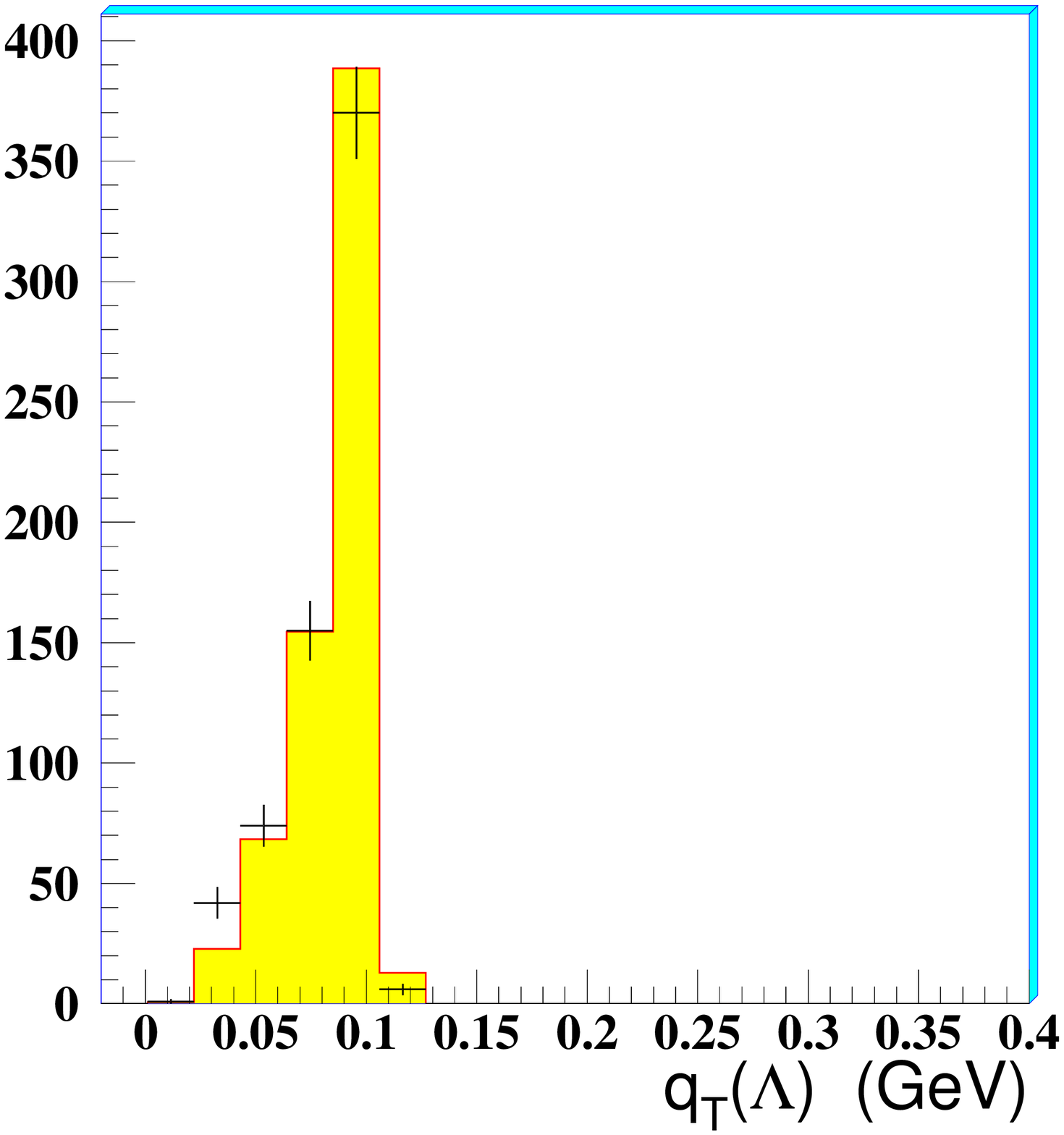}
\includegraphics[scale=0.185]{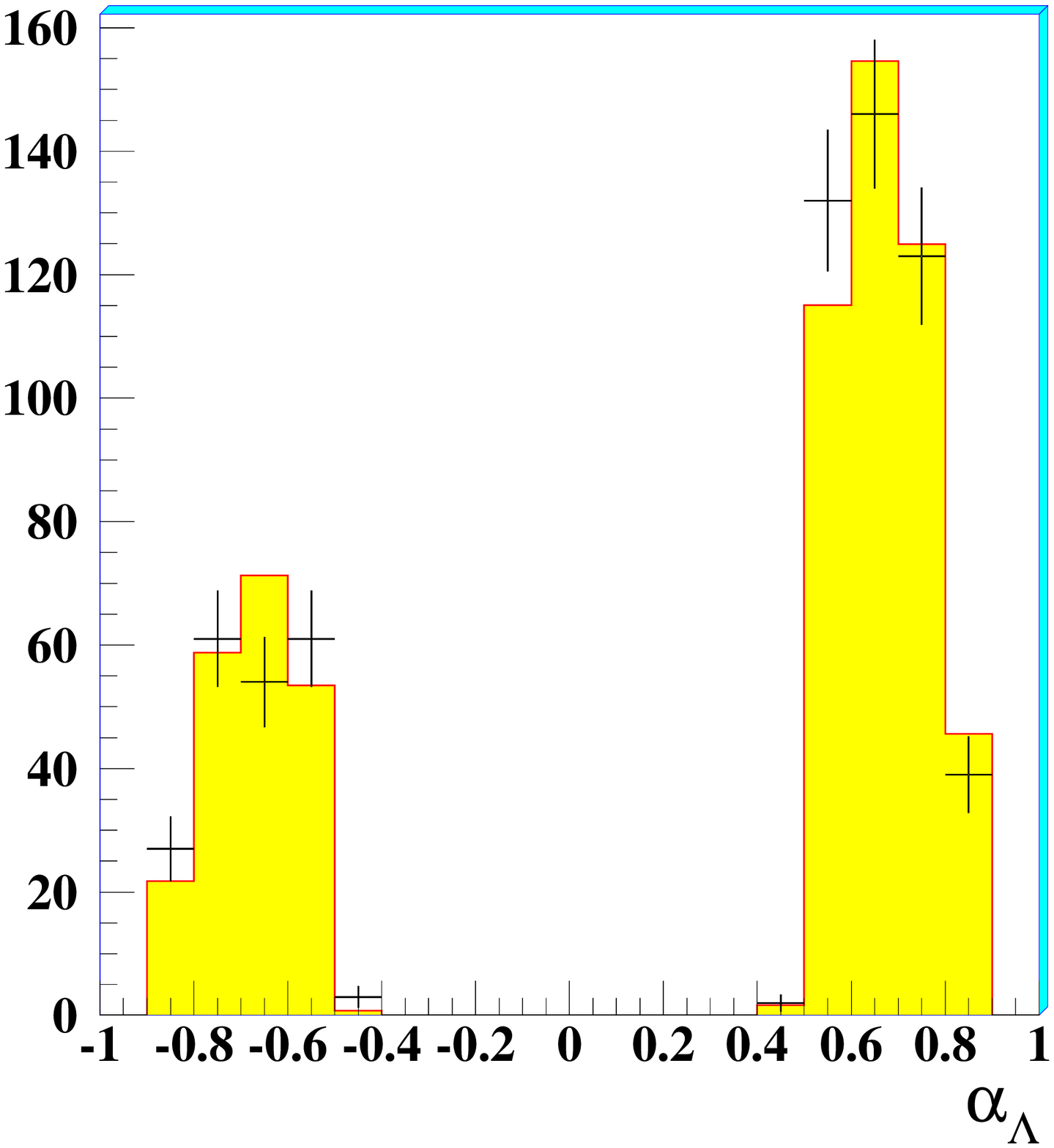}
\caption{Confronto tra varie distribuzioni delle \PgOm\ e \PagOp\ reali 
	 (punti con errori statistici o distribuzioni bidimensionali 
	 in chiaro) e le relative distribuzioni ottenute col Monte Carlo 
	 (distribuzioni in giallo).}
\label{OmComparison}
\end{center}
\end{figure}

In fig.~\ref{OmComparison} sono mostrati i confronti per le \PgOm\ e le \PagOp;  
guardando da sinistra verso destra, sono riportate le distribuzioni per: 
${b_y}_{cas}/\sigma_{y}$, ${b_z}_{cas}/\sigma_{z}$, $close_{cas}$, $x_{cas}$; 
$close_{V^0}$, $x_{V^0}$, $\phi_{cas}$, $\phi_{V^0}$;  
$\cos(\theta^*_{\Omega})$, $\cos(\theta^*_{\Lambda})$, $M(\Lambda,K)$, $M(p,\pi)$; 
$p_T$\ versus $y_{\Omega}$\ per i dati reali (in chiaro), 
$p_T$\ versus $y_{\Omega}$\ secondo il Monte Carlo (in giallo), $p_T$, $y_{\Omega}$;  
$q_T$\ versus $\alpha_{Arm}$\ per il decadimento delle $\Lambda$\ reali (in chiaro) 
e di quelle generate (in giallo), $q_T$, $\alpha_{Arm}$.  

\begin{figure}[p]
\begin{center}
\includegraphics[scale=0.185]{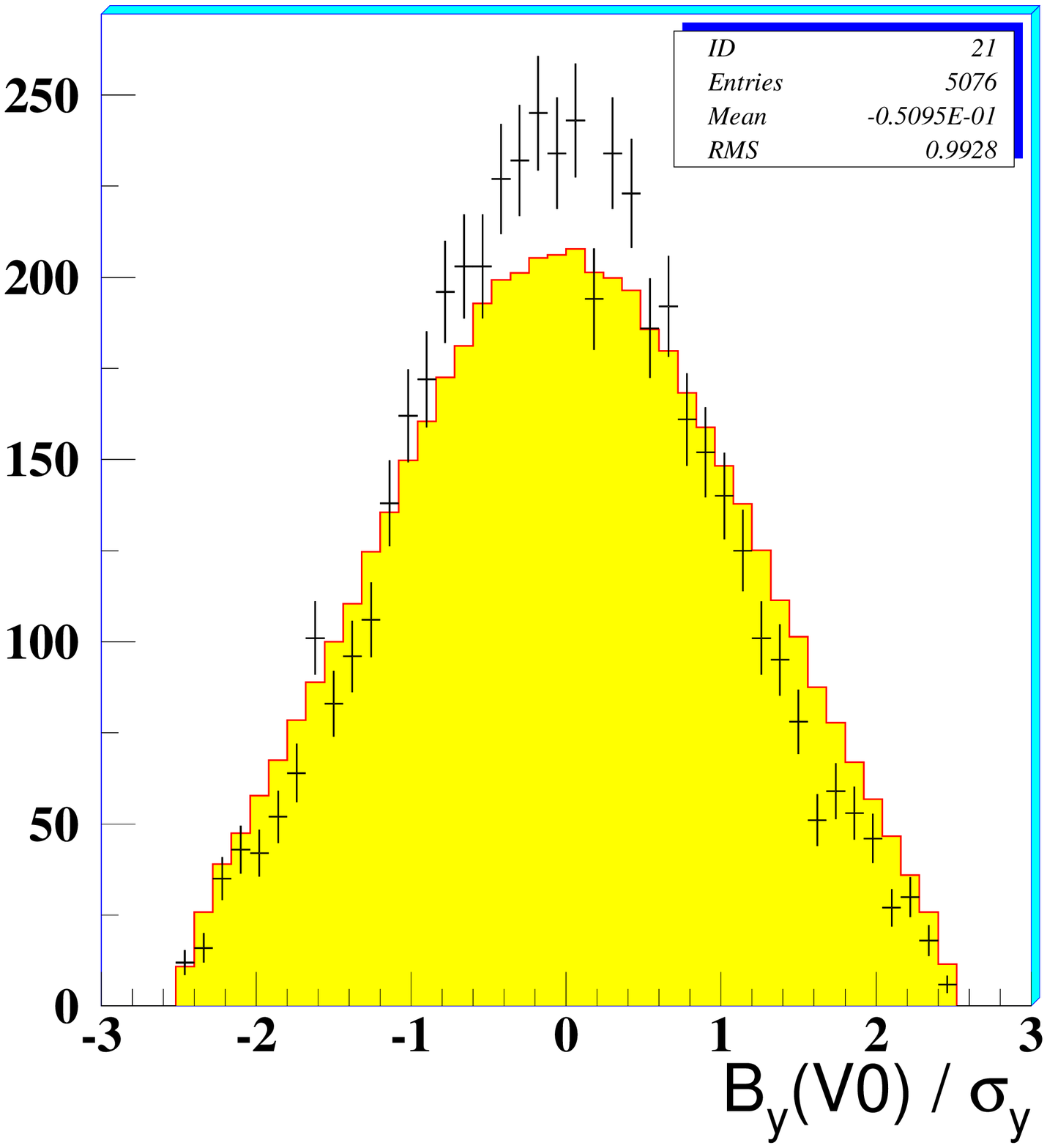}
\includegraphics[scale=0.185]{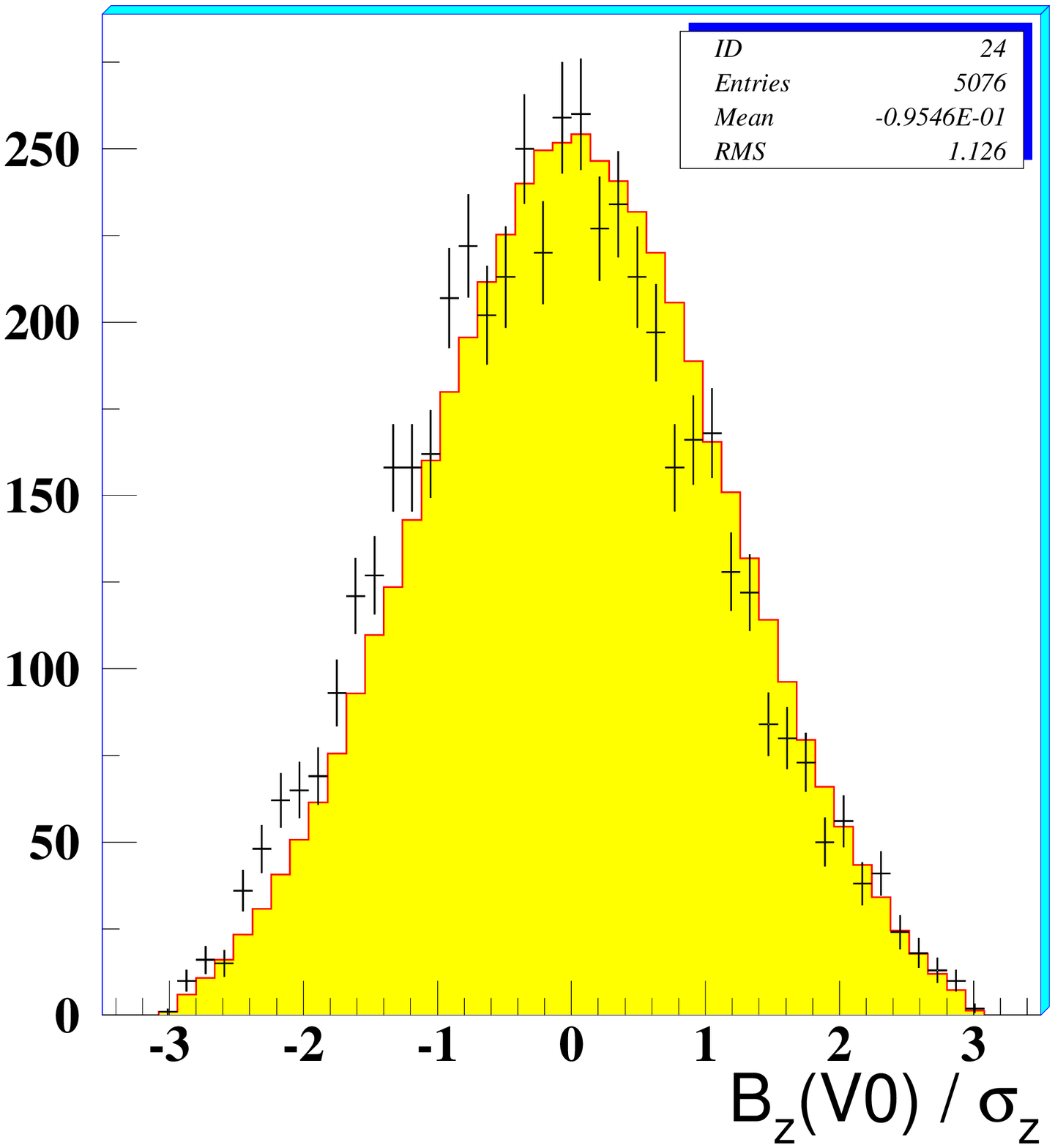}
\includegraphics[scale=0.185]{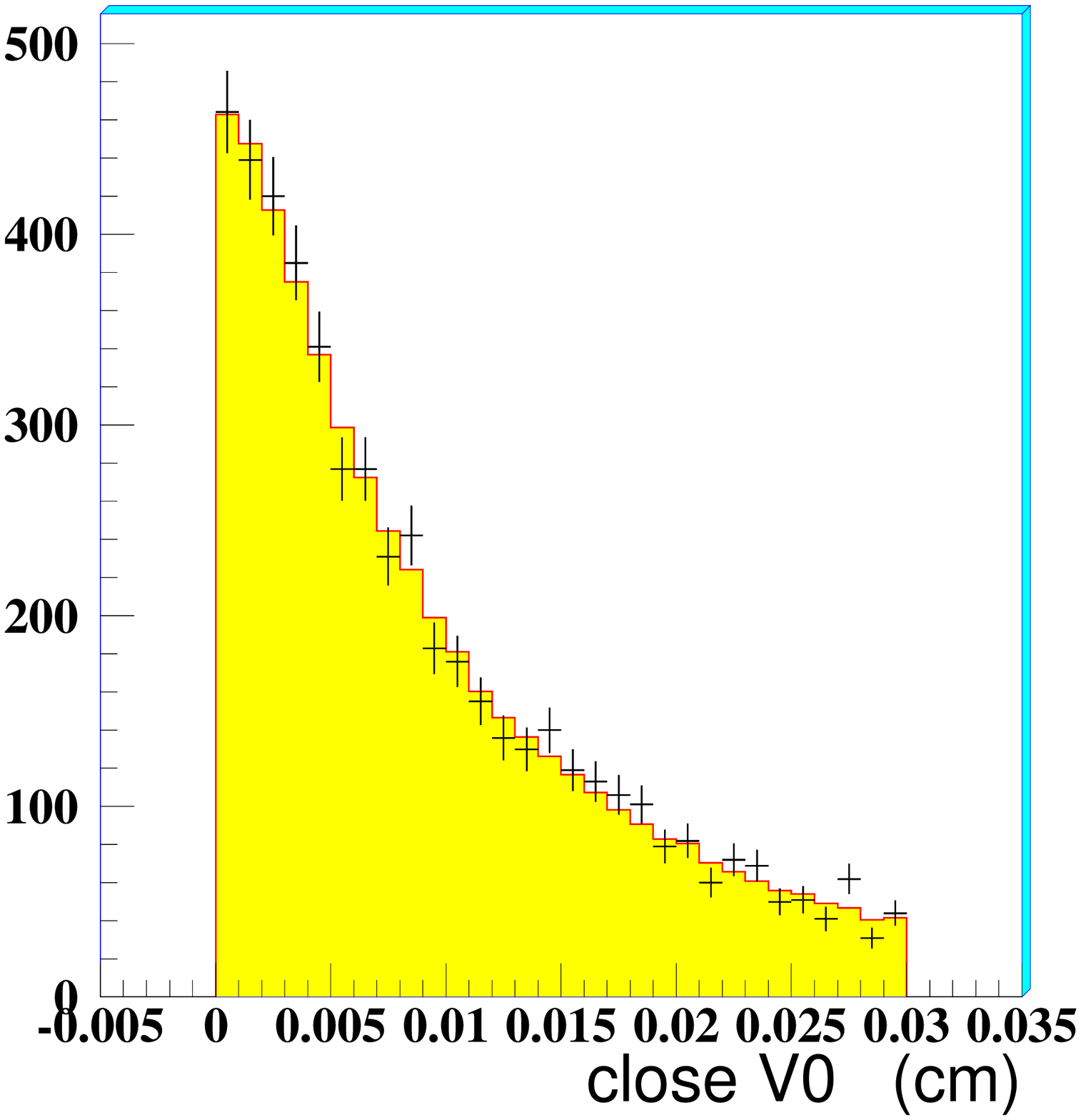}
\includegraphics[scale=0.185]{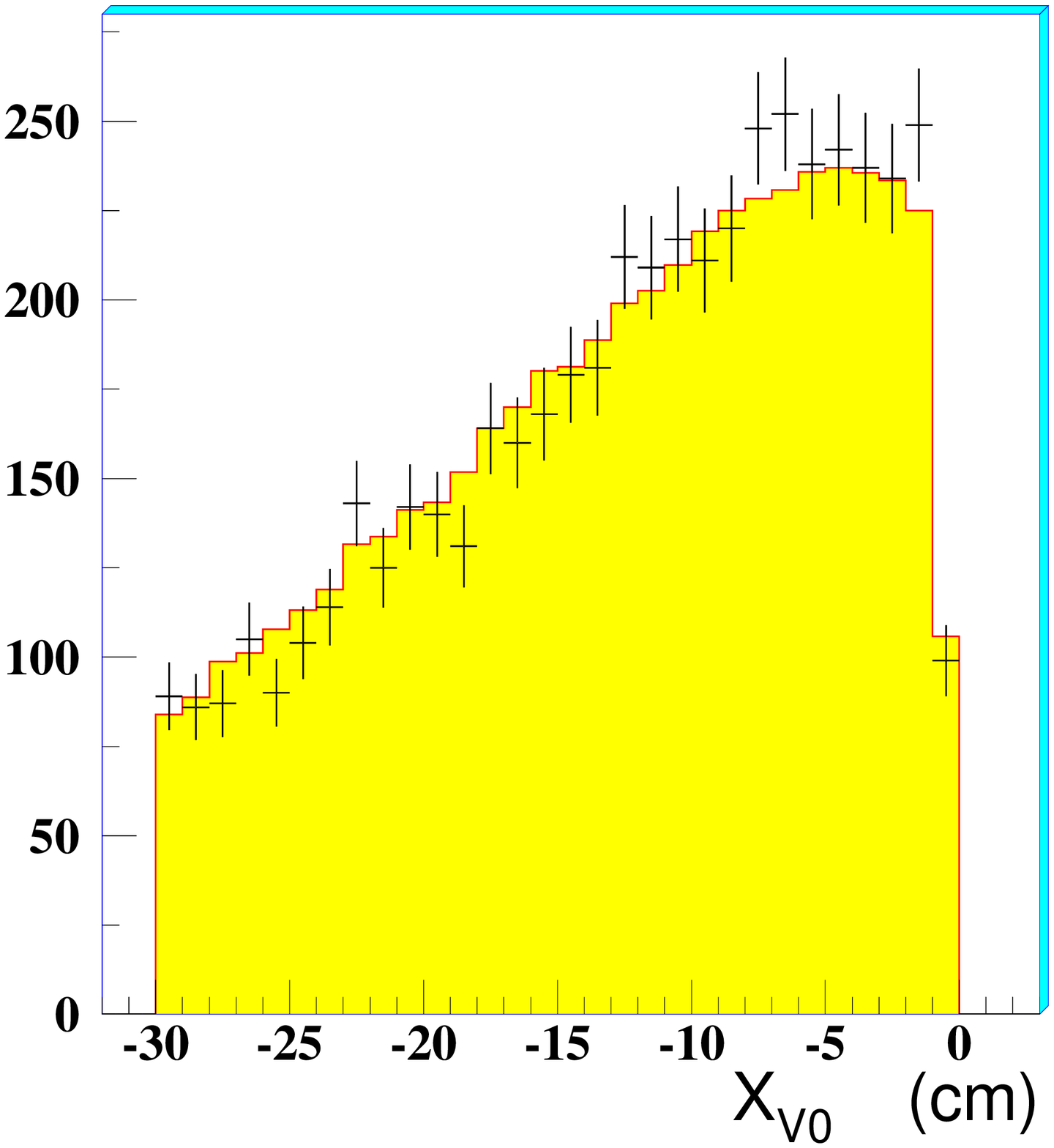} \\ 
\includegraphics[scale=0.185]{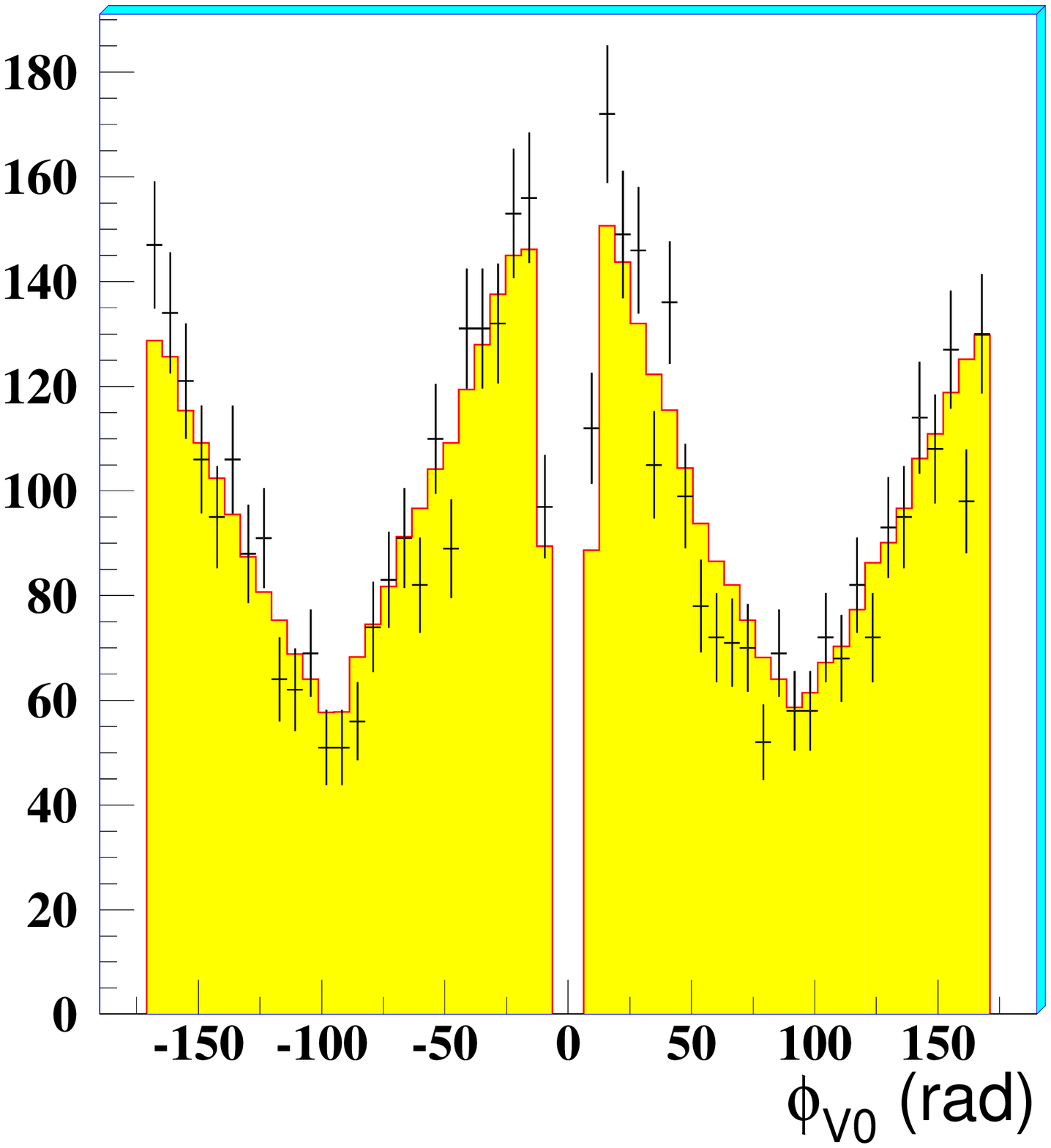}
\includegraphics[scale=0.185]{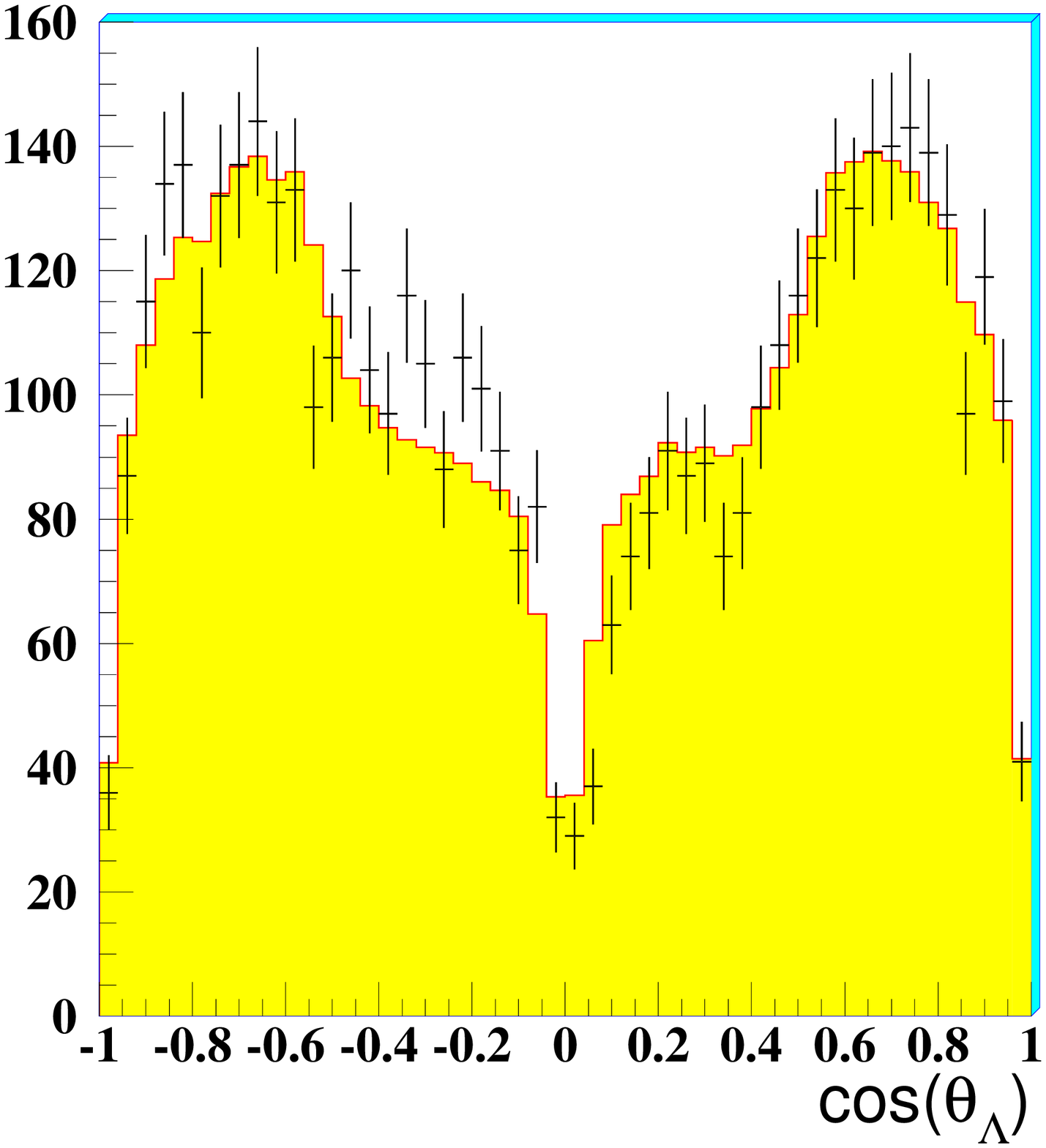}
\includegraphics[scale=0.185]{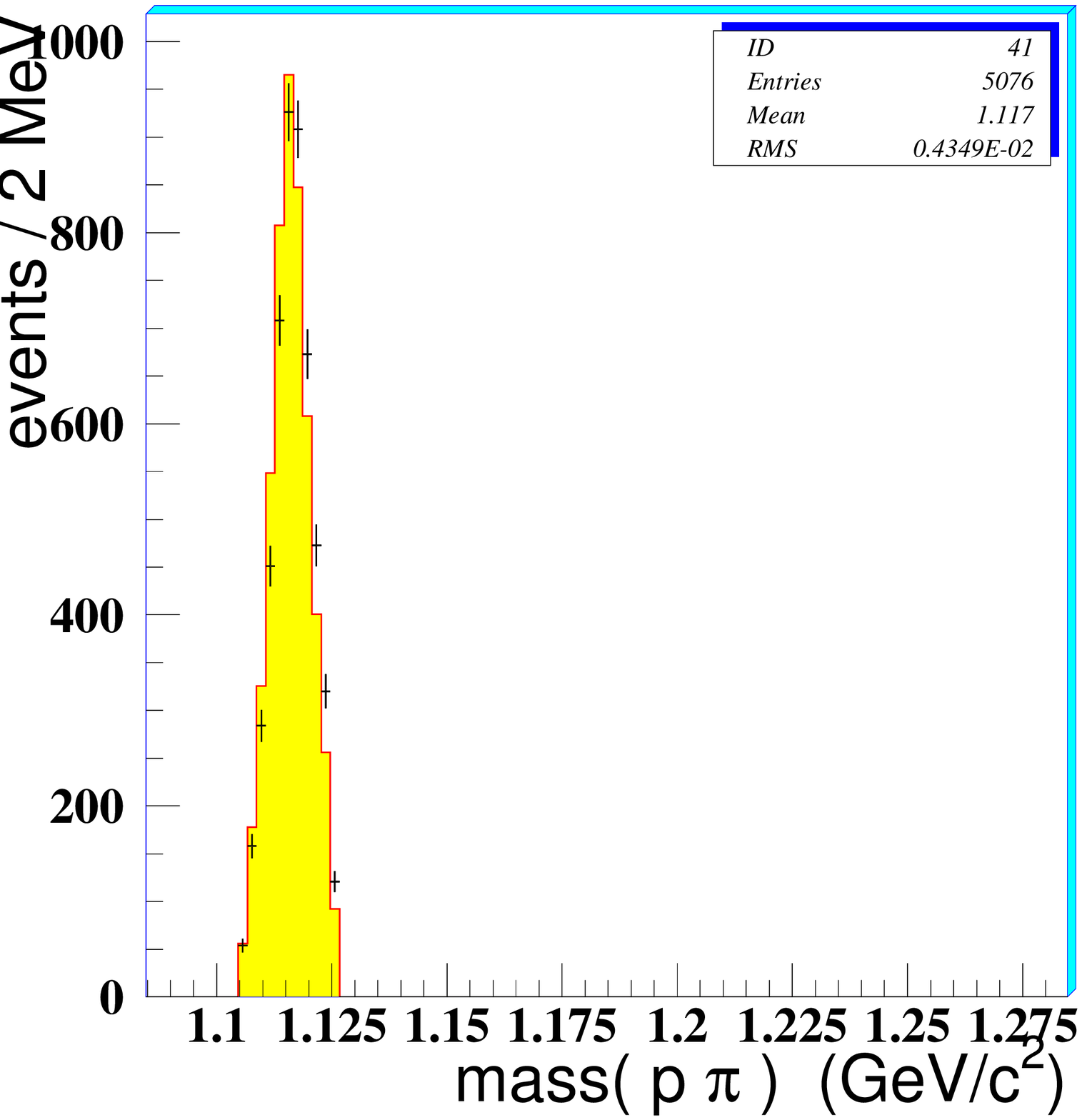}\\
\includegraphics[scale=0.185]{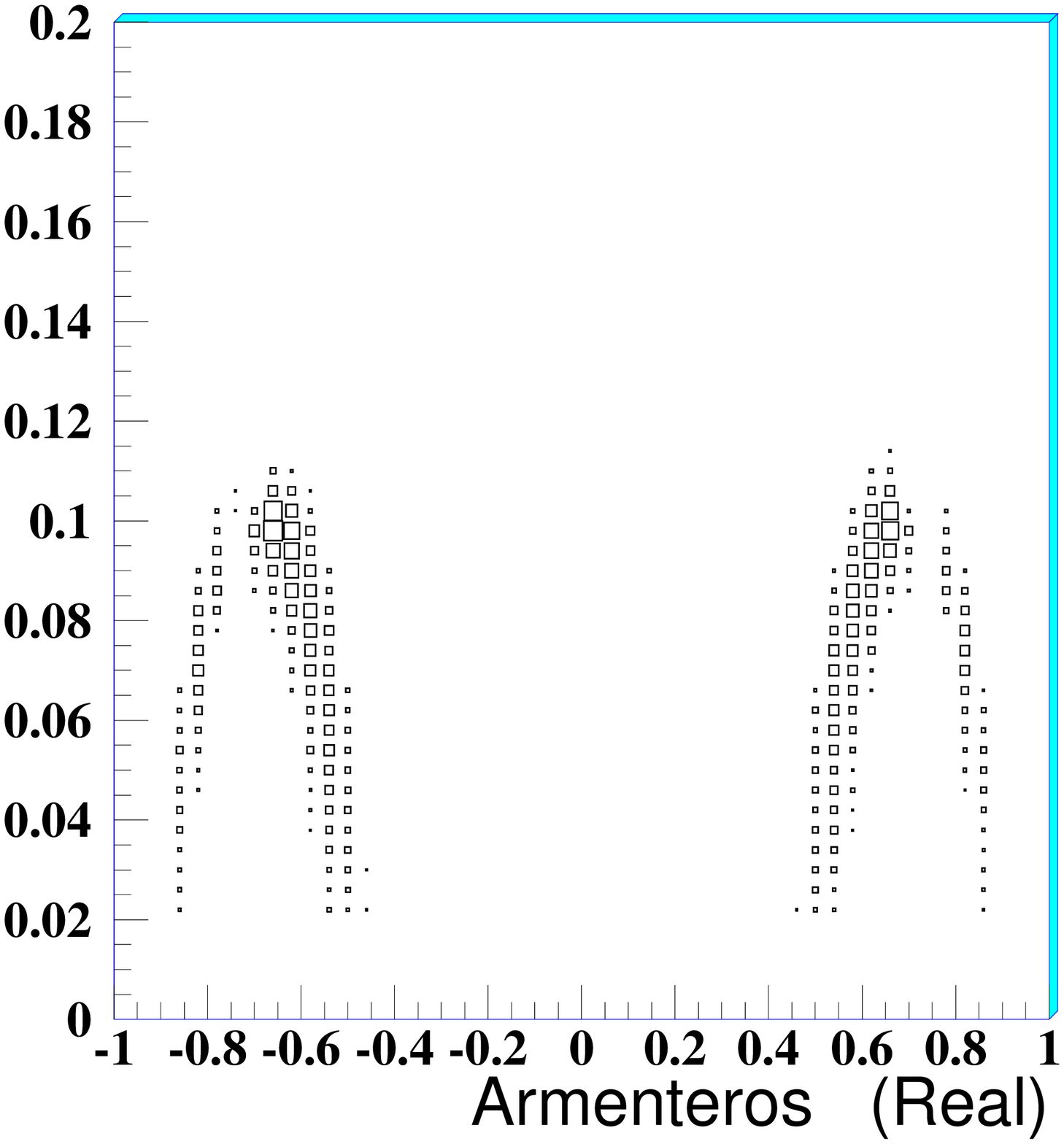}
\includegraphics[scale=0.185]{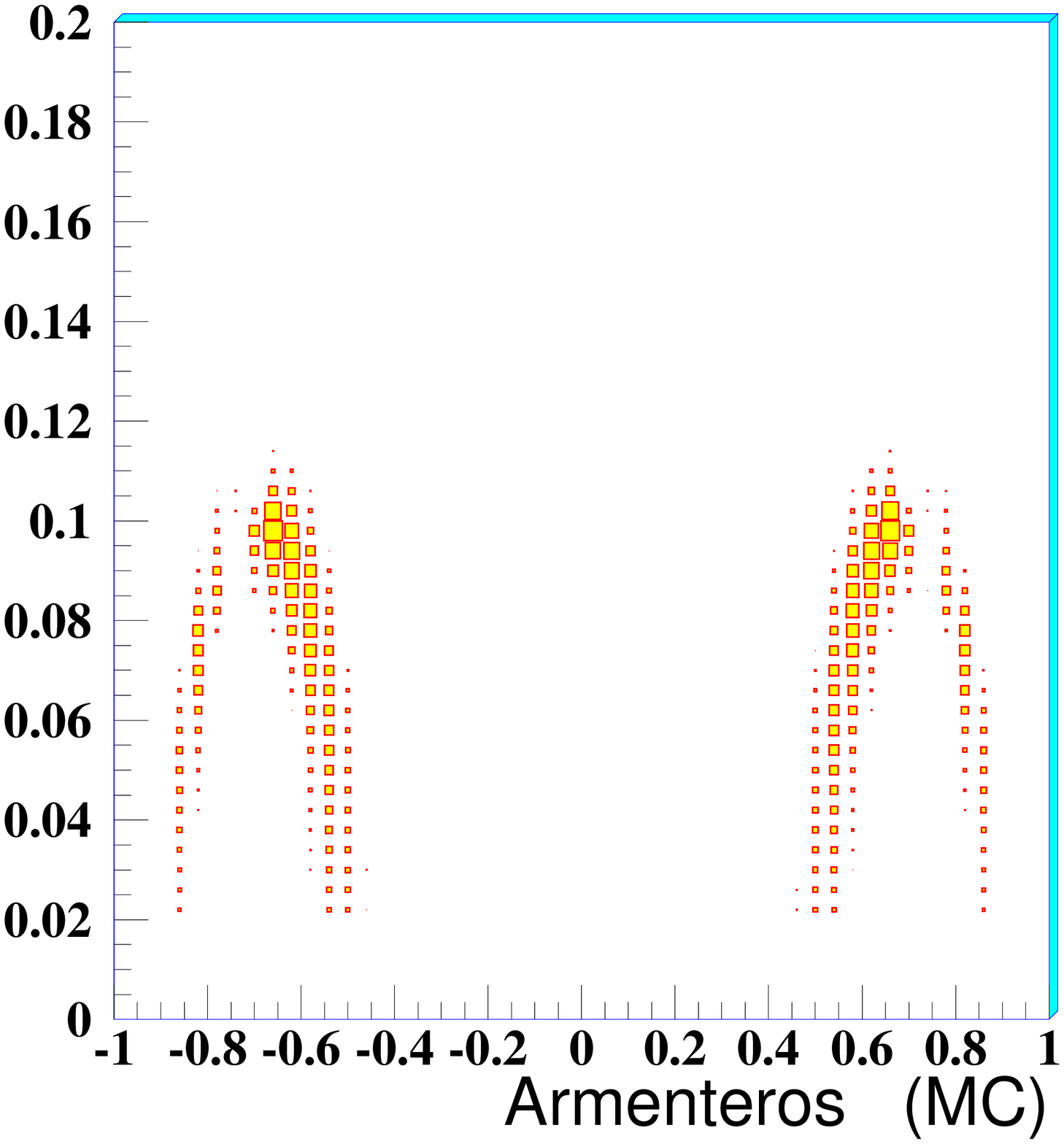}
\includegraphics[scale=0.185]{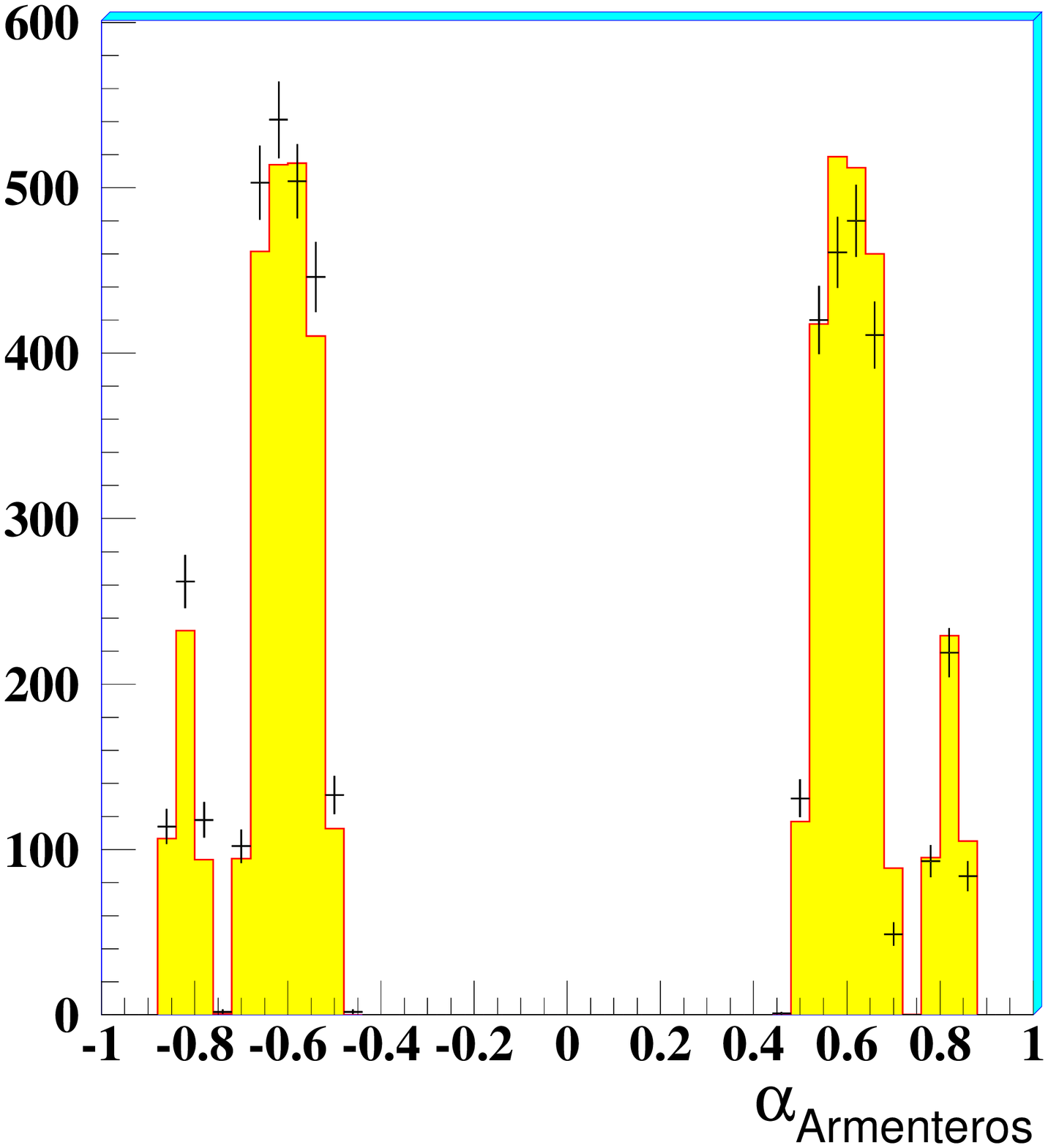}
\includegraphics[scale=0.185]{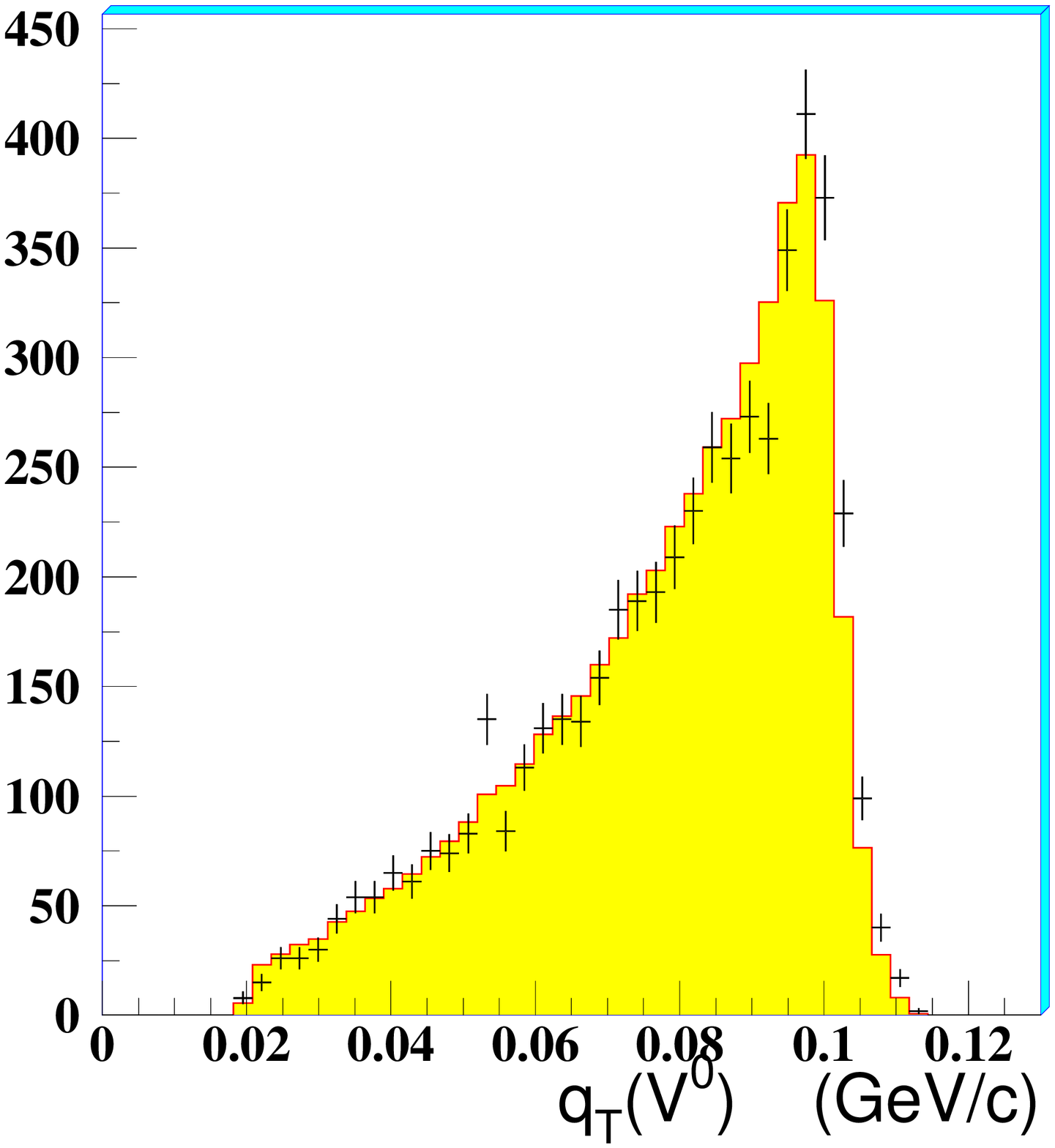}\\ 
\includegraphics[scale=0.185]{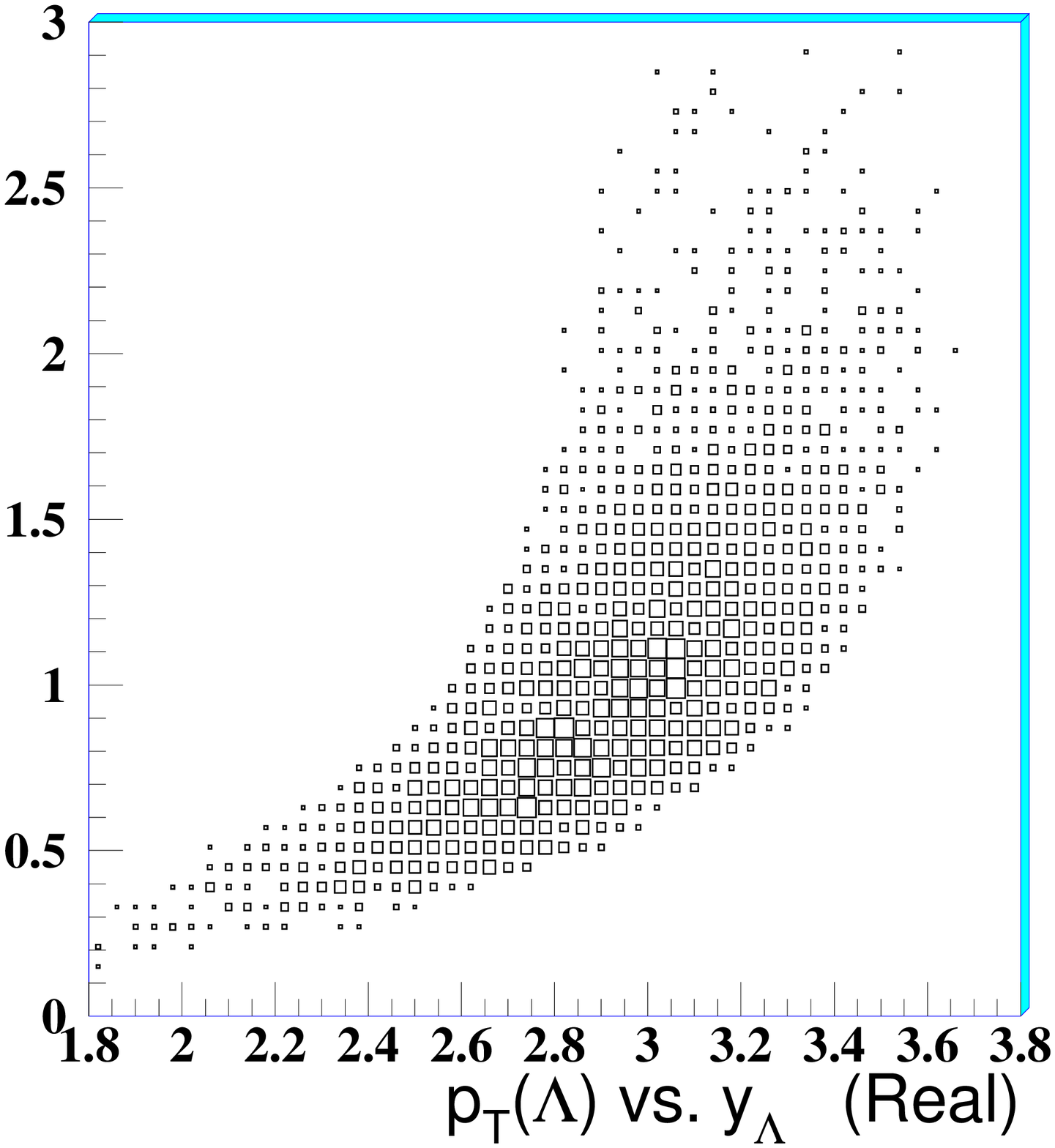}
\includegraphics[scale=0.185]{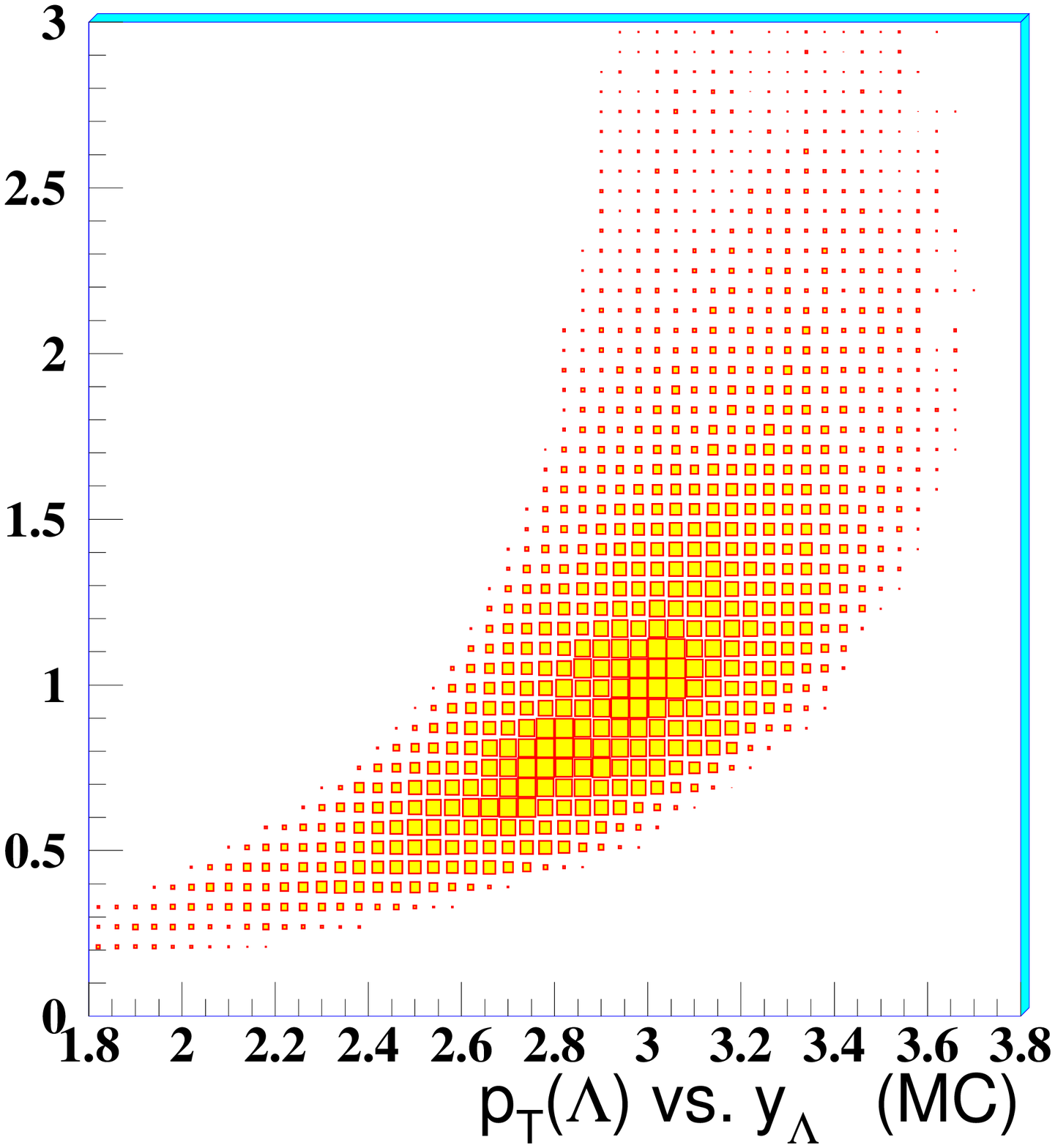}
\includegraphics[scale=0.185]{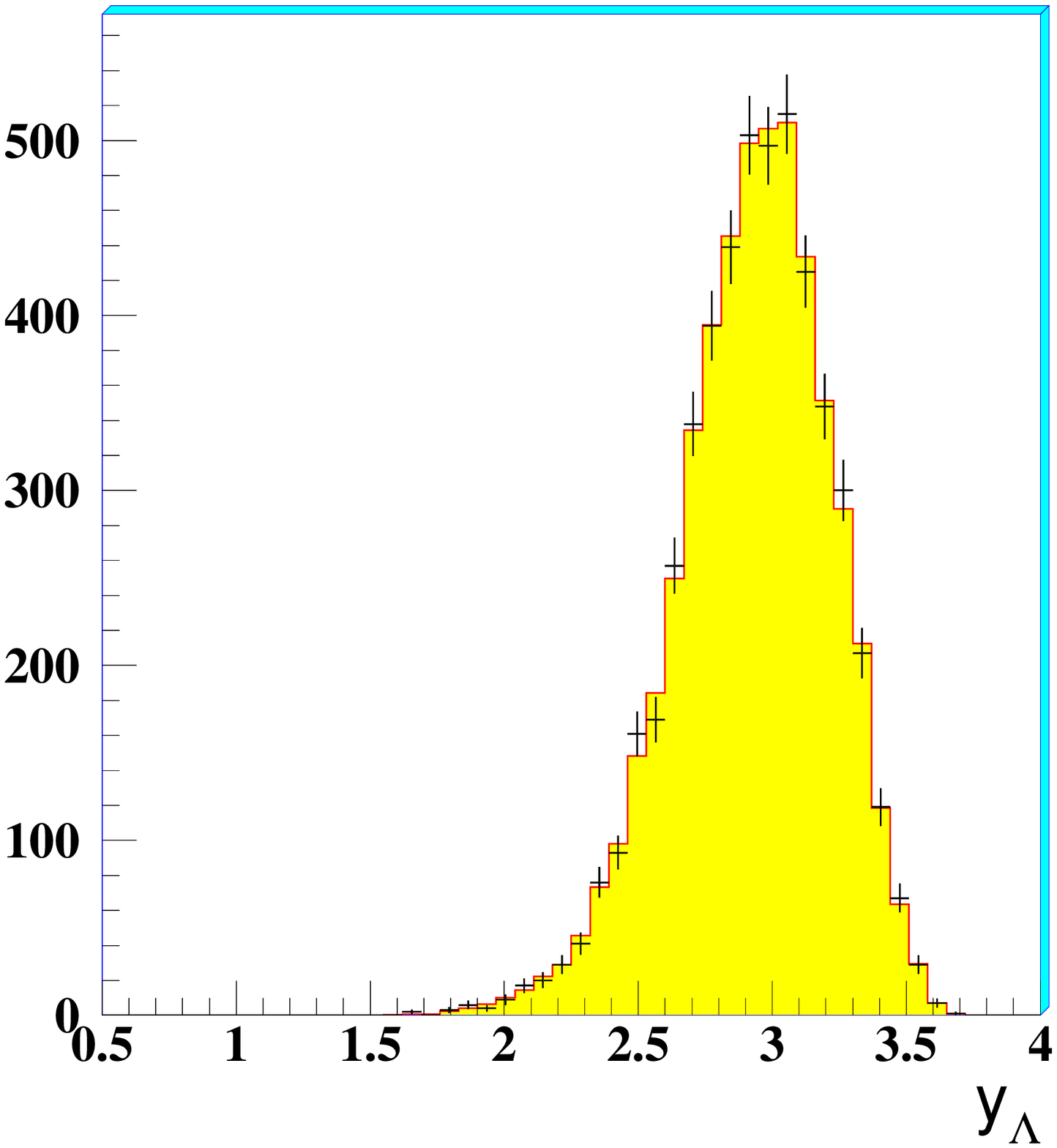}
\includegraphics[scale=0.185]{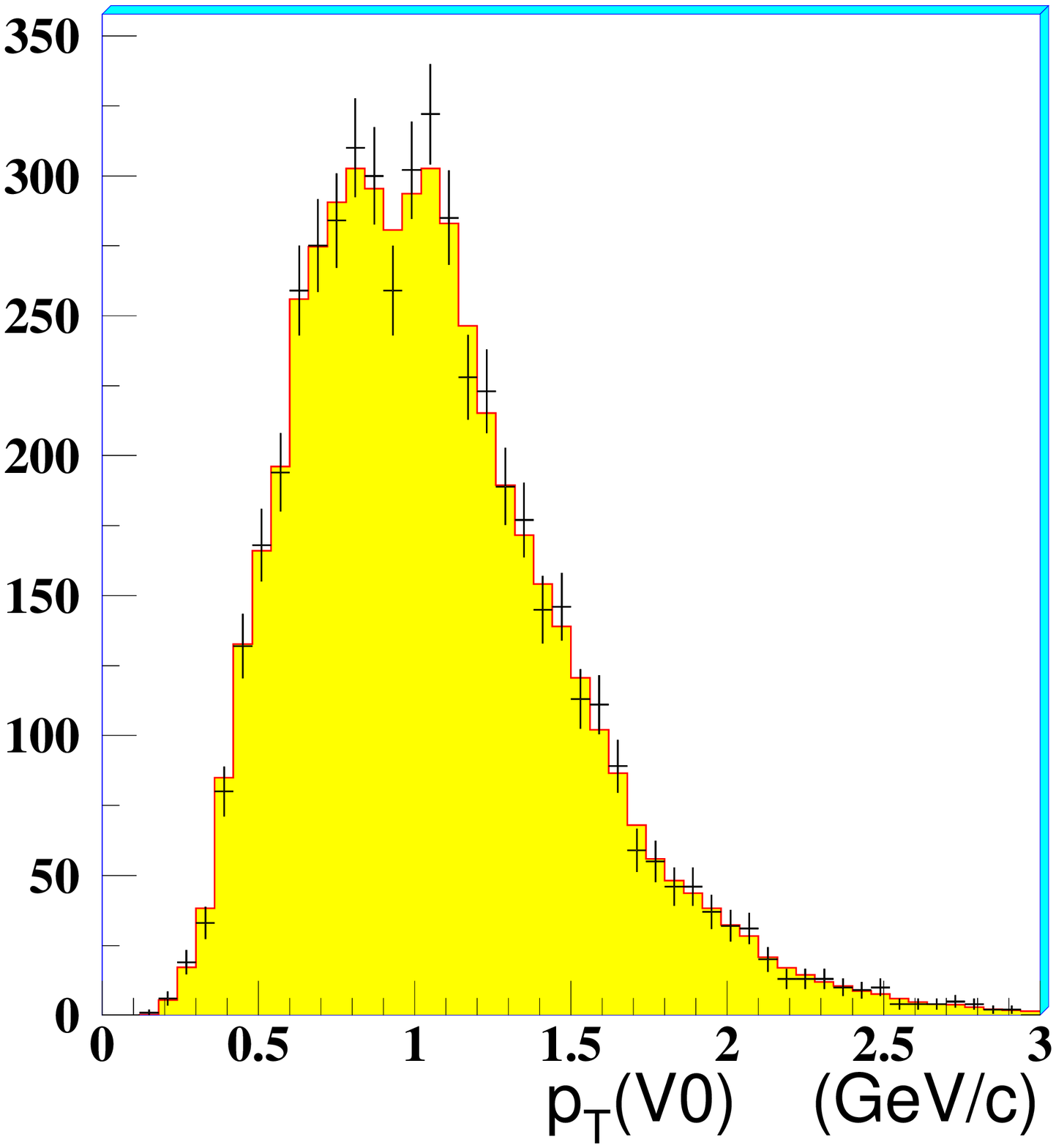}\\
\includegraphics[scale=0.185]{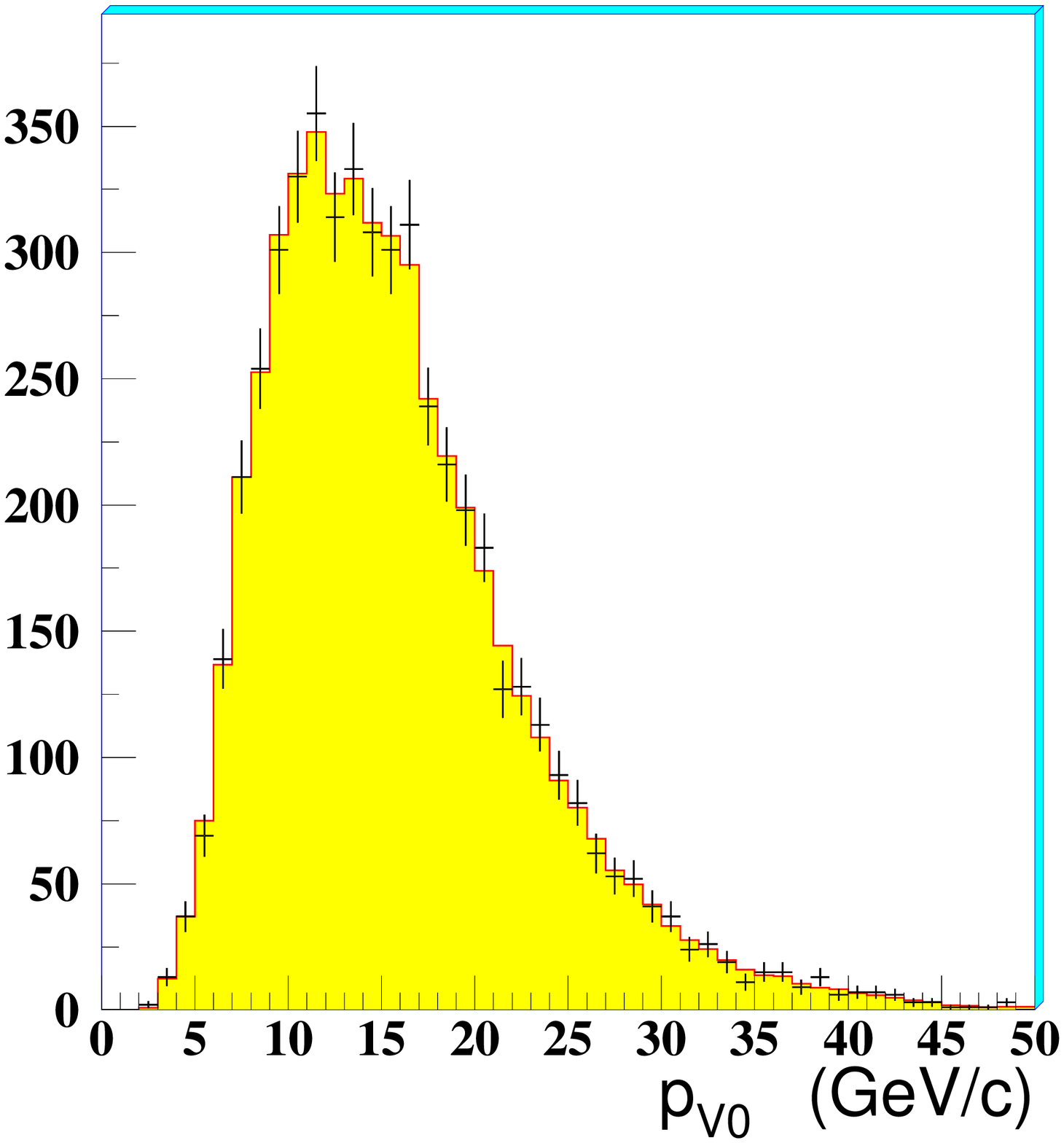}
\includegraphics[scale=0.185]{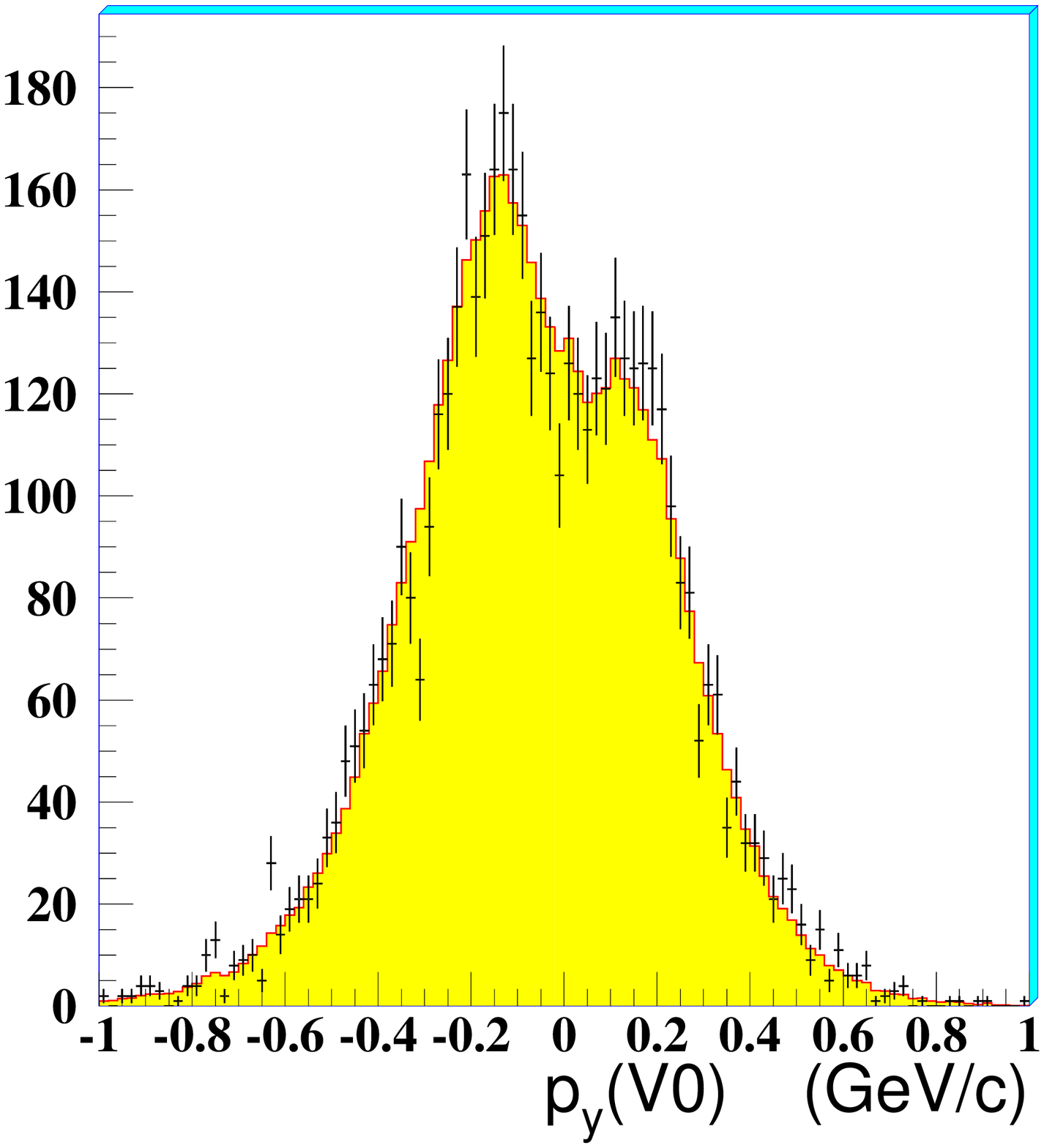}
\includegraphics[scale=0.185]{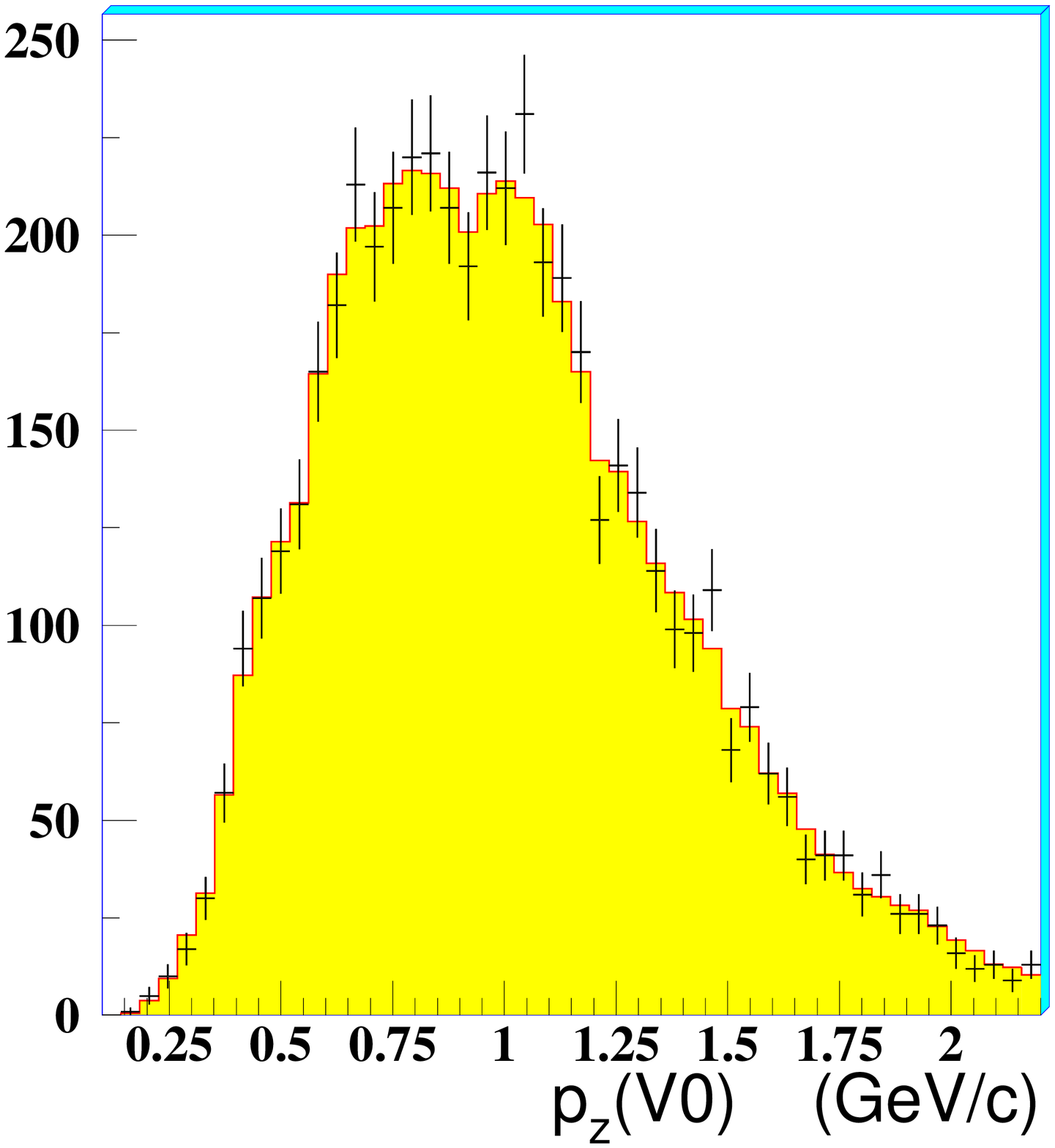}\\
\caption{Confronto tra varie distribuzioni delle \PgL\ e \PagL\ reali
         (punti con errori statistici o distribuzioni bidimensionali
         in chiaro) e le relative distribuzioni ottenute col Monte Carlo
         (distribuzioni in giallo).}
\label{LaComparison}
\end{center}
\end{figure}

In fig.~\ref{LaComparison} sono mostrati i dati relativi alle \PgL\ ed \PagL, 
in particolare le distribuzioni di:   
${b_y}_{V^0}/\sigma_{y}$, ${b_z}_{V^0}/\sigma_{z}$, $close_{V^0}$,   
$x_{V^0}$; $\phi_{V^0}$, $\cos(\theta_{\Lambda})$, massa invariante $M(p,\pi)$; 
il grafico di Armenteros per le $V^0$\ reali (in chiaro) e per quelle
del Monte Carlo (in giallo), $\alpha_{Arm}$, ${q_T}$; 
$p_T$\ versus $y_{\Lambda}$\ per i dati reali (in chiaro), 
$p_T$\ versus $y_{\Lambda}$\ per i dati del Monte Carlo (in giallo), 
rapidit\`a della $\Lambda$, $p_T$; modulo dell'impulso della $V^0$, 
$p_{y}$\ e $p_z$\ della $V^0$. 

\begin{figure}[p]
\begin{center}
\includegraphics[scale=0.185]{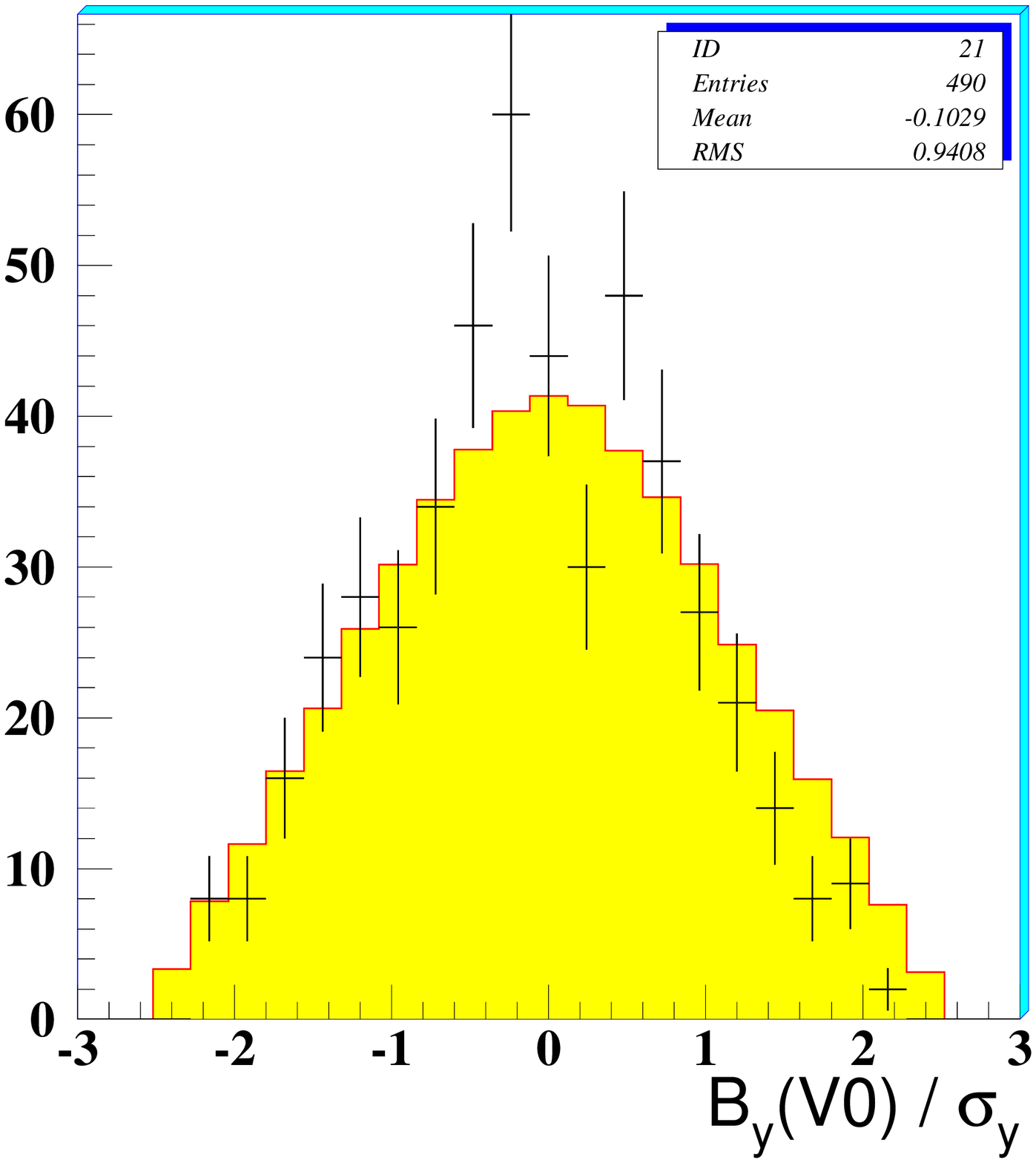}
\includegraphics[scale=0.185]{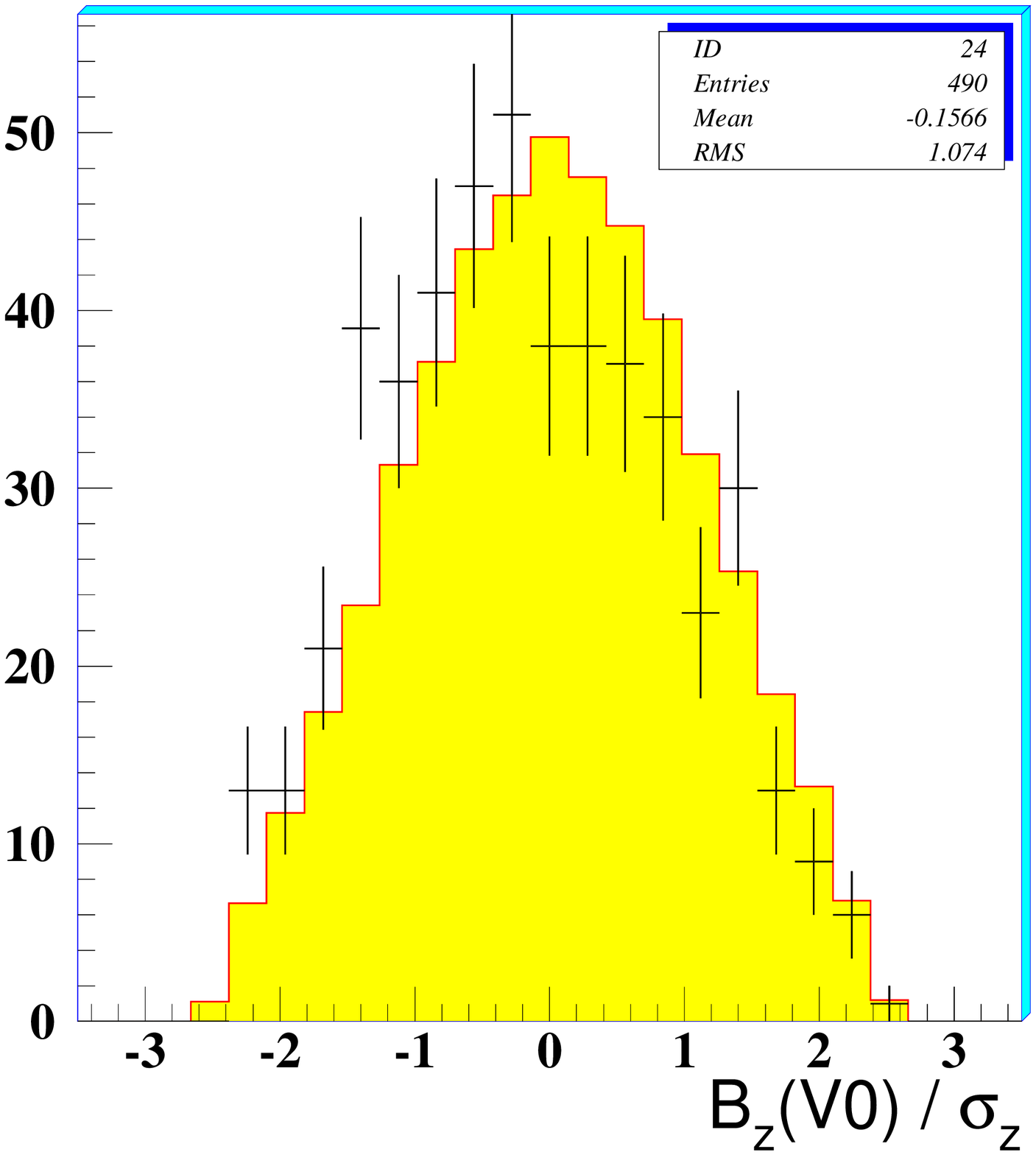}
\includegraphics[scale=0.185]{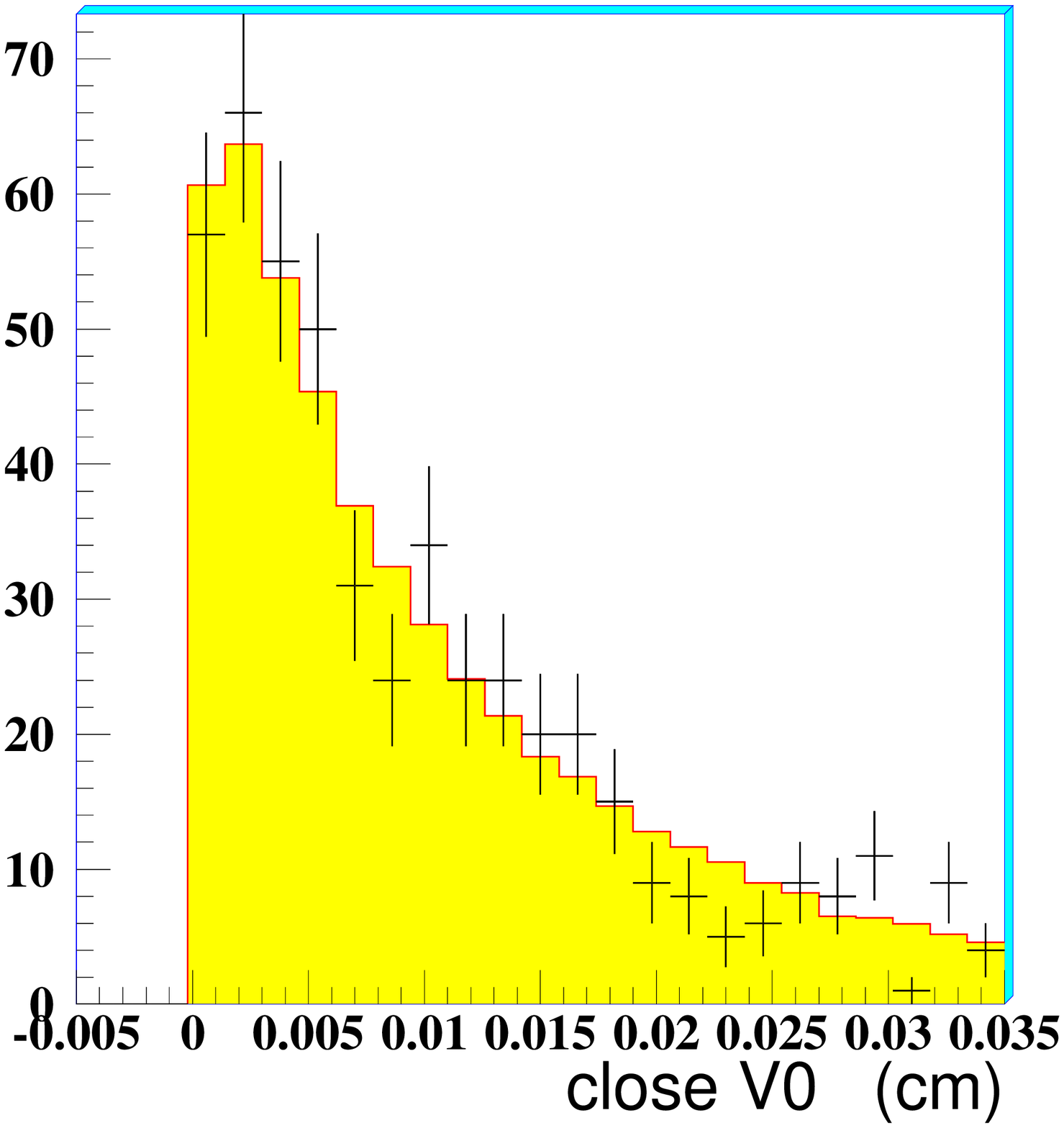}\\
\includegraphics[scale=0.185]{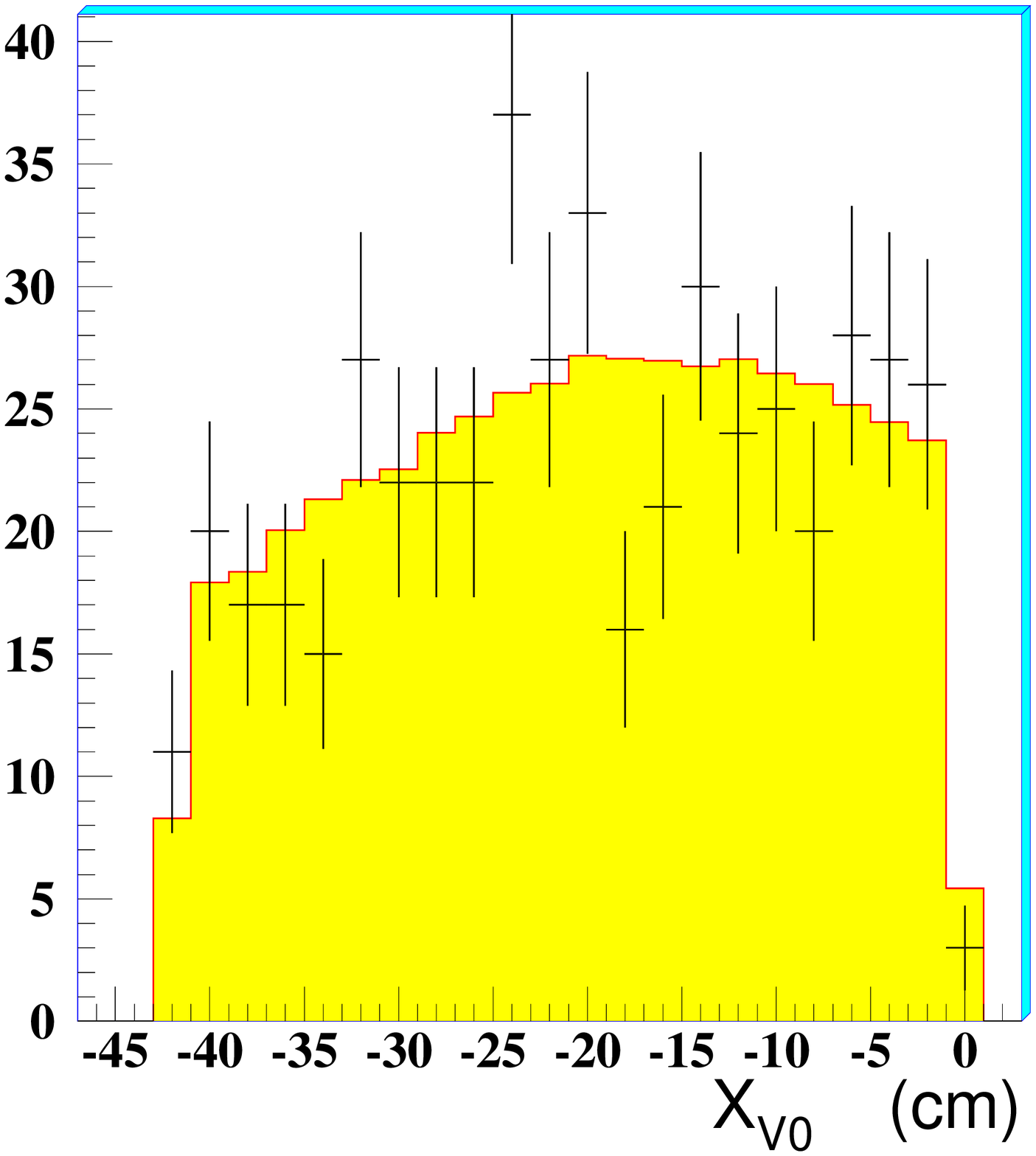} 
\includegraphics[scale=0.185]{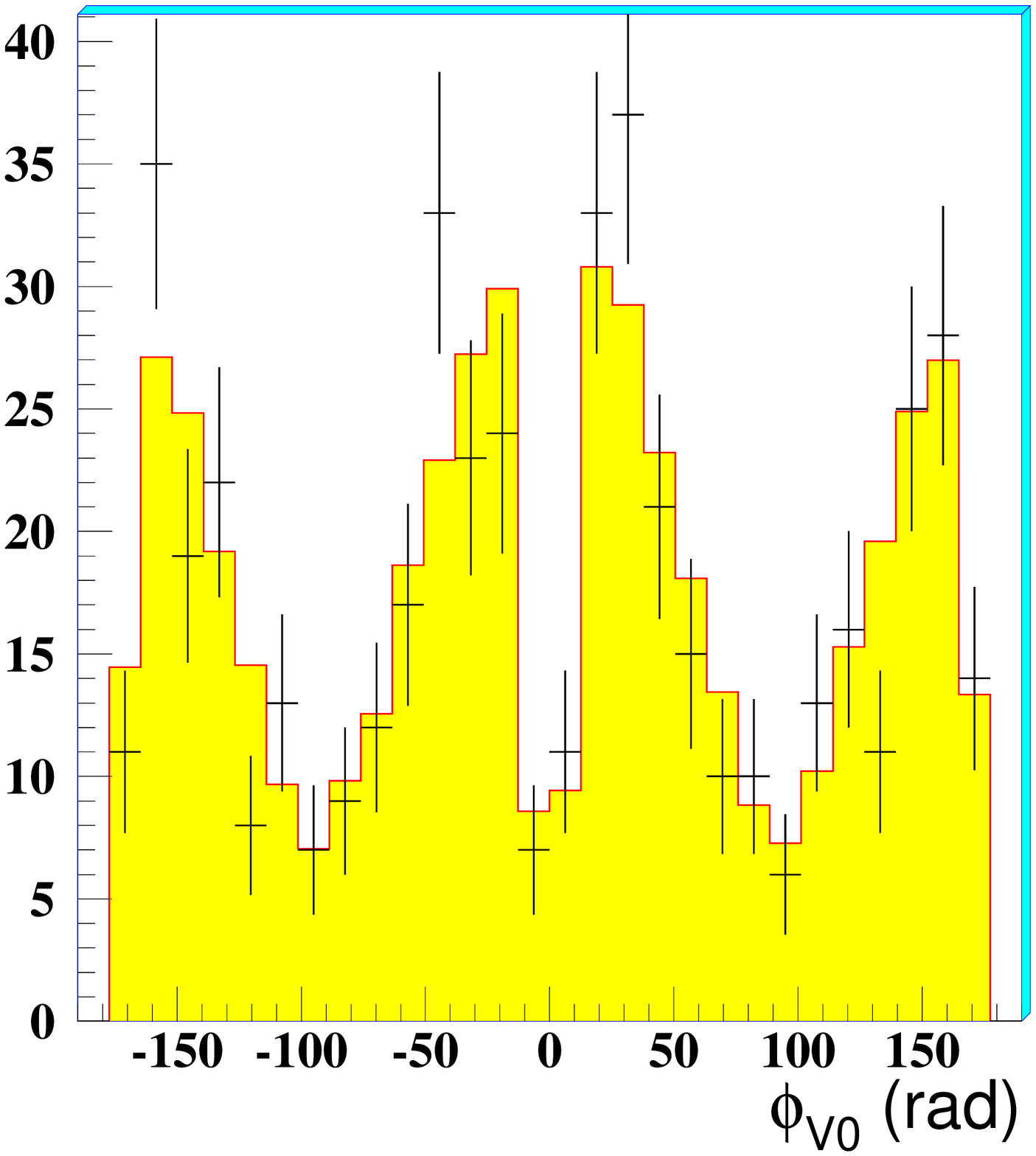}
\includegraphics[scale=0.185]{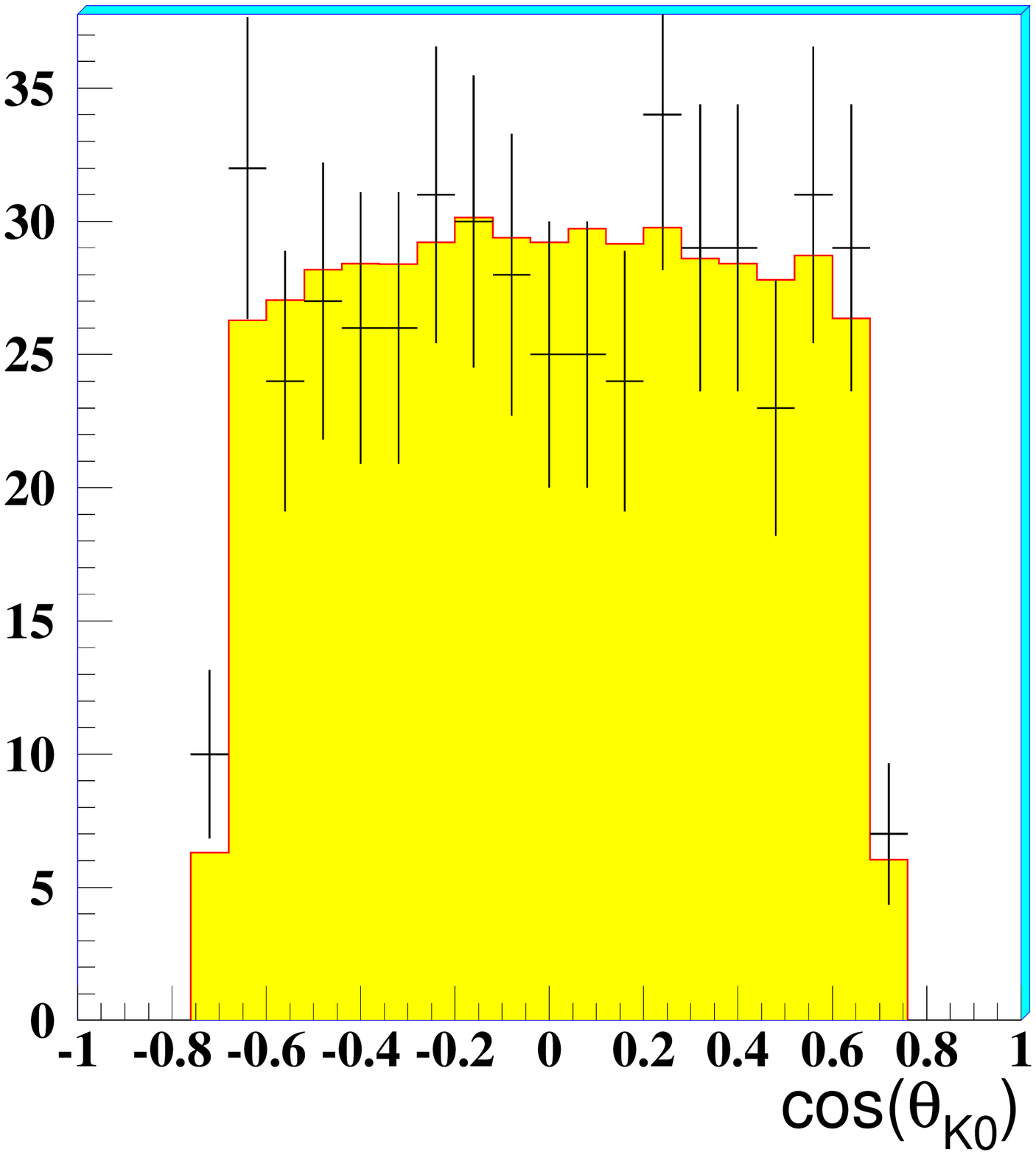}\\
\includegraphics[scale=0.185]{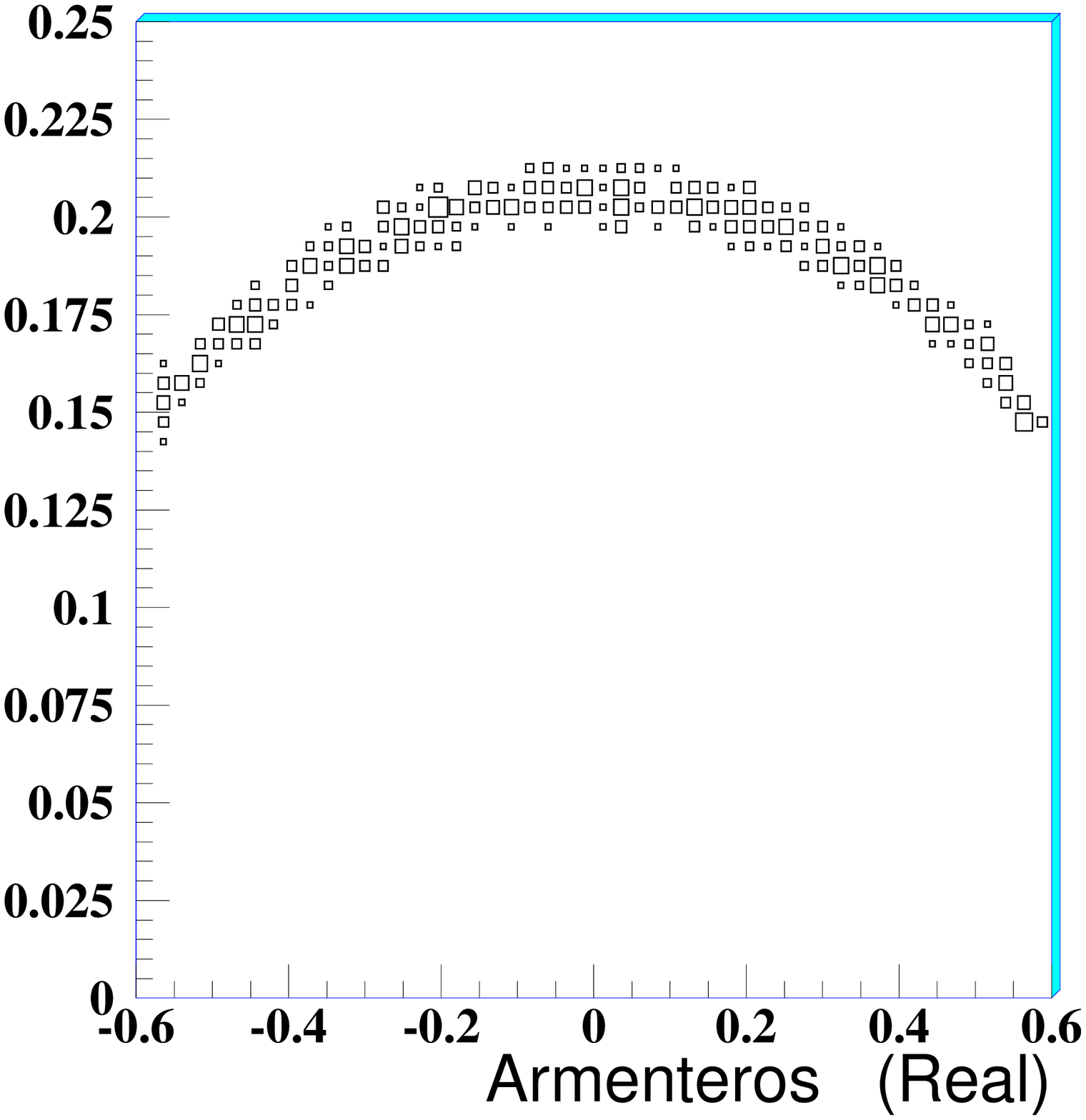}
\includegraphics[scale=0.185]{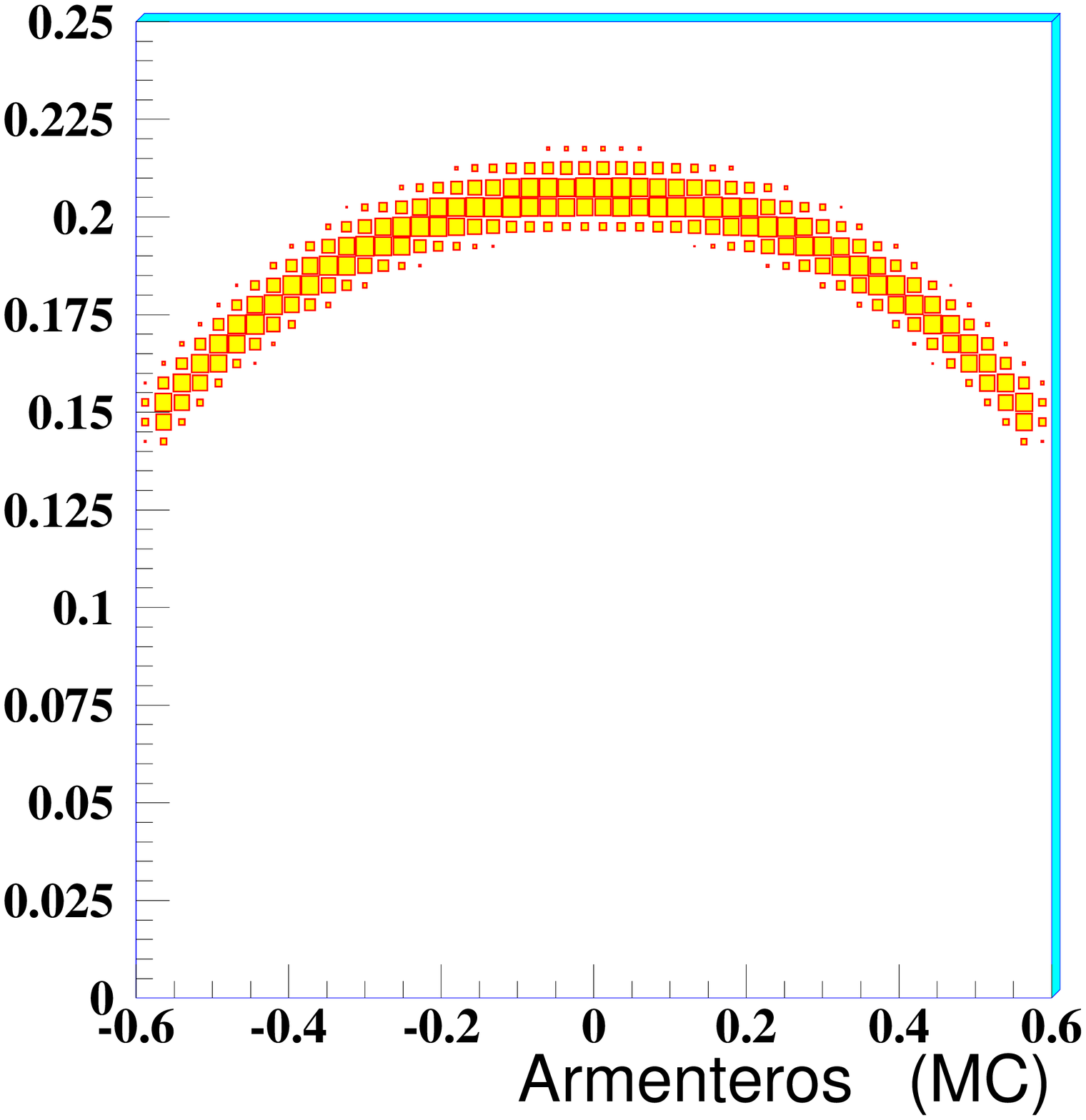}
\includegraphics[scale=0.185]{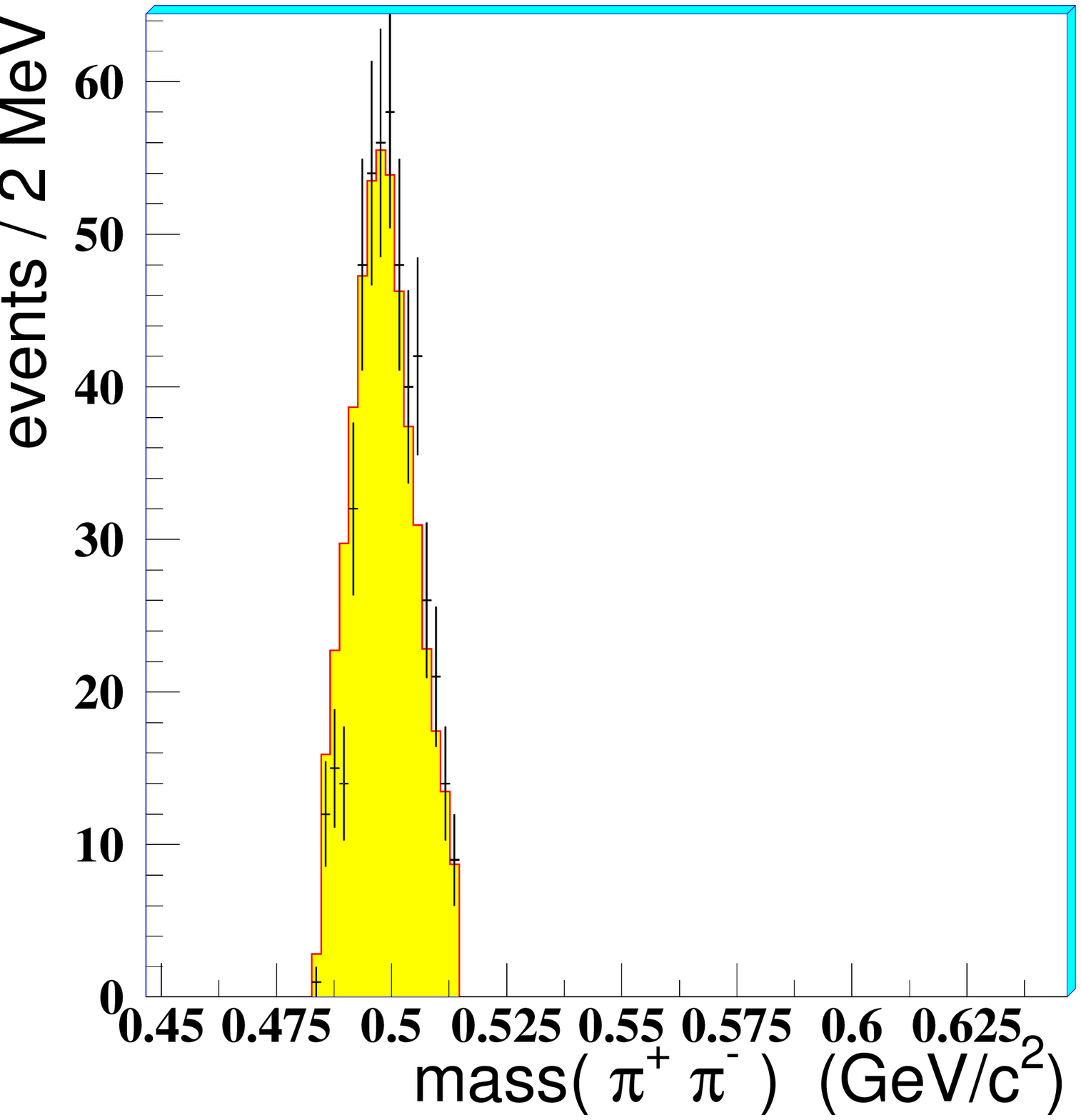}\\
\includegraphics[scale=0.185]{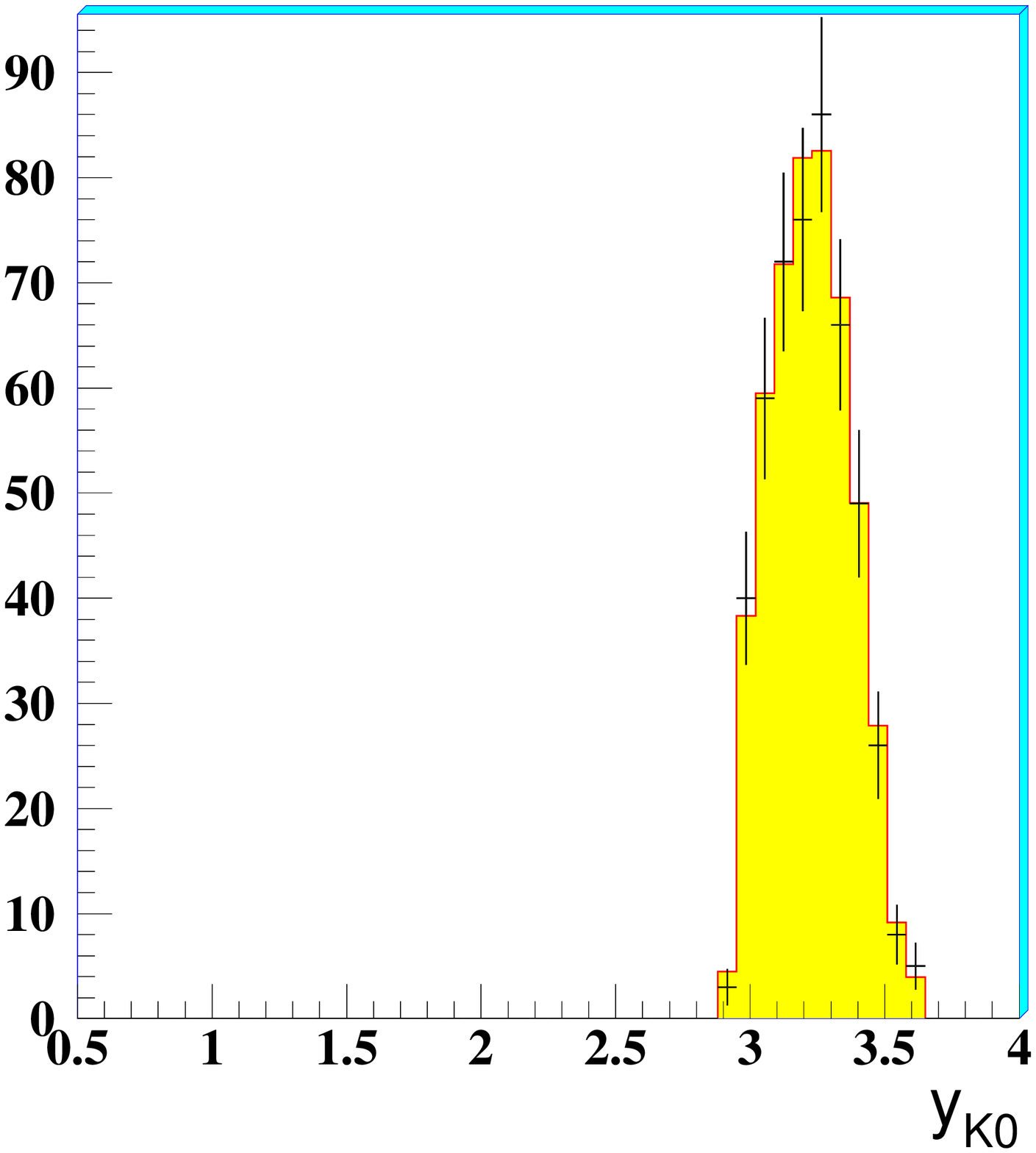}
\includegraphics[scale=0.185]{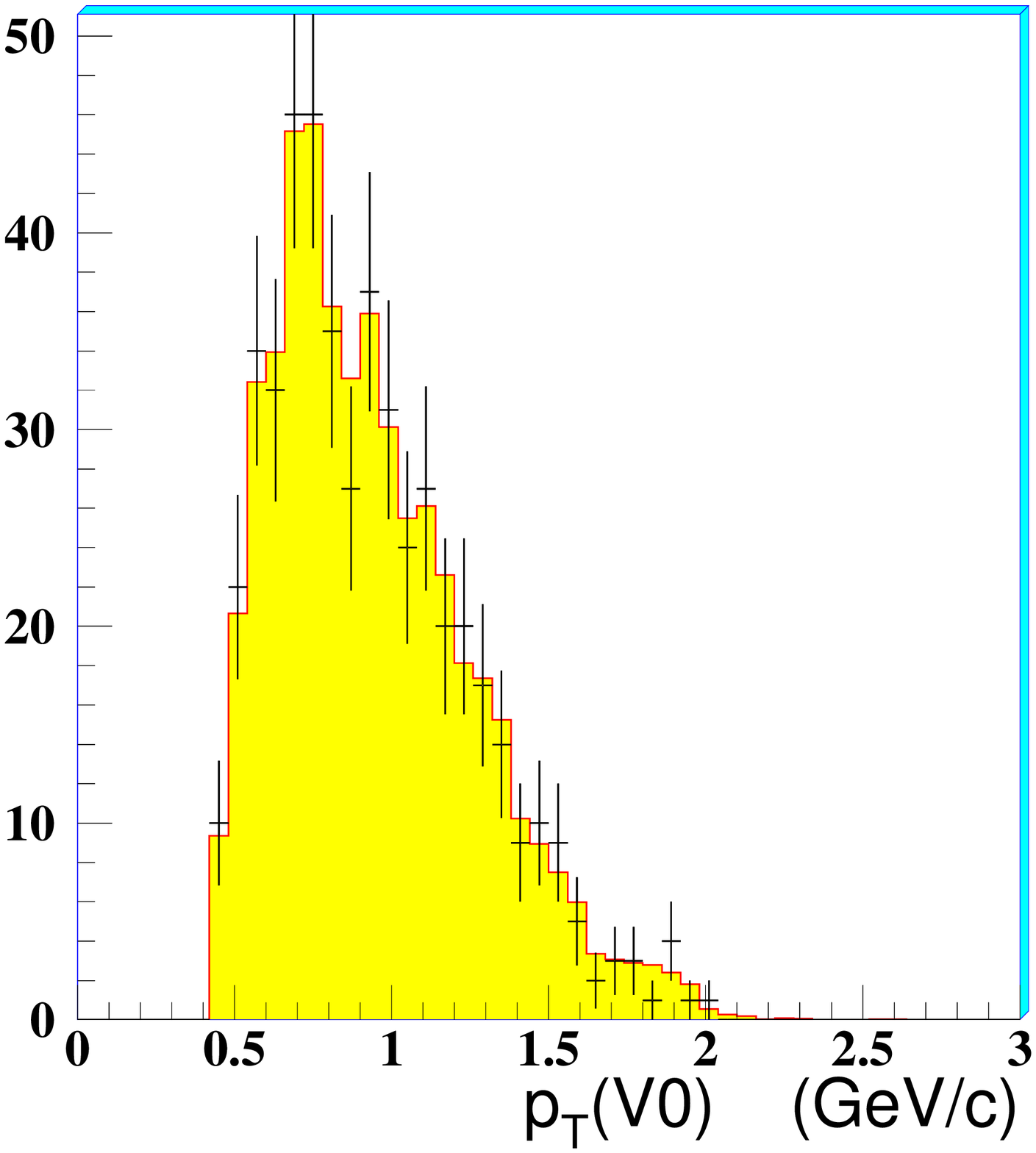}
\includegraphics[scale=0.185]{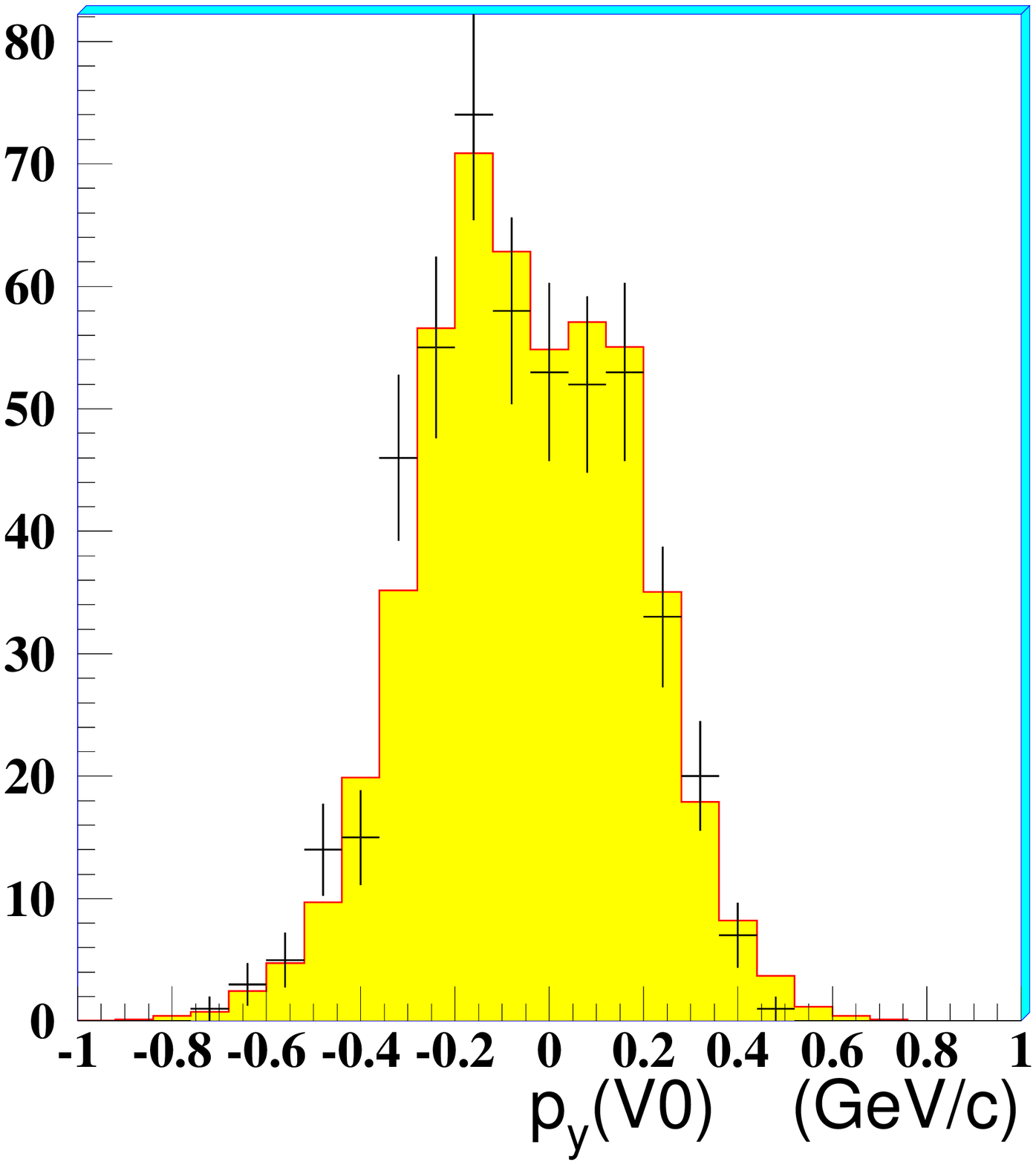}
\includegraphics[scale=0.185]{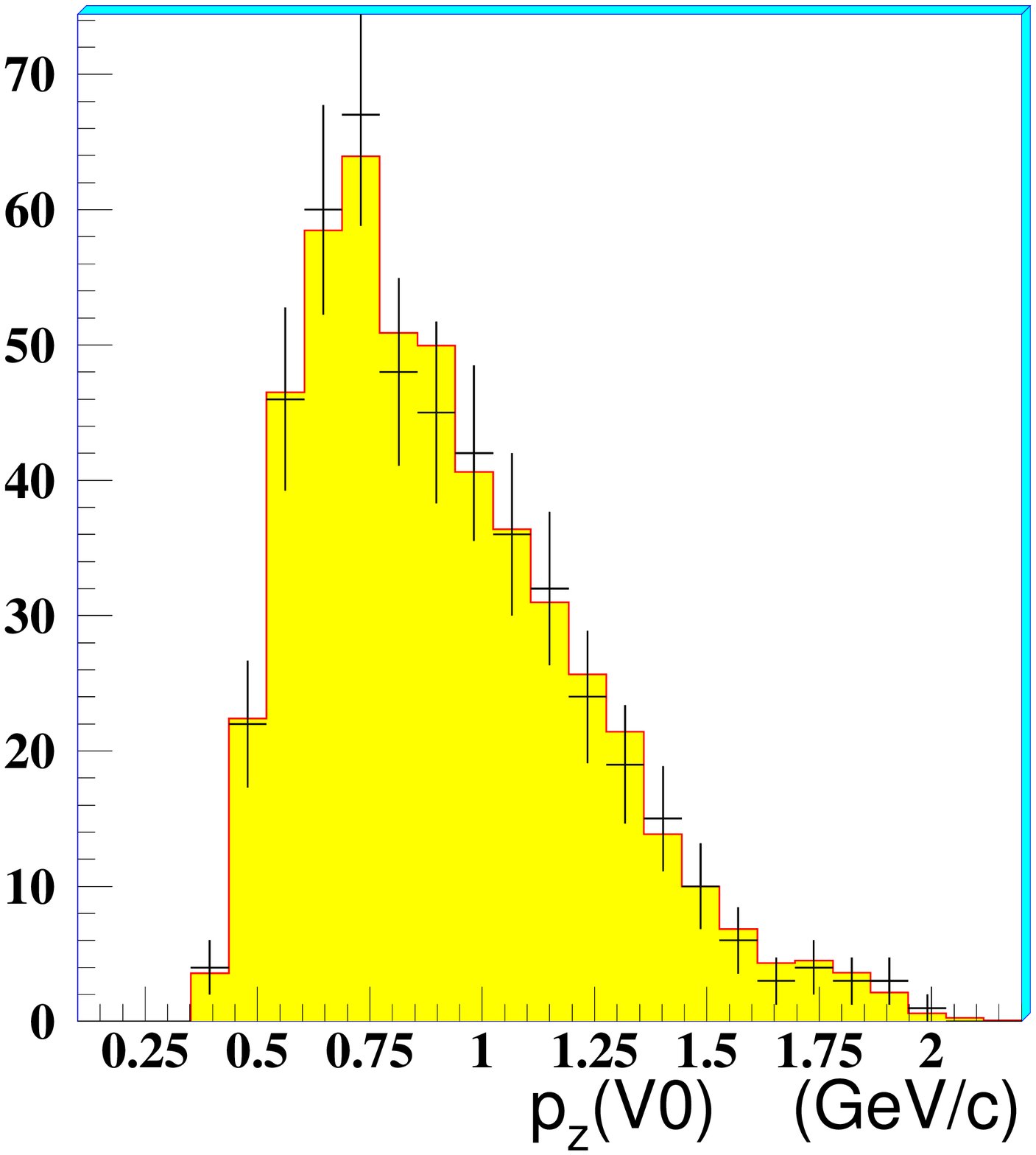}
\caption{Confronto tra varie distribuzioni dei \PKzS\ reali 
         (punti con errori statistici o distribuzioni bidimensionali
         in chiaro) e le relative distribuzioni ottenute col Monte Carlo
         (distribuzioni in giallo).}
\label{K0Comparison}
\end{center}
\end{figure}

In fig.~\ref{K0Comparison}, infine, sono mostrati i confronti relativi allo 
studio preliminare condotto sui \PKzS;   
col solito ordine da sinistra a destra, le distribuzioni mostrate si riferiscono a: 
${b_y}_{V^0}/\sigma_{y}$, ${b_z}_{V^0}/\sigma_{z}$, 
$close_{V^0}$; $x_{V^0}$, $\phi_{V^0}$, $\cos(\theta^*_{\PKzS})$; 
al grafico di Armenteros per i dati reali (in chiaro) e per le particlle 
Monte Carlo (in giallo), massa invariante $M(\pi^+,\pi^-)$; 
rapidit\`a della \PKzS, $p_T$, $p_y$\ e $p_{z}$\ della $V^0$.  
\newline
Le distribuzioni sono tutte riprodotte in modo eccellente, entro gli 
errori statistici, ad esclusione di quelle sui parametri d'impatto della 
particella primaria studiata. Per le diverse specie, si osserva infatti che 
le distribuzioni reali (maggiormente quelle in $y$) presentano un picco centrale pi\`u 
pronunciato rispetto a quelle del Monte Carlo; inoltre, 
%ma ci\`o si osserva solo per 
soprattutto per  
le $V^0$, la distribuzione della proiezione $z$\ dei dati reali \`e leggermente 
traslata verso valori negativi ($\approx 0.1 \sigma$\ per le $V^0$).  
Come spiegato all'inizio di questo paragrafo, nella simulazione le particelle 
vengono generate attorno al vertice {\em run per run}  secondo una gaussiana di 
larghezze pari alle $\sigma_{y},\sigma_{z}$\ di quel {\em run}. 
La fase successiva di ricostruzione 
introduce un ulteriore sbrodolamento, per cui la distribuzione 
bidimensionale (normalizzata alle $\sigma_{y},\sigma_{z}$)  
dei parametri d'impatto delle particelle Monte Carlo, 
$(b_{y}/\sigma_{y}$,$b_{z}/\sigma_{z})$,  
{\em ricostruiti} da ORHION \`e pi\`u larga della gaussiana 
%normale (cio\`e di $\sigma$\ unitaria) 
introdotta nella fase di {\em generazione} 
delle particelle. Inoltre, le particelle reali non sono misurate tutte 
con la stessa precisione, pertanto non \`e detto, a priori, che la 
distribuzione dei loro parametri d'impatto sia di forma gaussiana. 
\newline
A causa dei tagli ({\em ad hoc}) operati in fase di analisi, la discrepanza 
tra il Monte Carlo ed i dati reali per queste distribuzioni introduce 
un errore sistematico, di entit\`a pari alla differenza, tra gli eventi reali 
e quelli simulati, delle frazioni di eventi esclusi per via dei tagli, rispetto 
al totale. Poich\'e, per\`o, i tagli operati per tutte le specie di particelle 
non sono troppo selettivi ($2.5\div3 \, \sigma_{y(z)}$), tale errore sistematico 
\`e, al pi\`u, dell'ordine di qualche percento. 
%Poich\'e a livello di analisi, per tutte le particelle studiate, si sono operati 
%dei tagli ({\em ad hoc}) non troppo  selettivi, a $2.5 \div 3 \, \sigma_{y(z)}$, 
%l'errore sistematico introdotto dalla discrepanza tra il Monte 
%Carlo ed i dati reali per queste distribuzioni --- di entit\`a pari alla differenza, 
%tra gli eventi reali e quelli simulati, delle frazioni di particelle escluse dal taglio  
%(sulle code delle distribuzioni) rispetto al totale --- \`e comunque 
%dell'ordine di pochi percento soltanto, per tutte le specie. 
%Nel caso delle cascate, 
In particolare, per le cascate,  
in cui l'entit\`a dell'effetto \`e ulteriormente ridotto rispetto alle $V^0$, 
questo errore sistematico \`e trascurabile rispetto 
all'errore statistico.% sulla misura del tasso di produzione. 
\newline
Nel caso delle $V^0$, invece, per cui la discrepanza \`e pi\`u marcata e per cui 
gli errori statistici sono molto piccoli, diventa fondamentale valutare con precisione 
l'entit\`a dell'errore sistematico introdotto nella misura del tasso 
di produzione~\footnote{Non ci si apetta alcun errore sistematico nella misura 
degli spettri di massa trasversa in quanto non si osserva  correlazione tra il  
parametro d'impatto e le variabili $p_T$\ e rapidit\`a.}. 
\newline
Per quanto riguarda il Monte Carlo, \`e possibile  calcolare agevolmente 
la frazione degli eventi {\em generati}, che sono poi stati ricostruiti e che 
hanno superato tutti gli altri tagli dell'analisi, ma che vengono scartati  
per la sola azione del taglio sul parametro d'impatto. Nel caso delle \PgL\ e \PagL\ 
tale frazione \`e pari all'8.5\%, nel caso dei \PKzS\ al 10\%.  
\newline
Per i dati reali si presenta invece una complicazione, 
dovuta alla presenza del fondo residuo di $V^0$\ geometriche, concentrato 
soprattutto sulle code della distribuzione (bidimensionale) del parametro 
d'impatto. Tuttavia, disponendo di una buona descrizione del fondo geometrico, ottenuta 
col metodo di mescolamento degli eventi ({\em cfr. paragrafo 4.2}), diventa 
possibile stimare la frazione del ``segnale fisico'' (cio\`e $V^0$\ reali {\em meno} 
il fondo geometrico) rimossa dall'azione del taglio sul parametro di impatto, 
cos\`i come fatto per le particelle Monte Carlo. 
Nel caso delle \PgL\ e \PagL\ tale frazione \`e pari al 12.5\%, nel caso di 
\PKzS\ al 13.5\%. 
Considerando la differenza trovata tra il Monte Carlo e gli eventi reali, si conclude 
che le correzioni calcolate per le $\PgL$\ e le $\PagL$, e conseguentemente i tassi  
di produzione, sono sistematicamente sottostimate del 4\% a causa di questa discrepanza, 
mentre quelle dei \PKzS\ del 3.5\%. Si terr\`a conto di ci\`o includendo questi contributi 
negli errori sistematici.  
\section{Determinazione delle finestre di accettanza delle particelle strane}
Per ciascuna specie di particella strana, e differentemente per ciascuna interazione 
considerata,  
%(ed anche, a parit\`a di interazioni, differentemente al variare del 
%periodo di presa dati, cui corrisponde generalmente anche una variazione nella 
%disposizione sperimentale del telescopio e nelle efficienze dei diversi piani di 
%rivelatori),  
\`e opportuno definire una finestra di accettanza,  
nel piano rapidit\`a--impulso trasverso, pi\`u limitata rispetto al limite  
intrinseco fornito dall'accettanza geometrica del telescopio.   
Questa opportunit\`a pu\`o presentarsi anche a parit\`a di interazioni, per diversi 
periodi della presa dati, se si \`e avuta una variazione nella
disposizione sperimentale del telescopio e nelle efficienze dei diversi piani di
rivelatori.  
\newline
La necessit\`a di una simile ridefinizione ha due origini: ({\em i}) si vogliono 
escludere le particelle dirette verso i bordi geometrici del telescopio, 
per le quali si possono avere grandi fluttuazioni statistiche dovute alla 
minor accettanza entro il telescopio dei prodotti di decadimento, rispetto 
ad una particella che punti verso il centro del telescopio; 
({\em ii}) \`e preferibile  effettuare  la misura della sezione d'urto inclusiva  
di produzione 
%differenziale 
$\frac{{\rm d}^2N(m_T,y)}{{\rm d}m_T{\rm d}y} $\ 
entro una regione del piano $(y,p_T)$\ in cui  
l'efficienza globale di ricostruzione ed identificazione \`e  
sufficientemente uniforme.  Anche in tal caso 
i due aspetti --- quello proprio dell'accettanza geometrica e quello 
dell'efficienza --- sono interconnessi e difficilmente separabili: 
senza cercare di scinderli, si vogliono qui semplicemente escludere le regioni di 
spazio delle fasi cui corrisponderebbero correzioni molto elevate, rispetto 
a quelle medie, in quanto eventuali piccole imprecisioni sulla determinazione 
di tali correzioni influenzerebbero molto le distribuzioni che si vogliono 
misurare, specialmente per le particelle caratterizzate da pi\`u bassa 
statistica.  
\newline
Due approcci per la determinazione della miglior finestra di accettanza sono 
stati seguiti, il secondo dei quali, molto pi\`u adatto per i campioni di bassa 
statistica, \`e stato sviluppato solo di recente~\cite{BrunoOmegaAcc}, 
apposta per le $\Omega$.     
%~\cite{BrunoOmegaAcc} 
%e sostituir\`a presto il primo metodo. 
\subsubsection{Determinazione della finestra di accettanza considerando la 
		stabilit\`a dei risultati}
\begin{figure}[p]
\begin{center}
\includegraphics[scale=0.33]{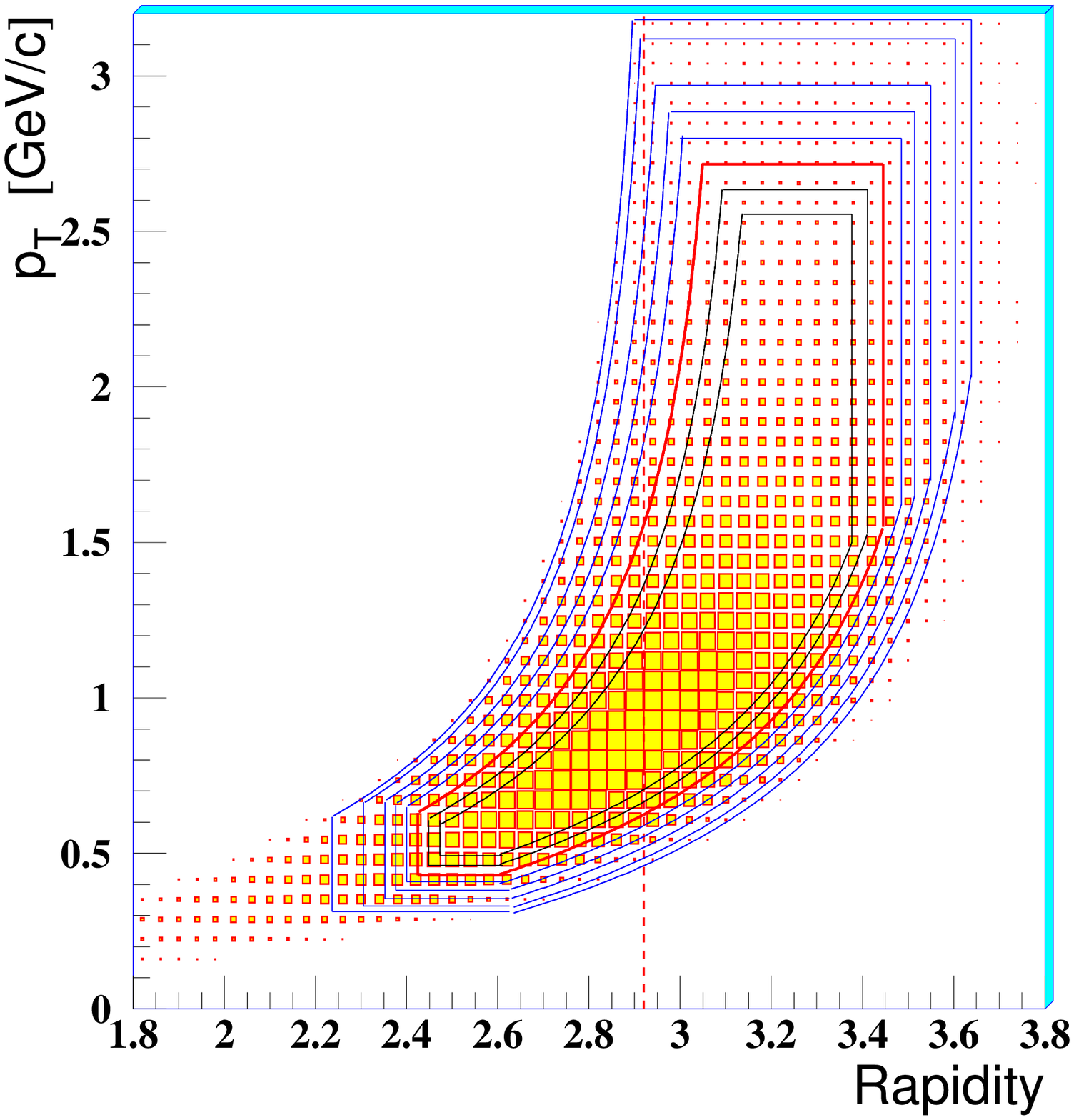}\\
\includegraphics[scale=0.28]{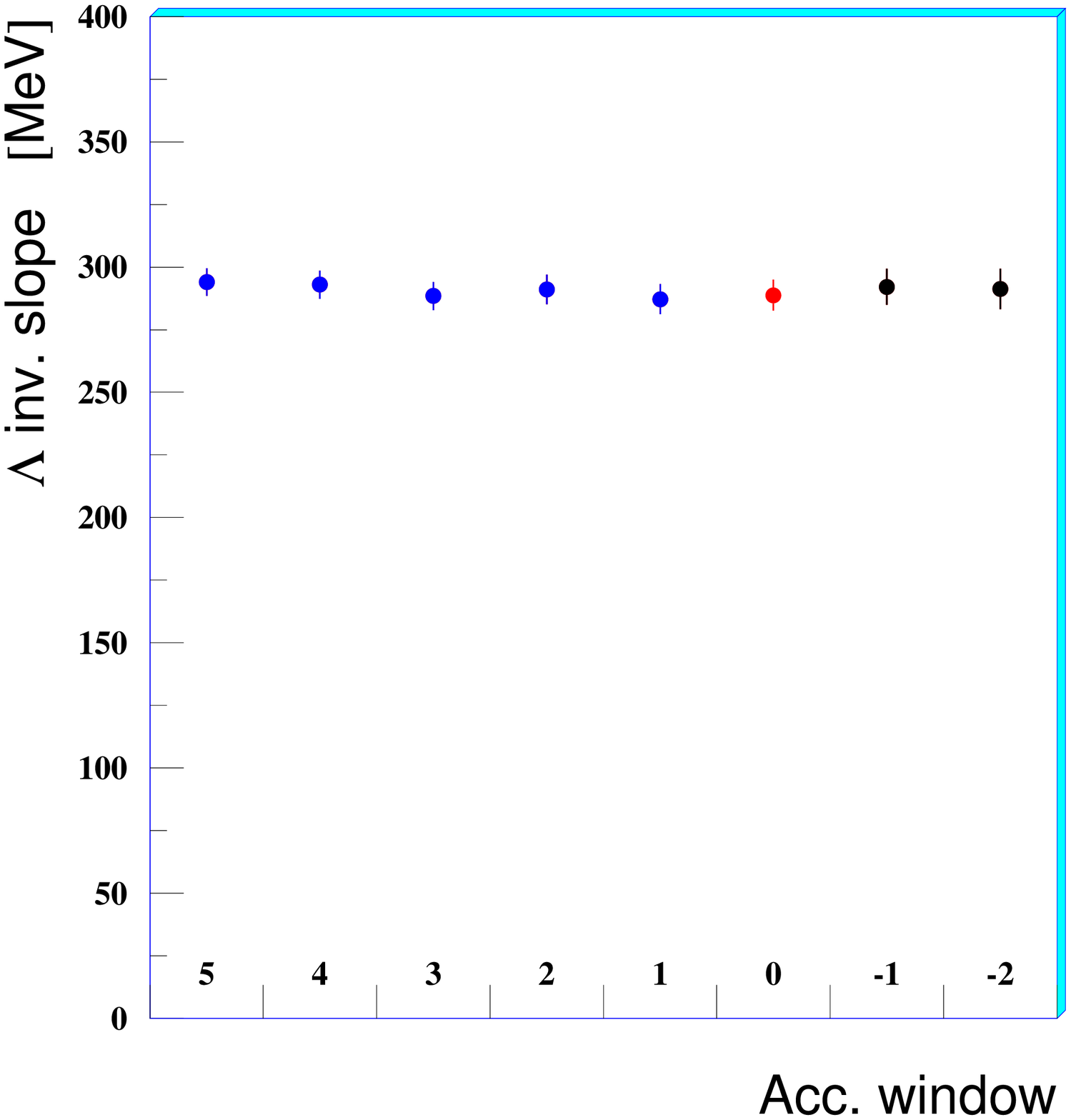}
\includegraphics[scale=0.28]{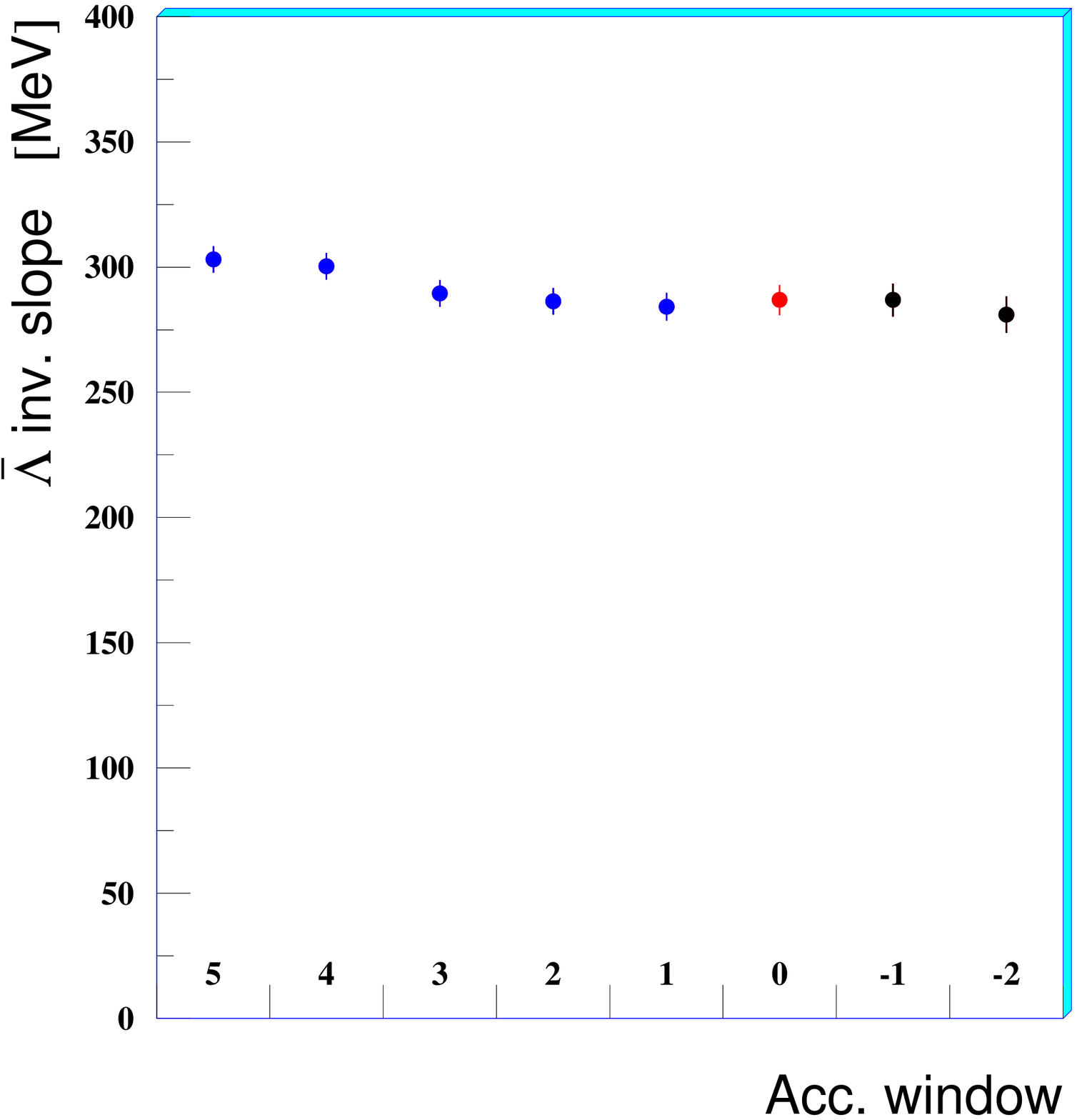}\\
\includegraphics[scale=0.28]{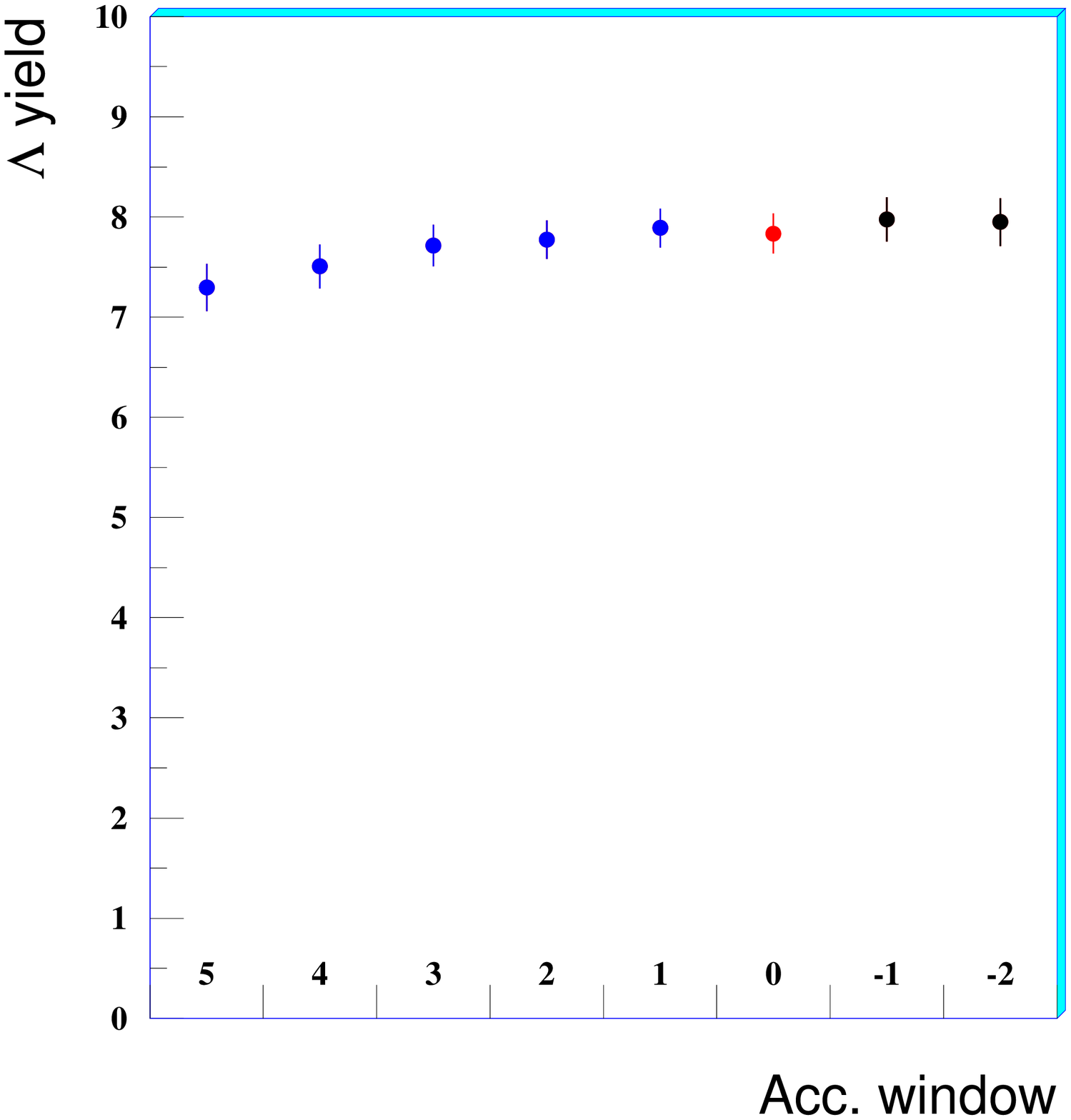}
\includegraphics[scale=0.28]{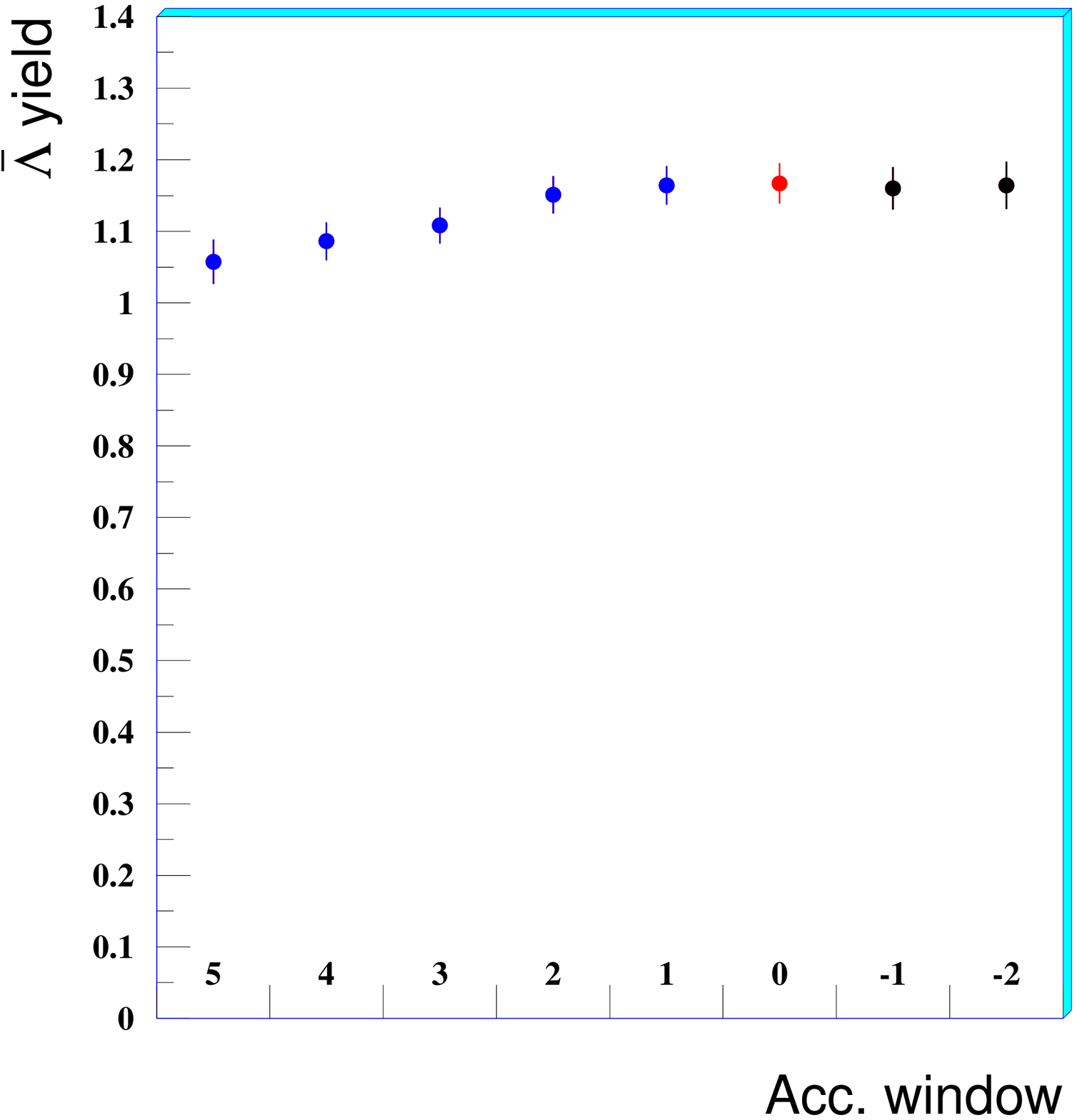}
\caption{Dipendenza della pendenza inversa ({\em ``inverse slope''}) 
	degli spettri di massa trasversa 
	e del tasso di produzione ({\em ``yield''}) estrapolato per \PgL\ e 
	\PagL\ dalla definizione della finestra di accettanza. Diverse
	finestre sono state definite, a partire da quelle pi\`u esterne 
	(in blu) cui sono state assegnate dei numeri interi decrescenti e 
	positivi, sino a quelle pi\`u interne (in nero) cui corrispondono 
	numeri interi negativi. Alla finestra scelta (numero $0$) \`e  
	assegnato il colore rosso.}  
\label{LambdaStability}
\end{center}
\end{figure}

\begin{figure}[p]
\begin{center}
\includegraphics[scale=0.28]{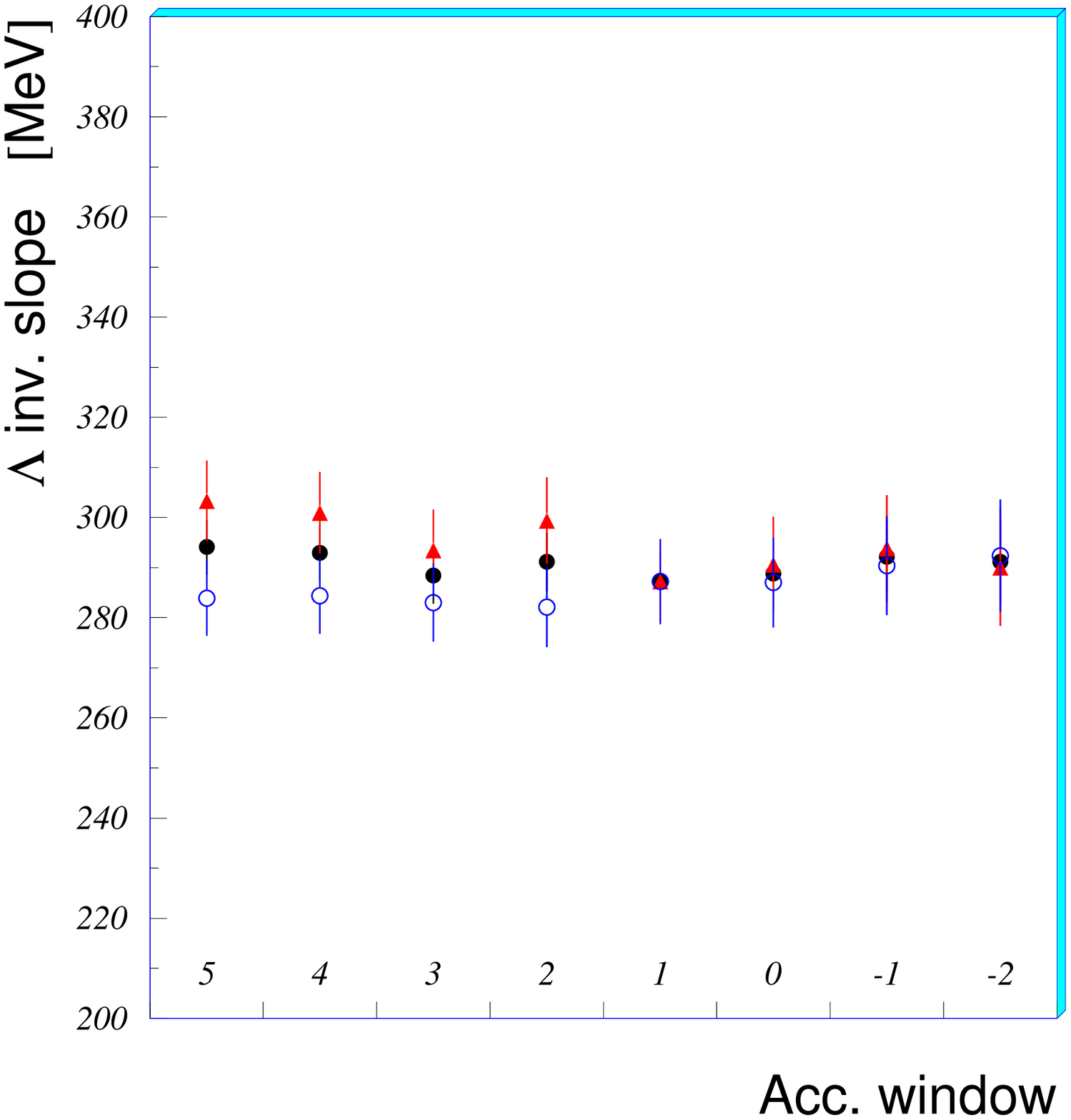}
\includegraphics[scale=0.28]{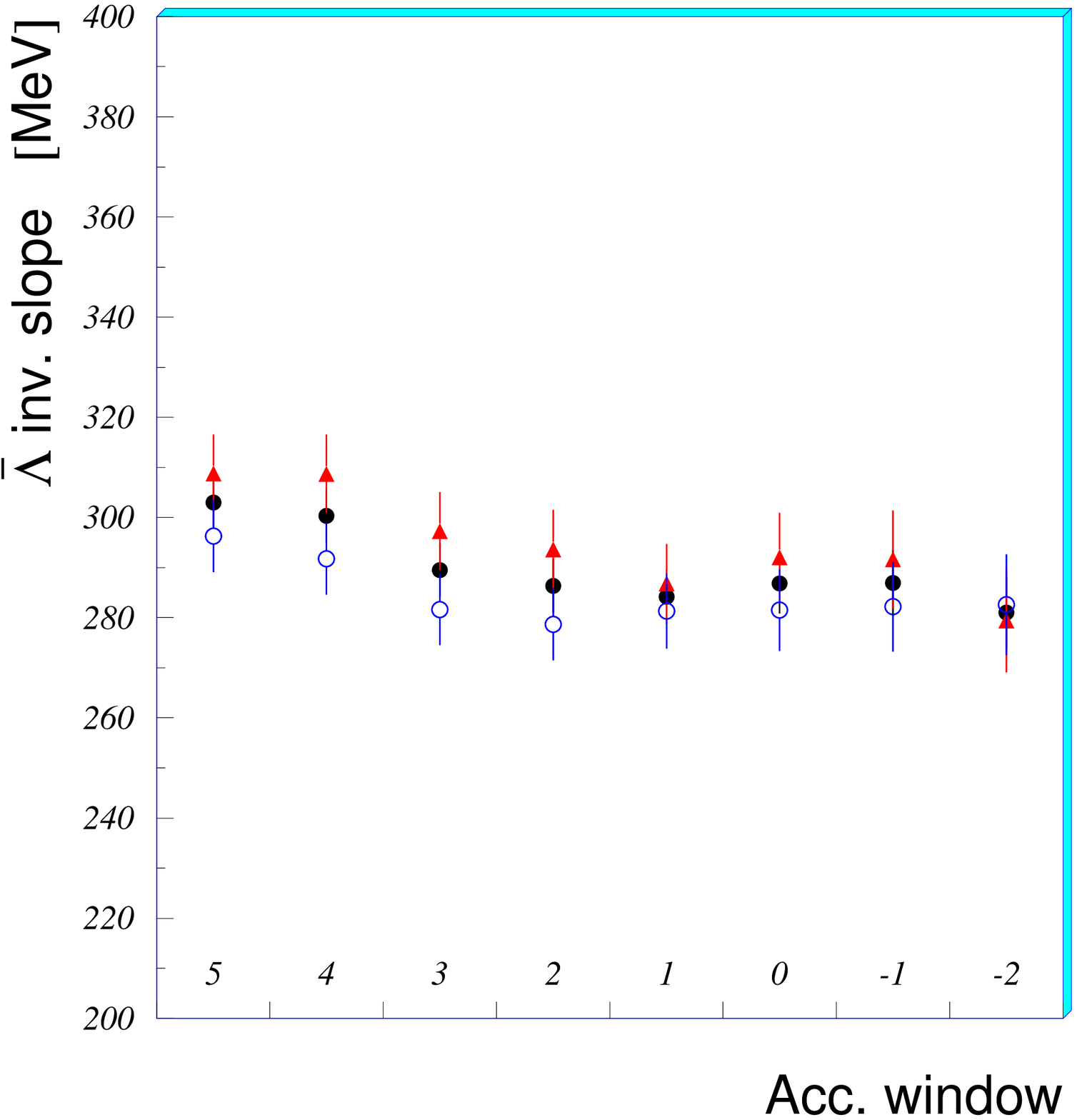}\\
\includegraphics[scale=0.28]{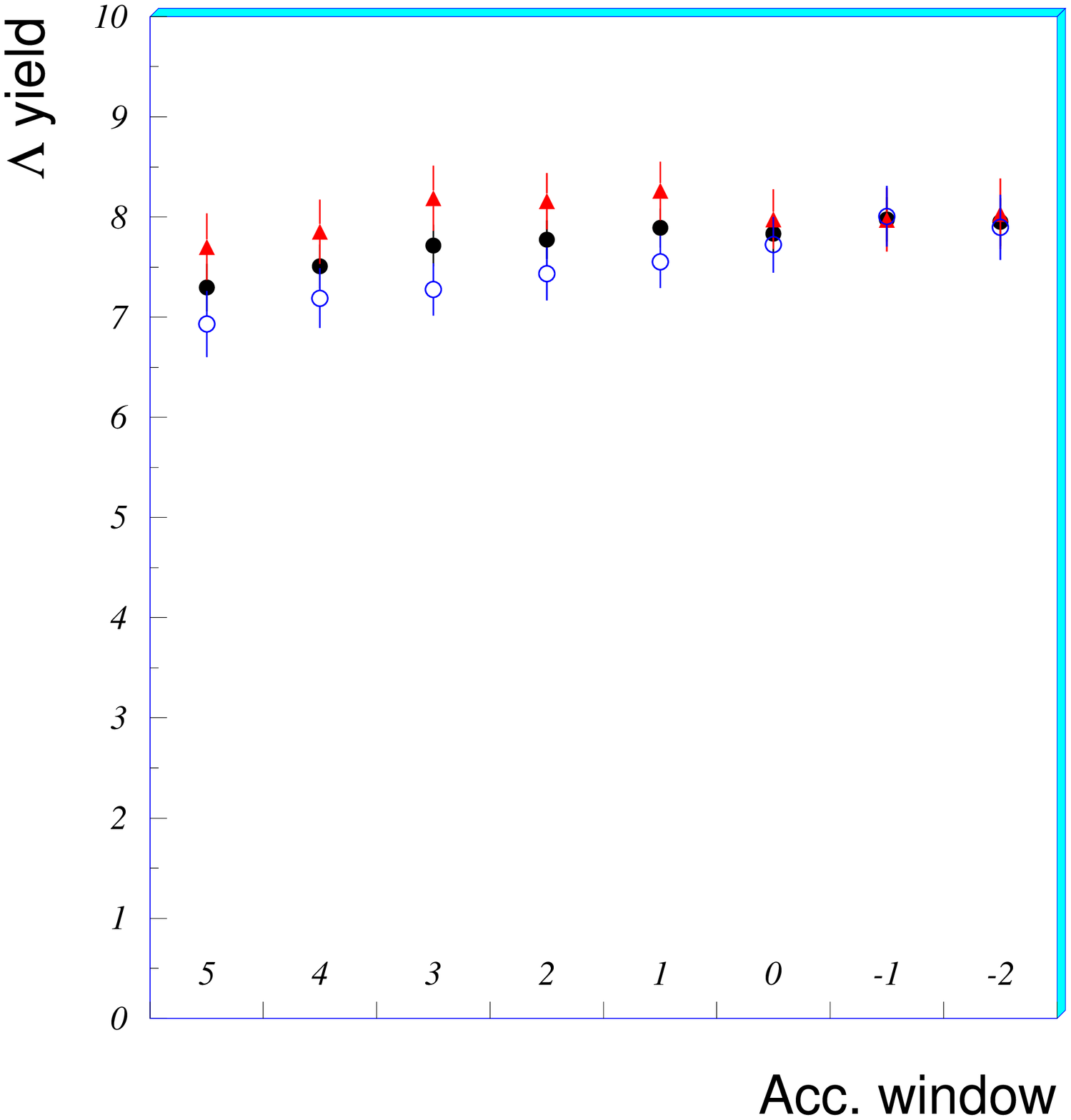}
\includegraphics[scale=0.28]{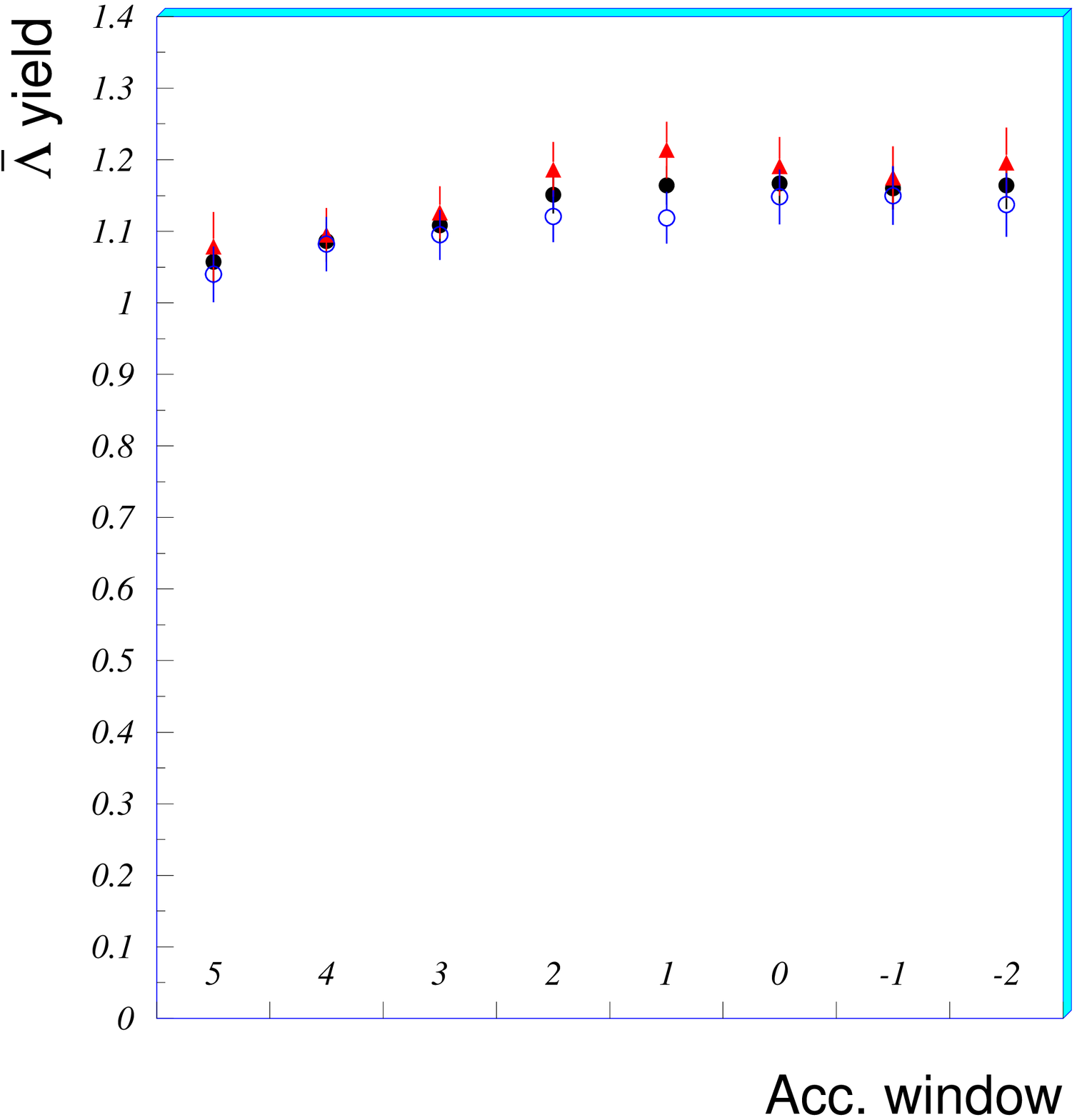}\\
\caption{Dipendenza delle {\em ``inverse slope''} (in alto) e degli {\em ``yield''} 
	 (in basso) per le \PgL\ (a sinistra) e le \PagL\ (a destra) in funzione 
	 delle finestre di accettanza, durante la prima met\`a (triangolini rossi) 
	 e la seconda met\`a (cerchietti blu vuoti) della presa dati e per 
	 l'intera presa dati (cerchietti neri pieni) del 1998.}
\label{LambdaPeriod}
\end{center}
\end{figure}

Il primo metodo sviluppato consiste nel calcolare le correzioni, come 
discusso nel {\em par. 4.3}, per tutte le particelle rivelate, e di definire 
inizialmente una finestra di accettanza fiduciale in modo da escludere quelle 
particelle per le quali il ``peso'' (la correzione) risulta superiore a circa 
$10$\ volte quello delle regioni pi\`u centrali, caratterizzate dai 
``pesi'' pi\`u bassi. A partire da questa  prima definizione di finestra 
di accettanza se ne considerano altre, pi\`u e meno estese, rispetto alle quali 
si studia la stabilit\`a dei risultati (tassi di produzione e spettri di 
massa trasversa). La finestra infine scelta \`e tale da cadere in una regione 
di stabilit\`a, rispetto alle altre definizioni, ed accettando quante pi\`u 
particelle possibile.  
\newline
In fig.~\ref{LambdaStability} \`e mostrato il risultato di questo tipo di studio 
nel caso delle \PgL\ e \PagL. Sono state definite otto finestre, numerate con gli 
interi dal $-2$\ al $+5$: ai numeri negativi corrispondono finestre via via 
meno estese e viceversa per i positivi. Sia per i tassi di produzione 
({\em ``yield''}), che sono estrapolati  
in una finestra di accettanza comune come si discuter\`a nel {\em par. 4.5}, sia per 
le pendenze inverse degli spettri di massa trasversa ({\em ``inverse slope''}), si 
riconoscono agevolmente dei {\em ``plateau''} che corrispondono alla regione 
di stabilit\`a desiderata. 
Il ginocchio di questi {\em ``plateau''} cade in corrispondenza della
finestra indicata col numero $+1$, ma un'analisi svolta in diversi periodi
di presa dati ha mostrato come nella finestra ``0'' i risultati
siano molto pi\`u stabili~\cite{BrunoLamStab}. I risultati di questo studio sono 
mostrati nella fig.~\ref{LambdaPeriod}, dove sono stati definiti due periodi, pari 
grosso modo alla prima ed alla seconda met\`a della presa dati del 1998, che si 
differenziano tra loro per un (improvviso) aumento (di circa il 35\%) della sezione 
del profilo, nella sola direzione $z$, del fascio di ioni di Pb incidente sul bersaglio:  
risulta evidente come  gli {\em yield} diventino compatibili solo 
a partire dalla finestra ``0'', sebbene il valor medio sia stabile gi\`a da 
quella ``+1''.  
La finestra $0$\
\`e dunque quella scelta per l'analisi delle \PgL\ e \PagL.  
\newline
Volendo tuttavia approfondire la comprensione delle differenze tra i due periodi, 
limitandosi a considerare solo quelle particelle che cadono entro la finestra 
``+1'' ma sono al di fuori di quella ``0'', si \`e appurato che esse sono essenzialmente  
imputabili alla regione prossima al bordo curvo inferiore della finestra, 
corrispondente ai piccoli angoli della particella emessa rispetto alla linea di fascio.  
Tale regione corrisponde alla parte dei rivelatori pi\`u prossimi al fascio, 
per cui l'efficienza \`e generalmente inferiore a causa del danneggiamento da 
radiazione. A parit\`a di impulso trasverso e di rapidit\`a, le variazioni nella 
posizione $z$\ del vertice primario modificano ovviamente l'accettanza geometrica 
di una particella, tuttavia i {\em ``pesi''} calcolati col Monte Carlo 
non riescono a correggere completamente per questo effetto, a conferma che 
le regioni prossime a bordi del telescopio sono le pi\`u critiche e vanno escluse.  

\begin{figure}[hbt]
\begin{center}
\includegraphics[scale=0.37]{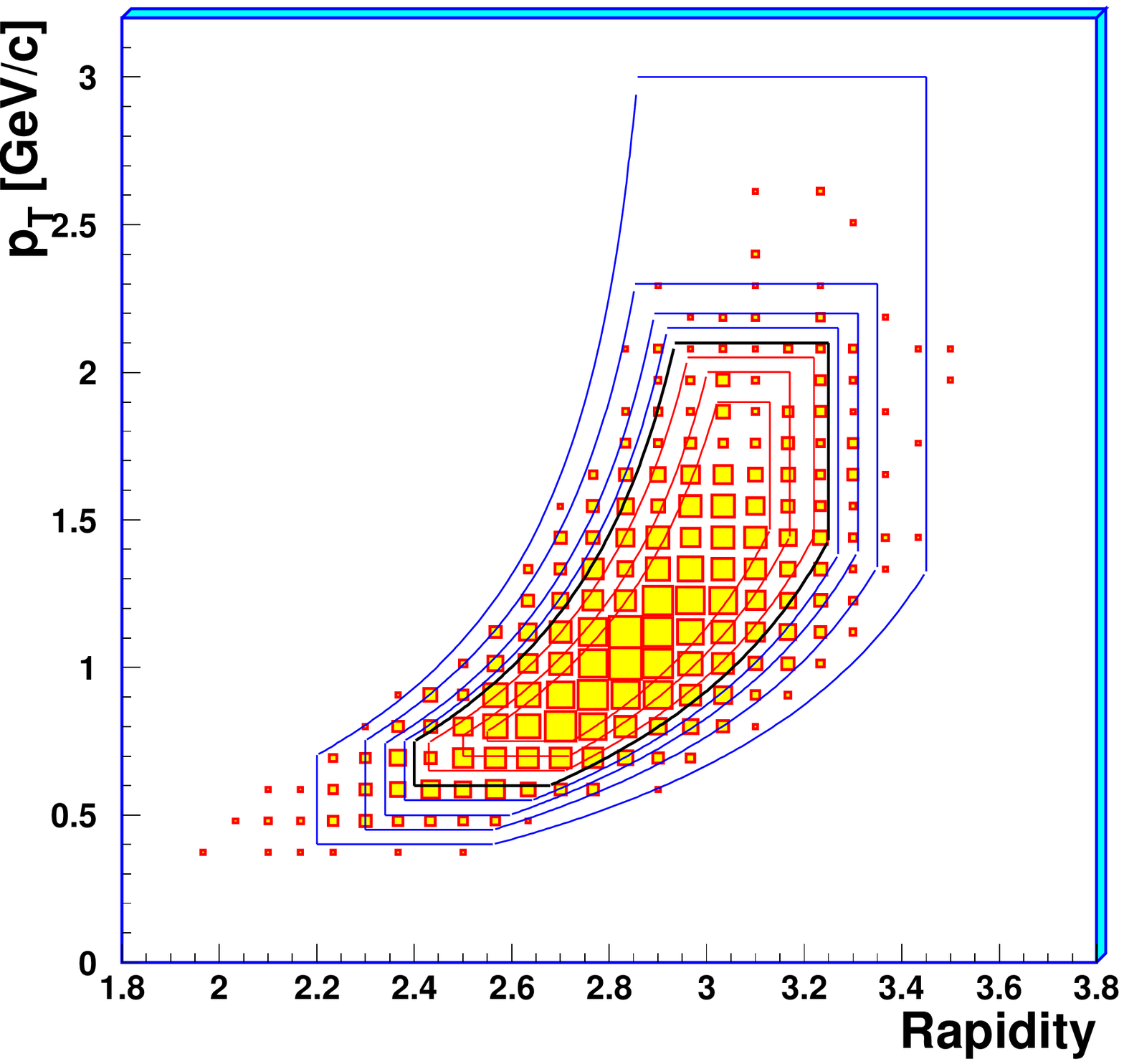}
\includegraphics[scale=0.37]{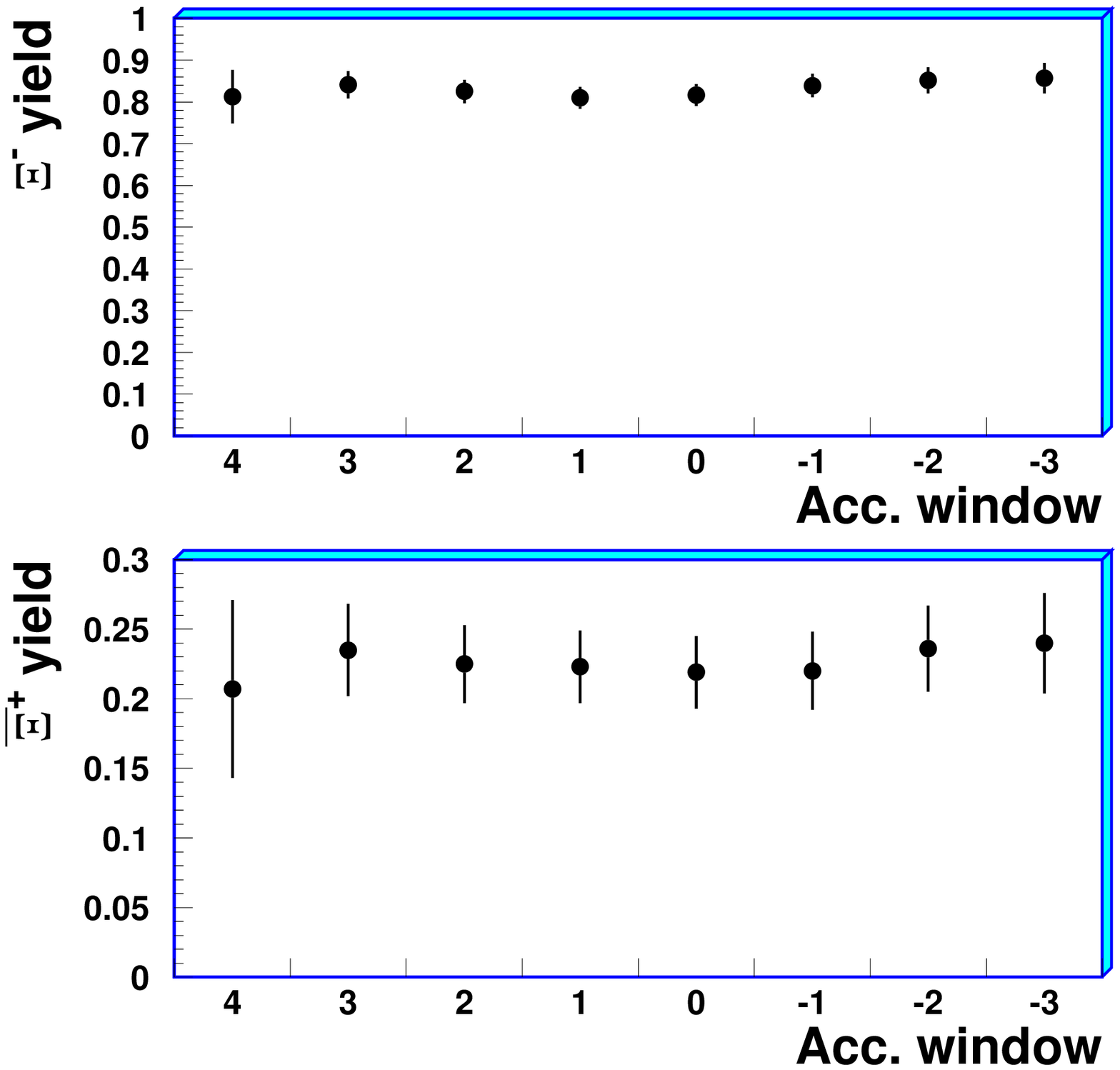}
\caption{Dipendenza del tasso di produzione ({\em ``yield''}) estrapolato 
	 delle particelle \PgXm\ e \PagXp\ dalla definizione della 
	 finestra di accettanza.  
	 Valori negativi sull'asse delle ascisse dei grafici di destra 
	 corrispondono alle finestre pi\`u interne (in rosso) sul grafico 
	 di sinistra, valori positivi a quelle pi\`u esterne (in blu). La 
	 finestra etichettata con $0$, in nero, \`e quella adottata.}
\label{XiStability}
\end{center}
\end{figure}

In fig.~\ref{XiStability} sono mostrati i grafici della stabilit\`a degli 
{\em ``yield''} delle \PgXm\ e delle \PagXp\ con la definizione della finestra 
di accettanza. 
La scelta adottata, corrispondente alla finestra ``0'', cade in una
regione molto stabile, entro gli errori statistici, sia per le \PgXm\ che per le
\PagXp. Sebbene la pi\`u bassa statistica non consenta ulteriori indagini, da un punto
di vista qualitativo, la maggior stabilit\`a dei risultati delle cascate rispetto a
quelli delle $\Lambda$\ \`e probabilmente dovuta alla maggior libert\`a
di distribuirsi nello spazio per i prodotti di decadimento della cascata
(in cui vi \`e un doppio decadimento) rispetto a quella concessa ai prodotti di
decadimento della $V^0$. Per una $V^0$, in altre parole, al crescere dell'angolo 
$\theta=\arccos\frac{p_T}{p}$, formato dall'impulso della $V^0$\ 
rispetto alla linea del fascio, la  probabilit\`a  che i suoi prodotti di 
decadimento siano accettati geometricamente entro il telescopio
aumenta molto pi\`u rapidamente rispetto al caso di una cascata. 

\begin{figure}[p]
\begin{center}
\includegraphics[scale=0.35]{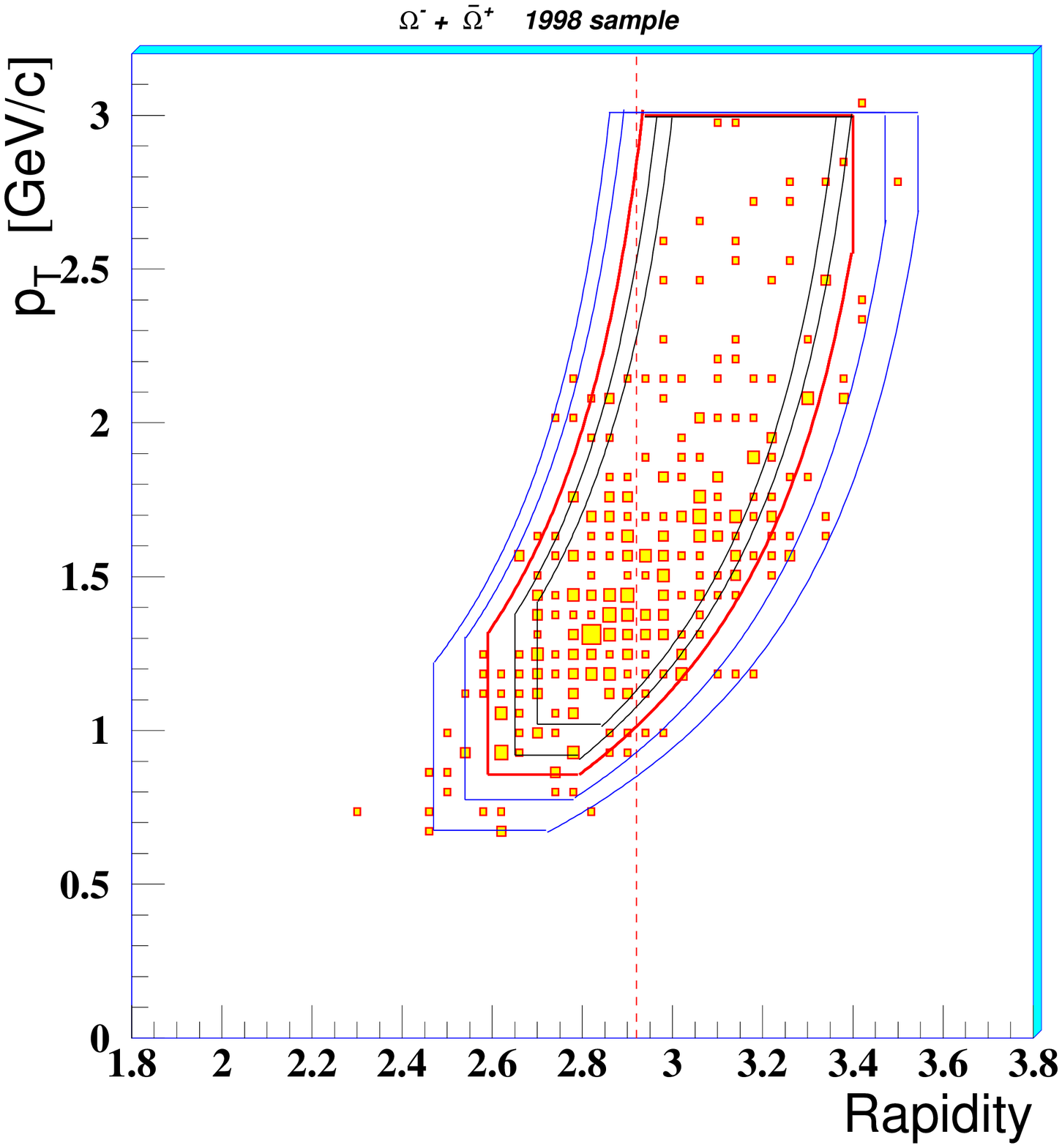} \\
\includegraphics[scale=0.30]{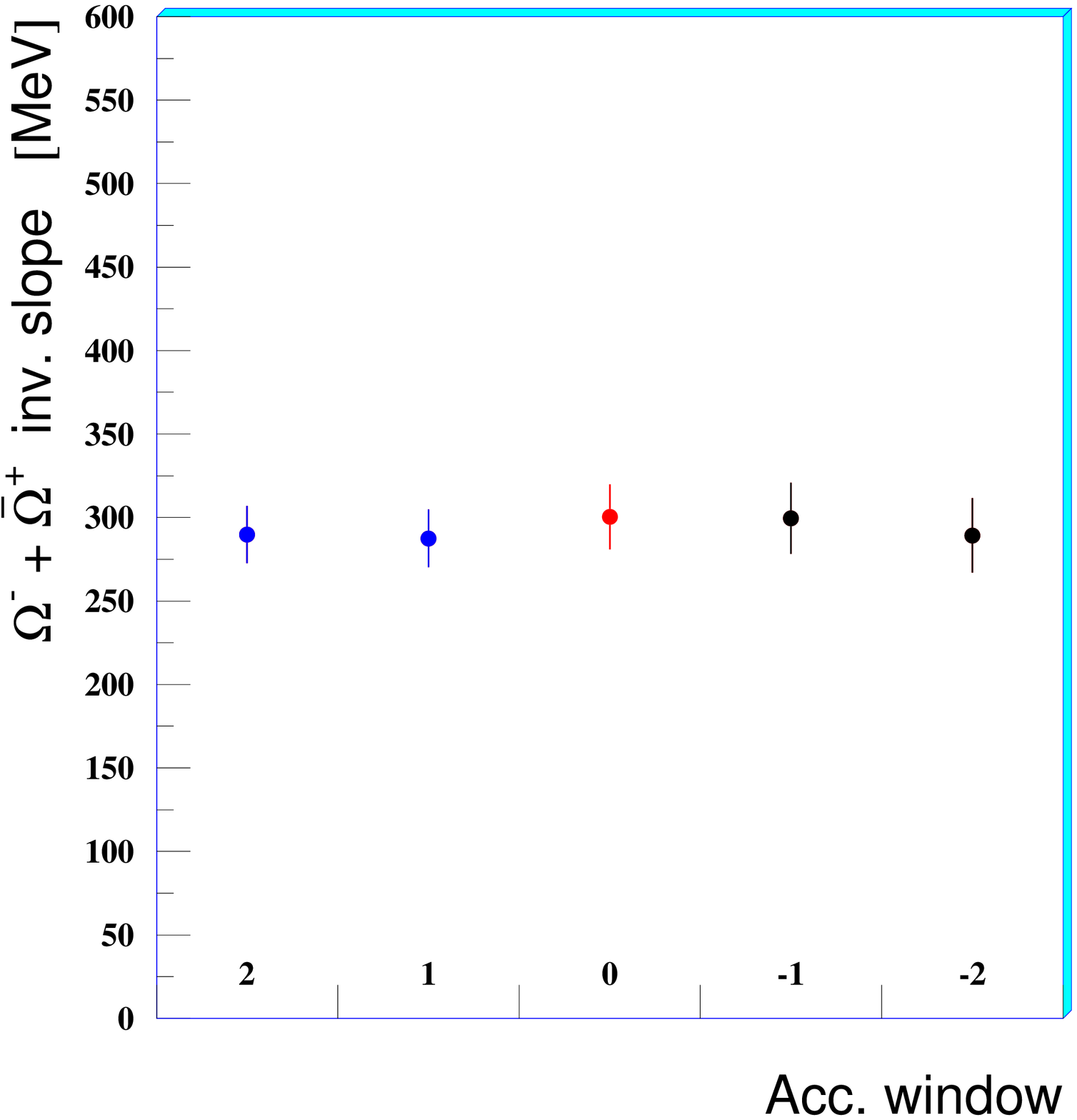}
\includegraphics[scale=0.30]{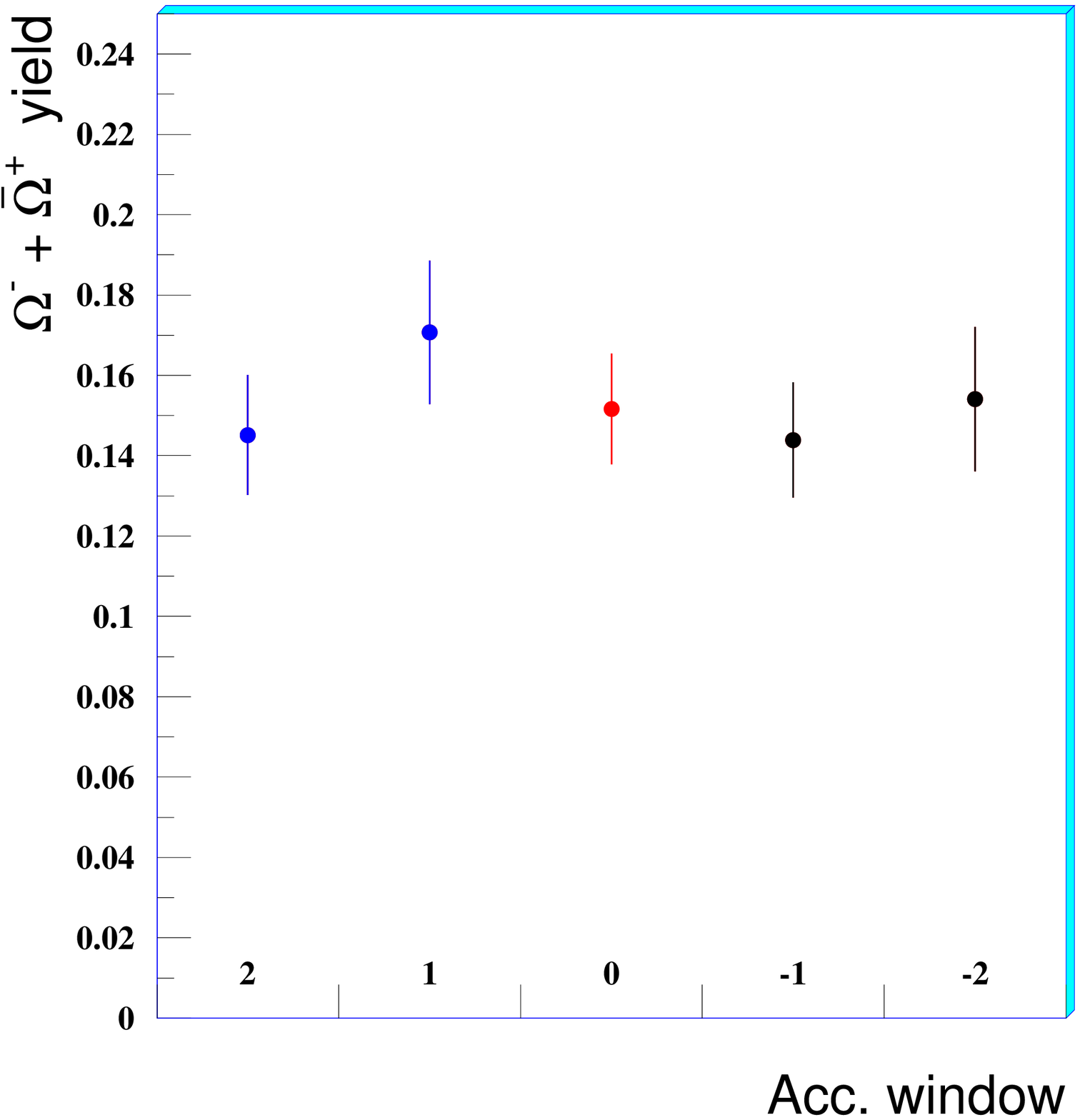}
\caption{Stabilit\`a delle pendenze inverse ({\em ``inverse slope''}) 
	 degli spettri di massa trasversa 
	 e dei tassi di produzione ({\em ``yield''}) estrapolati, 
	 al variare della definizione della finestra di accettanza per il 
	 campione di \PgOm\ e \PagOp\ dell'anno 1998. Valori negativi 
	 sull'asse delle ascisse corrispondono alle finestre pi\`u interne, 
	 valori positivi a quelle pi\`u esterne.}
\label{OmStability}
\end{center}
\end{figure}

In fig.~\ref{OmStability} \`e mostrato infine lo studio analogo condotto sul 
campione delle $\Omega$\ raccolte nell'anno 1998. Anche in tal caso, la scelta 
adottata d\`a confidenza nella stabilit\`a dei risultati, come per  
tutte le altre particelle.  
\newline
Questo approccio funziona bene quando la densit\`a di particelle nella 
finestra che si vuole definire \`e sufficientemente elevata; invece   
nel caso di campioni con basso numero di particelle, spesso non si \`e in grado 
di valutare in modo obiettivo se alcune zone sono spopolate perch\'e  
effettivamente 
%poco prodotte 
la produzione \`e inferiore 
in quella regione o 
perch\'e essa ha bassa efficienza.  
\subsubsection{Determinazione della finestra di accettanza per via Monte Carlo}
\begin{figure}[hbt]
\begin{center}
\includegraphics[scale=0.32]{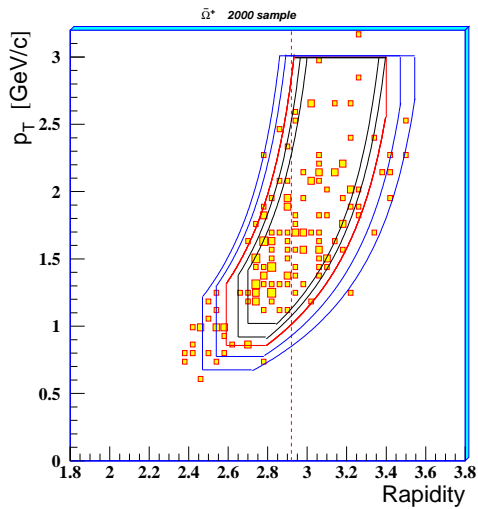}
\caption{Distribuzione nel piano $y_{\Omega}$--$p_T$\ delle \PagOp\ raccolte 
	 nella presa dati del 2000. Le finestre sovrapposte rappresentano 
	 possibili scelte per la definizione della regione fiduciale 
	 (si veda il testo per la discussione).}
\label{AOmWin2k}
\end{center}
\end{figure}

\begin{figure}[p]
\begin{center}
\includegraphics[scale=0.35]{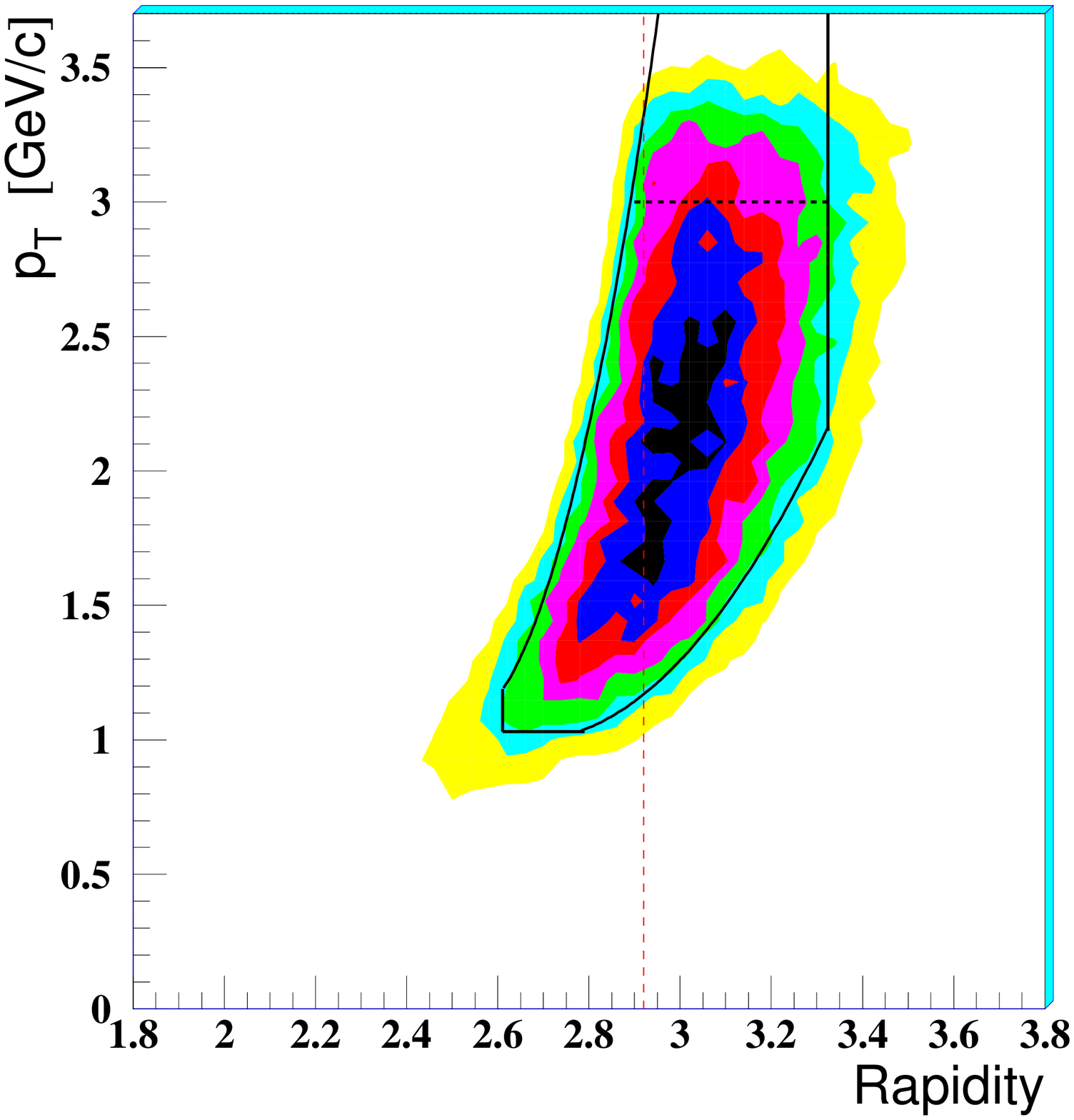}\\
\includegraphics[scale=0.58]{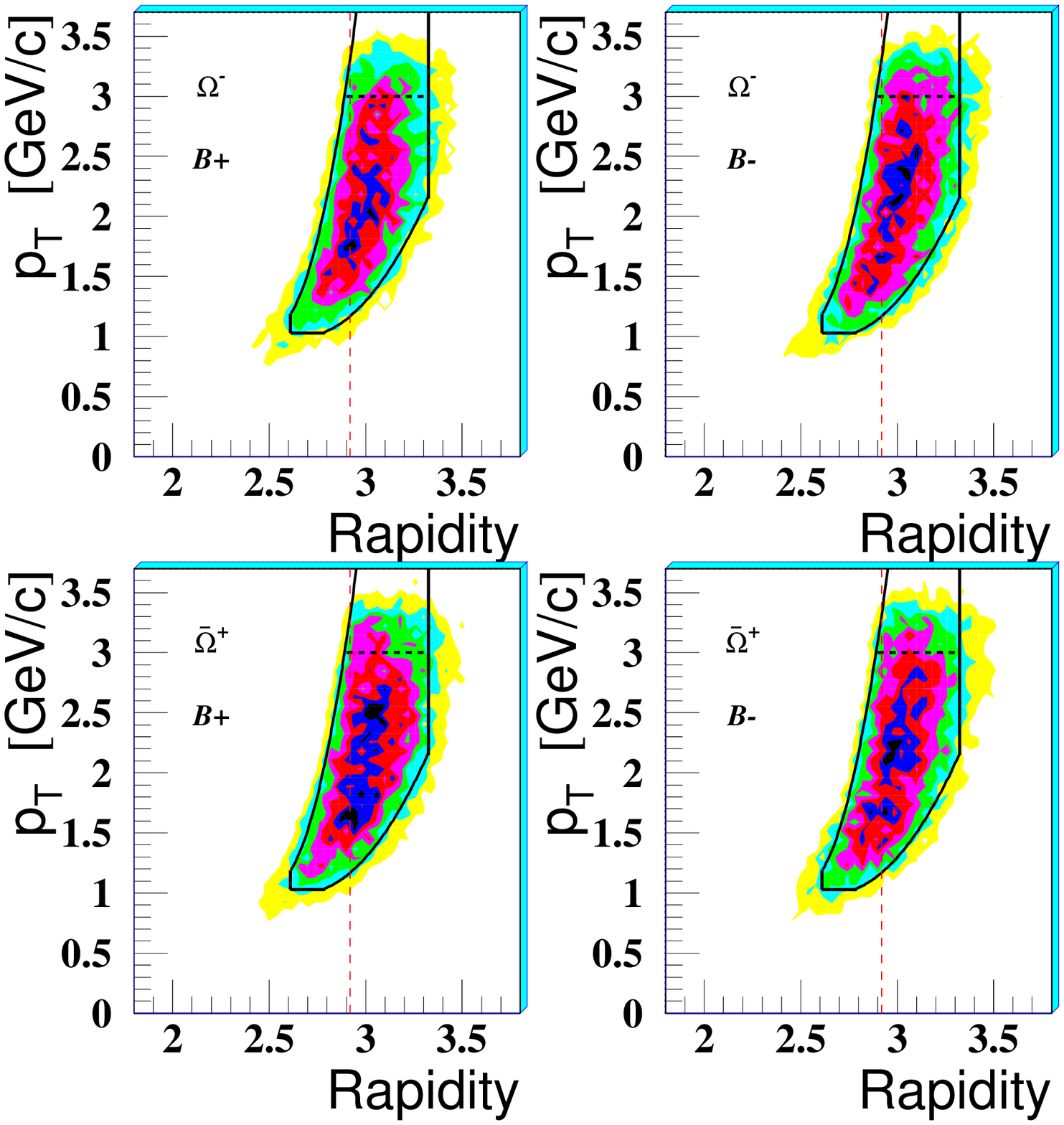}
\caption{Risultati per la distribuzione nel piano $(y,p_T)$ 
	 delle $\Omega$\ generate con la simulazione Monte Carlo 
	 nella disposizione sperimentale dell'anno 2000. Il grafico in alto si 
	 riferisce all'unione del campione di \PgOm\ ed \PagOp\ generate. 
	 I quattro grafici in basso alle diverse combinazioni 
	 \PgOm, \PagOp\ e polarit\`a del campo magnetico. 
	 La curva nera definisce la miglior finestra di accettanza determinata 
	 con questo studio, che ben si adatta a tutti i cinque grafici.}  
\label{OmWindowMC}
\end{center}
\end{figure}

\begin{figure}[p]
\begin{center}
\vspace{-1.2cm}
\includegraphics[scale=0.30]{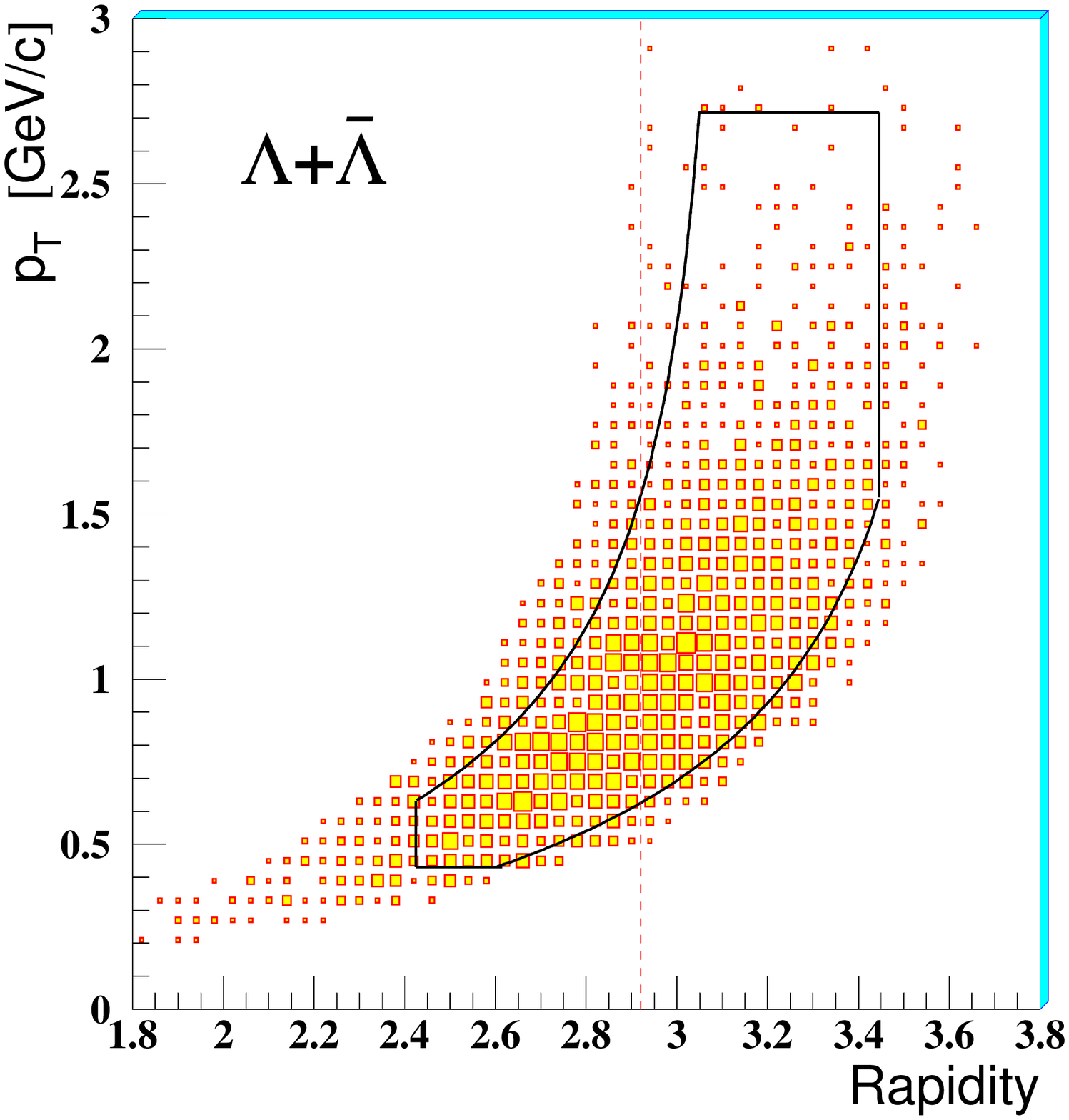}
\includegraphics[scale=0.30]{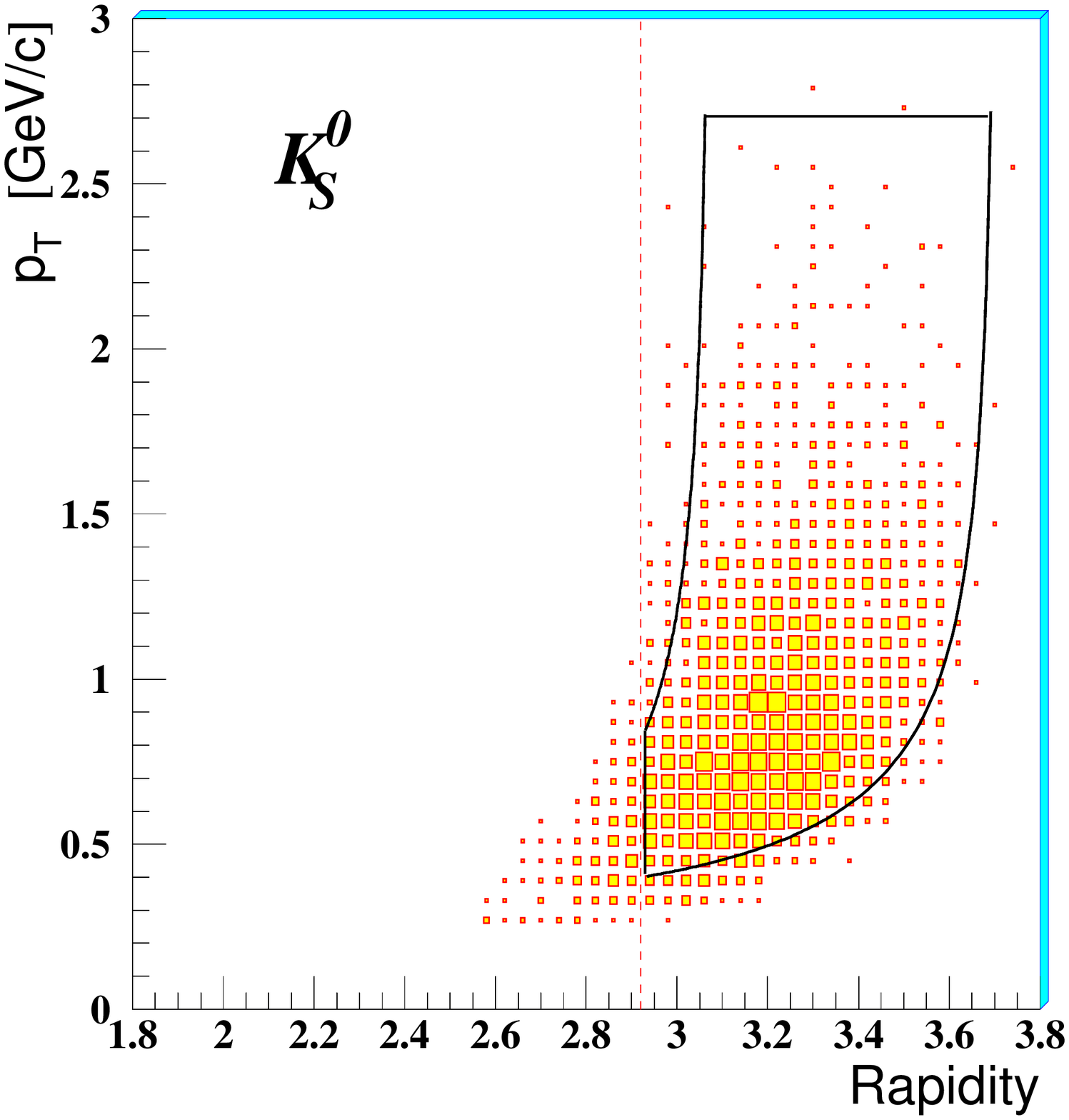}\\
\includegraphics[scale=0.30]{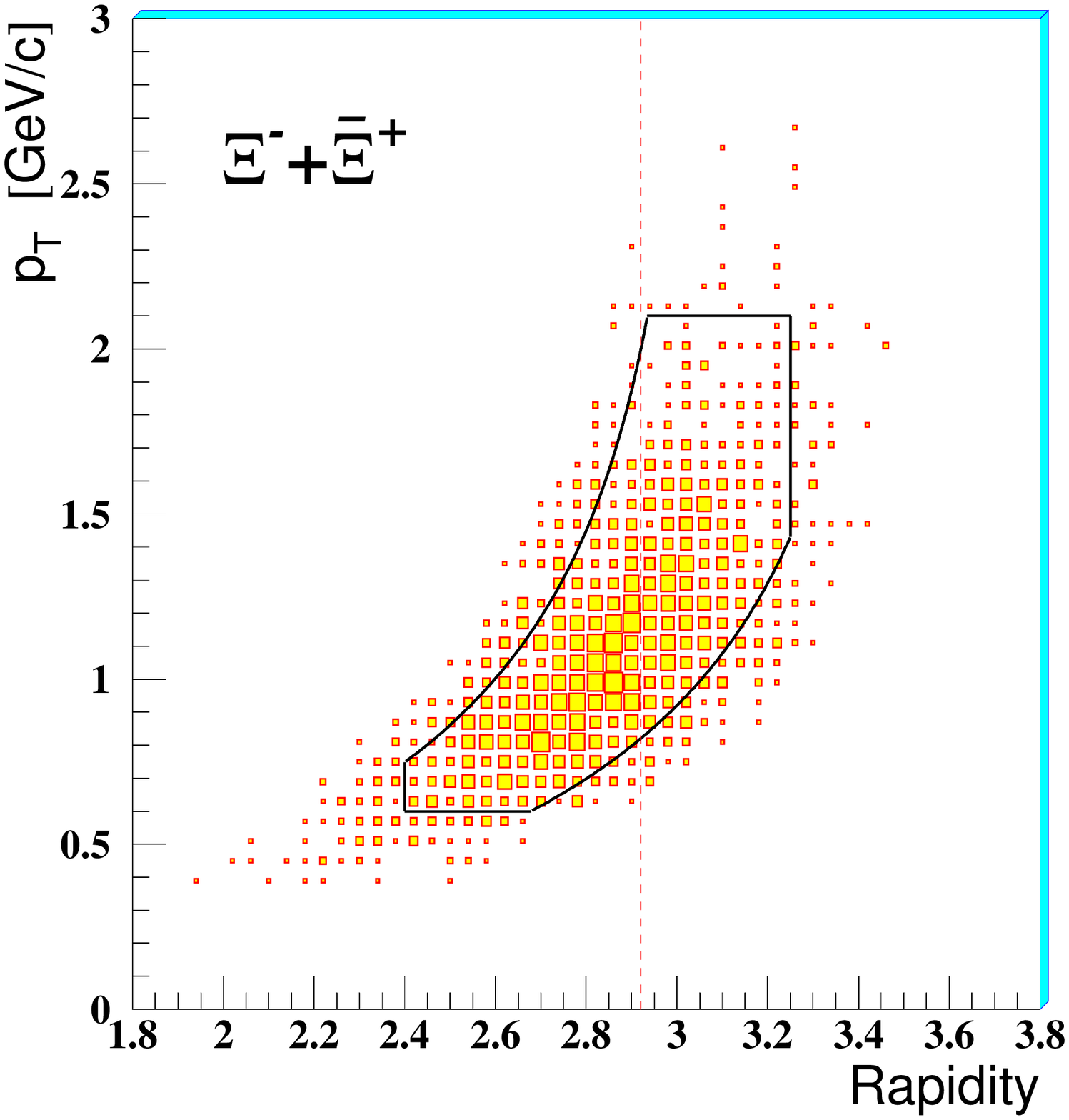}
\includegraphics[scale=0.30]{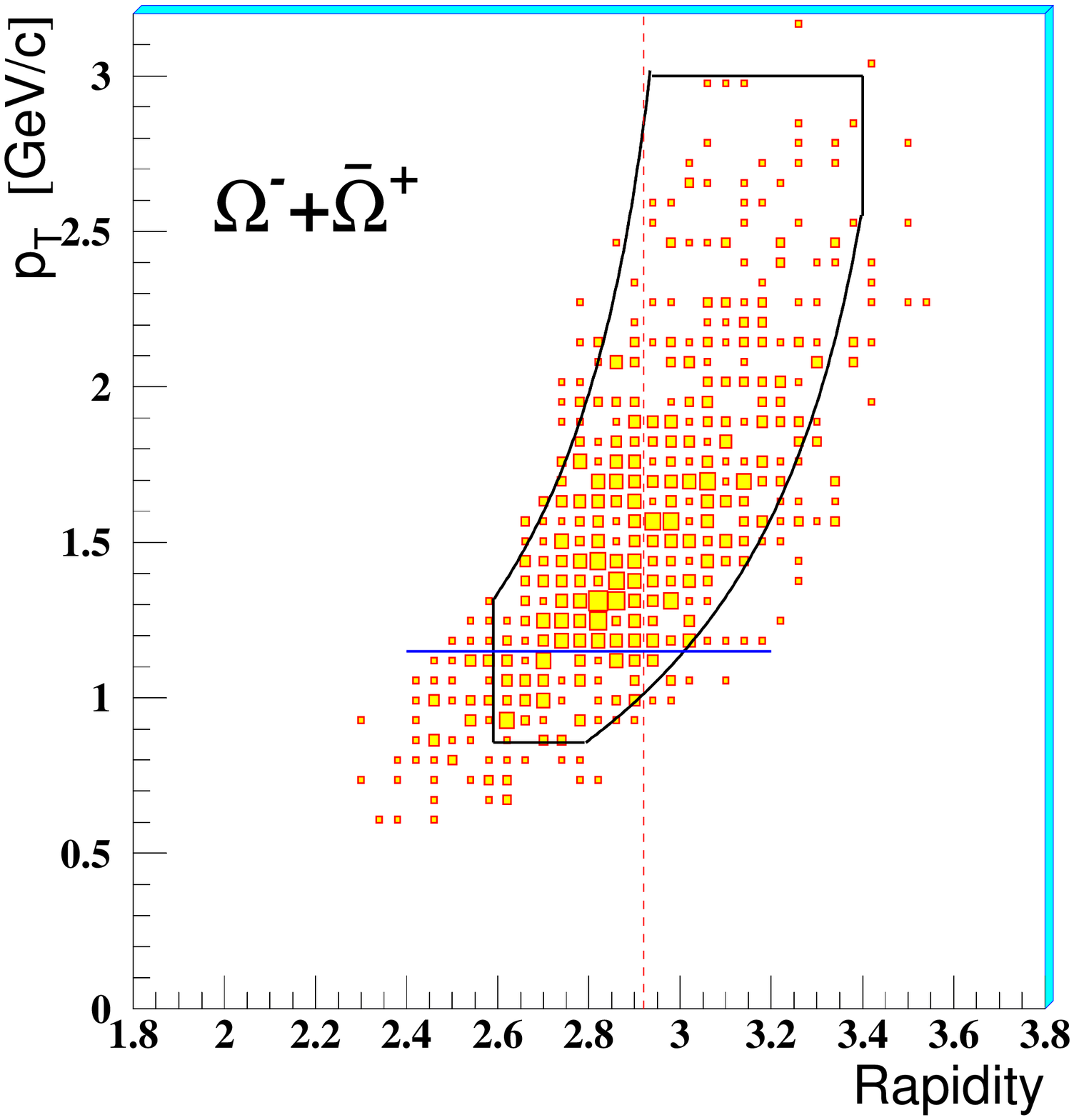}\\
\includegraphics[scale=0.38]{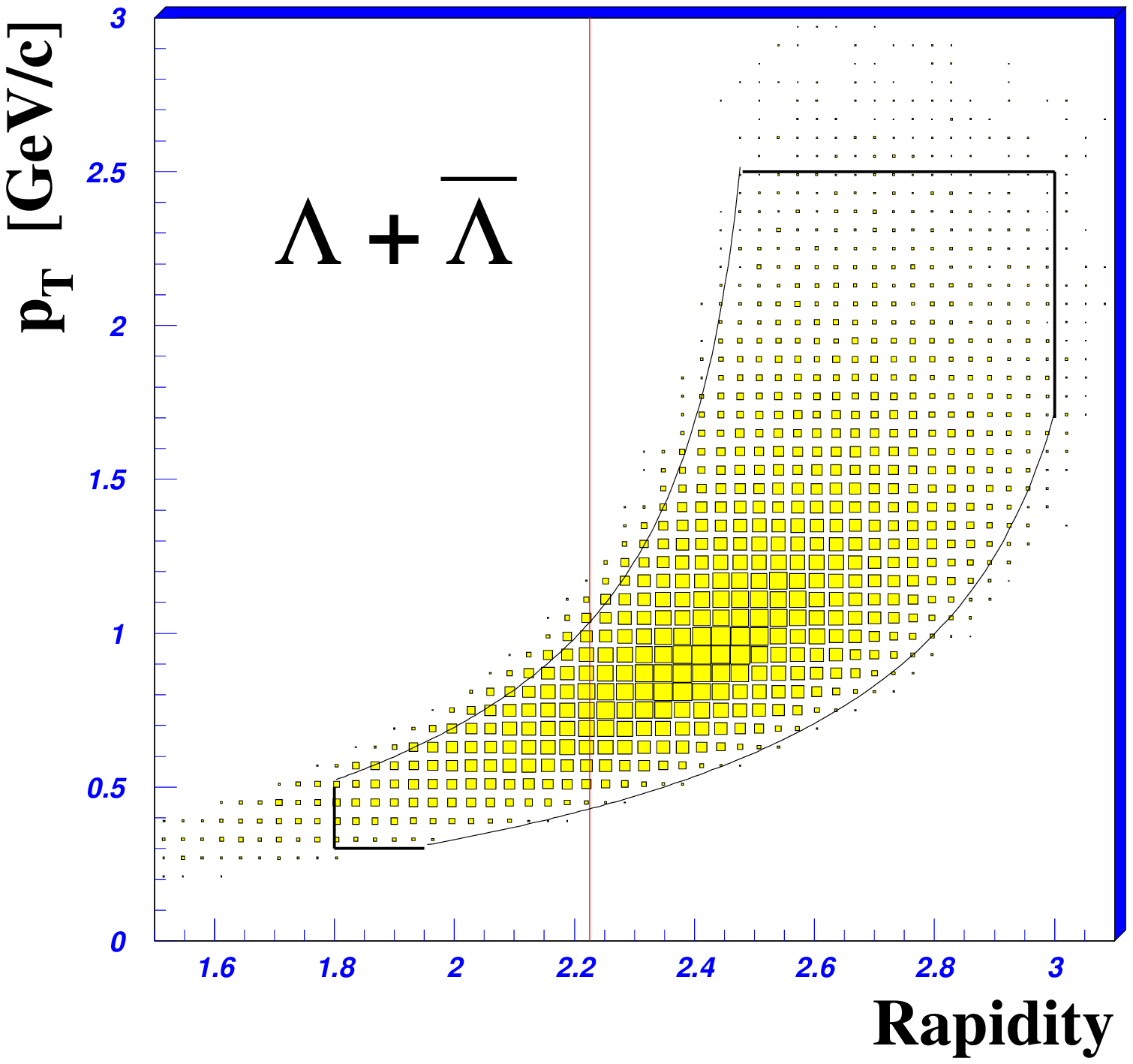}
\includegraphics[scale=0.38]{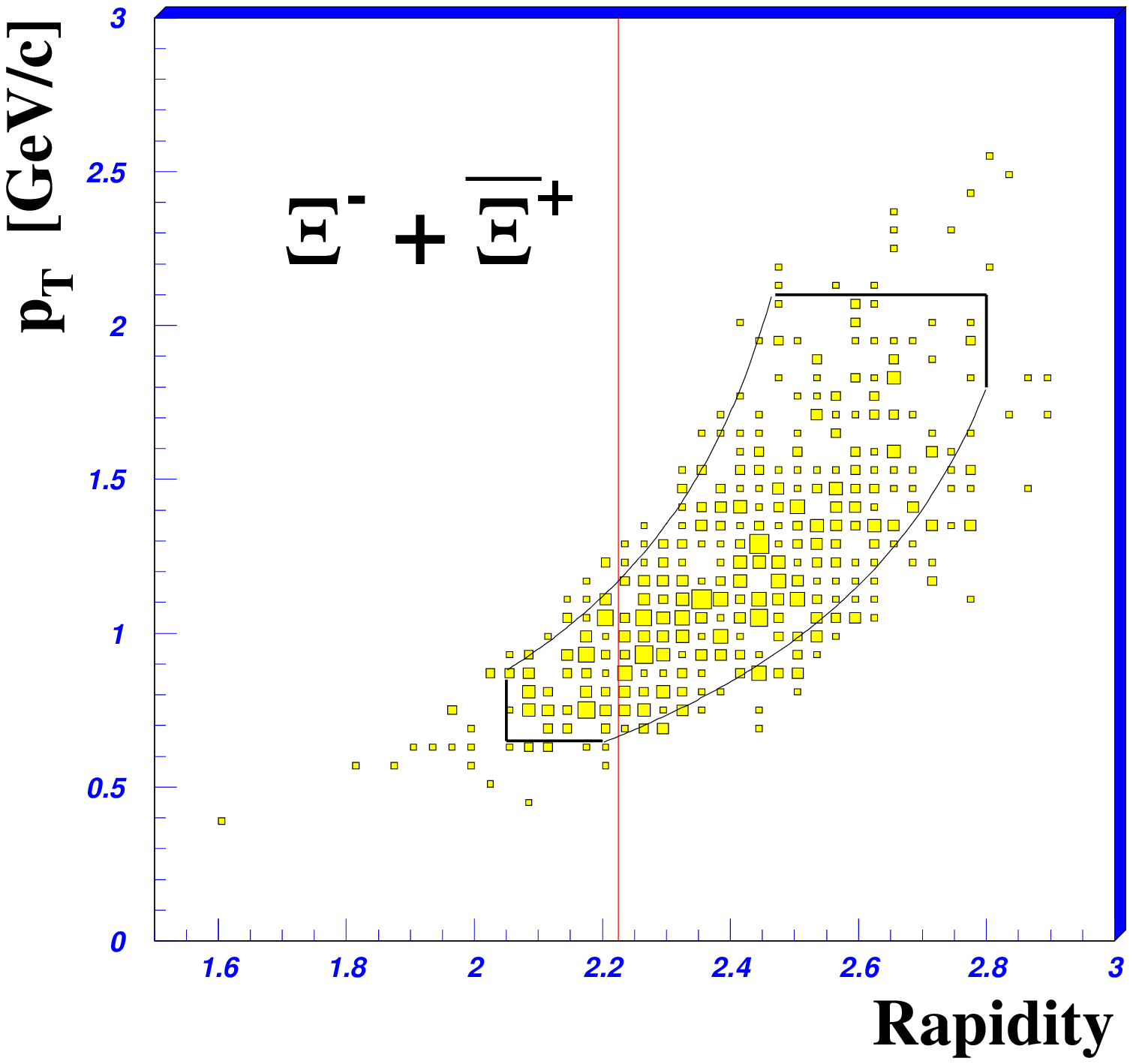}
\caption{Finestre di accettanza delle particelle strane nelle collisioni 
         Pb-Pb a 160 A GeV/$c$\ (i quattro grafici pi\`u in alto) e 
	 nelle collisioni Pb-Pb a 40 A GeV/$c$\ (i due grafici in basso). 
	 A tutti gli iperoni mostrati sono stati sovrapposti i relativi anti-iperoni. 
	 Le linee rosse verticali rappresentano la rapidit\`a del centro di massa.}
\label{AccWinFinal}
\end{center}
\end{figure}

Si consideri ad esempio il campione delle \PagOp\ raccolte nel 2000, che 
ha pur consistenza statistica superiore a quello del 1998, e la cui distribuzione 
nel piano $(y_{\Omega},p_T)$\ \`e mostrata in fig.~\ref{AOmWin2k}. 
Si immagini di definire diverse finestre di accettanza, come nel caso delle altre 
specie considerate, ad esempio quelle disegnate nella fig.~\ref{AOmWin2k}: 
si comprende facilmente che il passaggio dall'una all'altra \`e 
accompagnato da notevoli fluttuazioni statistiche dovute a quelle {\em poche} 
particelle, cui sono associate {\em elevate} correzioni, che vengono 
incluse od escluse. Inoltre, quando la densit\`a \`e bassa, nel definire 
la finestra di accettanza si pu\`o incrementare o diminuire la superficie 
della stessa, in modo non obiettivo, senza che il numero di particelle 
racchiuse al suo interno varii apprezzabilmente, ed il tasso di produzione  
varier\`a di conseguenza.  
Infine, con il precedente approccio, vi \`e il rischio di 
eliminare quelle regioni del piano in cui vi \`e una minor popolazione, 
escludendole semplicemente dalla finestra di accettanza, senza per\`o 
accertarsi se ci\`o \`e dovuto a bassa efficienza, nel qual caso 
l'esclusione pu\`o essere legittima, oppure perch\'e l\`i le particelle 
non vengono prodotte tanto abbondantemente quanto atteso, nel qual caso 
si perdono importanti informazioni. 
Potrebbe ad esempio nascere tale dubbio nel caso delle \PagOp\ 
nella regione intorno a $\left( y_{\Omega}\approx y_{cm} \, 
, \, p_T \approx 1.0\, {\rm GeV}/c\right)$, 
in cui si osserva uno spopolamento della distribuzione.   
\newline
Per superare queste difficolt\`a si sono generate delle \PgOm\ e delle 
\PagOp\ con il Monte Carlo,  
assumendo una distribuzione {\em uniforme} nel piano $y_{\Omega}$--$p_T$,  
nell'intervallo di valori di  $p_T$\  
$\left[0.5 \,{\rm GeV}/c \; , \; 3.4 \, {\rm GeV}/c \right] $. Mentre il limite 
inferiore dell'intervallo tiene conto del limite imposto  
dall'accettanza geometrica del telescopio, quello superiore \`e   
applicato solo perch\'e non si ha speranza di misurare le distribuzioni di 
$\Omega$\ reali oltre $p_T \approx 3$\ GeV/$c$, in quanto la loro produzione 
\`e (esponenzialmente) soppressa con $p_T$.  
Come nel caso del calcolo delle correzioni, 
gli {\em ``hit''} rilasciati dai prodotti carichi di decadimento sono stati 
introdotti in eventi di fondo, e quindi i nuovi eventi ibridi sono 
stati processati con ORHION. Si sono infine applicati i criteri di selezione 
dell'analisi per ricostruire il campione delle particelle generate.  
In tal modo si \`e ottenuta una misura della ``funzione di risposta'' 
$F(y_{\Omega},p_T)$\ del telescopio e dei programmi di ricostruzione 
per le particelle $\Omega$, funzione che fornisce la probabilit\`a di 
ricostruire una \PgOm\ od una \PagOp\ di dato impulso trasverso e rapidit\`a. 
In fig.~\ref{OmWindowMC} \`e mostrato il risultato della simulazione: 
muovendosi dal centro della finestra, in cui si ha la maggior efficienza 
(zona in nero), verso le regioni esterne di efficienza via via pi\`u bassa, 
ciascun cambiamento di colore corrisponde ad una diminuzione del 12.5\% 
rispetto all'efficienza massima. 
\`E quindi possibile definire delle curve 
di livello, sulle quali l'efficienza globale di rivelazione e ricostruzione 
\`e costante. Si ottiene in tal modo un criterio obiettivo per la definizione 
della finestra di accettanza; ad esempio nella fig.~\ref{OmWindowMC}, la finestra 
in nero \`e stata tracciata in modo tale da escludere le zone in  cui 
l'efficienza sia inferiore a circa il 30\% dell'efficienza massima.  
Questo metodo \`e stato sino ad ora sviluppato solo per le $\Omega$\  
nella configurazione 
%della presa dati 
del 2000, ma sar\`a presto esteso 
per rifinire la definizione della finestra di accettanza di tutti i campioni 
di bassa statistica  (cascate a $40$\ GeV/$c$, $\Omega$\ del 1998).  
\newline
Le finestre di accettanza definite per l'analisi delle diverse particelle 
prodotte nelle collisioni Pb-Pb a 160 e 40 A GeV/$c$\  
sono mostrate in fig.~\ref{AccWinFinal}. 
I risultati delle \PgOm\ ed \PagOp\ a 160 A GeV/$c$\  
che si presenteranno in questo capitolo sono stati ottenuti  
unendo il campione del 1998 e quello del 2000, assumendo la miglior  
finestra determinata per il solo campione del 1998, ma innalzando  
il valore minimo di $p_T$\ in corrispondenza del segmento blu  
in fig.~\ref{AccWinFinal} (al valore $1.15$\ GeV/$c$). 
La finestra delle $\Lambda$\ a 40 GeV  
\`e preliminare e verr\`a utilizzata solo per determinare il rapporto  
$\frac{\bar{\Lambda}}{\Lambda}$\ a questa energia. La finestra delle  
$\Omega$\ a 40 GeV \`e stata assunta, anch'essa in modo preliminare,  
coincidente con quella delle $\Xi$\ alla stessa energia, in attesa della  
determinazione per via Monte Carlo, in quanto vi rientrano tutte le 39  
particelle raccolte.  
\section{Determinazione della centralit\`a della collisione} 
\begin{figure}[p]
\begin{center}
\includegraphics[scale=0.36]{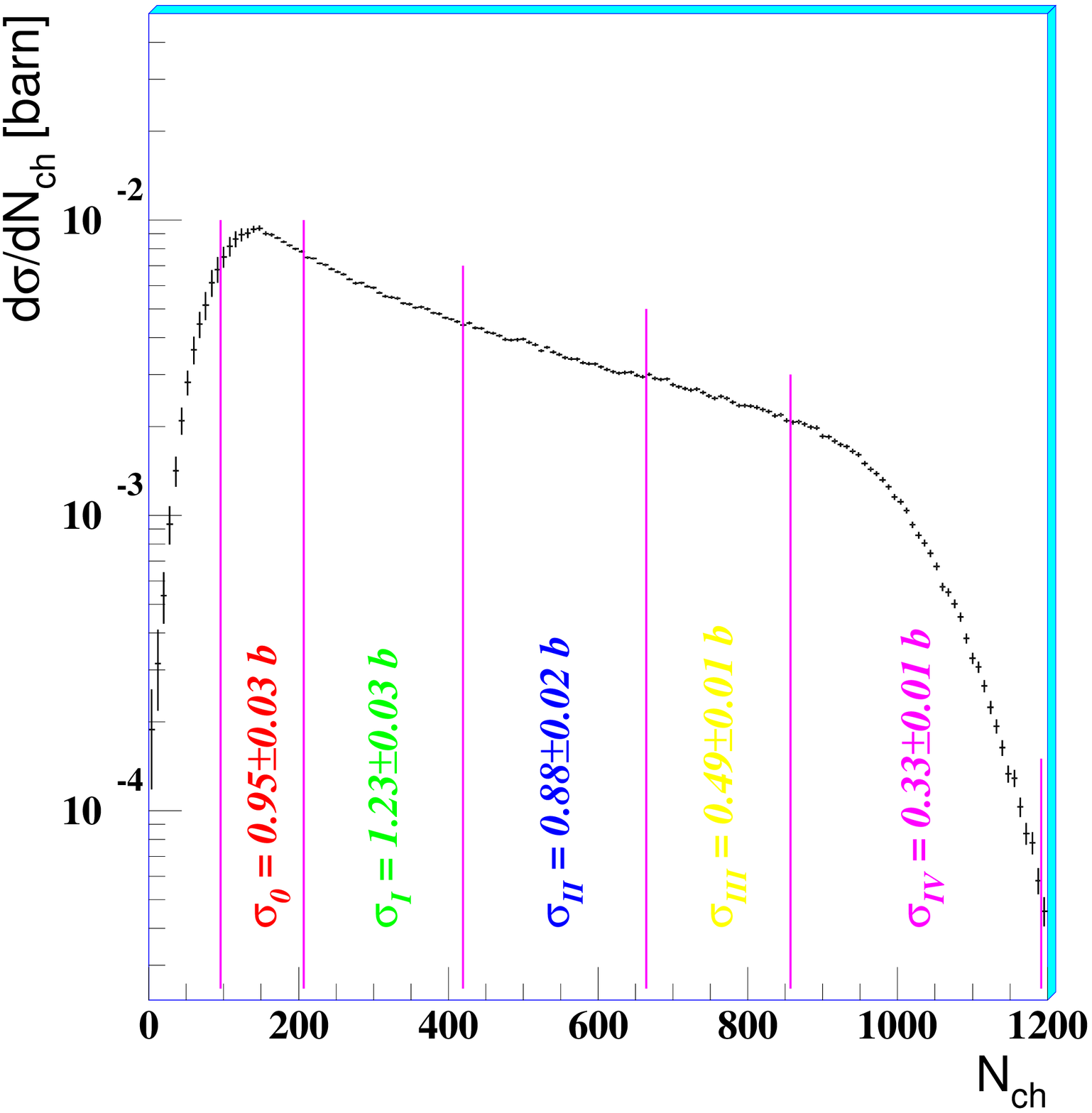}
\includegraphics[scale=0.43]{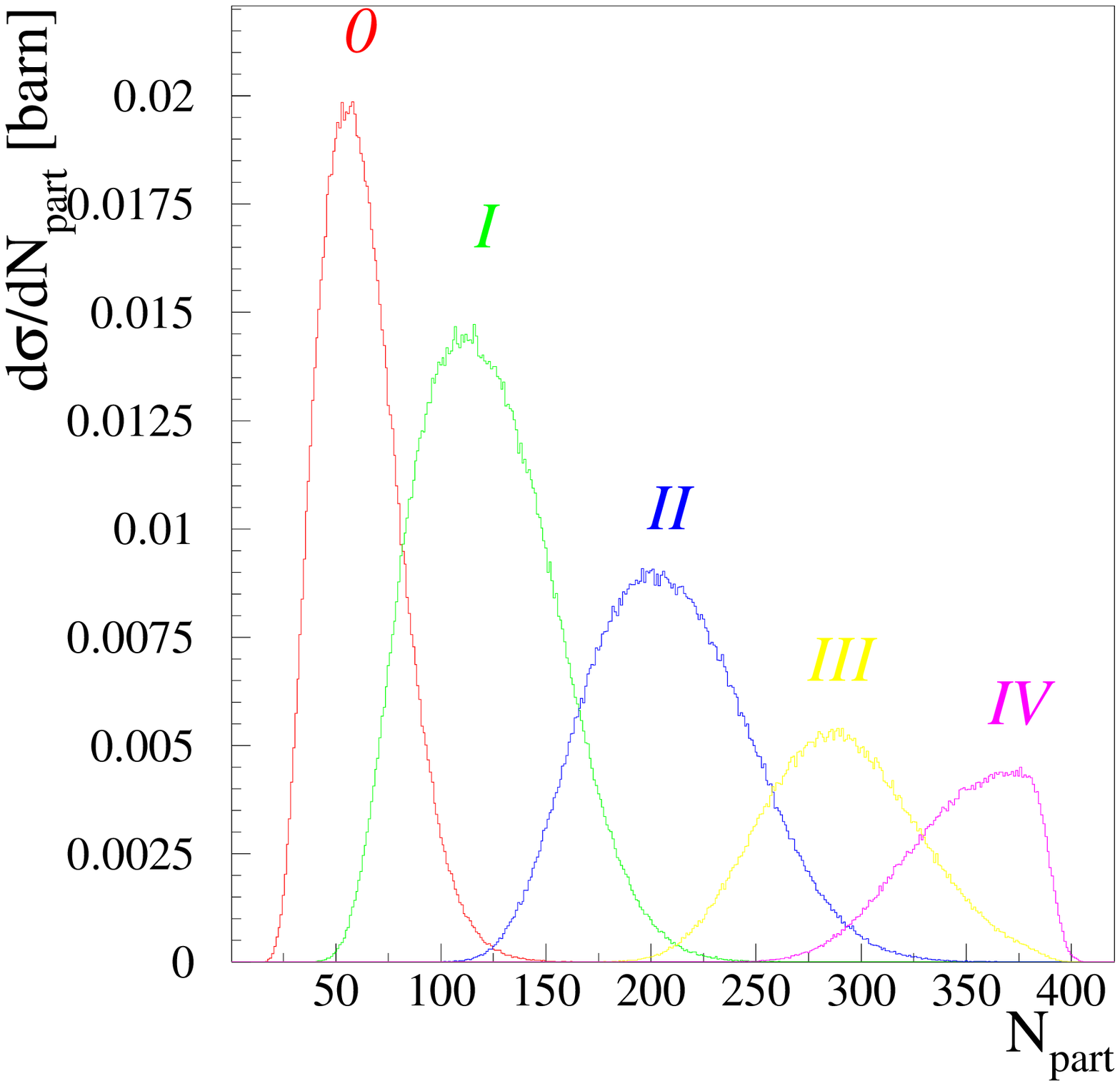}
\caption{{\em A sinistra:} 
	distribuzione della molteplicit\`a di particelle cariche nelle 
	collisioni Pb-Pb a 160 A GeV/$c$; i cinque intervalli in cui \`e divisa 
	la distribuzione definiscono le cinque classi di centralit\`a.  
        {\em A destra:} distribuzione del numero di nucleoni partecipanti nelle
        cinque classi di centralit\`a definite in termini di molteplicit\`a 
	di particelle cariche. 
%        La classe 0 corrisponde alle collisioni pi\`u periferiche, la classe IV a 
%	quelle pi\`u centrali.
         }
\label{CentralityPic}
\end{center}
\end{figure}

\begin{figure}[p]
\begin{center}
\includegraphics[scale=0.43]{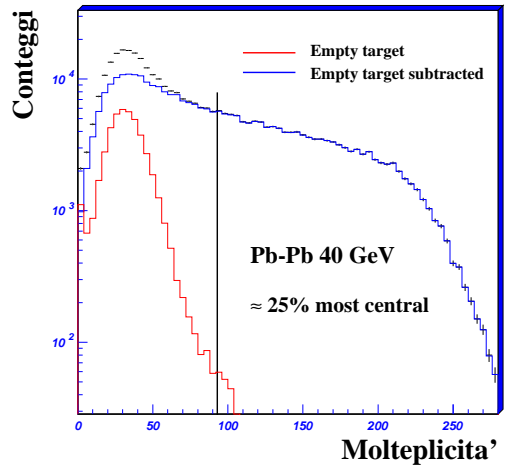}
\caption{Distribuzione della molteplicit\`a di {\em ``hit''} nell'interazione  
	 Pb-Pb a 40 A GeV/$c$\ col contributo della contaminazione di 
	 {\em ``empty target''} (curva in rosso). 
	 I risultati presentati in questa tesi si riferiscono al 25\% delle collisioni 
	 pi\`u centrali.} 
\label{multPb40}
\end{center}
\end{figure}

La descrizione dettagliata del metodo adoperato dall'esperimento NA57 
per la determinazione della centralit\`a della collisione \`e presentata   
in~\cite{WA97Centr}. Si ritiene utile riassumere qui brevemente i concetti 
pi\`u importanti ed i risultati che saranno poi utilizzati per analizzare    
la produzione di stranezza in funzione della centralit\`a della collisione.  
\newline
Nel {\em paragrafo 2.4.2} sono stati descritti i rivelatori di molteplicit\`a a 
micro-strip (MSD). La determinazione sperimentale della molteplicit\`a di 
particelle cariche prodotte in una collisione avviene a partire dal conteggio 
delle particelle cariche campionate dalle due stazioni di questi rivelatori 
negli intervalli di pseudo-rapidit\`a, rispettivamente, 
2 $<$\ $\eta$\ $<$\ 3 e  3 $<$\ $\eta$\ $<$\ 4. La copertura azimuthale \`e di 
circa il 30\% per ciascuna stazione. Ogni {\em ``strip''} fornisce 
un segnale analogico proporzionale al rilascio di energia in essa; la molteplicit\`a 
di {\em ``hit''} nei rivelatori viena calcolata con un algoritmo che tiene conto, 
oltre che del numero di {\em cluster}~\footnote{Un cluster \`e definito come l'insieme 
delle {\em ``strip''} contigue sopra soglia.}, anche dell'energia totale depositata,  
della sottrazione del rumore elettronico e di una piccola correzione per i doppi {\em ``hit''}. 
Si valuta quindi la contaminazione delle interazioni che avvengono in aria o negli altri 
materiali lungo la linea di fascio, considerando i {\em run} speciali eseguiti senza 
bersaglio (contaminazione di {\em ``empty target''});   
tale contaminazione viene sottratta dalla distribuzione sperimentale. 
Si considera ulteriormente una correzione dovuta ai raggi $\delta$\ 
(pi\`u precisamente alla coda di pi\`u alta energia di questi) 
prodotti principalmente nel passaggio in aria degli ioni che 
non hanno interagito nel bersaglio, 
dell'ordine dell'1\%. 
Per ottenere la molteplicit\`a di particelle cariche 
nell'intervallo di pseudo-rapidit\`a 2 $<$\ $\eta$\ $<$\ 4, si corregge infine, con 
una simulazione Monte Carlo, la distribuzione della molteplicit\`a di {\em ``hit''} 
per accettanza geometrica, risposta del rivelatore (cio\`e per efficienza, 
doppi {\em ``hit''}, condivisione di carica tra {\em ``strip''} contigue), interazioni 
secondarie e conversioni di raggi $\gamma$. 
\newline
Nel grafico di sinistra della fig.~\ref{CentralityPic} \`e mostrata la distribuzione 
di particelle cariche nelle collisioni Pb-Pb a 160 A GeV/$c$\ 
calcolata con il metodo appena descritto.  
In fig.~\ref{multPb40} \`e mostrata la distribuzione di molteplicit\`a 
di {\em ``hit''} misurata in collisioni Pb-Pb a 40 A GeV/$c$; il contributo 
dell'{\em empty target} corrisponde alla curva in rosso nella stessa 
fig.~\ref{multPb40}.  
Per entrambe le distribuzioni, la soppressione di eventi a bassa
molteplicit\`a \`e dovuta al trigger di centralit\`a operato dagli scintillatori
a petali.
\newline
Il trigger d'interazione $CU1$\ \`e studiato per selezionare i soli urti centrali tra 
i nuclei di piombo; la frazione della sezione d'urto anelastica totale $\sigma_I$\  
cui gli eventi selezionati dal trigger corrispondono pu\`o essere calcolata tramite 
la relazione: 
\begin{equation}
\frac{\sigma_{trig}^{exp}}{\sigma_I}=\frac{\lambda_I}{L}
 \left[ <\frac{CU1}{BEAM\, \overline{DT}}>  
%        <\frac{INT\, \overline{DT}}{CU1}> 
                                           \right]
\label{FracSect}
\end{equation}
dove $L$\ \`e lo spessore del bersaglio, $\lambda_I$\ \`e la lunghezza d'interazione 
anelastica ed il termine tra parentesi quadre contiene le medie, fatte su tutta la 
presa dati, dei conteggi dei corrispondenti livelli di trigger e misura la frequenza 
di interazione. Esso rappresenta il rapporto tra il numero di eventi 
selezionati ($CU1$) e quello degli ioni del fascio contati escludendo il tempo morto 
($BEAM\, \overline{DT}$). Risulta che la sezione d'urto 
selezionata $\sigma_{trig}^{exp}$\ vale $4.15 \pm 0.11$\ barn nelle collisioni 
Pb-Pb a 160 A GeV/$c$\ e $4.07 \pm 0.16$\ barn in quelle Pb-Pb a 40 A GeV/$c$,
corrispondendo quindi a circa il 60\% della sezione d'urto 
totale anelastica $\sigma_I$\ per entrambe le energie.  
%sia nelle collisioni a 160 GeV/$c$\ per nucleone sia in quelle a 40  GeV/$c$\ per nucleone.  
\newline
La distribuzione di molteplicit\`a viene descritta nell'ambito del modello dei 
nucleoni partecipanti ({\em cfr. paragrafo 1.4.2}). 
Si assume che %il valor medio della 
la molteplicit\`a di particelle 
cariche sia proporzionale 
(in media) 
al numero di nucleoni partecipanti, calcolato con il modello di Glauber  
({\em cfr. paragrafo 1.3.3} ed appendice A): 
\[ <N_{ch}> = q N_{part} \, .\]
Fissato il parametro d'impatto ${\rm {\bf b}}$\ dei due nuclei, il valor medio del 
numero di nucleoni partecipanti \`e fornito dall'eq.~\ref{1.30} e la dispersione 
del numero di particelle cariche prodotte \`e calcolabile come:
\begin{equation}
\sigma^2_{N_{ch}}=q^2\sigma^2_{N_{part}}+N_{part}\sigma^2_q+(\sigma^{exp}_{N_{ch}})^2
\label{dispersion}
\end{equation}
dove $\sigma_{N_{part}}$\ \`e la dispersione del numero di nucleoni partecipanti 
per fissato parametro d'impatto --- quantit\`a che viene valutata con una 
simulazione Monte Carlo del modello di Glauber --- 
e $\sigma_q$\ quella del numero di particelle cariche per nucleone partecipante. 
La dispersione $\sigma_q$\ viene valutata dalle informazioni sulle collisioni 
p-p e risulta proporzionale alla molteplicit\`a di particelle cariche 
(cio\`e $\sigma_q=aq$)~\cite{Alner}; all'energia dell'SPS e per $2\, <\, \eta\,< \,4$ 
risulta $a\simeq1$.  
L'ultimo addendo nel secondo membro dell'eq.~\ref{dispersion} \`e la dispersione sperimentale 
dovuta al campionamento nella misura di $N_{ch}$.
\newline
La sezione d'urto differenziale per collisioni anelastiche Pb-Pb, che 
viene approssimata ai dati sperimentali con una procedura di {\em ``best fit''}, assume 
la seguente espressione:
\begin{equation}
\frac{{\rm d}\sigma_{an}^{Pb-Pb}}{{\rm d}N_{ch}}(N_{ch})=
\int{ {\rm d}{\rm {\bf b}}[1-P(0,{\rm {\bf b}})] \times 
 \frac{1}{\sqrt{2\pi}\sigma_{N_{ch}}} 
 \exp \left\{ -\frac{\left[N_{ch}-q<N_{part}({\rm {\bf b}})>\right]^2}{2\sigma^2_{N_{ch}}} 
 \right\} }
\label{WNMfitFunction}
\end{equation}
dove il termine $[1-P(0,{\rm {\bf b}})] $\ \`e la probabilit\`a di avere almeno una 
collisione anelastica (nucleone-nucleone) con parametro di impatto ${\rm {\bf b}}$.  
Nella procedura di {\em ``best fit''} \`e possibile ricavare sia la costante 
di proporzionalit\`a $q$, pari al numero medio di particelle cariche prodotte 
per partecipante, sia la sezione d'urto selezionata dal trigger 
$\sigma_{trig}^{fit}$~\footnote{La sezione d'urto pu\`o anche essere fissata a priori 
ponendola pari a quella sperimentale ($\sigma_{trig}^{exp}$).}.  
Per le collisioni 
Pb-Pb a 160 A GeV/$c$\ risulta: $q=2.62\pm0.03^{+0,02}_{-0.05}$, 
$\sigma_{trig}^{fit}=4.42\pm0.14^{+0.15}_{-0.07}$\ barn; per quelle a 40 A GeV/$c$: 
$q=1.43\pm0.01$, $\sigma_{trig}^{fit}= 3.6\pm0.2$\ barn. 
In queste espressioni, il primo errore \`e 
quello di origine statistica, il secondo (se presente) di tipo sistematico.  
A 40 A GeV/$c$\ si \`e dunque trovata  
una differenza del 14\% tra la determinazione di $\sigma_{trig}$\ sperimentale 
e quella ricavata dal {\em ``best fit''} alla distribuzione di molteplicit\`a con il modello dei 
nucleoni partecipanti. \`E attualmente in corso una indagine per appurare l'origine 
di questa discrepanza.  
\newline
Nelle collisioni Pb-Pb a 160 A GeV/$c$\ sono state definite cinque classi 
di centralit\`a, corrispondenti a cinque intervalli contigui di molteplicit\`a 
di particelle cariche ($N_{ch}$), come mostrato in fig.~\ref{CentralityPic}; 
le quattro classi pi\`u centrali corrispondono 
con buona approssimazione alle quattro classi di centralit\`a definite 
nell'esperimento WA97 ({\em cfr. paragrafo 1.5.3}) e sono state indicate 
con gli stessi numeri romani $I$, $II$, $III$ e $IV$, dove la $IV$ classe 
\`e quella delle collisioni pi\`u centrali. In aggiunta, \`e stato possibile definire 
una nuova classe pi\`u periferica, cui corrisponde un numero medio  
di nucleoni partecipanti $<N_{part}>=62$, che verr\`a indicata col numero 
arabo $0$. Le distribuzioni del numero di nucleoni partecipanti per le diverse 
classi di centralit\`a sono mostrate nella parte destra della fig.~\ref{CentralityPic}. 
%\newline
In tab.~4.4 sono riassunti i risultati del modello dei nucleoni partecipanti   
per le diverse classi di centralit\`a nelle collisioni Pb-Pb a 160 A GeV/$c$.
\begin{table}[h]
\begin{center}
\begin{tabular}{|c|c|c|lc|c|} \hline
Classe& $N_{ch}$ & $\sigma^{exp}_{trig}$\ [barn] & $<N_{part}>$\ &{\em Larghezza}& $<b>$\ [fm]\\ \hline
 0   & $ 96.0<N_{ch}<206.6 $ & $ 0.95\pm0.03 $ & $ 62\pm4$ & $^{+22}_{-26}$ & $10.43\pm0.14$ \\
 I   & $206.6<N_{ch}<419.8 $ & $ 1.23\pm0.03 $ & $121\pm4$ & $^{+37}_{-42}$ & $ 8.57\pm0.12$ \\
 II  & $419.8<N_{ch}<664.0 $ & $ 0.88\pm0.02 $ & $209\pm3$ & $^{+41}_{-49}$ & $ 6.30\pm0.09$ \\
 III & $664.0<N_{ch}<857.1 $ & $ 0.49\pm0.01 $ & $290\pm2$ & $^{+38}_{-43}$ & $ 4.22\pm0.06$ \\
 IV  & $857.1<N_{ch}<1191.7$ & $ 0.33\pm0.01 $ & $349\pm1$ & $^{+41}_{-31}$ & $ 2.46\pm0.02$ \\ \hline
\end{tabular}
\end{center}
\caption{Intervalli di molteplicit\`a di particelle cariche ($N_{ch}$) e sezione  
	 d'urto sperimentale $\sigma^{exp}_{trig}$\  
	 dopo la sottrazione della contaminazione di {\em ``empty target''} 
	 nelle cinque classi di centralit\`a definite 
	 nell'interazione Pb-Pb a 160 A GeV/$c$. Per ciascuna classe 
	 \`e riportato il numero medio di nucleoni partecipanti alla collisione $<N_{part}>$,  
	 la {\em larghezza} del picco a met\`a altezza a destra ($+$) ed a sinistra ($-$)  
	 rispetto al valore $<N_{part}>$, ed il parametro d'impatto medio $<b>$.
\label{tab4.4}}
\end{table}
\newline 
I risultati che saranno presentati per le collisioni Pb-Pb a 40 A GeV/$c$\ 
si riferiscono alle collisioni pi\`u centrali che riguardano circa il 25 \% 
delle sezione d'urto anelastica $\sigma_I$\ e corrispondono alla parte di distribuzione  
di molteplicit\`a di {\em ``hit''} a destra del segmento verticale in nero nella 
fig.~\ref{multPb40}. 
Il numero medio di nucleoni partecipanti nella regione di centralit\`a cos\`i  
selezionata risulta pari a $<N_{part}> = 262 \pm 17 $. 
L'errore associato a questa determinazione, di tipo sistematico, 
\`e pari alla differenza tra i risultati di due differenti metodi di calcolo:  
nel primo  la frazione di sezione d'urto si ricava dalla procedura  
di {\em ``best fit''} alla distribuzione sperimentale e corrisponde 
quindi ad un ulteriore parametro libero,  
nel secondo tale frazione \`e quella misurata sperimentalmente, secondo  
l'eq.~\ref{FracSect}; in questo secondo metodo si prescinde quindi dalla 
relazione di stretta proporzionalit\`a tra la molteplicit\`a di particelle 
cariche ed il numero di nucleoni partecipanti.  
%\`e viene imposta nel calcolo dei nucleoni partecipanti.  
Si terr\`a conto della presenza di questo errore sistematico  
nella discussione dei risultati nel prossimo capitolo.  
\section{Calcolo della produzione di particelle strane} 
Una volta determinata la correzione da apportare a ciascuna particella 
identificata e la finestra di accettanza fiduciale, il tasso di produzione di 
una data specie all'interno della regione cinematica selezionata pu\`o  
essere calcolato sommando tali correzioni e normalizzando il risultato al 
numero di eventi raccolti. 
In formule: 
\begin{equation}
Y_{Acc\,Win}^{\bf j} = \frac{\sum_i w_i}{\overline{N}^{\bf j}}
\label{YieldInWind}
\end{equation}
dove la sommatoria delle correzioni $w_{i}$\ (fornite dall'eq.~\ref{Weight}) \`e 
estesa a tutte le particelle di un dato tipo che cadono all'interno delle corrispondenti 
regioni cinematiche e che provengono da eventi di una data classe ${\bf j}$\ di molteplicit\`a:  
ad esempio, per la IV classe di centralit\`a delle collisioni Pb-Pb a 160 A GeV/$c$,  
${\bf j} = IV $\ e  si considerando solo gli eventi per cui $857.1<N_{ch}<1191.7$\   
({\em cfr.} tabella 4.4). Nell'eq.~\ref{YieldInWind}, infine, il denominatore  
$\overline{N}^{\bf j}$\ \`e pari al numero di {\em trigger} di livello pi\`u alto 
$CU1$\ ({\em cfr. paragrafo 2.5.1}), corretto per la contaminazione di eventi 
di {\em ``empty target''} nel modo di seguito discusso. 
\newline
Durante la presa dati, al termine di ciascun {\em ``burst''} vengono registrate le  
informazioni dei contatori di fascio ({\em cfr. paragrafo 2.5.1}). Esse si riferiscono  
a quantit\`a integrate su un gran numero di proiettili incidenti, 
ed a partire da queste informazioni \`e possibile ad esempio calcolare il numero 
di ioni incidenti ($BEAM$) ed il numero di trigger di tipo $CU1$\  durante 
ciascun {\em ``burst''}. Tipicamente, nelle interazioni Pb-Pb a 160 A GeV/$c$\ in un 
{\em ``burst''} 
si contano $1.5\times10^5$\ $BEAM$\ e $2000$\ trigger $CU1$. Per ciascuna presa dati, 
viene anche estratto un campione rappresentativo (selezionando gli eventi uno 
ogni duecento), che non \`e necessario processare con ORHION, ma dei cui eventi si 
calcola la molteplicit\`a di particelle cariche $N_{ch}$. Tali eventi sono dunque 
un campione rappresentativo di tutti quelli selezionati dalla condizione $CU1$, ed 
a partire da essi \`e possibile calcolare la frazione di eventi della classe 
di molteplicit\`a ${\bf j}$, pari a   
\[ \frac{N^{\bf j}_{ref\, Pb\,1\%}}{\sum_{\bf j}{N^{\bf j}_{ref\, Pb\,1\%}}} \, , \]
dove la somma al denominatore si estende su tutte le centralit\`a raccolte 
con la condizione di trigger sperimentale $CU1$\ (il denominatore 
\`e semplicemente il numero di eventi totali nel campione di riferimento, il numeratore 
\`e pari al numero di eventi nel campione di riferimento ma della sola classe ${\bf j} $, 
entrambi ottenuti con il bersaglio di piombo).   
Servendosi 
poi dei {\em run} speciali eseguiti senza bersaglio (si parla di eventi di tipo 
{\em ``empty target''}), si considera il numero $N^{\bf j}_{Emp \, tgt}$\ di eventi 
selezionati dal trigger $CU1$, se pur in assenza di bersaglio, della classe di  
molteplicit\`a ${\bf j} $. Per calcolare la contaminazione {\em assoluta} 
delle interazioni non avvenute all'interno del bersaglio, durante la presa dati 
{\em con} il bersaglio, si \`e  moltiplicata la quantit\`a $N^{\bf j}_{Emp \, tgt}$\  
per il rapporto $ \frac{\sum{{BEAM}^{Pb\, 1\%}}}{\sum{{BEAM}^{Emp \, tgt}}} $\ 
tra il numero di $BEAM$\ contati durante la presa dati degli eventi col bersaglio 
(cui corrisponde il numero di trigger $\sum{CU1^{Pb\, 1\%}}$) 
e quello in assenza di bersaglio (cui corrisponde il numero di trigger 
$N^{\bf j}_{Emp \, tgt}$\ per la generica classe ${\bf j}$).   
Il fattore di correzione $\overline{N}^{\bf j}$\ ---  
che tiene dunque conto del numero di trigger raccolti per la classe ${\bf j} $\ 
e che \`e corretto per la contaminazione  
di eventi in cui l'interazione non ha avuto luogo nel bersaglio --- 
assume pertanto l'epressione:  
\begin{equation}
\overline{N}^{\bf j}=
\left[\sum{CU1^{Pb\, 1\%}} \cdot 
  \frac{N^{\bf j}_{ref\, Pb1\%}}{\sum_{\bf j}{N^{\bf j}_{ref\, Pb1\%}}}\right]  
 - \left[ N^{\bf j}_{Emp \, tgt}  
  \cdot \frac{\sum{{BEAM}^{Pb\, 1\%}}}{\sum{{BEAM}^{Emp \, tgt}}} \right]
\label{Normalization}
\end{equation}
\newline
L'errore sulla misura del tasso di produzione entro la finestra di accettanza \`e 
calcolato come 
\begin{equation}
\delta Y_{Acc\,Win}^{\bf j} =
 \frac{\sqrt{\sum_{i}{w_i^2+(\delta w_i)^2}}}{\overline{N}^{\bf j}} \, ,
\label{ErrYield}
\end{equation}
sommando cio\`e in quadratura l'errore statistico legato al numero di particelle presenti e 
l'errore sulla correzione dato dall'eq.~\ref{ErrWeight}; l'incertezza legata al numero 
di eventi selezionati risulta invece trascurabile. 
\newline
I valori calcolati per il tasso di produzione nella finestra di accettanza 
delle diverse particelle 
(eq.~\ref{YieldInWind}) e per l'errore ad esso associato (eq.~\ref{ErrYield}), 
non sono qui riportati ma verranno utilizzati nel {\em paragrafo 4.8},  
per calcolare una quantit\`a pi\`u significativa che non dipenda dalla particolare   
forma della finestra di accettanza in cui la misura \`e stata eseguita.   
\section{Distribuzioni di massa trasversa}
\begin{figure}[p]
\begin{center}
\includegraphics[width=0.45\textwidth,height=0.365\textheight]{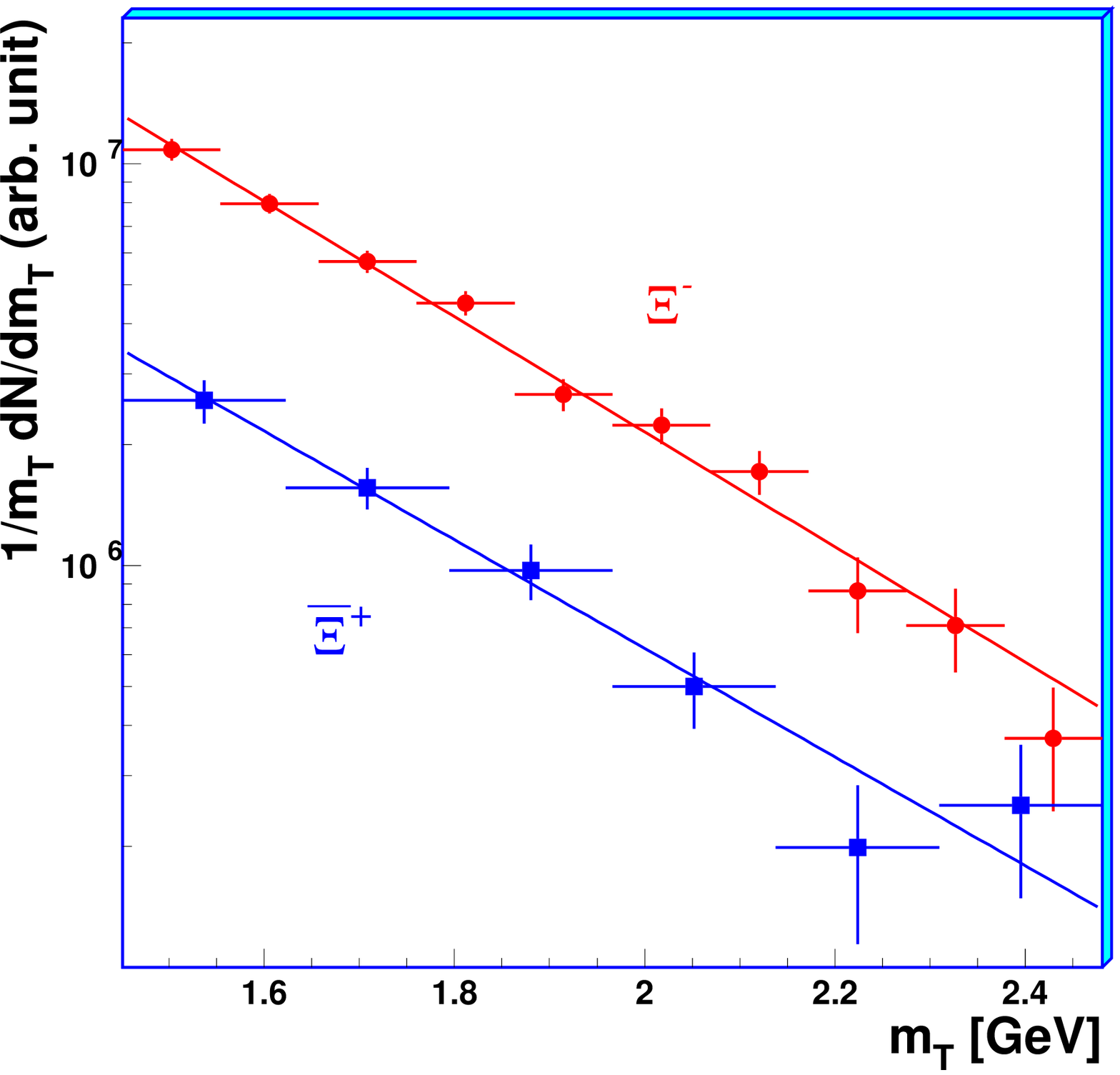}
\includegraphics[scale=0.37]{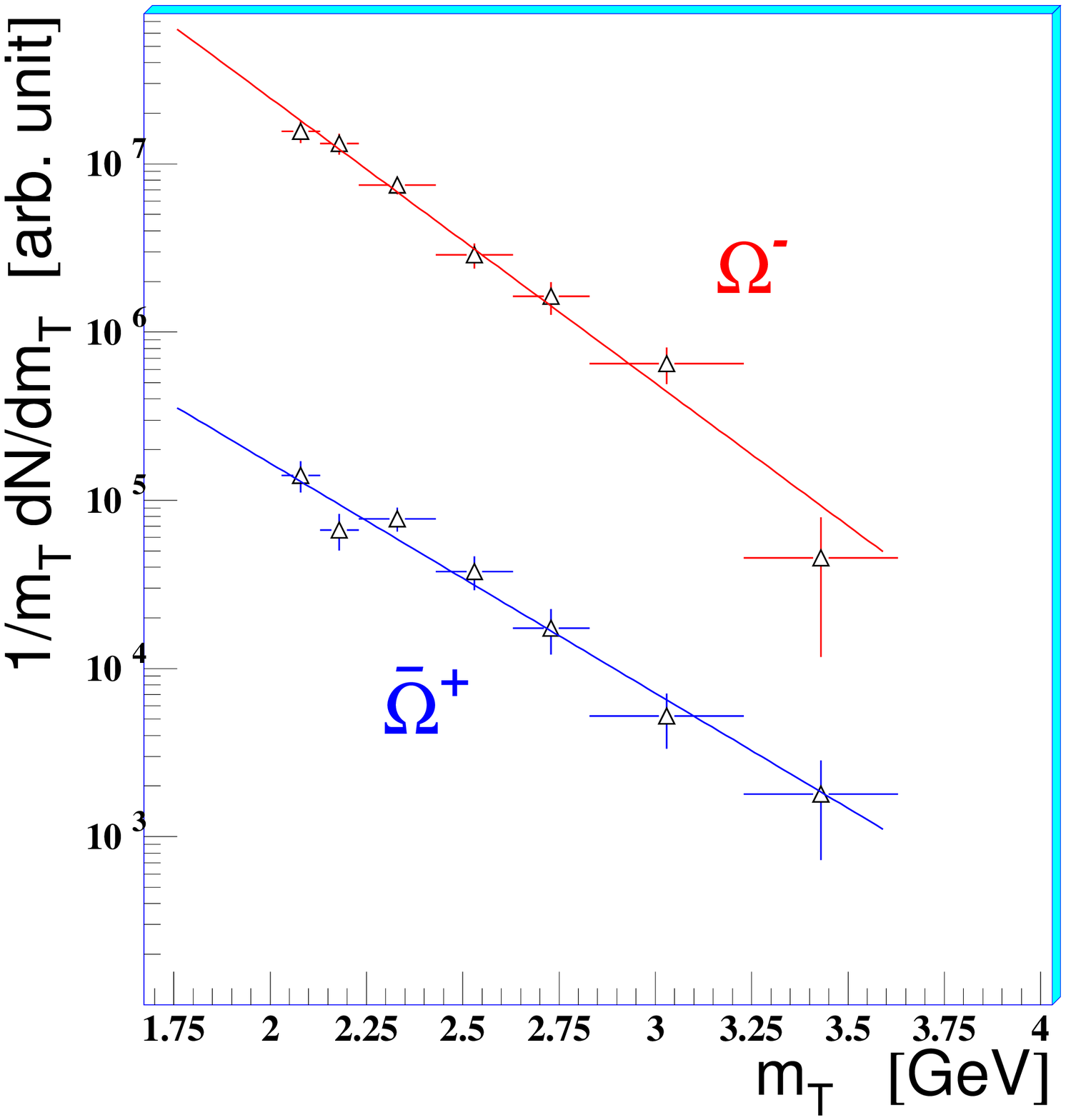}\\
\includegraphics[scale=0.37]{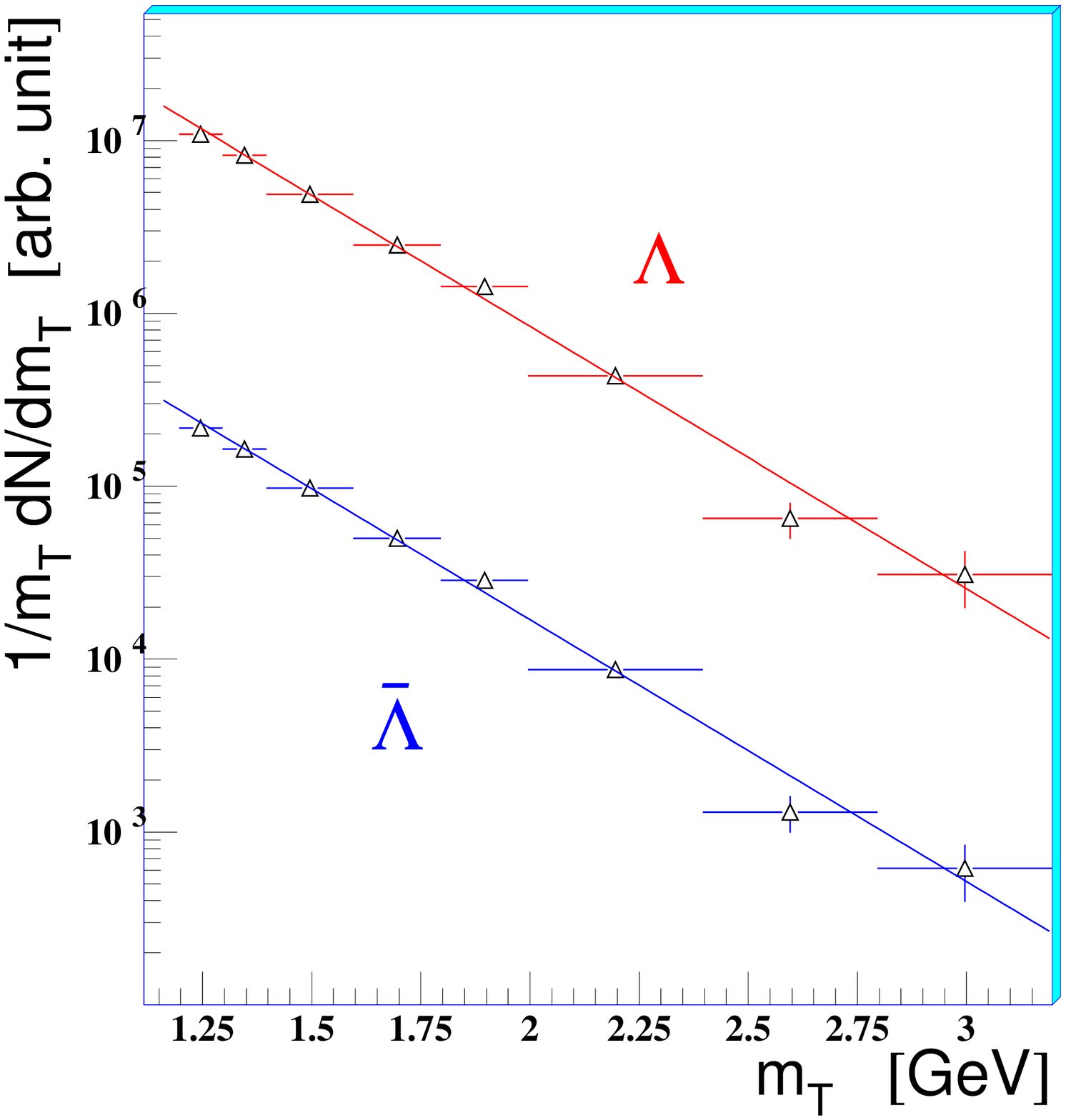}
\includegraphics[scale=0.37]{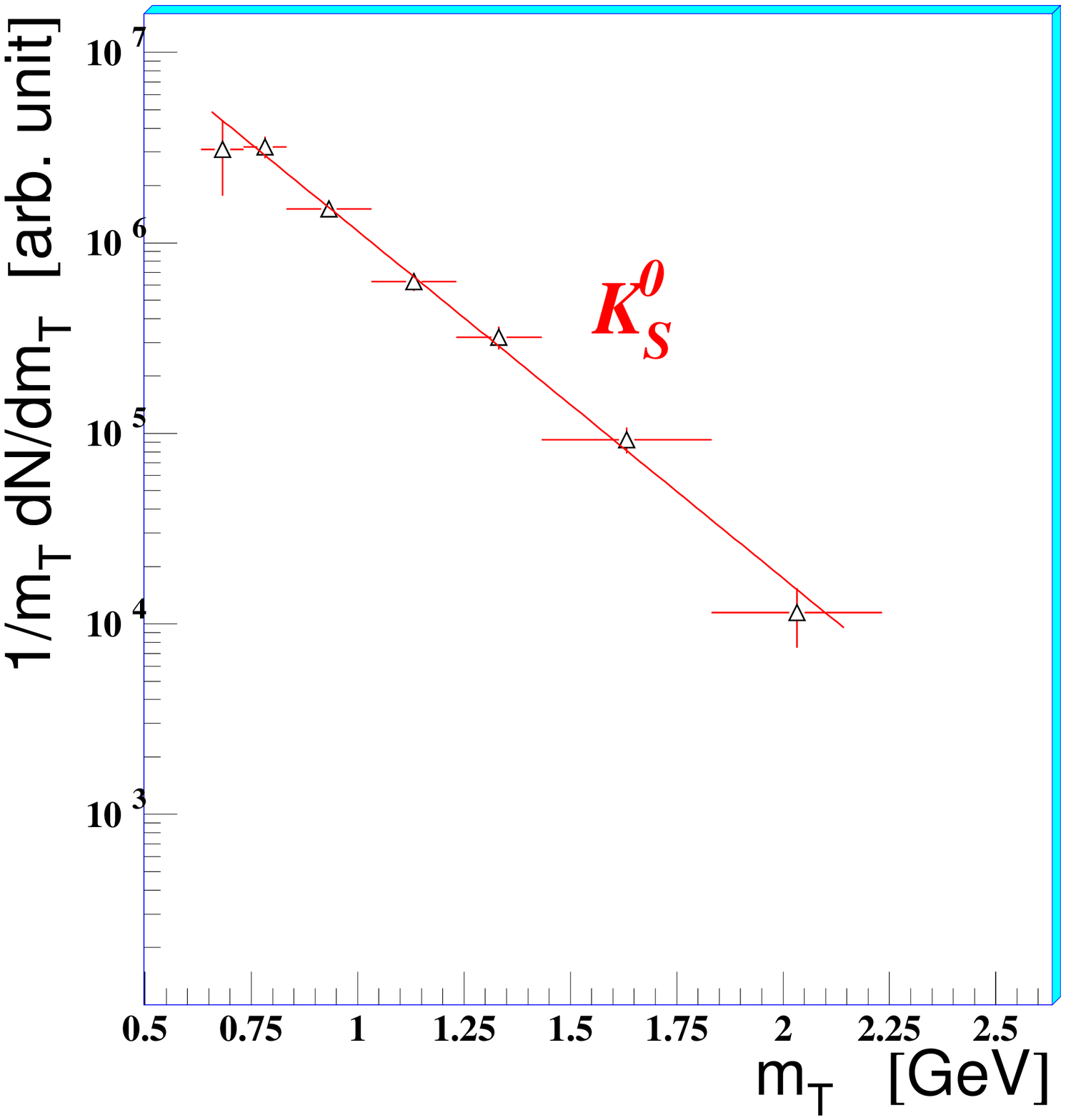}
\caption{Spettri di massa trasversa per le diverse particelle strane 
	 studiate dall'esperimento NA57 nelle collisioni Pb-Pb a 
	 160 A GeV/$c$. Le \PgOm\ e \PagOp\ sono ottenute unendo la 
	 statistica dell'anno 1998 a quella del 2000. Tutte le altre specie 
	 provengono dal solo campione del 1998. I risulati per i \PKzS\ 
	 sono ancora preliminari.}
\label{mTspectra160}
\end{center}
\end{figure}

Entro la finestra di accettanza determinata per ciascuna particella, 
\`e stato eseguito il {\em ``best fit''} della distribuzione sperimentale, 
corretta con la procedura descritta nel {\em paragrafo 4.3}, alla 
funzione:
\begin{equation}
\frac{{\rm d}^2N(m_T,y)}{{\rm d}m_T{\rm d}y} =f(y,m_T) = 
 \mathcal{A} m_T \exp\left[-\frac{m_T}{T_{a}}\right]
\label{DoubleDiff}
\end{equation}
Il {\em ``best fit''} \`e stato eseguito con il metodo della 
massima verosimiglianza, considerando la probabilit\`a congiunta 
di tutte le misure: 
\[ \mathcal{L}=\prod_{\substack{i=1}}^{\substack{N}}{f(y_i,{m_T}_i)}^{w_i} \, ,\]  
dove l'esponente $w_i$\ \`e pari al ``peso'' relativo alla $i$-esima particella 
misurata, di rapidit\`a $y_i$\ e massa trasversa ${m_T}_i$. In tal modo la singola 
misura di peso  $w_i$\ \`e trattata come $w_i$\ misure indipendenti di peso unitario.  
Conviene minimizzare l'opposto del logaritmo della funzione di verosimiglianza 
$\mathcal{L}$, in modo da trasformare il prodotto in una somma di logaritmi in cui 
gli esponenti $w_i$\ diventano fattori moltiplicativi:    
\begin{equation}
F = - \ln{\mathcal{L}} = 
 - \sum_{\substack{i=1}}^{\substack{N}}{w_i \ln{\left\{ f(y_i,{m_T}_i) \right\}}}
\label{Likely1}
\end{equation}
Si ricorda tuttavia che nel metodo della massima verosimiglianza, 
la funzione densit\`a di probabilit\`a $f$, da cui si assume che 
le misure sperimentali siano state campionate, deve essere normalizzata 
ad una costante ($I(T_a)=\iint_{Acc Win}{f(y,{m_T}) {\rm d}y {\rm d}m_T} = {\rm cost}$). 
Con questo metodo non \`e pertanto possibile ricavare anche la costante moltiplicativa  
$\mathcal{A}$\ che, nella procedura di minimizzazione, viene variata di volta in volta 
per soddisfare la condizione di normalizzazione.  Sviluppando l'eq.~\ref{Likely1} 
e volendo normalizzare correttamente la funzione densit\`a di probabilit\`a, si 
ricava infine l'espressione:  
\begin{equation}
F = \frac{2}{\overline{w}} \sum_{\substack{i=1}}^{\substack{N}}{
 w_i \cdot \left[ \ln\{{m_T}_i\} - \frac{{m_T}_i}{T_a} - \ln\{I(T_a)\} \right]
    }
\label{Likely2}
\end{equation}
che \`e stata  utilizzata per la minimizzazione.   
Nell'eq.~\ref{Likely2} la costante di normalizzazione $I(T_a)$\ \`e pari 
all'integrale 
%analitico 
della funzione $f$\ con $\mathcal{A}=1$\ ed il 
fattore $\frac{2}{\overline{w}}$, dove $\overline{w}$\ \`e il valor medio dei 
pesi, \`e introdotto per valutare in modo corretto l'errore 
del parametro $T_a$\ del {\em fit}.  
\newline
I risultati del {\em ``best fit''} ottenuti con il metodo della massima 
verosimiglianza appena descritto sono stati sistematicamente controllati, 
per confronto con quelli ottenuti dal {\em ``best fit''} con il metodo del 
$\chi^2$. 
I valori del $\chi^2$\ ridotto risultano distribuiti attorno ad uno, 
confermando la validit\`a dell'ipotesi di distribuzione esponenziale in 
$m_T$, secondo l'eq.~\ref{DoubleDiff}. 
\newline
In fig.~\ref{mTspectra160} sono mostrate le proiezioni in $m_T$\  
della distribuzione sperimentale $\frac{{\rm d}^2N(m_T,y)}{{\rm d}m_T{\rm d}y}$\ 
delle diverse particelle strane studiate nelle collisioni Pb-Pb a 160 A 
GeV/$c$\footnote{
Nell'eseguire la proiezione in $m_T$\ si tiene conto della forma della finestra  
di accettanza, applicando un fattore moltiplicativo a ciascun canale di $m_T$\
pari all'inverso dell'estensione in rapidit\`a della finestra di accettanza
in quel canale di $m_T$.}.  
Le rette sovrapposte sono proporzionali alla funzione esponenziale $m_T\exp(-m_T/T_a)$, 
con il parametro $T_a$\ ricavato dalla procedura di {\em ``best fit''}.  
Questi grafici danno ulteriore fiducia nella validit\`a dei risultati 
ottenuti.  
\newline 
In tab.~\ref{tab4.5} sono riassunti i risultati del {\em ``best fit''} 
per il parametro $T_a$\ delle particelle strane in collisioni Pb-Pb 
a 160 A GeV/$c$. Questi risultati saranno oggetto di discussione nel 
prossimo capitolo insieme a quelli relativi alle collisioni 
Pb-Pb a 40 A GeV/$c$, di seguito presentati.   
\begin{table}[h]
 \begin{center}
 \begin{tabular}{|c|c||c|c|c|c|c|}
 \hline
   & {\bf 0-IV} &
   {\bf 0} & {\bf I} &
   {\bf II} & {\bf III}
   & {\bf IV}   \\ \hline
   {\bf \PgL} & {289 $\pm$\ 7} &
   {238 $\pm$\ 19} & {276 $\pm$\ 13} &
   {282 $\pm$\ 12} & {308 $\pm$\ 15} & {306 $\pm$\ 16} \\
   {\bf \PagL} & {287 $\pm$\ 6} &
   {271 $\pm$\ 19} & {261 $\pm$\ 11} &
   {286 $\pm$\ 11} & {319 $\pm$\ 15} & {296 $\pm$\ 15} \\
%
%   {\bf \PKzS} & {235 $\pm$\ 8} & 
%   {188 $\pm$\ 19} &  {285 $\pm$\ 22} & {186 $\pm$\ 16} & 
%   {266 $\pm$\ 21} &  {228 $\pm$\ 19} \\
   \cline{3-7}
   {\bf \PKzS} & {235 $\pm$\ 8} &
   \multicolumn{5}{|c}{} \\
   {\bf \PgXm} & {303 $\pm$\ 11} &
   \multicolumn{5}{|c}{} \\
   {\bf \PagXp} & {321 $\pm$\ 23} &
   \multicolumn{5}{|c}{} \\
   {\bf \PgOm} & {280 $\pm$\ 16} &
   \multicolumn{5}{|c}{} \\
   {\bf \PagOp} & {324 $\pm$\ 29} &
   \multicolumn{5}{|c}{} \\
   \cline{1-2}
\end{tabular}
\end{center}
\caption{Parametro $T_a$\ (detto di {\em ``inverse slopes''}), espresso in MeV, ottenuto dal 
 {\em ``best fit''} con il metodo della massima verosimiglianza in tutto l'intervallo 
 ({\bf 0-IV}) e per le diverse classi di centralit\`a nelle collisioni Pb-Pb a 160 A GeV/$c$. 
 Gli errori sono solo quelli statistici.  
  \label{tab4.5}}
\end{table}
\subsection{Cascate in Pb-Pb a 40 A GeV/$c$}
\begin{figure}[tb]
\begin{center}
\includegraphics[scale=0.37]{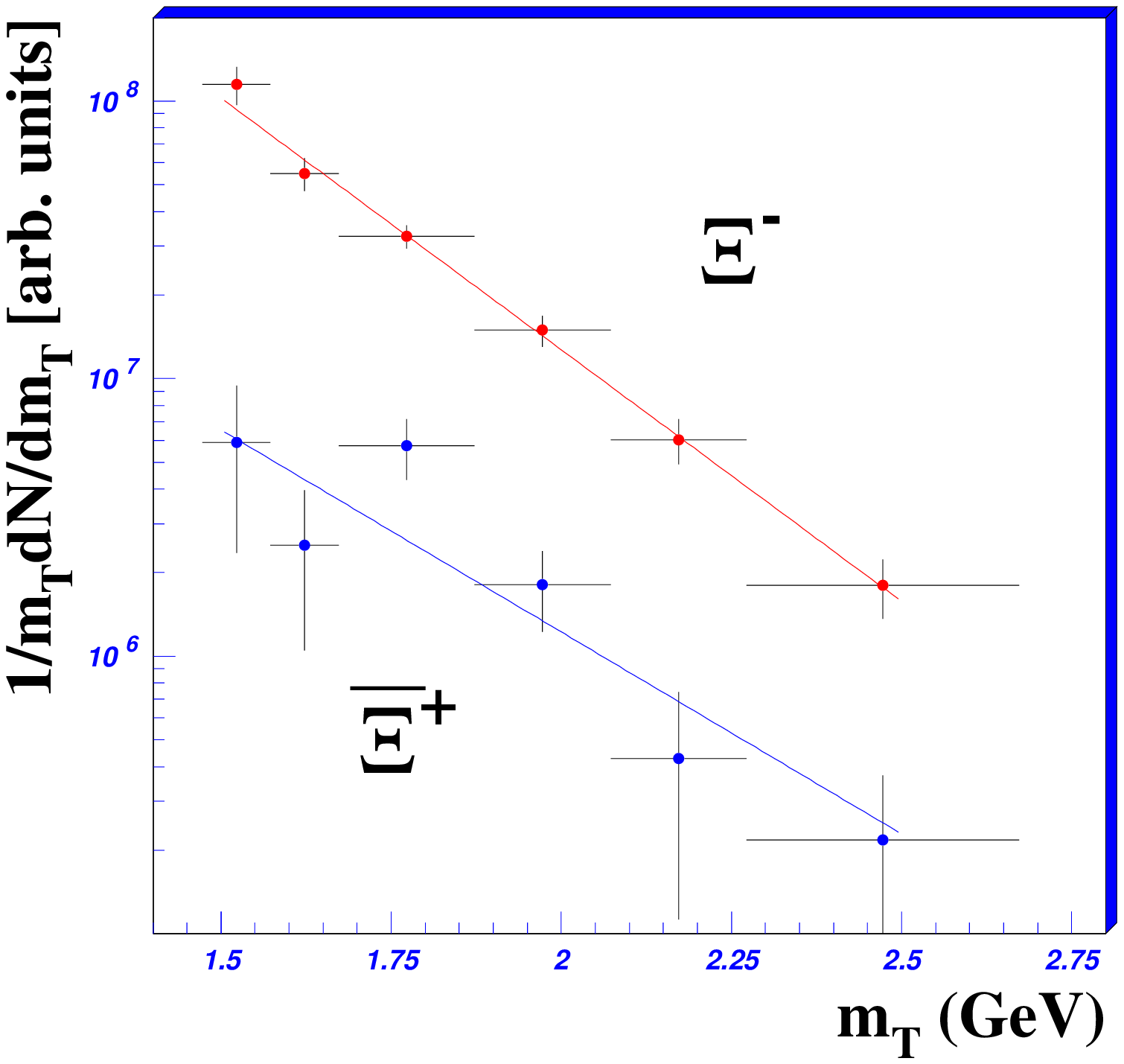}
\includegraphics[scale=0.37]{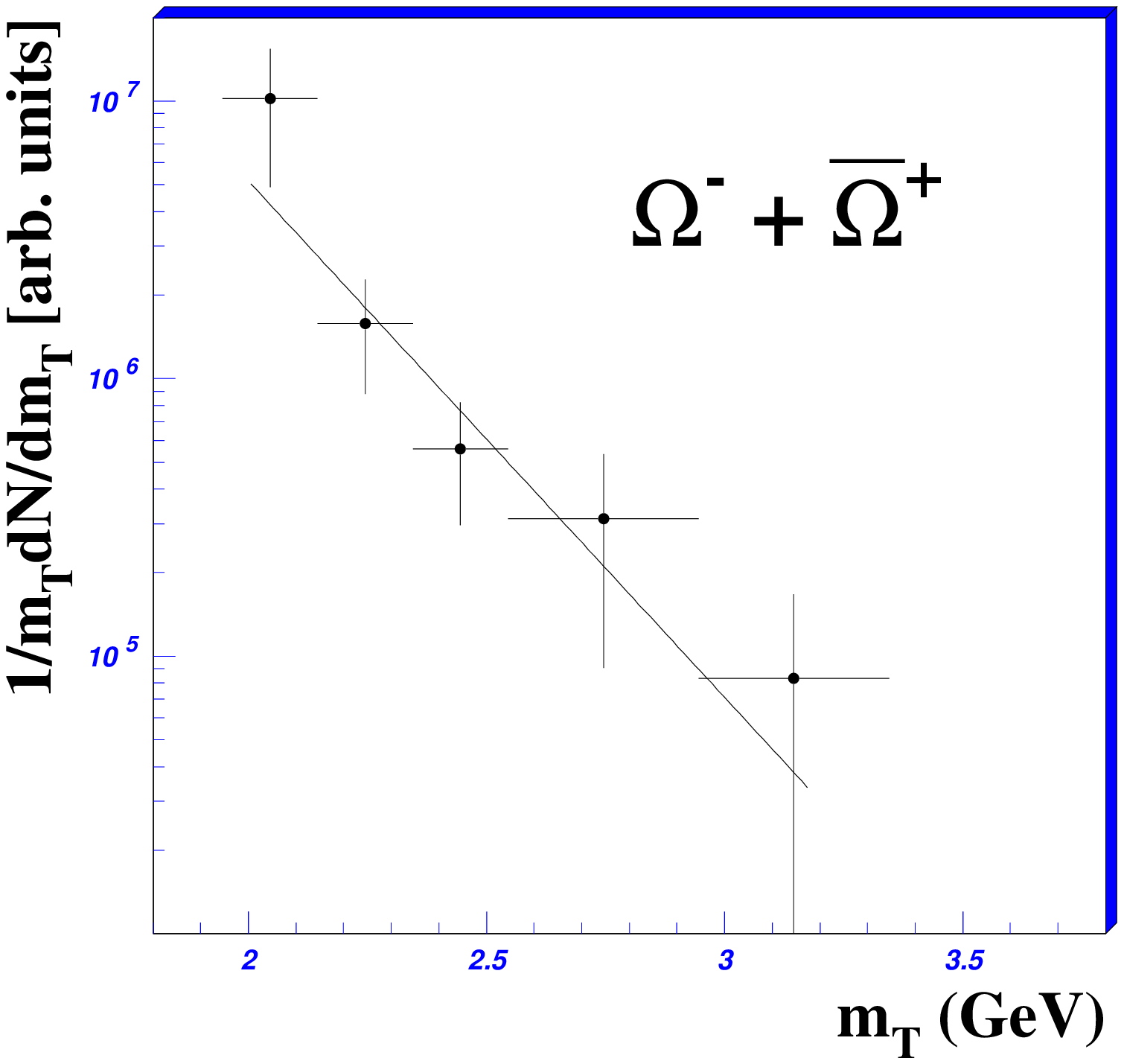}
\caption{Spettri di massa trasversa nelle collisioni Pb-Pb a 40 A GeV/$c$\ 
	 per le particelle \PgXm\ ed \PagXp\ (a sinistra) 
	 e per l'unione dei  campione di \PgOm\ ed \PagOp\ (a destra).} 
\label{mTspectra40}
\end{center}
\end{figure}

Lo stesso metodo di analisi \`e stato seguito nelle collisioni Pb-Pb 
40 A GeV/$c$\ in cui, sino ad ora, si sono studiate le sole 
cascate~\cite{EliaQM02}.   
In fig.~\ref{mTspectra40} sono mostrati gli spettri di massa trasversa  
per \PgXm, \PagXp\ e per il campione di \PgOm\ unito a quello di \PagOp.  
Nella tab.~\ref{tab4.6} sono riportati i valori del parametro $T_a$\ ricavati 
dalla procedura di {\em ``best fit''}.  
\begin{table}[h]
 \begin{center}
  \begin{tabular}{|c|c|}
   \hline
   {\bf \PgXm} & {228 $\pm$\ 13} \\
   {\bf \PagXp} & {280 $\pm$\ 63} \\
  {\bf \PgOm\ + \PagOp} & {326 $\pm$\ 66} \\ \hline
\end{tabular}
\end{center}
\caption{Parametro di {\em ``inverse slopes''} $T_a$\ (MeV) per le cascate 
   nelle collisioni Pb-Pb a 40 A GeV/$c$. Gli errori riportati sono solo 
   quelli statistici.   
\label{tab4.6}}
\end{table}
\section{Estrapolazione della produzione di particelle strane}
\begin{figure}[htb]
\begin{center}
%\includegraphics[scale=0.52]{cap4/yie_thesis.eps}
%\caption{Tasso di produzione delle particelle strane in funzione del numero di nucleoni 
%partecipanti in collisioni Pb-Pb a 160 A GeV/$c$.  
%Nel grafico di destra sono raggruppate le particelle che non condividono alcun quark di 
%valenza con quelli originariamente presenti nei nuclei incidenti (quark $u$\ e $d$).} 
\includegraphics[scale=0.52]{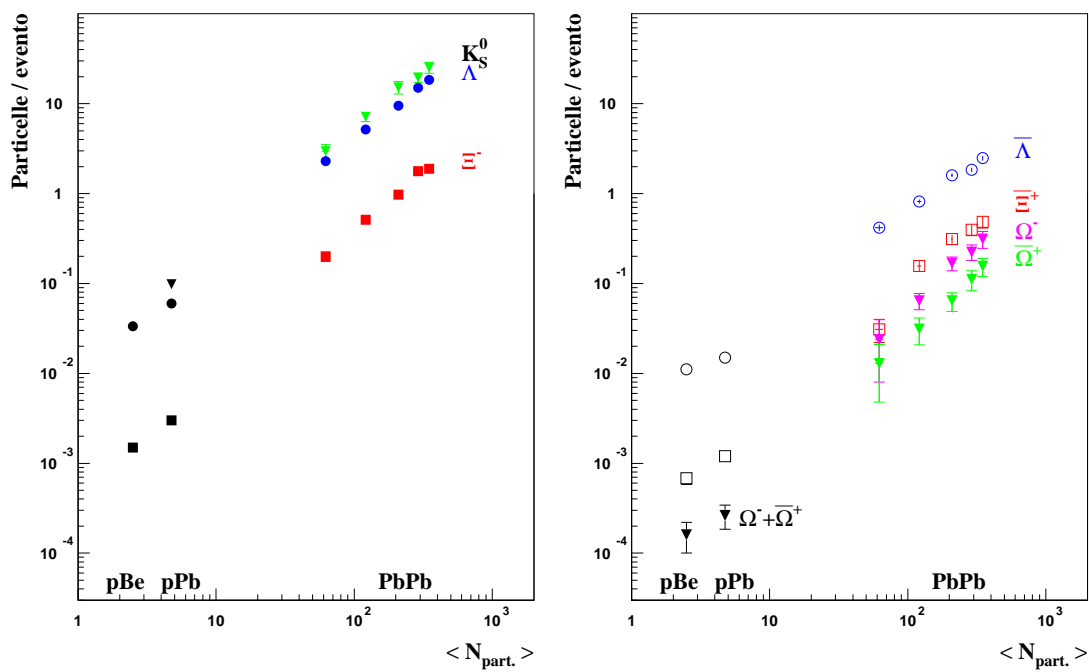}
\caption{Tasso di produzione delle particelle strane in funzione del numero di nucleoni
partecipanti in collisioni p-Be, p-Pb e Pb-Pb a 160 A GeV/$c$. 
I dati delle collisioni di riferimento p-Be e p-Pb sono quelli misurati dalla 
collaborazione WA97~\cite{WA97web,Rocco}.  
Nel grafico di destra sono raggruppate le particelle che non condividono alcun quark di
valenza con quelli originariamente presenti nei nuclei incidenti (quark $u$\ e $d$).
I punti p-Be e p-Pb delle $\Omega$\ di WA97 
si riferiscono al campione di \PgOm\ unito a quello di \PagOp.}  
\label{yieldThesis}
\end{center}
\end{figure}

L'eq.~\ref{YieldInWind} fornisce la misura del tasso di produzione 
delle diverse specie nella relativa finestra di accettanza cinematica.   
Volendo determinare una quantit\`a che non si riferisca 
alla particolare finestra di accettanza utilizzata, che \`e differente 
a seconda della particella considerata e dell'interazione studiata, in 
modo da rendere il risultato pi\`u significativo e 
direttamente confrontabile tra le varie particelle e con quelli di altri 
esperimenti,  
\`e preferibile estrapolare i risultati ad una regione ben definita ed uguale 
per tutte le specie di particelle. Si sceglie quindi 
un'unit\`a di rapidit\`a  
attorno al valore centrale ($y \in \left[y_{cm}-0.5 \, ,\, y_{cm}+0.5 \right]$) 
e l'intero spettro di massa trasversa  
($m_T \in \left[m,\infty \right]$):  
\begin{equation}
Y= I_{tot} = \int^{\infty}_{m}{\rm d}m_T \int_{y_{CM}-0.5}^{y_{CM}+0.5} 
  {\rm d}y \frac{{\rm d}^2 N }{{\rm d}m_T {\rm d}y}
\label{ExtrY}
\end{equation}
Nel calcolare il tasso di produzione estrapolato si pu\`o pensare di 
utilizzare la forma analitica 
della funzione  $ \frac{{\rm d}^2 N }{{\rm d}m_T {\rm d}y} $\  
data dall'eq~\ref{DoubleDiff}, che \`e stata determinata nel {\em ``best fit''} 
alla distribuzione sperimentale, entro la finestra di accettanza, come discusso 
nel paragrafo precedente. Si ricorda per\`o che il metodo della massima 
verosimiglianza, preferibile a quello del $\chi^2$, 
determina il solo parametro $T_a$\ ma non la normalizzazione assoluta  
$\mathcal{A}$~\footnote{Il {\em ``best fit''} con il metodo del $\chi^2$, fornendo  
invece sia  $T_a$\ che $\mathcal{A}$, permette di calcolare il tasso estrapolato 
direttamente per via analitica, con l'eq.~\ref{ExtrY}.}, 
e quindi, operativamente, il tasso di produzione estrapolato viene 
misurato come: 
\begin{equation}
Y= Y_{Acc\,Win} \cdot S = Y_{Acc\,Win} \cdot \frac{I_{tot}}{I_{Acc\,Win}}
\label{OperExtrY}
\end{equation}
dove il fattore di estrapolazione $S$\ \`e pari al rapporto tra gli integrali 
calcolati in maniera analitica, a partire dalla funzione determinata 
dal {\em ``best fit''} di massima verosimoglianza 
(ponendo arbitrariamente $\mathcal{A}=1$),   
sull'intero spettro di $m_T$\ entro una unit\`a di rapidit\`a (eq.~\ref{ExtrY}) e sulla 
finestra di accettanza cinematica  
($I_{Acc\,Win}=\iint_{Acc\,Win}\frac{{\rm d}^2 N }{{\rm d}m_T {\rm d}y}\, {\rm d}m_T{\rm d}y$).  
\newline
L'errore sul tasso estrapolato in questa nuova finestra \`e stato calcolato nel modo seguente: 
\begin{equation}
\delta Y = \left[ (S \cdot \delta Y_{Acc\,Win})^2 \, + \, (\delta S \cdot Y_{Acc\,Win})^2
                    \right]^{1/2} \, ,
\label{ErrExtrY}
\end{equation}
sommando cio\`e in quadratura l'errore $\delta Y_{Acc\,Win}$\ fornito 
dall'eq.~\ref{ErrYield} con 
quello $ \delta S = \frac{{\rm}d S}{{\rm}d T_a} \delta T_a$\ dovuto al fattore di 
estrapolazione. $ \delta T_a$\ indica l'errore nella determinazione del parametro $T_a$, 
calcolato durante la procedura di {\em ``best fit''}, e la derivata $\frac{{\rm}d S}{{\rm}d T_a}$\ 
\`e stata calcolata facendo variare di una quantit\`a molto piccola, 
nell'eq.~\ref{DoubleDiff}, la temperatura apparente risultante dal {\em ``best fit''}.  
\newline
I valori dei tassi di produzione estrapolati per le particelle strane prodotte in 
collisioni Pb-Pb a 160 e 40 A GeV/$c$\ sono riportate in tab.~\ref{tab4.7}.   
I valori per le collisioni a 160 GeV/$c$\ si riferiscono, in questa tabella, all'insieme 
delle cinque classi di centralit\`a ($0+I+II+III+IV$), quelli a 40 GeV/$c$\ all'unica 
classe definita nel {\em paragrafo 4.5} (pari alle interazioni pi\`u centrali per 
circa il 25\% della $\sigma_I$).  
Nel calcolo dell'estrapolazione del tasso di produzione 
delle cascate a 40 A GeV/$c$\ si \`e considerato un 
parametro $T_a$\ comune per le particelle e le rispettive antiparticelle, pari al 
valore ricavato eseguendo il {\em ``best fit''} sul campione di particelle unito 
a quello delle rispettive anti-particelle (cio\`e \PgXm+\PagXp\ ed \PgOm+\PagOp).    
Tutto ci\`o allo scopo di ridurre l'errore di estrapolazione associato a quello 
della determinazione del parametro $T_a$, molto elevato in particolare per 
le anti-particelle.  
I valori per le \PgOm\ e le \PagOp\ riportati nella tab.~\ref{tab4.7}, sono  
quelli presentati alla conferenza {\em ``Quark Matter 2002''}~\cite{ManzQM02} 
e sono stati calcolati con una {\em ``inverse slope''} $T_a=251\pm19$\ pari a 
quella misurata dall'esperimento WA97 per le \PgOm\ + \PagOp~\cite{mt_WA97}. 
Eseguendo l'estrapolazione con la $T_a$\ relativa al campione di \PgOm\ + \PagOp\ 
misurata da NA57 ($T_a=293\pm14$),  si ottengono dei tassi estrapolati per 
le $\Omega$\ inferiori del 18\% a quelli riportate nella tab.~\ref{tab4.7}, 
in conseguenza del pi\`u piccolo fattore di estrapolazione $S=5.86$, che si 
ottiene con la nuova $T_a$. 
\begin{table}[h]
\begin{center}
\begin{tabular}{|c|c|c|c|c|} \cline{2-3}
\multicolumn{1}{c}{ }  & \multicolumn{2}{|c|}{\bf Pb-Pb 160 A GeV/$c$}  \\ \cline{2-3}
\multicolumn{1}{c|}{ }  & $Y$    &  \multicolumn{1}{|c|}{$S$}    \\ \cline{1-3}
\PKzS & $9.0\pm0.6$ & \multicolumn{1}{c|}{$3.38$}   \\ \cline{4-5}
\PgL  & $7.8\pm0.2$   & $2.63$ & \multicolumn{2}{|c|}{\bf Pb-Pb 40 A GeV/$c$}  \\ \cline{4-5}
\PagL & $1.17\pm0.03$ & $2.63$ & \multicolumn{1}{|c|}{$Y$} & $S$ \\ \cline{4-5}
\PgXm & $0.819\pm0.026$ & $3.39$ & $1.21\pm0.09$   & $5.2$ \\ 
\PagXp& $0.220\pm0.014$ & $3.44$ & $0.076\pm0.017$ & $5.2$ \\ 
\PgOm & $0.118\pm0.011$ & $7.14$ & $0.071\pm0.036$ & $4.3$ \\ 
\PagOp& $0.054\pm0.007$ & $7.14$ & $0.020\pm0.010$ & $4.3$ \\ \hline
\end{tabular}
\end{center}
\caption{Tasso di produzione delle particelle strane e fattore $S$\ di estrapolazione 
	 nelle collisioni Pb-Pb a 160 e 40 A GeV/$c$.   
\label{tab4.7}}
\end{table}
\newline
In tab.~\ref{tab4.8} sono mostrati i tassi di produzione estrapolati per le 
particelle strane prodotte in collisioni Pb-Pb a 160 A Gev/$c$\ nelle cinque 
classi di centralit\`a definite in NA57. 
\newline 
Per comodit\`a sono anche riportati 
il numero medio di nucleoni partecipanti relativi a ciascuna classe 
ed i valori misurati dall'esperimento WA97 nelle collisioni p-Be alla stessa 
energia~\cite{WA97Nucl,WA97web}.   
\begin{table}[h]
\begin{center}
\footnotesize{
\begin{tabular}{|c|c|ccccc|} \hline
   & {\bf p-Be} & \multicolumn{5}{|c|}{{\bf Pb-Pb}} \\\hline\hline
$<N_{par}>$ & 2.5 & 62 & 121 & 209 & 290 & 349 \\ \hline
%\PKzS & --- &$3.68\pm0.67$&$6.16\pm0.73$&$19.3\pm3.0$&$17.7\pm2.2$&$26.0\pm3.6$\\\hline
 \PKzS & --- &$2.96\pm0.54$&$7.17\pm0.85$&$15.2\pm2.4$&$19.6\pm2.4$&$25.3\pm3.5$\\\hline
 \PgL  &$0.0344\pm0.0005$&$2.30\pm0.22$&$5.19\pm0.29$&$9.5\pm0.5$&$15.0\pm0.8$&$18.5\pm1.1$\\ \hline
 \PagL &$0.0111\pm0.0002$&$0.417\pm0.035$&$0.82\pm0.04$&$1.60\pm0.07$&$1.84\pm0.10$&$2.47\pm0.14$\\ \hline
 \PgXm &$0.0015\pm0.0001$&$0.200\pm0.024$&$0.51\pm0.04$&$0.98\pm0.06$&$1.78\pm0.11$&$1.88\pm0.14$\\ \hline
 \PagXp&$0.00068\pm0.0001$&$0.031\pm0.009$&$0.16\pm0.02$&$0.31\pm0.04$&$0.40\pm0.05$&$0.48\pm0.07$\\ \hline
 \PgOm &---&$0.024\pm0.016$&$0.064\pm0.013$&$0.17\pm0.03$&$0.22\pm0.04$&$0.31\pm0.07$\\ \hline
 \PagOp&---&$0.013\pm0.008$&$0.031\pm0.010$&$0.065\pm0.015$&$0.11\pm0.03$&$0.16\pm0.04$ \\ \hline
\PgOm+\PagOp&$0.00016\pm0.00006$&
		$0.037\pm0.017$&$0.095\pm0.017$&$0.23\pm0.03$&$0.33\pm0.05$&$0.47\pm0.07$\\\hline
\end{tabular}
}
\end{center}
\caption{Tasso di produzione delle particelle strane nelle collisioni p-Be (misurati da WA97) 
	 e Pb-Pb (misurati da NA57) a 160 A GeV/$c$\ in funzione del numero medio di nucleoni 
	 partecipanti alla collisione.
\label{tab4.8}}
\end{table}
L'estrapolazione per tutte le particelle \`e eseguita adoperando 
l' {\em ``inverse slope''} $T_a$\ ottenuta dall'unione delle cinque classi 
di centralit\`a, ad esclusione del caso delle \PgL\ e \PagL\ in cui \`e 
stato possibile adoperare i valori misurati in ciascuna classe. Inoltre, per le 
\PgOm\ e le \PagOp\ si \`e assunta nuovamente la temperatura misurata da WA97. 
L'andamento dei tassi di produzione in funzione del numero di nucleoni 
partecipanti \`e mostrato nella fig.~\ref{yieldThesis}.  
\newline
Questi risultati saranno oggetto di discussione nel prossimo capitolo. 
%\subsection{Calcolo degli errori sistematici di estrapolazione}
%\section{Considerazioni finali}

%% file: cap5/cap5.tex
\chapter{Discussione dei risultati} 
\section{Introduzione}
In questo capitolo si discuteranno i risultati ottenuti dall'analisi 
della produzione di particelle strane in collisioni Pb-Pb a 160 ed a 
40 A GeV/$c$\ discussa nel capitolo precedente. 
\newline
Si discuteranno i risultati sulle temperature 
apparenti~\footnote{In questo capitolo la temperatura apparente 
verr\`a indicata con $T_{app}$, anzich\'e con $T_a$, 
per renderla pi\`u agevolmente 
distinguibile dalla temperatura di {\em ``freeze-out''} $T$.},  
con particolare riferimento alla simmetria tra particelle e relative 
anti-particelle ed alla dipendenza dall'energia della collisione.  
\newline
Da un'analisi simultanea di tutti gli spettri di massa trasversa delle 
particelle strane si disaccoppier\`a la componente termica da quella 
dovuta al flusso trasverso di espansione collettiva. 
\newline
Saranno  presentati e discussi i rapporti di produzione delle particelle
strane.
\newline
Si studier\`a l'incremento della produzione di particelle strane in 
funzione della centralit\`a, rispetto 
a quanto atteso nelle collisioni adroniche in assenza di plasma, confrontando 
i dati delle collisioni di riferimento p-Be e p-Pb di WA97.  
%Si confronteranno quindi i risultati trovati con quelli dell'esperimento WA97 
%nelle collisioni Pb-Pb a 160 A GeV/$c$.
\newline
Si studier\`a infine la dipendenza dall'energia della collisione, confrontando 
i risultati con quelli dell'esperimento STAR al RHIC.
\section{Spettri di massa trasversa}
\subsection{Temperature apparenti in collisioni Pb-Pb a 160 A GeV/$c$}
\begin{figure}[tb]
\begin{center}
\includegraphics[scale=0.45]{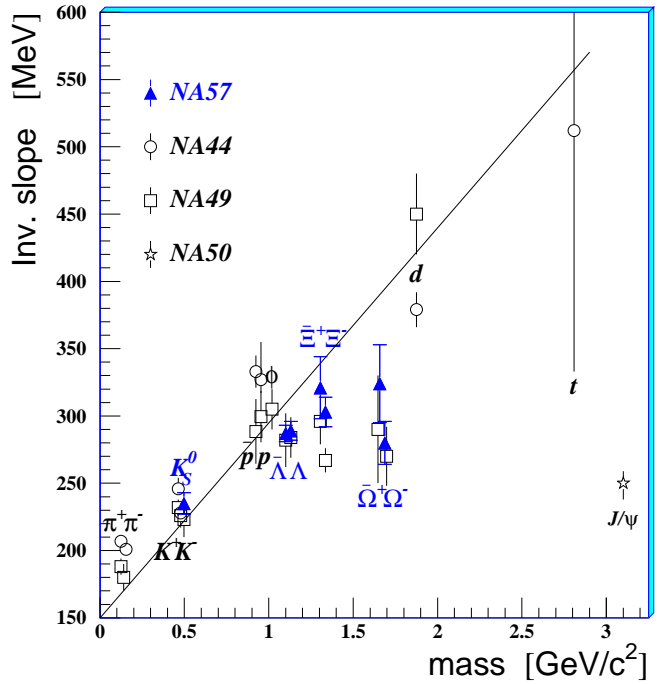}
\caption{Temperature apparenti misurate in collisioni Pb-Pb a 160 A 
         GeV/$c$\ in funzione della massa a riposo delle particelle. 
	 I triangoli blu sono i valori determinati per le particelle 
	 strane in quest'analisi; i cerchietti, i quadratini e le stelline 
	 nere sono,  
	 rispettivamente, i risultati degli esperimenti NA44, NA49 ed NA50
	 (si veda il testo per le referenze relative alle diverse specie). 
%	 Il punto ad $m=3.1$\ GeV/$c2$\ si riferisce all'osservazione di 
%	 NA50~\cite{JPSI1} di uno spettro termico per lo stato di 
%	 charmonio J/$\psi$.  
	 La linea rappresenta un'approssimazione lineare ai risultati 
	 forniti dal modello RQMD~\cite{RQMD} per le particelle non-strane.}  
\label{CompilaT}
\end{center}
\end{figure}

Lo studio degli spettri di impulso delle particelle prodotte in collisioni  
tra nuclei pesanti fornisce indicazioni circa il raggiungimento  
dell'equilibrio termico locale nella sorgente.  
\newline
In fig.~\ref{CompilaT} \`e mostrata una compilazione dei pi\`u recenti 
risultati sulle pendenze inverse degli spettri di massa trasversa 
ottenute da {\em ``best fit''} con la funzione esponenziale del tipo  
${\rm d}N/{\rm d}m_t \propto m_t \exp[-m_T/T_{app}]$. Le 
temperature apparenti ({\em ``inverse slope''})   
sono rappresentate in funzione della massa della particella considerata.  
\newline
I dati della collaborazione NA49 provengono dalle seguenti referenze: 
\Pgpm, \PKm\ e \PKp\ da~\cite{NA49K+K-PI-};  
\PKzS\ da~\cite{NA49K0}; 
\Pgpp, \Pp\ e \Pap\ da~\cite{NA49PI+PROT}; 
\Pgf\ da~\cite{NA49PHI}; 
\PgL\ e \PagL\ da~\cite{NA49LA}; 
\PgXm\ e \PagXp\ da~\cite{NA49XI}; 
\PgOm\ e \PagOp\ da~\cite{NA49OM}; 
deuterone da~\cite{NA49DEU}. 
\newline
Quelli della collaborazione NA44 
sono stati raccolti dalle seguenti referenze: 
\Pgpm, \Pgpp, \PKm, \PKp, \Pp\ e \Pap\ da~\cite{NA44remain}; 
deuterone e tritone da~\cite{NA44DEUTRI}.  
Il punto ad $m=3.1$\ GeV/$c^2$\ si riferisce alla determinazione di
NA50~\cite{JPSI1} di uno spettro termico per lo stato di charmonio J/$\psi$.  
\newline
L'errore riportato \`e il pi\`u grande tra quello statistico e  
quello sistematico, quando quest'ultimo viene valutato.  
\newline 
Si osserva in generale un aumento della temperatura apparente con 
la massa delle particelle. Questo comportamento \`e atteso in uno scenario 
in cui il moto termico sia accompagnato da un'espansione collettiva 
nella direzione trasversa del sistema, come discusso nel {\em paragrafo 1.4.2}. 
Nel limite non relativistico, cio\`e per $p_t \ll m_0$, si attende infatti la 
semplice relazione lineare $T_{app}=T+m_0<\beta^2_\perp>$, fornita dall'eq.~\ref{Tapp2}. 
Questa semplice relazione non permette per\`o di ricavare informazioni 
{\em quantitative} su $T$\ e $<\beta^2_\perp>$\ separatamente, in quanto solo per 
pochi punti degli spettri da cui sono state estratte le $T_{app}$\ si \`e 
ancora in regime non-relativistico. Pur con queste limitazioni, si osserva 
che le particelle strane si discostano rispetto dall'andamento di semplice  
linearit\`a, che in  fig.~\ref{CompilaT} \`e indicato servendosi 
di un'approssimazione lineare ai risultati del modello di 
produzione adronica RQMD~\cite{RQMD} a partire dalle sole particelle non strane.
%che \`e anche quello suggerito dal modello di produzione adronica RQMD~\cite{RQMD}, 
%la cui predizione \`e indicata dalla linea nella fig.~\ref{CompilaT}. 
Un simile comportamneto per le 
particelle strane \`e gi\`a stato osservato dalla collaborazione WA97~\cite{mt_WA97}.  
Ancora maggiore \`e lo scostamento dalla linearit\`a per la temperatura 
apparente della J/$\psi$, misurata da NA50.   
\newline 
A riguardo dei risultati del deuterone e del tritone, bisogna tuttavia 
osservare che queste particelle sono probabilmente prodotte successivamente, 
rispetto alle altre: dovrebbero infatti formarsi prevalentemente 
nelle interazioni dello stato finale tra nucleoni trasportati dal 
flusso.  
\newline
\`E incoraggiante osservare che i diversi esperimenti misurano temperature 
apparenti compatibili tra le stesse particelle   
(ad eccezione dei \Pgppm\ in cui per\`o vi \`e la complicazione  
dei decadimenti delle risonanze), e che si osservano valori 
generalmente compatibili tra le particelle e le rispettive antiparticelle.  
\subsection{{\em ``Freze-out''} termico} 
\begin{figure}[p]
\begin{center}
\includegraphics[scale=0.54]{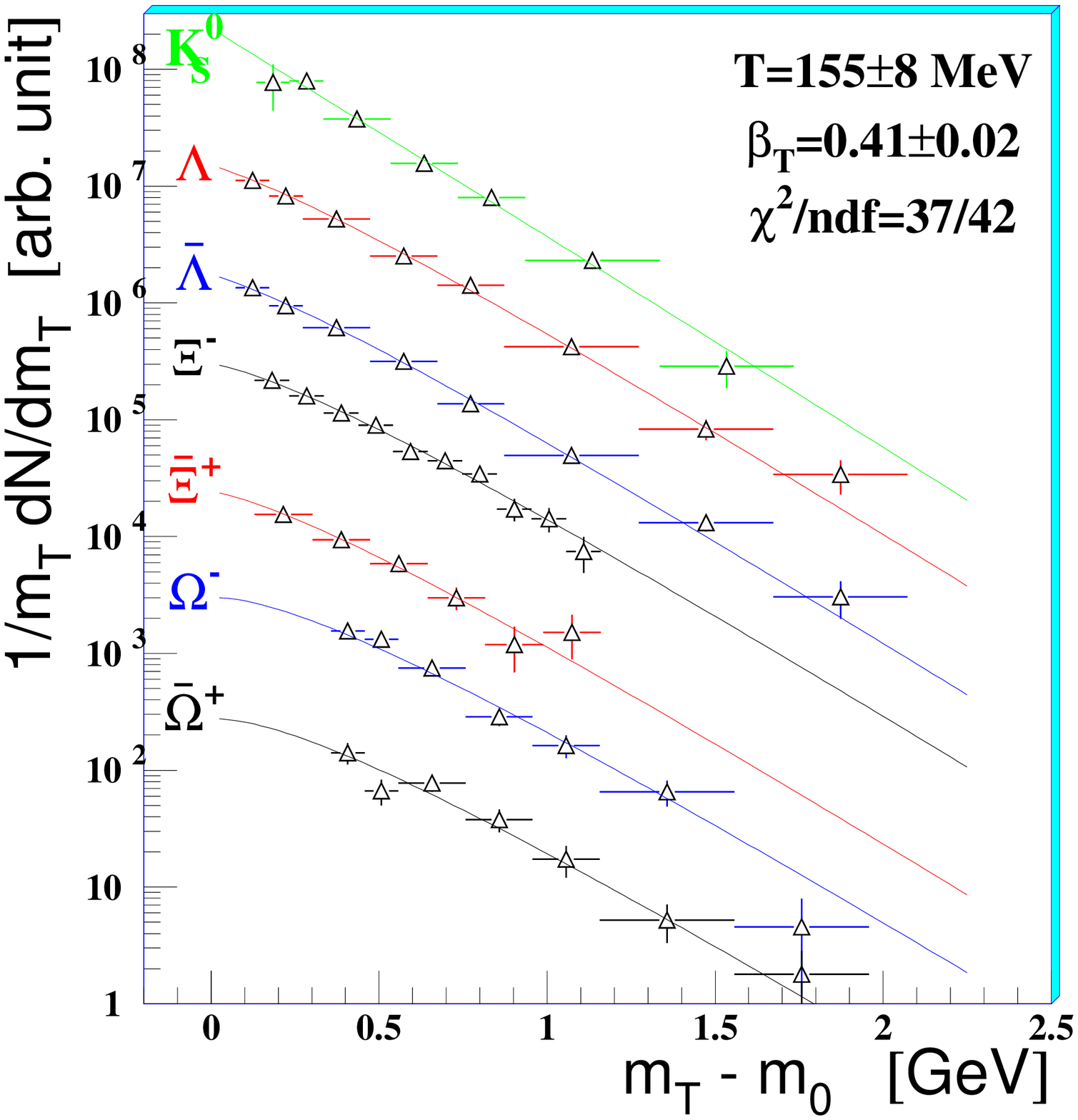}\\
\caption{Spettri di massa trasversa delle particelle strane 
         in collisioni Pb-Pb a 160 A GeV/$c$. 
	 Le curve sovrapposte sono il risultato del {\em ``best fit''} 
	 dell'eq.~\ref{Blast} ai punti sperimentali.}
\label{RadialFlow}
\includegraphics[scale=0.25]{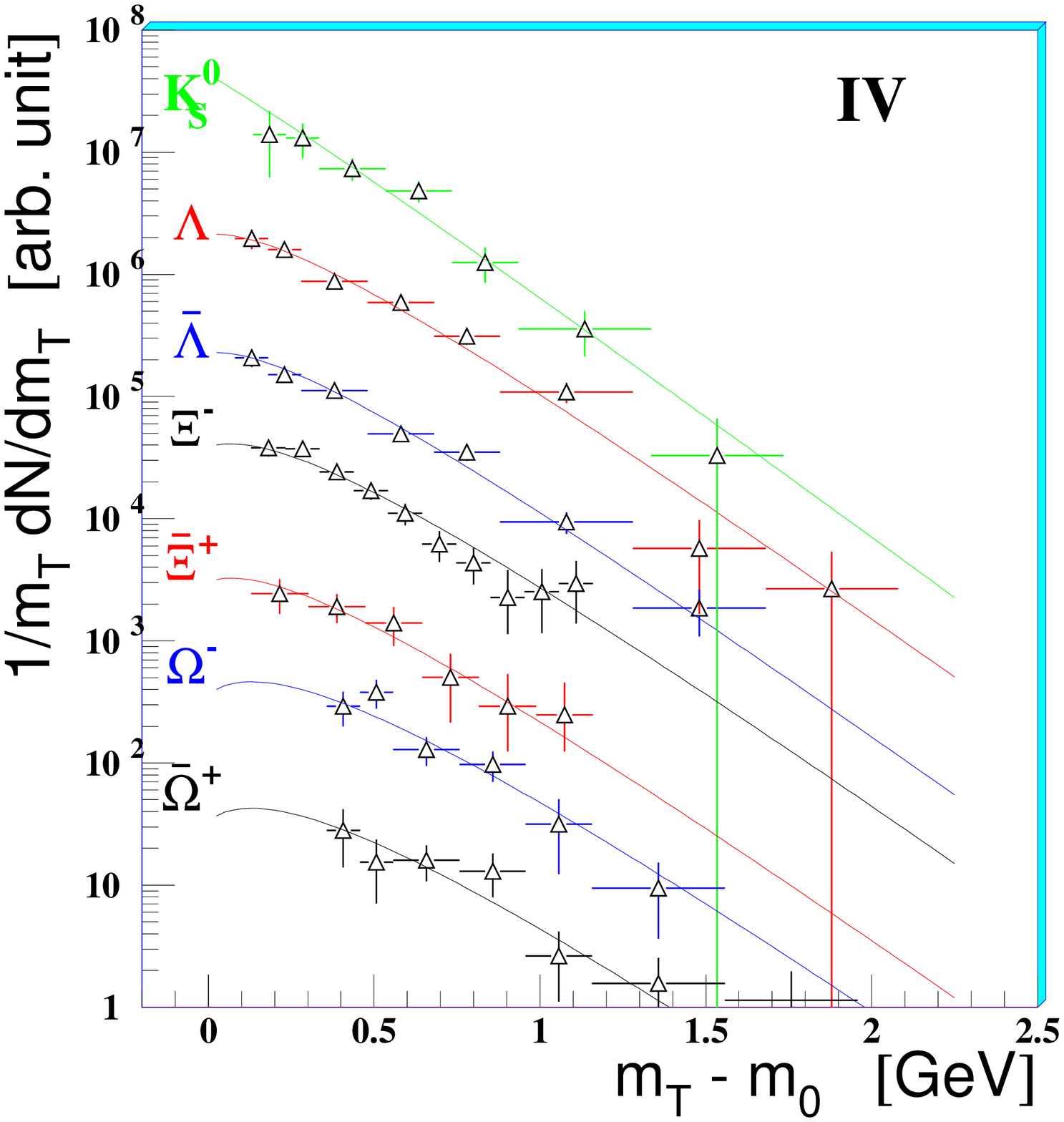}
\includegraphics[scale=0.25]{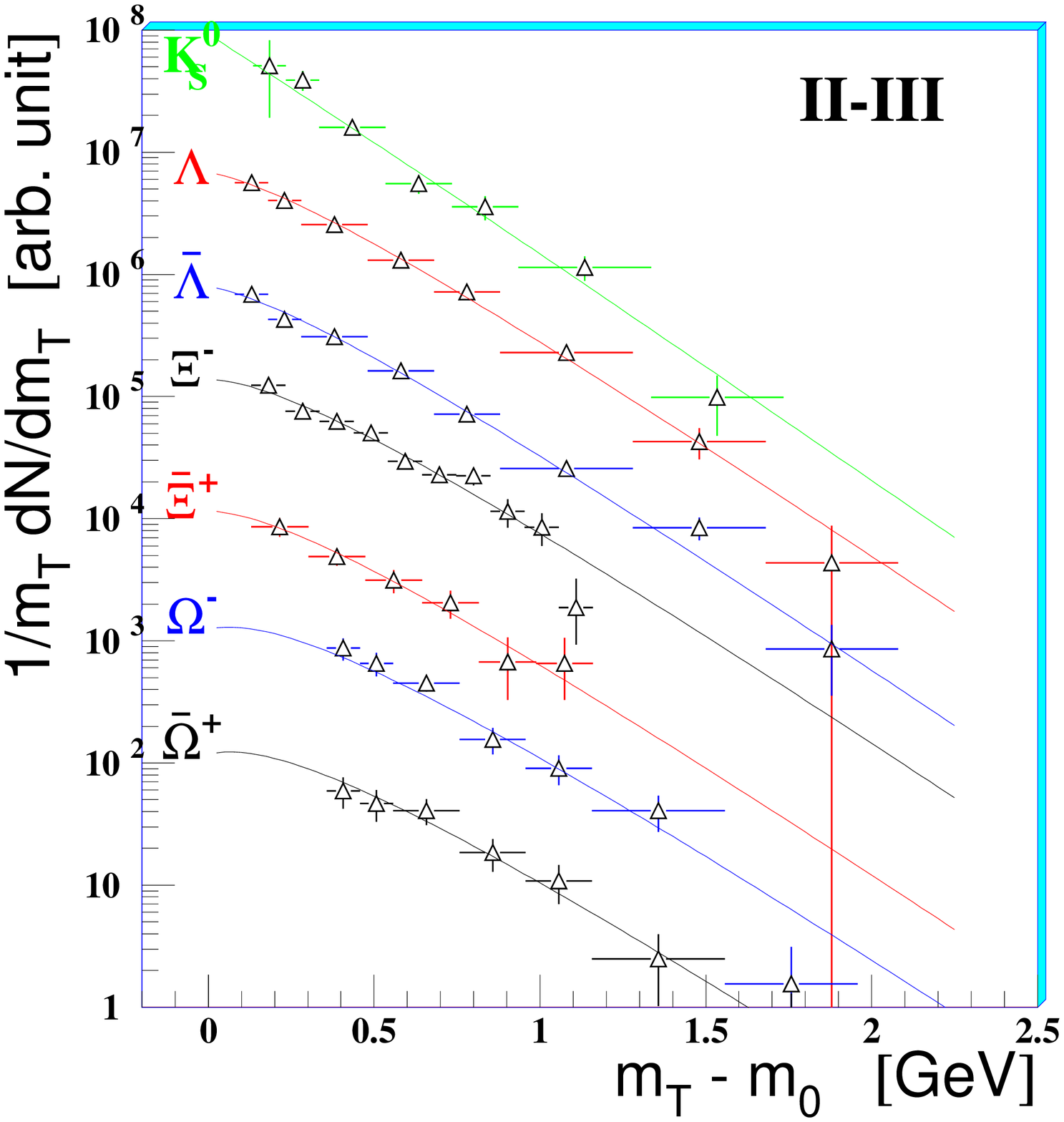}
\includegraphics[scale=0.25]{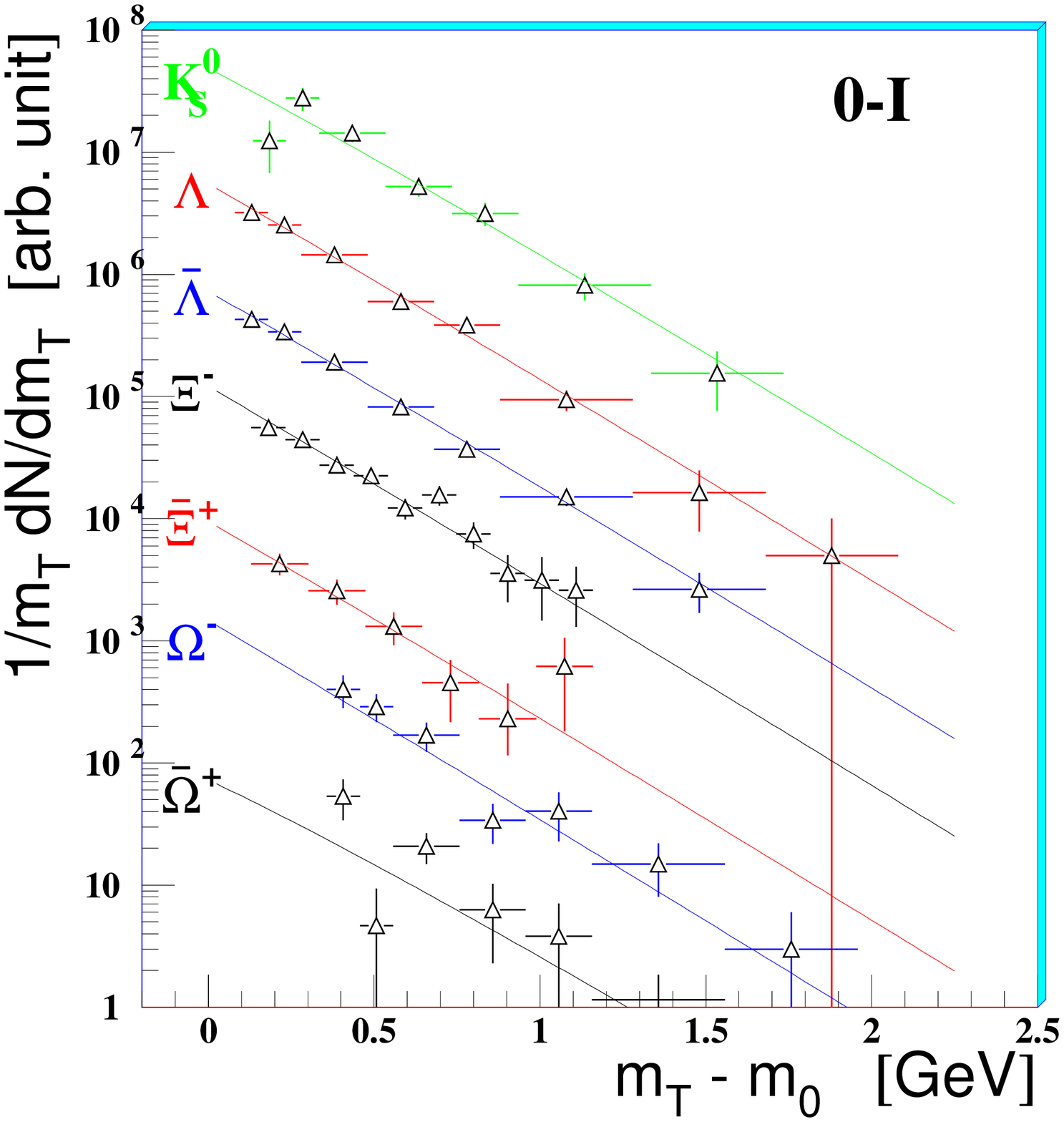}
\caption{Spettri di massa trasversa in collisioni  Pb-Pb a 160 A GeV/$c$ 
         per diverse classi di centralit\`a. Le curve sovrapposte sono il 
	 risultato del {\em ``best fit''} dell'eq.~\ref{Blast} 
	 ai punti sperimentali. Nella tab.~\ref{tab5.2} 
	 sono riportati i valori dei parametri del {\em ``best fit''}.}  
\label{RadialFlowMSD}
\end{center}
\end{figure}

Nella descrizione idrodinamica dell'evoluzione spazio-temporale delle 
collisioni, gli spettri di massa trasversa sono sensibili al flusso 
trasverso della materia adronica. 
\newline
Per caratterizzare il flusso, gli spettri di massa trasversa possono 
essere descritti dalla funzione~\cite{Blast1,Blast2}:
\begin{equation}
\frac{{\rm d}^2N}{m_T{\rm d}y{\rm d}m_T} \propto 
   m_T \cdot K_1\left(\frac{m_T \cosh \rho}{T} \right)
   \cdot I_0\left(\frac{p_T \sinh \rho}{T} \right)
\label{Blast}
\end{equation}
dove $K_1$\ ed $I_0$\ sono due funzioni di Bessel (modificate) e 
$\rho={\rm arctanh} \beta_\perp$.  Questa espressione \`e ricavata 
nell'appendice A della referenza ~\cite{Blast1}, a partire da un sorgente 
termica in cui prima si definisce 
(utilizzando ``l'angolo di {\em boost}'' $\rho={\rm arctanh} \beta_\perp$) 
un campo di velocit\`a trasversali,  
a simmetria azimuthale, nella fettina centrale della {\em ``fireball''} 
(cio\`e per $x=0$): ${u'}^\nu(\tilde{t},r,x=0) = (\cosh\rho,0,\vec{e}_r \sinh\rho)$. 
Si esegue quindi un {\em ``boost''} nella direzione longitudinale 
(con ``angolo di {\em boost''} $\eta$) per generare completamente 
il campo di velocit\`a~\footnote{Si ricorda che ``gli angoli di {\em boost}'' 
$\eta$\ e $\rho$\ sono pari al coseno iperbolico 
del fattore $\gamma$\ di Lorentz della rispettiva trasformazione.}:
\begin{equation}
u^{\mu}(\rho,\eta)=(\cosh{\rho}\cosh{\eta} , \cosh{\rho}\sinh{\eta}, \vec{e}_r \sinh{\rho}) .
\label{BlastSpeed}
\end{equation}
Questa descrizione permette la scelta di diversi profili per la velocit\`a del flusso 
trasverso; ad esempio la scelta $\beta_\perp(r)=\beta_s(\frac{r}{R})^2$\ \`e la  
pi\`u prossima alla soluzione idrodinamica~\footnote{In questa espressione si
assume una {\em ``fireball''} di raggio $R$\ in cui il flusso assume 
il massimo valore sulla superficie, dove \`e pari a $\beta_s$.}. 
L'espressione analitica data 
dall'eq.~\ref{Blast} si riferisce invece al caso pi\`u semplice di profilo uniforme: 
$\beta_\perp(r)=\beta_\perp={\rm cost}$.  
In~\cite{Blast1} si mostra come la forma del profilo influenzi poco i risultati 
dell'analisi.  
\newline
I parametri del modello sono dunque la temperatura $T$\ del {\em ``freeze-out''} e la 
velocit\`a del flusso trasverso $\beta_\perp$.  Si consideri lo spettro di massa trasversa 
di una data specie di particella; esso pu\`o venir descritto da un insieme di coppie 
di valori $(\beta_\perp,T)$. Tenendo conto degli errori associati a questa determinazione, 
ciascuno spettro di massa trasversa individua una regione fiduciale nel piano 
$\beta_\perp$--$T$. Soltanto l'analisi di pi\`u spettri, per particelle di diversa massa, 
permette quindi di separare in modo non ambiguo il contributo termico da quello 
associato al flusso trasverso.  
\newline
Un fit simultaneo della funzione~\ref{Blast} 
a tutti gli spettri di massa trasversa delle particelle strane 
misurati in questo lavoro per le collisioni Pb-Pb a 160 A GeV/$c$,  
fornisce i seguenti  valori:  
\begin{enumerate}
\item[ ] $T=155\pm8$\ MeV 
\item[ ] $\beta_\perp=0.41\pm0.02$ 
%\item[ ] con $\chi^2/ndf = 37/42$.
\end{enumerate}
con $\chi^2/ndf = 37/42$.  
In fig.~\ref{RadialFlow} \`e mostrato il 
risultato del {\em ``best fit''} per i diversi spettri di massa trasversa 
delle particelle strane.  
Il modello descrive in modo eccellente i dati sperimentali di tutte le 
particelle strane, $\Xi$\ ed $\Omega$\ incluse,  
perlomeno negli intervalli di $m_T$\ considerati.   
\`E stato suggerito da diversi autori~\cite{OmegaDecoup1,OmegaDecoup2} che 
le $\Xi$\ ed in particolare le $\Omega$\ possano disaccoppiarsi prima rispetto 
alle altre specie (cio\`e il {\em ``freeze-out''} sarebbe anticipato). I risultati 
qui trovati 
non danno indicazioni in questa direzione. Per quanto concerne le $\Omega$,  
assume quindi notevole interesse studiare il comportamento degli spettri 
sperimentali di massa trasversa per $m_T$\ prossimo ad $m_0$. \`E pertanto in 
corso uno studio in tal senso, volto ad estendere la regione fiduciale di 
accettanza verso valori i pi\`u piccoli possibile di $m_T$. Si ricordi che  
la regione a bassi valori di $m_T$\ \`e la pi\`u delicata per quel che riguarda 
il calcolo delle correzioni, come discusso nel {\em paragrafo 4.4}. 
\newline
Gli spettri di massa trasversa sono stati successivamente divisi in due gruppi: 
nel primo sono contenute quelle particelle che condividono almeno un quark di 
valenza in comune con quelli $u$\ e $d$\ originariamente presenti nei nuclei 
collidenti, nell'altro le particelle che non hanno alcun quark in comune. 
I \PKzS\ sono stati inclusi nel primo gruppo in quanto, come ``combinazione'' di 
\PKz$(d\bar{s})$ e \PaKz$(\bar{d}s)$,``contengono'' un $50\%$ di quark $d$.  
In tab.~\ref{tab5.1} sono riassunti i risultati del {\em ``best fit''} eseguito 
separatamente sugli spettri del primo gruppo e su quelli del secondo. 
\begin{table}[h]
\begin{center}
\begin{tabular}{|c|c|c|c|} \hline
                    & $T$\ (MeV)  &   $\beta_\perp$    & $\chi^2/ndf$ \\ \hline
\PKzS, \PgL, \PgXm  & $154\pm9$   &   $0.41\pm0.02$    & $15.1/19$    \\
\PagL, \PagXp, \PgOm, \PagOp &
                      $156\pm16$  &   $0.40\pm0.04$    & $23.2/21$    \\ \hline
\end{tabular}
\end{center}
\caption{Temperatura di {\em ``freeze-out''} e velocit\`a trasversa  
ottenuti dal {\em ``best fit''} dell'eq.~\ref{Blast} agli spettri 
di massa trasversa delle particelle strane.
Le particelle sono state divise tra quelle che contengono almeno un quark 
di valenza in comune con i nucleoni preesistenti entro i nuclei interagenti  
(quark $u$\ e $d$) e quelli che non hanno alcun quark in comune.  
%Il \PKzS, in quanto ``combinazione'' di \PKz$(d\bar{s})$ e \PaKz$(\bar{d}s)$, 
%\`e stato incluso tra i primi.  
\label{tab5.1}}
\end{table}
La temperatura di {\em ``freeze-out''} e la velocit\`a del flusso trasverso sono
compatibili entro gli errori sperimentali, confermando la simmetria tra 
il settore barionico e quello anti-barionico, gi\`a evidenziata dallo 
studio delle temperature apparenti.  
L'assenza di un'asimmetria negli spettri di massa trasversa tra i barioni e gli 
anti-barioni strani, appurata ad un elevato livello di precisione 
gi\`a nella misura delle temperature apparenti, rappresenta un'evidenza 
sperimentale molto importante. Suggerisce infatti un meccanismo di 
produzione comune per i barioni e gli anti-barioni strani,  come atteso 
nel caso in cui la sorgente primaria di queste particelle sia 
una ``fireball'' di QGP ({\em cfr. paragrafo 1.5.2}).  Per conservare questa 
simmetria, si deve anche escludere la possibilit\`a di sostanziali interazioni 
all'interno della materia in fase adronica che dovrebbe formarsi con 
l'adronizzazione dello stato di QGP. Ci\`o pu\`o avvenire~\cite{RafelskiBook} 
o per mezzo di un improvviso disaccoppiamento delle particelle 
dalla ``fireball'' di QGP, successivo  
ad un considerevole surraffreddamento, oppure per un processo di evaporazione degli 
adroni dalla superficie, sequenziale nel tempo, senza la formazione di una 
fase adronica dell'intero volume.  
Questa simmetria tra particelle ed anti-particelle non viene invece 
riprodotta dai modelli di trasporto, in cui si introducono i gradi 
di libert\`a degli adroni confinati.   
\newline
Un'altra possibile direzione d'indagine \`e la dipendenza dei parametri 
del {\em ``freeze-out''}  dalla centralit\`a della collisione. 
Le particelle sono quindi state raggruppate nuovamente  
con le anti-particelle e divise per classi di centralit\`a. Per ragioni 
di convenienza statistica, si sono raggruppate la classe di centralit\`a 
$0$\ con quella  $I$\ e la classe $II$\ con la $III$\ 
({\em cfr.} tab.~\ref{tab4.4}). In fig.~\ref{RadialFlowMSD} 
\`e mostrato il risultato del {\em ``best fit''} eseguito separatamente per le tre 
nuove classi di centralit\`a cos\`i definite.  
I risultati dei parametri del {\em ``best fit''}  sono riportati nella tab.~\ref{tab5.2}.  
\begin{table}[h]
\begin{center}
\begin{tabular}{|c|c|c|c|} \hline
            & $T$\ (MeV)  &   $\beta_\perp$    & $\chi^2/ndf$ \\ \hline
$IV$        & $132\pm13$  &   $0.46\pm0.03$    & $37.4/41$    \\
$II$--$III$ & $145\pm9 $  &   $0.43\pm0.02$    & $50.0/42$    \\
$0$--$I$    & $250\pm10$  &   $\approx 0\pm0.2$& $50.3/41$    \\ \hline
\end{tabular}
\end{center}
\caption{Dipendenza della temperatura di {\em ``freeze-out''} e della 
velocit\`a del flusso trasverso dalla centralit\`a della collisione.
I dati della classe di centralit\`a $II$\ e $III$\ e quelli delle 
classi $0$\ ed $I$\ ({\em cfr. paragrafo 4.5} ed in particolare la tab.~\ref{tab4.4})  
sono stati uniti per ridurre l'errore statistico.  
\label{tab5.2}}
\end{table}
Al diminuire della centralit\`a della collisione 
(cio\`e all'aumentare del parametro d'impatto) si osserva una diminuzione 
del flusso trasverso ed un aumento della temperatura finale del 
{\em ``freeze-out''}. 
Una temperatura pi\`u elevata pu\`o esser giustificata 
dalla meno intensa espansione del sistema, sino al  {\em ``freeze-out''}: 
espandendosi meno, la materia adronica si raffredderebbe  di meno. 
Nel caso delle collisioni pi\`u periferiche (classe $0$--$I$), la soluzione 
privilegiata \`e addirittura quella con flusso trasverso nullo e sola 
componente termica. 
\newline
Osservando le figg.~\ref{RadialFlow}~e~\ref{RadialFlowMSD} si deduce che, 
%secondo il modello, 
%all'aumentare del flusso trasverso (o al diminuire della temperatura) si ha 
%un pi\`u marcato appiattimento degli spettri, per valori di $m_t$\ prossimi 
%alla massa a riposo $m_0$\ delle particelle. 
%Inoltre l'effetto diventa pi\`u marcato all'aumentare di $m_0$.  
per masse trasverse prossime alla massa a riposo $m_0$\ delle particelle, 
l'espansione trasversa introduce una curvatura concava degli spettri di  
massa trasversa, tanto pi\`u accentuata quanto maggiore \`e la massa $m_0$. 
Ci\`o \`e evidente in modo particolare nella fig.~\ref{RadialFlowMSD}, 
confrontando le curve del modello 
determinate per la classe $0$--$I$, in cui non si osserva un 
apprezzabile flusso trasverso, con quelle per la classe $IV$\ in cui la velocit\`a 
del flusso trasverso \`e pari a circa la met\`a della velocit\`a della luce.  
\newline
I valori ricavati in questa analisi per la classe di maggior centralit\`a (la $IV$)  
sono compatibili con quelli quotati dalla collaborazione NA49~\cite{QM02NA49}, 
in un'analisi condotta in un analogo intervallo di centralit\`a 
(collisioni pi\`u centrali da circa il 5\% a circa il 10\% della sezione d'urto 
anelastica, a seconda dell'energia).  
L'esperimento NA49 studia infatti le collisioni Pb-Pb pi\`u centrali in funzione  
dell'energia della collisione (a 40, 80 e 160 A GeV/$c$) ma non ha  
osservato, in questo tipo di analisi, alcuna significativa dipendenza dall'energia della 
collisione~\cite{QM02NA49}. Anche NA49 non osserva asimmetria tra il {\em ``freeze-out''} 
delle particelle e quello delle anti-particelle, alle tre energie considerate.   
\newline
I risultati della  $IV$\ classe di centralit\`a sono anche compatibili con 
quelli determinati, per la stessa classe di centralit\`a, 
dall'analisi dell'interferometria HBT per le particelle di carica negativa  
(prevalentemente \Pgpm) sui dati di WA97. Quest'analisi sar\`a l'oggetto 
della discussione dell'ultimo capitolo. In quello studio si mostrer\`a 
ulteriormente una dipendenza dalla centralit\`a analoga, da  
un punto di vista {\em qualitativo}, a quella qui determinata.  

I risultati di questo studio sembrano suggerire, nel caso delle particelle  
di massa a riposo pi\`u elevata, un appiattimento degli  
spettri di massa trasversa per valori di $m_T$\ prossimi ad $m_0$.  
Nel calcolare i tassi di produzione si \`e estrapolata la misura eseguita 
nella finestra di accettanza di ciascuna particella all'intero spettro 
di $m_T$, secondo la funzione esponenziale dell'eq.~\ref{DoubleDiff}, nel modo  
discusso nel {\em paragrafo 4.8}. Supponendo pertanto che la relazione 
esponenziale venga violata per valori di $m_T$\ prossimi alla massa a 
riposo della particella, si sarebbero sistematicamente sottovalutati i tassi 
di produzione. 
\`E opportuno quindi stimare 
%l'errore introdotto nell'estrapolazione con la funzione esponenziale. 
di quanto variano i tassi di produzione estrapolando con la funzione 
definita dall'eq.~\ref{Blast}.  
La stima verr\`a fatta per le particelle di 
massa pi\`u elevata, le $\Xi$\ e le $\Omega$, in modo da ottenere un limite superiore 
nel caso delle altre particelle, in cui l'effetto, qualora fosse davvero presente,  
\`e molto minore. 
\newline
Si consideri il fattore di estrapolazione:
\begin{equation}
S'_{exp}=\frac{\int_{m_0}^{\infty}{\mathcal{A}m_T\exp{\left[-\frac{m_T}{T_{app}}\right]}{\rm d}m_T}}
        {\int_{m_T^{min}}^{\infty}{\mathcal{A}m_T\exp{\left[-\frac{m_T}{T_{app}}\right]}{\rm d}m_T}}
\label{OurExtrap}
\end{equation}
definito come il rapporto tra l'integrale sull'intero spettro di $m_T$\ 
della funzione esponenziale misurata entro la finestra sperimentale 
(dove la $T_{app}$\ \`e quella misurata sperimentalmente) e quello 
sull'intervallo $[m_T^{min}, \infty ]$\ della stessa funzione. 
Nell'eq.~\ref{OurExtrap} $m_T^{min}$\ \`e il valor minimo di $m_T$\ 
corrispondente al valor minimo di $p_T$\ nella finestra di 
accettanza: $m_T^{min}=\sqrt{(p_T^{min})^2 + m_{0}^2} $.  
\newline
Si indica invece con $S'_{blast}$\ il fattore di estrapolazione che si ottiene
a partire dallo spettro di massa trasversa descritto dall'eq.~\ref{Blast},
nella seguente forma:
\begin{equation}
S'_{blast}=\frac{\int_{m_0}^{\infty}{{\rm d}^2N}{/({\rm d}y{\rm d}m_T)}\,{\rm d}m_T}
          {\int_{m_T^{min}}^{\infty}{{\rm d}^2N}{/({\rm d}y{\rm d}m_T)}\,{\rm d}m_T} \, .
\label{BlastExtrap}
\end{equation}
Nell'eq.~\ref{BlastExtrap} i limiti d'integrazione sono gli stessi dell'eq.~\ref{OurExtrap},
ma si integra la funzione definita nel modello che prevede il moto collettivo trasverso
sovrapposto al moto termico
(cio\`e l'eq.~\ref{Blast}), anzich\'e la semplice funzione esponenziale.
\newline
Si osservi che le definizioni qui date per il fattore di
estrapolazione, eq.~\ref{OurExtrap} ed eq.~\ref{BlastExtrap}, differiscono da quella
usata nel {\em paragrafo 4.8} (eq.~\ref{OperExtrY}), in cui si opera anche
l'integrazione sulla variabile rapidit\`a nella finestra di
accettanza sperimentale.
\newline
Nel caso di \PgXm e \PagXp, $m_T^{min}$\ \`e pari a circa $1.45$\ GeV/$c^2$\  
e risulta $S'_{exp}=1.421$\ per le \PgXm\ ed $S'_{exp}=1.389$\ per le \PagXp; 
per entrambe le particelle risulta  $S'_{blast}=1.354$. La differenza tra i 
due possibili metodi di estrapolazione \`e del $5\%$\  nel caso delle 
\PgXm, e del $2.5\%$\ nel caso delle \PagXp.  
\newline
Nel caso delle $\Omega$,  in questa analisi $m_T^{min}$\ \`e pari a circa 
$2.03$\ GeV/$c^2$\ e risulta quindi $S'_{exp}=2.86$\ se 
calcolato con $T_{app}=293$\ MeV, che \`e il valore della temperatura 
apparente misurato per il campione di  
\PgOm+\PagOp. Il modello fornisce invece un fattore di estrapolazione pi\`u piccolo,  
$S'_{blast}=2.30$. Ci\`o vuol dire che un'estrapolazione a partire dai risultati 
ottenuti da questo modello fornirebbe un tasso estrapolato inferiore dell'20\% 
rispetto a quelli calcolati estrapolando con la funzione 
$\mathcal{A}m_T \exp(-m_T/T_{app})$.  
Si ricordi tuttavia che in questa analisi l'estrapolazione \`e stata eseguita 
servendosi della temperatura apparente misurata da WA97, come discusso 
nel {\em paragrafo 4.8}, il che ha condotto 
a dei tassi estrapolati superiori del 18\% rispetto a quelli che si avrebbero 
con la $T_{app}$\ misurata.  
%Risulta $S'_{exp}=3.50$\ se calcolato con la temperatura apparente misurata da WA97, 
%
\subsection{Dipendenza dall'energia della collisione} 
\begin{figure}[htb]
\begin{center}
\includegraphics[scale=0.45]{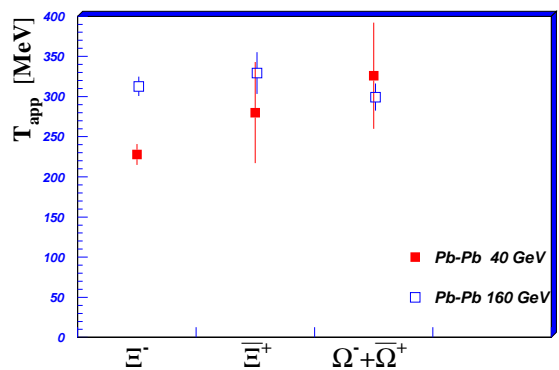}
\caption{Temperature apparenti misurate per le cascate nelle collisioni 
         Pb-Pb a 160 ed a 40 A GeV/$c$.}
\label{TempEnergy}
\end{center}
\end{figure}

Nelle collisioni Pb-Pb a 40 A GeV/$c$, disponendo allo stato attuale dell'analisi  
dei soli spettri di massa trasversa delle cascate ($\Xi$\ ed $\Omega$), non 
\`e ancora possibile separare in maniera quantitativa la temperatura di 
{\em ``freeze-out''} dal flusso trasverso. 
\newline
Le temperature apparenti possono per\`o fornire una prima indicazione.  
Per poter confrontare in maniera consistente i dati a 160 GeV/$c$\ con 
quelli a 40 GeV/$c$\ si sono raggruppate le tre classi di 
maggior centralit\`a definite a 160 GeV/$c$\ ($II$,$III$,$IV$). 
In tal modo, ad entrambe le energie, si considerano gli eventi 
pi\`u centrali corrispondenti a circa il 25\% della sezione 
d'urto anelastica.  
\newline
In fig.~\ref{TempEnergy} sono mostrate le temperature apparenti 
misurate per \PgXm, \PagXp\ e \PgOm+\PagOp\ in collisioni Pb-Pb a 
160 A GeV/$c$\ (quadretti vuoti blu) ed a 40 A GeV/$c$\ (quadretti pieni rossi). 
Entro gli errori sperimentali, anche a 40 A GeV/$c$, si osserva una stessa 
temperatura apparente per \PgXm\ e \PagXp.
\newline
Confrontando in funzione dell'energia, 
le \PagXp\ e le \PgOm+\PagOp, caratterizzate per\`o da un maggior errore statistico, 
sono tra loro compatibili, mentre per la temperatura apparente delle 
\PgXm, misurata con maggior precisione, 
si osserva una significativa ($\approx 6\, \sigma$) diminuzione a pi\`u 
bassa energia. Questo risultato pu\`o essere giustificato o da un minor flusso trasverso 
oppure, pi\`u semplicemente, dalla  minore energia iniziale disponibile nel sistema.  
\section{Rapporti di produzione}
La misura dei rapporti di produzione fornisce informazioni riguardanti l'equilibrio chimico 
nella produzione delle diverse specie di particelle. Nel {\em paragrafo 1.5} si \`e 
evidenziata l'importanza di stabilire il grado di saturazione nella produzione 
di stranezza, in quanto ci\`o potrebbe consentire di discriminare uno scenario di 
produzione adronica da uno che preveda la formazione di QGP.  
\newline
I rapporti di produzione, inoltre, vengono utilizzati nell'ambito di modelli termici,  
contenenti l'intera evoluzione dinamica delle collisioni, per determinare i 
parametri caratteristici della sorgente negli istanti immediatamente successivi 
all'urto, secondo un approccio di tipo statistico-termodinamico. 
Ad esempio, da un {\em ``best fit''} con uno di questi modelli ai 
rapporti di produzione misurati da WA97 per le particelle 
strane ed a quelli di molte altre particelle \`e stata ricavata  
la temperatura del {\em ``freeze-out chimico''} di circa 170 MeV~\cite{ModelBraun}.  
\subsubsection{Rapporti di produzione in Pb-Pb a 160 A GeV/$c$}
I rapporti di produzione riguardanti  \PKzS, $\Lambda$, $\Xi$\ ed $\Omega$\ 
prodotte in collisioni Pb-Pb a 160 A GeV/$c$ sono riportati in tab.~\ref{tab5.3}.  
\begin{table}[h]
\begin{center}
\begin{tabular}{|c|c||c|c|} \hline
 \multicolumn{4}{|c|}{\bf Pb-Pb a 160 A GeV/$c$} \\ \hline
 $\frac{\PgOm}{\PgXm}   $ & $0.144\pm0.014$ & $\frac{\PgXm}{\PgL} $ & $0.105\pm0.004$   \\
 $\frac{\PagOp}{\PagXp} $ & $0.245\pm0.035$ & $\frac{\PagXp}{\PagL} $ & $0.188\pm0.013$ \\ 
 $\frac{\PgOm+\PagOp}{\PgXm+\PagXp} $ & $0.166\pm0.013$ &
 $\frac{\PgXm+\PagXp}{\PgL+\PagL} $ & $0.116\pm0.004$  \\ \hline \hline
 $\frac{\PagL}{\PgL} $   & $0.150\pm0.005$  & $\frac{\PgL}{\PKzS}   $     & $0.87\pm0.06$ \\
 $\frac{\PagXp}{\PgXm} $ & $0.269\pm0.019$  & $\frac{\PagL}{\PKzS}  $     & $0.13\pm0.01$ \\
 $\frac{\PagOp}{\PgOm} $ & $0.46\pm0.07$    & $\frac{\PgL+\PagL}{\PKzS} $ & $0.997\pm0.07$ \\ \hline
\end{tabular}
\end{center}
\caption{Rapporti di produzione delle particelle strane nelle 
	collisioni Pb-Pb a 160 A GeV/$c$. Altri rapporti sono calcolabili 
	a partire dai valori in tab.~\ref{tab4.7}. 
\label{tab5.3}}
\end{table}
Essi sono stati ottenuti considerando i tassi di produzione calcolati 
nella regione di estrapolazione $p_T \, > \, 0$\ GeV/$c$\ e $|y-y_{CM}|<0.5$. 
Gli errori riportati sono solo statistici e non includono quelli di estrapolazione.  
\newline
Dai dati della tab.~\ref{tab5.3} si osserva un andamento crescente dei rapporti 
del tipo anti-particella/particella, in funzione del loro contenuto di stranezza.  
\subsubsection{Confronto tra i rapporti di produzione in p-Be, p-Pb e Pb-Pb a 160 A GeV/$c$}
Il calcolo della produzione di stranezza in interazioni tra nuclei pesanti assume 
particolare rilevanza se riferita a quella dovuta ad interazioni pi\`u semplici, 
quali quelle di tipo protone-nucleo. I rapporti di produzione delle particelle 
strane sono stati quindi confrontati con quelli misurati da WA97 nelle 
collisioni p-Pb~\cite{WA97web,Rocco} e p-Be~\cite{WA97web}, relativi 
alla stessa regione di estrapolazione. 
Nella tab.~\ref{tab5.5} sono riportate le ``variazioni'' dei rispettivi 
rapporti nel passaggio da collisioni p-Pb a collisioni Pb-Pb (prima colonna) e da 
collisioni p-Be a collisioni Pb-Pb (seconda colonna). Queste ``variazioni'' sono 
ottenute semplicemente dividendo tra loro i rapporti nelle collisioni Pb-Pb con
quelli p-Pb o p-Be.  
\begin{table}[h]
\begin{center}
\begin{tabular}{|c|c|c|} \hline
Rapporti & ${\rm E}\left( \frac{Pb-Pb}{p-Pb}\right)$ & ${\rm E}\left( \frac{Pb-Pb}{p-Be} \right)$ \\
 \hline
 $\frac{\PgOm}{\PgXm}   $ & $2.1\pm0.9$    &           \\
 $\frac{\PagOp}{\PagXp} $ & $  6\pm3  $    &           \\
 $\frac{\PgOm+\PagOp}{\PgXm+\PagXp} $ &  $2.8\pm1.0$ & $2.7\pm0.8$ \\
 \hline\hline
 $\frac{\PgXm}{\PgL}$    &  $2.06\pm0.18$  & $2.34\pm0.18$   \\
 $\frac{\PagXp}{\PagL}$  &  $2.35\pm0.34$  & $3.1\pm0.5$     \\
 $\frac{\PgXm+\PagXp}{\PgL+\PagL} $ & $2.07\pm0.16$ & $2.37\pm0.17$   \\
 \hline\hline
 $\frac{\PagL}{\PgL} $   & $0.58\pm0.05$   & $0.60\pm0.05$  \\
 $\frac{\PagXp}{\PgXm} $ & $0.71\pm0.11$   & $0.67\pm0.09$  \\
 $\frac{\PagOp}{\PgOm} $ & $2.2 \pm1.4 $   &    \\
 \hline
\end{tabular}
\end{center}
\caption{Variazioni nei rapporti di produzione degli iperoni nel passaggio 
         da collisioni Pb-Pb a collisioni p-Pb e p-Be. La variazione \`e 
	 calcolata dividendo il rapporto misurato in collisioni Pb-Pb 
	 con quello misurato in collisioni p-Pb o p-Be. 
	 I rapporti delle collisioni di riferimento p-A sono quelli 
	 misurati da WA97, come esposto nel testo.
\label{tab5.5}}
\end{table}
\newline
Si pu\`o notare che, per quanto riguarda i rapporti che coinvolgono particelle 
con diverso contenuto di stranezza,  
%(primi due blocchi orizzontali in tab.~\ref{tab5.5}), 
la variazione 
%corrisponde ad un incremento tanto maggiore quanto pi\`u grande \`e 
aumenta con 
il contenuto di stranezza delle particelle coinvolte. L'incremento, inoltre, risulta 
maggiore per le anti-particelle che per le particelle. Poich\'e nelle normali interazioni 
adroniche 
%tendono ad attenuare il contenuto (di stranezza) di 
\`e sfavorita la produzione di 
adroni multi-strani, 
ed ancor pi\`u quella di anti-barioni multi-strani, per via delle reazioni con scambio 
di {\em flavour}, l'incremento osservato non \`e compatibile con uno scenario di 
reazione coinvolgente unicamente collisioni tra adroni. Tale risultato \`e invece 
in linea con uno scenario che prevede la formazione  e l'emissione da una sorgente 
di materia deconfinata. 
\newline
Per quanto riguarda i rapporti del tipo anti-particella/particella, le variazioni 
riportate nella tab.~\ref{tab5.5} corrispondono ad una decrescita per le \PagL\ e 
le \PagXp, attribuibile a fenomeni di assorbimento degli anti-barioni che si 
manifestano in maniera maggiore nel sistema di collisione pi\`u massivo. Il 
rapporto relativo alle $\Omega$, al contrario, pur in presenza di un grande errore 
statistico, presenta un incremento nel passaggio da interazioni p-A a quelle Pb-Pb, 
probabilmente in seguito alla concomitanza di due fattori: il modesto assorbimento 
di \PgOm\ e \PagOp\ nel mezzo adronico (nel caso Pb-Pb) e la soppressione della 
produzione di barioni con tre unit\`a di stranezza in interazioni adroniche, 
a causa dell'alta soglia in massa (nel caso p-A). 
%Questa anomalia rafforza l'ipotesi 
%che le $\Omega$\ possano essere prodotte in collisioni Pb-Pb secondo un meccanismo 
%differente da quello in interazioni p-Pb. 
\subsubsection{Rapporti di produzione in Pb-Pb a 40 A GeV/$c$}
In tab.~\ref{tab5.4} sono calcolati i rapporti di produzione per $\Xi$\ ed $\Omega$\ 
in collisioni Pb-Pb a 40 A GeV/$c$. \`E anche riportato il valore del rapporto 
\PagL/\PgL, sebbene non ancora corretto per accettanza ed efficienza. Questo valore 
fornisce infatti gi\`a una attendibile indicazione, in quanto le correzioni sui rapporti 
anti-particella/particella sono piccole, in virt\`u della simmetria tra lato destro e 
sinistro dei rivelatori e dell'inversione della direzione del campo magnetico eseguita 
durante la presa dati.   
\begin{table}[h]
\begin{center}
\begin{tabular}{|c|c||c|c|} \hline
\multicolumn{4}{|c|}{\bf Pb-Pb a 40 A GeV/$c$} \\ \hline
 $\frac{\PagL}{\PgL} $   & $0.0248\pm0.0007$ & $\frac{\PgOm}{\PgXm}  $ & $0.06\pm0.03$ \\
 $\frac{\PagXp}{\PgXm} $ & $0.063\pm0.015$ & $\frac{\PagOp}{\PagXp} $ & $0.26\pm0.14$ \\
 $\frac{\PagOp}{\PgOm} $ & $0.3\pm0.2$   & $\frac{\PgOm+\PagOp}{\PgXm+\PagXp} $ & $0.07\pm0.03$ 
 \\ \hline
\end{tabular}
\end{center}
\caption{Rapporti di produzione delle particelle strane nelle
        collisioni Pb-Pb a 40 A GeV/$c$. Altri rapporti sono calcolabili
        a partire dai valori in tab.~\ref{tab4.7}.
\label{tab5.4}}
\end{table}
Questi risultati saranno commentati nello studio in funzione dell'energia della 
collisione, confrontando anche con i dati dell'esperimento STAR alla pi\`u alta 
energia di RHIC. 
\section{Produzione di stranezza in funzione della centralit\`a}
\begin{figure}[hbt]
\begin{center}
\includegraphics[scale=0.52]{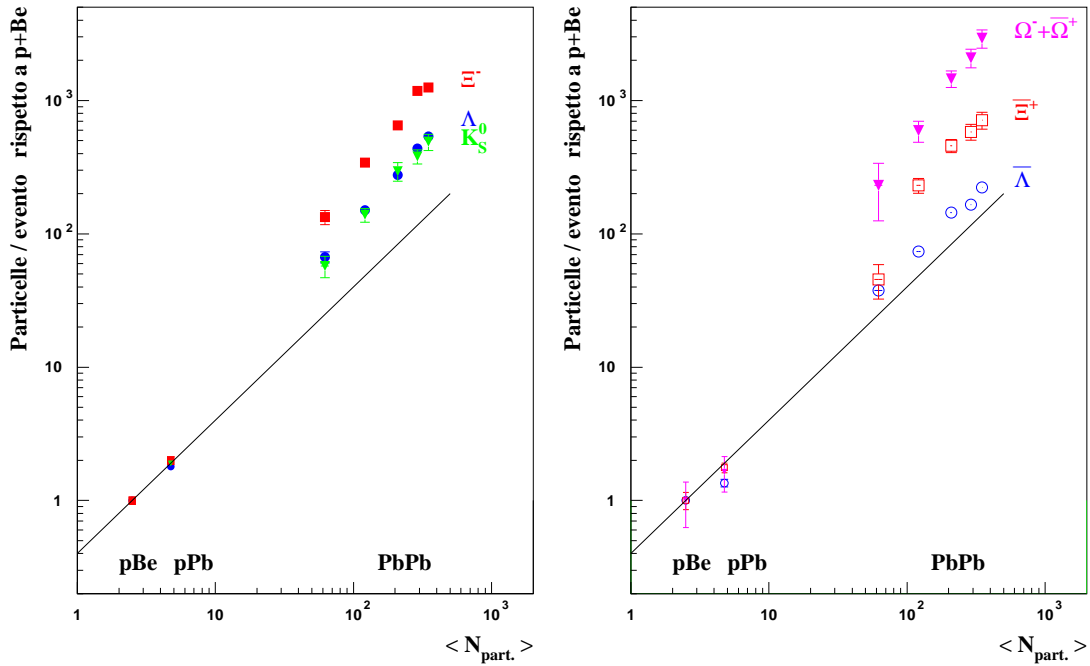}
\caption{Numero di particelle per evento, in unit\`a della corrispondente 
produzione osservata in collisioni p-Be, in funzione del numero di 
partecipanti. I punti delle interazioni di riferimento p-Be e p-Pb sono le 
misure della collaborazione WA97~\cite{WA97web,Rocco}.  
La linea continua indica la proporzionalit\`a col numero di partecipanti. 
I \PKzS\ sono stati riscalati in modo da disporre il punto relativo alle 
interazioni p-Pb sulla linea di proporzionalit\`a, non disponendo del tasso 
di produzione in collisioni p-Be.}
\label{xyz}
\end{center}
\end{figure}

In fig.~\ref{yieldThesis} \`e stata mostrata la dipendenza  
del numero di \PKzS, $\Lambda$, $\Xi$\ ed $\Omega$\ prodotte per evento 
(cio\`e i tassi di produzione estrapolati) in funzione del numero di partecipanti, 
nelle interazioni p-Be, p-Pb e Pb-Pb.   
%\newline
Le barre verticali d'errore in fig.~\ref{yieldThesis} indicano le sole incertezze 
statistiche.   
%e non includono i sistematici dovuti al {\em ``feed-down''} e quelli introdotti dalla procedura 
%di estrapolazione. Le barre orizzontali di errore indicano invece l'indeterminazione 
%sul numero medio di partecipanti in ciascuna classe.
\newline
In fig.~\ref{xyz}, la produzione di ciascun tipo di particella \`e stata scalata in modo 
da porre ad uno i relativi valori per interazioni p-Be. 
La produzione di iperoni risulta cos\`i espressa in unit\`a 
della corrispondente produzione per 
interazioni p-Be ed \`e confrontata con la linea continua proporzionale al numero 
di partecipanti, passante per il punto comune relativo all'interazione p-Be. 
Nel caso dei \PKzS, non disponendo del tasso di produzione in collisioni p-Be ma
di quello in collisioni p-Pb, si sono riscalati tutti i punti in modo che quello 
p-Pb si disponga sulla linea di proporzionalit\`a.  
\newline
Si pu\`o osservare che la produzione di tutte le particelle strane cresce con la 
centralit\`a molto pi\`u velocemente rispetto a quanto previsto dalla semplice 
proporzionalit\`a col numero di partecipanti (e quindi dal semplice effetto di 
sovrapposizione tra nucleoni interagenti) e questo incremento aumenta in 
funzione del contenuto di stranezza della particella, essendo maggiore per 
le $\Omega$\ rispetto alle $\Xi$, e per le $\Xi$\ rispetto alle $\Lambda$\   
ed ai \PKzS.   
L'incremento delle $\Omega$\ rispetto alla crescita col numero di partecipanti, in 
particolare, si manifesta con pi\`u di un ordine di grandezza. 
\subsection{Incremento nella produzione di stranezza rispetto alle collisioni p-A}
\begin{figure}[htb]
\begin{center}
\includegraphics[scale=0.52]{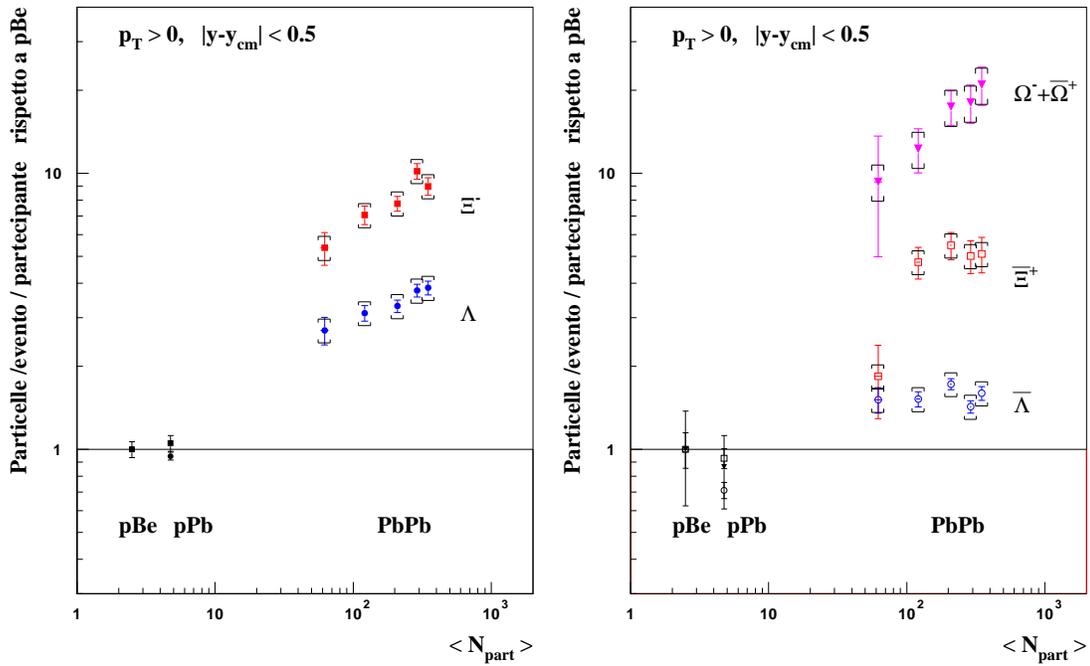}
\caption{Iperoni prodotti per evento e per partecipante normalizzati   
         a quelli dell'interazione p-Be, in funzione  del numero medio di 
	 nucleoni partecipanti.  
         Per le barre di errore si \`e adoperato il 
	 simbolo $_{\sqcup}^{\sqcap}$\ (in nero con le ``alette'' laterali) 
         % \( {\overset{\sqcap}{\underset{\sqcup}{ }}} \)
         per indicare l'errore di natura sistematica, quello $ _{-}^{-}$\ 
	 (nello stesso colore dei punti sperimentali, senza ``alette'')  
	 per indicare l'errore statistico.}
\label{EnhancQM02}
\end{center}
\end{figure}

\begin{figure}[hbt]
\begin{center}
\includegraphics[scale=0.42]{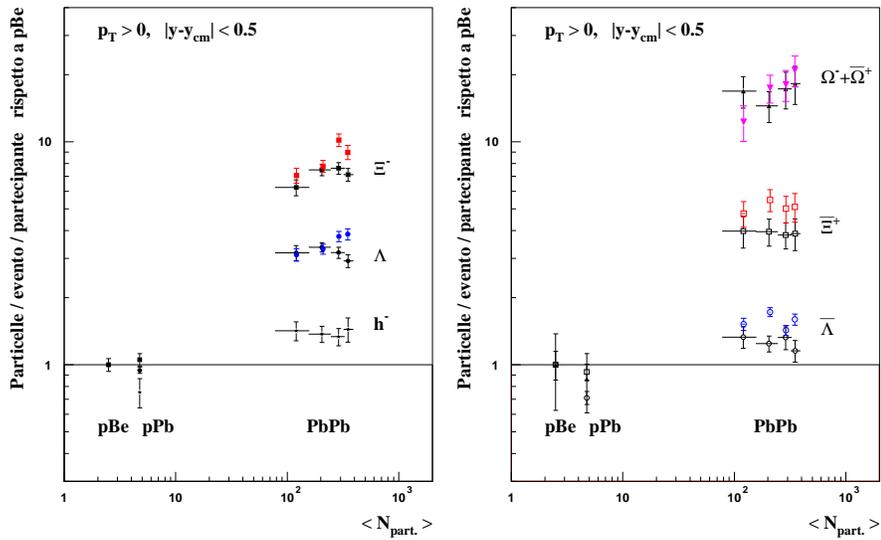}
\caption{Confronto dell'incremento della produzione degli iperoni
         in collisioni Pb-Pb a 160 A GeV/$c$\ con quello misurato da WA97
         nella regione di centralit\`a comune $N_{part}\ge100$.
         I dati di riferimento delle collisioni p-Be e p-Pb sono quelli di WA97.}
\label{WA97Comp}
\end{center}
\end{figure}

\begin{figure}[htb]
\begin{center}
\includegraphics[scale=0.45]{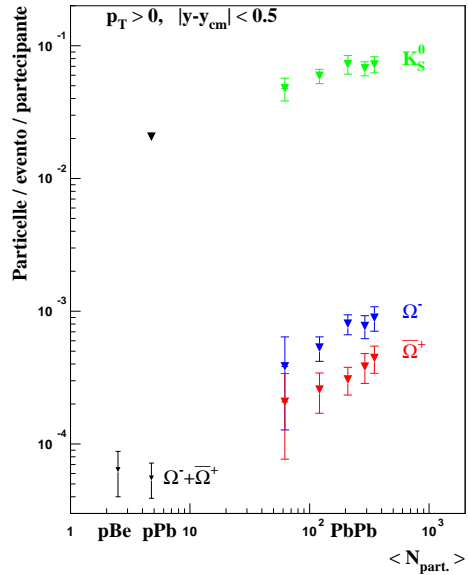}
\caption{Particelle prodotte per evento normalizzate al numero medio di 
nucleoni partecipanti in funzione del numero di partecipanti. I punti 
p-Be e p-Pb (misurati da WA97) per le $\Omega$\ si riferiscono al 
segnale di \PgOm+\PagOp.}
\label{OmegaK0s}
\end{center}
\end{figure}

\begin{figure}[p]
\begin{center}
\includegraphics[scale=0.50]{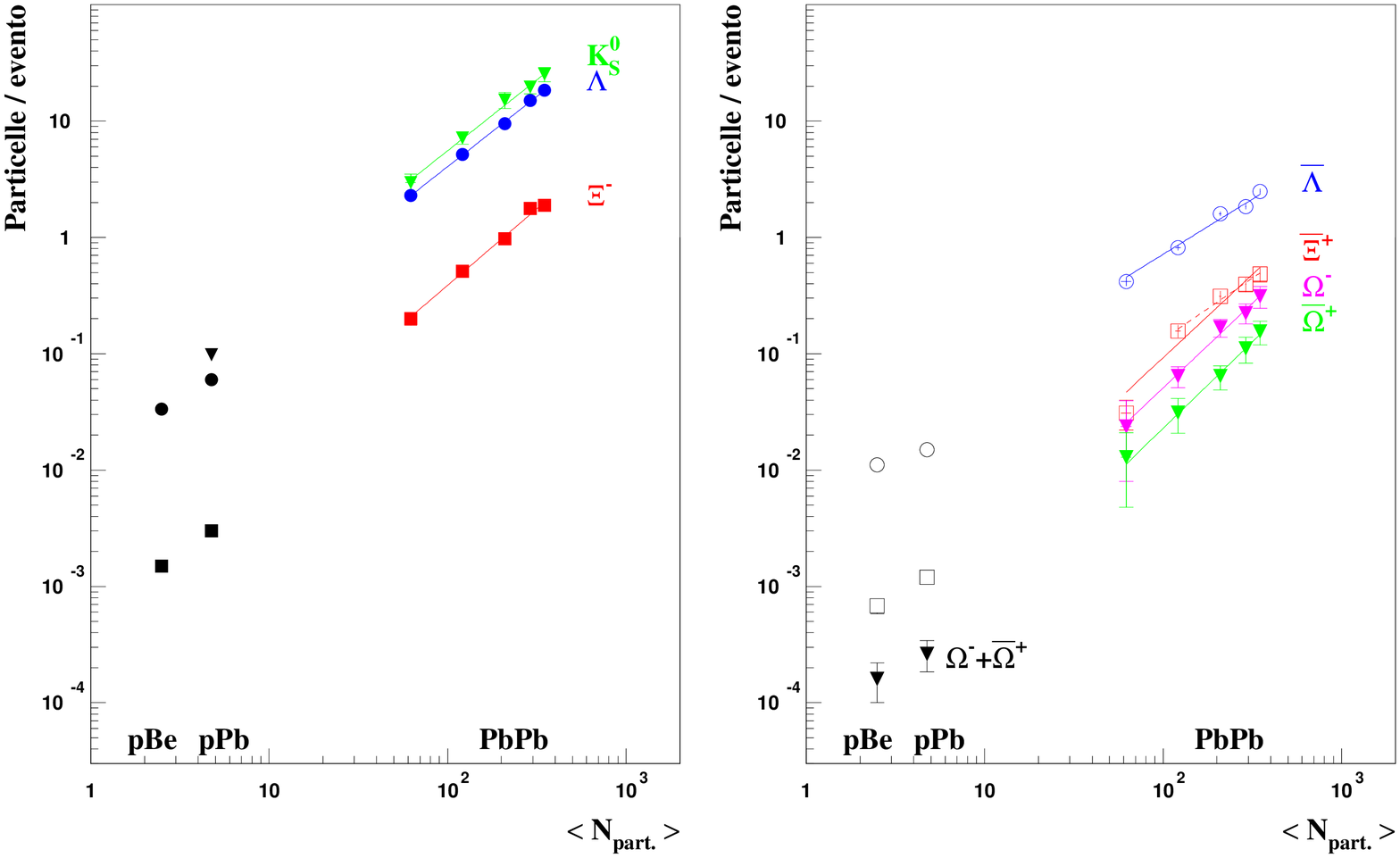}
\caption{{ {\em ``Best fit''} 
	della funzione potenza (${\rm cost}\, N_{part}^\beta$) ai tassi 
 	 di produzione delle particelle strane in funzione della 
     centralit\`a, per le sole collisioni Pb-Pb a 160 A GeV/$c$. 
	 I risultati del {\em ``best fit''} per l'esponente $\beta$\ sono  
	 riportati nella prima colonna della tab.~\ref{tab5.6}.
	 Per le \PagXp\ si \`e anche eseguito il {\em ``best fit''} escludendo il
	 punto di centralit\`a pi\`u periferica (curva tratteggiata).} 
\label{PowerLawFit_a}}
%\end{center}
%\end{figure}
%\begin{figure}[p]
%\begin{center}
\vspace{0.7cm}
\includegraphics[scale=0.50]{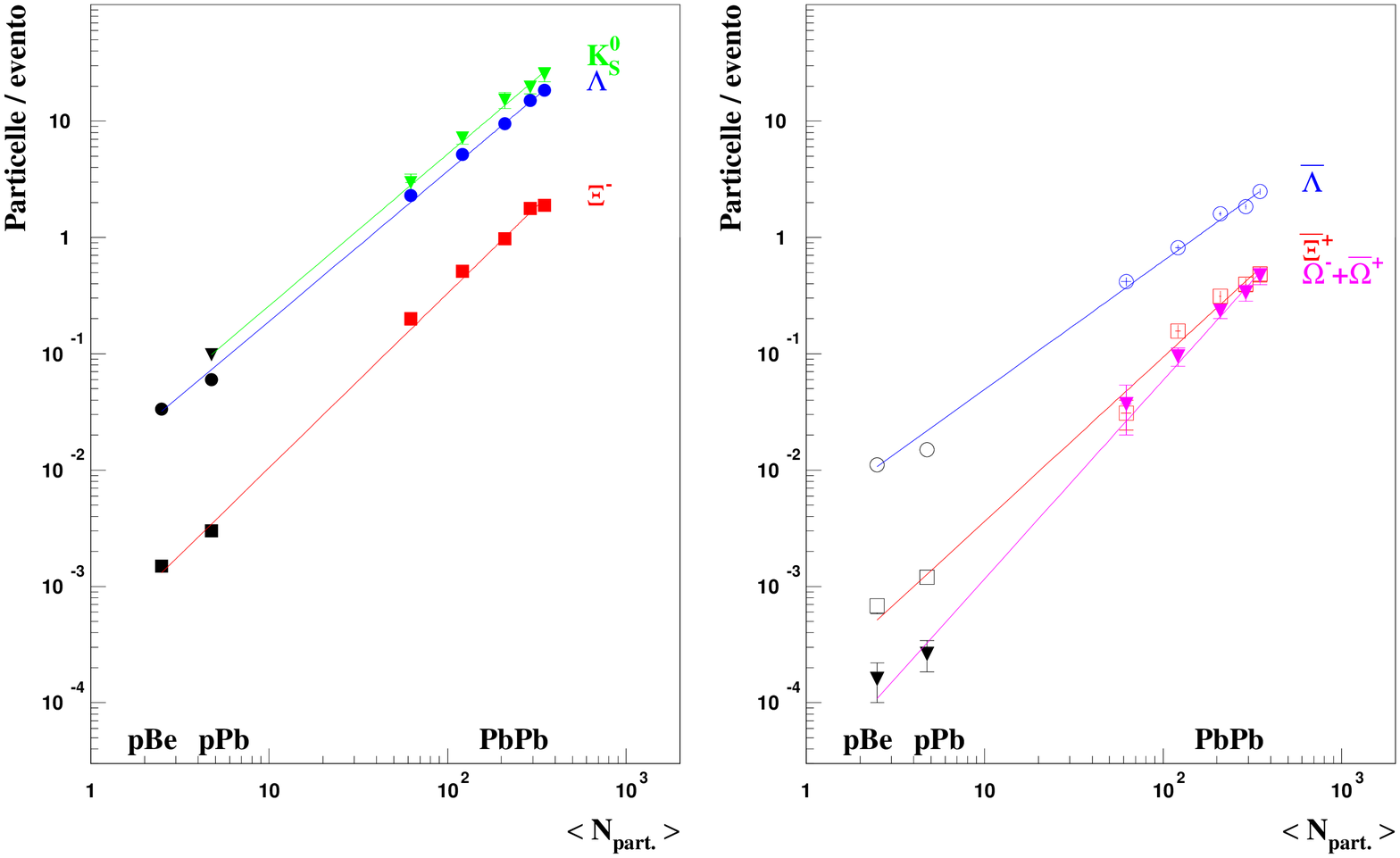}
\caption{{ {\em ``Best fit''} della funzione potenza ai tassi di produzione 
	  delle particelle strane in collisioni p-Be (non disponibile per i \PKzS), 
	  p-Pb e Pb-Pb. I risultati del {\em ``best fit''} sono riportati nella seconda 
	  colonna della tab.~\ref{tab5.6}.}
\label{PowerLawFit_b}}
\end{center}
\end{figure}

Per poter valutare in maniera quantitativa l'incremento 
della produzione di particelle strane, rispetto alla normale sovrapposizione 
di collisioni calcolata (in termini di nucleoni partecipanti) a partire dalle 
interazioni di riferimento (p-A), si definisce la quantit\`a:  
\begin{equation}
E = \left( \frac{<Yield>}{<N_{part}>}\right)_{{\rm Pb-Pb}} /
    \left( \frac{<Yield>}{<N_{part}>}\right)_{{\rm p-Be}}  \, .
\label{Enanch}
\end{equation}
Nell'eq.~\ref{Enanch} il numeratore \`e pari al tasso di produzione 
nell'intervallo di centralit\`a considerato (eq.~\ref{OperExtrY}), 
diviso il numero medio di partecipanti in tale intervallo, numero 
calcolato col modello di Glauber come discusso nel {\em paragrafo 4.5};  
il denominatore \`e il tasso 
di produzione misurato in collisioni p-Be diviso il numero 
medio di partecipanti ($\simeq2.5$) in tali collisioni.  
In fig.~\ref{EnhancQM02} \`e mostrato l'andamento dell'incremento di 
produzione (specifico), ora definito, in funzione del numero di nucleoni 
partecipanti alla collisione.  
Oltre agli errori statistici \`e riportata anche la stima attuale di   
quelli sistematici, in cui si tiene anche conto dell'errore di  
estrapolazione.   
%({\em qui dovr\`o spiegare meglio !}).  
\newline
Tutti i punti rappresentativi della quantit\`a $E$\ si trovano al di sopra della 
linea continua orizzontale, corrispondente ad assenza di incremento. 
Gli incrementi risultano crescenti con il contenuto di stranezza della particella. 
Il massimo incremento si osserva per la particella $\Omega$, con maggior 
contenuto di stranezza, e corrisponde ad un fattore circa 20 per la classe 
di maggior centralit\`a. 
Questi risultati risultano difficilmente interpretabili in un contesto di 
produzione adronica ed un incremento cos\`i pronunciato della produzione 
degli stati \PgOm$(sss)$\ e \PagOp$(\bar{s}\bar{s}\bar{s})$\ nel passaggio 
da interazioni protone-nucleo a quelle nucleo-nucleo, presente anche dopo 
aver considerato l'effetto di sovrapposizione dei nucleoni interagenti, indica 
chiaramente la possibilt\`a che le particelle possano essere prodotte da una 
sorgente di quark e gluoni liberi e deconfinati.
% Confronto con WA97 qui
\newline
In fig.~\ref{WA97Comp} \`e mostrato un confronto tra questi risultati (punti  
colorati) 
e quelli di WA97 (punti in bianco e nero), nell'intervallo di centralit\`a 
coperto %da WA97, pari a $N_{part}\ge100$, 
da WA97 ($N_{part}\ge100$) 
corrispondente con ottima 
approssimazione alle quattro classi di maggior centralit\`a di NA57. 
Considerando globalmente i risultati,  
\`e cos\`i confermato l'incremento della produzione di stranezza nelle 
collisioni Pb-Pb a 160 A GeV/$c$\ trovato da WA97, rispetto a quanto 
atteso dalla sovrapposizioni di ``normali'' (cio\`e in assenza di plasma) 
collisioni nucleone-nucleone.
%approssimate sperimentalmente nel miglior modo possibile dalle collisioni p-Be.
\`E altres\`i confermata la stessa gerarchia nell'incremento della produzione
di iperoni.
%: l'incremento aumenta con l'aumentare del contenuto di stranezza dell'iperone.  
\newline
Entrando pi\`u nei dettagli, risulta che i tassi di produzione 
misurati da WA97 sono inferiori di un 10-20\% per tutte le specie; inoltre, 
almeno per le \PgL\ e le \PgXm, si osserva una dipendenza dalla 
centralit\`a significativamente differente. 
Questa differenze sono state investigate e comprese~\cite{Kristin,Quercigh} 
e sono pricipalmente imputabili al modo non del tutto soddisfacente  
con cui si correggeva per un'istanbilit\`a del fascio di piombo utilizzato 
da WA97 (i dati dei protoni non presentano invece questo problema).  
\newline
Come gi\`a discusso nel primo capitolo, 
considerando {\em globalmente} le indicazioni fornite dai diversi esperimenti coinvolti 
nella ricerca del QGP all'SPS ed  ``in primis'' quelle qui illustrate  
sulla produzione di stranezza e 
quelle sulla soppressione della produzione di stati di charmonio,  
%diventa fuori da ogni ragionevole dubbio affermare che 
si pu\`o affermare, con ragionevole certezza, che 
nelle collisioni Pb-Pb pi\`u centrali venga formato il QGP. 
Se si accetta questo scenario,  la domanda che subito si pone diventa la seguente: 
in che maniera avviene la transizione tra il regime di produzione in 
ambiente adronico (collisioni p-Be e p-Pb) e quello in presenza di QGP 
(collisioni Pb-Pb pi\`u centrali)~?
%\`E interessante notare come la crescita degli incrementi con il contenuto 
%di stranezza sia pi\`u  sostenuta per le antiparticelle 
\newline
Gli incrementi di tutte le specie, ad eccezione delle \PagL, mostrano una 
crescita col numero di nucleoni partecipanti. 
\newline
Nel caso delle \PagXp\ si osserva una decrescita improvvisa nella classe di centralit\`a pi\`u 
periferica: passando dalla classe $I$\ ($<N_{part}>=121$) alla classe 
 $0$\ ($<N_{part}>=62$) il tasso di produzione per partecipante decresce 
di un fattore $2.6$, corrispondente ad un effetto pari a $3.5\, \sigma$. 
Le sole \PagXp\  suggerirebbero dunque un improvviso raggiungimento 
delle condizioni per la formazione del QGP tra i $60$\ ed i $120$\ partecipanti. 
Sar\`a interessante vedere se questo risultato verr\`a confermato studiando il 
campione di $\Xi$\ dei dati dell'anno 2000. 
%
%{\ em here comment on global increase with respect to the line, but we observe
%         deviation from linearity within the five Pb-Pb points.}
\newline
Considerando le altre specie, \PgXm, \PgL, \PgOm+\PagOp, si osserva un incremento 
crescente con maggior continuit\`a.   
Ci\`o \`e ugualmente vero anche per le \PgOm\ e le \PagOp\ considerate 
separatamente, e per i \PKzS, come risulta considerando la 
dipendenza dei tassi di produzione per
partecipante dalla centralit\`a~\footnote{La normalizzazione ai dati
del p-Be, di cui non si dispone, introduce solamente un fattore di scala comune
per tutti i punti, senza modificare gli andamenti.}, mostrata nella fig.~\ref{OmegaK0s}.
In questa seconda prospettiva si \`e eseguita una
procedura di {\em ``best fit''} assumendo una dipendenza del tipo
$N_{part}^\beta$, 
applicata inizialmente ai punti relativi alla sola 
produzione in collisioni Pb-Pb, come mostrato in fig.~\ref{PowerLawFit_a}. 
I risultati del  {\em ``best fit''} forniscono i valori 
dell'esponente $\beta$\ riportati nella prima colonna della tab.~\ref{tab5.6}. 
Quindi si \`e ripetuta la procedura del {\em ``best fit''}, includendo anche 
i punti delle interazioni di referimento p-A, come mostrato nella 
fig.~\ref{PowerLawFit_b}.    
I risultati  dei nuovi esponenti $\beta$, cos\`i determinati,
sono riportati nella seconda colonna della tab.~\ref{tab5.6}.  
\begin{table}[h]
\begin{center}
\begin{tabular}{|c|lr|lr|} \hline 
 Particelle    & $\beta$\ (Pb-Pb) & $\chi^2/ndf$ & 
                    $\beta$\  (p-A, Pb-Pb) & $\chi^2/ndf$ \\ \hline\hline
 \PKzS 		&$1.22\pm0.19$&	$0.6/3$ & $1.30\pm0.02$ & $1.3/4$ \\ \hline
 \PgL		&$1.21\pm0.05$&	$0.9/3$ & $1.29\pm0.01$ & $55/5$   \\ \hline
 \PagL		&$1.00\pm0.07$&	$8.6/3$  & $1.11\pm0.01$ & $63/5$   \\ \hline
 \PgXm 		&$1.32\pm0.09$&	$4.8/3$ & $1.50\pm0.01$ & $20/5$ \\ \hline
 \PagXp		&\begin{tabular}{l}
                 $1.43\pm0.19$\\ 
		 $1.05\pm0.19$
		 \end{tabular}& 
		             \begin{tabular}{r}
				$9.1/3 $\\
				$0.6/2 $
			     \end{tabular}
		 		& $1.41\pm0.03$ & $12/5$ \\ \hline
 \PgOm+\PagOp   &$1.46\pm0.23$& $0.35/3$& $1.71\pm0.05$ & $3.4/5$ \\ \hline
 \PgOm 		&$1.45\pm0.28$& $0.62/3$\\ \cline{1-3}
 \PagOp 	&$1.49\pm0.36$& $0.19/3$\\ \cline{1-3}
\end{tabular}
\end{center}
\caption{Valori di {\em ``best fit''} dell'esponente $\beta$\ e del $\chi^2$\ 
         ridotto nell'approssimazione di una dipendenza 
	 dei tassi di produzione dal numero di nucleoni partecipanti  
	 secondo una funzione di potenza ($\propto N_{part}^\beta$). 
	 Nella prima colonna si sono aprrossimati i soli cinque 
	 punti dell'interazione Pb-Pb, nella seconda anche quelli delle 
	 interazioni di riferimento p-Be e p-Pb. 
	 Nel {\em ``best fit''} ai soli punti Pb-Pb delle \PagXp, si \`e anche   
	 provato ad escludere il punto di minor centralit\`a (classe $0$), 
	 l'unico che si discosta significativamente da una semplice dipendenza
	 lineare.  
\label{tab5.6}}
\end{table}
\newline
Dai risultati di questo studio si pu\`o affermare che una 
descrizione con la legge di potenza sull'intero intervallo di collisioni 
non sembra una scelta appropriata. Dai valori del $\chi^2$\ si deduce infatti 
che questa descrizione pu\`o essere rigettata con 
un livello di confidenza del 97.5\% per tutte le particelle ad esclusione dei 
\PKzS\ e delle $\Omega$. 
Queste ultime (le $\Omega$)  
sono la specie con l'errore  statistico associato pi\`u elevato, 
%ed i  risultati sui \PKzS\ qui presentati provengono da un'analisi 
%preliminare condotta su un piccolo numero di particelle.
per i \PKzS\ bisogna tener presente che si dispone di un solo punto sperimentale (p-Pb)  
relativo alle interazioni di riferimento e  che i risultati qui presentati provengono 
da un'analisi preliminare condotta su un piccolo numero di particelle. 
Per tutte le altre specie questi risultati suggeriscono un {\em brusco} 
salto passando dalle collisioni p-A a quelle Pb-Pb. 
\newline
Limitandosi a considerare l'approssimazione con la legge di potenza 
sui soli punti delle collisioni Pb-Pb, tale descrizione si adatta 
molto meglio ai risultati sperimentali. Gli esponenti $\beta$, ad 
esclusione di quello relativo alle \PagL, sono tutti significativamente 
superiori ad uno. L'eliminazione del punto di minor centralit\`a (la 
classe $0$) modifica apprezzabilmente il risultato del {\em ``best fit''}
solo nel caso delle \PagXp. 
%\newline
Ci\`o vuol dire che l'incremento della produzione di stranezza~\footnote{
Per incremento si intende la definizione data nell'eq.~\ref{Enanch}.} non 
satura completamente con il numero di partecipanti, anche nelle collisioni 
Pb-Pb pi\`u centrali corrispondenti ad $N_{part} \apprge 120$, ma continua ad 
aumentare. La dipendenza \`e tanto pi\`u rapida quanto maggiore \`e 
il contenuto di stranezza della specie considerata: 
il valore dell'esponente passa da  
$\beta \approx 1.2$\ per le particelle con una unit\`a di stranezza (\PgL\ e \PKzS)  
a $\beta \approx 1.45$\ per le particelle con tre unit\`a di stranezza  
(\PgOm\ e \PagOp). 

\noindent
La conclusione pi\`u significativa che si trae da questo studio \`e pertanto  
la seguente: 
passando dalle collisioni p-A a quelle Pb-Pb si osserva un salto 
talmente brusco che neanche una legge di potenza riesce 
a raccordare in maniera continua; considerando le sole collisioni Pb-Pb la 
dipendenza dal numero di partecipanti diviene sostanzialmente continua ed 
aumenta all'aumentare del contenuto di stranezza.

%Per poter discriminare tra i due scenari proposti diventa fondamentale ridurre gli 
\noindent
Per poter trarre delle conclusioni ancor pi\`u precise, si dovrebbe tentare 
di ridurre in maniera significativa gli errori statistici e, simultaneamente,
mantenere un buon controllo di quelli sistematici. 
Per quanto riguarda le $V^0$\ (\PgL\ e \PKzS), la procedura da seguire \`e quella 
di implementare metodi alternativi di correzione per accettanza ed efficienza, 
in modo da poter utilizzare interamente il campione di particelle raccolte. 
Venedo alle cascate, per quel che riguarda le $\Xi$\ si raddoppier\`a (perlomeno) la 
consistenza statistica del campione di particelle, mentre per le $\Omega$\ si 
pu\`o solo pensare di seguire un'altro approccio: rinunciare alla purezza del 
campione, con dei criteri di selezione molto meno selettivi. In tal modo 
diventa per\`o necessario disporre di una buona descrizione del fondo, con 
metodi, ancora da sviluppare, analoghi a quello discusso nel 
{\em paragrafo 4.2} a proposito delle $V^0$.
\newline
Il metodo d'indagine pi\`u efficace sarebbe per\`o quello 
di esplorare l'intervallo di centralit\`a compreso tra $<N_{part}>=62$\
(classe 0 delle interazioni Pb-Pb) ed $<N_{part}>=4.5$ (interazioni p-Pb).
La Collaborazione NA57 ha quindi avanzato la richiesta di studiare le collisioni  
di un sistema pi\`u leggero del Pb-Pb, quale l'In-In ({\em cfr. paragrafo 2.7}).  
\section{Produzione di stranezza in funzione della energia della collisione}
\begin{figure}[htb]
\begin{center}
\includegraphics[scale=0.45]{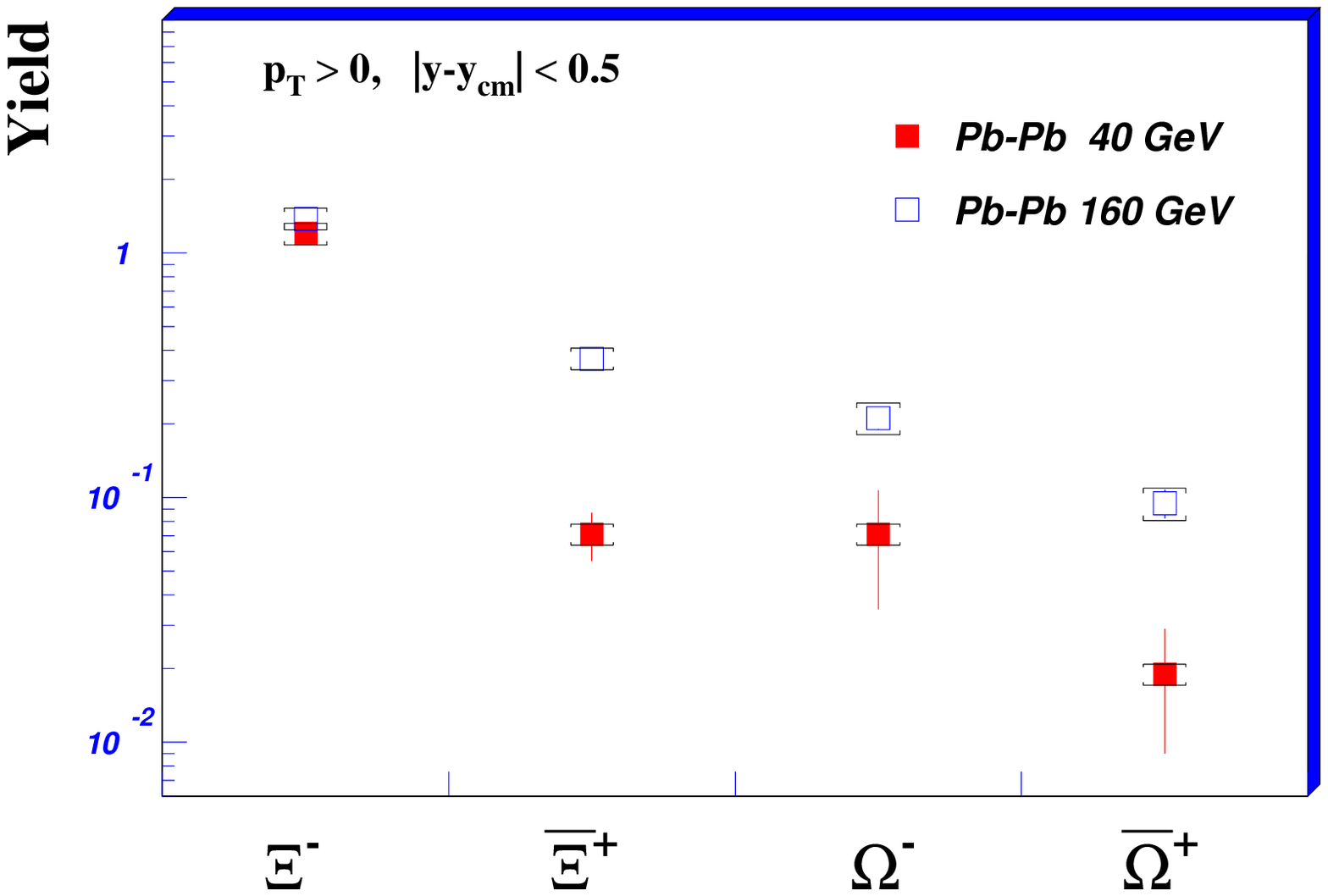}
\includegraphics[scale=0.45]{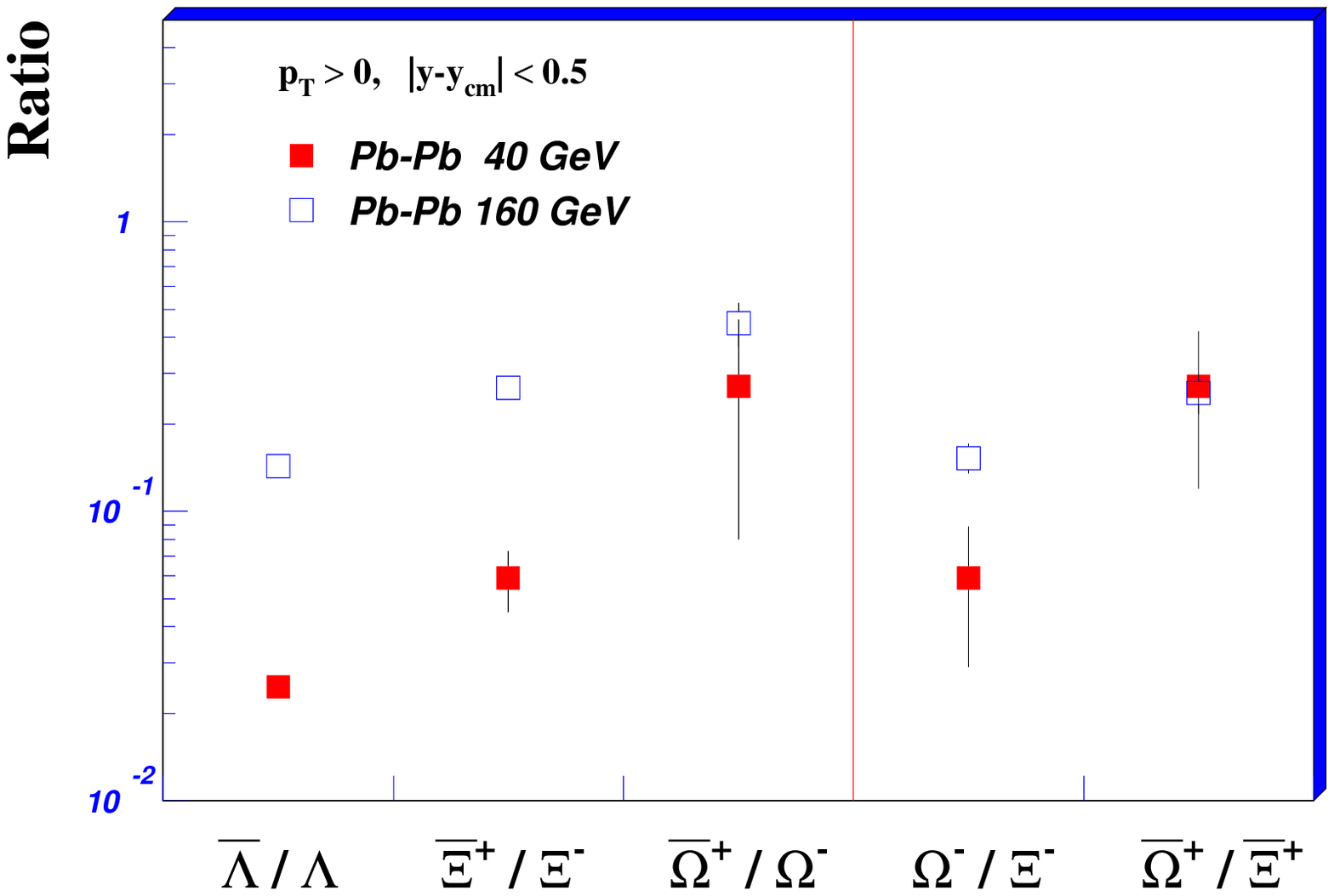}
\caption{{\em A sinistra:}
         confronto tra il tasso di produzione ({\em ``yield''})di
         $\Xi$\ ed $\Omega$\ nelle collisioni Pb-Pb a 160 A GeV/$c$ e quello
	 nelle collisioni  Pb-Pb a 40 A GeV/$c$.
	 \newline {\em A destra:}
         rapporti di produzione degli iperoni strani in collisioni Pb-Pb 
         a 160 e 40 A GeV/$c$.}
\label{RatEnergy}
\end{center}
\end{figure}

In fig.~\ref{RatEnergy} \`e mostrato il confronto tra la produzione di 
$\Xi$\ ed $\Omega$\  in collisioni Pb-Pb a 160 A GeV/$c$\ e quella in collisioni 
Pb-Pb a 40 A GeV/$c$. Per questo confronto, i tassi di produzione 
nelle collisioni a 160 GeV/$c$\ sono stati ricalcolati considerando 
unicamente le tre classi di maggior centralit\`a (II-III-IV); in tal modo, 
ad entrambe le energie, si considerano le collisioni pi\`u centrali relative 
a circa il $25\%$\ della sezione d'urto analestica. Nella fig.~\ref{RatEnergy} 
sono anche mostrati i rapporti di produzione calcolabili a partire da 
questi tassi di produzione, ed il rapporto \PagL/\PgL, che ancora deve 
essere corretto per accettanza ed efficienza come discusso nel 
{\em paragrafo 5.3}. 
\newline
Dalla fig.~\ref{RatEnergy} si deduce che  la produzione di 
\PgXm\ \`e circa la stessa a 40 ed a 160 GeV, mentre le \PagXp, \PgOm\ e \PagOp\ 
vengono prodotte meno abbondantemente a pi\`u bassa energia. Questo andamento 
\`e giustificato tenendo conto che, a pi\`u bassa energia, si attende una 
maggiore densit\`a barionica: essa favorisce la produzione di iperoni che 
condividono almeno un quark di valenza con i nucleoni collidenti. La maggior 
densit\`a barionica \`e conseguenza del maggior frenamento dei nuclei dopo 
l'urto %che si ha 
a pi\`u bassa energia, come discusso nel {\em paragrafo 1.4.2}. 
Ad energie molto pi\`u elevate (come ad esempio al RHIC), invece, i due nuclei collidenti 
diventano l'uno ``trasparente'' all'altro (nel senso discusso nel {\em paragrafo 1.4.2}) 
e la densit\`a barionica della {\em ``fireball''} approssima zero.   
\newline 
Dal grafico dei rapporti di produzione, si osserva che, andando da 160 a 40 GeV, 
il rapporto anti-iperone/iperone diminuisce. L'effetto maggiore si ha nel caso 
delle $\Lambda$\ ma \`e quasi nullo (entro gli errori) per le $\Omega$. 
Per i rapporti $\Omega$\ su $\Xi$, si osserva una riduzione di un 
fattore tre nel caso di \PgOm/\PgXm, mentre il rapporto \PagOp/\PagXp\ \`e 
quasi uguale, come gi\`a osservato commentando i tassi di produzione.  

Per poter stabilire se si ha gi\`a incremento nella produzione di particelle
all'energia di 40 GeV/$c$\ per nucleone, ed in quale quantit\`a, bisogna
attendere la disponibilit\`a dei dati di riferimento p-Be alla stessa energia.

\begin{figure}[htb]
\begin{center}
\includegraphics[scale=0.45]{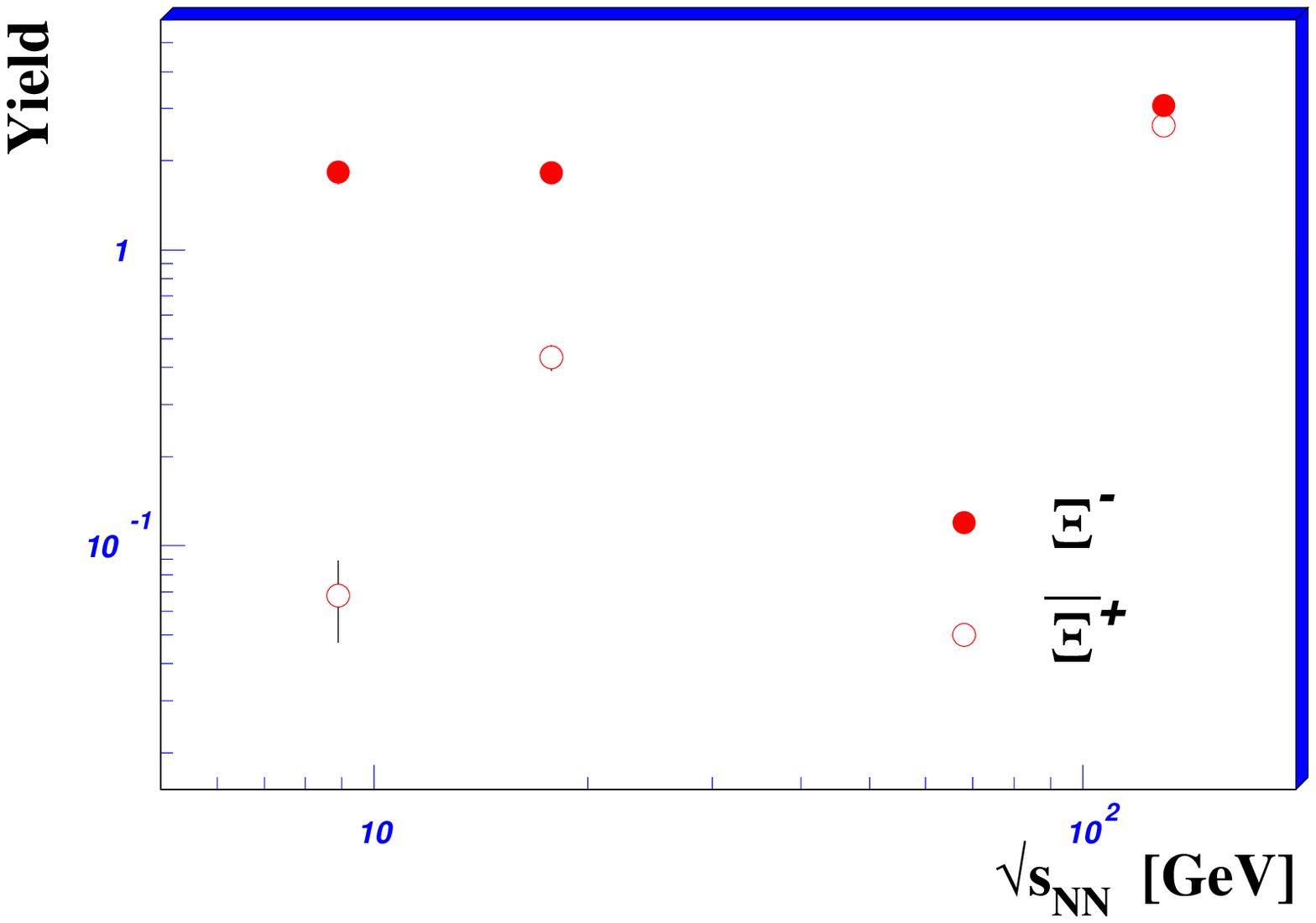}
\includegraphics[scale=0.45]{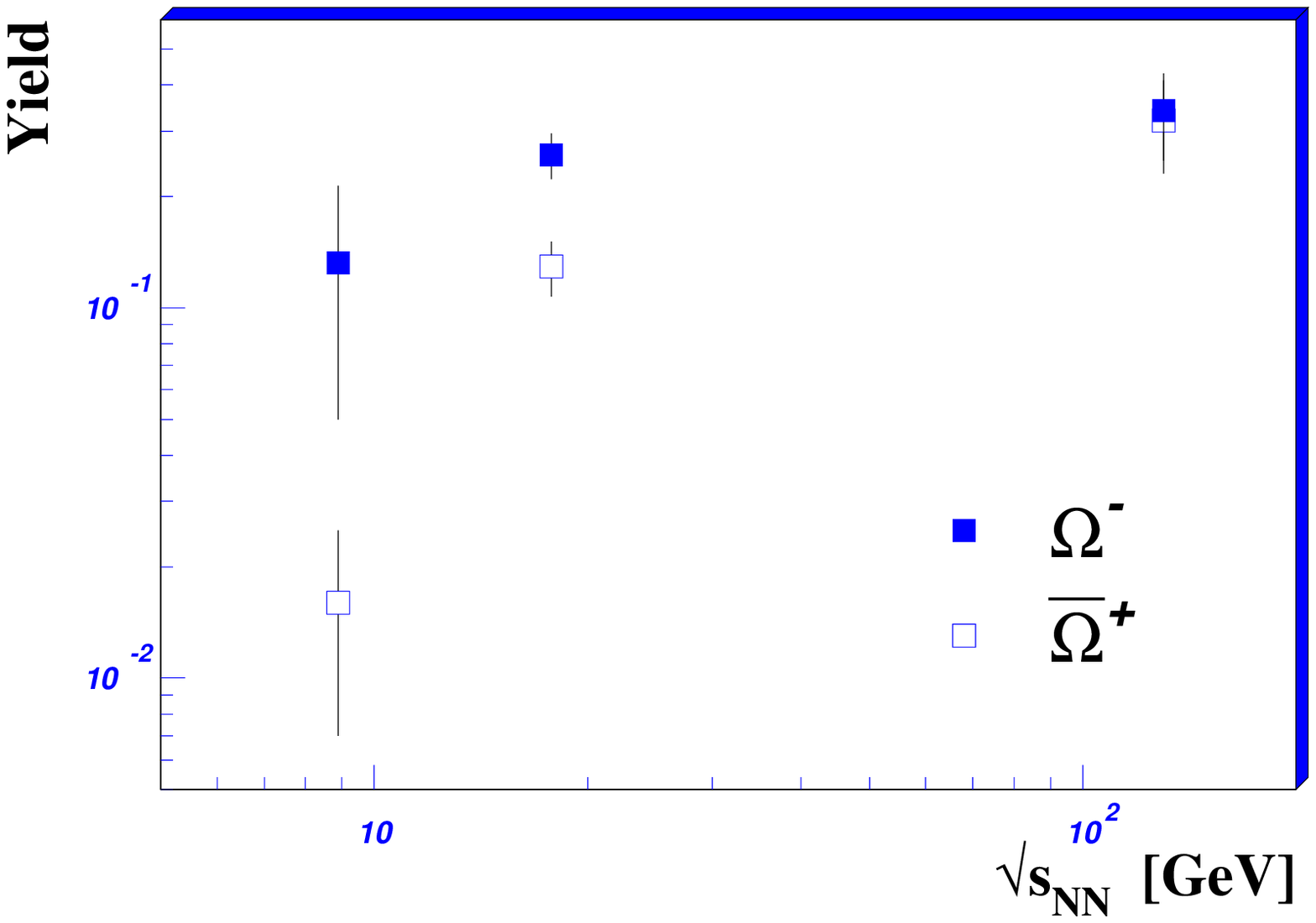}\\
\includegraphics[scale=0.55]{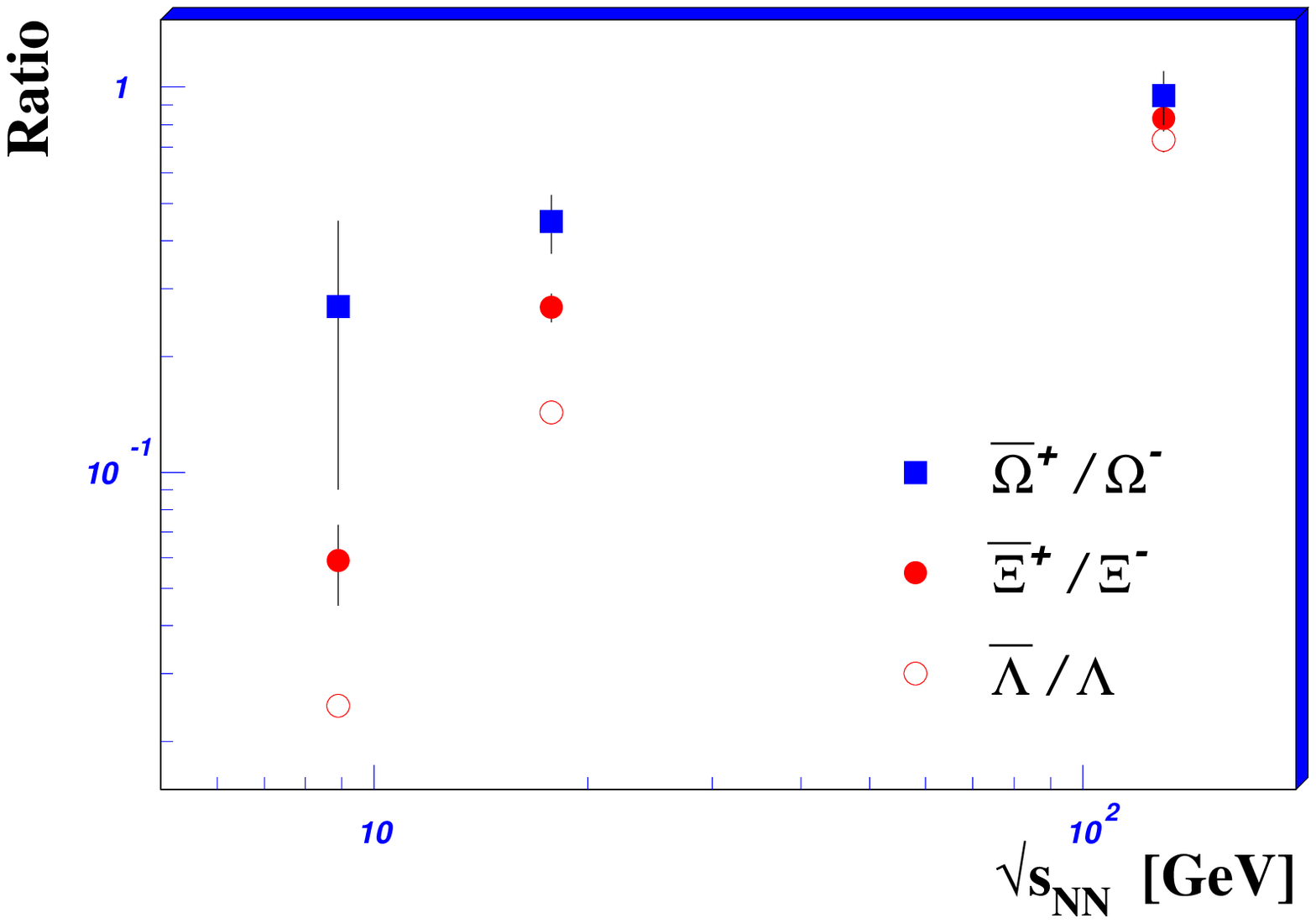}
\caption{{\em In alto}: confronto tra le rese di produzione delle $\Xi$\ e 
         delle $\Omega$\ misurate nelle collisioni Pb-Pb a 40 e 160 
	 A GeV/$c$\ e quelle misurate da STAR~\cite{STAR} nelle  
	 collisioni Au-Au a $\sqrt{s_{NN}}=130$\ GeV.  
	 \newline
	 {\em In basso}: andamento dei rapporti di produzione anti-iperone/iperone
	 in funzione dell'energia nel centro di massa. I dati a pi\`u alta 
	 energia sono quelli pubblicati da STAR~\cite{STAR}.}
\label{AllEnergy}
\end{center}
\end{figure}

Nella fig.~\ref{AllEnergy} si confrontano i tassi di produzione misurati ad 
entrambe le energie con i primi risultati dell'esperimento STAR al RHIC. 
I risultati di STAR si riferiscono alle collisioni Au-Au pi\`u centrali per il 
14\% della sezione d'urto anelastica~\cite{STAR}. 
Per confronto, i tassi di produzione di NA57 sono stati ricalcolati, sia
a 160 GeV/$c$\ che a 40 GeV/$c$, in un'intervallo di centralit\`a corrispondente
all'unione delle due classi pi\`u centrali definite a  160 GeV/$c$\ (III-IV). Ci\`o
corrisponde alle collisioni pi\`u centrali relative a circa il 15\% della sezione
d'urto anelastica. 
All'aumentare dell'energia della collisione, i tassi di produzione delle \PgXm\ e 
delle \PgOm\ crescono molto meno rapidamente di quelli delle relative 
anti-particelle. In particolare, all'energia di RHIC, i tassi di produzione di 
particelle e relative  anti-particelle sono circa le stesse. Questo 
comportamento \`e conseguenza, come detto, della minor densit\`a barionica 
della {\em ``fireball''} all'aumentare dell'energia della collisione.  
Ci\`o \`e evidente considerando i rapporti di produzione, anch'essi mostrati 
nella fig.~\ref{AllEnergy}. Nel calcolare i rapporti di produzione, al fine di 
ridurre l'incertezza statistica, si sono considerate nuovamente le tre classi 
di maggior centralit\`a, corrispondenti alle collisioni pi\`u centrali 
per il $25\%$\ della sezione d'urto anelastica, dove la dipendenza 
dalla centralit\`a \`e simile  per particelle ed anti-particelle. 
Dalla fig.~\ref{AllEnergy} si deduce che 
tutti i rapporti anti-iperone/iperone aumentano con l'energia della collisione, 
approssimando il valore unitario all'energia di RHIC, valore 
atteso per la produzione di stranezza nella fase di QGP in un'ambiente a densit\`a 
barionica nulla. Inoltre, la dipendenza 
dall'energia \`e pi\`u debole per le particelle con pi\`u elevato contenuto 
di stranezza.  

%% file: cap6/cap6.tex
\chapter{Studio della dinamica di espansione 
	in collisioni Pb-Pb a 160 $A$ GeV/$c$\ 
	con l'interferometria HBT}  
\section{Introduzione}
Come accennato nel {\em paragrafo 1.4.3} la tecnica dell'interferometria 
di intensit\`a permette di ottenere importanti informazioni circa le 
dimensioni geometriche del sistema allo stato del {\em freeze-out} 
e sulla dinamica  di espansione %che lo ha condotto 
sino a tale stato.  
\newline
Per una {\em ``review''} con gli ultimi sviluppi della tecnica di analisi ``HBT'' 
nella fisica degli ioni pesanti, sia sotto l'aspetto teorico che sperimentale, 
si consiglia la referenza~\cite{Tomasik}.   
Essa comprende un'ampia rassegna e discussione dei risultati sperimentali
dall'AGS al RHIC.  
\newline
In questo capitolo si descriveranno i risultati di un'analisi eseguita   
sui dati raccolti nell'anno 1995 dall'esperimento WA97  
relativi all'interazione Pb-Pb a 160 A GeV/$c$.  
\newline
Per i dettagli dell'analisi si rimanda alla referenza~\cite{HBTpaper}, che \`e 
allegata, per comodit\`a, al termine di questa tesi.  
Dopo una discussione sui fondamenti teorici del metodo, 
%seguir\`a una breve descrizione dell'apparato sperimentale di WA97 e 
si discuteranno i risultati principali dell'analisi svolta.   
\section{L'interferometria HBT nella fisica degli ioni pesanti} 
La tecnica dell'interferometria di intensit\`a (o del secondo ordine), 
sviluppata da R.Hanbury-Brown e R.Q.Twiss negli anni Cinquanta per l'astronomia 
e la radio astronomia~\cite{Hanbury}, fu riscoperta nella fisica delle particelle, 
in maniera del tutto indipendente, qualche anno pi\`u tardi da G.Goldhaber, 
S.Goldhaber, W.Lee e A.Pais~\cite{Goldhaber}. Fin d'allora tale tecnica \`e 
indicata coi nomi di ``interferometria HBT'', ``effetto GGLP'' o 
``correlazione di Bose Einstein'' se applicata a bosoni. 
\newline
Il metodo fa leva sugli effetti, prodotti dalla propriet\`a di simmetria 
(bosoni) od antisimmetria (fermioni) degli stati di pi\`u particelle identiche, 
sulla densit\`a di stati nello spazio delle fasi. 
In astronomia si studiano coppie di fotoni, mentre in fisica nucleare e delle 
particelle elementari si utilizzano generalmente coppie di pioni 
(adesso anche tripletti e quadrupletti~\cite{3pioni}\cite{4pioni}),  
o coppie di nucleoni, di kaoni o di iperoni dello stesso tipo.  
\subsection{Correlazione per sorgenti caotiche}
\begin{figure}[tb]
  \centering
  \resizebox{0.98\textwidth}{!}{%
  \includegraphics{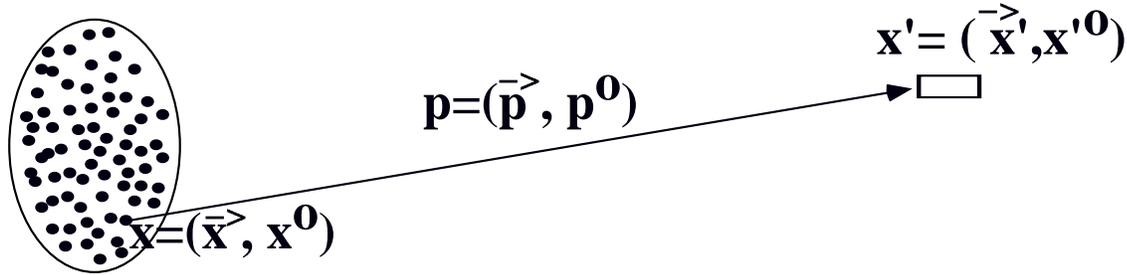}}
 \caption{Un pione di impulso $p$\ \`e emesso da un punto-sorgente $x$\ di
  una sorgente estesa ed \`e rivelato nel punto $x'$.}
 \label{sorgente1}
\end{figure}

Si consideri una sorgente estesa di particelle di una data specie, che 
supporremo bosoni (ad esempio \Pgpm). La trattazione per fermioni identici 
procede sulla stessa falsariga e non si ripeter\`a: l'unica differenza compare 
allorquando si introduce la propriet\`a di simmetria 
(antisimmetria per i fermioni) per la funzione d'onda che descrive le due 
particelle identiche. 
\newline
La sorgente estesa emette 
un \Pgpm\ di quadrimpulso \( p = (\vec{p},p^{0}) \) dal punto 
\( x = (\vec{x},x^{0}) \) dello spazio tempo, osservato da un rivelatore in
\( x' = (\vec{x'},{x'}^{0}) \), come schematizzato in fig.~\ref{sorgente1}. 
\`{E} conveniente utilizzare, nella trattazione che 
segue, il sistema centro di massa della sorgente; la scelta di un tale 
sistema \`e pi\`u ovvia per le collisioni di due particelle identiche ai 
collisionatori, mentre negli esperimenti a bersaglio fisso la sorgente \`e 
in moto rispetto ai rivelatori. I benefici di tale scelta discendono dal fatto 
che le distribuzioni di impulso per particella singola e per coppie, 
\( P(p) \) e \( P(p_{1},p_{2}) \), non dipendono dalle coordinate del 
rivelatore \( x' \), ma da quelle della sorgente  \( x \).
La risoluzione spaziale e temporale nella misura di \(p\) e \(x'\) non 
permette di determinare il punto di origine del pione nella sorgente, 
ma soltanto di stabilire 
che un pione \`e emesso da un punto qualsiasi della sorgente estesa. 
Vi sono molti punti-sorgente \(x\) da cui il pione pu\`o aver origine, 
che soddisfano tutti la relazione approssimata, 
che descrive la traiettoria classica del pione, 
\begin{equation}
 \vec{x'} - \vec{x} \approx \frac{\vec{p}}{p^0}(t'-t) 
 \label{classic}
\end{equation}
dove \( \frac{\vec{p}}{p^0} = \frac{d\vec{x'}}{dt'} \) \`e la velocit\`a 
del \Pgpm.
L'ampiezza di probabilit\`a per la propagazione del pione da \(x\) ad \(x'\), 
calcolata con il metodo dell'integrale di percorso di Feynman, \`e data 
da:
\begin{equation}
 \psi(p\,:\,x \rightarrow x') = \sum_{tutti\ i\ percorsi} e^{iS(percorso)}
 \label{eq:path}
\end{equation} 
dove: 
\[ S(percorso)=-m \int{\sqrt{1-\left(\frac{d\vec{x'}}{dt'}\right)^2} \,dt'} \] 
\`e ``l'azione'' relativistica per il pione, 
il cui valore dipende dal percorso nello spazio-tempo. 
Tutti i percorsi nella eq.~\ref{eq:path} hanno 
estremi di integrazione fissati; il contributo dominante della somma viene 
dalla traiettoria classica, descritta approssimativamente 
dall'eq.~\ref{classic}. 
I contributi provenienti dagli altri percorsi, variando 
notevolmente e con segno a caso, tendono a cancellarsi a vicenda. \`{E} 
ragionevole quindi approssimare la eq.~\ref{eq:path} col contributo del solo 
termine classico; si assume altres\`i che il pione si propaghi con 
probabilit\`a nulla di attenuazione od assorbimento 
\[ \psi(p\,:\,x \rightarrow x') \approx e^{iS(traiettoria\ classica)} \]
Valutando l'azione lungo la traiettoria classica si trova: 
\[S(traiettoria\ classica)\approx p\cdot(x-x') \approx p^{0}(t-t')-\vec{p}\cdot(\vec{x}-\vec{x'}) \]
e pertanto, in tale approssimazione, la eq.~\ref{eq:path} si riduce alla 
\begin{equation}
 \psi(p\,:\,x \rightarrow x') \approx e^{ip\cdot(x-x')}
 \label{eq:planewave}
\end{equation}
Nel seguito, pur servendosi di tale approssimazione, si utilizzer\`a il 
simbolo \( = \) anzich\`e \( \approx \) per semplicit\`a di notazione.
La eq.~\ref{eq:planewave} fornisce l'ampiezza di probabilit\`a per un \Pgpm\  
(non interagente) di propagarsi da \( x \)\  ad\  \( x' \)\  con impulso~ \( p \). 
L'ampiezza di probabilit\`a per la sua produzione \`e descritta, invece, da un 
modulo \( A(p,x) \), che si pu\`o assumere positivo, ed una fase 
\( \phi(x) \). L'ampiezza di probabilit\`a, per la produzione di \Pgpm\ in 
\( x \) e la sua propagazione in \( x' \) \`e data quindi da 
\begin{equation}
\Psi(p\,:\,x \rightarrow x')=A(p,x)e^{i\phi(x)}\psi(p\,:\,x \rightarrow x')=
   A(p,x)e^{i\phi(x)}e^{ip\cdot(x-x')}
 \label{eq:cinque}
\end{equation}
L'ampiezza totale per la produzione da un qualsiasi punto-sorgente \`e dunque 
\begin{equation}
 \Psi(p\,:\, \left\{ \begin{array}{c} 
            tutti\ i \\ 
            punti\ x 
                 \end{array} \right\} 
                                     \rightarrow x') 
                      =\sum_{x} A(p,x)e^{i\phi(x)}e^{ip\cdot(x-x')}
\label{eq:sei}
\end{equation}
La somma sui punti-sorgente richiede la conoscenza della densit\`a \(\rho(x)\) 
di questi entro la sorgente. 
%Nota \(\rho(x)\), 
Se si suppone continua la distribuzione dei punti nella sorgente, 
la sommatoria \(\sum_{x}\) si 
pu\`o riscrivere come integrale su \(x\) 
\begin{equation}
 \sum_{x}{\ldots} \longrightarrow \int{\rho(x)}\,dx\, \ldots
\label{eq:sum-int}
\end{equation}
La distribuzione di impulsi per singola particella, \(P(p)\), che fornisce la 
densit\`a di probabilit\`a, per un pione di impulso \(p\), di essere prodotto 
entro la sorgente e di giungere al rivelatore, \`e pari al modulo quadro della 
\(\Psi\)
\begin{equation}
 P(p)=|\Psi|^{2}=|\sum_{x}{ A(p,x)e^{i\phi(x)}e^{ip \cdot(x-x')}}|^{2}
 =|\sum_{x}{ A(p,x)e^{i\phi(x)}e^{ip \cdot x}}|^{2}
 \label{singleparticle}
\end{equation}
I risultati fin qui ottenuti valgono per una qualsiasi sorgente; si consideri 
ora una {\em sorgente caotica}, che si definisce tale se la fase \(\phi(x)\)\ 
dipende in modo del tutto arbitrario dalla coordinata \(x\)\ (funzione 
``random''): essa non presenta, cio\`e, alcuna continuit\`a nei suoi valori al 
variare di \(x\).
Conviene in tal caso riscrivere la eq.~\ref{singleparticle} come
\begin{equation}
 P(p)=\sum_{x}{A(p,x)^{2}}+\sum_
                    {\substack{x,y\\ x \neq y }}
                 {A(p,x)A(p,y)e^{i\phi(x)}e^{-i\phi(y)}e^{ip\cdot(x-y)}}
 \label{formula8}
\end{equation}
Tenendo conto della caratteristica di caoticit\`a, che si riflette nel 
comportamento ``random'' di \(\phi(x)\), la seconda somma 
d\`a contributo nullo, poich\'e gli addendi nella somma, 
avendo modulo che varia 
lentamente con \(x\)\ e fase fluttuante rapidamente, si cancellano a vicenda.
Pertanto la eq.~\ref{formula8} si riscrive, in termini della 
densit\`a \(\rho(x)\), come:
\begin{equation}
 P(p)=\int{A^2(p,x)\,\rho(x)\,dx}
 \label{eqq10}
\end{equation}
Conviene confrontare quest'ultima espressione con le propriet\`a della 
funzione di distribuzione dello spazio delle fasi \(f(p,x)\), nota come 
funzione di emissione o densit\`a di Wigner, per la quale vale: 
\begin{equation}
 P(p)=\int{f(p,x)\,dx}
 \label{eqf(x)}
\end{equation}
Pertanto, per una sorgente caotica, vale la relazione 
\( f(p,x)=A^2(p,x)\,\rho(x)\), che permette di ricavare \(A(p,x)\) 
a partire da 
\(f(p,x)\), noto \(\rho(x)\)
\begin{equation}
 A(p,x)=\sqrt{\frac{f(p,x)}{\rho(x)}}
 \label{eq12.b}
\end{equation}

\begin{figure}[tb]
  \centering
  \resizebox{0.98\textwidth}{!}{%
  \includegraphics{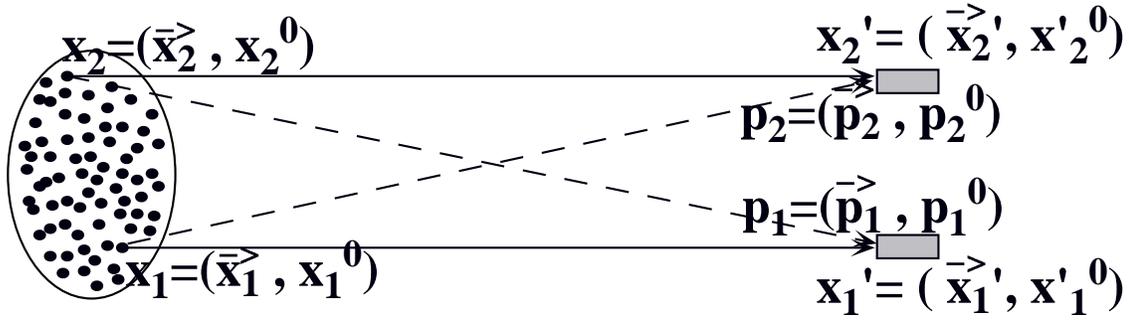}}
 \caption{Due pioni, emessi da una sorgente estesa, sono rivelati nei punti 
  $x_1'$\ e $x_2'$\ con impulsi pari, rispettivamente, a $p_1$\ e $p_2$. 
  Il pione di impulso $p_1$\ ($p_2$) pu\`o provenire sia dal punto-sorgente 
  $x_1$\ che da $x_2$.}
 \label{sorgente2}
\end{figure}

Si vuole ora calcolare la distribuzione \(P(p_{1},p_{2})\)\ per una coppia di 
\Pgpm di impulsi \(p_{1}\)\ e \(p_{2}\). Si consideri un \Pgpm\ di quadrimpulso 
\(p_{1}\) rivelato nel punto \(x'_{1}\) ed un secondo \Pgpm\ di impulso 
\(p_{2}\) e rivelato in \(x'_{2}\); nel caso in cui il primo pione si 
propaghi da \(x_{1}\) ed il secondo da \(x_{2}\), come descritto 
in fig.~\ref{sorgente2} 
dalle linee continue, l'ampiezza di probabilit\`a vale  
\[ \psi(p_{1}\,:\,x_{1} \rightarrow x'_{1})\psi(p_{2}\,:\,x_{2} \rightarrow x'_{2}) \approx e^{ip_{1}\cdot(x_{1}-x'_{1})} e^{ip_{2}\cdot(x_{2}-x'_{2})} \] 
nella stessa approssimazione usata in precedenza.
L'ampiezza di probabilit\`a per la produzione in \(x_{i}\) e la successiva 
propagazione fino ad \(x'_{i}\), per la coppia di \Pgpm\ di impulso \(p_{i}\) 
\((i=1,2)\), \`e data da 
\[ A(p_{1},x_{1})e^{i\phi(x_{1})} A(p_{2},x_{2})e^{i\phi(x_{2})} 
\psi(p_{1}\,:\,x_{1} \rightarrow x'_{1}) \psi(p_{2}\,:\,x_{2} 
\rightarrow x'_{2}) \] 
Tuttavia il \Pgpm\ di impulso  \(p_{1}\), rivelato in \(x'_{1}\), pu\`o anche 
provenire da \(x_{2}\), quando l'altro proviene da \(x_{1}\). 
In questo secondo 
caso l'ampiezza di probabilit\`a vale:
\[ A(p_{1},x_{2})e^{i\phi(x_{2})} A(p_{2},x_{1})e^{i\phi(x_{1})} 
\psi(p_{1}\,:\,x_{2} \rightarrow x'_{1}) \psi(p_{2}\,:\,x_{1} 
\rightarrow x'_{2})\]
L'ampiezza di probabilit\`a, a causa dell'indistinguibilit\`a di una coppia 
di bosoni identici, deve essere simmetrica nello scambio degli indici 
che permettono di distinguere tra i due bosoni. In tal caso, gli unici indici 
che distinguono i due pioni sono le coordinate spazio-temporali dei 
punti-sorgente \(x_{1}\)\ ed \(x_{2}\), in quanto si assume di misurare gli 
impulsi \(p_{1}\)\ e \(p_{2}\)\ nei punti \(x'_{1}\)\ ed \(x'_{2}\).
L'ampiezza di probabilit\`a deve essere quindi simmetrica nello scambio di 
\(x_{1}\)\ ed \(x_{2}\)
\begin{equation}
\begin{split}
 \frac{1}{\sqrt{2}}\{& A(p_{1},x_{1})e^{i\phi(x_{1})}
   A(p_{2},x_{2})e^{i\phi(x_{2})}e^{ip_{1}\cdot(x_{1}-x'_{1})}
    e^{ip_{2}\cdot(x_{2}-x'_{2})} + \\ 
 & A(p_{1},x_{2})e^{i\phi(x_{2})}A(p_{2},x_{1})e^{i\phi(x_{1})}
 e^{ip_{1}\cdot(x_{2}-x'_{1})} e^{ip_{2}\cdot(x_{1}-x'_{2})}\} \\
 & \equiv e^{i\phi(x_{1})}e^{i\phi(x_{2})}\Phi(p_{1},p_{2}\,:\,x_{1}x_{2} 
        \rightarrow x'_{1}x'_{2})
\end{split}
\label{ampli}
\end{equation}
dove \(\Phi(p_{1},p_{2}\,:\,x_{1}x_{2} \rightarrow x'_{1}x'_{2}) \) \`e la 
parte di ampiezza di probabilit\`a non dipendente da \(\phi\)
\begin{multline}
\label{Phi}
  \Phi(p_{1},p_{2}\,:\,x_{1}x_{2}\rightarrow x'_{1}x'_{2})=\\
 \frac{1}{\sqrt{2}}\{A(p_{1},x_{1})A(p_{2},x_{2})
   e^{ip_{1}\cdot(x_{1}-x'_{1})}e^{ip_{2}\cdot(x_{2}-x'_{2})} +\\ 
  A(p_{1},x_{2})A(p_{2},x_{1})e^{ip_{1}\cdot(x_{2}-x'_{1})}
                               e^{ip_{2}\cdot(x_{1}-x'_{2})} \} 
\end{multline}
I due pioni possono esser prodotti in altri punti-sorgente; l'ampiezza totale 
\`e la somma delle ampiezze per tutte le combinazioni dei punti-sorgente. 
Pertanto l'ampiezza di probabilit\`a per due pioni identici di essere 
prodotti dalla sorgente e quindi rivelati in \(x'_{1}\)\ ed \(x'_{2}\), 
con impulsi pari a, rispettivamente, \(p_{1}\)\ ed \(p_{2}\), \`e data da:
\begin{equation}
 \Psi(p_{1}p_{2}\,:\,\left\{ \begin{array}{c} 
            tutti\ i \\ 
            punti\ x_{1} x_{2}  
                 \end{array} \right\} 
                                     \rightarrow x'_{1} x'_{2})=
 \sum_{x_{1},x_{2}}{e^{i\phi(x_{1})}e^{i\phi(x_{2})}
     \Phi(p_{1},p_{2}\,:\,x_{1}x_{2}\rightarrow x'_{1}x'_{2})} 
 \label{amplitot}
\end{equation}
La distribuzione di probabilit\`a \(P(p_{1},p_{2})\), pari al 
modulo quadro della eq.~\ref{amplitot}, vale quindi
\begin{equation}
 P(p_{1},p_{2})=\frac{1}{2!} \left|\Psi(p_{1}p_{2}\,:\,\left\{ 
      \begin{array}{c} 
            tutti\ i \\ 
            punti\ x_{1} x_{2}  
                 \end{array} \right\} 
                                     \rightarrow x'_{1} x'_{2}) \right|^{2}
\label{sedici}
\end{equation} 
I risultati ottenuti nelle eq.~\ref{ampli},~\ref{amplitot},~\ref{sedici}
 sono del tutto generali e validi per lo studio della correlazione degli 
impulsi tanto per sorgenti caotiche che per quelle coerenti.
Per una {\em sorgente caotica}, sfruttando ancora la natura 
``random'' delle fasi, sviluppando la eq.~\ref{sedici} e separando i 
termini che contengono \(\phi\), si ha 
\begin{equation}
 \label{diciass}
\begin{split}
& P(p_{1},p_{2})  = \frac{1}{2} \sum_{x_{1},x_{2}} \{ 
 \Phi^{*}(p_{1},p_{2}\,:\,y_{1}y_{2} \rightarrow x'_{1}x'_{2})_{
  \left|\substack{y_1=x_1\\ y_2=x_2} \right.} 
   \Phi(p_{1},p_{2}\,:\,x_{1}x_{2}\rightarrow x'_{1}x'_{2}) \;+\\
 & \quad \quad\ \quad  \quad \quad
  \Phi^{*}(p_{1},p_{2}\,:\,y_{1}y_{2}\rightarrow x'_{1}x'_{2})_{
 \left|\substack{y_1=x_2\\ y_2=x_1} \right.}  
 \Phi(p_{1},p_{2}\,:\,x_{1}x_{2}\rightarrow x'_{1}x'_{2}) \} \\
 & \quad \quad \quad \;+ \frac{1}{2}\sum_{ 
  \substack{x_1,x_2,y_1,y_2\\ \{x_1,x_2\} \neq \{y_1,y_2 \} }}
  \{ e^{i\phi(x_{1})+i\phi(x_{2})}
   \Phi(p_{1},p_{2}\,:\,x_{1}x_{2}\rightarrow x'_{1}x'_{2}) \cdot\\
 & \quad \quad \quad \quad \quad \quad \quad \quad \quad \quad \quad \quad
   e^{-i\phi(y_{1})-i\phi(y_{2})}
  \Phi^{*}(p_{1},p_{2}\,:\,y_{1}y_{2}\rightarrow x'_{1}x'_{2}) 
  \}
 \end{split}
\end{equation}
I due termini nella prima sommatoria della eq.~\ref{diciass} sono uguali per 
la simmetria di scambio e, per sorgenti caotiche, le somme sui termini che 
contengono le fasi danno contributo nullo. Pertanto si pu\`o riscrivere
\[ P(p_{1},p_{2}) = \sum_{x_{1},x_{2}} \left|
  \Phi(p_{1},p_{2}\,:\,x_{1}x_{2}\rightarrow x'_{1}x'_{2}) \right|^{2} \]
e, cambiando le sommatorie in integrali secondo la prescrizione  
dell'eq.~\ref{eq:sum-int}, si ottiene
\[ P(p_{1},p_{2}) = \int \left|
  \Phi(p_{1},p_{2}\,:\,x_{1}x_{2}\rightarrow x'_{1}x'_{2}) \right|^{2} 
 \, \rho(x_{1})\rho(x_{2}) \, dx_{1} \,dx_{2}. \] 
Includendo la eq.~\ref{Phi}, quest'ultima espressione diventa
\begin{equation}
 \begin{split}
 \label{venti}
 P(p_{1},p_{2})& = \int A^{2}(p_{1},x_{1}) \,\rho(x_{1})\,dx_{1}
     \int A^{2}(p_{2},x_{2})\,\rho(x_{2})\,dx_{2}\\
   & \quad + \int A(p_{1},x_{1})A(p_{2},x_{1})
             e^{i(p_{1}-p_{2}) \cdot x_{1}}\,\rho(x_{1})\,dx_{1}\\
   &  \quad  \quad  \quad  \quad 
             \int A(p_{2},x_{2})A(p_{1},x_{2})
             e^{i(p_{2}-p_{1}) \cdot x_{2}}\,\rho(x_{2})\,dx_{2}
 \end{split}
\end{equation}
Utilizzando infine la eq.~\ref{eqq10}, la funzione di distribuzione degli 
impulsi per due particelle identiche si esprime come
\begin{equation}
 \label{ventuno}
 P(p_{1},p_{2})=P(p_{1})\cdot P(p_{2})\,+\, \left|
 \int e^{i(p_{1}-p_{2})\cdot x}\rho(x)A(p_{1},x)A(p_{2},x)\,dx \right|^{2}
\end{equation}
Conviene, a questo punto, introdurre la funzione di densit\`a effettiva, 
\( \rho_{eff}(x;p_{1},p_{2}) \), per riscrivere la eq.~\ref{ventuno}\ come:
\begin{equation}
 \label{ventidue}
 P(p_{1},p_{2})=P(p_{1})\cdot P(p_{2}) \left( 1\;+\; 
  \left|\int e^{i(p_{1}-p_{2}) \cdot x} \rho_{eff}(x;p_{1},p_{2}) \,dx
  \right|^{2} \right)
\end{equation}   
dove 
\begin{equation}
 \label{ventitre}
 \rho_{eff}(x;p_{1},p_{2})\;=\;\frac{\rho(x)A(p_{1},x)A(p_{2},x)}
                                 {\sqrt{P(p_{1})P(p_{2})}}
\end{equation}
In termini della funzione di distribuzione dello spazio delle fasi, la 
\(\rho_{eff}(x;p_{1},p_{2}) \)\ si esprime come
\begin{equation}
 \label{ventiquattro}
 \rho_{eff}(x;p_{1},p_{2})\;=\;\frac{\sqrt{f(p_{1},x)f(p_{2},x)}}
            {\sqrt{\int f(p_{1},x_{1})\,dx_{1} \int f(p_{2},x_{2})\,dx_{2}}}
\end{equation}
La trasformata di Fourier di \(\rho_{eff}(x;p_{1},p_{2}) \)\ viene indicata 
con
\begin{equation}
 \label{venticinque}
 \tilde{\rho}_{eff}(q;p_{1}p_{2})\;=\;\int \, e^{iq \cdot x}
       \rho_{eff}(x;p_{1}p_{2})\,dx
\end{equation}
dove \( q=p_{1}-p_{2} \)\ . La funzione di distribuzione \(P(p_{1},p_{2}) \)\ 
dipende dunque dalla trasformata di Fourier della \( \rho_{eff} \);
dalle eq.~\ref{venticinque},~\ref{ventidue} si ottiene infatti
\begin{equation} 
 \label{ventisei}
 P(p_{1},p_{2})=P(p_{1})\cdot P(p_{2}) \left( 1\;+\;
      \left| \tilde{\rho}_{eff}(q;p_{1}p_{2}) \right|^{2} \right)
\end{equation}\\
Si definisce {\em funzione di correlazione} 
\( C_{2}(p_{1},p_{2}) \)\ il rapporto
\begin{equation} 
 \label{ventisette}
 C_{2}(p_{1},p_{2})=\frac{P(p_{1},p_{2})}{P(p_{1})\cdot P(p_{2}).}
\end{equation}
Esso vale, per una sorgente caotica:
\begin{equation} 
 \label{ventotto}
 C_{2}(p_{1},p_{2})=1+ \left| \tilde{\rho}_{eff}(q;p_{1}p_{2}) \right|^{2}
\end{equation}
Pertanto, per una sorgente caotica estesa, la funzione di correlazione 
\( C_{2}(p_{1},p_{2}) \)\ \`e direttamente collegata alla trasformata di 
Fourier della densit\`a effettiva, secondo la eq.~\ref{ventotto}, e 
pu\`o essere 
utilizzata per misurare la configurazione dello spazio delle fasi della 
sorgente al momento dell'emissione delle particelle.\\
\newline
Come si esporr\`a nei paragrafi successivi, in molte applicazioni 
si parametrizza la densit\`a effettiva con una distribuzione gaussiana 
\begin{equation} 
 \label{trentuno}
 \rho_{eff}(x;p_{1}p_{2})=\frac{N}{4\pi^{2}R_{x}R_{y}R_{z}\sigma_{t}}
     exp\left\{-\frac{x^{2}}{2R_{x}^{2}}-\frac{y^{2}}{2R_{y}^{2}}
     -\frac{z^{2}}{2R_{z}^{2}}-\frac{t^{2}}{2\sigma_{t}^{2}}\right\} 
\end{equation}
dove la costante di normalizzazione, \( N(p_{1},p_{2})\), e le 
deviazioni standard \(R_{x}\),\(R_{y}\),\(R_{z}\)\ e \(\sigma_{t}\), 
dipendono da \(p_{1}\)\ e \(p_{2}\).  
Il coefficiente \(N\)\ si pu\`o prendere in modo tale che 
\begin{equation} 
 \label{trentadue}
 N(p_{1},p_{2})= \int \rho_{eff}(x;p_{1},p_{2}) \, dx
\end{equation}
Per la disuguaglianza di Schwartz, si ha anche
\begin{equation} 
 \label{trentatre}
 \int \sqrt{f(p_{1},x)f(p_{2},x)}\,dx  \leq 
 \sqrt{\int f(p_{1},x_{1})\,dx_{1} \int f(p_{2},x_{2})\, dx_{2}}
\end{equation}
e quindi, per le eq.~\ref{ventiquattro},~\ref{trentatre}
\[ \int \rho_{eff}(x;p_{1},p_{2})\, dx= N(p_{1},p_{2}) \leq 1 \]
dove l'uguaglianza vale per \(p_{1}=p_{2} \).
Si ha dunque \( N(q=0)=1 \).
La trasformata di Fourier della \(\rho_{eff}(x;p_{1},p_{2})\), 
eq.~\ref{venticinque}, \`e pari, in questa particolare parametrizzazione, a
\begin{equation} 
 \label{trentaquattro}
 \tilde{\rho}_{eff}(q;p_{1}p_{2})=N\,exp \left\{ 
  -\frac{R_{x}^{2}q_{x}^{2}}{2}-\frac{R_{y}^{2}q_{y}^{2}}{2}
  -\frac{R_{z}^{2}q_{z}^{2}}{2}-\frac{\sigma_{t}^{2}q_{t}^{2}}{2}\right\}
\end{equation}
La funzione di correlazione diventa, nella parametrizzazione  
gaussiana,
\begin{equation}
 \label{correl_gauss}
 C_{2}(p_{1},p_{2}) = C_{2}(q;p_{1}p_{2})= 1 +
                  N\,exp \left\{-R_{x}^{2}q_{x}^{2}-R_{y}^{2}q_{y}^{2}-
R_{z}^{2}q_{z}^{2}-\sigma_{t}^{2}q_{t}^{2}\right\} 
\end{equation}
I risultati relativi alla parametrizzazione gaussiana possono essere usati 
per analizzare dati sperimentali sulla correlazione degli impulsi. 
Bisogna enfatizzare che i parametri \(N\), \(R_{i}\)\ e  \(\sigma_{t}\)\ 
in queste espressioni, sono funzioni di \(p_{1}\)\ e \(p_{2}\). Soltanto 
in casi speciali questa dipendenza viene a mancare. Questo accade  
quando \(A(p,x)\)\ \`e indipendente da \(x\), e \(f(p,x)\)\ si 
fattorizza in \( f(p,x)=A^{2}(p)\rho(x)\). In tal caso si ha \(
\rho_{eff}(x)=\rho(x)\)\ e \(P(p)=\int f(p,x)\,dx=A^{2}(p)\), ponendo \(
\int \rho(x)\,dx=1 \); inoltre \( C_{2}(p_{1},p_{2})=C_{2}(q) \)\ \`e 
dato semplicemente dalla trasformata di Fourier della distribuzione 
spazio-temporale della sorgente.
Pertanto una parametrizzazione gaussiana, con parametri indipendenti da 
\(p_{1}\)\ e \(p_{2}\), fornisce una descrizione rigorosa per una sorgente 
statica, ma \`e chiaramente approssimativa per una sorgente in cui il centro 
della distribuzione dei punti sorgente di pioni di fissato impulso  
cambia al variare del valore dell'impulso, come nel 
caso della produzione di particelle accompagnata da espansione collettiva 
della sorgente.\\
Un'approssimazione ottenuta da Pratt~\cite{Pratt}, per la \(
\rho_{eff}(x;p_{1},p_{2})\), valida quando \(p_{1}\simeq p_{2}\), si 
ottiene espandendo \(f(p,x)\)\ rispetto a \(f(K,x)\), dove 
\(K=\frac{p_{1}+p_{2}}{2} \)
\begin{align}
 f(p_{1},x)& =f(K,x)+\frac{\partial f(p,x)}{\partial x^{\mu}}_{
 \left| \scriptstyle p=K \right.}\frac{q^{\mu}}{2}+
 \frac{\partial^{2} f(p,x)}{2!\partial x^{\mu}\,\partial x^{\nu} }_{
 \left| \scriptstyle p=K \right.} \frac{q^{\mu}q^{\nu}}{4} \nonumber\\
f(p_{2},x)& =f(K,x)-\frac{\partial f(p,x)}{\partial x^{\mu}}_{
 \left| \scriptstyle p=K \right.}\frac{q^{\mu}}{2}+
 \frac{\partial^{2} f(p,x)}{2!\partial x^{\mu}\,\partial x^{\nu} }_{
 \left| \scriptstyle p=K \right.} \frac{q^{\mu}q^{\nu}}{4} \nonumber
\end{align}  
Trascurando termini del second'ordine in \(q^{\mu}\), poich\'e 
\[ \sqrt{f(p_{1},x)f(p_{2},x)}=f(K,x)+ \mathcal{O} (q^{\mu}q^{\nu}) , \]
la densit\`a effettiva vale:
\begin{equation}
 \label{ultima}
 \rho_{eff}(x;p_{1},p_{2})\simeq \frac{f(K,x)}
            {\sqrt{\int f(p_{1},x_{1})\,dx_{1} \int f(p_{2},x_{2})\,dx_{2}}}
\end{equation}
\subsection{Sorgenti coerenti e parzialmente coerenti: parametro 
            di caoticit\`a} 
Da questa derivazione dettagliata della funzione di 
correlazione \`e possibile capire l'origine della correlazione per una 
sorgente caotica.  Per questo tipo di sorgente, la funzione di fase 
\(\phi(x)\)\ non ha andamento regolare e lo sviluppo della somma nella 
distribuzione di probabilit\`a pu\`o essere completato separando i 
termini in cui la fase ``random'' rende nullo il contributo alla 
probabilit\`a, da quelli in cui la fase non compare. Uno dei termini che 
non \`e eliminato proviene dalla imposta simmetria di scambio richiesta 
per bosoni identici, ed \`e il termine proporzionale alla trasformata di 
Fourier di una funzione della %funzione di 
distribuzione dello spazio delle fasi \(f(k,x)\).  
\`{E} dunque la simmetria di scambio per particelle 
identiche l'origine della correlazione tra gli impulsi, per una sorgente 
caotica. \newline
Che tipo di correlazione \`e attesa per una sorgente estesa 
coerente? 
\newline
Una sorgente coerente ha una fase di produzione \(\phi(x)\)\ che \`e una 
funzione regolare delle coordinate \(x\)\ dei punti sorgente. 
\`E facile ricavare~\cite{BrunoTesi,Wong55} che per una sorgente coerente 
riulta 
\(P(k_{1},k_{2})= P(k_{1})P(k_{2})\), senza alcuna correlazione. 
\newline
Per l'analisi dei dati sperimentali, conviene spesso introdurre un 
{\em parametro di caoticit\`a}\ \(\lambda\)\ che modifica la funzione di 
correlazione, eq.~\ref{correl_gauss}, nella forma: 
\begin{equation}
 \label{eq10}
  C_{2}(q)= 1 + \lambda exp \left\{-R_{x}^{2}q_{x}^{2}
   -R_{y}^{2}q_{y}^{2}-R_{z}^{2}q_{z}^{2}-\sigma_{t}^{2}q_{t}^{2}\right\}
   \end{equation}
Questa parametrizzazione interpola tra il caso di sorgente coerente, 
corrispondente a \(\lambda=0 \), e quello di sorgente completamente 
caotica, per cui \(\lambda=1 \). 
\newline 
Il parametro \(\lambda\)\ non dipende solo dal livello di coerenza intrinseca  
della sorgente, ma anche \cite{Boal}\ dalla contaminazione dovuta ad altre 
particelle interpretate erroneamente come pioni, dal numero di 
punti-sorgente, da decadimenti di risonanze, da processi di ``scattering'' 
multiplo, dalla media sul parametro di impatto, dalla risoluzione 
sperimentale finita e, non ultima, dall'espansione della sorgente. 
\newline
In particolare, \`e stato mostrato~\cite{sette,otto,NA49HBT} che le particelle 
erroneamente identificate influenzano unicamente il parametro $\lambda$\ 
(riducendone il valore) e non i ``raggi HBT'' $R_i$\ (e $\sigma_{t}$) da 
cui si estraggono le informazioni sulla geometria della sorgente. 
Pertanto in molte analisi sperimentale si considereranno tutte le coppie 
$h^-$-$h^-$\ di particelle di carica negativa, non identificate, 
ma costituite prevalentemente da coppie \Pgpm-\Pgpm. 
Questo approccio \`e stato seguito in questo lavoro, in quanto non \`e 
possibile identificare in WA97 particelle primarie che non decadono.  
\subsection{Approssimazioni di continuit\`a e di ``on shell''}
Conviene riesprimere la funzione di correlazione, eq.~\ref{ventotto}, in 
termini della sola funzione di emissione $ f(p,x)$, utilizzando 
l'approssimazione di Pratt (eq.~\ref{ultima}) nell'eq.~\ref{ventiquattro} 
\begin{equation}
 C_{2}(\vec{q},\vec{K}) = 1+ \frac{\left|\int f(K,x)e^{iqx}\,dx\right|^2}
                            {\int f(p_1,x_1)\,dx_1\int f(p_2,x_2)\,dx_2}  
 \label{quarantadue}
\end{equation}
Nell'ulteriore approssimazione, nota come ``approssimazione di continuit\`a'', 
la regolarit\`a della dipendenza da $ p $\ della funzione di emissione 
permette di porre:
\begin{equation}
 f(K - \frac{1}{2}q,x_1)f(K + \frac{1}{2}q,x_2) \approx 
 f(K,x_1)f(K,x_2)
\end{equation}
In~\cite{42} si mostra come gli errori introdotti da questa approssimazione 
siano trascurabili per le tipiche funzioni di emissione degli adroni. 
Utilizzando l'approssimazione di continuit\`a, la funzione di correlazione 
assume l'espressione:
\begin{equation}
 C_{2}(\vec{q},\vec{K})\approx 1+ \frac{\left|\int f(K,x)e^{iqx}\right|^2\,dx}
                                   {\left|\int f(K,x)\right|^2\,dx}  
 \label{quarantatre}
\end{equation}
La funzione di emissione $f(K,x)$\ dipende in generale dal quadrivettore $K$, 
dove $K^0=\frac{1}{2}(E_1+E_2)$. In molte applicazioni si usa un'ulteriore 
approssimazione, detta di ``on shell'', in cui:
\begin{equation}
 f(K^0,\vec{K},x)\approx f(E_K,\vec{K},x) \quad\quad {\rm dove:} \quad 
       E_K=\sqrt{m^2+\vec{K}^2}
\end{equation} 
Anche in tal caso le correzioni, che possono essere calcolate 
sistematicamente, risultano essere trascurabili~\cite{42}. 
\subsection{Condizione di shell di massa}
Sebbene la funzione di correlazione sia ottenuta come trasformata di Fourier 
della $f(K,x)$, come indicato dalla eq.~\ref{quarantatre}, la funzione di 
emissione non pu\`o essere ricostruita in maniera univoca dal correlatore. 
Ci\`o \`e dovuto al fatto che le particelle rivelate obbediscono alla 
condizione di ``shell di massa'' $p_1^2=p_2^2=m_1^2=m_2^2$, e pertanto: 
\begin{equation}
 K\cdot q=p_1^2-p_2^2=m_1^2-m_2^2=0
 \label{quarantaquattro}
\end{equation}
il che implica che solo tre delle quattro componenti dell'impulso 
relativo $q$\ sono cinematicamente indipendenti. Quindi la dipendenza da 
$q$\ di 
$C_2(q,K)$\ permette di sondare solo tre delle quattro direzioni indipendenti 
dello spazio-tempo. \`{E} inevitabile quindi introdurre una dipendenza dai 
modelli nel ricostruire la $f(K,x)$. Questa dipendenza pu\`o essere rimossa 
fornendo ulteriori informazioni non provenienti dalla correlazione tra 
particelle identiche. L'eq.~\ref{quarantaquattro} suggerisce di 
utilizzare la correlazione tra particelle non identiche, con diverse 
combinazioni di masse. Particelle non identiche non presentano ovviamente 
correlazione di natura statistica, ma la correlazione proveniente 
dall'interazione nello stato finale contiene tuttavia informazioni sulla 
funzione di emissione. Attualmente la correlazione tra particelle non 
identiche \`e oggetto di intensa attivit\`a di ricerca~\cite{99}\cite{100}
\cite{101}\cite{168}\cite{156}\cite{114}.
\subsection{Parametrizzazione gaussiana della funzione di correlazione}
La funzione di correlazione \`e generalmente parametrizzata con una funzione 
gaussiana nelle componenti della differenza tra gli impulsi delle due 
particelle. Questa discende ovviamente da una parametrizzazione della 
funzione di emissione anch'essa di tipo gaussiano. 
Si esprima dunque $f(K,x)$\ come 
\begin{equation}
 f(K,x)=N(K)f(K,\bar{x}(K))\exp\left[{-\frac{1}{2}
      \tilde{x}^{\mu}(K)B_{\mu\nu}\tilde{x}^{\nu}(K)}\right]
 \label{quarantacinque}
\end{equation}
dove $B_{\mu\nu}$\ \`e un tensore simmetrico $3 \times 3$\ che descrive 
la curvatura della funzione di correlazione intorno a $\vec{q}=0$. 
Le coordinate $\tilde{x}_{\mu}$\ sono definite riferendosi a $\bar{x}(K)$, 
il ``centro effettivo della sorgente'' per bosoni emessi con impulso 
$\vec{K}$\ ~\cite{41}\cite{89}\cite{171}:
\begin{equation}
 \tilde{x}^{\mu}(K)=x^{\mu} - \bar{x}^{\mu} \quad \quad 
  \bar{x}^{\mu}(K)=<x^{\mu}>(K)
\label{quarantasei}
\end{equation}
dove $ <\ldots>$\ indica una media sulla funzione di emissione:
\[ <g>(K)=\frac{\int g(x)f(K,x)\,dx}{\int f(K,x)\,dx} \]
La scelta 
\begin{equation}
 (B^{-1})_{\mu\nu}(K)=<\tilde{x}_{\mu}\tilde{x}_{\nu}>(K) 
\label{superciccia}
\end{equation}
assicura che la funzione di emissione gaussiana (eq.~\ref{quarantacinque}) abbia 
le stesse larghezze quadratiche medie (nello spazio-tempo) della funzione 
di emissione completa. Inserendo l'eq.~\ref{quarantacinque} nella relazione 
fondamentale, eq.~\ref{quarantadue}, si ottiene la seguente 
forma gaussiana per la funzione di correlazione:
\begin{equation}
 C_2(\vec{q},\vec{K})=1 + 
  \exp\left[-q_{\mu}q_{\nu}<\tilde{x}^{\mu}\tilde{x}^{\nu}>(\vec{K})\right]
 \label{quarantotto}
\end{equation}
in cui si sono assunte le approssimazioni di continuit\`a e di ``on shell'', 
che permettono di esprimere le varianze spazio-temporali 
$<\tilde{x}^{\mu}\tilde{x}^{\nu}> $\ come funzioni del solo $\vec{K}$. 
Poich\'e la funzione di correlazione dipende solo dalle distanze relative 
$\tilde{x}_{\mu}$, essa non fornisce alcuna informazione circa la posizione 
assoluta $\bar{x}(\vec{K})$ del centro della sorgente nello spazio-tempo.
La funzione di correlazione, eq.~\ref{quarantotto}, d\`a accesso alle 
larghezze quadratiche medie della ``sorgente effettiva'' di particelle di 
impulso $\vec{K}$. In generale queste larghezze non caratterizzano 
l'estensione totale della regione di collisione, ma piuttosto la 
``regione di omogeneit\`a''~\cite{153}, cio\`e le dimensioni della regione 
da cui vengono emesse con maggiore probabilit\`a coppie di particelle con 
impulso $\vec{K}$. Le varianze spazio-temporali coincidono con l'estensione 
totale della sorgente solo quando la funzione di emissione non presenta 
correlazioni tra impulso e posizione e si fattorizza come
\[ f(K,x)=A^2(K)\rho(x) \] \\
Per confrontare l'eq.~\ref{quarantotto} con i dati sperimentali  
bisogna eliminare una delle quattro componenti di $q_{\mu}$\ per mezzo 
della condizione di shell di massa.
A seconda della scelta delle tre componenti indipendenti, si ottengono 
diverse parametrizzazioni gaussiane; le pi\`u accreditate sono quella 
cartesiana e quella di Yano-Koonin-Podgoretski\u{i} (YKP). 
\subsubsection{Parametrizzazione cartesiana}
\begin{figure}[htb]
  \centering
  \resizebox{0.98\textwidth}{!}{%
  \includegraphics{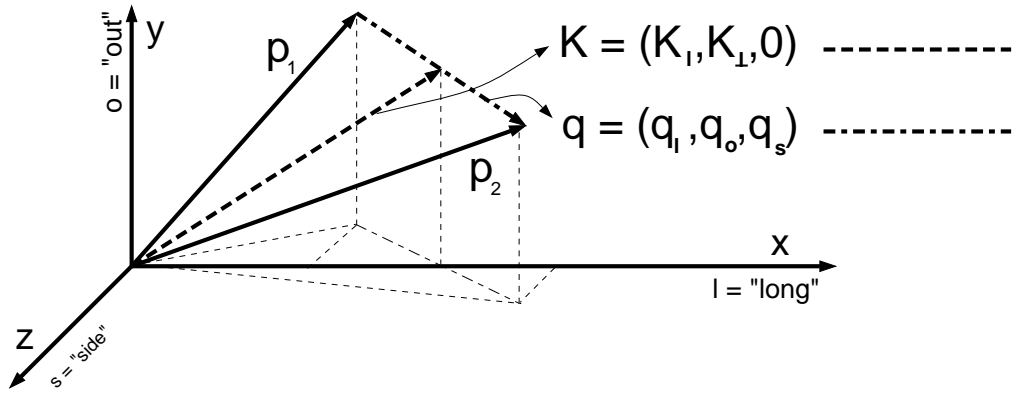}}
 \caption{ 
  Sistema di coordinate {\em osl}: la direzione longitudinale
  ({\em long}) \`e parallela all'asse del fascio. 
 Nel piano ad esso perpendicolare, la direzione {\em out} \`e presa 
 parallela alla componente trasversa dell'impulso della coppia 
 $\vec{K}_{\bot}$; la restante componente 
  cartesiana individua la direzione {\em side}.}
 \label{osl.fig}
\end{figure}

Nella parametrizzazione cartesiana si adopera il sistema di coordinate di 
riferimento {\em osl}, illustrato in fig.~\ref{osl.fig}, basato sulle tre 
componenti cartesiane spaziali della differenza d'impulso 
$q_o$~({\em out}), $q_s$~({\em side}), 
$q_l$~({\em long}). 
La componente temporale \`e eliminata per mezzo della condizione di shell di 
massa (eq.~\ref{quarantaquattro}):
\begin{align}
 & q_0=\vec{\beta}\cdot\vec{q} \quad\quad\quad\quad 
 \vec{\beta}=\frac{\vec{K}}{K^0}\\
 & \vec{\beta}=(\beta_l,\beta_{\bot},0) \quad\quad
 \text{nel sistema {\em osl}}
\end{align}
In generale, nel sistema di riferimento del laboratorio, 
$C_2(\vec{q},\vec{K})$\ dipende non solo da $K_{\bot}$\ e $K_l$, ma anche 
dall'orientazione azimutale $\Phi$\ dell'impulso trasverso della coppia 
$\vec{K_{\bot}}$. Questo angolo non compare esplicitamente nel 
sistema~{\em osl} che \`e orientato differentemente per ogni coppia. 
Si deve dunque definire $\Phi$\ rispetto a qualche direzione 
indipendente dalla 
particolare coppia considerata nel sistema laboratorio, ad esempio rispetto 
al parametro di impatto $\vec{b}$. Per collisioni centrali ($\vec{b}=0$) 
la regione di collisione presenta simmetria azimutale e le funzioni di 
emissione e di correlazione non dipendono da $\Phi$. Nel sistema {\em osl} 
l'indipendenza da $\Phi$\ si manifesta in una simmetria per riflessione 
rispetto alla direzione {\em side}~\cite{40}:
\begin{align}
 & f_{lab}(K_{\bot},\Phi,K_l;x)=f_{lab}(K_{\bot},\Phi+\delta\Phi,K_l;x)
  \Longleftrightarrow \nonumber \\
 & f_{osl}(K_{\bot},K_l;x_l,x_s,x_o,t)=f_{osl}(K_{\bot},K_l;x_l,-x_s,x_o,t)
\end{align}
Per la simmetria per riflessioni $x_s \rightarrow -x_s$\ della funzione 
di emissione, si ha $\bar{y}=<y>=0$\ e le tre varianze spazo-temporali 
$<\tilde{x}_{\mu}\tilde{x}_{\nu}>(\vec{K})$\ lineari in $\tilde{y}$\ sono 
nulle. Il tensore simmetrico $B_{\mu\nu}(\vec{K})$\ ha quindi solo sette 
componenti indipendenti non nulle. Esse si combinano per formare quattro 
parametri $R^2_{ij}(\vec{K})$\ non nulli che caratterizzano la funzione di 
correlazione~\cite{41}\cite{89}:
\begin{align}
 & R^2_s(\vec{K})=<\tilde{x}_s^2>(\vec{K}) \\
 & R^2_o(\vec{K})=<(\tilde{x_o}-\beta_{\bot}\tilde{t})^2>(\vec{K}) \\
 & R^2_l(\vec{K})=<(\tilde{x_l}-\beta_{l}\tilde{t})^2>(\vec{K}) \\
 & R^2_{ol}(\vec{K})=<(\tilde{x_o}-\beta_{\bot}\tilde{t})
     (\tilde{x_l}-\beta_{l}\tilde{t})>(\vec{K}) \\
 & R^2_{os}(\vec{K})=0 \\
 & R^2_{sl}(\vec{K})=0 
\end{align}
Nel sistema {\em osl}, nell'ipotesi di simmetria azimutale, la funzione di
correlazione assume quindi l'espressione 
\begin{equation}
\label{osl-eq}
 C_2(\vec{q},\vec{K})=1+\lambda \exp\left[
 -R^2_oq^2_o-R^2_sq^2_s-R^2_lq^2_l-2|R_{ol}|R_{ol}q_lq_o\right]
\end{equation}
che sar\`a utilizzata per riprodurre i dati sperimentali\footnote{
Nel caso di collisioni non centrali, la simmetria azimutale \`e 
ristabilita quando si considera un insieme molto numeroso di 
collisioni caratterizzato da una distribuzione ``random'' del 
parametro di impatto $\vec{b}$.}. 
\subsubsection{Parametrizzazione di Yano-Koonin-Podgoretski\u{i}}
La condizione di shell di massa, $K \cdot q = 0$, permette di scegliere 
differentemente le tre componenti indipendenti di $q$. 
La parametrizzazione di Yano-Koonin-Podgoretski\u{i} (YKP), che 
presuppone una regione di collisione a simmetria azimutale\footnote{Vedi  
nota 1.}, utilizza le  
componenti $q_{\bot}=\sqrt{q_{o}^2 + q_s^2}$, $q^0$, $q_{\parallel}$\ e la 
funzione:
\begin{multline}
 \label{YKPeq}
 C_2(\vec{q},\vec{K})=1+\lambda \exp \left[
 -R^2_{\bot}(\vec{K})q^2_{\bot} -R^2_{\parallel}(\vec{K})
 (q^2_{\parallel}-(q^0)^2)   \right.  \\
 \left.-(R^2_0(\vec{K})+R^2_{\parallel}(\vec{K}))(q\cdot U(\vec{K}))^2 \right]
\end{multline}
In questa espressione $ U(\vec{K})$\ \`e una quadrivelocit\`a con 
la sola componente spaziale non nulla nella direzione longitudinale:
\begin{align}
 & U(\vec{K})=\gamma(\vec{K})(1,v_{yk}(\vec{K}),0,0)   \nonumber \\
 & \gamma(\vec{K})=\frac{1}{\sqrt{1-v_{yk}^2(\vec{K})}} \nonumber
\end{align}
Poich\'e $(q^2_l-(q^0)^2)$, $(q\cdot U(\vec{K}))^2$\ e $q^2_{\bot}$\ si 
comportano come scalari a seguito di trasformazioni di Lorentz (``boost'') nella 
direzione longitudinale, i tre parametri $R^2_{\bot}(\vec{K})$, 
$R^2_0(\vec{K})$\ e $R^2_{\parallel}(\vec{K})$\ sono invarianti a seguito 
di tali trasformazioni. I loro valori non dipendono quindi dalla velocit\`a 
longitudinale del sistema di riferimento. 
Il parametro $v_{yk}(\vec{K})$, noto come velocit\`a di Yano-Koonin, \`e 
strettamente legato alla velocit\`a della sorgente~\cite{180}. La 
corrispondente rapidit\`a 
\begin{equation}
 Y_{YK}(\vec{K})=\frac{1}{2} \ln \left(\frac{1+v_{yk}(\vec{K})}{1-v_{yk}(\vec{K})}
                  \right)
\label{rapYkn}
\end{equation}
si trasforma in maniera additiva a seguito di ``boost'' longitudinali. 
Le relazioni tra i parametri introdotti e le varianze spazio-temporali 
$<\tilde{x}^{\mu}\tilde{x}^{\nu}>$\ sono date da~\cite{180}\cite{79}:
\begin{align}
 & R^2_{\bot}(\vec{K})=R^2_{s}(\vec{K})=<\tilde{y}^2>(\vec{K}) \\
 & R^2_{0}(\vec{K})=A-v_{yk}C \\
 & R^2_{\parallel}(\vec{K})=B-v_{yk}C  \\
 & v(\vec{K})=\frac{A+B}{2C}\left(1-\sqrt{1-\left(\frac{2C}{A+B}\right)^2}
               \right)
\end{align}
dove $ \vec{\beta}=\vec{K}/K^0$\ e, con la notazione 
$\tilde{\xi}=\tilde{y}+i\tilde{z}$, 
\begin{align}
 & A=<(\tilde{t}-\frac{\tilde{\xi}}{\beta_{\bot}})^2>(\vec{K}) \nonumber \\
 & B=<(\tilde{x}-\frac{\beta_{l}}{\beta_{\bot}}\tilde{\xi})^2>(\vec{K}) 
                                 \nonumber \\
 & C=<(\tilde{t}-\frac{\tilde{\xi}}{\beta_{\bot}})
  (\tilde{x}-\frac{\beta_{l}}{\beta_{\bot}}\tilde{\xi})>(\vec{K})  \nonumber
\end{align}
\subsection{Approccio sperimentale}
Nell'eq.~\ref{ventisette}, la definizione di $C_2$\ corrisponde alla scelta 
$\mathcal{N}=1$\ del fattore di normalizzazione nella definizione 
pi\`u generale della funzione di correlazione:
\begin{equation}
 C_2=\mathcal{N}\frac{P(p_1,p_2)}{P(p_1)P(p_2)}.
 \label{generale}
\end{equation}
$P(p_1,p_2)$\ e $P(p)$, che qui conviene riscrivere in termini 
delle sezioni d'urto differenziali come: 
\begin{align}
 & P(p)=E\frac{1}{\sigma}\frac{d\sigma_{\pi}}{d\vec{p}} \nonumber \\
 & P(p_1,p_2)=E_1E_2\frac{1}{\sigma}
  \frac{d\sigma_{\pi\pi}}{d\vec{p}_1d\vec{p}_2} \nonumber 
\end{align}
sono normalizzati a:
\begin{align}
 & \int P(p) \frac{d^3p}{E} =<\hat{N}> \nonumber \\
 & \int P(p_1,p_2) \frac{d^3p_1}{E_1} \frac{d^3p_2}{E_2}=<\hat{N}(\hat{N}-1)>
 \nonumber
\end{align}
In queste espressioni $<\hat{N}>$\ \`e il valore di aspettazione 
dell'operatore numero $\hat{N}$.\\ 
Da un punto di vista sperimentale conviene normalizzare separatamente 
all'unit\`a il numeratore  ed il denominatore della eq.~\ref{generale}, 
il che fornisce \[ \mathcal{N}=\frac{<\hat{N}>^2}{<\hat{N}(\hat{N}-1)>} \]
Questa scelta permette infatti di intendere la funzione di correlazione 
come il fattore che collega la sezione d'urto differenziale per due pioni 
del mondo reale (cio\`e in presenza della simmetria di Bose-Einstein), con 
una ideale, in cui la correlazione di Bose-Einstein non esiste,
\begin{equation}
d\sigma_{\pi\pi}^{BE}/d\vec{p}_1d\vec{p}_2=C_2(\vec{q},\vec{K})
      d\sigma_{\pi\pi}^{NBE}/d\vec{p}_1d\vec{p}_2
\end{equation}
\newline
%Il problema \`e quindi quello di ottenere una distribuzione 
%costruita a partire da particelle 
Per calcolare il denominatore della funzione di correlazione 
(eq.~\ref{generale}), occorre quindi costruire una distribuzione a 
partire da particelle 
\begin{itemize}
\item con la stessa massa dei pioni
\item con la stessa energia ed impulso delle coppie reali
\item estratte da eventi con la stessa molteplicit\`a e lo stesso stato 
      finale
\item con le stesse correlazioni di origine dinamica (decadimenti di 
        risonanze)
\item senza la correlazione di Bose-Einstein.
\end{itemize}
Vi sono diversi modi di costruire, dai dati sperimentali, questa distribuzione 
di  riferimento. 
\begin{enumerate}
\item Il metodo pi\`u semplice \`e quello di prendere coppie \Pgpm\ \Pgpp 
 dallo stesso evento. Energia ed impulso sono conservate e la correlazione 
 di Bose-Einstein sparisce. Tuttavia correlazioni di diversa origine vengono 
 introdotte poich\'e l'interazione coulombiana \`e 
 ora attrattiva e molte coppie   
 provengono da decadimenti di risonanze.    
\item Un altro approccio possibile \`e quello di rimescolare le componenti 
      trasverse dell'impulso dei pioni. Questo non aggiunge nuove 
      correlazioni ed il quadrimpulso dell'evento \`e conservato. Ovviamente 
      vi \`e violazione per i singoli pioni ma, poich\'e generalmente 
      $|p_t| \ll E$, si ha $E^2-p^2 \approx m_{\pi}^2$\ per le nuove 
      particelle. Il problema principale \`e che la correlazione di 
      Bose-Einstein non \`e completamente rimossa, specialmente nella 
      direzione longitudinale.
\item Prendendo pioni identici da eventi diversi non si introducono nuove 
      correlazioni. Si viola per\`o la conservazione dell' energia e 
      dell'impulso. Questo effetto si riduce combinando eventi con stato 
      finale simile e diventa trascurabile per i tipici stati finali delle 
      interazioni tra ioni pesanti, caratterizzati da un'elevata 
      molteplicit\`a. 
      La correlazione di Bose-Einstein \`e in tal caso quasi del tutto 
      eliminata~\cite{Zajc}. 
\end{enumerate}
\subsection{Un modello per una sorgente in espansione}
Come esposto nei paragrafi precedenti, la condizione di shell di massa 
rende l'interpretazione dei raggi HBT 
dipendente dal modello con cui si parametrizza la funzione di emissione. 
Un modello alquanto generale e flessibile \`e stato sviluppato da 
S. Chapman et al.~\cite{WU5}. La funzione di emissione ha la forma, nel 
sistema di unit\`a naturali (c=1):   
\begin{equation}
f(K,x)=\frac{M_{\bot}\cosh(\eta-Y)}{(2\pi)^3\sqrt{2\pi(\Delta\tau)^2}}
       \exp\left[-\frac{K \cdot u(x)}{T} \right]
       \exp\left[-\frac{(\tau-\tau_0)^2}{2(\Delta\tau)^2}  
     -\frac{r^2}{2R^2_G}-\frac{(\eta-\eta_0)^2}{2(\Delta\eta)^2} \right]
\label{modelf}
\end{equation}
dove le coordinate nelle direzioni longitudinali e 
temporali sono parametrizzate dalla rapidit\`a (spazio-temporale) 
$\eta=0.5\ln[(t+x)/(t-x)]$\ e dal tempo proprio longitudinale 
$\tau=\sqrt{t^2-x^2}$, quelle trasversali da $r=\sqrt{y^2+z^2}$. 
L'impulso della coppia \`e specificato dalla rapidit\`a della 
coppia $Y=0.5\ln[(1+\beta_l)/(1-\beta_l)]$\ e dalla massa trasversa 
$M_{\bot}=\sqrt{m^2+K_t^2}$, avendo assunto $K^0=E_K=\sqrt{m^2+\vec{K}^2}$\  
(approssimazione di ``on shell''). 
$u^{\mu}$\ parametrizza infine la velocit\`a del flusso nella forma:
\[ u^{\mu}(x)=\left(\cosh\eta_l(\tau,\eta)\cosh\eta_t(r),
 \sinh\eta_l(\tau,\eta)\cosh\eta_t(r),\frac{y}{r}\sinh\eta_t(r),
 \frac{z}{r}\sinh\eta_t(r) \right). \]
Il primo fattore dell'eq.~\ref{modelf} specifica la forma 
dell'iper-superficie al ``freeze-out''.  
Il secondo codifica l'ipotesi di 
equilibrio termico locale sovrapposto ad un'espansione collettiva: si 
assume una rapidit\`a del flusso longitudinale indipendente da $\tau$\  
($\eta_l(\tau,\eta)=\eta$), cio\`e un profilo (di Bj{\o}rken~\cite{Bjo83}) 
$v_l=x/t$\ nella direzione longitudinale, ed un profilo lineare di 
intensit\`a $\eta_f$\ per la rapidit\`a di flusso trasverso: 
$\eta_t(r)=\eta_f \frac{r}{R_{G}}$. 
L'ultimo fattore ha invece un'interpretazione puramente geometrica. 
\newline
I parametri del modello sono dunque l'intensit\`a del flusso trasverso 
$\eta_f$\ in unit\`a di rapidit\`a o, equivalentemente, il valor medio 
della {\em velocit\`a} del flusso trasverso $<\beta_T>$\ 
(la velocit\`a e la rapidit\`a del flusso sono legate dalla relazione 
$\beta_\perp = \tanh \eta_t$; nel seguito ci si riferir\`a a 
$<\beta_T>$, che verr\`a indicato pi\`u semplicemente come 
$\beta_T$, piuttosto che ad $\eta_f $),  
la temperatura al freeze-out $T$, 
il raggio geometrico trasverso $R_G$,  
il tempo proprio di freeze-out $\tau_0$\ e la durata dell'emissione 
$\Delta\tau$\ (ossia la dispersione di $\tau$\ attorno a $\tau_0$), 
il centro della distribuzione di rapidit\`a della  
sorgente $\eta_0$\ e la sua larghezza gaussiana $\Delta\eta$.
%
%\section{L'esperimento WA97 all'SPS}
%Nella fig.~\ref{WA97picture}
%\begin{figure}[h]
%\begin{center}
%\includegraphics[scale=0.50]{cap6/WA97_pict2rc.eps}
%\caption{ }
%\label{WA97picture}
%\end{center}
%\end{figure}
%
\subsection{Estrazione dei parametri della sorgente dalla studio dei ``raggi HBT''}
Come discusso nei paragrafi precedenti, nel caso di una sorgente statica i raggi
HBT estratti dal {\em ``best fit''} alla distribuzione misurata sperimentalmente 
forniscono direttamente le dimensioni geometriche della sorgente. Nel caso 
di una sorgente in espansione questa semplice interpretazione non \`e pi\`u valida 
ed i raggi HBT dipendono dall'impulso della coppia $K$. 
Per comprendere questo aspetto, %riferendosi alla fig.~\ref{scketch}, 
si consideri un osservatore che guardi una regione di collisione che si trova in 
uno stato di rapida espansione. 
\begin{figure}[hbt]
\begin{center}
\includegraphics[scale=0.50]{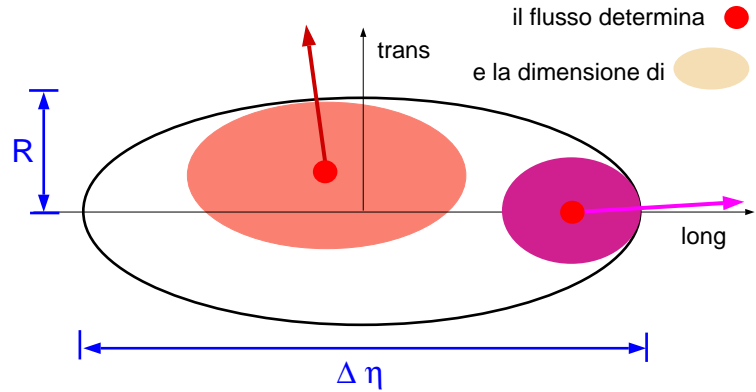}
\caption{Schema della regione di emissione delle particelle con espansione dinamica  
 collettiva. L'ampiezza e l'orientazione del momento medio della coppia $\vec{K}$\ 
 fungono come filtro di lunghezze d'onda: al variare di $\vec{K}$, le misure HBT 
 sono sensibili a diverse parti della regione di collisione.}
\label{scketch}
\end{center}
\end{figure}
Alcune regioni della sorgente si muovono verso l'osservatore e lo spettro 
d'impulso delle particelle emesse da queste regioni risulter\`a 
``spostato verso il blu'' per effetto Doppler ({\em ``blue-shift''});  
altre regioni della sorgente si allontanano dall'osservatore ed appariranno 
quindi ``spostate verso il rosso'' ({\em ``red-shift''}). Se l'osservatore 
guarda la regione di collisione con un filtro di lunghezze d'onda, 
osserver\`a soltanto una parte della regione di collisione. Con 
l'espressione coniata da Sinyukov~\cite{153} si dir\`a che l'osservatore guarda 
una ``regione di omogeneit\`a''.  
\newline
Nell'interferometria HBT, il ruolo del filtro di lunghezze d'onda \`e 
espletato dall'impulso della coppia $\vec{K}$. La direzione di $\vec{K}$\ 
corrisponde alla direzione da cui la sorgente viene osservata. Il modulo di 
$\vec{K}$\ caratterizza la velocit\`a centrale $\beta=\vec{K}/K_0$\ di quella 
parte della sorgente che \`e osservata attraverso il filtro di lunghezza d'onda. 
La situazione \`e schematizzata nella fig.~\ref{scketch}; al variare della direzione 
e del modulo di $\vec{K}$, regioni differenti della sorgente vengono misurate. 
La posizione del centro di una regione di omogeneit\`a $<x_{\mu}>$\ 
({\em cfr.} eq.~\ref{quarantasei}) dipende da $\vec{K}$\ e si trova generalmente 
tra il centro della sorgente e l'osservatore. Nell'approssimazione gaussiana, la 
regione di omogeneit\`a \`e descritta con un ellissoide quadrimensionale 
(nello spazio-tempo) centrato attorno a $<\tilde{x}_{\mu}>$\ e caratterizzato 
(localmente) da una funzione di emissione $f(\vec{K},x)$, 
secondo l'eq.~\ref{quarantacinque}.  Le larghezze di questa regione di 
omogeneit\`a corrispondono alle varianze spazio-temporali 
$<\tilde{x}_{\mu}\tilde{x}_{\nu}>(\vec{K})$,   
fornite dall'eq.~\ref{superciccia}. Pertanto i raggi HBT danno accesso alle varianze 
spazio-temporali (cio\`e alle larghezze della regione di omogeneit\`a); essi non 
dipendono invece dal ``centro effettivo'' della sorgente $<x_{\mu}>(\vec{K})$\ per 
particelle di impulso $\vec{K}$.  

Le informazioni sulla geometria al {\em ``freeze-out''} e sulla dinamica 
di espansione 
della sorgente possono essere ricavate dallo studio della dipendenza dei raggi HBT 
dall'impulso della coppia. 
L'interpretazione non \`e per\`o univoca per quanto discusso nel {\em paragrafo 6.2.3}, 
ma \`e indispensabile riferirsi a qualche modello.  
\newline
A partire dal modello proposto in questo paragrafo (eq.~\ref{modelf}), si possono 
ricavare~\cite{153,WU5} le relazioni seguenti che permettono di estrarre i parametri 
della sorgente:  
\begin{equation}
R_{\perp}(K_t,Y_{\pi\pi})=
R_G \left[ 1+M_{t}\frac{\beta^2_{\perp}}{T}\cosh(Y_{YK}-Y_{\pi\pi})\right]
 ^{-\frac{1}{2}}
\label{Rtransv}
\end{equation}

\begin{equation}
R_{\parallel}(K_t,Y_{\pi\pi})= \tau_o[\frac{M_t}{T}\cosh(Y_{YK}-Y_{\pi\pi}) -
  \cosh^{-2}(Y_{YK}-Y_{\pi\pi}) +1/(\Delta\eta)^2]^{-\frac{1}{2}}
\label{Rperp}
\end{equation}

\begin{equation}
 \Delta\tau \, \simeq R_0  \quad {\rm ad \; elevati}\;  K_t
\label{eq:Deltatau}
\end{equation}

\begin{equation}
(\Delta \eta)^2=(\Delta y)^2-\frac{T}{m_T}
 \label{eq:Deltaeta}
\end{equation}
In queste espressioni $R_{\perp}$, $R_{\parallel}$\ ed $R_0$\ sono i raggi HBT 
della parametrizzazione di Yano-Koonin (eq.~\ref{YKPeq}); $K_t$\ e $Y_{\pi\pi}$\ 
sono la componente trasversale dell'impulso medio e la rapidit\`a media della coppia  
\begin{align}
K_{t} & =\frac{1}{2}\sqrt{(p_{y1}+p_{y2})^2+(p_{z1}+p_{z2})^2} \, ;\nonumber \\
Y_{\pi\pi} & =\frac{1}{2}\log\frac{E_1+E_2+p_{x1}+p_{x2}}
{E_1+E_2-p_{x1}-p_{x2}} 
\label{KtAndY}
\end{align}
ed $M_t=\sqrt{K_t^2+m_{\pi}^2}$.  
Nell'eq.~\ref{eq:Deltaeta} $\Delta y$\ \`e pari alla larghezza della distribuzione 
di rapidit\`a dei $\pi^-$, quantit\`a misurabile dagli esperimenti di grande accettanza, 
e permette di ottenere la larghezza della distribuzione di rapidit\`a della 
sorgente $\Delta\eta$. Questa quantit\`a, cui WA97 non ha accesso, \`e stata 
misurata da NA49 che ha determinato il valore 
$\Delta\eta=1.3 \pm 0.1$~\cite{NA49DeltaEta}.   
Dall'eq.~\ref{Rtransv} \`e possibile ricavare il raggio geometrico trasverso $R_G$\ 
ed il rapporto $\frac{\beta_{\perp}^2}{T}$. Dallo studio della distribuzione di massa 
trasversa di singola particella si ottiene un'ulteriore informazione sulle due 
variabili $\beta_{\perp}$\ e $T$, come prescritto dall'eq.~\ref{Tapp1} che per 
convenienza si riscrive: 
\[ T_{app} = T \sqrt{\frac{1+\beta_\perp}{1-\beta_\perp}} \, .\]
Dalla dipendenza del raggio $R_{\parallel}$\ da $(K_t,Y_{\pi\pi})$\ (eq.~\ref{Rperp}) 
si ricava il valor medio del tempo (proprio) di {\em ``freeze-out''} $\tau_0$.  
La durata del {\em ``freeze-out''} 
$\Delta\tau$\ si ricava dal valore del raggio $R_0$\ ad elevati $K_t$.  
Infine si possono estrarre informazioni sull'espansione longitudinale dallo 
studio della dipendenza di $v_{yk}$\ (od equivalentemente dalla sua 
rapidit\`a $Y_{YK}$, eq.~\ref{rapYkn}) dalla rapidit\`a media della coppia $Y_{\pi\pi}$.  
\section{Discussione dei risultati}
Ci si soffermer\`a inizialmente sull'aspetto dell'analisi che riguarda il 
calcolo delle correzioni per l'interazione coulombiana tra le particelle della  
coppia poich\`e il metodo che \`e stato sviluppato a tal fine rappresenta 
una novit\`a. Inoltre si sono confrontati tra loro i diversi approcci 
generalmente seguiti per calcolare questo tipo di correzione.  
Per tutti gli altri aspetti tecnici dell'analisi, 
quali la correzione dovuta alla risoluzione di coppie di tracce o la procedura di 
{\em ``best fit''} delle correlazioni misurate sperimentalemente,  
si rimanda alla pubblicazione acclusa (cio\`e alla referenza~\cite{HBTpaper}).  
\subsection{Correzione per interazione coulombiana}
L'interazione coulombiana tra due particelle di carica elettrica dello stesso
segno introduce un'anticorrelazione in quanto le due particelle interagenti
vengono accelerate l'una rispetto all'altra.  Il calcolo delle correzioni 
\`e di notevole importanza in quanto gli effetti dell'interazione coluombiana non 
sono trascurabili rispetto a  quelli dell'interferenza di Bose-Einstein ed 
agiscono sulla funzione di correlazione entrambi a piccoli valori dell'impulso 
relativo e su scale confrontabili. I diversi metodi utilizzati per calcolare 
le correzioni possono quindi modificare i risultati dei ``raggi HBT'' e 
l'andamento della loro dipendenza dall'impulso medio della 
coppia $\vec{K}$.  
\newline
L'interazione coulombiana tra due particelle cariche \`e descritta 
dalla funzione d'onda relativa della coppia, espressa in termini della funzione 
ipergeometrica confluente $F$:  
\begin{align}
\nonumber
\Psi^{coul}_{\vec{q}/2}(\vec{r}) = & \Gamma(1+i\eta)e^{-\frac{1}{2}\pi\eta}
   e^{\frac{i}{2}\vec{q}\cdot\vec{r}} F(-i\eta;1;z_{-})  \\
   z_{\pm}= &\frac{1}{2}(qr\pm\vec{q}\cdot\vec{r})=\frac{1}{2}qr(1\pm\cos\theta) 
\label{Hyper}
\end{align}
dove $r=|\vec{r}|$, $q=|\vec{q}|$\ e $\theta$\ denota l'angolo tra i due vettori 
$\vec{r}$\ e $\vec{q}$. Il parametro di Sommerfeld  $\eta=\alpha/(v_{rel}/c)$\ 
dipende dalla massa della particella $m$\ e dall'intensit\`a della costante  
di accopiamento elettromagnetica $\alpha$. Conviene indicare: 
\begin{equation}
\eta_{\pm}=\pm \frac{e^2}{4\pi}\frac{\mu}{q/2}=\pm \frac{me^2}{4 \pi q}
\nonumber
\end{equation}
dove $\mu$\ \`e pari alla massa ridotta ed il segno $+$\ ed il segno $-$ si riferiscono, 
rispettivamente, a coppie di carica opposta e di ugual carica. 
Se le particelle vengono emesse da una sorgente {\em statica} 
a distanza relativa iniziale $r$\ con probabilit\`a $f_{stat}(\vec{r},\vec{K})$, 
la correlazione per l'interazione coulombiana  
\`e pari alla media del modulo quadro della funzione d'onda eq.~\ref{Hyper}:
\begin{equation}
C_2^{Coul}(\vec{q},\vec{K})=\int {\rm d}^3r \, f_{stat}(\vec{r},\vec{K}) 
     |\Psi^{coul}_{\vec{q}/2}(\vec{r})|^2 \, .
\label{CoulCorr}
\end{equation}
La dipendenza da $\vec{K}$\ viene generalmente trascurata nel calcolare le 
correzioni coulombiane. 

La maniera pi\`u semplice e pi\`u frequentemente adoperata per ``correggere''
gli effetti introdotti dall'interazione coulombiana tra le due particelle
che contribuiscono alla correlazione sperimentale prescrive di includere
il cosiddetto fattore di Gamow; questo \`e pari al modulo quadro della
funzione d'onda $\Psi^{coul}_{\vec{r}/2}(0)$,
soluzione dell'equazione di Schrodinger a due corpi e valutata a separazione
radiale nulla.
Si assume infatti implicitamente una sorgente di tipo puntiforme:
$f_{stat}(\vec{r})=\delta^{(3)}(\vec{r})$.
La correlazione dovuta all'interazione coulombiana fornita dall'eq.~\ref{CoulCorr}
diventa quindi il fattore  
\begin{equation}
G(\eta)=\left| \Psi^{coul}_{\vec{q}/2}(0) \right|^2 = \frac{2\pi\eta}{e^{2\pi\eta}-1}
\label{Gamow}
\end{equation}
Nei primi studi la funzione di correlazione di Bose-Einstein per particelle cariche
identiche era calcolata dividendo quella misurata per il fattore di Gamow:
\[ C_{corr}(\vec{q},\vec{K}) = C_{mis}(\vec{q},\vec{K}) / G(\eta_{-}). \]
La correzione di Gamow sovrastima tuttavia le correzioni da apportare, in
quanto assume una sorgente puntiforme;  
nel caso di interazioni tra nuclei pesanti, caratterizzate da sorgenti
di estesi volumi, l'imprecisione diviene intollerabile~\cite{Gamow-bad1}. 
Inoltre, in un ambiente ad elevata molteplicit\`a quale \`e
quello prodotto in
collisioni relativistiche tra ioni pesanti, effetti di schermaggio possono
ridurre l'interazione dello stato finale fino a quando le particelle non
si sono allontanate sufficientemente dal vertice dell'interazione.

Un notevole miglioramento si pu\`o ottenere calcolando il fattore di correzione 
per una sorgente statica di dimensioni finite, come suggerito da diversi  
autori~\cite{Prat86,Bowler,Baym}. Tipicamente si assume una distribuzione 
gaussiana, del tipo $f_{stat}(\vec{r})\propto\exp[-\vec{r}^2/4R^2]$. 
Il fattore di correzione e la funzione di correlazione corretta diventano quindi,  
rispettivamente:  
\begin{align}
 & F^{stat}_{corr}(\vec{q}) = 
   \int {\rm d}^3 r f_{stat}(\vec{r})|\Psi^{coul}_{\vec{q}/2}(\vec{r})|^2 
 \label{usami} \\
 & C_{corr}(\vec{q},\vec{K})=C_{mis}(\vec{q},\vec{K}) / F^{stat}_{corr}(\vec{q}) 
 \label{usami2}
\end{align}

Nel modello classico suggerito da Baym e Braun-Munzinger~\cite{Baym}
si tiene conto degli effetti dell'interazione coulombiana ed anche di quelli 
di schermaggio del mezzo nucleare 
trascurando l'interazione coulombiana delle particelle fino a 
quando queste distano per meno di una certa quantit\`a $r_0$, 
ed includendo la sola interazione
coulombiana relativa quando la loro separazione supera $r_0$. 
La differenza di impulso iniziale, $\vec{q}_0$, e quella finale misurata,
$\vec{q}$, sono quindi legate dalla relazione:
\[ \frac{(\vec{q}/2)^2}{2\mu}=\frac{(\vec{q}_0/2)^2}{2\mu} \pm
 \frac{e^2}{r_0}  \]
dove il segno $+$\ vale per particelle di stesso segno di carica, quello
$-$\ per coppie di carica opposta.   
%e $\mu$\ \`e la massa ridotta delle due particelle.
La funzione di correlazione per due particelle viene quindi modificata dallo
Jacobiano $|d^3q_0/dq^3|=q_0/q$\ e diventa:
\begin{equation}
 C(\vec{q})=\left|\frac{d^3q_0}{dq^3} \right| C_0(\vec{q})=C_0(\vec{q})
                \sqrt{1 \mp \frac{2\mu e^2}{r_o(q/2)^2}}
\label{toy-equation}
\end{equation}
Per calcolare il valore di $r_0$\ necessario per
correggere i dati sperimentali
nella correlazione tra particelle di ugual carica, si pu\`o studiare la
correlazione tra particelle di carica opposta (\Pgpm\ e \Pgpp) 
per le quali la correlazione non ha origine dalla statistica quantistica, ma
essenzialmente dall'interazione nello stato finale.  
Assumendo che la correlazione abbia solo origine nell'interazione 
coulombiana ($C_0=1$) si parametrizza la distribuzione di coppie di carica 
opposta con la funzione: 
 \[ \sqrt{1 + \frac{2\mu e^2}{r_o(q/2)^2}}. \]
Noto $r_0$, si corregge quindi il contributo che ciascuna coppia apporta alla funzione
di correlazione tra particelle identiche moltiplicando per il fattore:
\begin{equation} 
\frac{1}{\sqrt{1 - \frac{m_{\pi} e^2}{r_o(q/2)^2}}} \, .
\label{BaymCorr}
\end{equation}

I tre metodi esposti in questa breve panoramica sono tutti, per diversi aspetti, 
insoddisfacenti. Delle limitazioni intrinseche del semplice 
fattore di Gamow si \`e gi\`a discusso; 
i metodi che assumono una sorgente estesa e statica sono altres\`i 
criticabili nello studio delle collisioni tra ioni ultra-relativistici 
in cui si attende un elevato flusso collettivo della materia nucleare da cui 
sono emesse le particelle. Il modello di  Baym e Braun-Munzinger ha delle 
attrattive, in quanto si basa sulla misura sperimentale della correlazione 
tra particelle di carica opposta (che quindi pu\`o tener anche 
conto di effetti strumentali altrimenti non apprezzabili), ma anche molte limitazioni. 
In primo luogo non \`e detto che trascurare completamente l'interazione tra le 
particelle della coppia per $r<r_0$\ tenga conto correttamente dello schermaggio 
elettrico dell'ambiente circostante. Inoltre, la correzione \`e la stessa per tutte 
le componenti di $\vec{q}$, dipendendo solo dal modulo: ci\`o vuol dire che si 
assume implicitamente simmetria spaziale sferica  per la sorgente e nessuna 
dipendenza dal tempo (cio\`e la sorgente \`e ancora di tipo statico). 

\begin{figure}[htb]
\begin{center}
\includegraphics[scale=0.20]{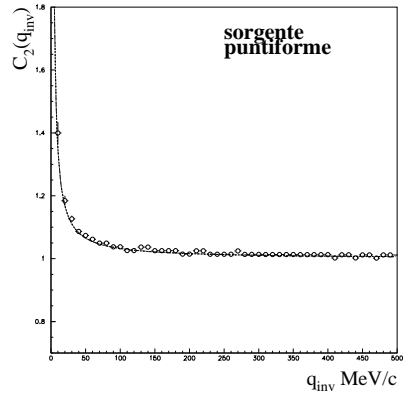}
\caption{Risultati del calcolo con il Monte Carlo 
 per la correlazione dovuta all'interazione coulombiana nel caso di sorgente 
 puntiforme (cerchietti) con sovrapposta la predizione del fattore di 
 Gamow (curva continua).}
\label{CoulomCorrTest}
\end{center}
\end{figure}

Si \`e sviluppato pertanto un nuovo metodo nel quale si calcolano le correzioni per 
interazione coulombiana assumendo che l'emissione delle particelle avvenga da una sorgente 
in espansione, secondo lo stesso modello fornito dall'eq.~\ref{modelf} che viene 
poi impiegato in fase di interpretazione dei risultati sperimentali.  
Il fattore di correzione viene dunque calcolato a partire 
dall'eq.~\ref{usami}, in cui $f_{stat}$\ viene sostituita 
dalla funzione dell'eq.~\ref{modelf} che assume una sorgente in espansione.  
Operativamente il calcolo non \`e eseguito in maniera analitica ma con una 
tecnica Monte Carlo in cui si generano particelle secondo tale funzione di 
emissione (eq.~\ref{modelf}) e per ciascuna coppia si calcola la 
funzione d'onda $\Psi^{coul}_{\vec{q}/2}(\vec{r})$. Il metodo Monte Carlo \`e 
stato verificato applicandolo al caso di una sorgente puntiforme, parametrizzata 
con la funzione $f_{point}=\delta^{(3)}(\vec{r})\delta(t-t_0)$\ 
in cui tutte le particelle sono emesse simultaneamente da un unico punto dello spazio, 
e confrontando i risultati con quanto prescritto dal fattore di Gamow.  
In fig.~\ref{CoulomCorrTest} \`e mostrato il risultato di questa verifica: la curva 
continua rappresenta la correlazione calcolata con il fattore di Gamow (eq.~\ref{Gamow}), 
i punti con le barre di errore il risultato del Monte Carlo per la correlazione 
di origine coulombiana nel caso di sorgente puntiforme. L'ottimo accordo tra 
le due distribuzioni d\`a fiducia nel metodo.  

\begin{figure}[htb]
\begin{center}
\includegraphics[scale=0.36]{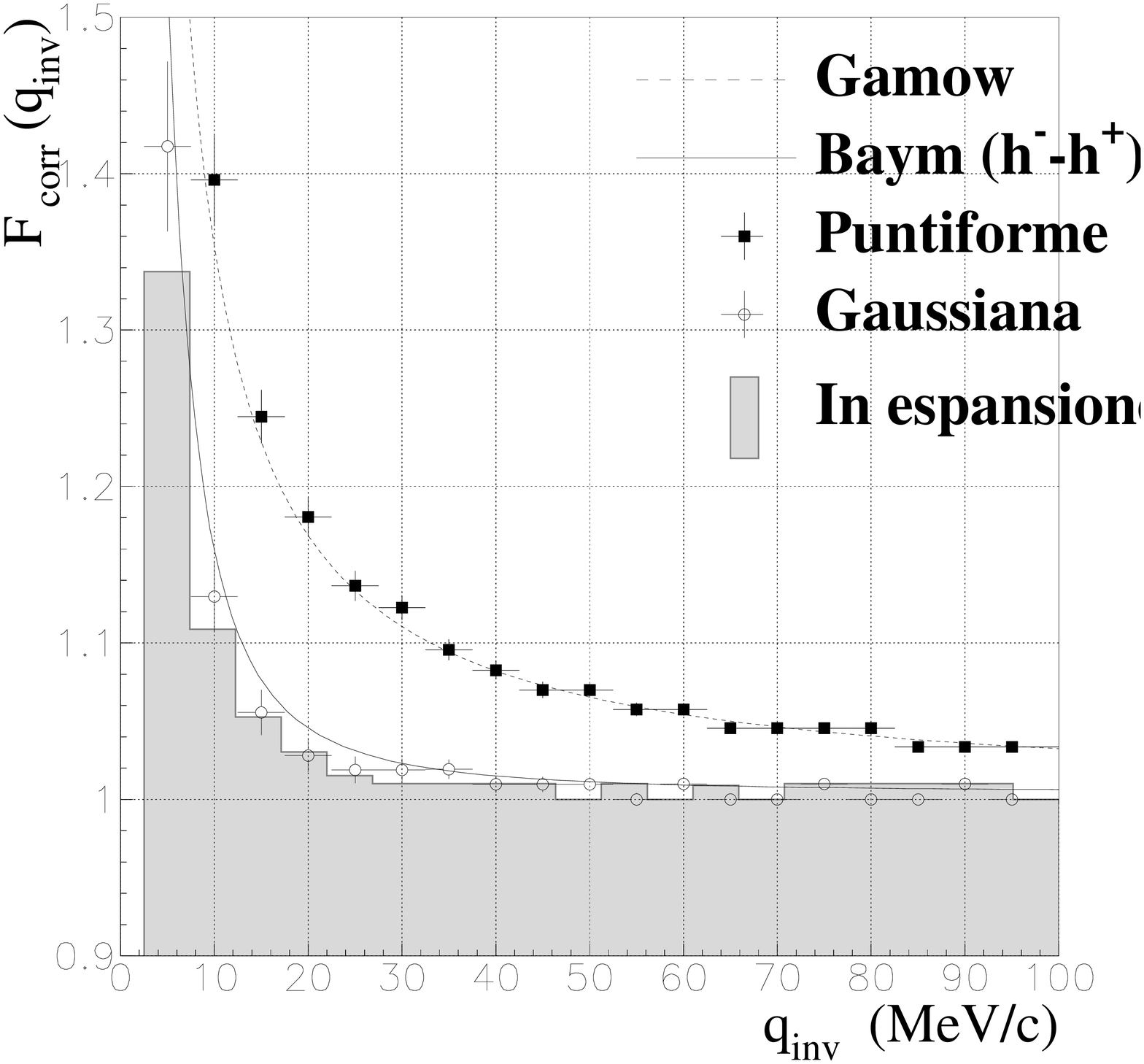}
\caption{Risultati delle correzioni per interazione coulombiana 
 secondo diversi metodi di calcolo. Si veda il testo per il loro significato.}
\label{CoulomCorr}
\end{center}
\end{figure}

Tuttavia, venendo al caso della funzione di emissione (eq.~\ref{modelf}), vi \`e 
il problema che ancora non si conoscono i 
parametri del modello ($T$,$\beta_\perp$,$R_G$,$\tau_0$,$\Delta\tau$,$\Delta\eta$),  
che sono le quantit\`a che si desidera misurare con l'analisi HBT. \`E dunque 
necessario procedere in modo iterativo:  
\begin{itemize}
\item
si parte da una correzione coulombiana iniziale ragionevole, quale 
quella ottenuta col metodo di Baym e Braun-Munzinger, a partire dalla misura 
della correlazione tra particelle di carica opposta, come si vedr\`a; 
\item
si calcolano dei valori iniziali per i parametri del modello, a partire 
dallo studio della dipendenza dei raggi HBT, estratti dalla funzione di 
correlazione cos\`i corretta, nel modo discusso in dettaglio nella 
pubblicazione allegata (cio\`e la referenza~\cite{HBTpaper}); 
\item 
si calcola nuovamente la correzione coulombiana, utilizzando adesso il 
nuovo metodo implementato, che assume la sorgente in espansione con i 
parametri ottenuti nella fase precedente; si 
calcolano nuovamente i raggi HBT per la funzione di correlazione cos\`i 
corretta, ed a partire da questi si ricalcolano i nuovi  
parametri del modello ($T$,$\beta_\perp$,$R_G$,$\tau_0$,$\Delta\tau$,$\Delta\eta$); 
\item
si ripete la procedura fino a quando non si ottiene 
la convergenza dei valori dei parametri. 
\end{itemize}
La convergenza \`e stata raggiunta
gi\`a dopo la prima iterazione, suggerendo cos\`i che la correzione iniziale del 
metodo di Baym e Braun-Munzinger e quella finale calcolata per una sorgente in
espansione sono tra loro abbastanza simili.  
\newline
In fig.~\ref{CoulomCorr} sono mostrati i risultati finali per le correzioni 
$F(\vec{q})$\ calcolate nel modo appena descritto e, per confronto, le correzioni 
calcolate con gli altri approcci descritti in precedenza.  
L'istogramma in grigio \`e il risultato del metodo sviluppato che suppone una 
sorgente in espansione. Esso \`e quello che fornisce le correzioni pi\`u 
piccole. Il metodo di Baym e  Braun-Munzinger prescrive una correzione 
secondo la curva a tratto continuo nella fig.~\ref{CoulomCorr}. Per calcolare 
questa curva si \`e studiata la correlazione tra le particelle di carica opposta, 
ottennendo un raggio effettivo $r_0=6.1 \pm 0.5$\ fm, compatibile con quello misurato 
da NA49~\cite{NA49HBT}, e si \`e calcolata la funzione fornita dall'eq.~\ref{BaymCorr}. 
I cerchietti vuoti si riferiscono al caso della sorgente statica gaussiana 
(eq.~\ref{usami}, eq.~\ref{usami2}) di raggio $R=4$\ fm, valore frequentemente  
incontrato in letteratura.  
Nella fig.~\ref{CoulomCorr} sono anche riportati i risultati del fattore di Gamow 
e della sorgente puntiforme, descritti in riferimento alla fig.~\ref{CoulomCorrTest}. 
\`E evidente come il fattore di Gamow sovrastimi in modo significativo le correzioni; 
il metodo con la sorgente gaussiana statica e quello dello studio della correlazione 
tra coppie di carica opposta (secondo il modello di Baym e Braun-Munzinger) 
sono tra loro in buon accordo. Entrambi sovrastimano per\`o le correzioni nel caso 
di una sorgente in espansione, soprattutto a piccoli valori di $q_{inv}$, che sono 
quelli pi\`u critici per lo studio delle correlazioni.
\subsection{Studio della dipendenza dei raggi HBT dall'impulso della 
coppia $\vec{K}$}
\begin{figure}[t]
  \centering
  \resizebox{0.58\textwidth}{!}{%
  \includegraphics{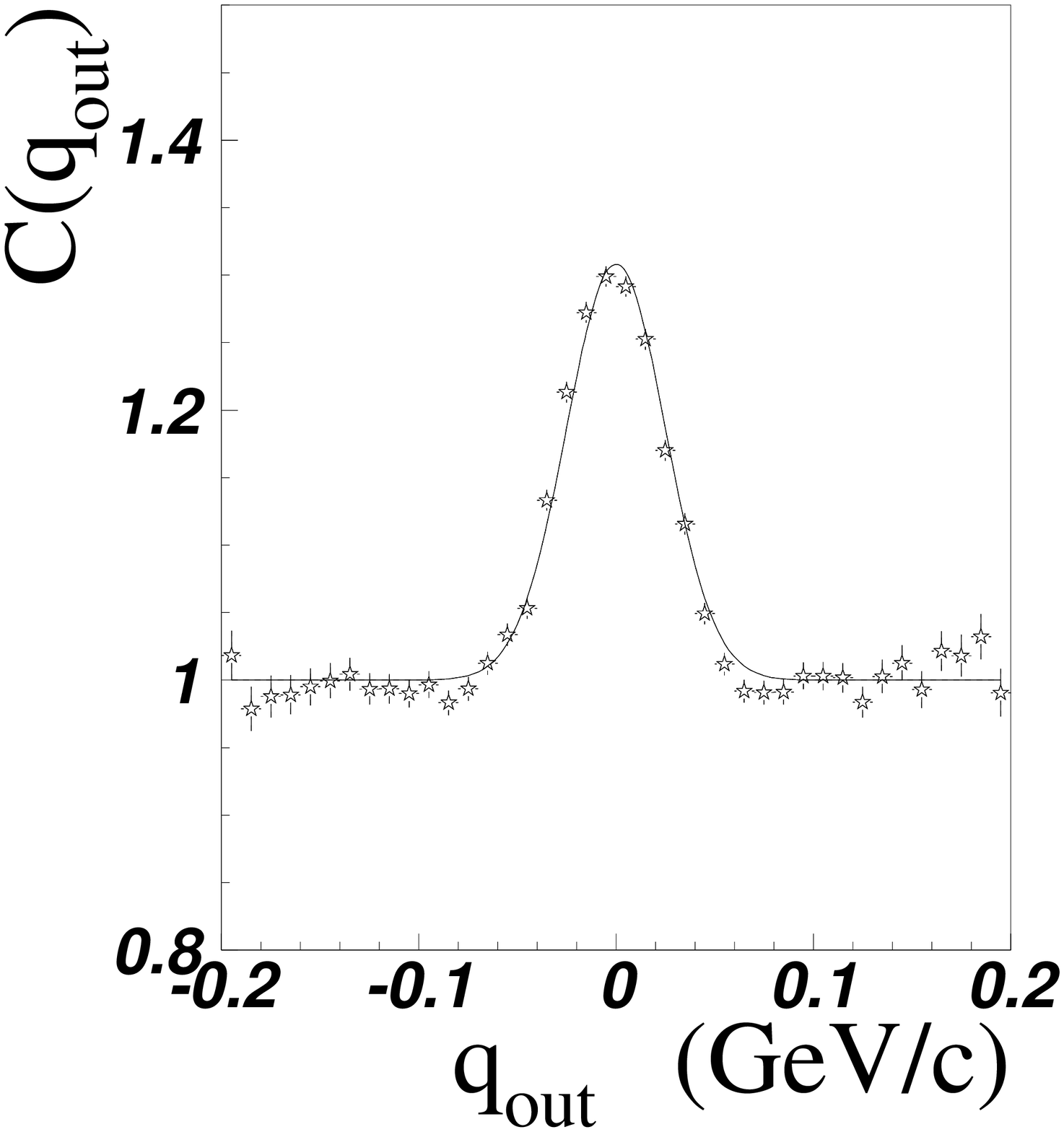}
  \includegraphics{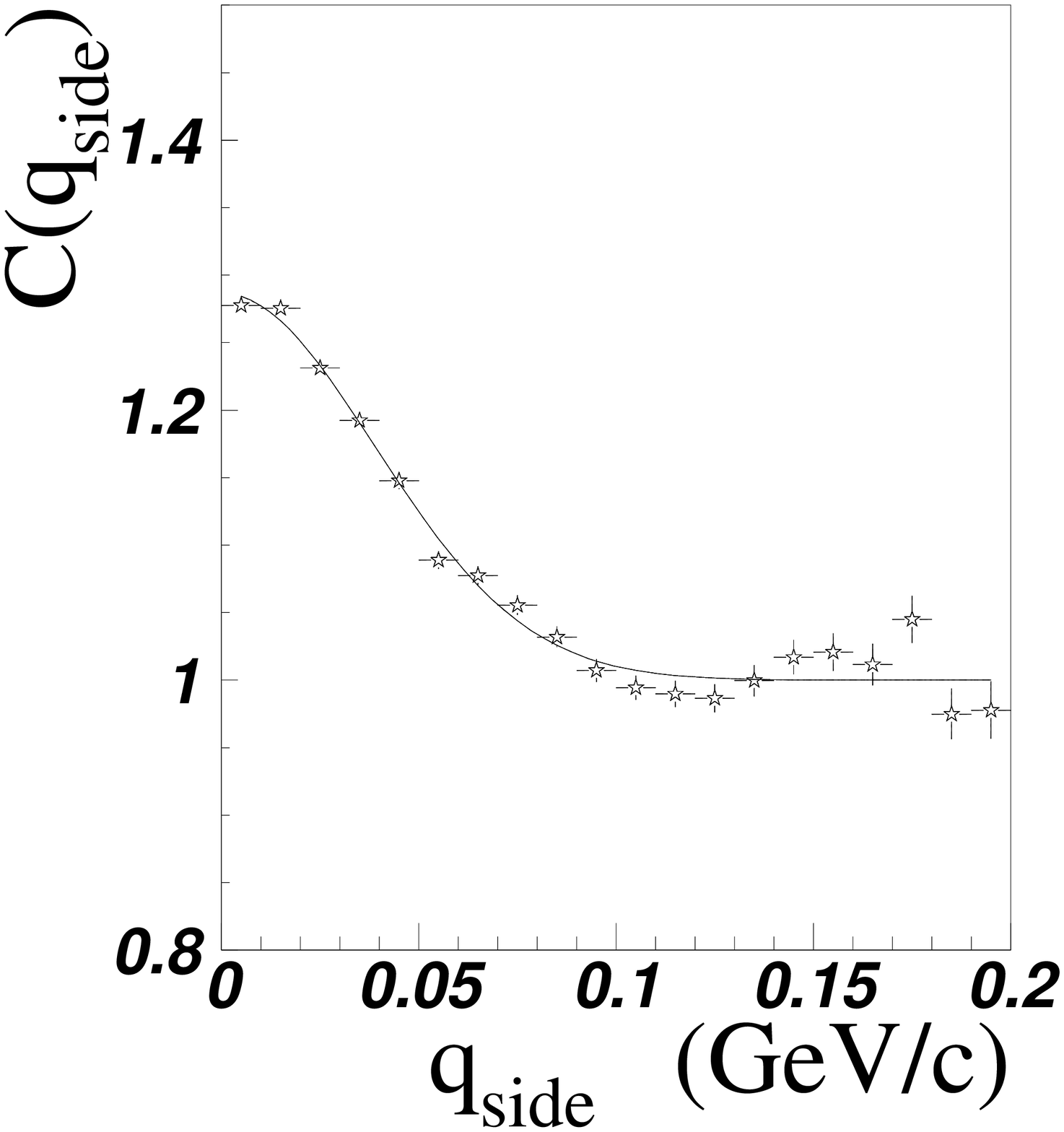}
  \includegraphics{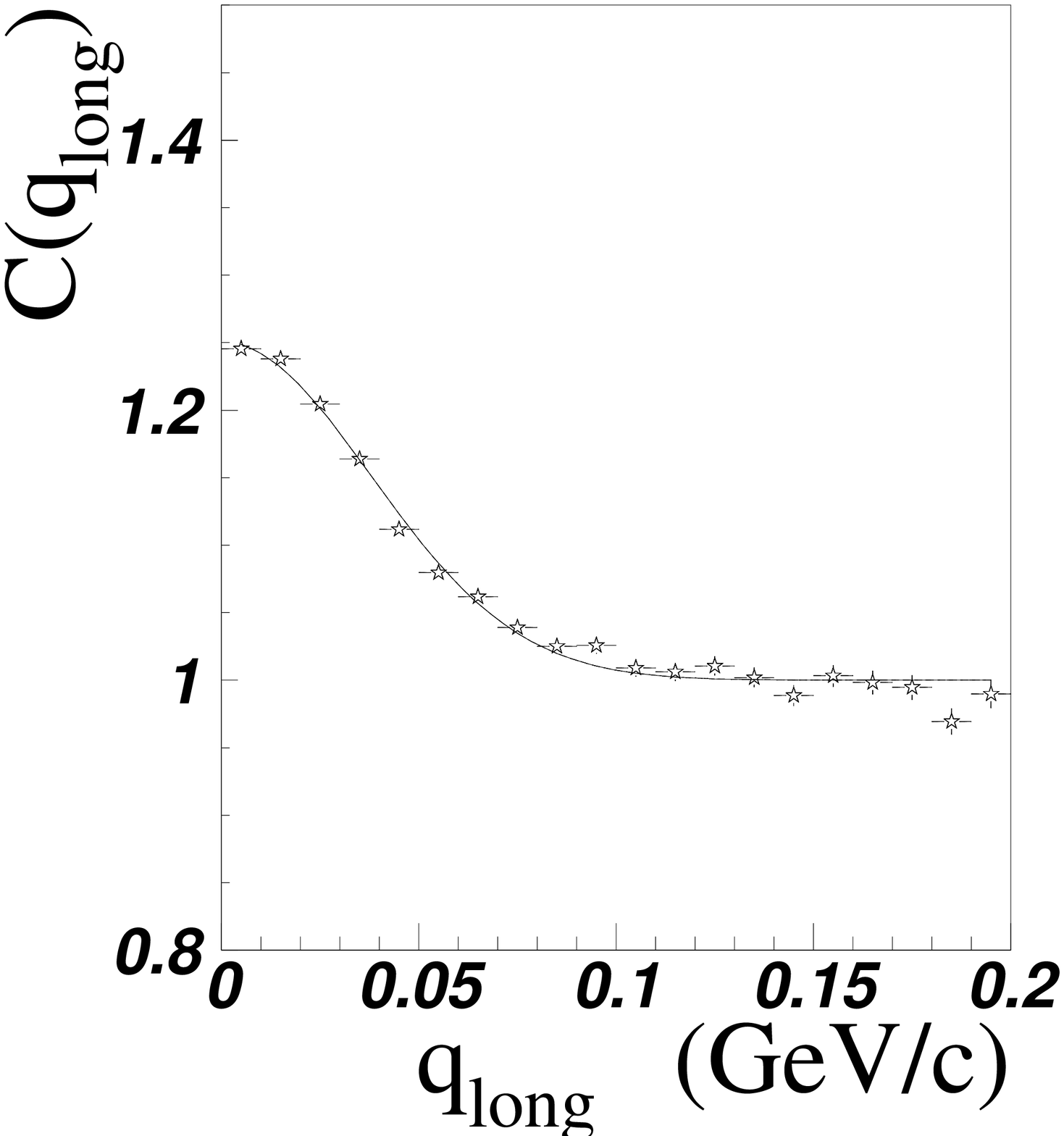}}\\
  \resizebox{0.58\textwidth}{!}{%
  \includegraphics{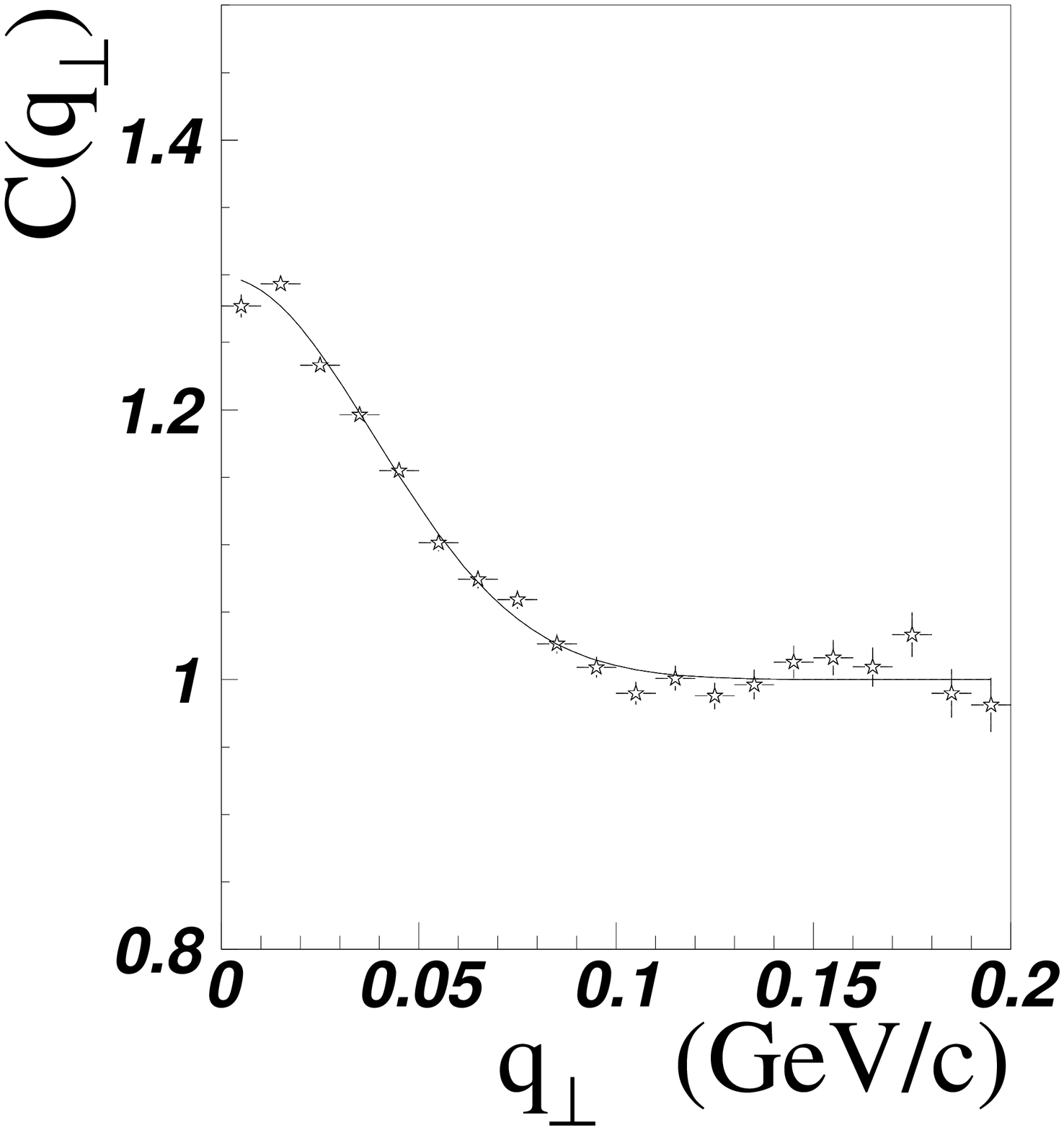}
  \includegraphics{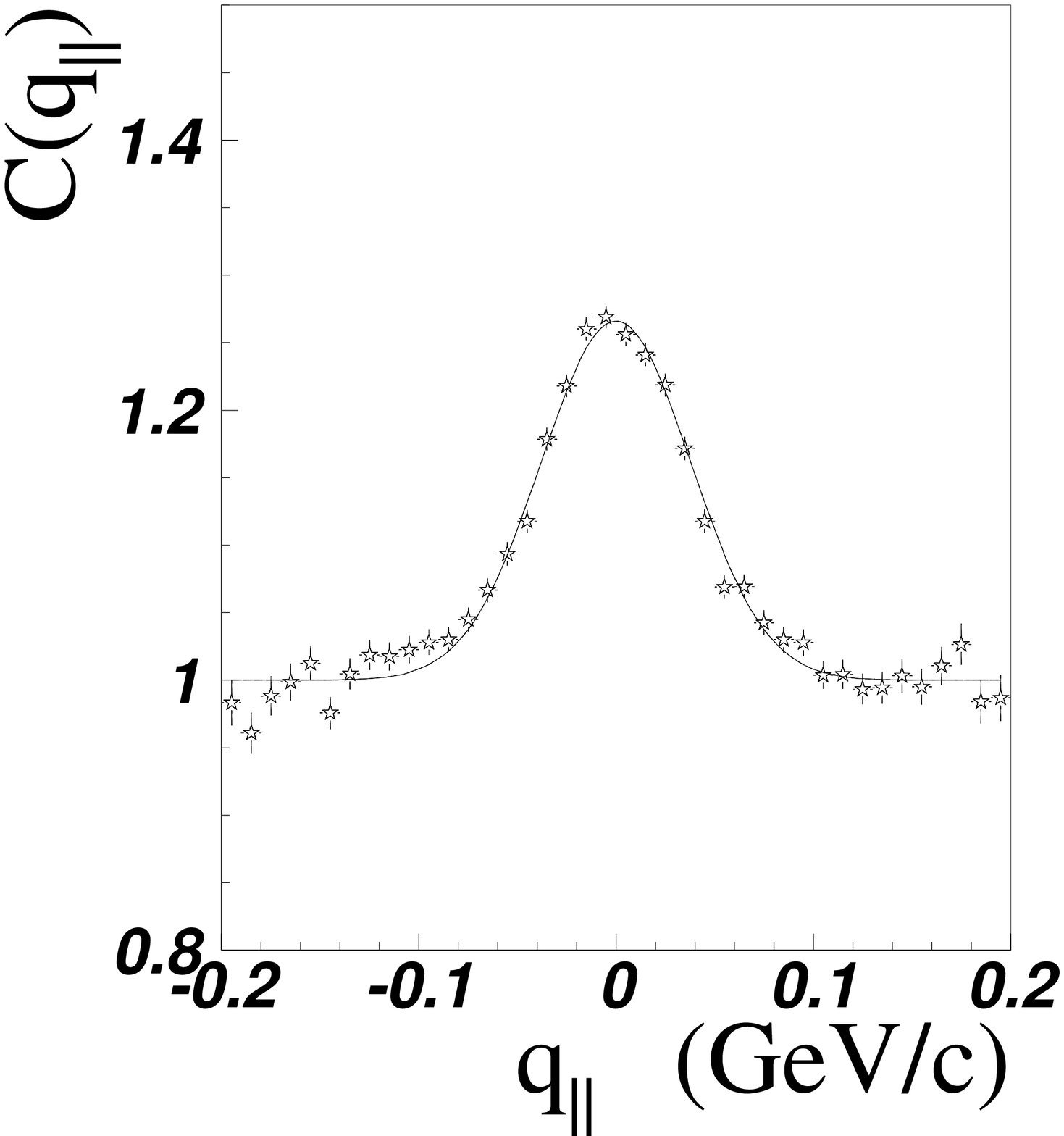}
  \includegraphics{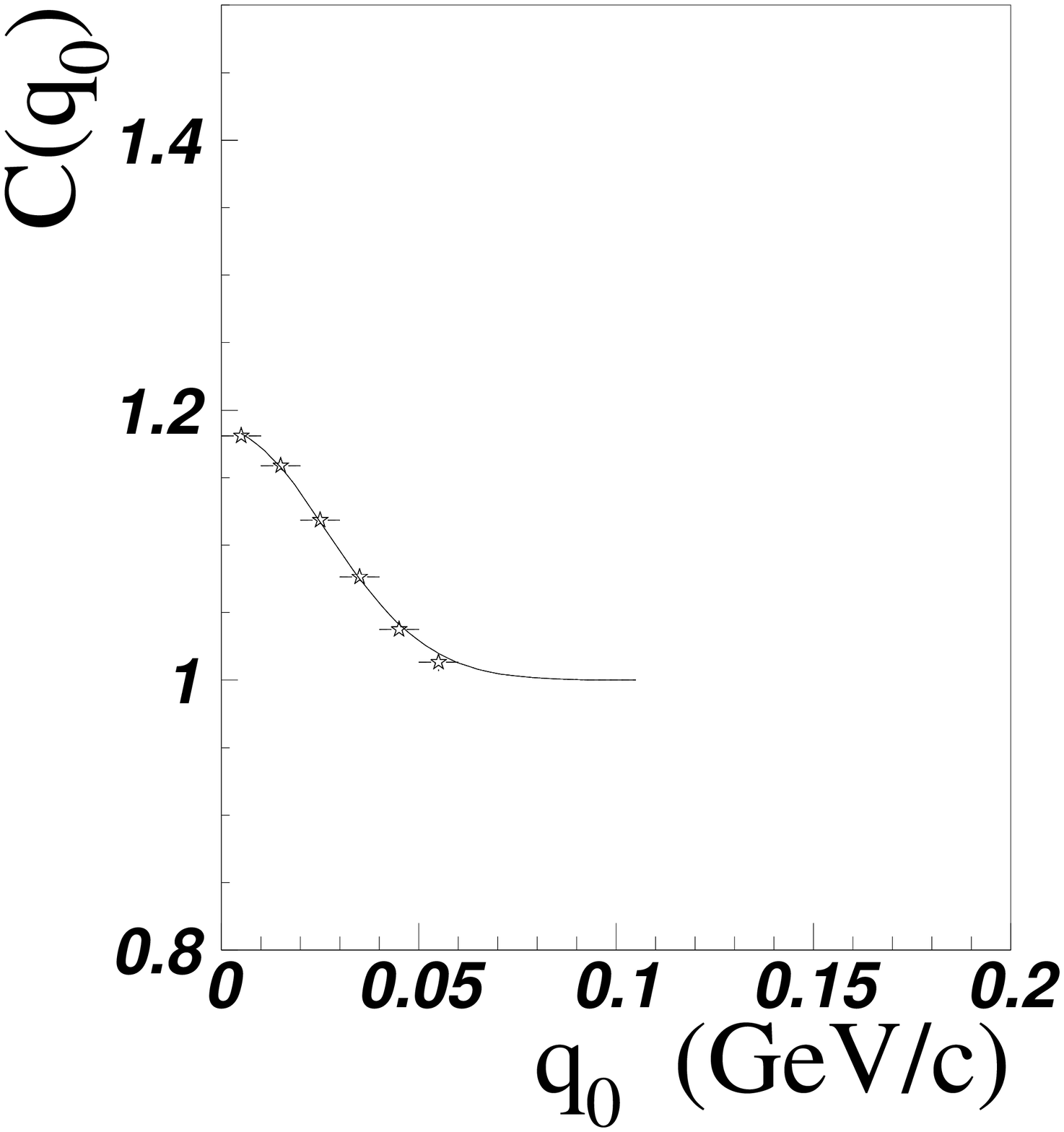}}
  \caption{{\rm Proiezioni della funzione di correlazione per coppie 
  $h^-$-$h^-$\ misurata adoperando la parametrizzazione cartesiana 
  (in alto) e quella YKP (in basso)  per coppie di rapidit\`a 
  $ 2.55  <  Y_{\pi\pi}  <  3.65 $\ ed impulso trasverso 
  $ 0.23  <  K_t  <  0.39 \, {\rm GeV}/c$.}}
  \label{projec}
 \end{figure}

Per poter studiare la dipendenza dei raggi HBT dall'impulso medio della 
coppia $\vec{K}$, si \`e suddivisa la finestra di accettanza, definita per 
questo studio nel piano $(Y_{\pi\pi},K_t)$, nei 20 intervalli 
indicati nella tab.~\ref{accept}.   
\begin{table}[h]
\begin{center}
\begin{tabular}{||l|c|c|c|c||} \hline \hline
\multicolumn{1}{|c|}{ } & \multicolumn{4}{c|}
{\bf $Y_{\pi\pi} \, \longrightarrow$} \\ \hline
& $\begin{array}{cc} 0.910 \\ 3.11 \end{array} $ &
$\begin{array}{cc} 0.915 \\ 3.28 \end{array} $ &
$\begin{array}{cc} 0.915 \\ 3.42 \end{array} $ &
$\begin{array}{cc} 0.925 \\ 3.60 \end{array} $  \\ \cline{2-5}
& $\begin{array}{cc} 0.660 \\ 3.10 \end{array} $ &
$\begin{array}{cc} 0.665 \\ 3.28 \end{array} $ &
$\begin{array}{cc} 0.665 \\ 3.42 \end{array} $ &
$\begin{array}{cc} 0.670 \\ 3.60 \end{array} $  \\ \cline{2-5}
 {\bf $\uparrow $} &
$\begin{array}{cc} 0.535 \\ 3.05 \end{array} $ &
$\begin{array}{cc} 0.535 \\ 3.23 \end{array} $ &
$\begin{array}{cc} 0.540 \\ 3.37 \end{array} $ &
$\begin{array}{cc} 0.540 \\ 3.56 \end{array} $  \\ \cline{2-5}
 {\bf $K_t$}  &
 $\begin{array}{cc} 0.435 \\ 3.03 \end{array} $ &
 $\begin{array}{cc} 0.440 \\ 3.23 \end{array} $ &
 $\begin{array}{cc} 0.440 \\ 3.37 \end{array} $ &
 $\begin{array}{cc} 0.445 \\ 3.55 \end{array} $  \\ \cline{2-5}
 & $\begin{array}{cc} 0.315 \\ 2.86 \end{array} $ &
 $\begin{array}{cc} 0.330 \\ 3.08 \end{array} $ &
 $\begin{array}{cc} 0.335 \\ 3.22 \end{array} $ &
 $\begin{array}{cc} 0.345 \\ 3.42 \end{array} $  \\ \hline \hline
\end{tabular}
\resizebox{0.40\textwidth}{!}{%
    \includegraphics{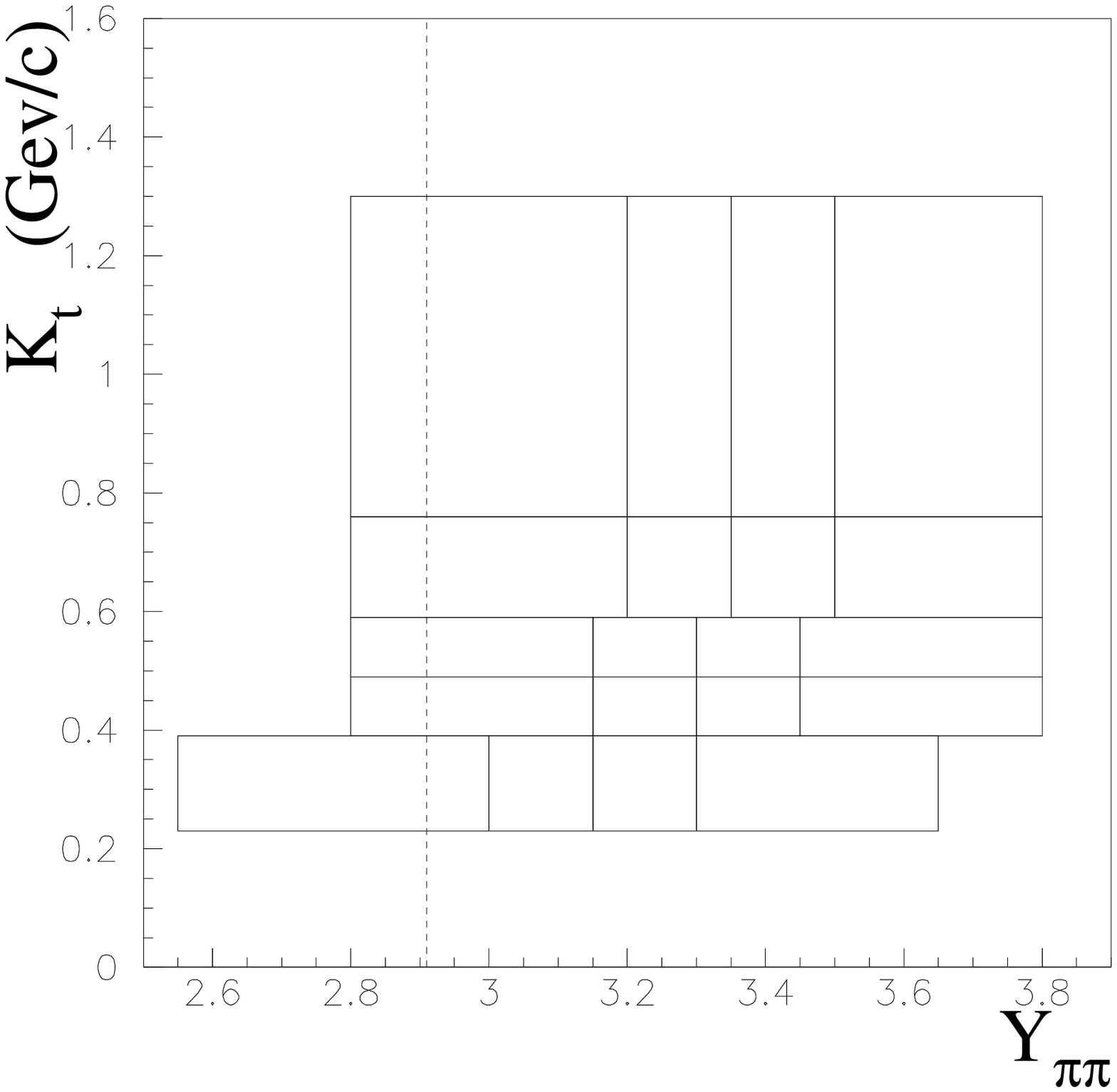}}
\caption{Finestra di accettanza per le coppie di particelle di carica 
         negativa nel piano $(Y_{\pi\pi},K_t)$\ (a destra). 
	 Per ciascuno dei venti 
	 rettangoli in cui si \`e suddivisa la finestra di accettanza 
	 sono riportati il valor medio della distribuzione di $K_t$\ 
	 (numeri in alto in ciascuna cella) e quello della 
	 distribuzione di $Y_{\pi\pi}$\ (numeri in basso). La linea tratteggiata 
	 in figura indica il valore di rapidit\`a centrale.  
\label{accept}}
\end{center}
\end{table}
La funzione di correlazione \`e stata quindi misurata in ciascuno dei 
venti rettangoli cos\`i definiti, sia nella parametrizzazione cartesiana 
(eq.~\ref{osl-eq}) sia in quella di Yano-Koonin (eq.~\ref{YKPeq}).  I 
raggi estratti dalla parametrizzazione cartesiana sono stati sistematicamente 
adoperati per controllare quelli dell'altra parametrizzazione, come spiegato 
in dettaglio in~\cite{HBTpaper}. 
In fig.~\ref{projec} sono mostrate le proiezioni della funzione di correlazione 
misurata nell'intervallo a valori di $K_t$\ pi\`u piccoli.

L'analisi \`e stata ripetuta in ciascuna delle quattro classi di centralit\`a 
definite nell'esperimento WA97 ({\em cfr. paragrafo 1.5.3} ed in particolare 
la tab.~1.5).  
In tal modo si \`e realizzato uno studio della dinamica di espansione 
della sorgente in funzione della centralit\`a della collisione.  

\subsubsection{Temperatura e dimensione trasversa al {\em ``freeze-out''}, 
       intensit\`a del flusso trasverso collettivo} 
\begin{figure}[htb]
\begin{center}
\resizebox{0.64\textwidth}{!}{%
\includegraphics{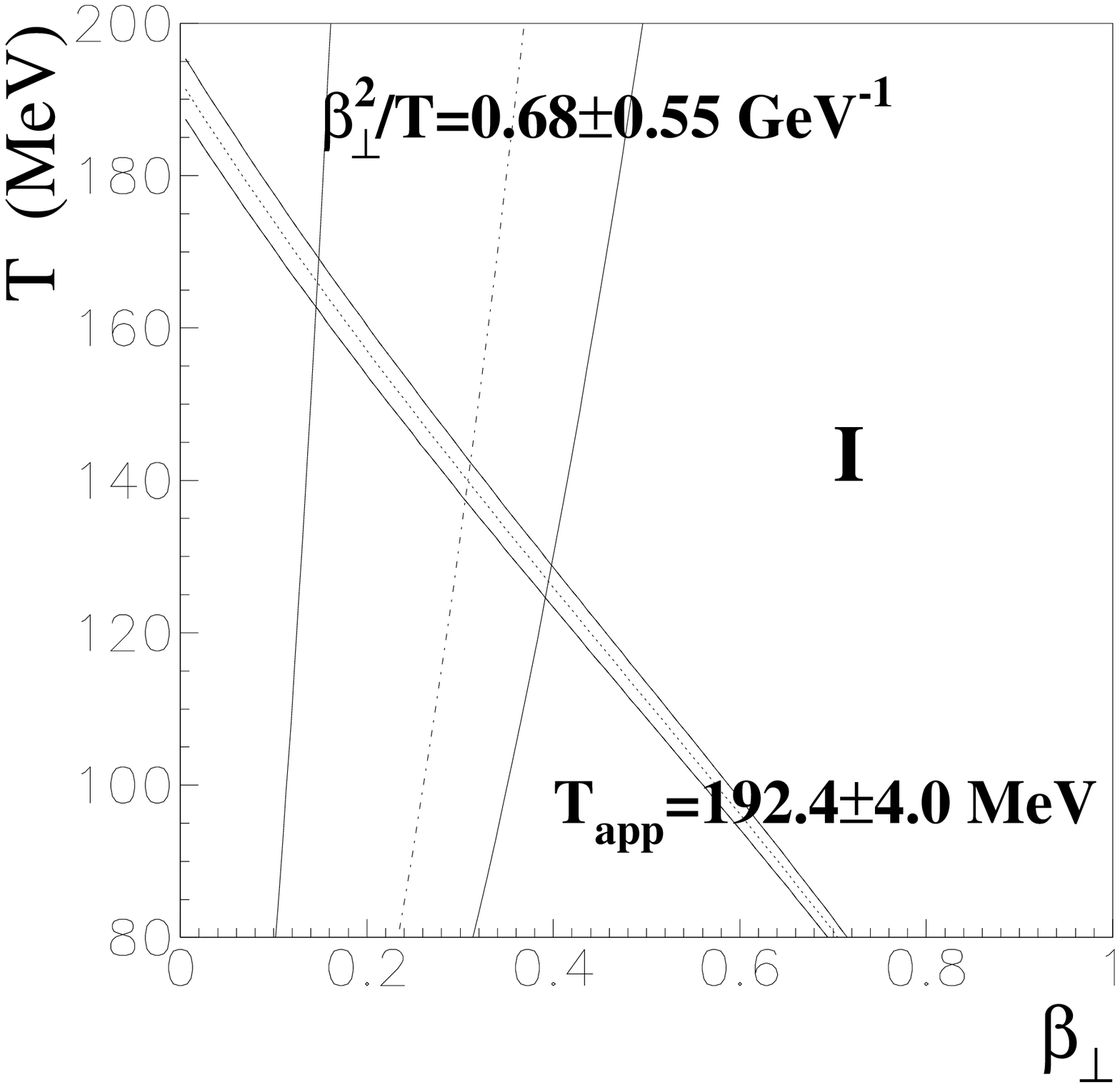}
\includegraphics{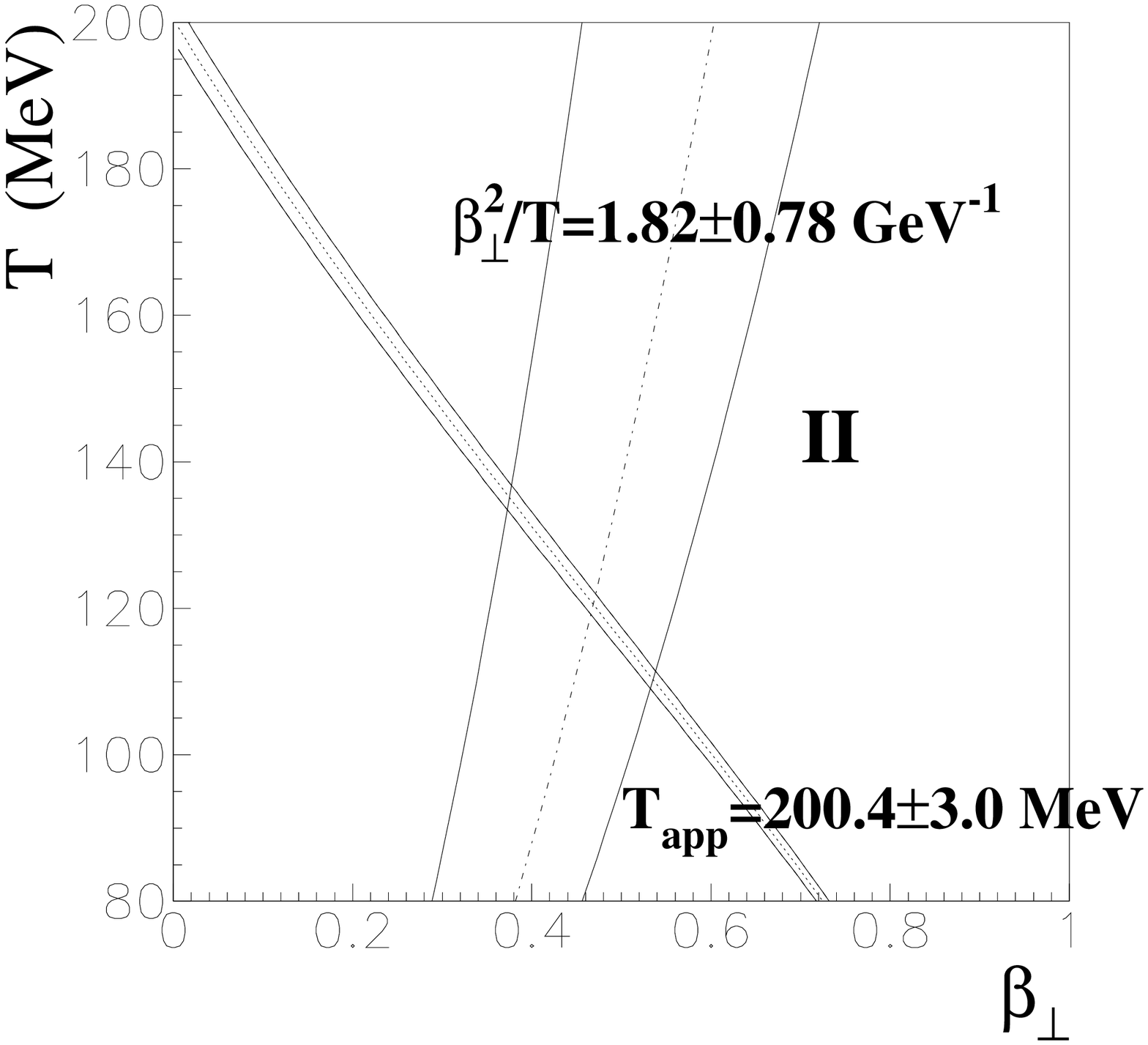}}\\
\resizebox{0.64\textwidth}{!}{%
\includegraphics{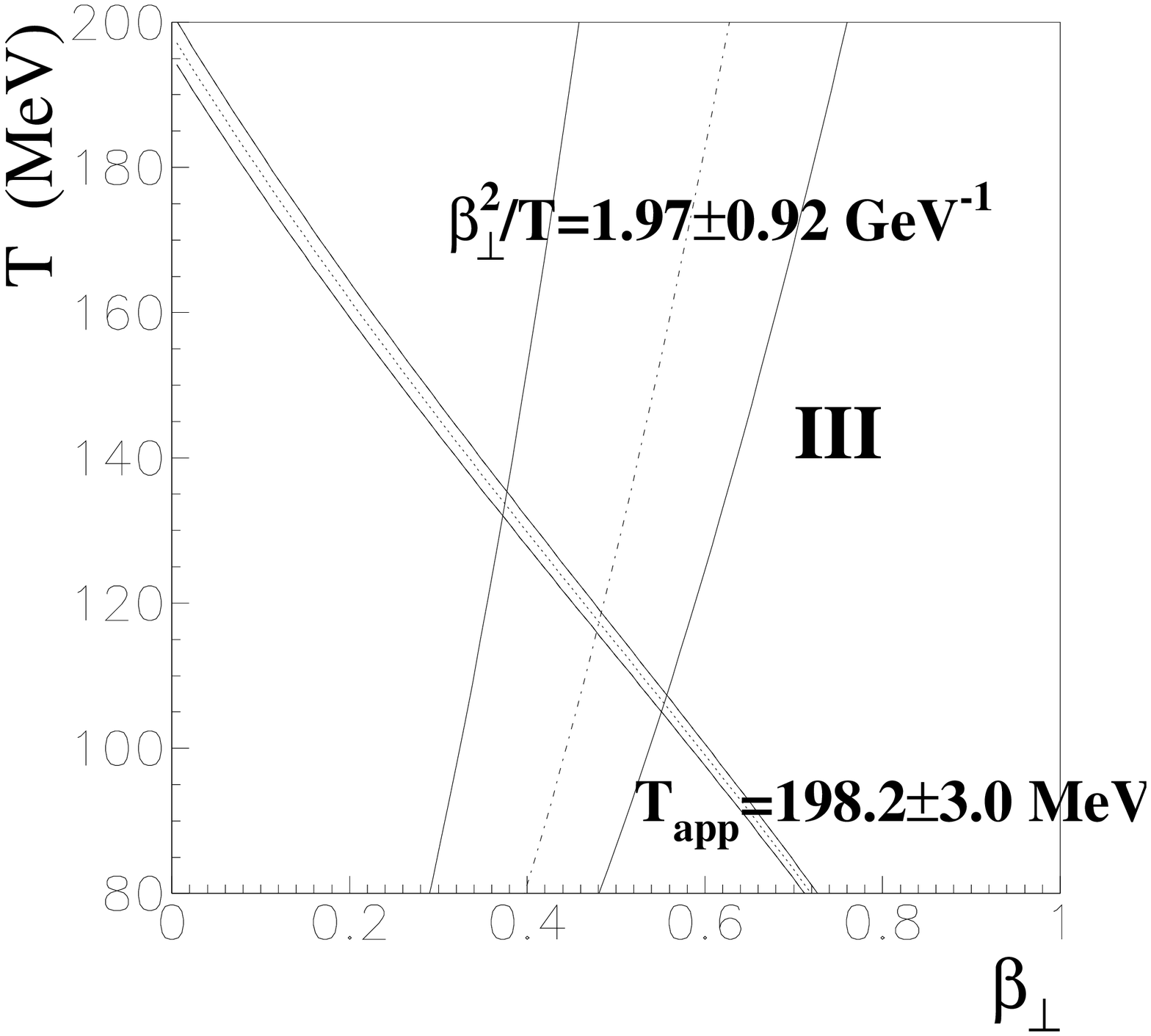}
\includegraphics{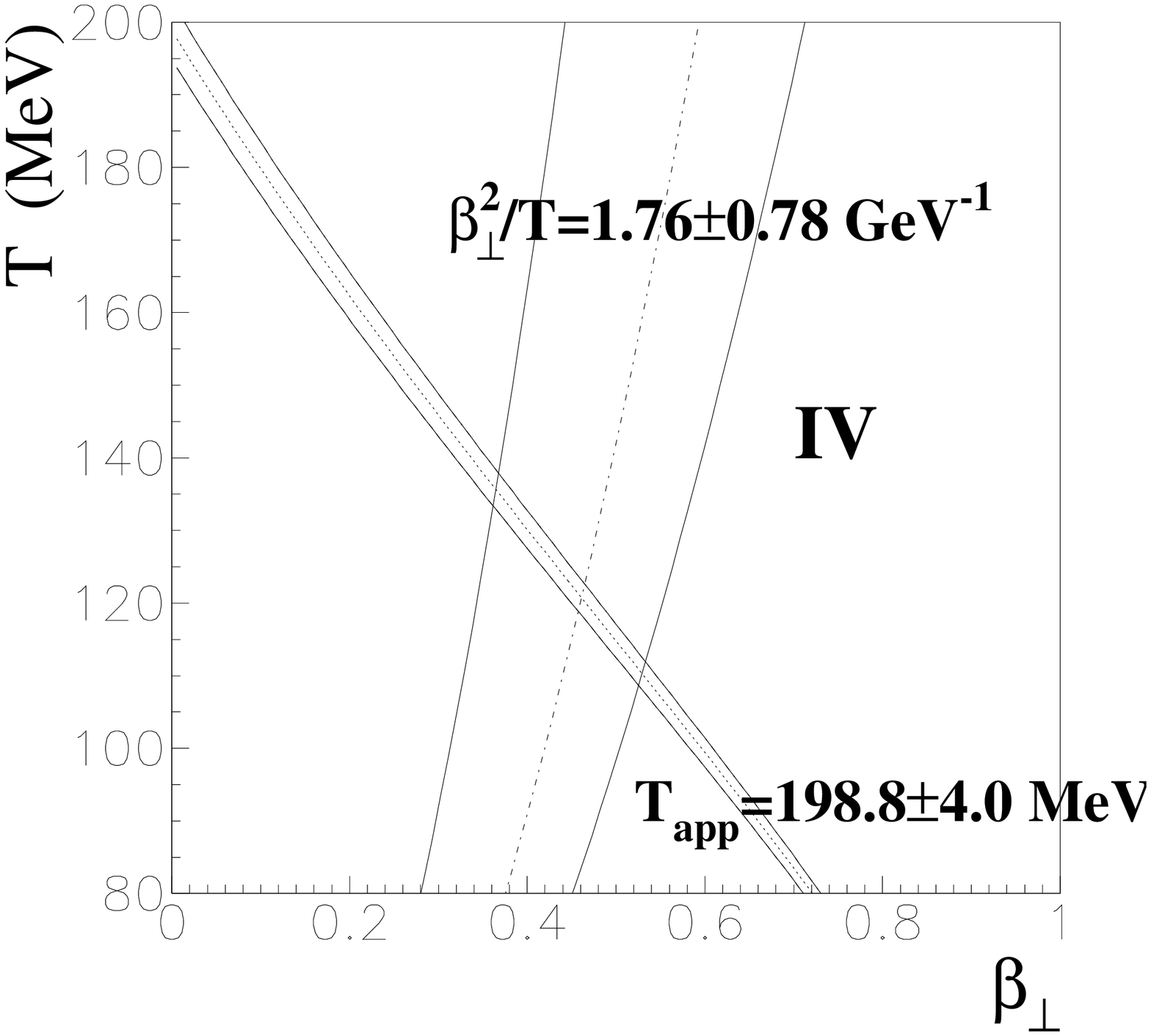}}
\caption{Regioni permesse per la coppia di variabili $\beta_\perp$\ 
         (velocit\`a del flusso trasverso) e $T$\ 
	 (temperatura di {\em ``freeze-out''}), 
	 in funzione della centralit\`a della collisione. La classe 
	 I corrisponde alle collisioni pi\`u periferiche, quella IV alle 
	 pi\`u centrali.}
\label{TransvExp}
\end{center}
\end{figure}

Dal {\em ``best fit''} ai raggi sperimentali $R_\perp$\ misurati nei venti 
intervalli di $(Y_{\pi\pi},K_t)$\ con la funzione definita dall'eq.~\ref{Rtransv} 
si sono estratti il raggio trasverso $R_G$\ ed il rapporto $\beta_\perp^2/T$. 
Combinando l'informazione sul rapporto $\beta_\perp^2/T$\ con quella 
della temperatura apparente calcolata dagli spettri di massa trasversa 
(eq.~\ref{Tapp1}) \`e possibile disaccoppiare le due quantit\`a. Ci\`o \`e 
mostrato nella fig.~\ref{TransvExp}, in cui per ciascuna classe di centralit\`a, 
si delimitano delle regioni fiduciali nel piano $(\beta_\perp,T)$. I valori 
adoperati per $T_{app}$\ sono quelli pubblicati in~\cite{mt_WA97}.  
I valori estratti per le quantit\`a $R_G$,  $\beta_\perp$\ e $T$\ sono 
riportati  nella tabella~\ref{tab6.2} riassuntiva e saranno discussi in seguito.
\subsubsection{Espansione longitudinale} 
\begin{figure}[p]
\begin{center}
    \resizebox{0.60\textwidth}{!}{%
    \includegraphics{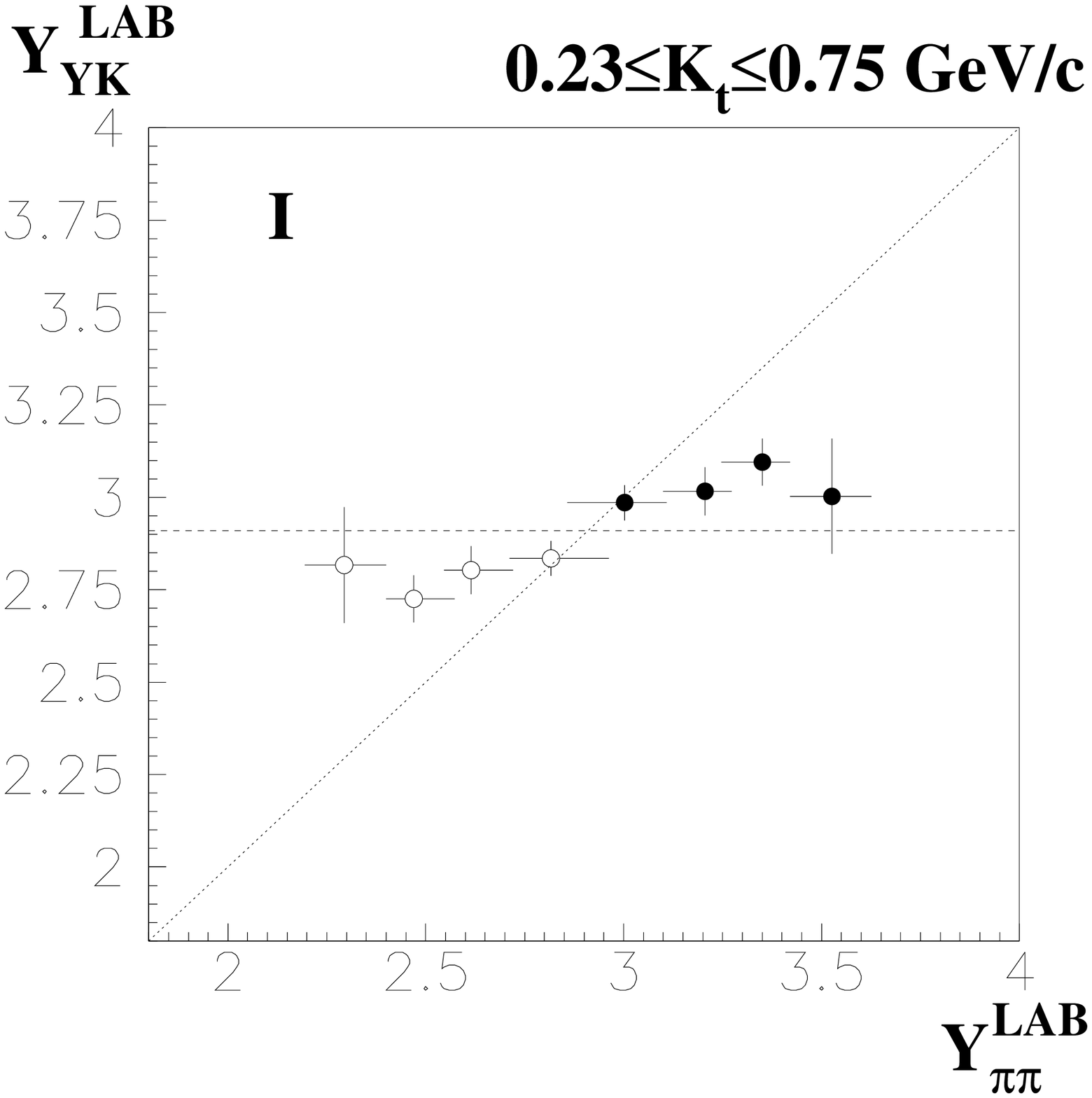}
    \includegraphics{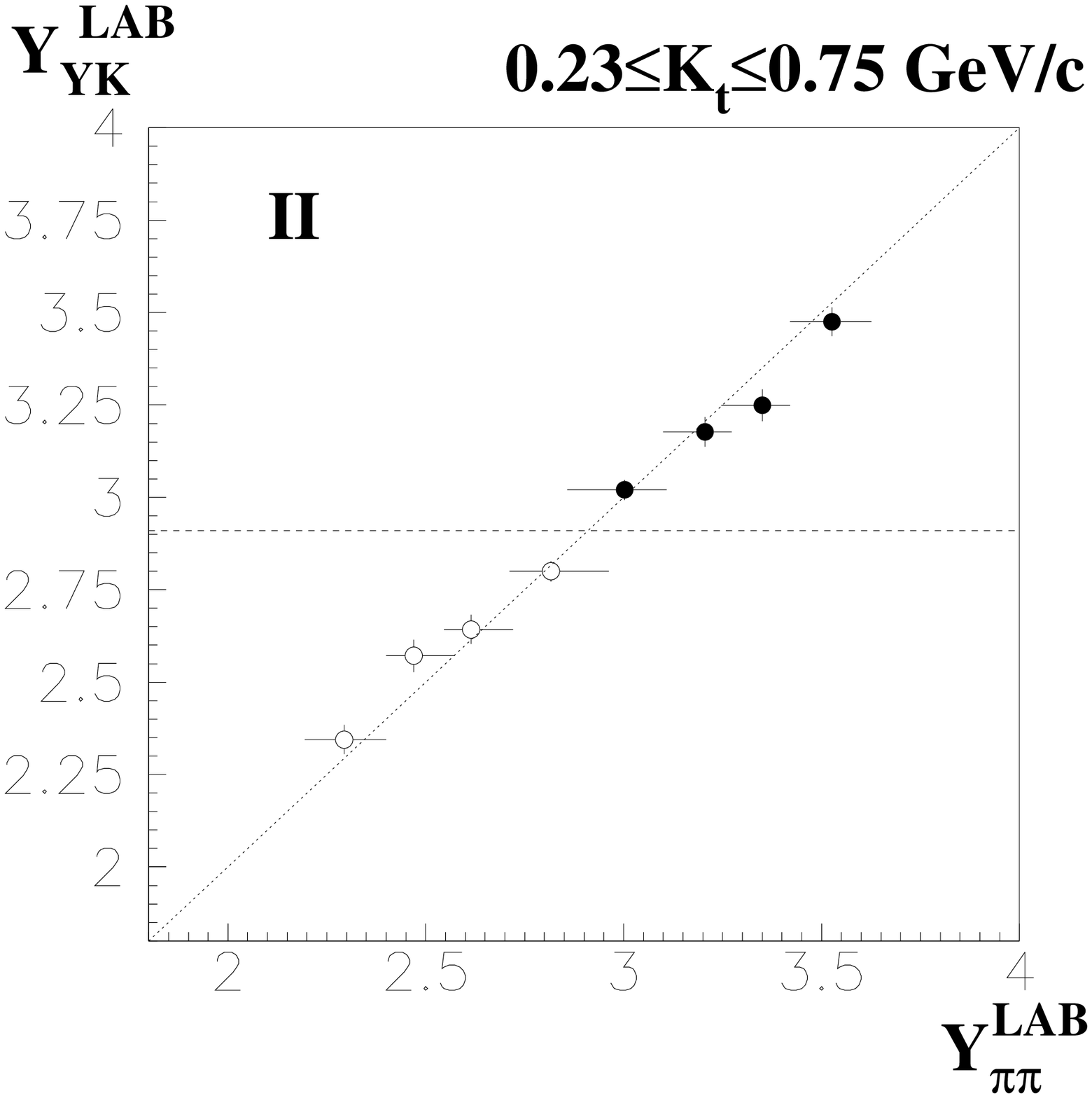}}\\
    \resizebox{0.60\textwidth}{!}{%
    \includegraphics{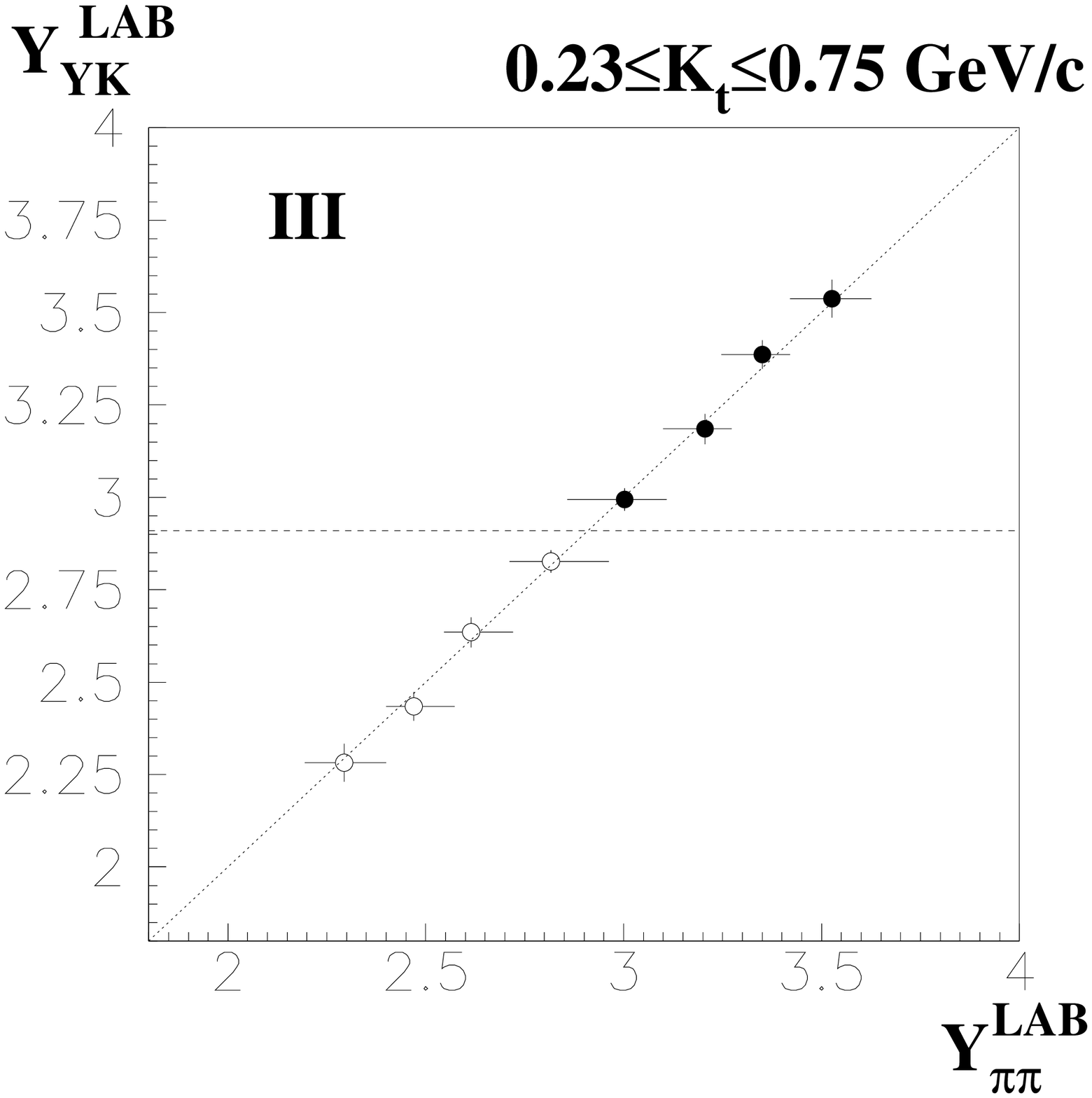}
    \includegraphics{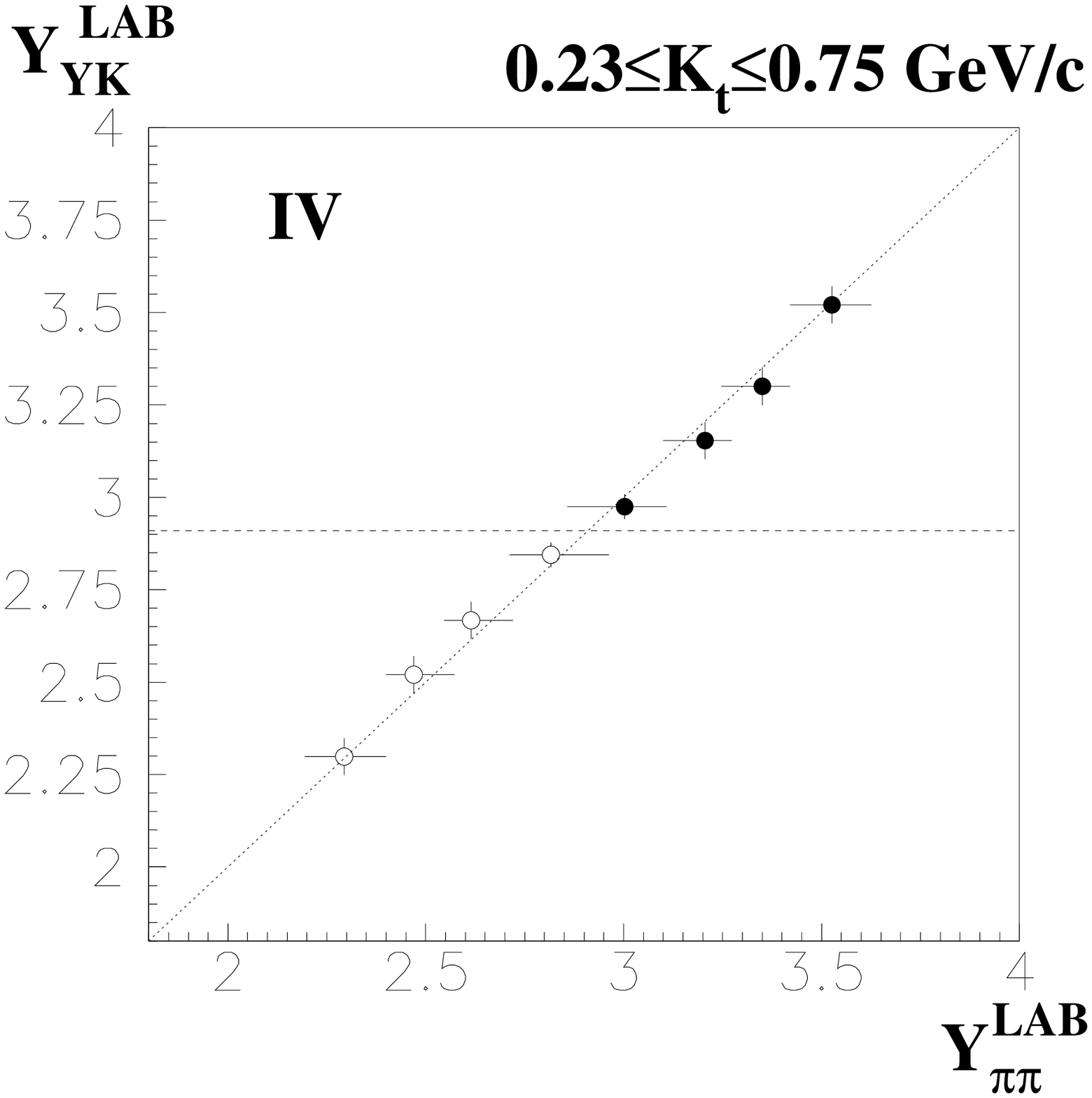}}
\caption{Dipendenza di $Y_{YK}$\ dalla rapidit\`a della coppia per le diverse 
         classi di centralit\`a. I cerchietti pieni sono i dati sperimentali, 
	 quelli vuoti sono i dati riflessi rispetto a $y_{cm}$.}
\label{LongExp1}
\vspace{1.0cm}
%\end{center}
%\end{figure}
%\begin{figure}[p]
%\begin{center}
%\resizebox{0.28\textwidth}{!}{%
%\includegraphics{cap6/tmpv_all.eps}}\\
\resizebox{0.50\textwidth}{!}{%
\includegraphics{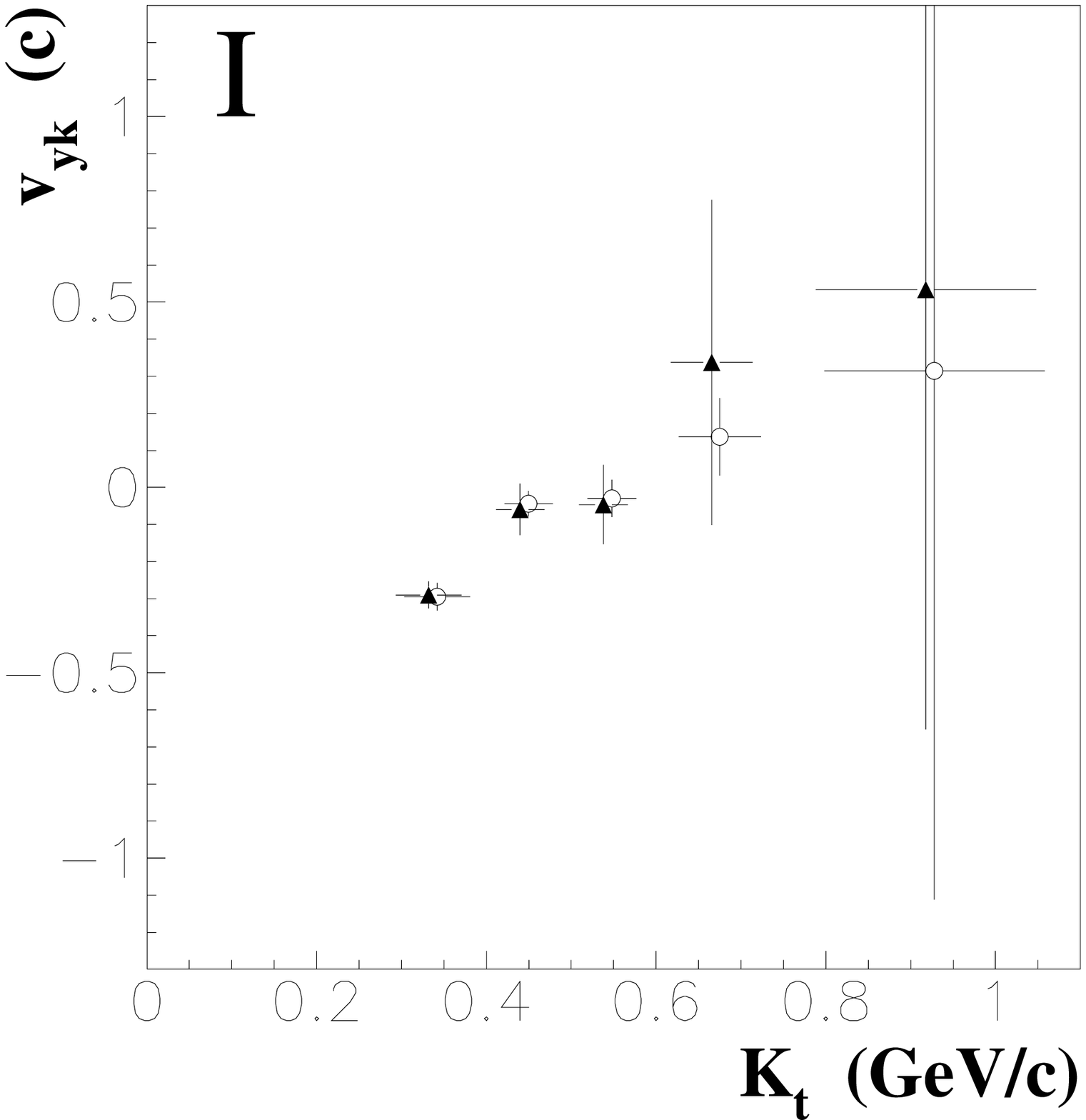}
\includegraphics{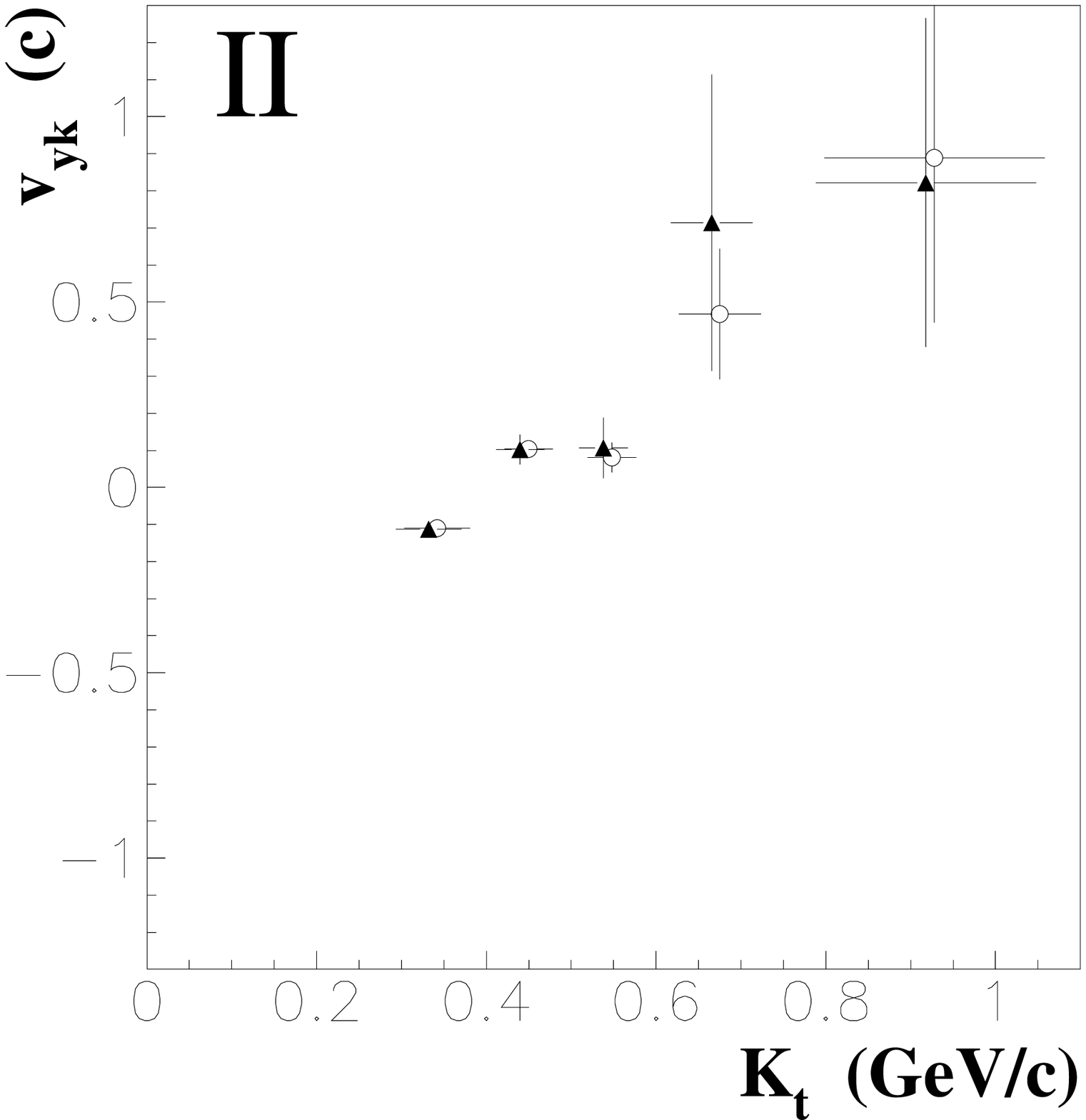}}\\
\resizebox{0.50\textwidth}{!}{%
\includegraphics{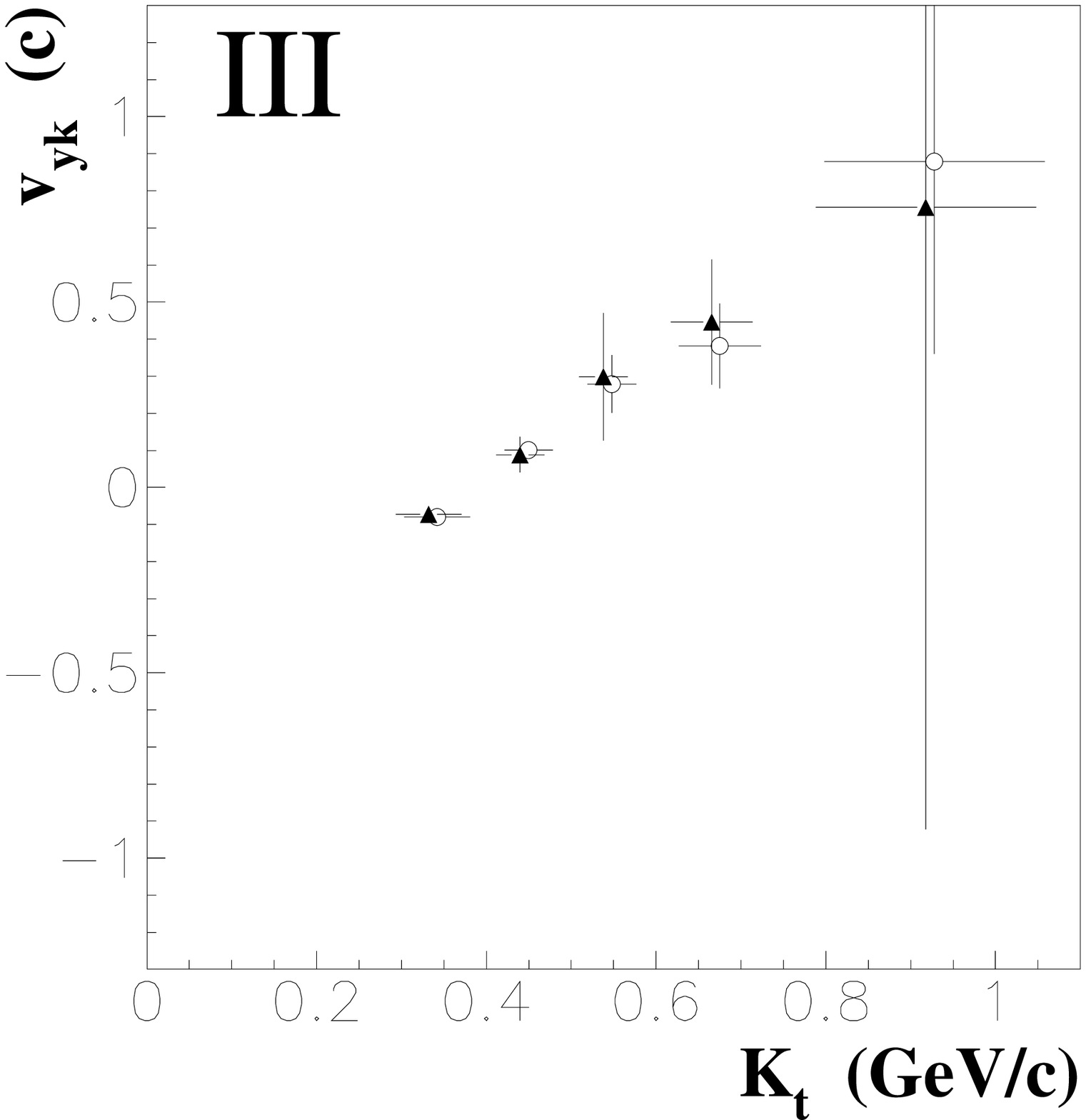}
\includegraphics{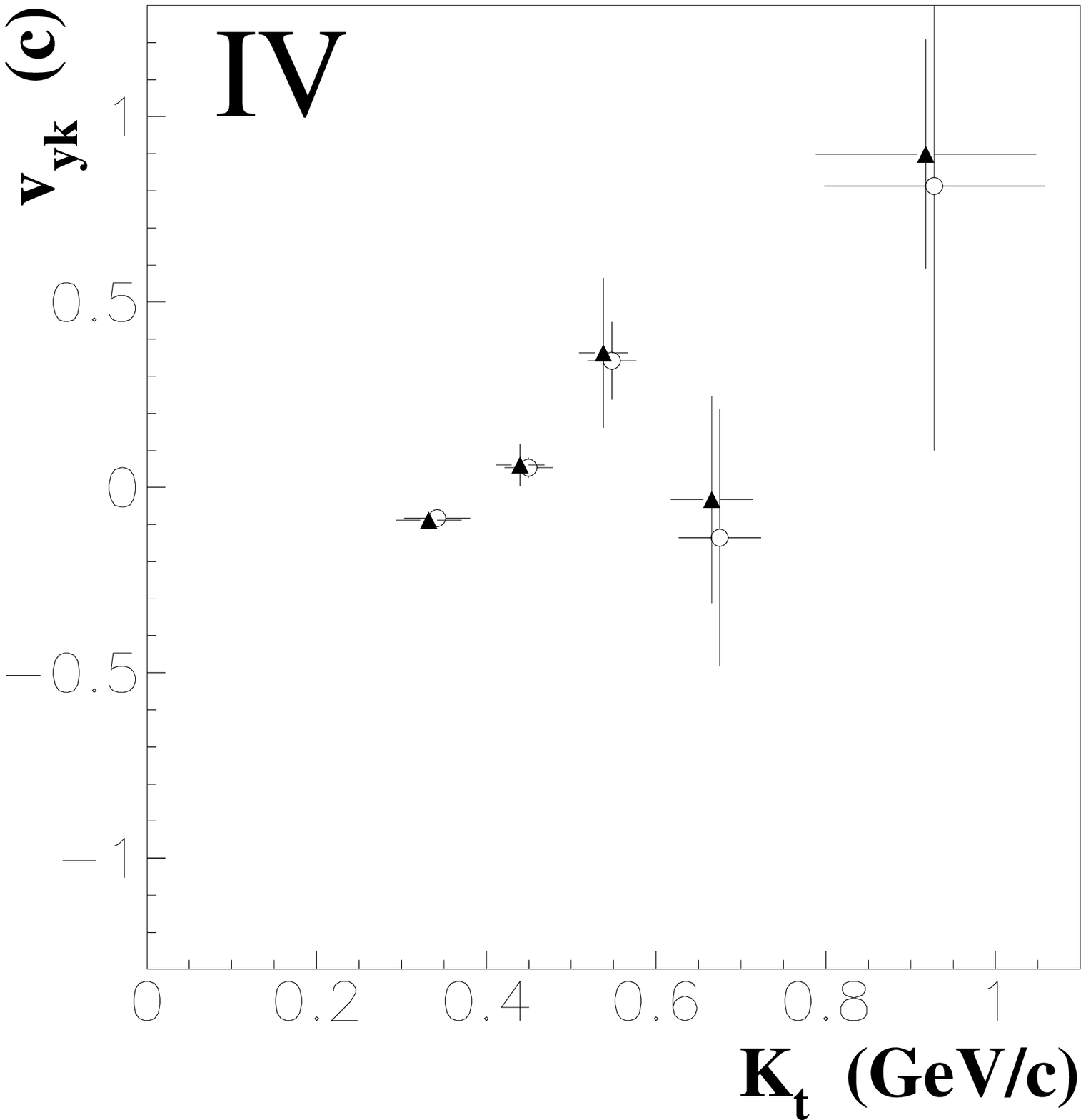}}
\caption{Dipendenza della velocit\`a di Yano-Koonin da $K_t$. }
\label{LongExp2}
\end{center}
\end{figure}

In fig.~\ref{LongExp1} \`e mostrata la dipendenza della rapidit\`a di $Y_{YK}$\ 
(eq.~\ref{rapYkn}) dalla rapidit\`a della coppia, per le quattro classi di 
centralit\`a. Entrambe le rapidit\`a sono qui calcolate nel riferimento del 
laboratorio.  
\`E stato suggerito~\cite{79} che la dipendenza della rapidit\`a $Y_{YK}$\ 
(eq.~\ref{rapYkn}) dalla rapidit\`a della coppia fornisce una misura diretta 
dell'espansione longitudinale della sorgente, ben separata dalla dinamica nella 
direzione trasversale. In particolare, per una sorgente statica (nella 
direzione longitudinale) $Y_{YK}$\ dovrebbe esser indipendente da $Y_{\pi\pi}$\ 
(linee orizzontali in fig.~\ref{LongExp1}),  mentre per una sorgente 
(in espansione) invariante per {\em ``boost''} di Lorentz nella direzione 
longitudinale le due variabili dovrebbero essere completamente correlate 
(linee a $45^o$\ nella fig.~\ref{LongExp1}). Si attende invece soltanto 
una debole dipendenza di $Y_{YK}$\ (od equivalentemente della 
relativa velocit\`a $v_{yk}$) da $K_t$.  
\newline
I dati sperimentali indicano pertanto un'espansione longitudinale molto 
meno intensa per la classe di centralit\`a pi\`u periferica. 
\newline
Studiando la correlazione tra  la velocit\`a di Yano-Koonin (a partire dalla quale 
si calcola $Y_{YK}$, secondo l'eq.~\ref{rapYkn}) e l'impulso trasverso 
della coppia $K_t$, si osserva, contrariamente alle previsioni del modello,
una forte dipendenza di $v_{yk}$\ da $K_t$, 
come \`e mostrato in fig.~\ref{LongExp2}. Eseguendo un {\em ``best fit''} ai 
dati sperimentali con la relazione lineare $v_{yk}=a+bK_t$, si trovano delle 
pendenze $b$\ simili tra le diverse classi di centralit\`a (con un valore di 
circa $1.5 [{\rm GeV}c^{-2}]^{-1}$), mentre l'intercetta $a$\ \`e diversa 
per la classe I ($a\approx -0.8c$), rispetto alle classi II-III-IV 
($a\approx-0.6c$). La relazione lineare diventa non fisica ($v_{yk}>c$) 
per $K_t>\frac{c-a}{b} \sim 1 \, {\rm GeV}c^{-1}$, e pertanto si attende 
una saturazione di $v_{yk}$, per elevati valori di $K_t$, ad un valore 
prossimo alla velocit\`a della luce.  
Tutto ci\`o implica che i due semplici scenari suggeriti come riferimenti 
discutendo la fig.~\ref{LongExp1} (quello di sorgente statica o di sorgente 
in rapida espansione longitudinale) non sono del tutto validi, ma bisogna tener 
conto del particolare intervallo di $K_t$\ considerato. Per elevati $K_t$, 
infatti, le linee a $45^o$\ della fig.~\ref{LongExp1} possono essere 
oltrepassate, mentre riducendo il valore di $K_t$\ i punti sperimentali 
si spostano verso la curva orizzontale.
Tuttavia la conclusione, tratta precedentemente, che l'espansione nella 
direzione longitudinale \`e meno intesa per la classe I \`e ancora valida:  
{\em (i)} i quattro grafici in fig.~\ref{LongExp1} sono infatti ottenuti per valori 
di $K_t$\ distributi (con stesso valor medio) nello stesso intervallo 
$[0.23\, , \, 0.75]\, {\rm GeV}/c$; {\em (ii)} l'intercetta $a$\ assume un valore 
pi\`u piccolo (o meglio meno negativo) per la sola classe I.
\subsubsection{Durata del {\em ``freeze-out''}: $\Delta\tau$}
\begin{figure}[htb]
\begin{center}
\resizebox{0.55\textwidth}{!}{\includegraphics{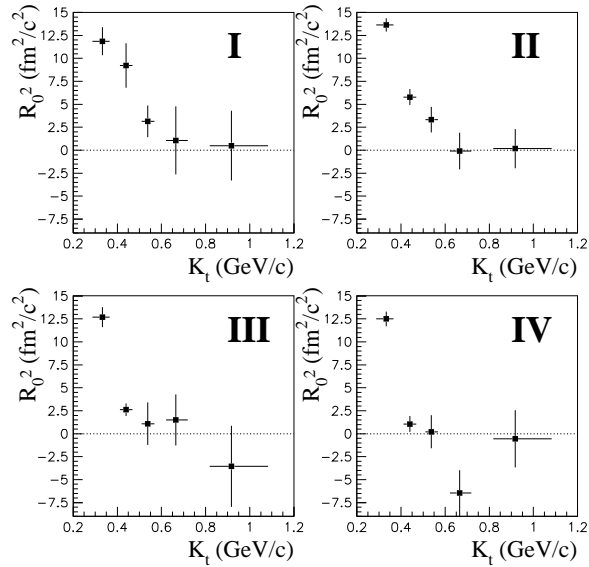}}
\caption{Dipendenza del parametro $R_0^2$\ estratto dal {\em ``best fit''} delle 
         funzioni di correlazione dall'impulso $K_t$, per le diverse 
	 classi di centralit\`a.}
\label{ExpDuration}
\end{center}
\end{figure}

In fig.~\ref{ExpDuration} \`e mostrata la dipendenza del quadrato del 
``raggio'' $R_0$\ dall'impulso trasverso della coppia, per le quattro 
classi di centralit\`a.  
Questi grafici sono stati ottenuti integrando sull'intero spettro di $Y_{\pi\pi}$\ 
allo scopo di ridurre l'errore statistico nella determinazione di $R_0^2$, che \`e 
il parametro con l'errore associato pi\`u elevato. Ci\`o \`e lecito in quanto non 
si \`e osservata dipendenza di $R_0^2$\ da $Y_{\pi\pi}$ 
({\em cfr.} pubblicazione~\cite{HBTpaper}). 
\newline
\`E preferibile misurare il parametro $R_0^2$, piuttosto che la sua radice 
quadrata, in quanto per alcuni modelli di emissione esso potrebbe assumere 
anche valori negativi~\cite{Tomasik1998}. Questo \`e il caso, ad esempio, delle 
sorgenti  {\em opache}, per le quali \`e previsto  un forte 
riassorbimento delle particelle dalla materia nucleare. Per tali sorgenti 
l'emissione delle particelle avviene dunque prevalentemente dalla 
superficie (processo di evaporazione), come nel caso dei fotoni emessi 
dal Sole.  I risultati trovati suggeriscono, per tutte le classi di centralit\`a,  
un {\em ``freeze-out''} quasi istantaneo (cio\`e $\Delta\tau \approx 0$) da parte 
di una sorgente trasparente.  
\subsubsection{Durata dell'espansione: tempo proprio del {\em ``freeze-out''} 
 $\tau_0$}
L'ultimo raggio estratto dal {\em ``best fit''}, $R_{\parallel}$, permette di 
ricavare il tempo proprio del {\em ``freeze-out''} $\tau_0$, 
tramite l'eq.~\ref{Rperp}.  
I valori del parametro $\tau_o$\ ricavati dal {\em ``best fit''} 
della funzione~\ref{Rperp} ai raggi $R_{\parallel}$\ misurati in 
funzione di $(K_t,Y_{\pi\pi})$\  
sono riportati nella tab.~\ref{tab6.2}, per le quattro 
classi di centralit\`a. Nuovamente, la classe I presenta una differente 
dinamica rispetto alle altre classi: il {\em ``freeze-out''} avviene dopo soli 
$3.7 \, {\rm fm}/c$, mentre per le altre classi si ha dopo i $5 \,{\rm fm}/c$.  
\subsection{Discussione finale}
Nella tab.~\ref{tab6.2} sono riassunti i risultati dello studio della 
correlazione HBT in funzione della centralit\`a della collisione.  
\begin{table}[h]
\begin{center}
%{\footnotesize
\begin{tabular}
{||p{0.7in}|c|c|c|c||} \hline \hline
   & IV & III & II & I \\ \hline
   $R_G ({\rm fm})$  & $5.1 \pm 0.6$ & $5.0 \pm 0.6$ & $4.6 \pm 0.4$
                 & $3.2 \pm 0.3$   \\ \hline
      $T ({\rm MeV})$   & $120^{+15}_{-11}$ & $117^{+16}_{-11}$ &
                $121^{+15}_{-11}$ & $140^{+26}_{-13}$  \\ \hline
   $\beta_{\perp} (c)$  & $0.46^{+0.07}_{-0.10}$ & $0.48^{+0.08}_{-0.11}$ &
                     $0.47^{+0.07}_{-0.10}$ & $0.30^{+0.09}_{-0.16}$
                     \\ \hline
$\tau_0 ({\rm fm}/c)$& $5.6\pm 0.2$ & $5.6 \pm 0.2$ &
               $5.1\pm 0.2$ & $3.7 \pm 0.2$   \\ \hline
 $\Delta\tau({\rm fm}/c)$ &
   \multicolumn{4}{c||}{$\approx 0$\ ad elevati $K_t$} \\ \hline
   Espansione longitud.
   & \multicolumn{3}{c|}{veloce} & lenta \\ \hline \hline
\end{tabular}%}
\end{center}
\caption{Parametri della sorgente di pioni per le collisioni Pb-Pb a 160 A GeV/$c$\ 
nelle quattro classi di centralit\`a di WA97.
\label{tab6.2}}
\end{table}
Questi dati forniscono implicitamente una descrizione dinamica della sorgente 
di particelle. 
\newline
Volendo verificare se questa descrizione sia autoconsistente,  
si pu\`o confrontare la radice della larghezza quadratica 
media bi-dimensionale \(R_{rms}^{freeze\, out}=\sqrt{2} R_{G}\)\ con l'analoga 
quantit\`a valutata per un nucleo freddo di piombo:
\begin{equation}
R_{rms}^{Pb}=\sqrt{\frac{2}{5}} 1.2 A^{1/3} \simeq 4.5 \, {\rm fm}
\label{ColdPb}
\end{equation}
Per le collisioni pi\`u centrali (classi III e IV), il sistema si espande di un 
fattore di $\sim 1.6$\ o, equivalentemente, di $2.65$\ fm nella direzione trasversa. 
Se la velocit\`a del flusso trasverso \`e pari a $\beta_\perp \simeq 0.47\, c$\ 
durante tutta l'espansione, nell'intervallo di tempo  pari a 
$\tau_0 \simeq 5.6 \, {\rm fm} c^{-1}$\ la materia nucleare percorre 
$\tau_0\beta_{\perp} \, \simeq 2.6$\ fm nella direzione trasversa. Questo 
valore \`e consistente con la precedente stima fornita dalla differenza 
$R_{rms}^{freeze out} - R_{rms}^{Pb}$. Pertanto, la descizione HBT \`e 
auto-consistente. 
\newline
Per collisioni pi\`u periferiche (classi I e II), la regione 
di sovrapposizione iniziale dei nuclei di piombo deve essere pi\`u piccola 
di quanto fornito dall'eq.~\ref{ColdPb}, ma una stima realistica di tale 
sovrapposizione non \`e immediata. Tuttavia \`e ancora possibile, partendo 
da questi dati, risalire al momento della collisione. Se al {\em ``freeze-out''} 
la radice della larghezza quadratica media bi-dimensionale 
nella direzione trasversa \`e pari a $4.5$\ fm ($6.5$\ fm) per la classe I 
(classe II), tenendo conto di un'espansione totale di 
$\tau_0\beta_\perp = 1.1$\ fm ($2.4$\ fm), \`e possibile stimare la larghezza 
iniziale della sorgente nella direzione trasversa pari a $3.4$\ fm ($4.1$\ fm).  
\newline
Non \`e possibile fornire una descrizione quantitativa dell'espansione 
nella direzione longitudinale. Qualitativamente si pu\`o affermare che 
l'espansione longitudinale diventa pi\`u intensa all'aumentare 
della centralit\`a della collisione (cio\`e per parametri di impatto 
pi\`u piccoli).  Inoltre i risultati ottenuti suggeriscono un salto 
improvviso passando dalla classe I alla classe II. 
\newline
Per quel che riguarda lo sviluppo temporale della collisione, si \`e dedotto 
che l'emissione dei pioni \`e un processo improvviso ($\Delta\tau \approx 0$), 
che inizia dopo $\sim 5.5\, {\rm fm}c^{-1}$\ ($3.7\, {\rm fm}c^{-1}$) nel 
caso delle collisioni pi\`u centrali (periferiche). Il processo di 
emissione avviene dall'intero volume e non \`e dominato dalla superficie. 
Ci\`o ricorda il processo di disaccopiamento dei fotoni nell'universo 
primordiale.  
\newline
\`E lecito supporre che il disaccopiamento dei pioni sia conseguenza 
del rapido raffreddamento e della diluizione della densit\`a barionica. 
Nei modelli termici~\cite{ModelBraun,Becca}, la temperatura del 
{\em ``freeze-out'' chimico} ($\approx 170 \, {\rm MeV}$) necessaria 
per descrivere i rapporti di produzione delle diverse particelle prodotte 
\`e considerevolmente pi\`u elevata della temperatura del 
{\em ``freeze-out'' termico}, qui determinata. Ci\`o implica che l'abbondanza 
relativa delle diverse specie raggiunge l'equilibrio prima del termine 
delle interazioni (forti) che nell'ultima fase sarebbero pertanto solo 
di tipo elastico ({\em ``freeze-out'' termico}).   
L'ultima fase dell'espansione sarebbe cos\`i caratterizzata da una diluizione 
volumetrica della densit\`a barionica senza alcuna apprezzabile ricombinazione 
delle particelle.  

%% file: concl.tex
\chapter*{Conclusioni}
\addcontentsline{toc}{chapter}{Conclusioni}
I risultati esposti in questa tesi hanno permesso di raggiungere una pi\`u 
approfondita comprensione della dinamica delle collisioni tra nuclei di piombo 
accelerati dall'SPS del CERN. 
I principali argomenti d'indagine hanno riguardato: 
\begin{itemize}
\item[{\em (i)}] lo studio della produzione di particelle strane nelle 
collisioni nucleari all'energia dell'SPS come evidenza sperimentale della 
transizione di fase nello stato di QGP;  
\item[{\em (ii)}] lo studio della dinamica di espansione in funzione della 
centralit\`a delle collisioni Pb-Pb a 160 A GeV/$c$\ con le  
distribuzioni di massa trasversa delle particelle strane e 
con l'interferometria HBT tra particelle di carica negativa 
(prevalentemente pioni). 
\end{itemize}
%newline with tab.

%newline with tab.
Per quanto riguarda la produzione di particelle strane, si sono 
considerati i dati raccolti dall'esperimento NA57 nelle collisioni 
Pb-Pb a 160 ed a 40 A GeV/$c$. 
Pi\`u in dettaglio, l'analisi svolta ha riguardato %: 
lo studio delle caratteristiche di produzione delle particelle ed 
anti-particelle con una, due o tre
unit\`a di stranezza: \PKzS, \PgL, \PagL, \PgXm, \PagXp, \PgOm\ e \PagOp,  
con particolare riguardo alla dipendenza dalla centralit\`a della collisione e 
dall'energia del proiettile.  

\noindent
Si sono calcolati i tassi di produzione delle $V^0$\ (\PKzS, \PgL\ e \PagL) 
e delle cascate (\PgXm, \PagXp, \PgOm\ e \PagOp) a 160 A GeV/$c$\ 
e quelli delle cascate a 40 A GeV/$c$. A partire da queste quantit\`a 
si sono anche ricavati i rapporti di produzione 
(elementi indispensabili dei modelli termici) che hanno fornito le prime 
indicazioni significative sulla produzione di stranezza. 
\newline
Nelle collisioni a 160 A GeV/$c$\ \`e stato possibile confrontare 
con delle interazioni di riferimento p-Be e p-Pb, alla stessa energia, 
raccolte precedentemente dall'esperimento WA97.   
All'energia di 40 A GeV, l'esperimento NA57 ha completato (nel mese di 
dicembre del 2001)  
la raccolta degli eventi in interazioni p-Be, attualmente in fase di ricostruzione.
\newline
I risultati sulla produzione di stranezza in funzione della centralit\`a 
(misurata dal numero di partecipanti) suggeriscono il seguente scenario (a 160 GeV):   
le particelle strane sono prodotte nella regione centrale di rapidit\`a 
in maniera crescente con la centralit\`a pi\`u rapidamente di 
quanto previsto dalla proporzionalit\`a col numero di partecipanti.  
%\newline
Questo incremento risulta pi\`u pronunciato all'aumentare del contenuto di 
stranezza e raggiunge un fattore 20 nel caso delle $\Omega$ nella classe 
di maggior centralit\`a. 
\newline
Questi risultati sono di difficile  
interpretazione in termini di produzione in fase adronica, data l'alta 
soglia in massa delle reazioni di produzione dei barioni multi-strani, e 
suggeriscono un diverso meccanismo di produzione delle particelle strane, 
le cui caratteristiche sono quelle attese per la fase di QGP.  
\newline
Volendo determinare in quale maniera, in termini di 
%volumi coinvolti nella collisione, 
nucleoni partecipanti, 
si passi dal regime di interazioni adroniche (collisioni p-Be e p-Pb) 
a quello di QGP, supposto per le collisioni Pb-Pb pi\`u centrali, si sono tratte 
le seguenti conclusioni: 
\begin{itemize}
\item[-]
 Considerando le sole \PagXp\ si \`e osservato un repentino incremento, 
 corrispondente ad un effetto pari a $3.5$\ $\sigma$, passando  
 da $<N_{part}> = 62$\ (classe $0$) ad $<N_{part}> = 121$\ (classe~$I$). 
 L'incremento satura quindi per le collisioni pi\`u centrali. 
\item[-] 
 Limitatamente alle sole collisioni Pb-Pb, 
 considerando tutte le altre specie ad esclusione delle \PagL\ --- compatibili 
 con un incremento che abbia ormai raggiunto un valore di saturazione --- si osserva 
 che l'incremento aumenta con maggior continuit\`a al crescere della centralit\`a; 
 la dipendenza secondo una funzione potenza ($Y \propto N_{part}^\beta$), 
 escludendo i punti relativi alle 
 collisioni p-A, fornisce una buona descrizione. 
 Il valore dell'esponente $\beta$\ aumenta con il 
 contenuto di stranezza della specie considerata passando da  
 $\beta \approx 1.2$\ per le particelle con una unit\`a di 
 stranezza (\PgL\ e \PKzS) 
 a $\beta \approx 1.5$\ per le particelle con tre unit\`a di stranezza  
 (\PgOm\ e \PagOp). 
\item[-]
 Passando
 dalle collisioni p-A a quelle Pb-Pb si osserva una crescita talmente
 brusca che neanche una dipendenza secondo una legge di potenza 
% ($Y \propto N_{part}^\beta$)
 riesce a raccordare con continuit\`a.
\end{itemize}
Per poter trarre argomentazioni conclusive sul modo in cui si 
passa dal regime adronico, proprio delle collisioni p-A,  
a quello di QGP, supposto per le collisioni Pb-Pb pi\`u centrali, 
%studiando la produzione di particelle strane, 
si avverte la necessit\`a 
di esplorare l'intervallo di centralit\`a compreso tra $<N_{part}>=62$\ 
(classe 0 delle interazioni Pb-Pb) 
ed $<N_{part}>=4.5$ (interazioni p-Pb). 
La Collaborazione NA57 ha quindi avanzato la richiesta di studiare le collisioni 
di un sistema pi\`u leggero del Pb-Pb, quale l'In-In.  

\noindent
Lo studio della produzione di stranezza in funzione dell'energia della 
collisione \`e stato eseguito confrontando i risultati delle collisioni 
Pb-Pb a 40 ed a 160  A GeV/$c$\ 
(cio\`e, in termini di energia nel centro di massa nucleone-nucleone, rispettivamente, 
$\sqrt{s_{NN}}=8.8$\ GeV e $\sqrt{s_{NN}}=17.3$\ GeV)  
e quelli delle collisioni Au-Au determinati dalla collaborazione STAR   
all'energia del RHIC ($\sqrt{s_{NN}}=130$\ GeV).  
\newline
I risultati sui tassi di produzione di $\Xi$\ ed $\Omega$\ nelle collisioni Pb-Pb a 
40 A GeV/$c$\ indicano una maggior densit\`a barionica nella regione della collisione 
rispetto ai 160 A GeV/$c$. 
\newline
I rapporti di produzione anti-iperone/iperone (per $\Lambda$, $\Xi$\ ed $\Omega$) 
aumentano con l'energia passando 
dall'energia di NA57 a quella di STAR, dove approssimano il valore unitario.  
La dipendenza dall'energia dei  rapporti \`e via via pi\`u debole per le particelle 
con maggior contenuto di stranezza.

Il secondo campo d'indagine ha riguardato lo studio della dinamica di  
espansione del sistema Pb-Pb dopo la collisione, valutando in particolare la 
temperatura di completo disaccoppiamento degli adroni dello stato finale 
({\em ``freeze-out} termico'') e la velocit\`a di espansione della 
{\em ``fireball''}, seguendo due diversi approcci: in uno si utilizzano le 
distribuzioni di massa trasversa delle particelle strane studiate, con un 
{\em ``best-fit''} globale nel contesto di un modello termico; nell'altro si 
studiano gli effetti di correlazione tra bosoni identici, 
essenzialmente pioni (interferometria HBT).   
\newline
Lo studio degli spettri di massa trasversa delle diverse particelle strane 
identificate ha evidenziato il raggiungimento di un equlibrio termico 
di tipo locale della materia interagente formatasi in seguito alla collisione. 
\newline
Le prime indicazioni qualitative fornite dallo studio 
--- in diversi esperimenti --- delle temperature apparenti 
estratte da tali distribuzioni, in funzione della massa delle particelle, 
suggeriscono la presenza di un moto collettivo di espansione nella direzione trasversa 
sovrapposto al moto termico.  
\newline
Lo studio {\em simultaneo} di tutti gli spettri a 160 A GeV/$c$\ effettuato in questo   
lavoro ha permesso di disaccoppiare 
in modo quantitativo la componente termica da quella dovuta al flusso trasverso.  
Si sono potute ricavare la temperatura del ``{\em freeze-out} termico'' e la 
velocit\`a media del flusso trasverso. Nel caso delle collisioni Pb-Pb pi\`u centrali 
si \`e misurata una temperatura di {\em ``freeze-out''} pari a circa $130$\ MeV ed una 
velocit\`a media di espansione collettiva trasversa pari a circa la met\`a della 
velocit\`a della luce.  
\newline
Tutte le particelle strane considerate,  $\Omega$\  
incluse,  rientrano   in questa descrizione.  
%$<\beta_\perp>=0.46$
\newline
Il ``{\em freeze-out} termico'' %cos\`i determinato  
ha le stesse caratteristiche, entro gli errori sperimentali, per le particelle che 
condividono almeno un quark di valenza in comune con quelli preesistenti nei 
nucleoni collidenti e per quelle che non ne condividono alcuno.
\newline
Lo studio della dipendenza degli spettri di massa trasversa dalla centralit\`a della  
collisione ha portato alla seguente conclusione: al diminuire della centralit\`a 
diminuisce l'intensit\`a del flusso trasverso collettivo ed aumenta la temperatura 
di {\em ``freeze-out''}. Per le due classi di collisioni pi\`u periferiche ($0$--$I$) 
la soluzione privilegiata \`e addirittura quella di assenza di moto collettivo 
trasverso ($\beta_\perp \approx 0$) e massima componente termica ($T\approx 250$\ MeV).   

\noindent
Lo studio delle temperature apparenti relative agli 
spettri di massa trasversa delle cascate ($\Xi$\ ed $\Omega$) a 40 
GeV/$c$\ ha fornito le seguenti indicazioni: 
\begin{itemize}
\item[-]
anche a 40 GeV/$c$ si misura, entro gli errori sperimentali, una 
stessa temperatura apparente per particelle (\PgXm) ed anti-particelle 
(\PagXp);
\item[-]
la temperatura apparente della \PgXm\ (quella %cui \`e associata 
misurata con 
la minor indeterminazione statistica) risulta significativamente 
inferiore a quella 
misurata a 160 GeV/$c$. Ci\`o pu\`o esser dovuto ad una minore intensit\`a 
del flusso trasverso ed alla minor energia iniziale del sistema.  
\end{itemize}

\noindent
Come secondo metodo d'indagine per lo studio della dipendenza  
dalla centralit\`a della dinamica di espansione delle collisioni Pb-Pb a 160 A GeV/$c$\ 
si \`e utilizzata la tecnica dell'interferometria HBT. Si sono analizzati i dati raccolti 
dalla collaborazione WA97 relativi alle particelle di carica negativa, costituite 
prevalentemente da \Pgpm.  
\newline
\`E stata sviluppata una nuova procedura per il calcolo delle correzioni per l'interazione  
coulombiana che tiene conto %in maniera consistente 
delle caratteristiche della sorgente  
da cui vengono emesse le particelle (forma e dimensione, dinamica di espansione, etc.).   
Questo metodo \`e stato confrontato %in modo sistematico 
con gli altri utilizzati  
tradizionalmente: il confronto ha mostrato che gli altri metodi tendono a 
sovrastimare le correzioni per piccoli valori della differenza d'impulso 
tra le due particelle (cio\`e nella regione pi\`u critica per lo studio delle 
correlazioni). 
\newline
Lo studio della correlazione \`e stato condotto in maniera indipendente in due 
diverse parametrizzazioni (quella cartesiana e quella di YKP), sia nel calcolo 
delle correzioni (per l'interazione coulombiana e per 
la risoluzione di coppie di tracce) sia nella procedura di {\em ``best fit''} 
alle funzioni di correlazione. Le due parametrizzazioni hanno fornito risultati 
tra loro compatibili.  
\newline
Dallo studio della dipendenza dei ``raggi HBT'' dall'impulso medio della coppia 
si sono ricavate le seguenti quantit\`a in funzione della centralit\`a delle 
collisioni:
\begin{itemize}
\item[-] il raggio trasverso gaussiano (bidimensionale) al {\em ``freeze-out''}; 
\item[-] la durata dell'espansione del sistema sino al {\em ``freeze-out''}; 
\item[-] la durata del {\em ``freeze-out''};  
\item[-] informazioni sull'espansione longitudinale; 
\item[-] la temperatura di {\em ``freeze-out''} e  
%\item[-] 
	 la velocit\`a media dell'espansione trasversa~\footnote{Queste ultime due 
         quantit\`a sono state disaccoppiate l'una dall'altra 
	 dall'analisi simultanea dello spettro 
	 di massa trasversa di singola particella degli $h^-$.}.  
\end{itemize}
I risultati ottenuti per la temperatura e per la velocit\`a del flusso nella 
direzione trasversa per la classe di collisioni pi\`u centrali (la $IV$) 
sono compatibili, entro gli errori sperimentali, con quelli determinati 
dallo studio dei soli spettri di massa trasversa delle particelle strane 
nella stessa classe di centralit\`a.  
%($T\approx$, $\beta_\perp$). 
Questa compatibilit\`a, relativa ai risultati di due diversi campioni di particelle  
(gli $h^-$, costituiti prevalentemente da \Pgpm, e le particelle strane),  
con due diversi metodi di analisi (l'interferometria HBT e 
lo studio degli spettri di massa trasversa per particelle di diversa massa) 
ed in due differenti esperimenti (WA97 ed NA57), d\`a  %ulteriore valenza ai  
fiducia nei 
risultati presentati.  
Nelle altre classi di centralit\`a, raggruppate in modo diverso, 
il confronto non \`e immediato. La dipendenza qualitativa dalla centralit\`a  
d\`a per\`o indicazioni ancora in accordo tra loro: 
l'intensit\`a del flusso trasverso diminuisce 
al diminuire della centralit\`a della collisione e  
contemporaneamente aumenta la temperatura finale di {\em ``freeze-out''}. 
\newline
Considerando tutti i parametri estratti dall'interferometria HBT, la pi\`u importante   
conclusione che si trae da questa analisi \`e la seguente: la classe di minor 
centralit\`a studiata (la $I$) presenta una dinamica di espansione 
{\em significativamente} differente da quella delle classi di maggior centralit\`a.  
Il sistema si espande di meno sia nella direzione longitudinale sia in quella trasversale, 
la dimensione trasversa al {\em ``freeze-out''} \`e minore 
(consistentemente con la minor intensit\`a dell'espansione), la durata  
dell'espansione \`e minore, la temperatura finale \`e pi\`u elevata.  
I parametri della sorgente per le due classi di collisioni pi\`u centrali
(III e IV) sono tra loro molto simili, suggerendo cos\`i lo stesso stato
geometrico e dinamico per queste classi. La classe II presenta dei parametri
intermedi tra le classe I e III-IV; la sua dinamica \`e tuttavia molto
pi\`u prossima a quella delle classi III-IV.  

Alla luce di questi risultati, considerando anche le indicazioni
degli altri esperimenti dell'SPS ed in modo particolare quelle di NA50
a proposito della soppressione della produzione del 
mesone $J/\Psi$, si pu\`o quindi affermare,
con ragionevole certezza, che nelle collisioni Pb-Pb pi\`u centrali
(a 160 A GeV/$c$) venga formato un nuovo stato della materia che ha molte
delle propriet\`a previste per il QGP.  In seguito il sistema si adronizza ed 
evolve sino al {\em ``freeze-out} termico'' secondo un'espansione ``esplosiva'', con velocit\`a 
pari a circa la meta della velovit\`a della luce, raffreddandosi 
ad una temperatura di circa 120-130 MeV.  

%% file: appendici/glauber.tex
\appendix
\chapter{Calcolo di alcune quantit\`a col modello di Glauber}
\begin{figure}
  \centering
  \resizebox{0.75\textwidth}{!}{%
  \includegraphics{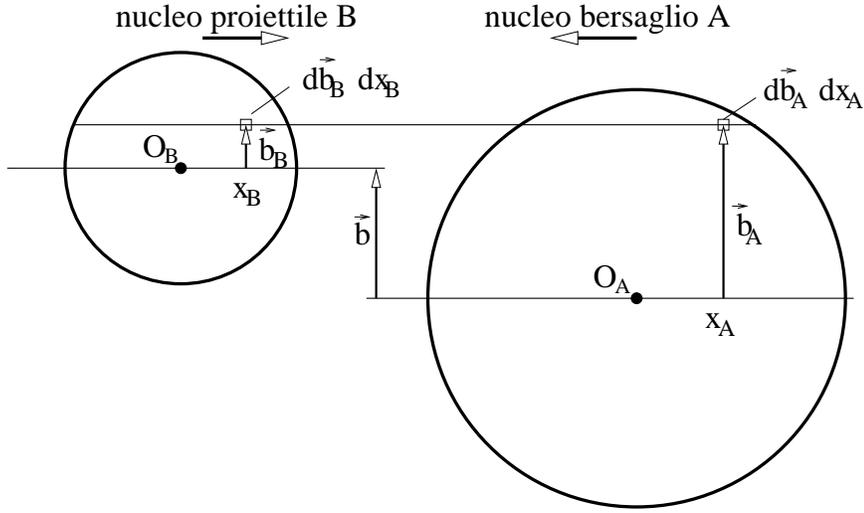}}
 \caption{Schema della collisione con parametro di impatto $ {\bf b} $\ 
  tra un nucleo proiettile $B$\ ed un nucleo bersaglio $A$.}
\label{collisione}
\end{figure}

\index{Calcolo di alcune quantit\`a col modello di Glauber}
Sia $  t({\bf b})\,d{\bf b} $\ la probabilit\`a di avere 
un'interazione anelastica nucleone--nucleone quando i due nucleoni sono situati ad un 
parametro d'impatto relativo $ {\bf b} $\ entro l'elemento di area 
trasversa $ d{\bf b} $. La funzione $ t({\bf b}) $\ \`e chiamata  
{\em funzione di spessore}\ nucleone--nucleone ed \`e normalizzata come 
\begin{equation}
 \int t({\bf b})\, d{\bf b}  \; = \; 1 
\label{1.22}
\end{equation}
Per collisioni tra nucleoni non polarizzati, $ t({\bf b}) $\  dipende 
dal modulo del parametro di impatto $ b $. 
Se $\sigma_{in}$\ \`e la sezione d'urto anelastica nucleone-nucleone, 
la probabilit\`a di un'interazione anelastica fra due nucleoni con 
parametro d'impatto relativo ${\bf b}$\ \`e $ t({\bf b}) \sigma_{in}$. 
Nella collisione tra un nucleo proiettile B con un nucleo bersaglio A, 
schematizzata in fig.~\ref{collisione}, sia 
$ \rho_{B}({\bf b}_{B},x_{B})\,d{\bf b}_{B}\,dx_{B} $\ la 
probabilit\`a di trovare un nucleone nell'elemento di volume 
$ d{\bf b}_{B}\,dx_{B} $\ del nucleo B, alla posizione 
$ ({\bf b}_{B},x_{B}) $. Si osservi che 
$ \rho_{B}({\bf b}_{B},x_{B})  $, con la normalizzazione 
\begin{equation}
 \int\rho_{B}({\bf b}_{B},x_{B})\, d{\bf b}_{B}\,dx_{B}  \; = \; 1 ,
\label{1.23}
\end{equation}
\`e l'usuale funzione di densit\`a nucleare divisa per il numero di nucleoni 
del nucleo. In modo simile si definisce la probabilit\`a di trovare un 
nucleone del nucleo A alla posizione $ ({\bf b}_{A},x_{A}) $. \\
La probabilit\`a che avvenga una collisione anelastica tra due nucleoni,  
in un urto tra i nuclei con parametro di impatto $ {\bf b} $, viene 
definita come la funzione di spessore nucleo-nucleo 
$ T({\bf b}) $\ moltiplicata 
per la sezione d'urto anelastica nucleone--nucleone $ \sigma_{in} $
\begin{equation} 
 T({\bf b}) \sigma_{in} \; = \;
 \int 
    T_{A}({\bf b}_{A}) T_{B}({\bf b}_{B}) 
    t \left[{\bf b} - ({\bf b}_{A} - {\bf b}_{B}) \right] 
        \sigma_{in}\,  d{\bf b}_{A}\,d{\bf b}_{B}
\label{1.24}
\end{equation}
dove le funzioni di spessore $ T_{i}({\bf b}_{i}) $, $ i\,=\,A,B $, 
sono definite come 
\begin{equation} 
 T_{i}({\bf b}_{i}) \; = \; \int 
    \rho_{i}({\bf b}_{i},x_{i}) \, dx_{i}\,
\label{1.25}
\end{equation}
In virt\`u delle equazioni ~\ref{1.22}, ~\ref{1.23}, le funzioni di 
spessore nucleare $ T_{A}({\bf b}_{A}) $, $ T_{B}({\bf b}_{B}) $\ 
e $ T({\bf b}) $\ sono anch'esse normalizzate all'unit\`a. \\
Applicando le definizioni introdotte, si pu\`o scrivere la probabilit\`a che 
per una collisione ad un dato parametro di impatto $ {\bf b} $\  si  
abbiano  esattamente {\em n}\ collisioni anelastiche nucleone--nucleone
\begin{equation}
 P(n,{\bf b})\;=\; \left(\begin{array}{c} AB \\ n \end{array} \right)
            \left[ T({\bf b})\sigma_{in} \right]^{n}
            \left[ 1 - T({\bf b})\sigma_{in} \right]^{AB - n}
\label{eq1.26}
\end{equation}
dove il primo fattore del termine di destra rappresenta il numero di 
possibili estrazioni di {\em n} collisioni tra le possibili {\em AB}\   
combinazioni nucleone--nucleone, il secondo fornisce la probabilit\`a di 
avere esattamente {\em n} interazioni ed il terzo fattore d\`a la  
probabilit\`a di avere $ AB\,-\,n $\ ``non-interazioni''. 
Sommando la ~\ref{eq1.26} da $ n\,=\,1 $\ ad {\em AB}, si ottiene la 
probabilit\`a di avere almeno un'interazione anelastica in una collisione 
con parametro d'impatto $ {\bf b} $ 
\begin{equation}
 \frac{d\sigma_{in}^{AB}({\bf b})}{d{\bf b}}\;=\;
 \sum_{n=1}^{AB}P(n,{\bf b})\;=\;
  1 \, - \, \left[ 1 - T({\bf b})\sigma_{in} \right]^{AB}
 \label{1.27}
\end{equation}
il cui integrale fornisce la sezione d'urto anelastica totale 
$ \sigma_{in}^{AB}({\bf b}) $
\begin{equation}
  \sigma_{in}^{AB}({\bf b}) \; = \; 
  \int \sum_{n=1}^{AB} P(n,{\bf b}) \, d{\bf b} \;=\; 
  \int \left\{
  1 \, - \, \left[ 1 - T({\bf b})\sigma_{in} \right]^{AB} \right\}\, d{\bf b}
 \label{1.28}
\end{equation}
Altre due quantit\`a di interesse che possono essere calcolate in modo 
analogo sono il numero medio di collisioni binarie, $ N_{coll} $, ed il 
numero medio di partecipanti, $ N_{part} $:
\begin{align}
 \label{1.29}
  <N_{coll}({\bf b})> & = \sum_{n=1}^{AB}n\,P(n,{\bf b}) \\
  <N_{part}({\bf b})> & = A \; \int 
  T_{A}({\bf s}) \left\{1-\left[1-T_{B}({\bf s}-{\bf b})
   \sigma_{in}\right]^{B}\right\}\, d^{2}{\bf s} \nonumber \\
   & \quad \quad +  B \; \int 
  T_{B}({\bf s}-{\bf b}) \left\{1-\left[1-T_{A}({\bf s})
   \sigma_{in}\right]^{A}\right\} \, d^{2}{\bf s}
   \label{1.30}
\end{align}
Per ``partecipante'' si intende un nucleone che subisca {\em almeno} 
un'interazione anelastica con un altro nucleone del sistema.  

%% file: appendici/V0cinematic.tex
\chapter{Geometria~del~decadimento~delle~$V^0$}
\begin{figure}[hbt]
\begin{center}
 \includegraphics[scale=0.38]{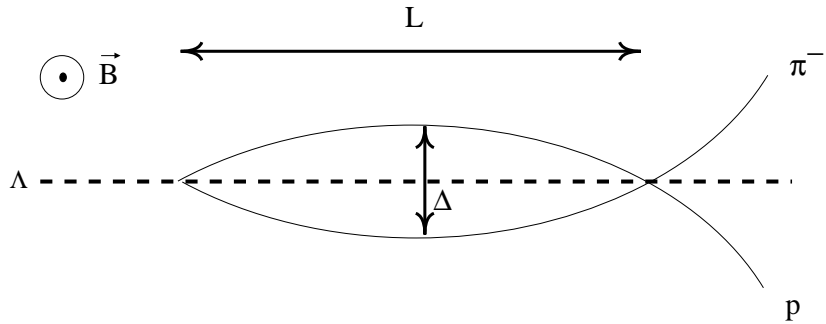}
 \caption{Topologia di decadimento ``cowboy'' per una $\Lambda$.}
\label{Arco}
\end{center}
\end{figure}

\index{Geometria~del~decadimento~delle~$V^0$}
In fig.~\ref{Arco} \`e mostrata la proiezione nel piano $xy$\ del 
decadimento di una $V^0$\ (nel caso specifico, quello di una $\Lambda$) nella 
configurazione ``cowboy''. Si consideri la distanza $L$\ tra il vertice 
della $V^0$\ ed il secondo punto di intersezione delle tracce di decadimento, 
la massima distanza trasversa $\Delta$\ tra le due tracce di decadimento lungo $L$. 
%Riferendosi alla fig.~\ref{Arco2}, 
Per una particella di carica $e$\ ed impulso $p$\ in moto entro un campo 
magnetico $B$\ perpendicolare alla sua traiettoria, il raggio di curvatura 
$R$\ si ricava dalla relazione: 
\begin{equation}
R= \frac{p}{eB}
\label{Radius}
\end{equation}
\begin{figure}[hbt]
\begin{center}
\includegraphics[scale=0.25]{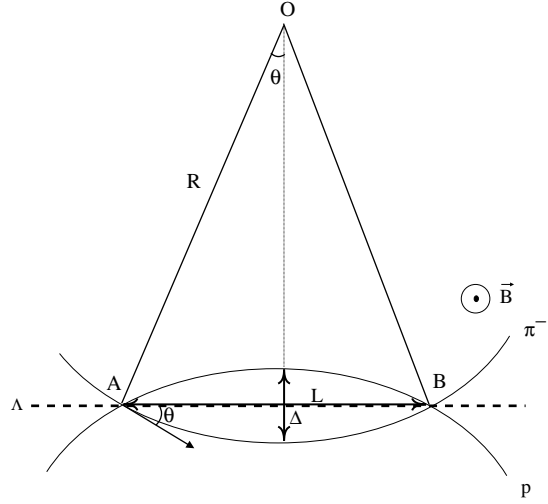}
\caption{Schema della geometria del decadimento di una $\Lambda$\ nella
         configurazione ``cowboy''.}  
\label{Arco2}
\end{center}
\end{figure}
Dalla fig.~\ref{Arco2} si deduce che $ \sin\theta = \frac{L}{2R} $;  per 
piccoli angoli $\theta$, risulta $\sin\theta \simeq \tan\theta = 
\frac{q_T}{q_L}$, dove $q_T$\ e $q_L$\ sono, rispettivamente, le 
componenti trasversali e longitudinali dell'impulso dei prodotti di 
decadimento rispetto alla linea di volo della $V^0$.  
\newline
Poich\'e le tracce di decadimento ricostruite nell'esperimento NA57 
dispongono tipicamente di impulsi di parecchi GeV/$c$, mentre il valore massimo 
del loro impulso trasverso nel riferimento della $V^0$\ vale 
$q_{T}^{MAX}=0.1$\ GeV/$c$\ per le $\Lambda$\ ed $\bar{\Lambda}$, e 
$q_{T}^{MAX}=0.2$\ GeV/$c$\ per i $K_S^0$, risulta $q_T \ll p$, per 
cui $q_L \simeq p$. Utilizzando dunque queste approssimazioni, 
si pu\`o riscrivere l'eq.~\ref{Radius} come:
\begin{equation}
L \simeq 2 \left( \frac{p}{eB} \right) \left(\frac{q_T}{p} \right) = \frac{2q_T}{eB}
\label{Corda}
\end{equation}
Tale distanza risulta dunque invariante per {\em boost} di Lorentz 
longitudinali ed il suo valore massimo \`e caratteristico della particella 
che decade. La massima distanza $L$, espressa in metri, risulta pari a:
\begin{equation}
L_{MAX} \simeq  \frac{2 \,q_T^{MAX}({\rm GeV}/c)}{e\,B({\rm Tesla})}
 \left(\frac{10^9e}{c} \right) = %\frac{2q_T}{eB} = 
 \frac{2 \, q_T^{MAX}({\rm GeV}/c)}{0.2998 \, B({\rm Tesla})}
\label{LMax}
\end{equation}
Nel caso di un campo magnetico di intensit\`a pari a $B = 1.4$\ Tesla, 
quale quello del magnete GOLIATH usato in NA57, risulta che: 
\begin{itemize}
\item[] $
L_{MAX} \approx \; 50 \;{\rm cm}  \quad {\rm per \; \Lambda \; ed \; \bar{\Lambda}} $
\item[] $
L_{MAX} \approx \; 100 \; {\rm cm} \quad {\rm per \; K_S^0} $
\end{itemize}
Per quanto riguarda la distanza $\Delta$, dalla fig.~\ref{Arco2} risulta:
\begin{equation}
\Delta = (R_1 -R_1 \cos\theta_1) + (R_2 -R_2 \cos\theta_2)
\label{Delta}
\end{equation}
dove gli indici 1 e 2 si riferiscono alle due tracce di decadimento. 
Dalla relazione~\ref{Radius}, considerando che 
$\cos\theta = \frac{q_L}{p}$, si ricava:
\begin{equation}
(R_i -R_i \cos\theta_i) = \frac{p_i}{eB}
\left( 1 - \frac{q_{L_i}}{p_i} \right) = \frac{1}{eB}
\left( \sqrt{q^2_T+q^2_{L_i}} - q_{L_i} \right)
\label{NoUse}
\end{equation}
Indicando con $p^*$, $E_i^*$\ e $\theta^*$\ le quantit\`a valutate nel 
sistema centro di massa, \`e possibile esprimere le componenti dell'impulso 
delle particelle di decadimento nel sistema del laboratorio come:
\begin{equation}
\begin{split}
q_{T_i} = & (-1)^i p^* \sin\theta^* \\
q_{L_i} = & \gamma  \left[ \beta E^*_i + (-1)^i p^* \cos\theta^* \right] .
\label{Transfo}
\end{split}
\end{equation}
Considerando il decadimento con massimo impulso trasverso, corrispondente a 
$\theta^*=\frac{\pi}{2}$\ e sostituendo l'eq.~\ref{Transfo} nell'eq.~\ref{NoUse}, 
si ottiene: 
\begin{equation}
(R_i -R_i \cos\theta_i) = \frac{1}{eB}\left[ 
 \sqrt{{p^*}^2 +(\gamma^2 - 1){E^*_i}^2} -\sqrt{\gamma^2 - 1}E^*_i
 \right]
\label{NoUse2}
\end{equation}
Pertanto $\Delta$\ dipende, attraverso $\gamma$, dall'impulso della particella 
che decade e, attraverso $E^*_i$\ e $p^*$, dalle masse delle particelle 
coinvolte nel decadimento. Esprimendo $E^*_i$\ ($p^*$) in GeV (GeV/$c$), risulta:
\begin{equation}
(R_i -R_i \cos\theta_i){\footnotesize (metri)} = 
\frac{1}{0.2998\, B{\footnotesize (Tesla)}}\left[
 \sqrt{{p^*}^2 +(\gamma^2 - 1){E^*_i}^2} -\sqrt{\gamma^2 - 1}E^*_i
 \right]
\label{NoUse3}
\end{equation}
Svolgendo il calcolo per un fattore $\gamma=10$, corrispondente ai tipici impulsi 
delle particelle di decadimento, e per $B=1.4$\ Tesla, si ottiene:
\begin{itemize}
\item[] $
\Delta(\gamma=10) = \; 0.9 \;{\rm cm}  \quad {\rm per \; \Lambda \; ed \; \bar{\Lambda}} $\item[] $
\Delta(\gamma=10) = \; 3.9 \; {\rm cm} \quad {\rm per \; K_S^0} $. 
\end{itemize}

%% file: biblio.tex
%

%% file: aknow.tex
\chapter*{{\em Ringraziamenti}}
{\em Vorrei spendere qualche parola per le persone che mi sono state 
vicine professionalmente e sotto il profilo umano in questi anni.  
Sono molto grato al mio tutore, il Prof. Bruno Ghidini, che mi ha  
instancabilmente seguito e sostenuto in ogni mia attivit\`a.  
Ringrazio quindi il Dr. Emanuele Quercigh, per aver letto con grande  
attenzione questa tesi e per tutti i consigli, gli insegnamenti ed il 
tempo dedicatomi.   
Voglio quindi esprimere la mia gratitudine e stima per il gruppo di 
Bari degli ``Ioni Pesanti'': un grazie di cuore, in ordine sparso, ad Alfredo 
Loconsole, Rosanna Fini, Vito Manzari, Domenico Elia, Rocco Caliandro,  
Vito Lenti e tutti gli altri.   
Sono altres\`i grato a tante persone con cui ho collaborato, le prime che  
mi sovvengono sono Adam Jacholkowski, Ladislav \v{S}\'andor e Nicola Carrer, ma 
sono sicuro di averne dimenticate tante altre.  
Questa \`e anche un'ottima occasione per ringraziare una lunga lista 
di persone cui devo molto: la mia famiglia, Annamaria e tutti i miei amici.} 

%% file: new.bbl
\begin{thebibliography}{9}
\addcontentsline{toc}{chapter}{Bibliografia}
%
%% Ref. del capitolo 1
%
\bibitem{r1_2} C. Quigg, {\em ``Gauge Theories of the Strong, Weak and Electromagnetic
Interactions''}, Benjamin-Cummings, Reading (1983).
\bibitem{r1_1} S.L. Glashow, Nucl. Phys. 22 (1961) 579;\\
S. Weinberg, phys. Rev. Lett. 19 (1967) 1264;\\
A. Salam, Proc. 8th Nobel Symposium , Stockholm 1968, ed. N. Svartholm 
(Almqvist and Wiksells, Stockholm 1968), p. 367.
\bibitem{NOW} M.C. Gonzalez-Garcia and Y. Nir, {\em ``Developments in Neutrino Physics''}, 
 hep-ph/0202058 (2001). \\
 G.G. Raffelt, {\em ``Neutrino Masses in Astroparticle Physics''}, astro-ph/0207220, (2002).
\bibitem{r1_7} Results presented at the XXXth International Conference on
High Energy Physics, Osaka (2000).
\bibitem{BagModel} A. Chodes et al., Phys. Rev. D9 (1974) 3471.  
\bibitem{Hax80} W.C. Haxton e L. Heller, Phys. Rev. D22 (1980) 1198.
\bibitem{Has81} P. Hasenfratz et al., Phys. Lett. B95 (1981) 199.
\bibitem{Sch86} D. Schram, Proc. 2nd ESO-CERN Symposium, Gerching (1986) 269.
\bibitem{Gle91} N.K. Glendenning et al., LBL-30645 (1991)
%\bibitem{OrderTransit} Y. Iwasaki et al., Phys. Rev. D53 (1996) 6443.
%\bibitem{OrderTransit2} S. Aoki et al., Nucl. Phys. B73 (1999) 459.
% Collisioni nucleari ultra-relativistiche
\bibitem{Glauber} R.J. Glauber in
{\em ``Lectures in Theoretical Physics vol.1''},
(Interscience, New York, 1959) 315.
\bibitem{Wong} C.Y. Wong, in
{\em ``Introduction to High--Energy Heavy--Ion Collision''},
(World Scientific, Singapore, 1994), 349--264.
\bibitem{Carrer18} D. Kharzeev, C. Louren\c{o}, M. Nardi e H. Satz,
 Z. Phys. C 74 (1997) 307.
\bibitem{Wong19} C. Y. Wong, in
 {\em ``Introduction to High--Energy Heavy--Ion Collision''},
 (World Scientific, Singapore, 1994), 275--277.
% QCD su reticolo 
\bibitem{Wil74} K.G. Wilson, Phys. Rev. D14 (1974) 2455.
\bibitem{Sat84} H. Satz, Nucl. Phys. A418 (1984) 447c.  
\bibitem{Condensate} K. Kanaya, to be published in the Proceedings of QM2002 (2002). \\
        http://alice-france.in2p3.fr/qm2002/Transparencies/22Plenary/Kanaya.ppt
\bibitem{OrderTransit} Y. Iwasaki et al., Phys. Rev. D53 (1996) 6443.
\bibitem{OrderTransit2} S. Aoki et al., Nucl. Phys. B73 (1999) 459.
\bibitem{Chirale} P.~Koch, Introduction to chiral symmetry, nucl-th/9512029 (1995).
\bibitem{QuarkMass} J. Gasser and H. Leutwyler, Phys. Rep. 87 (1982) 77.
\bibitem{QuarkCost} J.F. Donoghue, Annu Rev. Nucl. Part. Sci. 39 (1989) 1.
%Rapidity distribution
\bibitem{Fer50} E. Fermi, Prog. Theor. Phys. 5 (1950) 570.
\bibitem{Lan53} L.D. Landau, Izv. Akad. Nauk. SSSr 17 (1953) 51.
\bibitem{Bjo83} J.D. Bj{\o}rken, Phys. Rev. D27 (1983) 140.
\bibitem{Mcl82} L. McLerran, Phys. Rep. 88 (1982) 379.
\bibitem{NetBaryon49} H. Appelsh\"{a}user et al., Phys.Rev.Lett. 82 (1999) 2471.
\bibitem{BRHAMS} I.G Bearden et al.,  nucl-ex/0112001, to be 
      published in Phys. Rev. Lett (2002).
\bibitem{mt_WA97} F. Antinori et al., The Eur. Phys. J. C14 (2000) 633. 
\bibitem{Carr26} H. de Vries, C.W. de Jager and C. de Vries, 
 {\em Atomic Data and Nuclear Data Tables}, 36 (1987) 495.
% Transverse energy distrbution
\bibitem{WA98PRC65} M.M. Aggarwal et al., Phys. Rev C65 (2001) 054912.\\
		    M.M. Aggarwal et al., Eur. Phys. J. C18 (2001) 651.
\bibitem{VENUS} K. Verner, Phys. Rep. 232 (1993) 87.     
\bibitem{EtPHENIX} K. Adcox et al., Phys. Rev. Lett. 87 (2001) 52301.
\bibitem{Bla96} J.P. Blaizot, J.Y. Ollitrault, Phys. Rev. Lett. 77 (1996) 1703.
\bibitem{Abbot} T.Abbot et al., Phys.Rev. C63 (2001) 064602.
% Multiplicity
\bibitem{WA97Centr} F. Antinori et al., Eur. Phys. J. C18 (2000) 57.
%Transverse mass distribution
\bibitem{Cooper-Frye} E. Schnedermann, J. Sollfrank and U. Heinz, 
In {\em ``Particle production in highly excited matter''}, H. Gutbrod and 
J. Rafelski editors, volume B303 of NATO ASI Series Physics (1993) 175.
\bibitem{28} U. Heinz, In {\em `` Hot hadronic matter, theory and 
experiment''}, H. Gutbrod and J. Rafelski editors, 
volume B346 of NATO ASI Series Physics (1995) 413. 
\bibitem{Carrer29} U. Heinz, preprint nucl-th/9801050 (1998).  
\bibitem{RQMD} H. van Hecke, H Sorge and Nu Xu, Phys. Rev. Lett. 81 (1998) 5764.
% HBT intro
\bibitem{Hanbury} R. Hanbury-Brown and R.Q. Twiss, Phil.Mag. 45 (1954) 633.
\bibitem{Goldhaber} G. Goldhaber, S. Goldhaber, W. Lee and A. Pais,
 Phys. Rev. 120 (1960) 300.
%Charmonium suppression
\bibitem{Satz} T. Matsui and H. Satz, Phys. Lett. B178 (1986) 416.  
\bibitem{Schr} S. Jacobs, M.G. Olsson and C.Suchyta, Phys. Rev. D33 (1986) 3338.
\bibitem{Dixit} V.V. Dixit, Mod. Phys. Lett. A5 (1990) 227.
\bibitem{DebyePQCD} D. Gross, R.D. Pisarski and L.G. Yaffe, Rev. Mod. Phys. 53 (1981) 43.
\bibitem{65} U. Heller, F. Karsch and J. Rank, Phys. Lett. B355 (1995) and 
              Phys. Rev. D57 (1998) 1438. 
\bibitem{JPSI1} M.C. Abreun et al., Phys. Lett. B499 (2001) 85.	       
\bibitem{NormalAbsor} S. Gavin, M. Gyulassy and A. Jackson, Phys. Lett. B207 (1998) 257.
\bibitem{JPSI2} M.C. Abreun et al., Phys. Lett. B477 (2000) 28. 
\bibitem{JPSI3} M.C. Abreun et al., Phys. Lett. B521 (2001) 195. 
\bibitem{Pro1}  H. Satz, Rep. Prog. Phys.  63 (2000) 1511. 
\bibitem{Pro2} J.P. Blaizot, P.M Dinh and J.Y. Ollitrault, Nucl. Phys., A698 (2002) 579.
\bibitem{Capella} N. Armesto and A. Capella, Phys. Lett., B430 (1998) 23.
\bibitem{Gavin} S. Gavin and R. Vogts, Phys. Rev. Lett. 78 (1997) 1006.
%photons
\bibitem{Kap93}   J. Kapusta, P. Lichard and D. Seibert, Phys. Rev. D47 (1993) 4171. 
\bibitem{PhotEm1} K. Kajantie and H.I. Miettinen, Z. Phys. C9 (1981) 341.
\bibitem{PhotEm2} F. Halzen and H.C. Liu, Phys. Rev. D25 (1982) 1842.
\bibitem{PhotEm3} B. Sinha, Phys. Lett. B128 (1983) 91.
\bibitem{PhotEm4} L.D. McLerran and T. Toimela, Phys. Rev. D31 (1985) 545.
\bibitem{Kaputsa} J.Kaputsa, P. Lichard and D. Seibert, Phys. Rev. D44 (1991) 2774.
\bibitem{Aurenche} P. Aurenche, F. Gellis, H. Zaraket and R. Kobes, Phys. Rev. D58 (1998).
\bibitem{Srivastava} D.K. Srivastava, Eur. Phys. J. C10 (1999) 487.
\bibitem{Ake90} T. {\AA}kesson et al., Z. Phys C64 (1990) 369.
\bibitem{Alb91} R. Alber et al., Z. Phys. C64(1994) 195.
\bibitem{Agg96} M.M. Aggarwal et al., Nucl. Phys. A610 (1996) 200. 
\bibitem{WA98}  M.M. Aggarwal et al., Phys Rev. Lett. 85 (2000) 3595.
\bibitem{PHENIXQM02} K. Reygers et al., to be published in the Proceedings of QM2002 (2002). \\ 
        http://alice-france.in2p3.fr/qm2002/Transparencies/22Parallel1/Reygers.pdf
% DiLeptons
\bibitem{Ruu91} P.V. Ruuskanen, Nucl. Phys. A525 (1991) 255.
\bibitem{CERESpA} G. Agakichiev et al., Eur. Phys. J. C4 (1998) 231.
\bibitem{CERESSAu}  G. Agakichiev et al., Phys. Rev. Lett. 75 (1995) 1272
\bibitem{CERES160} B. Lenkeit et al., Nucl. Phys. A661 (1999) 23
\bibitem{CERES40} H. Appelsh\"{a}user et al., Nucl. Phys. A698 (2002) 253.
\bibitem{ReviewDilep} R. Rapp and J. Wambach, Adv. Nucl. Phys. 25 (2000) 1.
\bibitem{Brown-Rho} G.E. Brown and M. Rho; hep-ph/0103102 and earlier 
                    references. 
\bibitem{DiLepThe2} R. Rapp, G. Chanfray and J. Wambach, 
                    Nucl. Phys. A617 (1997) 472.
% Fluctuation
\bibitem{NA49pID} S. Afanasiev et al., Nucl. Meth. A430 (1999) 210.
\bibitem{Fluct1} H. Appelsh\"{a}user et al., {\em Event-by-Event fluctuations 
of the kaon to pion ratio in central Pb-Pb collision at 158 GeV per Nucleon} 
preprint hep-ex/0009053 (2000).
\bibitem{Fluct2}  H. Appelsh\"{a}user et al., Phys. Lett. B459 (1999) 679.
% Stranezza
\bibitem{Raf82-86} J. Rafelski and B. M\"{u}ller, Phys. Rev. Lett. 48 (1982) 1066.\\
J. Rafelski and B. M\"{u}ller, Phys. Rev. Lett. 56 (1986) 2334. 
\bibitem{Raf86} P. Koch, B. M\"{u}ller and J. Rafelski, Phys. Rep. 142 (1986) 167.
\bibitem{Raf92} J. Rafelski, Phys. Lett. B262 (1991) 333. \\
                J. Rafelski, Nucl. Phys. A544 (1992) 279.
\bibitem{Raf96} J. Rafelski, J. Letessier and A. Tounsi, 
                Acta Phys. Pol. B27 (1996) 1035.
\bibitem{Raf2000} J. Letessier and J. Rafelski, nucl-th/0003014 (2000) 36 pagg.
\bibitem{Bilic} N. Bili\'c, J. Cleymans, I. Dadi\'c and D. Hislop, 
	        Phys. Rev. C52 (1995) 401.
\bibitem{Koch} P. Koch and J. Rafelski, Nucl. Phys. A444 (1985) 678
\bibitem{Letessier} J. Letessier, A. Tounsi and J. Rafelski, Phys. Lett. B390 (1997) 363.
\bibitem{WA97Nucl} F. Antinori et al., Nucl. Phys. A661 (1999) 130.
\bibitem{WA97Eur} F. Antinori et al., Eur. Phys. J. C11 (1999)  79.
\bibitem{WA97PhysLett} E. Andersen et al., Phys. Lett. B433 (1998) 209.
\bibitem{WA97Lietava} R. Lietava et al., J. Phys. B25 (1999) 181.
\bibitem{NA49PLB444} H. Appelsh\"{a}user et al., Phys. Lett. B444 (1998) 523.
\bibitem{WA85PLB447} F. Antinori et al., Phys. Lett. B447 (1999) 178.
\bibitem{NA35} T. Alber et al., Z. Phys. C64 (1994) 195.
\bibitem{WA85-WA94} D. Evans et al., J. Phys. G: Nucl. Phys. 25 (1999) 209.
\bibitem{NA44} H. Boggild et al., Phys. Rev. C59 (1999) 328.
\bibitem{WA97PLB449} E. Andersen et al., Phys. Lett. B449 (1999) 401.
\bibitem{NA49SiKler} F. Sikl\'er et al., Nucl. Phys. A661 (1999) 45.
\bibitem{NA44PLB471} I. Bearden et al., Phys. Lett. B471 (1999) 6.
\bibitem{NA52} S. Kabana et al., Nucl. Phys. A661 (1999) 370. \\
               S. Kabana et al., J. Phys G: Nucl. Part. Phys. 25 (1999) 217.
\bibitem{Cle91} J. Cleymans, Nucl. Phys. A525 (1991) 205.
\bibitem{Pisarski} E. Breaten and R.D. Pisarski, Nucl. Phys. B337 (1990) 569.
\bibitem{Altherr93} T. Altherr and D. Seibert, Phys. Lett. B313 (1993) 149 \\
		    T. Altherr and D. Seibert, Phys. Rev.  C49 (1994) 1684.
\bibitem{Biro} T.S. Bir\'o, P. L\'evai and B. M\"{u}ller, Phys. Rev. D42 (1990) 3078.
\bibitem{Ukawa} A. Ukawa, Nucl. Phys. A498 (1989) 227.
\bibitem{WA97}  http://wa97.web.cern.ch/WA97/WelcomeOld.html
\bibitem{PressRelease} http://www.cern.ch/CERN/Announcements/2000/NewStateMatter 
%                      \\ si veda anche U.~Heinz, M.~Jacob, nucl-th/00022042.
%
%%% Ref del capitolo 2
%
\bibitem{NA57p} NA57 proposal, CERN/SPSLC/96-40 (1996).  
\bibitem{WA85}  http://www.ep.ph.bham.ac.uk/exp/WA85/
\bibitem{WA94}  http://www.ep.ph.bham.ac.uk/exp/WA94/ \\
		S. Abatzis et al., J. Phys. G: Nucl. Part. Phys. 23 (1997) 1857-1864.
\bibitem{OMEGA} {\em The CERN OMEGA spectrometer, 25 years of physics}, 
		CERN Yellow Report, cernrep/97-02, 
		edito da M.R.M.~Jacob and E.~Quercigh (1997).
\bibitem{SPS} N. Angert et al. Preprint CERN 93-01, (1993). 
\bibitem{WA97p} WA97 proposal, CERN/SPSLC/91-29 (1991) 263.
\bibitem{RD19} E.H.M Heijne et. al., Nucl. Instr. and Meth. A349 (1994) 138. \\
               F. Antinori et. al.,  Nucl. Instr. and Meth. A360 (1995)  91.
\bibitem{Omega2} M. Campbell et al., Nucl. Instr. and Meth. A342 (1994) 52.
\bibitem{Omega3} E.H.M Heijne et al., Nucl. Instr. and Meth. A383 (1996) 55.
%\bibitem{VanDeVen} P. van de Ven, {\em ``Cemtrality dependence of \Lam production 
% in Pb-Pb collisions'' } Ph.D Thesis (2001).
\bibitem{DeRijke} P. C. de Rijke, {\em ``Performance of the microstrip detectors 
in the NA57 experiment for the lead-99 run''}, Student thesis, Universiteit Utrecht, 
The Netherlands (2001).
\bibitem{trigger1}  F. Antinori et al., {\em A new VME trigger processor for the 
 NA57 experiment}, Proceedings of ''Electronics for LHC Experiment'',  
 London (1997) 364-368.  
\bibitem{trigger2}  F. Antinori et al., {\em The NA57 trigger processor}, 
 Proceedings of ''Electronics for LHC Experiment'',  Roma (1998) 364-368. 
\bibitem{NA57p1} NA57 proposal, CERN/SPSLC/96-40 (1996) 4-5. 
\bibitem{HPSS} {\em CERN Introduction to HPSS}, \\
      http://tilde-dmac3.web.cern.ch/~dmac3/hpss/hpss\_cernintro/hpss\_cernintro.htm
\bibitem{CASTOR} The CASTOR PROJECT
	       http://cern.ch/it-div-ds/HSM/CASTOR/
\bibitem{ORHION} J.C. Lasalle and A. Michalson, {\em ORHION User's guide and 
        reference manual}, WA97 internal note, 1994. 
\bibitem{SPLINE} H. Wind, Nucl. Instr. and Meth. 115 (1974) 431. 
\bibitem{NA57Mem} F. Antinori et al., {\em NA57 Status Report and Future Plans} 
             CERN/SPSC 2002-012, SPSC/M680 (2002). 
%
%% Referenze del capitolo 3
%
\bibitem{Pod54} J. Podolansky e R. Armenteros, Phil. Mag. 45 (1954) 13.
\bibitem{Omega98Bruno} G.E. Bruno, Riunione di Collaborazione NA57, 
	CERN 11-12 Luglio 2002.
\bibitem{XiCarrer} N. Carrer, Riunione di Collaborazione NA57, 
	CERN 28-29 Giugno 2001.
\bibitem{BrunoMoriond} G.E. Bruno et al., to be published in the Proceeding of 
        {\em The XXXVII Rencontres de Moriond on QCD and High Energy 
	Hadronic Interactions},  hep-ex/0207047 (2002).
\bibitem{BrunoBKG} G.E. Bruno, Riunione di Collaborazione NA57, 
	CERN 07-08 Febbraio 2002 (2002).
\bibitem{EliaQM02} D. Elia et al., Proceedings of the Quark Matter 2002 Conference - 
	Nantes, France 18-24 July 2002 to be published on Nuclear Physics A.
%
%% Referenze del capitolo 4
%
\bibitem{GEANT3} R. Brun et al., GEANT3, CERN program library Q123.
\bibitem{BrunoOmegaAcc} G.E. Bruno, Riunione di Collaborazione NA57, 
	CERN 23-24 Settembre 2002.
\bibitem{BrunoLamStab} G.E. Bruno, Riunione di Collaborazione NA57, 
	CERN 23-24 Maggio 2002. 
\bibitem{Alner} G.J. Alner et al., Phys. Lett. B167 (1986) 476.
\bibitem{ManzQM02} V. Manzari et al., Proceedings of the Quark Matter 2002 Conference -
        Nantes, France 18-24 July 2002 to be published on Nuclear Physics A.
\bibitem{WA97web} http://wa97.web.cern.ch/WA97/Data/TableQM99.html  
\bibitem{Rocco} R.~Caliandro, Tesi di Dottorato, Universit\`a di Bari.
%
%% Referenze del capitolo 5
%
\bibitem{NA49K+K-PI-} S.~V. Afanasiev et al., nucl-ex/0205002, 
		   to appear in Phys. Rev. C33 (2002).
\bibitem{NA49K0}  S. Margetis et al., J. Phys. G: Nucl. Part. Phys. 25 (1999) 189.
\bibitem{NA49PI+PROT} S.V.~Afanasiev et al., Nucl. Phys. A610 (1996) 188.
\bibitem{NA49PHI} R.A.~Barton et al., Phys. G: Nucl. Part. Phys. 27 (2001) 367.
\bibitem{NA49LA} T. Alber et al., J. Phys. G: Nucl. Part. Phys. 23 (1997) 1817.
\bibitem{NA49XI} S.V.~Afanasiev et al., Phys.Lett. B538 (2002) 275.
\bibitem{NA49OM} M.~van~Leeuwen for the NA49 Collaboration at Quark Matter 2002 Conference -
        Nantes, France 18-24 July 2002,  
http://alice-france.in2p3.fr/qm2002/Transparencies/20Plenary/VanLeeuwen.pdf 
\bibitem{NA49DEU} S.V.~Afanasiev et al., Phys. Lett., B : 486 (2000) no.1-2, pp.22-8
\bibitem{NA44remain} I.G. Bearden et al., arXiv:nucl-ex/0202019; 
 submitted to Phys. Rev. C.
\bibitem{NA44DEUTRI} I.G. Bearden et al., Eur. Phys. J. C23 (2002) 237.
\bibitem{Blast1} E.~Schnedermann, J. Sollfrank and U.~Heinz, Phys. Rev. C48 (1993) 2462.
\bibitem{Blast2} E.~Schnedermann and U.~Heinz, Phys. Rev. C50 (1994) 275  
		  [arXiv:hep-ex/9402018].
\bibitem{OmegaDecoup1} H.~van~Hecke, H.~Sorge and N.~Xu, Phys. Rev. Lett. 81 (1998) 5764
			[arXiv:nucl-th/9804035].
\bibitem{OmegaDecoup2} A.~Dumitri, S.A.~Bass, M.~Bleicher, H.~Stocker and W.~Greiner, 
		       Phys. Lett. B460 (1999) 411 [arXiv:nucl-th/9901046].
\bibitem{RafelskiBook} J.~Letessier and J.~Rafelski, in 
		      {\em ``Hadrons and Quark-Gluon Plasma''} edito da 
		      Cambridge University Press (2002) pagg. 148-154.
\bibitem{QM02NA49} M~van~Leeuwen et al.; Proceedings of the Quark Matter 2002 Conference -
        Nantes, France 18-24 July 2002 to be published on Nuclear Physics A. 
	[arXiv:nucl-ex/0208014].
\bibitem{ModelBraun} P. Braun-Munzinger, I. Heppe and J. Stachel, Phys. Lett. B465 (1999) 15. 
\bibitem{Kristin} K.~Fanebust et al.,  J. Phys. G:  Nucl. Part. Phys. 28(2002) 1607
\bibitem{Quercigh} E.~Quercigh, ``Comments on systematic effects in Heavy Ion Experimets'', 
		   Acta Physica Polonica N.12 Vol.33 (2002).
\bibitem{STAR} J. Castillo et al., J. Phys. G: Nucl. Part. Phys. 28 (2002) 1987.\\
               B. Hippolyte, PhD thesis, Universite de Strasbourg.
%
%% Referenze del capitolo 6
%
% HBT
\bibitem{Tomasik} B.~Tomasik and U.A. Wiedemann, in "Quark Gluon Plasma 3", 
eds. R.C.Hwa and X.-N.Wang, World Scientific (to be published)  [arXiv:hep-ph/0210250].
\bibitem{HBTpaper} F. Antinori et al., J. Phys. G 27 (2001) 2325.
\bibitem{3pioni} H. B{\o}ggild et al., NA44 Collaboration,
  Nucl. Phys. A 638 (1998) 471c.
\bibitem{4pioni} C. Adler er al., STAR Collaboration, 
  to be submitted to Phys. Rev. C. 
\bibitem{Pratt} S. Pratt, Phys. Rev. Lett. 53 (1984) 1219.  
\bibitem{BrunoTesi} G.E.~Bruno, Tesi di Laurea, Universit\`a di Bari. 
\bibitem{Wong55}  C.Y. Wong, in
{\em ``Introduction to High--Energy Heavy--Ion Collision''},
(World Scientific, Singapore, 1994), 451-456.
\bibitem{Boal} D.~Boal, C.K.~Gelbke e B.K.~Jennings, Rev. Mod. Phys. 62 (1990) 553.
\bibitem{sette} A.~Bamberger et al., Phys. Lett. B203 (1988) 320. \\
                T.~Humanic et al., Z. Phys. C38 (1988) 79.
\bibitem{otto} T.~Alber et al., Z. Phys. C66 (1995) 77.\\
               T.~Alber et al., Phys. Rev. Lett. 74 (1995) 1303.
\bibitem{NA49HBT} H. Appelsh\"{a}user et al., Eur. Phys. J. C (1998) 661.
\bibitem{42} S. Chapman , P. Scotto e U. Heinz, Heavy Ion Physics 1 (1995) 1.
\bibitem{99} R. Lednicky e V.L. Lyuboshitz, Sov. J. Nucl. Phys. 35 (1982) 770.
\bibitem{100} R. Lednicky e V.L. Lyuboshitz, Heavy Ion Physics 3 (1996) 93.
\bibitem{101} R. Lednicky, V.L. Lyuboshitz, B. Erazmus e D. Nouais,
Phys. Lett. B 373 (1996) 30.
\bibitem{168} S.A. Voloshin, R. Lednicky, S. Panitkin e N. Xu,
Phys. Rev. Lett. 79 (1997) 4766.
\bibitem{156} S. Soff  et al., J. Phys. G 23 (1997) 2095.
\bibitem{114} D. Miskowiec, in {\em CRIS'98: Measuring the size of things in
 the Universe: HBT interferometry and heavy ion physics},
  ed. da S. Costa et al., (World Scientific, Singapore, 1998), nucl-ex/9808003.
\bibitem{41} S. Chapman , P. Scotto e U. Heinz,
  Phys. Rev. Lett. 74 (1995) 4400.
\bibitem{89} M.Herrmann e G. F. Bertsch, Phys. Rev. C 51 (1995) 328.
\bibitem{171} U.A. Wiedemann, P. Scotto e U. Heinz,
  Phys. Rev. C 53 (1996) 918.
\bibitem{153} Yu. Sinyukov, in {\em ``Hot Hadronic Matter: Theory and
   Experiment}, ed. da J. Letessier et al. (Plenum, New York, 1995), p. 309.
\bibitem{40} S. Chapman e  U. Heinz, Phys. Lett. B 340 (1994) 250.
\bibitem{180} Y.-F. Wu, U. Heinz, B. Tomasik e U.A. Wiedemann,
  Eur. Phys. J. C 1 (1998) 599.
\bibitem{79} U. Heinz, B. Tomasik, U.A. Wiedemann e Y.-F. Wu,
  Phys. Lett. B 382 (1996) 181.
\bibitem{Zajc} W.A. Zajc et al., Phys. Rev. C 29 (1984) 2173.
\bibitem{WU5} S. Chapman, J. R. Nix, U. Heinz, Phys. Rev. C 52 (1995) 2694.
\bibitem{Gamow-bad1} T. Alber et al., Nucl. Phys. A 590 (1995) 453c.
\bibitem{Prat86} S. Pratt, Phys. Rev. D33 (1986) 72.
\bibitem{Bowler} M.G. Bowler , Phys. Lett. B270 (1991) 69.
\bibitem{Baym} G.~Baym and P. Braun-Munzinger, Nucl. Phys. A610 (1996) 286.  
\bibitem{NA49DeltaEta} P. Jones et al.,  Nucl. Phys. A610 (1996) 188.
\bibitem{Tomasik1998} H.~Heiselberg and A.P.~Vischer, Eur. Phys. J. C2 (1998) 593. \\
         B. Tomasik and U. Heinz, preprint nucl-th/9805016 (1998).
\bibitem{Becca} F.~Beccatini, M.~Ga\'{z}dzicki M and  J.~Sollfrank  
                Eur. Phys. J.  C5 (1998) 143.
%
%
\end{thebibliography}
